%% file: MIT-Thesis.tex
\DocumentMetadata{ 
	pdfstandard = a-2b,
	pdfversion  = 1.7,
	lang		= en-US,
}

\documentclass[twoside]{mitthesis} 
\usepackage{listings}

\usepackage[version=4]{mhchem}

\usepackage{lipsum}
\IfPackageAtLeastTF{lipsum}{2021/09/20}{\setlipsum{auto-lang=false}}{}

\counterwithout{footnote}{chapter}

\usepackage[style=ieee,maxbibnames=10,sorting=none]{biblatex}

\usepackage{booktabs}
\usepackage{array}

\newcommand*\YourName{Bruno S. Scheihing Hitschfeld}

\hypersetup{%
	pdftitle={MIT thesis template},
	pdfkeywords={\YourName, Massachusetts Institute of Technology, MIT},
	pdfauthor={\YourName},
	pdfauthortitle={Professor of Mechanical Engineering},
	pdfcaptionwriter={\YourName},
	pdfurl={https://lienhard.mit.edu},
	pdfcontactemail={lienhard@mit.edu},
	pdfcontactaddress={77 Massachusetts Avenue, Room 3-166},
    pdfcontactcity={Cambridge, MA},
    pdfcontactpostcode={02139},
    pdfcontactcountry={USA},
    pdfcontacturl={https://lienhard.mit.edu},
    pdfsubject={Template for writing MIT theses with the mitthesis class},
	colorlinks=true,
	linkcolor=Blue3,
	citecolor=Blue3,
	urlcolor=violet,
	filecolor=red, 
 	pdfborder={0 0 0},
	bookmarksnumbered=true,
	bookmarksopen=true,
	bookmarksopenlevel=1,
	pdfpagelayout=SinglePage,
	pdfdisplaydoctitle=true,
	pdfstartview=Fit,
	pdfnewwindow=true,
}
\AtEndDocument{
	\hypersetup{pdfcopyright={Copyright © \DegreeYear\ by \YourName. \PDFRightsText},}
}
	
\usepackage{braket}
\usepackage{bbm}
\usepackage{bm}
\usepackage{slashed}
\usepackage{amssymb}

\newcommand*\diff{\mathop{}\!\mathrm{d}}

\newcommand{\nn}{\nonumber}

\newcommand{\be}{\begin{eqnarray}}
\newcommand{\ee}{\end{eqnarray}}
\newcommand{\ma}{\mathrm}
\newcommand{\ml}{\mathcal}
\newcommand{\bs}{\boldsymbol}

\newcommand{\T}{\mathcal{T}}
\newcommand{\W}{\mathcal{W}}

\def \bes {\begin{subequations} }
\def \ees {\end{subequations}}

\def \pd {\partial}
\def \g {\gamma}
\def \a {\alpha}
\def \b {\beta}
\def \vp {{\bm p}}
\def \no {\nonumber}

\def\O{{\mathcal O}}

\def\p{{\bs p}}
\def\q{{\bs q}}
\def\k{{\bs k}}

\def\D{{\mathbb{D}}}
\def\Tr{{\rm Tr}}
\def\tr{{\rm tr}}

\def\C{{\mathcal C}}

\def\I{{\mathcal I}}

\def \lcb {l_\text{Cb}}
\def \lcbdot {\dot{l}_\text{Cb}}

\def \pT {p_{\perp}}
\def \pz {p_{z}}

\def \tA {A}

\def \ta {a}
\def \s {\sigma}

\def \sE {{\cal E}}

\def \tq {\widetilde{q}}
\def \tqB {\widetilde{q}_B}

\def \ad {\text{ad}}

\def \als {\alpha_{S}}
\def \betas {\beta_{S}}
\def \gas {\gamma_{S}}

\def \As {A_{S}}
\def \Bs {B_{S}}
\def \Cs {C_{S}}
\def \qs {q_{S}}
\def \le {\left}
\def \ri {\right}

\begin{document}


\title{Emergence, Formation and Dynamics of Hot QCD Matter}



\Author{Bruno Sebastian Scheihing Hitschfeld}{Department of Physics}[B.Sc. Physics, Universidad de Chile, 2017][M.Sc. Physics, Universidad de Chile, 2019]

\Degree{Doctor of Philosophy}{Department of Physics}

\Supervisor{Krishna Rajagopal}{William A.M. Burden Professor of Physics}

\Acceptor{Lindley Winslow}{Professor of Physics}{Associate Department Head}


\DegreeDate{September}{2024}

\ThesisDate{June 10, 2024}

%
%
\CClicense{CC BY-NC-ND 4.0}{https://creativecommons.org/licenses/by-nc-nd/4.0/}
%

\maketitle
	\cleardoublepage




\begin{abstract}
	\input{abstract.tex}
\end{abstract}
	\cleardoublepage

\include{acknowledgments.tex}%
	\cleardoublepage



\setcounter{tocdepth}{3}
\setcounter{secnumdepth}{3}

\tableofcontents
\listoffigures
\listoftables



\include{introduction.tex}
\include{overview.tex}

\include{quarkonium-v2.tex}
\include{hydrodynamization.tex}
\include{outlook.tex}


\appendix
\include{appendixa.tex}
\include{appendixb-v2.tex}
\include{appendixc.tex}


{\raggedright
\printbibliography[title={References},heading=bibintoc]
}



\end{document}

%% file: abstract.tex
%
%


Understanding the dynamics of Quantum Chromodynamics (QCD) in quantitative detail is one of the main frontiers in particle physics. While the last century gave us the formulation of the theory of nuclear interactions, QCD, as well as that of the rest of visible matter encoded in the Standard Model of Particle Physics, much remains to be understood. In particular, the hot QCD matter produced in high energy collisions of heavy ions presents a unique challenge to theory and phenomenology due to the vast number of different phenomena that take place in such a collision, and even more so because it is an out-of-equilibrium process. In this thesis, we make progress in two concrete directions in the vast landscape of hot QCD physics. The first one is quarkonium transport inside quark-gluon plasma (QGP), the high temperature phase of QCD. Over the past two decades it has been realized that a significant fraction of quarkonium suppression in high energy heavy ion collisions comes from dynamic dissociation and recombination processes, instead of static screening of the interaction potential as originally proposed by Matsui and Satz. Our contribution is the formulation of the precise correlation functions in QCD at finite temperature that describe the dissociation and recombination processes of heavy quarkonium in QGP, as well as their calculation in weakly coupled QCD and strongly coupled $\mathcal{N}=4$ supersymmetric Yang-Mills theory. We also formulate the Euclidean version of these correlation functions so that they may be calculated using Lattice QCD techniques. In this way, our results provide the necessary ingredients to carry out an analysis of the suppression of $\Upsilon$ states in heavy ion collisions in terms of the parameters of the QCD lagrangian. 
The second contribution we make is the development of tools to understand the process of hydrodynamization in QCD kinetic theory and their application to a simplified description where only a subset of the QCD scattering mechanisms are included. By doing this, we learn that the process of hydrodynamization in this theory, and specifically, how memory of the initial condition is lost, follows the recently proposed Adiabatic Hydrodynamization scenario.
Concretely, hydrodynamization proceeds through a sequential process in which a monotonously shrinking set of low-energy states dominate the dynamics, where the opening of an energy gap relative to the ground state(s) signals the start of each stage of this process. The hydrodynamic attractor is reached when only one low-energy state remains as the ground state, and the system approaches local thermal equilibrium following the adiabatic evolution of this low-energy state.

%% file: acknowledgments.tex


\chapter*{Acknowledgments}
\addcontentsline{toc}{chapter}{\protect\textbf{Acknowledgments}}


\textit{To my family,}

First and foremost, I wish to thank my parents for their unconditional and continuous support over all of my life and studies, without whom I would certainly not have been able to be here, submitting this thesis to the Department of Physics at MIT.

I can only try to describe in words the full extent of their support and encouragement throughout my upbringing, knowing full well that my description will pale in comparison. The endless conversations I had with my father, Rodrigo, about any imaginable subject, with him always thoroughly answering every time I asked ``why?'' to the point my curiosity had been satiated, regardless of whether I was 4, 9, or 16 years old. His unwavering support and aid in anything I needed to do, often well beyond what I considered reasonable once I was old enough to appreciate it, with a completely selfless dedication to me and my sister, Irma, is something I will forever be grateful for and never forget. The extraordinary disposition of my mother, Nancy, to teach and play with us growing up, over uncountably many hours, with endless patience in dealing with any state of mind we were in, from the happiest day to the most extreme tantrum, always making sure that we grew in experience, knowledge and intelligence, will never disappear from my memory.

Thank you both, mom and dad, for everything you did for me.

Thank you, Irma, for continuing to remind me that not everything in life needs to be undertaken with such a serious attitude as I often choose to have.

\vspace{0.5cm}

\noindent
\textit{to my mentors and collaborators,}

For the last five years, I am grateful to have enjoyed the support of my thesis advisor, Krishna Rajagopal, in all and every activity I set out to do in connection with my PhD and beyond. The sheer number of opportunities I have had to connect with people, to present my work at more conferences I can count, as well as to start scientific collaborations, would not have been possible without your support and guidance. I have greatly benefited from your ability to spot what is an important question worth spending my time on, and what is a question that needs to be made sharper before starting a research project. I hope to have learned enough from you so that I can one day transmit your wisdom and insight to the next generation.

I also want to thank Xiaojun Yao, my most frequent collaborator, the scientist with whom I have had most numerous enlightening discussions throughout my PhD studies. These discussions have often enabled me to see past what is known, and make a contribution of my own to our collective understanding of the fundamental building blocks of matter all around us.

Thanks to Jasmine Brewer, Jamie Karthein, Govert Nijs, Rachel Steinhorst, Urs Wiedemann, and Yi Yin, for countless stimulating discussions about exciting questions in the field of heavy-ion physics and the many-body physics of QCD.

Thank you to my mentors and collaborators at Universidad de Chile, Gonzalo Palma and Fernando Lund. You have continued to inspire me and energized me to seek answers to questions beyond the scope of my main research field, maintaining my curiosity across all of the spectrum of theoretical physics.

\vspace{0.5cm}

\noindent
\textit{and to my friends and colleagues,}

Finally, I wish to thank all of my friends and fellow graduate students for the full life I have lived here in Boston and in my travels through Europe, especially during the last year as I prepared this thesis and decided on the next steps in my life.
Zhiquan, Rachel, Artur, Wenzer, Patrick, Josh, Dominika, Elba, Rebecca, David, Sam, Nico, Marianne, Federica, Dana, Adriana, Nina, Andrea, Arjun, Hannah, Janice, Xo\'an, Sergio, Thomas, Cari, Kyle, Atakan, Anjie, Aasmund, Dimitra, Pierre, Sean, Lisa, Wentao, Wenjie, Rikab, Ryan, Mati, Anna, Jean, Raji, Angus, John, Annie, Sa\'ul, Alexander, Yitian, Manu, Richard, Nicole, and Stella, thank you for all the little and big things you did for me every day, from the smallest talk at the local coffee machine, to the latest of late night conversations at any and all bars and brewpubs in town.

\vspace{0.5cm}

\hspace{\fill} \textit{Thank you all.}

%% file: introduction.tex

\chapter{Introduction}

The last century was host to several revolutions in our understanding of the Universe. For the first time in the history of humanity, what was previously thought of as indivisible, the atom, was revealed to have subatomic structure, comprised of a nucleus of positive electric charge and a cloud of electrons around it, as in Rutherford's model~\cite{Rutherford1911}. 
This was demonstrated in a series of experiments by Geiger and Mardsen~\cite{Geiger:1908ian,Geiger-Mardsen1909,Geiger:1910ian,Geiger-Mardsen1913}, which, at the same time as it advanced our understanding of the atom, demonstrated the usefulness of perhaps the most fruitful method to reveal the microscopic structure of matter in our Universe to date: collide particles, and by doing so at increasingly higher energies, probe the structure of matter at smaller and smaller length scales. Since the early 1900s, collider technology has evolved to a point where nowadays protons can be collided at $0.99999999$ times the speed of light, with its latest feat being the discovery of the Higgs boson at the Large Hadron Collider at CERN by the ATLAS~\cite{ATLAS:2012yve} and CMS~\cite{CMS:2012qbp} collaborations.

The advances that were needed to get us to this point were as groundbreaking on the theory frontier as on the experiment frontier. To fully understand the hydrogen atom, let alone helium and all the heavier atoms and isotopes, the development of a brand new theory, quantum mechanics~\cite{Heisenberg1925,Dirac1925,Born:1925mph,Born:1926uzf,Schrodinger1926}, was paramount. Although its earliest formulations allowed for a correct description of the hydrogen atom by describing the wavefunction of an electron in the electrostatic potential of a proton, it was clear that a complete microscopic theory of matter needed to include (at least) protons, electrons, and photons as degrees of freedom. This led to the development of Quantum Electrodynamics (QED) in the late 1920s~\cite{Dirac1927}. The main new feature that this theory presented was that the number of particles in a physical process need not be conserved, so as to describe the empirical fact that light (photons) can be absorbed and emitted by matter. As it turned out, the number of electrons and protons is also not conserved, demonstrated by the observation of $\beta$-decay and its explanation by Enrico Fermi in his theory of weak interactions~\cite{Fermi1934}. These were the earliest examples of quantum field theories, the language in which we nowadays describe the fundamental constituents of the Universe. 

However, the formulation of a complete theory of nuclear interactions had to wait until the 1970s. The sheer number of particles/resonances in the spectrum provided by the strong interactions, as well as the richness and complexity of their dynamics and structure made it so that enough empirical data had to be gathered first. In fact, even the name we nowadays give to matter comprised of quarks and gluons (the constituents of protons and neutrons), \textit{hadrons}, only entered the mainstream of particle physics in 1962~\cite{Okun:1962kca}. At that time, quarks and gluons had not been discovered yet; the proposal that quarks might be the microscopic constituents of mesons and baryons only came in 1964 by the hand of Murray Gell-Mann~\cite{Gell-Mann:1964ewy} and George Zweig~\cite{Zweig:1964jf,Zweig:1964ruk}. 

It took yet another step in the late 1960s along the lines of the Geiger-Mardsen experiments to conclusively reveal that protons and neutrons had internal structure. This time, a team of MIT and Stanford Linear Accelerator Center (SLAC) scientists used a process called \textit{deep inelastic scattering} (DIS), by which electrons were accelerated to sufficiently high energies such that they were able to disintegrate nucleons through scattering with what would be revealed as their point-like constituents. While the particles participating in the collision were fundamentally the same as 60 years before, the results demonstrated that protons and neutrons have microscopic structure~\cite{Bloom:1969kc,Breidenbach:1969kd}. Furthermore, the results of this experiment showed that these constituents were essentially free~\cite{Friedman:1972sy} for the purposes of these collisions by confirming a theoretical prediction known as \textit{Bjorken scaling}~\cite{Bjorken:1968dy,Bjorken:1969ja}.

The fact that quarks could only be probed directly by means of high-energy scattering, and that they would behave as essentially free, point-like particles in this limit, put together with the fact that at low energies they must condense into nuclei and become unobservable as individual particles posed a significant theoretical challenge. This challenge was met by David Gross and Frank Wilczek~\cite{Gross:1973id,Gross:1973ju,Gross:1974cs} and David Politzer~\cite{Politzer:1973fx} with their discovery of \textit{asymptotic freedom} in a non-Abelian gauge theory called Quantum Chromodynamics (QCD), which, crucially, was able to feature the quarks necessary to describe the low-energy hadron spectrum via Gell-Mann and Zweig's model. Asymptotic freedom is the property of QCD by which interactions become weaker and weaker as the energy of the particles participating in the collision is driven to higher and higher values (more precisely, the value of the coupling $g$, to be introduced in the next paragraph, decreases with increasing energy of the participating particles). This explained the fact that the constituents of protons and neutrons observed at SLAC behaved as almost free particles, and conversely, provided a mechanism by which quarks could be bound in nuclei at lower energies and never seen as free particles due to the greater strength of QCD interactions at such scales. It also signalled the need for the existence of additional force carriers, known as gluons, which were soon discovered later on that decade~\cite{Barber:1979yr,PLUTO:1979dxn,JADE:1979rke,TASSO:1979zyf}.

All of this, and more, can be accounted for by the relatively simple Lagrangian
\begin{equation} \label{eq:QCD-lagrangian}
    \mathcal{L} = - \frac14 F_{\mu \nu}^a F^{\mu \nu a} + \sum_I \bar{\psi}_I \left( i \slashed{D} - m_I  \right) \psi_I \, ,
\end{equation}
where $F_{\mu \nu} \equiv \frac{i}{g} [D_\mu, D_\nu]$ is the non-Abelian field strength, $D_\mu \equiv \partial_\mu - i g A_\mu^a T^a$ is the gauge-covariant derivative, $A_\mu^a$ is the non-Abelian gauge field, $\psi_I$ are Dirac spinors describing each quark flavor $I$, $g$ is the strong coupling constant, and a slash $\slashed{D} \equiv \gamma^\mu D_\mu $ denotes a contraction with the Dirac matrices $\gamma^\mu$. The index $I$ labels the quark species, which in QCD runs over $u,d,s,c,b,t$ (up, down, strange, charm, bottom, and top quarks). A detailed discussion of all of these fields and quantities is nowadays readily available in quantum field theory textbooks, e.g.,~\cite{Weinberg:1995mt,Weinberg:1996kr,peskin1995introduction,Schwartz:2014sze,Srednicki:2007qs}.
Supplemented with empirical data, the characteristic energy scale at which QCD interactions become uncontrollably strong to admit a description in terms of quarks and gluons as weakly interacting particles can be derived directly from this Lagrangian, and it is usually denoted by $\Lambda_{\rm QCD} \approx 200 \, {\rm MeV}$. In practice, calculations based on a small coupling expansion break down even at higher energies. $\Lambda_{\rm QCD}^{-1}$ plays a crucial role in low-energy hadronic physics as it determines the typical size of hadrons, and in turn also in high-energy collisions as it determines the scale at which the fragmentation of quarks and gluons becomes non-perturbative and leads to hadronic matter.

\begin{figure}[t]
    \centering
    \includegraphics[width=0.7\textwidth]{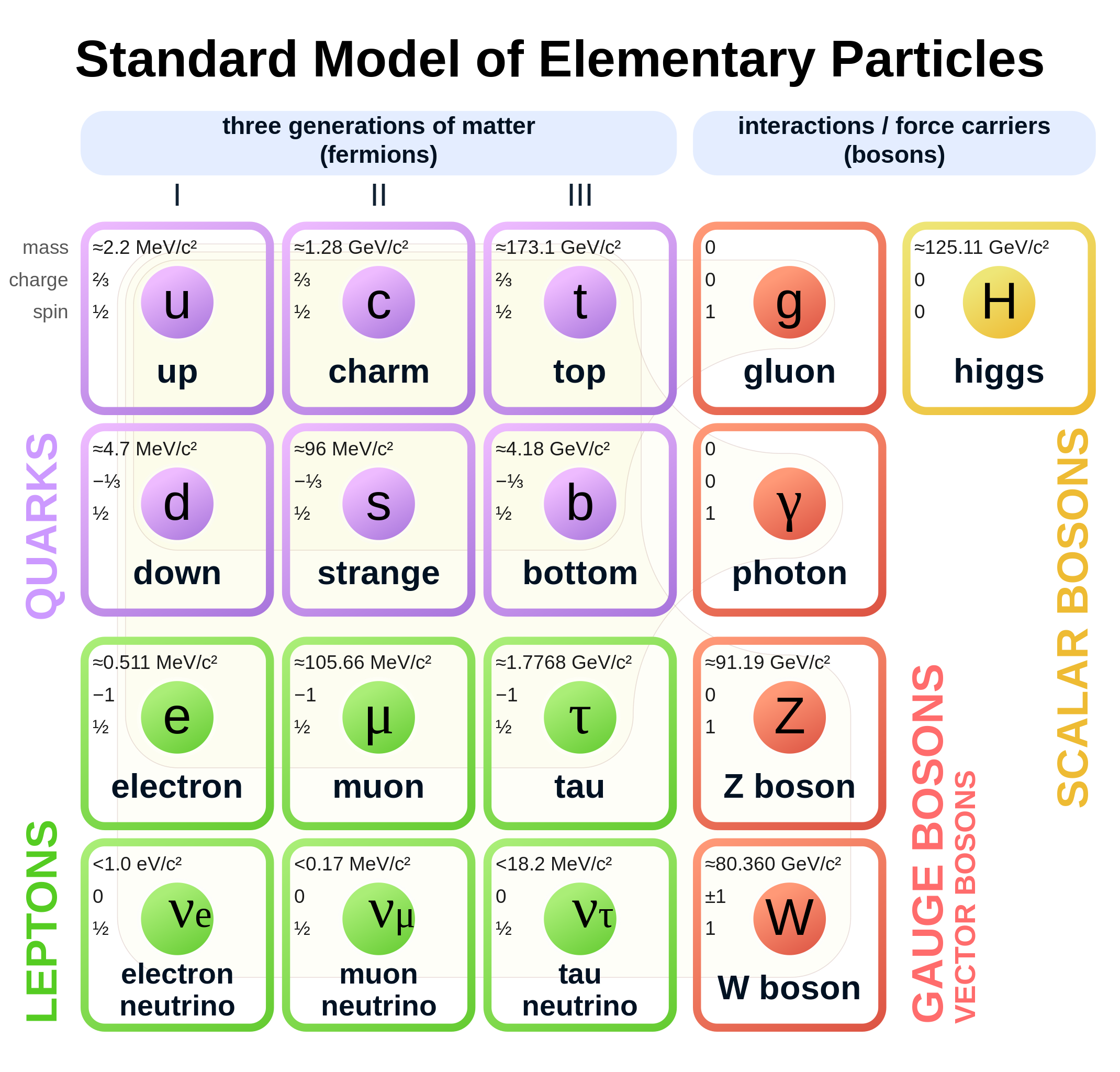}
    \caption{The Standard Model of Particle Physics, with all of the particles that have been discovered so far in Earth-based experiments. These particles constitute all of visible matter around us. This image has been released into the public domain by its author, which we reproduce from \href{https://commons.wikimedia.org/wiki/File:Standard_Model_of_Elementary_Particles.svg}{Wikimedia Commons} {\footnotesize https://commons.wikimedia.org/wiki/File:Standard\_Model\_of\_Elementary\_Particles.svg }. The particles that interact via the strong force, i.e., those that appear in the QCD Lagrangian~\eqref{eq:QCD-lagrangian}, are the six types of quarks ($u,d,s,c,b,t$) and the gluons.
    }
    \label{fig:standard-model}
\end{figure}

Nowadays, QCD is but one of the many ingredients that go into the Standard Model of Particle Physics, summarized in Figure~\ref{fig:standard-model}. While fundamental puzzles still remain to be addressed, such as the origin of neutrino masses or the nature of dark matter and dark energy (of which none will be discussed with any level of detail here), which almost certainly will require that we discover and understand new physical phenomena, whenever it has been possible to carry out a calculation within the Standard Model to compare with empirical data, its success in describing every single Earth-based experiment up to the present point in human history has been nothing short of astonishing.

This does not mean that the Standard Model is fully understood in quantitative detail. In fact, out of all the matter in the Standard Model, and even though it constitutes some of the most common particles in our Universe (protons and neutrons), QCD still remains its most challenging part, mostly due to the fact that perturbation theory in terms of its fundamental degrees of freedom becomes inapplicable at the energy scales relevant for the study of most hadrons. In many cases, it is our understanding of QCD itself which limits our ability to interpret experimental results, even when the observable at hand is not a QCD object. For example, the biggest theoretical uncertainty in the determination of the anomalous magnetic dipole moment of the muon comes from QCD dynamics~\cite{Wittig:2023pcl}. Even in the regime where perturbation theory is applicable, i.e., on the collider physics side at high energies, the largest uncertainties in theoretical calculations also come from QCD.

In fact, an open problem in theoretical physics is to understand quantitatively the nature of the mechanism that confines quarks and gluons into nucleons, i.e., protons and neutrons, as well as other hadrons, making it impossible to measure them as asymptotic states in collider experiments. This property is, fittingly, known as \textit{confinement}. At the very least, shedding light into this mechanism would allow for more precise calculations of the hadron spectrum by informing Lattice QCD calculations, a field theoretical framework that allows one to evaluate QCD properties numerically in a controlled, systematically improvable setting (to the extent that it already provides an accurate description of the hadron spectrum for more than a decade now~\cite{MILC:2009mpl,BMW:2008jgk,BMW:2014pzb,PACS-CS:2013vie}), and one to which we will return to later on this thesis. At the very optimistic best, understanding the confinement mechanism in full detail would allow us to write down an explicit formula that determines the proton mass in terms of parameters in the microscopic Lagrangian of QCD (i.e., in terms of $g$ and the quark masses $m_I$ determined at a reference energy scale or renormalization scheme). This formula is, as of yet, unknown.

\begin{figure}[t]
    \centering
    \includegraphics[width=0.8\textwidth]{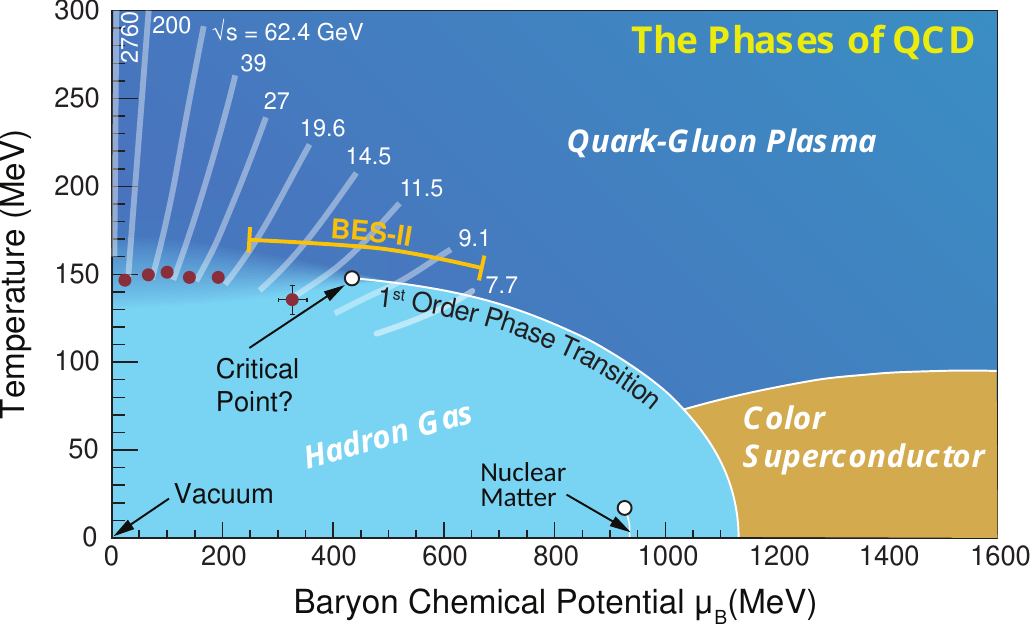}
    \caption{Sketch of the QCD phase diagram, assuming a (conjectured) critical endpoint that marks the beginning of the separation between the quark-gluon plasma (QGP) phase and the hadron gas phase as the baryon chemical potential is increased. The white lines that start in the QGP phase represent a sketch of the path in the phase diagram that some heavy ion collisions traverse, each at its corresponding collision energy. The leftmost line (2760 GeV) represents an instance of a collision at the LHC, while the ones at lower energies in this figure represent collisions at RHIC. Adapted from Ref.~\cite{An:2021wof}.}
    \label{fig:phase-diagram}
\end{figure}

Furthermore, there are also qualitative questions about QCD that are still open. Given its richness in matter content, QCD has a highly nontrivial phase diagram, which is still far from completely understood. In Figure~\ref{fig:phase-diagram} we show what is nowadays perhaps the most widespread expectation for its features as a function of temperature $T$ and baryon chemical potential $\mu_B$. Other than the zero chemical potential line $\mu_B = 0$, which is well-understood via numerical calculations as well as decades of heavy-ion collision experiments, and extrapolations from there up to $\mu_B/T \sim 2$, not much is known, even qualitatively. The critical endpoint, although theoretically well-motivated~\cite{Rajagopal:1992qz,Stephanov:1996ki,Stephanov:1998dy,Berges:1998rc,Halasz:1998qr,Stephanov:2006zvm,Yin:2018ejt} has so far escaped conclusive detection~\cite{STAR:2020tga,STAR:2021fge}. The final results of the experimental search from the Beam Energy Scan II (BES-II) program at the STAR experiment at the Relativistic Heavy Ion Collider (RHIC) at Brookhaven are, as of the time of writing of this thesis, still under analysis. At larger chemical potentials, asymptotic freedom eventually allows for tractable calculations where perturbation theory becomes reliable, and a color superconductor phase is expected~\cite{Alford:1997zt,Rapp:1997zu,Alford:1998mk,Son:1998uk} (although the number of different color superconducting phases that exist as a function of baryon chemical potential is still unknown).

However, despite the fact that QCD, written as a quantum field theory in terms of quarks and gluons is famously intractable, that has not stopped theoretical progress. Rather, it has channeled it through the development of so-called Effective Field Theories (EFTs). An EFT is a quantum field theory in its own right, designed to describe a smaller set of physical processes than the complete quantum field theory that it approximately describes, and is characterized by one (or more) expansion parameters, also called power-counting parameters, which ensure that the corresponding approximations are under quantitative control. Depending on how one constructs these theories, one calls them either `top-down' or `bottom-up', corresponding to whether they are obtained by integrating out the degrees of freedom of a more complete theory explicitly or not, respectively. Out of the theories we have mentioned so far, an example of an EFT is the Fermi theory of weak interactions, which is the low-energy limit of the electroweak sector of the Standard Model, and can be considered as a `top-down' effective theory. In the case of `bottom-up' effective theories, a ``microscopic''  Lagrangian is not used, either because it is not known or because it is hard to carry out calculations to which the EFT could be matched.
By construction, EFTs have limited regimes of applicability, usually related to the kinematics of a given physical process under consideration, which can greatly simplify the theoretical analysis required to obtain a quantitative answer/prediction for a physical observable in that process by, e.g., assuming that the ratio between two energy scales is small.

Given the plethora of energy scales present in QCD, different effective theories have been developed to tackle physical processes in each regime. 
\begin{itemize}
    \item At low energies, below the energy scale of strong interactions $\Lambda_{\rm QCD}$, the most successful effective theory has been Chiral Perturbation Theory~\cite{Weinberg:1978kz,Gasser:1983yg,Gasser:1984gg}, which allows one to calculate the dynamics of pions and kaons (the lightest mesons), and can even be extended to describe nucleons at low energies. This is a `bottom-up' effective theory, as no direct derivation from the QCD lagrangian exists where the quark and gluon degrees of freedom are integrated out into meson and baryon fields explicitly. As such, the coefficients in the Lagrangian are fixed by matching them to experimental measurements of the masses of hadrons. Once these coefficients are fixed, this Lagrangian is predictive for calculating scattering amplitudes of hadrons at low energies, and most phenomena at energy scales well below $\Lambda_{\rm QCD}$.
    \item Conversely, at asymptotically high energies there are times when asymptotic freedom is enough to provide perturbative control to calculations of certain processes. However, even high-energy processes in QCD often require to account for contributions that come from smaller energies, because the kinematics of QCD allows (and, in fact, mandates) it. A prime example of an effective theory constructed for precisely this situation is Soft-Collinear Effective Theory (SCET)~\cite{Bauer:2000ew,Bauer:2001ct,Bauer:2000yr,Bauer:2001yt,Bauer:2002uv,Bauer:2002nz}, which allows to systematically separate (factorize) the ``hard,'' perturbatively calculable contributions, from the ``soft'' or collinear contributions to a physical process, which often require the use of resummation techniques for them to be evaluated (alternatively, one can use the predictive power of this factorization to determine the soft contributions from experimental data). State-of-the-art predictions for collider experiments use this theory to treat the high-energy particles that come out of a collision, including, but not exclusively, $p p \,\,{\rm or} \,\, e^+ e^- \to {\rm dijets}$, $pp \to H$, and $B$ hadron decays (in conjunction with HQET, see below).
    \item At high temperatures $T \gg \Lambda_{\rm QCD}$, one might think that the fact that QCD is asymptotically free means that perturbation theory is reliable and no re-summation or effective description is needed. This is certainly true for particles with energy/momentum $p$ of the order of the temperature itself. However, because of the interactions of QCD (and most other interacting theories), another relevant scale emerges, given by $g T$, which at weak coupling defines the characteristic (inverse) length at which (color-electric) soft modes get screened inside a hot plasma. This scale emerges as a result of the coupling between hard modes ($p \sim T$) and all the rest of modes in the theory, and thus singles out ``soft'' modes with $p \sim g T$, which means that a precise description of the soft modes of the theory requires to re-sum contributions from the hard sector of the theory. This is most easily achieved by constructing an effective theory that does this automatically. Such EFT is called Hard Thermal Loop (HTL) effective theory. Its first developments were presented in~\cite{Pisarski:1988vd,Braaten:1989mz,Braaten:1989kk,Braaten:1990az,Taylor:1990ia,Frenkel:1989br}, and its effective action was developed shortly thereafter~\cite{Braaten:1991gm,Frenkel:1991ts}. Any observable that depends on the dynamics of the soft modes at high temperatures requires the use of this re-summation. This includes thermodynamic quantities such as the free energy~\cite{Andersen:2002ey}. Recent examples of calculations that use this re-summation are the computation of the thermodynamic pressure at two-loop accuracy in $\mathcal{N}=4$ SYM~\cite{Du:2020odw} and the next-to-leading order self-energy of photons in a hot QED medium~\cite{Gorda:2022fci}. As an additional remark on finite temperature QCD, we note that $g T$ is not the only emergent scale at finite temperature: color-magnetic modes only start being screened at a scale defined by $g^2 T$. This eventually generates a breakdown of perturbation theory~\cite{Linde:1980ts,Gross:1980br} because diagrams with an arbitrary power of coupling constants need to be re-summed to obtain accurate fixed-order results as a function of the same coupling constant. An example of this is that the last perturbatively calculable term of the QCD thermodynamic pressure is of order $g^6 \ln 1/g$~\cite{Kajantie:2002wa}, and the calculation of the $\mathcal{O}(g^6)$ contributions remains an open problem where ongoing efforts are still underway~\cite{Navarrete:2022adz} (a hierarchy of EFTs is needed at this order to reliably account for the non-perturbative effects~\cite{Braaten:1995jr}).
    \item Another example of an effective theory with widespread applications is Heavy Quark Effective Theory (HQET)~\cite{Georgi:1990um}, which describes the charm $c$ and bottom $b$ quarks taking advantage of the fact that their masses $m_c \approx 1.27 \, {\rm GeV}$, $m_b \approx 4.65 \, {\rm GeV}$ are much larger than $\Lambda_{\rm QCD}$. If one is only interested in sufficiently low-energy processes, then heavy quark number is conserved, at least for times less than the time that it takes for them to decay via the electroweak interactions to lighter quarks ($\sim 10^{-20} \, {\rm s}$), which is orders of magnitude longer than the characteristic time scale of QCD interactions ($\sim 10^{-24}\, {\rm s}$). This theory gives predictive power to calculations involving bottom and charmed hadrons, by expanding in inverse powers of the heavy quark mass and thus greatly simplifying the number of operators that need to be considered for any given observable of interest in this context. For example, it is possible to systematically calculate the masses of $B$ mesons (containing a $b$ quark) and their decay rates by organizing the calculation as an expansion in powers of $\Lambda_{\rm QCD}/m_b$, with a finite number of additional parameters at each order in the expansion, which can then be determined by experimental measurements of the properties of a few of these hadrons and be predictive for all the rest. In the same way, one can also use this EFT to calculate properties of $D$ mesons (containing a $c$ quark), although the corrections in the expansion parameter $\Lambda_{\rm QCD}/m_c$ are larger than for $B$ mesons.
    \item A further step of integrating out degrees of freedom can be made from HQET if one is interested in bound state problems of such heavy particles, taking advantage of the fact that in many such problems the heavy quarks move slowly $v/c \ll 1$ relative to the frame in which the total spatial momentum of the system vanishes, with $v$ serving as an expansion parameter. The EFT that emerges after doing so is called Non-Relativistic QCD (NRQCD)~\cite{Caswell:1985ui,Bodwin:1994jh,Soto:2006zs}, or Non-Relativistic EFT in general. For the specific case of QCD, it allows for systematic calculations of the properties of bottomonium ($b\bar{b}$) bound states (as well as of many other hadrons), which will be of great interest to us later on in this thesis, but it also allows for a systematic approach to precision calculations regarding the hydrogen atom (in an effective theory called NRQED), as both the electron and the proton are non-relativistic in this system.
    \item Finally, as the last EFT we will mention, and the main conceptual framework on which a part of this thesis is based during Chapter~\ref{ch:quarkonia-in-QGP}, if one is interested in the dynamics of bound states of heavy quarks, then one can further integrate out the degrees of freedom of QCD that generate a potential interaction between heavy quark pairs and encode that information in single-particle Hamiltonians, by which, in essence, a Schr\"odinger equation is derived for the dynamics of the heavy quark pair system, and the only remaining degrees of freedom that are left with explicit dynamics are the internal ones (color, spin) together with their relative position, in addition to the light QCD degrees of freedom with which the heavy pair may interact. Such a theory is called potential Non-Relativistic QCD (pNRQCD)~\cite{Pineda:1997bj,Brambilla:1999xf,Brambilla:2004jw}, and has been extensively applied to the study of the dynamics of quarkonium bound states, especially so in the presence of quark-gluon plasma (QGP), a subject in which we will deeply immerse ourselves later on this thesis.
\end{itemize}

Furthermore, first-principles calculations of static QCD properties have also made steady progress over the past few decades. By discretizing spacetime and evaluating the QCD path integral on a Euclidean lattice, it is nowadays possible to calculate these properties directly from the fundamental degrees of freedom of QCD with unprecedented precision~\cite{Joo:2019byq,Cirigliano:2019jig,Detmold:2019ghl,USQCD:2022mmc}. The lattice spacing $a$ provides a natural regulator that allows for systematically improvable calculations by taking the limit $a \to 0$ while holding fixed the physical quantities to which the theory is calibrated, which then has predictive power for all other observables in the theory. The thermodynamic properties of QCD at $T>0$, $\mu_B = 0$ are also directly calculable~\cite{Borsanyi:2010cj,Borsanyi:2013bia,HotQCD:2014kol}. The drawbacks to this method are i) the massive computational resources needed to carry out a calculation and ii) that the determination of real time observables suffers from intrinsic limitations coming from the fact that said calculations are carried out in Euclidean spacetime (as opposed to Minkowski spacetime, which is the mathematical description of the spacetime we live in at distances where gravity has no significant effect). Overcoming both of them is, of course, an active area of research nowadays.

On the other hand, the many-body physics of QCD in out-of-equilibrium settings has mostly been explored with other effective descriptions, each derived from QCD for a subset of observables in a given kinematic regime. We distinguish these from the EFTs we described above because these descriptions forego some of the quantum aspects of QCD, and while they can be accurate in their respective regimes of validity for the observables they are engineered to describe, they include uncontrolled approximations that do not allow one to recover the quantum dynamics of the system. Examples include classical-statistical simulations in the color glass condensate framework~\cite{McLerran:1993ni,Mueller:1999wm,Iancu:2001md,Iancu:2002xk} (valid at large occupation numbers), effective kinetic theory~\cite{Arnold:2002zm} (valid at weak coupling, when particles can be treated as point-like with short-range interactions), and a hydrodynamic description (when only the dynamics of conserved quantities, e.g., energy and charge densities, are of interest). In the latter, a relativistic fluid description is given to QCD matter in local thermal equilibrium, and all that needs to be specified are the transport coefficients encoding corrections to ideal hydrodynamics.

Finally, another approach to gain insight on QCD physics is to resort to different quantum field theories that share some degree of similarity with it, and carry out calculations in those theories in order to get estimates of analog quantities. Examples of theories that model low-energy aspects of QCD, such as confinement or chiral symmetry breaking, include the Nambu--Jona-Lasinio model~\cite{Nambu:1961fr,Nambu:1961tp}, the Schwinger model~\cite{Schwinger:1962tp} and the Gross--Neveu model~\cite{Gross:1974jv}, among many others. Commonly used models to explore the high-temperature phase of Yang-Mills theories are the supersymmetric versions of such theories, because they also exhibit a phase of deconfined color-charged matter. In particular, due to its high degree of symmetry, $\mathcal{N}=4$ supersymmetric Yang-Mills theory ($\mathcal{N}=4$ SYM) allows one to carry out many calculations that are intractable in other theories. Even calculations at strong coupling are nowadays feasible in this theory due to the discovery of the AdS/CFT correspondence~\cite{Maldacena:1997re,Maldacena:1998im}, whereby expectation values of operators can be calculated using a holographic description of the theory that only requires one to solve classical equations of motion in a higher-dimensional spacetime, provided the theory is in the limit $N_c \gg 1$ and $N_c g^2 \gg 1$ (in this order). This duality has been particularly useful to gain intuition about the physics of QCD at finite temperature just above the transition temperature $T_c$, where QGP is a strongly coupled fluid. For instance, one can calculate expectation values of Wilson loops at zero and finite temperature~\cite{Maldacena:1998im,Rey:1998ik,Liu:2006he}, correlation/response functions from which transport coefficients can be extracted~\cite{Policastro:2001yc,Policastro:2002se,Policastro:2002tn,Kovtun:2004de,Casalderrey-Solana:2006fio,Casalderrey-Solana:2007ahi}, and even hydrodynamics has a natural formulation in terms of the dynamics of the metric in a higher dimensional spacetime~\cite{Bhattacharyya:2007vjd}. For a comprehensive review on applications of AdS/CFT to QGP physics, see~\cite{Casalderrey-Solana:2011dxg}.

Progress in our understanding of QCD physics benefits from all of the above methods, more often than not in complementary ways. Among all of the experiments that have been conducted on Earth, perhaps none of them tests our understanding of QCD in all of the ways we just discussed as comprehensively as heavy ion collisions, where their complementarity is used to the greatest possible degree.

\paragraph{Heavy Ion Collisions} \hspace{\fill}

\begin{figure}[t]
    \centering
    \includegraphics[width=0.8\textwidth]{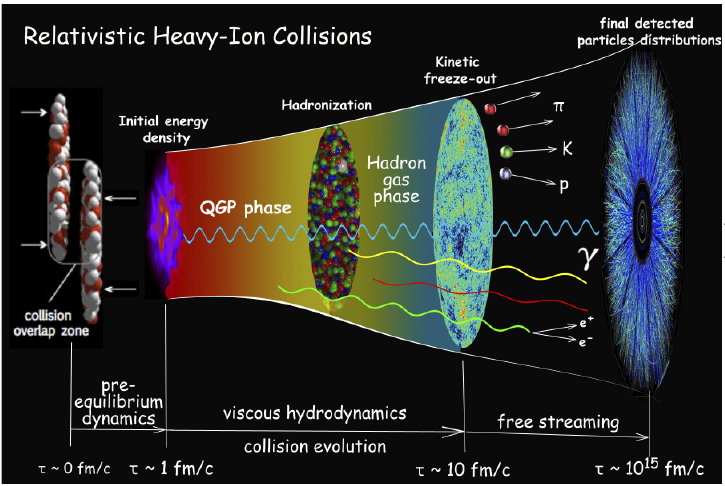}
    \caption{Sketch of the stages that QCD matter undergoes in a Heavy-Ion collision. Figure credit: Paul Sorensen and Chun Shen, extracted from~\cite{Heinz:2013wva}.}
    \label{fig:heavy-ion-collision}
\end{figure}

Heavy Ion Collisions (HICs) are collider experiments where two heavy nuclei, usually lead (Pb) or gold (Au), are collided at ultra-relativistic energies with the purpose of studying the dynamics of QCD matter in extreme conditions (for a review see~\cite{Busza:2018rrf}). They provide a unique window into the dynamics of the quantum fields that comprise matter all around us in their ``deconfined'' phase, where quarks and gluons break free from their nuclear bindings and interact to form the most perfect fluid ever discovered, quark-gluon plasma (QGP). However, this phase of matter, where quarks and gluons are heated up to temperatures greater than $T_c = 155 \, {\rm MeV} \sim \Lambda_{\rm QCD}$, only exists for a finite amount of time in each HIC, and to properly describe the data obtained by measuring the final state of the debris of the collision in terms of the underlying theory, a precise understanding of each stage of the evolution of QCD matter after the collision is necessary. Figure~\ref{fig:heavy-ion-collision} shows a schematic representation of these stages.

Due to the complexity of QCD dynamics, and the fact that the temperature reached in a HIC is not nearly high enough to make perturbation theory quantitatively controlled, its theoretical description requires the use of (multiple) EFT(s). Each different stage and physical process in a HIC is thus analyzed via different effective descriptions, or even models when derivations or calculations that relate an effective theory with the microscopic theory are unfeasible or unavailable. As we will see, the latter is not at all uncommon, and one sometimes has to resort to doing calculations in theories other than QCD, such as supersymmetric $\mathcal{N}=4$ Yang-Mills, to get numerical estimates of QCD quantities in strongly coupled regimes.

\begin{figure}
    \centering
    \includegraphics[width=0.51\textwidth]{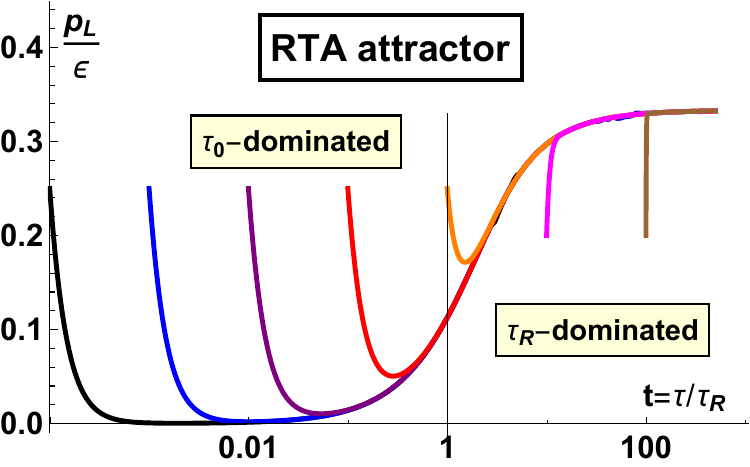}
    \includegraphics[width=0.51\textwidth]{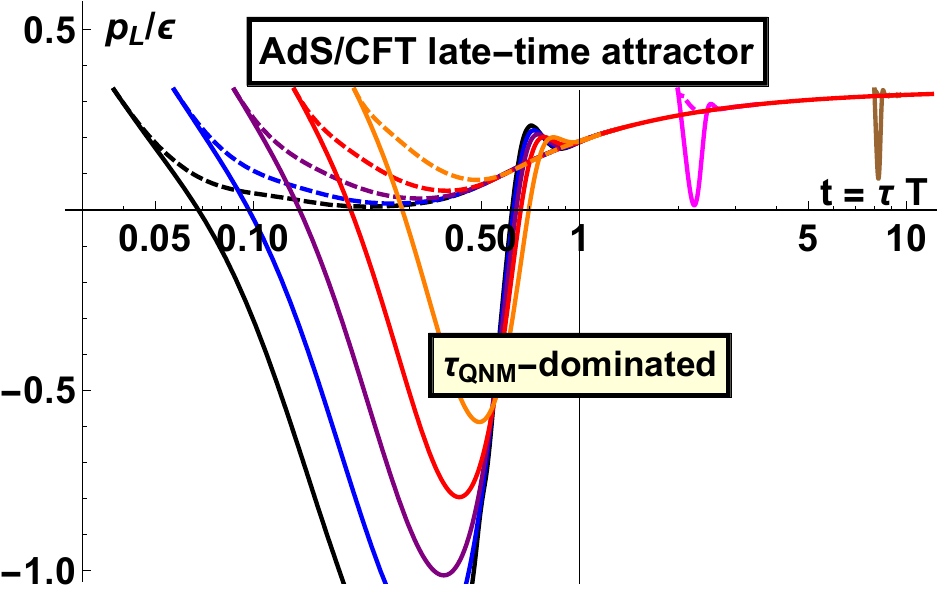}
    \includegraphics[width=0.51\textwidth]{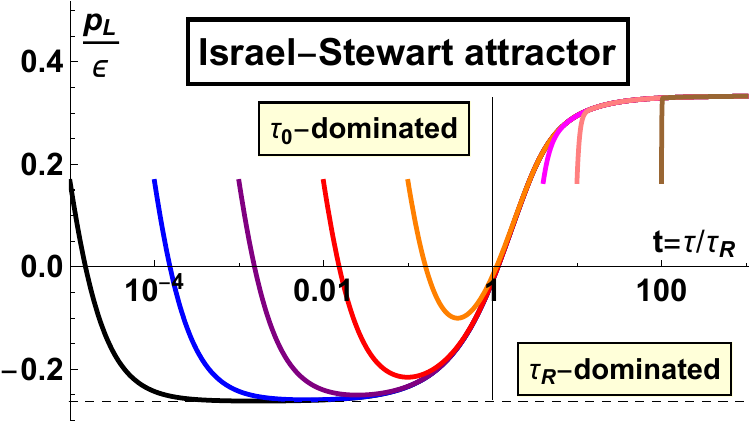}
    \caption{Attractor curves for the ratio between longitudinal pressure and energy density in different theory descriptions of the hydrodynamization process in a heavy-ion collision. Figure reproduced from~\cite{Kurkela:2019set}.}
    \label{fig:attractors}
\end{figure}

A rough summary of the phenomenological description of quark-gluon plasma in HICs is as follows:
\begin{enumerate}
    \item Pre-equilibrium (initial stage) dynamics: After the collision, highly excited QCD matter evolves out of equilibrium. The initial state, two large atomic nuclei, is disintegrated into deconfined quarks and gluons, and their interactions begin to drive the plasma towards local equilibrium, a process known as \textit{hydrodynamization}. Historically, the early success of hydrodynamic modelling in describing HIC data in the early 2000s pointed to rapid hydrodynamization of the plasma formed in the collisions~\cite{Kolb:2000fha,Kolb:2000sd,Kolb:2001qz,Huovinen:2001cy,Heinz:2001xi,Heinz:2002un}. However, in those days the estimates for the thermalization time of QCD coming from kinetic theory~\cite{Baier:2000sb,Arnold:2002zm} were too long to explain this rapid hydrodynamization. This remained as a puzzle for a few years, until holographic calculations via the AdS/CFT correspondence allowed to establish rapid hydrodynamization in strongly coupled $\mathcal{N}=4$ Yang-Mills theory~\cite{Chesler:2008hg,Chesler:2009cy,Chesler:2010bi,Heller:2011ju,Heller:2012km}, in a typical time scale given by $1/T$, where $T$ is the temperature after hydrodynamization. Qualitatively, this provides a sensible estimate because QGP at temperatures reached in HICs is a strongly coupled fluid, meaning that it is possible to obtain insights on its hydrodynamization process (that of a strongly-coupled Yang-Mills plasma) by solving classical equations of motion in a higher-dimensional spacetime that describe an analog strongly coupled plasma. That being said, nowadays it is possible to describe this process of hydrodynamization using effective descriptions of QCD, without resorting to other theories. A common approach is to describe the initial condition and its dynamics for during the first $0.1 \, {\rm fm}/c$ (and at most up to $\tau \sim 0.3 \, {\rm fm}/c$ after the collision) via the CGC framework~\cite{McLerran:1993ni,Mueller:1999wm,Iancu:2001md,Iancu:2002xk}, where the non-Abelian gauge field is treated as classical, and then switch to a kinetic theory description until local thermal equilibrium is reached. It has been shown that even though it is an intrinsically weakly coupled description, kinetic theory provides a reasonable estimate of the hydrodynamization time in a HIC~\cite{Kurkela:2015qoa} provided that the coupling is chosen at a realistic intermediate value, and also similar to the AdS/CFT estimates at intermediate couplings. None of these descriptions fully accounts for the quantum field dynamics of QCD at realistic couplings, but they do provide quantitative estimates as well as insights about the qualitative features of its hydrodynamization process. In fact, a universal feature of the process of hydrodynamization across all of these descriptions is the presence of out-of-equilibrium attractors~\cite{Kurkela:2019set} (see Figure~\ref{fig:attractors}), i.e., dynamically preferred trajectories in the phase space of the system that its evolution approaches starting from generic initial conditions, which characterize how the loss of memory to the initial condition and subsequent hydrodynamization takes place in each theory.
    \item QGP in local thermal equilibrium: Once local thermal equilibrium is reached, the number of degrees of freedom one has to follow to describe the evolution of the bulk of the system decrease dramatically. It is in this setup that the ultimate EFT, hydrodynamics, shines. Once local thermal equilibrium has been reached, the only modes (collective excitations of the system) that live long enough for them to affect the bulk dynamics are those associated to the dynamics of conserved quantities, such as energy density or charge densities. Hydrodynamics is the theory that describes the dynamics of these modes, and it is the ultimate EFT in the sense that the low energy limit of \textit{any} many-body system (quantum or not) is a hydrodynamic description of its conserved quantities. 
    Different fluids are characterized by different constitutive relations and transport coefficients, which should be calculated from the microscopic description of the fluid or measured from experiment (both if one intends to test the microscopic description with experimental data). Therefore, the fluid description of QGP features numerical quantities that need to be defined and calculated in QCD. At weak coupling, this has been done using perturbation theory together with a kinetic description~\cite{Arnold:2000dr,Arnold:2003zc}. Static quantities, such as the equation of state, can be calculated from Lattice QCD at finite temperature (for reviews see~\cite{Petreczky:2012rq,Ratti:2018ksb}), but transport coefficients can be notoriously hard to calculate with this method. This is because transport coefficients are defined as the low frequency limit of correlation functions which are, in turn, defined as functions of real time, and to determine them from Euclidean calculations one has to carry out an analytic continuation which is mathematically well-defined only if one has access to a dense set of points where the calculation is carried out in imaginary time, which is not feasible unless further theoretical input is given. Therefore, estimates of their values have often had to be found outside of QCD. In fact, as it has long been realized~\cite{Shuryak:2003xe,Muller:2006ee,Shuryak:2008eq}, and it continues to be verified, e.g., according to parameter inference carried out using Bayesian analysis methods by \textit{Trajectum}~\cite{Nijs:2020ors} and JETSCAPE~\cite{JETSCAPE:2020shq}, the closest theoretical estimate we have of the shear viscosity of QGP at temperatures around $\Lambda_{\rm QCD}$ has been provided by the AdS/CFT correspondence, by which it was found that the specific shear viscosity equals $\eta/s = 1/(4\pi)$ in strongly coupled $\mathcal{N}=4$ Yang-Mills theory~\cite{Policastro:2001yc}. In this way, insight on the properties of QGP has been gained in complementary ways, using experimental data, first-principles QCD methods, as well as holography as input to the hydrodynamic description of QGP. In a HIC, a QGP will exist until chemical freeze-out, i.e., when interactions cease to maintain QCD matter in its deconfined phase, which (in a central collision) takes place around $\tau \sim 10 \, {\rm fm}/c$ after the collision, when the temperature has dropped to $T \sim T_c \approx 155 \, {\rm MeV}$.
    \item Hadronization, hadron gas phase, and kinetic freezeout: Once the temperature drops below the QCD crossover transition temperature $T_c \approx 155 \, {\rm MeV}$~\cite{Borsanyi:2010bp,Bazavov:2011nk}, QCD matter starts becoming confined again and its microscopic description is better thought of as a hadron gas instead of the deconfined QCD matter that is QGP. A common description of this stage is the Hadron Resonance Gas (HRG) model. The idea behind this model is to take the measured spectrum of hadrons and resonances and construct the thermodynamic partition function from these states. While the earliest versions of this model date back to works by Hagedorn~\cite{Hagedorn:1965st,Hagedorn:1980kb}, current versions of the model also incorporate long-range (van der Waals) interactions~\cite{Samanta:2017yhh,Vovchenko:2020lju}. In this approximation, one can recalculate the equation of state and transport coefficients with which the final state of the QGP can be propagated until kinetic freeze-out, when particles stop colliding with each other and simply free-stream away from the point where the collision took place (possibly decaying to other particles before being observed in a detector). The kinetic freeze-out temperature is usually around $T \sim 100 \, {\rm MeV}$.
\end{enumerate}
As may be apparent from this description, the observable quantities that are measurable at a particle detector are the hadrons (or their decay products) that free-stream after kinetic freeze-out has happened. As such, there is no direct probe of the evolution of QGP in the earlier stages, other than what can be constrained as possible through simulations and demanding consistency with empirical observations. This is nowadays done through Bayesian analysis of data based on models that can be simulated in a computer (e.g., JETSCAPE~\cite{Kauder:2018cdt} or \textit{Trajectum}~\cite{Nijs:2020roc}). These can be sensitive to the initial condition given to the model, and as such, qualitative and quantitative understanding of the pre-equilibrium dynamics is a necessity, which is still an active area of research nowadays. In fact, motivated by this, as we will discuss in Chapter~\ref{ch:hydrodynamization-in-HIC}, in this thesis we will seek to establish \textit{how} out-of-equilibrium QCD matter loses memory to the initial condition and thus explain \textit{why} a hydrodynamic QGP can be formed so quickly ($\tau \sim 1 \, {\rm fm}$) after the collision took place, even in weakly coupled settings. Concretely, we follow up on the proposal of~\cite{Brewer:2019oha} and propose that the out-of-equilibrium attractors that drive models of QCD matter towards hydrodynamization can be accurately described by the instantaneous eigenstate(s) of the effective Hamiltonian of the theory (i.e., the operator that generates time evolution) when it is written in an ``adiabatic frame'', which we will specify how to find. In this frame, it is manifest that a large part of the degrees of freedom of the system are very short-lived, thus making the low energy eigenstates of the effective Hamiltonian the only ones that are relevant to describe the dynamics of the plasma. In this way, we provide much needed intuition on the mechanisms that drive out-of-equilibrium systems to hydrodynamics. We expect the insights gained by understanding these qualitative aspects of hydrodynamization will be helpful for i) simplifying calculations of the pre-hydrodynamic stage of a HIC and ii) understanding the process of hydrodynamization and thermalization in broader many-body theory settings.

\paragraph{Hard and Electromagnetic Probes of QGP} \hspace{\fill}

Notwithstanding our discussion in the preceding paragraph, the debris of the locally equilibrated matter that comprised the QGP created in the collision is not the only way that QGP can be probed. While it is certainly unfeasible to carry out a measurement analog to that of the Geiger-Mardsen experiments or the DIS process, where an external probe is made to collide directly with the object to be studied, it is possible to carry out measurements on other matter produced by the collision, and compare with analog measurements conducted in collision systems where no QGP is formed, such as proton-proton ($p$-$p$) collisions. There are two large classes of such probes:
\begin{itemize}
    \item ``Hard'' probes: particles or matter produced with such high energy that they will never come into local thermal equilibrium with the QGP created in the collision. Some of the most prominent examples include ``jets,'' i.e., highly energetic beams of matter made up of either heavy or light particles whose propagation is modified relative to vacuum due to the presence of a QGP medium, and heavy hadrons (with at least one heavy quark in them) whose final yields at the detector change due to the interactions with the QGP.
    \item Electromagnetic probes: due to the comparatively weak coupling of electromagnetism, photons and electroweak bosons that later decay into lepton pairs can also be used to obtain information about the dynamics of a HIC. They are produced at one point in time during the collision and do not interact further with QCD matter, thus providing a clean, calculable process to compare data with theory. However, their weak coupling to the QGP is also a disadvantage compared to hard probes made of QCD matter, whose dynamics get modified much more strongly and thus provide richer information about how the QGP responds to such probes.
\end{itemize}

EFTs are, again, an invaluable tool to analyze the processes involving hard and electromagnetic probes of QGP in HICs, and especially so for hard probes made up of QCD matter, where calculations directly from the QCD Lagrangian are unwieldy. In this context,
\begin{itemize}
    \item SCET provides an invaluable tool to study jets. Even if there is an additional energy scale involved in probing QGP, its temperature $T$, this one is of the same order of magnitude as $\Lambda_{\rm QCD}$ and therefore a scale separation still exists and allows for controlled calculations. However, there are additional complications in comparison to jets in vacuum, such as the Landau-Pomeranchuk-Migdal (LPM) effect, by which repeated interactions with the medium can cause destructive interference and suppress the radiation spectrum of a jet. Applications of SCET to in-medium jet propagation include momentum broadening calculations~\cite{DEramo:2010wup,DEramo:2012uzl,Benzke:2012sz,Vaidya:2020cyi} and jet substructure calculations~\cite{Vaidya:2020lih}, among many others.
    \item HQET allows one to study the dynamics of open heavy quarks in a thermal medium. Again, since the additional energy scale given by the temperature is of the same order as $\Lambda_{\rm QCD}$, one may integrate out the high-energy degrees of freedom of QCD in the same way as in vacuum. A classic application of HQET in the context of HICs is the formulation of the heavy quark diffusion coefficient, first carried out in~\cite{Moore:2004tg,Casalderrey-Solana:2006fio,Caron-Huot:2009ncn}. By doing so, one can formulate precisely the quantity that needs to be calculated in the low-energy quantum field theory, which may then be calculated using other techniques (such as lattice QCD~\cite{Altenkort:2023oms} or even holographic methods~\cite{Casalderrey-Solana:2006fio}). In this sense, it is often said that the  calculation ``factorizes,'' something which was only achievable by means of the separation of energy scales that the EFT makes manifest.
    \item NRQCD and pNRQCD allow one to make quantitative statements about the formation of heavy hadrons after the QGP freezes out, and in particular, pNRQCD allows one to describe the in-medium dynamics of heavy quarkonium. In recent years, a substantial body of work starting from this EFT has been carried out with the goal of describing quarkonium suppression in HICs~\cite{Brambilla:2016wgg,Brambilla:2017zei,Brambilla:2020qwo,Brambilla:2021wkt,Brambilla:2022ynh,Brambilla:2023hkw,Yao:2017fuc,Yao:2018nmy,Yao:2018sgn,Yao:2020eqy,Yao:2020xzw}. We will build on these works and refine this description later on in this thesis. This was needed because, so far, the description of quarkonium in QGP had been mostly driven by our understanding of weakly coupled physics. In Chapter~\ref{ch:quarkonia-in-QGP} we rigorously formulate for the first time the non-perturbative objects that need to be calculated in the low-energy quantum field theory (QCD with only light quarks) to describe quarkonium transport in QGP, even in the regime where the plasma itself is strongly coupled, as is the case in HICs at RHIC and the LHC. Supplemented with knowledge of the quarkonium spectrum, these are measurable properties of QGP that can be determined (or at least constrained) from quarkonium suppression data in HICs.
\end{itemize}
We will devote special attention to this last bullet point in this thesis. 

Historically, bound states of heavy quarks were proposed by Matsui and Satz~\cite{Matsui:1986dk} as probes of QCD deconfinement. Their heuristic picture was that in the presence of deconfined QCD matter, the attractive potential in the singlet channel between a heavy quark pair would be ``screened'' at a length scale proportional to the inverse temperature of the plasma. Therefore, excited states of quarkonium that would be bound in vacuum due to the confinement property of QCD would ``melt'' in a hot QGP. As such, a suppression of the final quarkonium abundances ($J/\psi$ and $\Upsilon$, respectively $c\bar{c}$ and $b\bar{b}$) was expected in HICs relative to $p$-$p$ collisions.

\begin{figure}
    \centering
    \includegraphics[width=0.49\textwidth]{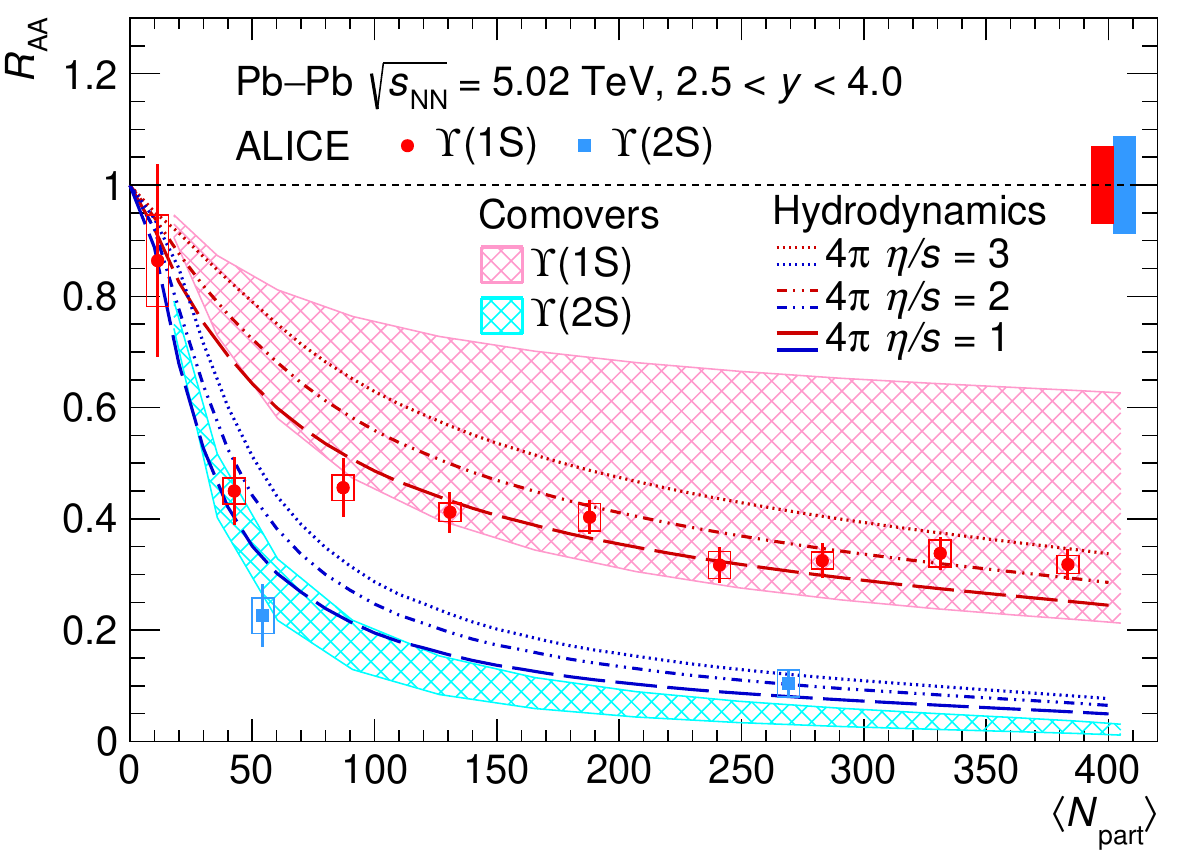}
    \includegraphics[width=0.49\textwidth]{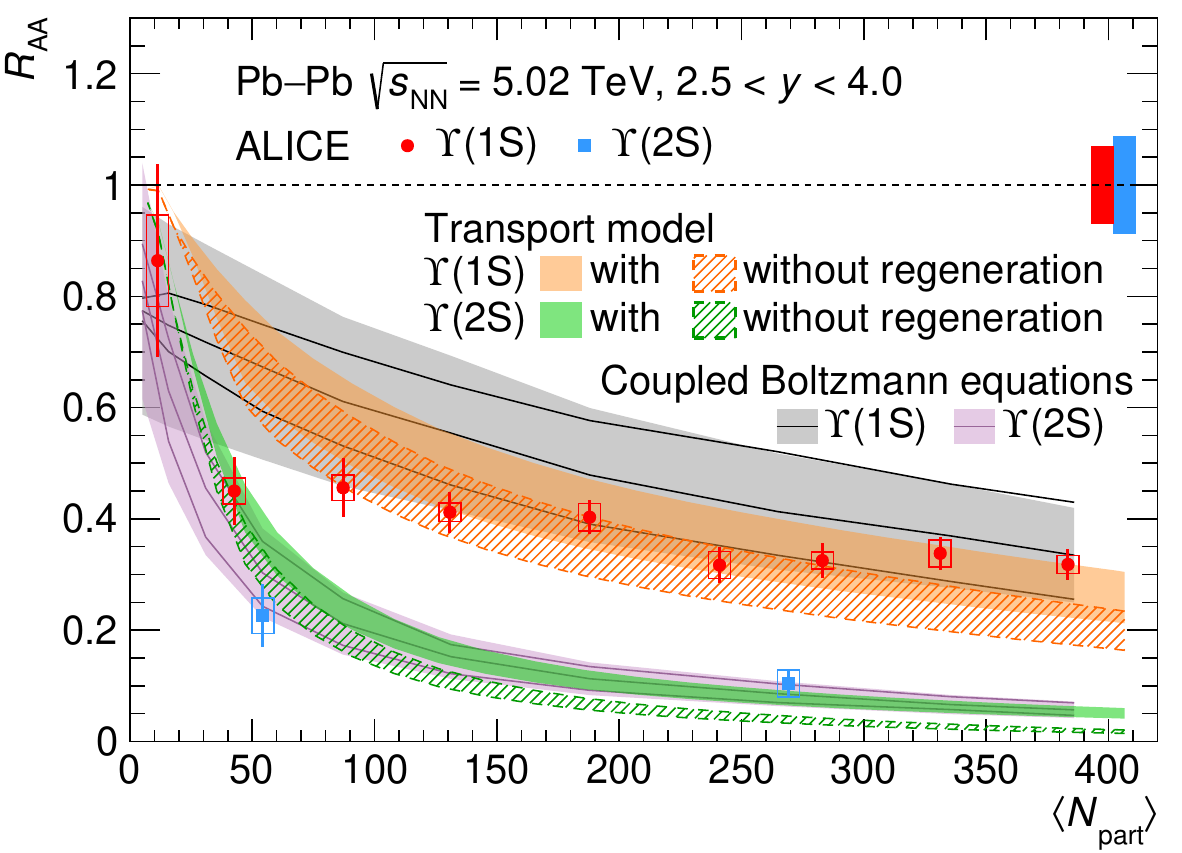}
    \includegraphics[width=0.49\textwidth]{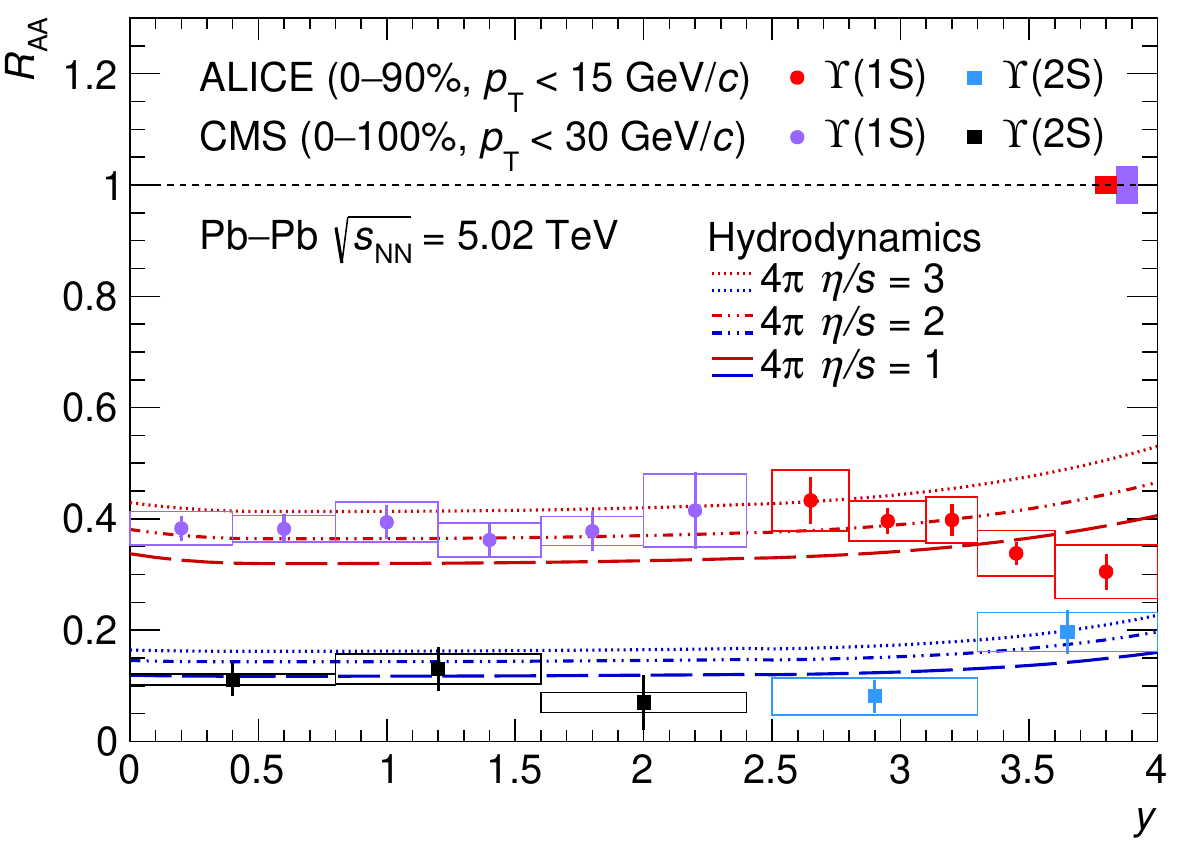}
    \includegraphics[width=0.49\textwidth]{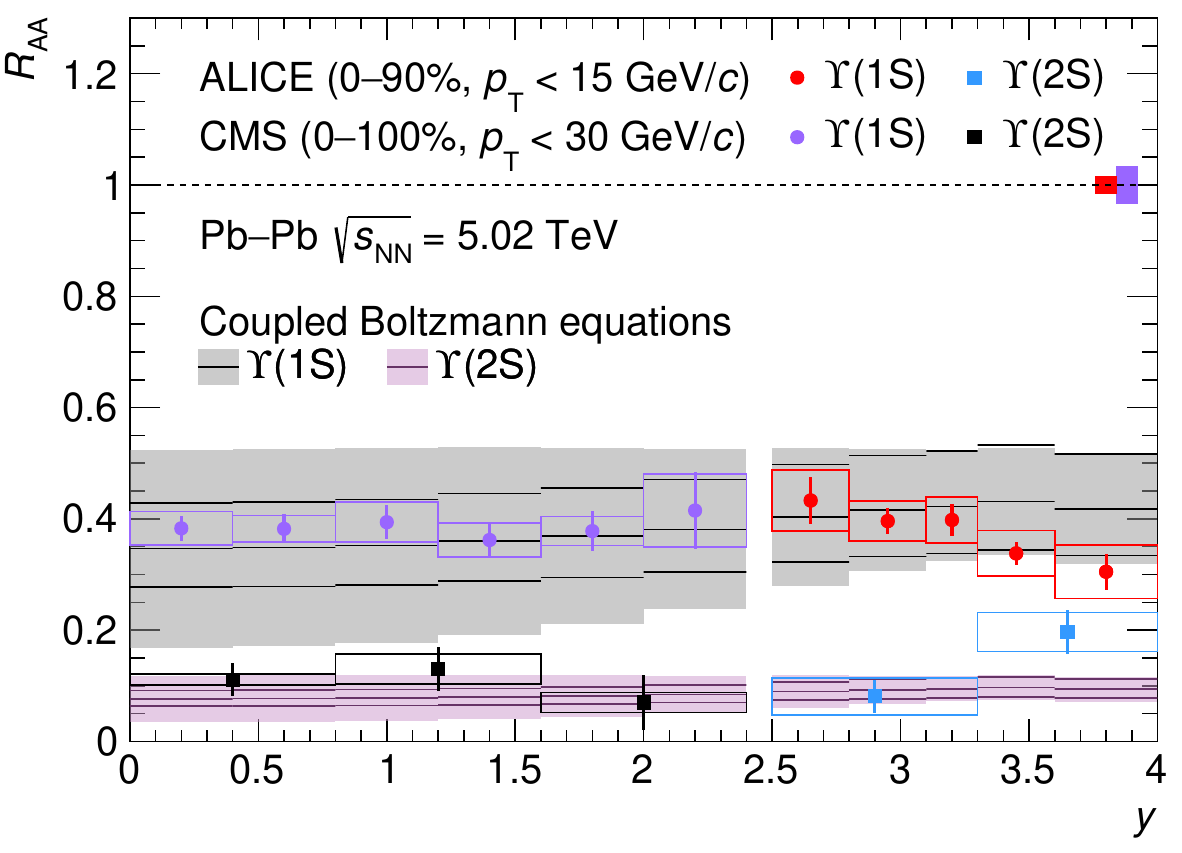}
    \includegraphics[width=0.49\textwidth]{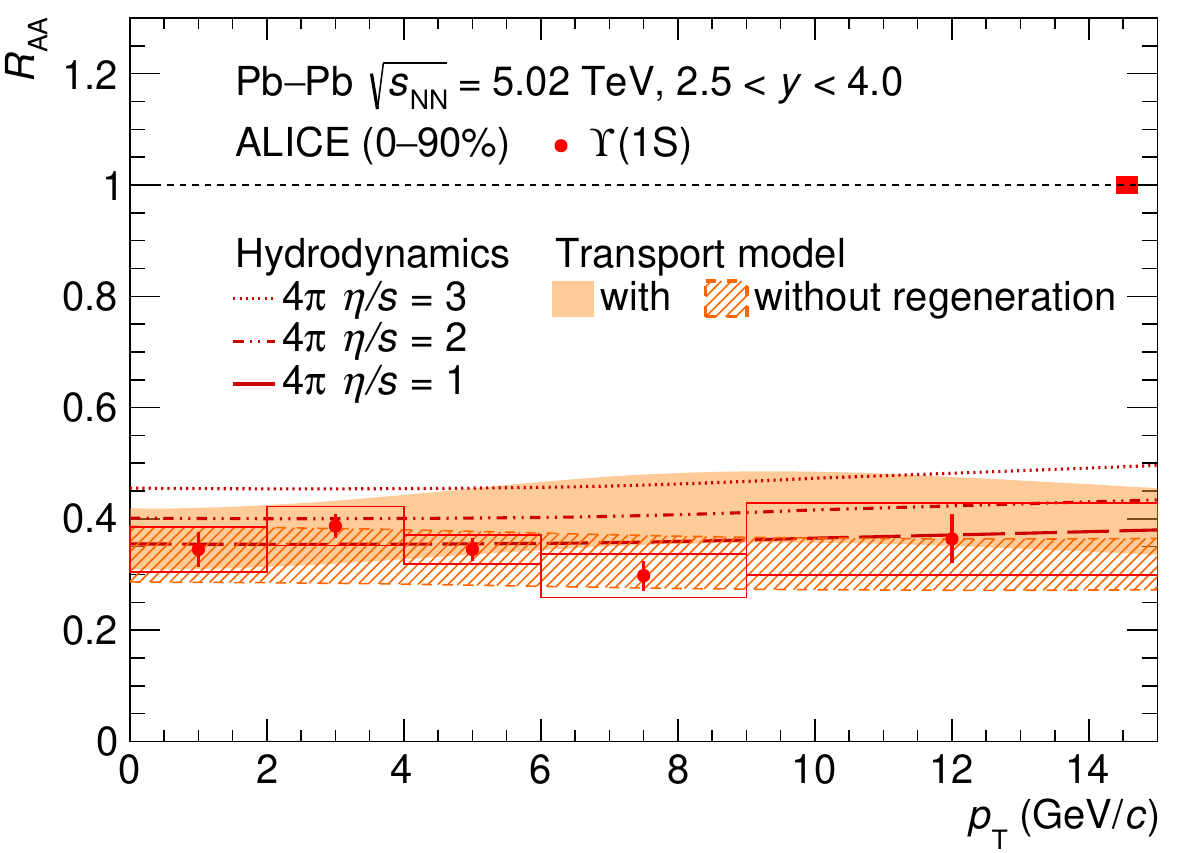}
    \caption{Top panels: nuclear modification factor $R_{AA}$ of $\Upsilon(1S)$ and $\Upsilon(2S)$ as a function of average number of participants. Middle panels: nuclear modification factor $R_{AA}$ of $\Upsilon(1S)$ and $\Upsilon(2S)$ as a function of rapidity. Bottom panel: nuclear modification factor $R_{AA}$ of $\Upsilon(1S)$ as a function of transverse momentum. The results are compared with several models. Figures reproduced from~\cite{ALICE:2020wwx}.}
    \label{fig:Quarkonium-Suppression-ALICE}
\end{figure}

\begin{figure}
    \centering
    \includegraphics[width=0.49\textwidth]{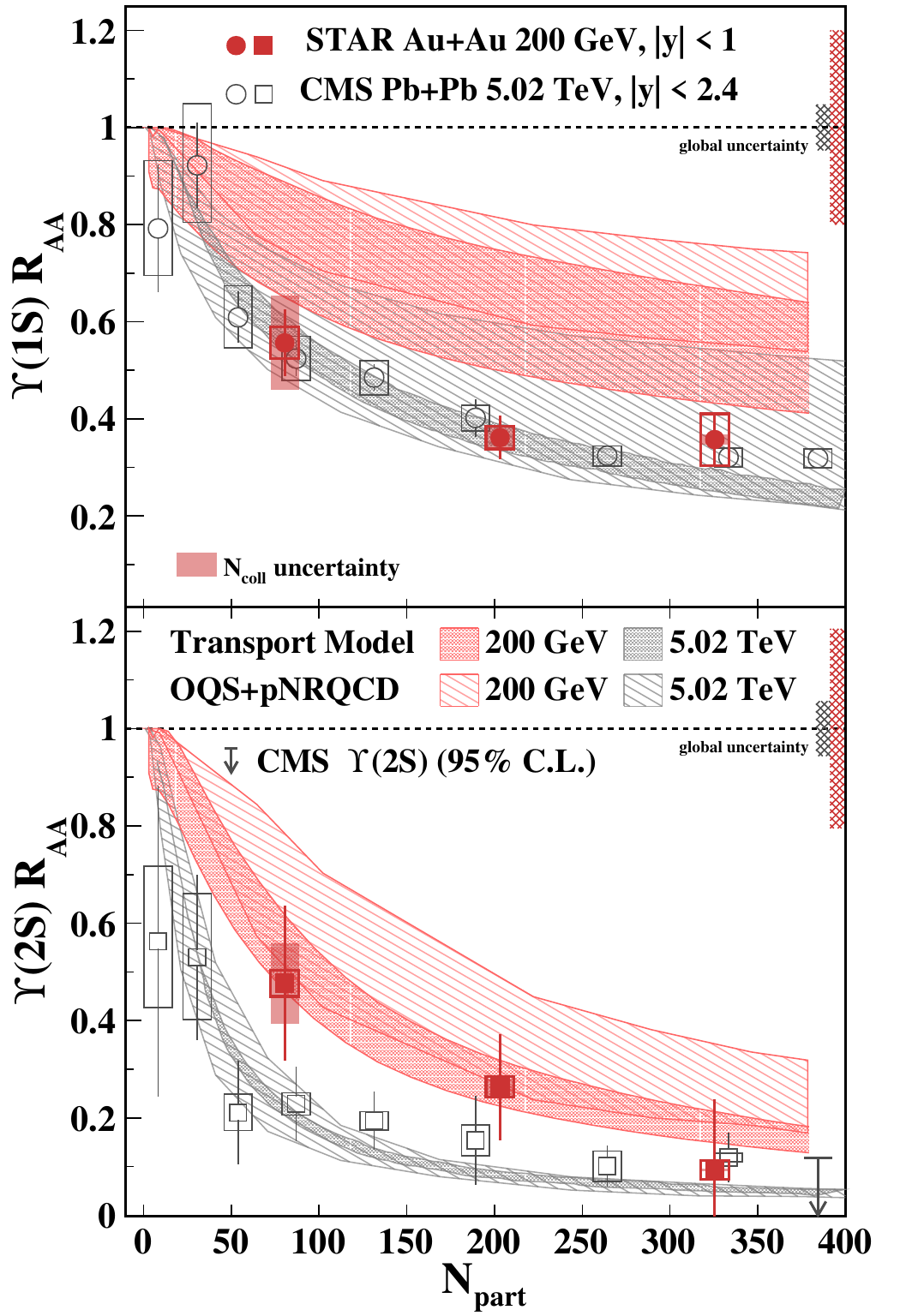}
    \includegraphics[width=0.49\textwidth]{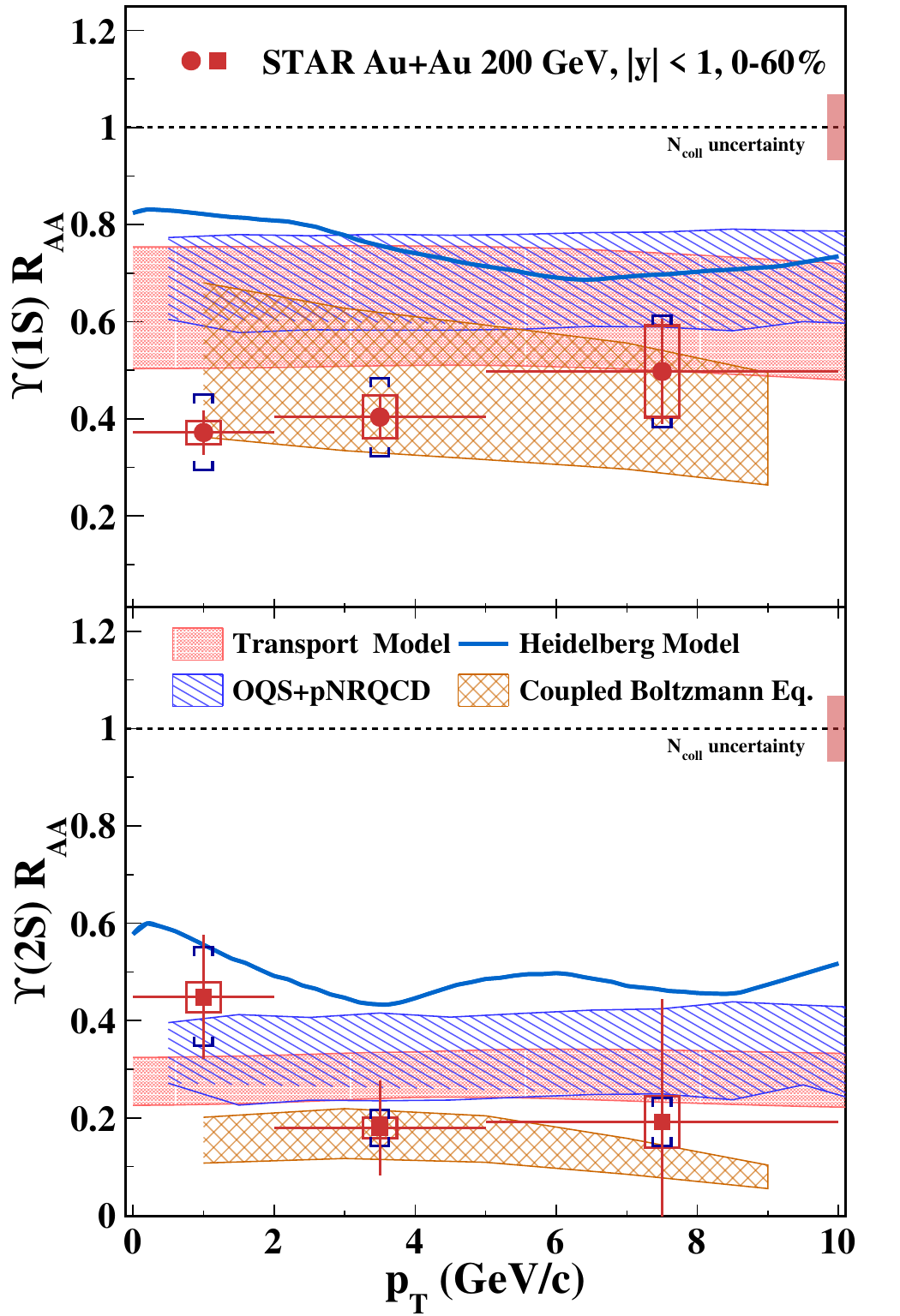}
    \caption{Left panels: nuclear modification factor $R_{AA}$ of $\Upsilon(1S)$ and $\Upsilon(2S)$ as a function of average number of participants. Right panels: nuclear modification factor $R_{AA}$ of $\Upsilon(1S)$ and $\Upsilon(2S)$ as a function of transverse momentum. The results are compared with several models. Figures reproduced from~\cite{STAR:2022rpk}.}
    \label{fig:Quarkonium-Suppression-STAR}
\end{figure}

Nowadays, quarkonium suppression has been abundantly measured. This is true for both $J/\psi$ states~\cite{STAR:2013eve,STAR:2019fge,ALICE:2015nvt,PHENIX:2006gsi,CMS:2016mah} and for $\Upsilon$ states~\cite{STAR:2013kwk,PHENIX:2014tbe,ALICE:2020wwx,STAR:2022rpk,CMS:2023lfu}. We will focus on the latter, because due to the large mass of the $b$ quark, the EFT description we employ and develop in this thesis is under better quantitative control for $b\bar{b}$ states than for $c\bar{c}$ states. We display the $\Upsilon$ suppression ratios reported in~\cite{ALICE:2020wwx} and~\cite{STAR:2022rpk} in Figures~\ref{fig:Quarkonium-Suppression-ALICE} and Figure~\ref{fig:Quarkonium-Suppression-STAR}, respectively, and their comparison to several models (see the references in the captions for details thereof). The quantity plotted in all of the panels is called the nuclear modification factor $R_{AA}$, and represents the ratio of the number of $\Upsilon$ states measured in Pb-Pb collisions to the $\Upsilon$ production cross-section in $p$-$p$ collisions in the same kinematic regime times the average number of inelastic nucleon-nucleon collisions in the Pb-Pb collision. As we can see from Figures~\ref{fig:Quarkonium-Suppression-ALICE} and~\ref{fig:Quarkonium-Suppression-STAR}, the theory uncertainty is generally larger than the experimental uncertainty. As such, there is a need for a first-principles theoretical description where the uncertainties can be reduced systematically.

\begin{figure}
    \centering
    \includegraphics[width=0.52\textwidth]{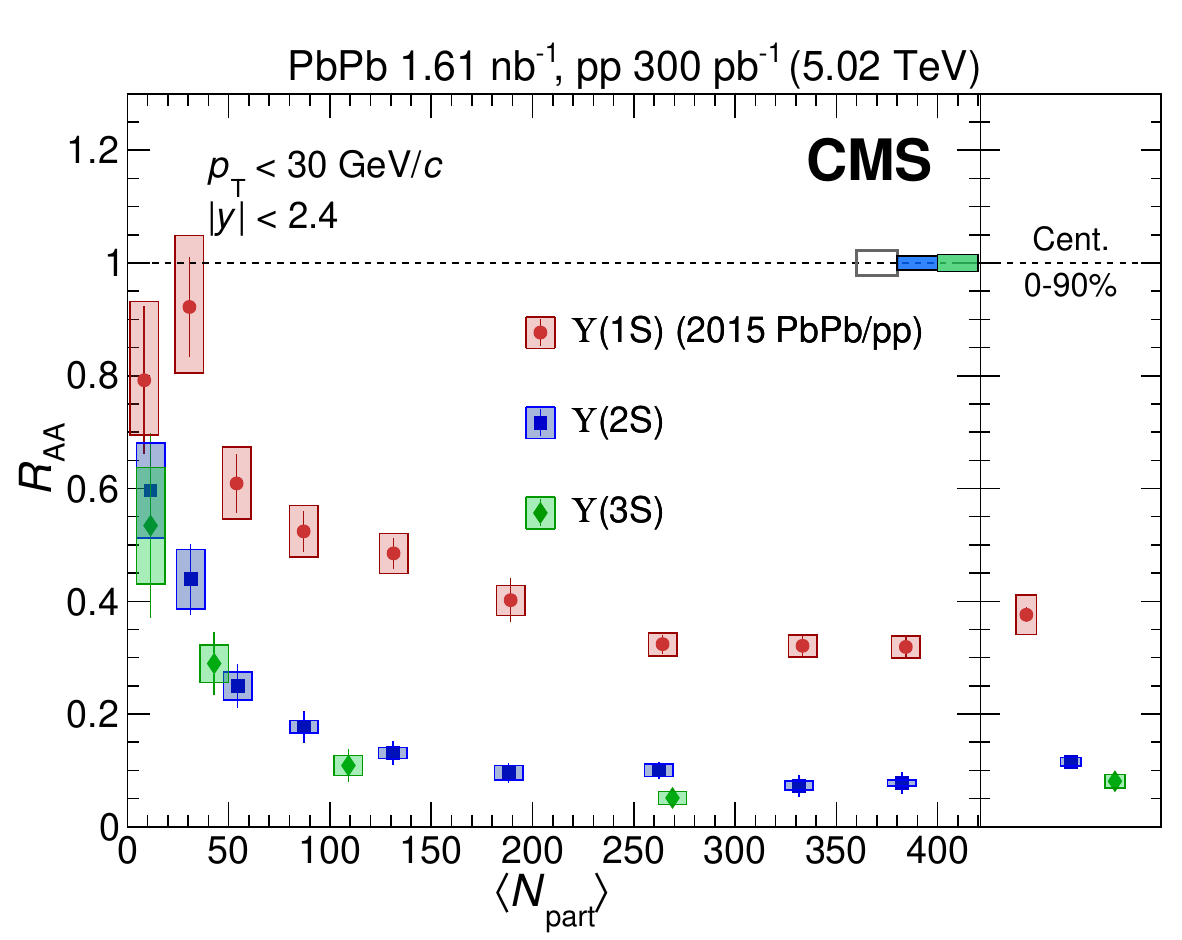}
    \includegraphics[width=0.47\textwidth]{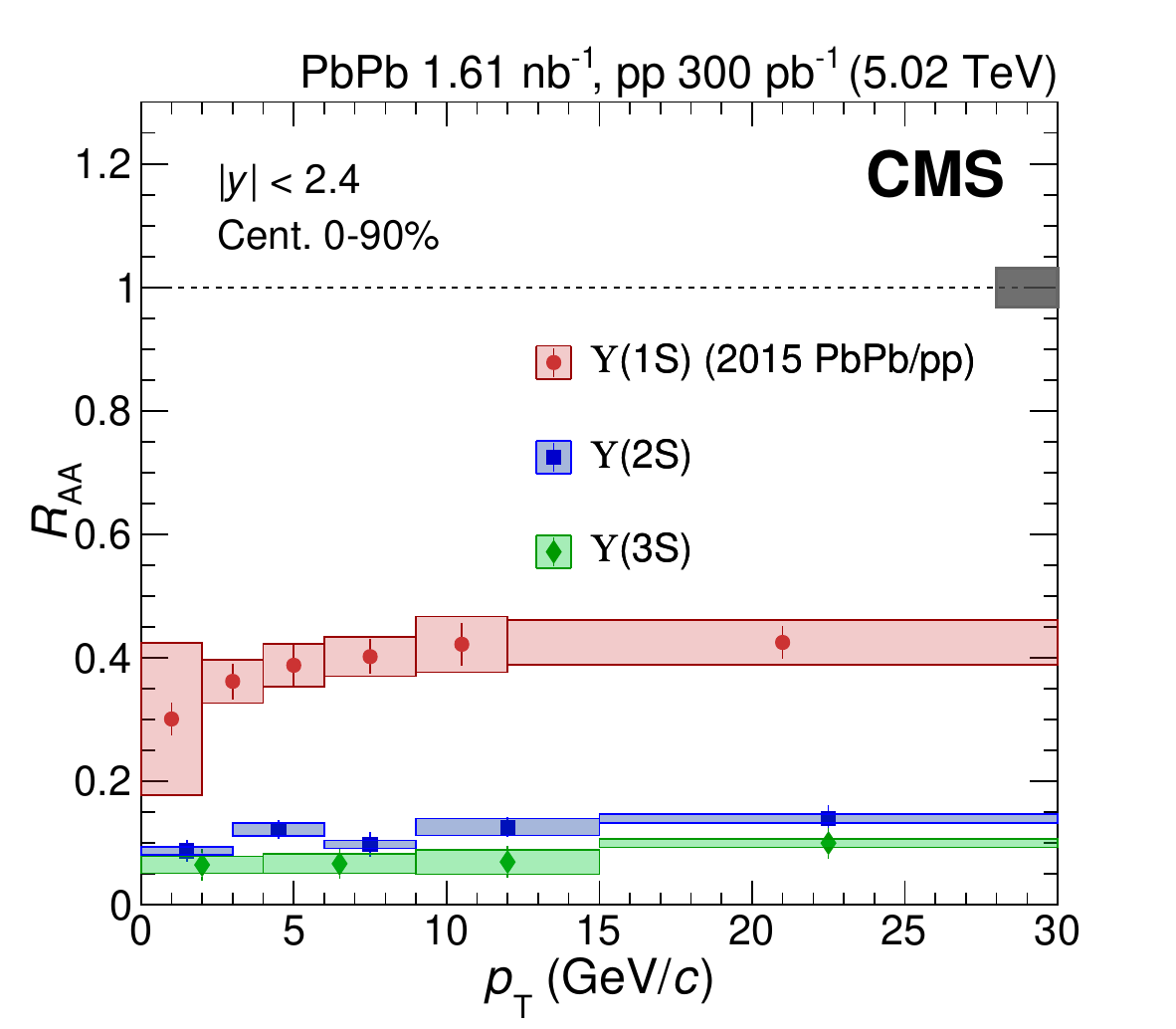}
    \caption{Left panel: nuclear modification factor $R_{AA}$ of $\Upsilon(1S)$, $\Upsilon(2S)$, and $\Upsilon(3S)$ as a function of average number of participants. Right panels: nuclear modification factor $R_{AA}$ of $\Upsilon(1S)$, $\Upsilon(2S)$, and $\Upsilon(3S)$ as a function of transverse momentum. Figures reproduced from~\cite{CMS:2023lfu}.}
    \label{fig:Quarkonium-Suppression-CMS}
\end{figure}

Furthermore, recent measurements by the CMS collaboration~\cite{CMS:2023lfu} have verified sequential that quarkonium suppression takes place up to the $\Upsilon(3S)$ state (see Figure~\ref{fig:Quarkonium-Suppression-CMS}), i.e., that $R_{AA}[\Upsilon(1S)] > R_{AA}[\Upsilon(2S)] > R_{AA}[\Upsilon(3S)]$, meaning that there is a wealth of data that informs and places sharp constraints on our theoretical description of quarkonium dynamics in QGP.

As it has turned out, the picture provided by Matsui and Satz has proven to be insufficient to describe data. Dissociation and recombination, processes by which a heavy quark pair in a color singlet state can transition to and from a color octet state (the other possible state for a $Q\bar{Q}$ pair: $3 \otimes 3 = 1 \oplus 8$), essentially by scattering off quarks or gluons from the QGP, can significantly influence the final abundances of quarkonium. In fact, recent Lattice QCD results~\cite{Bala:2021fkm} of the $Q\bar{Q}$ (free) energy from Wilson loop expectation values suggest that screening may play no role at all at temperatures of $T \sim 200 \, {\rm MeV}$ and lower, and those results also cast doubt on whether screening provides a significant effect on the $Q\bar{Q}$ interaction potential at higher temperatures relevant for HICs at RHIC or the LHC.

As such, a framework that can account for dissociation and recombination of quarkonium inside QGP is necessary to describe quarkonium suppression data. Furthermore, to interpret the data in terms of the underlying microscopic theory, one needs to calculate the effect of these processes from QCD. To make progress in calculations, EFTs provide an invaluable tool. And, indeed, a significant amount of progress has happened in this direction in the last ten years through the application of pNRQCD to describe quarkonium suppression~\cite{Brambilla:2016wgg,Brambilla:2017zei,Brambilla:2020qwo,Brambilla:2021wkt,Brambilla:2022ynh,Brambilla:2023hkw,Yao:2017fuc,Yao:2018nmy,Yao:2018sgn,Yao:2020eqy,Yao:2020xzw}. However, despite the substantial body of literature on the subject, the precise formulation of the correlation functions that need to be calculated non-perturbatively (or measured experimentally) to describe the dynamics of in-medium quarkonium at realistic coupling strengths and temperatures was not available until the work described in this thesis was published.

In summary, the research presented in this thesis advances our understanding of hot QCD matter in two crucial ways:
\begin{itemize}
    \item Dynamics of quarkonium in QGP: We present the first studies formulating the precise QCD quantities to which quarkonium dissociation and recombination in QGP is sensitive, and as such, provide solid ground to relate experiment with the parameters of the microscopic theory, QCD. We expect that phenomenological studies that use our results will be able to shed light into the nature of the confinement-deconfinement transition in a dynamical setting, as well as give further insight into the thermal correlations of hot QCD matter that make QGP the most perfect fluid in nature, and last but not least, its response to hard probes embodied in quarkonia. 
    \item Thermalization and hydrodynamization of QCD matter: We present the first studies showing that the hydrodynamization process of a simplified version of QCD kinetic theory can be described by a monotonously dwindling set of adiabatically evolving states, thus explaining \textit{why} attractor solutions emerge and memory of the initial condition is lost and \textit{how} hydrodynamics is approached from a highly out-of-equilibrium initial condition, concluding at a point in time where only one adiabatically evolving state remains, which captures entirely and uniquely the properties of a locally equilibrated plasma.
\end{itemize}

The remaining chapters of this thesis are organized as follows: 
\begin{itemize}
    \item In Chapter~\ref{ch:overview} we enumerate where each part of this thesis can be found in published literature, as well as mention other work carried out by the author of this thesis that is not covered here.
    \item In Chapter~\ref{ch:quarkonia-in-QGP} we formulate the gauge theory correlation functions that describe quarkonium transport in a thermal medium (Section~\ref{sect:open}), and calculate them, firstly at weak coupling in QCD (Section~\ref{sect:nlo}), secondly at strong coupling in $\mathcal{N} = 4$ SYM (Section~\ref{sec:strong-coupling}), as well as formulate them in Euclidean QCD (Section~\ref{sec:latticec}), paving the way for a lattice QCD determination of quarkonium transport properties. We close this chapter by laying out the prospects for a phenomenological application of these correlators (Section~\ref{sec:pheno}).
    \item In Chapter~\ref{ch:hydrodynamization-in-HIC} we demonstrate that the Adiabatic Hydrodynamization scenario, as proposed in~\cite{Brewer:2019oha}, is realized in a kinetic theory description of QCD. Even though the collision kernel studied here is, admittedly, incomplete for a full QCD description, the terms that are present have a direct connection to the microscopic scattering mechanisms of quarks and gluons in out-of-equilibrium QCD matter. As such, this is a significant step forward in our understanding of the hydrodynamization/thermalization process of QCD, because we now have the ability to systematically isolate the degrees of freedom at early times that will survive the process of hydrodynamization. 
    \item Finally, in Chapter~\ref{ch:outlook} we give our conclusions, as well as our outlook for the next steps after this work, open problems, and the challenges in the road ahead.
\end{itemize}

In summary, by refining our theoretical understanding of out-of-equilibrium QCD dynamics, this thesis lays the groundwork for subsequent phenomenological studies that will connect QCD theory and experiment more directly than ever before. We also hope this thesis will motivate continuing theoretical work on these topics.

%% file: overview.tex

\chapter{Overview} \label{ch:overview}

This thesis covers work that was carried out to advance our understanding of Hot QCD matter in two directions: 1) the propagation of pairs of heavy quarks through QGP and what information from the QGP is imprinted in their dynamics and onto HIC observables, and 2) the dynamics of the process of hydrodynamization that describes how the initial state of a HIC, i.e., two highly Lorentz-contracted nuclei, quickly reaches local thermal equilibrium and becomes a hydrodynamic QGP.

The purpose of this Chapter is to describe and provide a quick reference to the published work on which this thesis is based.

\section{Chapter 3: Dynamics of Quarkonia in Quark-Gluon Plasma}

This chapter is based on five papers and one conference proceedings that explore the dynamics of heavy quark pairs in the presence of a thermal environment.

\paragraph{Non-Abelian electric field correlator at NLO for dark matter relic abundance and quarkonium transport~\cite{Binder:2021otw}} \hspace{\fill}

\textit{In collaboration with Tobias Binder, Kyohei Mukaida and Xiaojun Yao. This work is described in Section~\ref{sect:nlo}, excluding Section~\ref{sec:axial-gauge}.}

We perform a complete next-to-leading order calculation of the non-Abelian electric field correlator in a SU($N_c$) plasma, which encodes properties of the plasma relevant for heavy particle bound state formation and dissociation, and is different from the correlator for the heavy quark diffusion coefficient. The calculation is carried out in the real-time formalism of thermal field theory and includes both vacuum and finite temperature contributions. By working in the $R_\xi$ gauge, we explicitly show the results are gauge independent, infrared and collinear safe. The renormalization group equation of this electric field correlator is determined by that of the strong coupling constant. Our next-to-leading order calculation can be directly applied to any dipole singlet-adjoint transition of heavy particle pairs. For example, it can be used to describe dissociation and (re)generation of heavy quarkonia inside the quark-gluon plasma well below the melting temperature, as well as heavy dark matter pairs (or charged co-annihilating partners) in the early universe.

\paragraph{Gauge Invariance of Non-Abelian Field Strength Correlators: The Axial Gauge Puzzle~\cite{Scheihing-Hitschfeld:2022xqx}} \hspace{\fill}

\textit{In collaboration with Xiaojun Yao. This work is described in Section~\ref{sec:axial-gauge}.}

Many transport coefficients of the quark-gluon plasma and nuclear structure functions can be written as gauge invariant correlation functions of non-Abelian field strengths dressed with Wilson lines. We discuss the applicability of axial gauge $n \cdot A = 0$ to calculate them. In particular, we address issues that appear when one attempts to trivialize the Wilson lines in the correlation functions by gauge fixing. We find it is always impossible to completely remove the gauge fields $n \cdot A$ in Wilson lines that extend to infinity in the $n$ direction by means of gauge transformations. We show how the obstruction appears in an explicit example of a perturbative calculation, and we also explain it more generally from the perspective of the path integral that defines the theory. Our results explain why the two correlators that define the heavy quark and quarkonium transport coefficients, which are seemingly equal in axial gauge, are actually different physical quantities of the quark-gluon plasma and have different values. Furthermore, our findings provide insights into the difference between two inequivalent gluon parton distribution functions.

\paragraph{Chromoelectric field correlator for quarkonium transport in the strongly coupled $\mathcal{N} = 4$ Yang-Mills plasma from AdS/CFT~\cite{Nijs:2023dks}} \hspace{\fill}

\textit{In collaboration with Govert Nijs and Xiaojun Yao. This work is described in Section~\ref{sec:strong-coupling}, excluding Section~\ref{sec:corr-flow}.}

Previous studies have shown that a gauge-invariant correlation function of two chromoelectric fields connected by a straight timelike adjoint Wilson line encodes crucial information about quark-gluon plasma (QGP) that determines the dynamics of small-sized quarkonium in the medium. Motivated by the successes of holographic calculations to describe strongly coupled QGP, we calculate the analog gauge-invariant correlation function in strongly coupled $\mathcal{N}=4$ supersymmetric Yang-Mills theory at finite temperature by using the AdS/CFT correspondence. Our results indicate that the transition processes between bound and unbound quarkonium states are suppressed in strongly coupled plasmas, and moreover, the leading contributions to these transition processes vanish in both the quantum Brownian motion and quantum optical limits of open quantum system approaches to quarkonia.

\paragraph{Real time quarkonium transport coefficients in open quantum systems from Euclidean QCD~\cite{Scheihing-Hitschfeld:2023tuz}} \hspace{\fill}

\textit{In collaboration with Xiaojun Yao. This work is described in Section~\ref{sec:lattice}.}

Recent open quantum system studies showed that quarkonium time evolution inside the quark-gluon plasma is determined by transport coefficients that are defined in terms of a gauge invariant correlator of two chromoelectric field operators connected by an adjoint Wilson line. We study the Euclidean version of the correlator for quarkonium evolution and discuss the extraction of the transport coefficients from this Euclidean correlator, highlighting its difference from other problems that also require reconstructing a spectral function, such as the calculation of the heavy quark diffusion coefficient. Along the way, we explain why the transport coefficient $\gamma_{\rm adj}$ differs from $\gamma_{\rm fund}$ at finite temperature at $O(g^4)$, in spite of the fact that their corresponding spectral functions differ only by a temperature-independent term at the same order. We then discuss how to evaluate the Euclidean correlator via lattice QCD methods, with a focus on reducing the uncertainty caused by infrared renormalons in determining the renormalization factor nonperturbatively.

\paragraph{Generalized Gluon Distribution for Quarkonium Dynamics in Strongly Coupled $\mathcal{N}=4$ Yang-Mills Theory~\cite{Nijs:2023dbc}} \hspace{\fill}

\textit{In collaboration with Govert Nijs and Xiaojun Yao. This work is described in Sections~\ref{sec:corr-flow} and~\ref{sec:weak-strong}.}

We study the generalized gluon distribution that governs the dynamics of quarkonium inside a non-Abelian thermal plasma characterizing its dissociation and recombination rates. This gluon distribution can be written in terms of a correlation function of two chromoelectric fields connected by an adjoint Wilson line. We formulate and calculate this object in $\mathcal{N}=4$ supersymmetric Yang-Mills theory at strong coupling using the AdS/CFT correspondence, allowing for a nonzero center-of-mass velocity $v$ of the heavy quark pair relative to the medium. The effect of a moving medium on the dynamics of the heavy quark pair is described by the simple substitution $T \to \sqrt{\gamma} \, T$, in agreement with previous calculations of other observables at strong coupling, where $T$ is the temperature of the plasma in its rest frame, and $\gamma = (1 - v^2)^{-1/2}$ is the Lorentz boost factor. Such a velocity dependence can be important when the quarkonium momentum is larger than its mass. Contrary to general expectations for open quantum systems weakly coupled with large thermal environments, the contributions to the transition rates that are usually thought of as the leading ones in Markovian descriptions vanish in this strongly coupled plasma. This calls for new theoretical developments to assess the effects of strongly coupled non-Abelian plasmas on in-medium quarkonium dynamics. Finally, we compare our results with those from weakly coupled QCD, and find that the QCD result moves toward the $\mathcal{N}=4$ strongly coupled result as the coupling constant is increased within the regime of applicability of perturbation theory. This behavior makes it even more pressing to develop a non-Markovian description of quarkonium in-medium dynamics.

\paragraph{Quarkonium transport in weakly and strongly coupled plasmas~\cite{Nijs:2023bok}} \hspace{\fill}

\textit{Ongoing work in collaboration with Govert Nijs and Xiaojun Yao. This work is described in Section~\ref{sec:pheno}, and corresponds to a contribution to the proceedings of the 30th International Conference on Ultra-relativistic Nucleus-Nucleus Collisions (Quark Matter 2023).}

We report on progress in the nonperturbative understanding of quarkonium dynamics inside a thermal plasma. The time evolution of small-size quarkonium is governed by two-point correlation functions of chromoelectric fields dressed with an adjoint Wilson line, known in this context as generalized gluon distributions (GGDs). The GGDs have been calculated in both weakly and strongly coupled plasmas by using perturbative and holographic methods. Strikingly, the results of our calculations for a strongly coupled plasma indicate that the quarkonium dissociation and recombination rates vanish in the transport descriptions that assume quarkonium undergoes Markovian dynamics. However, this does not imply that the dynamics is trivial. As a starting point to explore the phenomenological consequences of the result at strong coupling, we show a calculation of the $\Upsilon (1S)$ formation probability in time-dependent perturbation theory. This is a first step towards the development of a transport formalism that includes non-Markovian effects, which, depending on how close the as of yet undetermined nonperturbative QCD result of the GGDs is to the strongly coupled $\mathcal{N}=4$ SYM result, could very well dominate over the Markovian ones in quark-gluon plasma produced at RHIC and the LHC.

\section{Chapter 4: Dynamics of Hydrodynamization and Emergence of Hydrodynamics in Heavy-Ion Collisions}

This chapter is based on two papers that explore the hydrodynamization process of QCD matter, with focus on how QGP forms in heavy ion collisions.

\paragraph{Scaling and adiabaticity in a rapidly expanding gluon plasma~\cite{Brewer:2022vkq}} \hspace{\fill}

\textit{In collaboration with Jasmine Brewer and Yi Yin. This work is described in Section~\ref{sec:intro-AH-BSY} through~\ref{sec:sum-adiab-early}.}

In this work we aim to gain qualitative insight on the far-from-equilibrium behavior of the gluon plasma produced in the early stages of a heavy-ion collision. It was recently discovered~\cite{Mazeliauskas:2018yef} that the distribution functions of quarks and gluons in QCD effective kinetic theory (EKT) exhibit self-similar “scaling” evolution with time-dependent scaling exponents long before those exponents reach their pre-hydrodynamic fixed-point values. In this work we shed light on the origin of this time-dependent scaling phenomenon in the small-angle approximation to the Boltzmann equation. We first solve the Boltzmann equation numerically and find that time-dependent scaling is a feature of this kinetic theory, and that it captures key qualitative features of the scaling of hard gluons in QCD EKT. We then proceed to study scaling analytically and semi-analytically in this equation. We find that an appropriate momentum rescaling allows the scaling distribution to be identified as the instantaneous ground state of the operator describing the evolution of the distribution function, and the approach to the scaling function is described by the decay of the excited states. That is to say, there is a frame in which the system evolves adiabatically. Furthermore, from the conditions for adiabaticity we can derive evolution equations for the time-dependent scaling exponents. In addition to the known free-streaming and BMSS fixed points, we identify a new “dilute” fixed point when the number density becomes small before hydrodynamization. Corrections to the fixed point exponents in the small-angle approximation agree quantitatively with those found previously in QCD EKT and arise from the evolution of the ratio between hard and soft scales.

\paragraph{Adiabatic Hydrodynamization and the Emergence of Attractors: a Unified Description of Hydrodynamization in Kinetic Theory~\cite{Rajagopal:2024lou}} \hspace{\fill}

\textit{In collaboration with Krishna Rajagopal and Rachel Steinhorst. This work is described in Section~\ref{sec:adiab-beyond-scaling}.}

``Attractor" solutions for the pre-hydrodynamic, far-from-equilibrium, evolution of the matter produced in relativistic heavy ion collisions have emerged as crucial descriptors of the rapid hydrodynamization of quark-gluon plasma (QGP). Adiabatic Hydrodynamization (AH) has been proposed as a framework with which to describe, explain, and predict attractor behavior that draws upon an analogy to the adiabatic approximation in quantum mechanics. In this work, we systematize the description of pre-hydrodynamic attractors in kinetic theory by showing how to use the AH framework to identify these long-lived solutions to which varied initial conditions rapidly evolve, demonstrating the robustness of this framework. In a simplified QCD kinetic theory in the small-angle scattering limit, we use AH to explain both the early- and late-time scaling behavior of a longitudinally expanding gluon gas in a unified framework. In this context, we show that AH provides a unified description of, and  intuition for, all the stages of what in QCD would be bottom-up thermalization, starting from a pre-hydrodynamic attractor and ending with hydrodynamization. We additionally discuss the connection between the notions of scaling behavior and adiabaticity and the crucial role of time-dependent coordinate redefinitions in identifying the degrees of freedom of kinetic theories that give rise to attractor solutions. The tools we present open a path to the intuitive explanation of how attractor behavior arises and how the attractor evolves in all stages of the hydrodynamization of QGP in heavy ion collisions.


\section{Work not covered in this thesis}

During and before my time at MIT, I have also published works in the fields of inflationary cosmology and condensed matter physics. 

\subsection{The generation of primordial non-Gaussianity in multi-field inflation}

\paragraph{Landscape tomography through primordial non-Gaussianity~\cite{Chen:2018uul}} \hspace{\fill}

\textit{In collaboration with Xingang Chen, Gonzalo A. Palma, Walter Riquelme and Spyros Sypsas.}

In this paper, we show how the structure of the landscape potential of the primordial Universe may be probed through the properties of the primordial density perturbations responsible for the origin of the cosmic microwave background anisotropies and the large-scale structure of our Universe. Isocurvature fields—fields orthogonal to the inflationary trajectory—may have fluctuated across the barriers separating local minima of the landscape potential during inflation. We analyze how this process could have impacted the evolution of the primordial curvature perturbations. If the typical distance separating consecutive minima of the landscape potential and the height of the potential barriers are smaller than the Hubble expansion rate parametrizing inflation, the probability distribution function of isocurvature fields becomes non-Gaussian due to the appearance of bumps and dips associated with the structure of the potential. We show that this non-Gaussianity can be transferred to the statistics of primordial curvature perturbations if the isocurvature fields are coupled to the curvature perturbations. The type of non-Gaussian structure that emerges in the distribution of curvature perturbations cannot be fully probed with the standard methods of polyspectra; instead, the probability distribution function is needed. The latter is obtained by summing all the n-point correlation functions. To substantiate our claims, we offer a concrete model consisting of an axionlike isocurvature perturbation with a sinusoidal potential and a linear derivative coupling between the isocurvature and curvature field. In this model, the probability distribution function of the curvature perturbations consists of a Gaussian function with small superimposed oscillations reflecting the isocurvature axion potential.

\paragraph{Reconstructing the Inflationary Landscape with Cosmological Data~\cite{Chen:2018brw}} \hspace{\fill}

\textit{In collaboration with Xingang Chen, Gonzalo A. Palma and Spyros Sypsas.}

We show that the shape of the inflationary landscape potential may be constrained by analyzing cosmological data. The quantum fluctuations of fields orthogonal to the inflationary trajectory may have probed the structure of the local landscape potential, inducing non-Gaussianity (NG) in the primordial distribution of the curvature perturbations responsible for the cosmic microwave background (CMB) anisotropies and our Universe’s large-scale structure. The resulting type of NG (tomographic NG) is determined by the shape of the landscape potential, and it cannot be fully characterized by 3- or 4-point correlation functions. Here we deduce an expression for the profile of this probability distribution function in terms of the landscape potential, and we show how this can be inverted in order to reconstruct the potential with the help of CMB observations. While current observations do not allow us to infer a significant level of tomographic NG, future surveys may improve the possibility of constraining this class of primordial signatures.

\paragraph{Non-Gaussian CMB and LSS statistics beyond polyspectra~\cite{Palma:2019lpt}} \hspace{\fill}

\textit{In collaboration with Xingang Chen, Gonzalo A. Palma and Spyros Sypsas.}

Cosmic inflation may have led to non-Gaussian initial conditions that cannot be fully parametrised by 3- and/or 4-point functions. In this work, we discuss various strategies to search for primordial non-Gaussianity beyond polyspectra with the help of cosmological data. Our starting point is a generalised local ansatz for the primordial curvature perturbation $\zeta$ of the form $\zeta = \zeta_G + F_{\rm NG}(\zeta_G)$, where $\zeta_G$ is a Gaussian random field and $F_{\rm NG}$ is an arbitrary function parametrising non-Gaussianity that, in principle, could be reconstructed from data. Noteworthily, in the case of multi-field inflation, the function $F_{\rm NG}$ can be shown to be determined by the shape of tomographic sections of the landscape potential responsible for driving inflation. We discuss how this generalised local ansatz leads to a probability distribution functional that may be used to extract information about inflation from current and future observations. In particular, we derive various classes of probability distribution functions suitable for the statistical analysis of the cosmic microwave background and large-scale structure.

\subsection{Thermal transport resistivity in solids due to phonon scattering by dislocations}

\paragraph{Scattering of phonons by quantum dislocations segments in an elastic continuum~\cite{PhysRevB.99.214102}} \hspace{\fill}

\textit{In collaboration with Fernando Lund.}

A canonical quantization procedure is applied to elastic waves interacting with pinned dislocation segments via the Peach-Koehler force. The interaction Hamiltonian, derived from an action principle that classically generates the Peach-Koehler force, is a power series of creation and annihilation operators. The leading term is quadratic, and keeping only this term the observable quantities of scattering processes are computed to all orders in perturbation theory. The resulting theory is characterized by the magnitude of $kL$, with $k$ the wavenumber of an incident phonon. The theory is solved for arbitrary $kL$, and different limits are explored. A significant result at this level is the scattering cross section for phonons by dislocation segments. This cross section has a much richer structure than the linear-in-frequency behavior that is inferred from scattering by an infinite, static, dislocation. When many dislocations are present, an effective mass operator is computed in the Weak and Independent Scattering Approximation. The contribution of the cubic terms is computed to leading order in perturbation theory, allowing for a comparison of the scattering of a phonon by excited dislocations and three-phonon scattering, as well as studying the dependence of scattering amplitudes on the temperature of the solid. It is concluded that the effect of dislocations will dominate for relatively modest dislocation densities. Finally, the full power series of the interaction Hamiltonian is considered. The effects of quantum corrections, i.e., contributions proportional to Planck's constant, are estimated, and found to be controlled by another wavenumber-dependent parameter $k d_q$, where $k d_q$ is a length proportional to $\sqrt{\hbar}$. The possibility of using the results of this paper in the study of the phononic thermal properties of two- and three-dimensional materials is noted and discussed.

\paragraph{The scattering of phonons by infinitely long quantum dislocations segments and the generation of thermal transport anisotropy in a solid threaded by many parallel dislocations~\cite{nano10091711}} \hspace{\fill}

\textit{In collaboration with Fernando Lund.}

A canonical quantization procedure is applied to the interaction of elastic waves --phonons-- with infinitely long dislocations that can oscillate about an equilibrium, straight line, configuration. The interaction is implemented through the well-known Peach-Koehler force. For small dislocation excursions away from the equilibrium position, the quantum theory can be solved to all orders in the coupling constant. We study in detail the quantum excitations of the dislocation line, and its interactions with phonons. The consequences for the drag on a dislocation caused by the phonon wind are pointed out. We compute the cross-section for phonons incident on the dislocation lines for an arbitrary angle of incidence. The consequences for thermal transport are explored, and we compare our results, involving a dynamic dislocation, with those of Klemens and Carruthers, involving a static dislocation. In our case, the relaxation time is inversely proportional to frequency, rather than directly proportional to frequency. As a consequence, the thermal transport anisotropy generated on a material by the presence of a highly-oriented array of dislocations is considerably more sensitive to the frequency of each propagating mode, and therefore, to the temperature of the material.

%% file: quarkonium-v2.tex

\chapter{Dynamics of Quarkonia in Quark-Gluon Plasma} \label{ch:quarkonia-in-QGP}

Strongly coupled systems, such as superconductors, topological insulators, cold atoms in optical lattices and neutron stars, usually exhibit complex behavior.
Historically, studying them has led to many breakthroughs in our understanding of matter.
One particular example in high energy nuclear physics is quark-gluon plasma (QGP) created in relativistic heavy ion collisions (HICs). In these experiments, two heavy nuclei are accelerated to almost the speed of light and then collide. Shortly after the collision, a hot and dense droplet of QGP is created that only lasts for a tiny fraction of a second ($10^{-22}$ s)\@. The short lifetime of the QGP created in HICs  makes it very challenging to measure its properties directly, and so indirect probes have been primarily used.
The microscopic nature of QGP at different energy scales
is studied by combining experimental measurements, phenomenological studies and theoretical calculations at weak and strong coupling.

A useful probe of QGP involves quarkonium~\cite{Mocsy:2013syh,Chapon:2020heu}\@, a bound state of a heavy quark-antiquark ($Q\bar{Q}$) pair. Low-lying quarkonium species ($J/\psi, \psi(2S)$ and $\Upsilon(1S), \Upsilon(2S), \Upsilon(3S)$) have binding energies smaller than their inverse sizes, and are more deeply bound the smaller their size is (i.e., the mass of the state is smaller). Thus, different quarkonium species can probe QGP at multiple scales. For a long time, it was believed that the suppression of quarkonium production in HICs probes the Debye screening of (the real part of) the $Q\bar{Q}$ potential~\cite{Matsui:1986dk,Karsch:1987pv}\@. However, systematic studies using thermal field theory showed that in addition to the Debye screening, the in-medium $Q\bar{Q}$ potential also develops a thermal imaginary part\footnote{Whether or not the dissociation rate is the expectation value of the imaginary part of the potential depends on the definition of the potential, i.e., at which scale each relevant process happens.}~\cite{Laine:2006ns,Beraudo:2007ky}\@, which is a reflection of quarkonium dissociation. When the temperature of QGP is low enough that a particular $Q\bar{Q}$ bound state can exist, the inverse process of dissociation, i.e., regeneration, also occurs and plays a crucial role in charmonium production~\cite{Thews:2000rj,Andronic:2003zv,Andronic:2007bi}\@. Many phenomenological studies of quarkonium suppression have shown that the dynamical processes of dissociation and regeneration are, if not more important than, as important as the Debye screening~\cite{Krouppa:2015yoa,Du:2017qkv,Yao:2020xzw,Brambilla:2022ynh,Song:2023zma}\@.

\begin{figure}[t]
\centering
\includegraphics[width=0.75\textwidth]{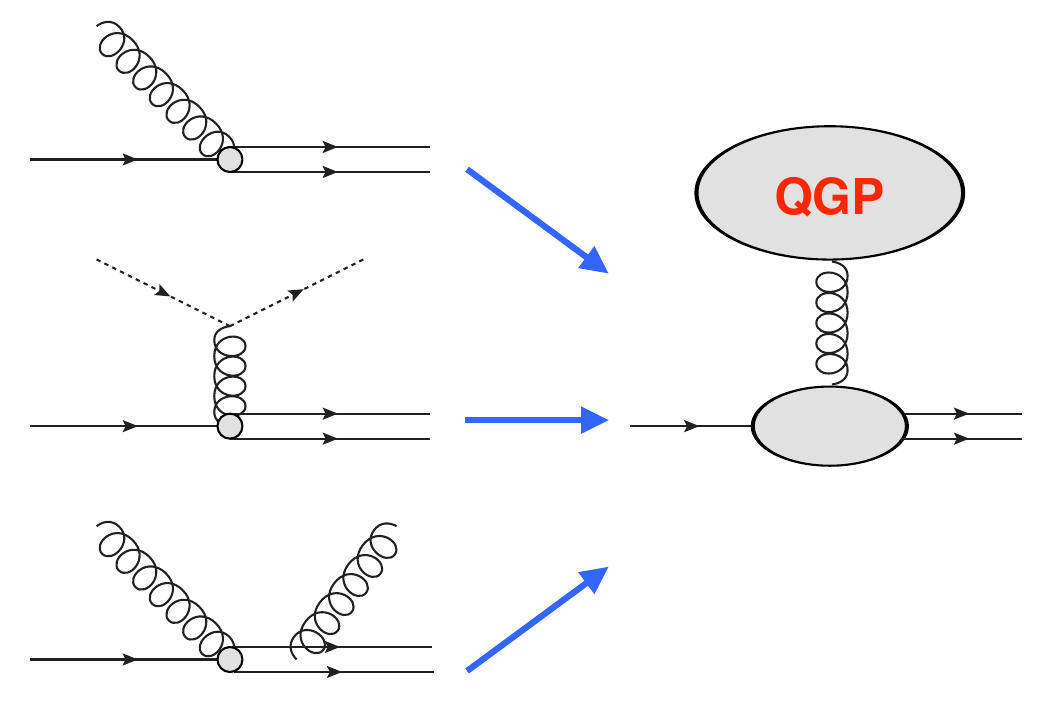}
\caption{A few perturbative Feynman diagrams for quarkonium dynamics (left) and its nonperturbative generalization (right) at leading order (dipole) in the multipole expansion. The single (double) solid line indicates a $Q\bar{Q}$ bound (unbound) state and the dotted line represents a light quark. The effective operator on the right is a chromoelectric field dressed with a timelike adjoint Wilson line. The physically measurable correlation functions of two of these effective operators are called Generalized Gluon Distributions (GGDs), as explained in the main text.}
\label{fig:f_g}
\end{figure}

The understanding of dynamical processes for quarkonium can be dated back to the early work by Peskin and Bhanot~\cite{Peskin:1979va,Bhanot:1979vb}, where they studied perturbatively the scattering process $g+(Q\bar{Q})_b \leftrightarrow Q+\bar{Q}$ (the subscript $b$ indicates a bound state), as shown in Fig.~\ref{fig:f_g}\@, in which the gluon is on shell. By convoluting the scattering amplitude squared with the Bose-Einstein distribution $n_B$ for the gluon, one can obtain the dissociation rate~\cite{Grandchamp:2001pf,Wong:2004zr,Rapp:2009my,Brambilla:2011sg} and the regeneration rate~\cite{Wong:2004zr,Yao:2017fuc} if QGP were a free gas of quarks and gluons. These studies have been generalized to the case of a weakly interacting gas in which the gluon mediating the $t$-channel $2\leftrightarrow3$ scattering processes ($q/g+(Q\bar{Q})_b \leftrightarrow q/g+Q+\bar{Q}$) is virtual \footnote{Other non-$t$-channel processes such as those shown in Fig.~\ref{fig:f_g} also contribute, which is a requirement of gauge invariance~\cite{Yao:2018sgn}. A complete set of diagrams at $g^4$ can be found in~\cite{Yao:2018sgn,Binder:2021otw}\@.}~\cite{Brambilla:2013dpa,Yao:2018sgn}\@. 
However, it is well known that at temperatures around $\Lambda_{\rm QCD}$, QGP is a strongly coupled fluid. This is the regime where most regeneration occurs and the binding energy cannot be neglected. 
Therefore, it is important to find the nonperturbative generalization of the Peskin-Bhanot and related higher order processes. 

With recent developments combining potential nonrelativistic QCD and open quantum systems~\cite{Akamatsu:2011se,Akamatsu:2014qsa,Katz:2015qja,Brambilla:2016wgg,Brambilla:2017zei,Blaizot:2017ypk,Kajimoto:2017rel,Blaizot:2018oev,Yao:2018nmy,Akamatsu:2018xim,Miura:2019ssi,Sharma:2019xum,Yao:2020eqy,Akamatsu:2021vsh,Brambilla:2021wkt,Miura:2022arv,Brambilla:2022ynh,Xie:2022tzs,alalawi2023impact} (see recent reviews~\cite{Rothkopf:2019ipj,Akamatsu:2020ypb,Sharma:2021vvu,Yao:2021lus}), a factorization formula was constructed for the dissociation and recombination of small-size quarkonium states~\cite{Yao:2020eqy}\@. At linear order in the multipole expansion, 
the dissociation and recombination rates are factorized into a nonrelativistic part that only involves the wavefunctions of the $Q\bar{Q}$ pair, which can be obtained from solving Schr\"odinger equations, and two Generalized Gluon Distributions (GGDs), which is the effective distribution of quasi-gluons from the medium that the $Q\bar{Q}$ pair absorbs or radiates. On the RHS of Fig.~\ref{fig:f_g} we show the effective operator whose correlation functions give rise to the GGDs, that we will rigorously define and calculate in what follows. As we will show later, provided knowledge of the quarkonium spectrum, its production mechanisms, and the thermal history of QGP in a heavy-ion collision, one can use quarkonium suppression data to measure (or at least constrain) the GGDs as nonperturbative properties of QGP, in the same way that parton distribution functions (PDFs) are measured in DIS experiments~\cite{Collins:1989gx}.  

In this chapter, we present the first complete calculation of the GGDs at NLO in weakly coupled QCD (Section~\ref{sect:nlo}), as well as the first nonperturbative study of it in $\mathcal{N}=4$ supersymmetric Yang-Mills theory using the AdS/CFT correspondence (Section~\ref{sec:strong-coupling}). We compare the two results and discuss what would be needed to calculate them using lattice (Euclidean) QCD methods (Section~\ref{sec:latticec}). Surprisingly, our findings suggest that a small-size $Q\bar{Q}$ pair weakly interacting with a strongly coupled plasma is an exception to the general expectation~\cite{Breuer:2002pc} that the dynamics of an open quantum system weakly coupled with a large thermal environment can be described by Markovian processes, and therefore, that the existing transport formalisms need to be generalized to include this regime. We give our outlook for studies of quarkonium dynamics in strongly coupled plasmas in Section~\ref{sec:pheno}.

\section{Effective field theory for quarkonia: pNRQCD}

Three energy scales are used to describe heavy quarkonium in vacuum: the heavy quark mass $M$, the inverse of quarkonium size $\frac{1}{r}$ and the binding energy $|E_b|$\@. Nonrelativistically, these three scales form a hierarchy $M\gg \frac{1}{r} \sim Mv \gg |E_b| \sim Mv^2$, where $v$ is the typical relative speed of the heavy quarks in the bound state. Depending on where the plasma temperature $T$ fits into the hierarchy, we may have different descriptions of quarkonium in-medium dynamics. The hierarchy $M\gg T$ is always true, since the highest temperature achieved in current heavy ion collision experiments is on the order of $500$\,MeV, which is smaller than the charm quark mass $M_c\approx1.3$\,GeV and much smaller than the bottom quark mass $M_b\approx4.2$\,GeV\@. 
Furthermore, because the inverse size $1/r$ of the low-lying states of quarkonium is always larger than their binding energy and the temperature of the QGP produced in HICs, the only question of phenomenological interest is how does $T$ compare with $E_b$, and we may always use $M\gg \frac{1}{r}\gg |E_b|, T$\@.

In the aforementioned hierarchy, the interaction can be described by \emph{potential nonrelativistic QCD} (pNRQCD) \cite{Pineda:1997bj,Brambilla:1999xf,Brambilla:2004jw}, which is an effective field theory for nonrelativistic two-body states. If the scale $\frac{1}{r}\sim Mv$ is perturbative, pNRQCD can be constructed perturbatively, under systematic nonrelativistic and multipole expansions, which are expansions in terms of $v$ and $r T$ respectively. The former allows the construction of the bound $Q\bar{Q}$ states as nonrelativistic objects, and the latter to approximate these states as pointlike objects from the point of view of a QGP environment with temperature $T$. At leading order in the nonrelativistic expansion and linear order in the multipole expansion, the Lagrangian density for the \emph{subsystem} consisting of a $Q\bar{Q}$ pair and its \emph{interaction} with the environment via the gauge field is given by:
\begin{align}
\ml{L}_\ma{pNREFT}^{(S,I)} =& \int \diff^3r\, \Tr\Big[  \ma{S}^{\dagger}(i\partial_0-H_{\rm s})\ma{S} + \mathrm{Adj}^{\dagger}( iD_0-H_\mathrm{adj} )\mathrm{Adj}  \nn\\
\label{eq:lagr}
&- V_A( \ma{Adj}^{\dagger}\bs r \cdot g{\bs E} \ma{S} + \ma{h.c.})  - \frac{V_B}{2}\ma{Adj}^{\dagger}\{ \bs r\cdot g\bs E, \ma{Adj}  \} +\cdots \Big]\,.
\end{align}
In this expression, curly brackets $\{\cdot, \cdot \}$ denote an anticommutator. The subsystem degrees of freedom include the singlet configuration of the $Q$-$\bar{Q}$ pair, $\ma{S}({\bs x}_{\rm cm}, \bs r, t)$, and the adjoint configuration $\ma{Adj}({\bs x}_{\rm cm}, \bs r, t)$,%
where the center-of-mass (c.m.) position of the heavy quark pair is denoted by ${\bs x}_{\rm cm}$  while $\bs r$ is the relative position (which is to be integrated over at the level of the Lagrangian density because, as seen from the point of view of the QGP environment, it is an internal degree of freedom of the $Q\bar{Q}$ subsystem). These operators contain the two-body bound and scattering states. The chromoelectric fields $\bm{E}$ and the gauge field in $D_0$ appearing in $\ml{L}_\ma{pNREFT}^{(S,I)}$ are all evaluated at the center-of-mass (c.m.) position ${\bs x}_{\rm cm}$\@.
At linear order in the multipole expansion, these states interact through the (non-Abelian) \emph{electric dipole operator}, $\bm{r} \cdot g \bm{E}$, shown in the second line of the Lagrangian density. Concretely, this operator describes bound state formation and dissociation.

The singlet and adjoint two-body fields are expressed as $N_c \times N_c$ matrices:
\begin{align}
  \mathrm{S} = \frac{{\bs 1}_{N_c}}{\sqrt{N_c}} S\,, 
  \quad \mathrm{Adj} = \frac{T^a}{\sqrt{T_F}} \mathrm{Adj}^a \,.
\end{align}
Here the identity matrix of size $N_c \times N_c$ is ${\bs 1}_{N_c}$,
an SU($N_c$) generator acting on the fundamental representation is $T^a$,
and its normalization is $\Tr (T^a T^b) = T_F \delta^{ab}$, where $T_F = 1/2$.
A summation over spin indices is implicit in the trace.
Under the ultrasoft\footnote{The gauge degrees of freedom are ``ultrasoft'' because those are the ones that couple to the composite singlet and octet fields, and due to their low energy scale cannot resolve the separation between the heavy quarks.} gauge transformation of $e^{ig \theta^a T^a}$, they transform as
\begin{align}
  S ({\bs x}_{\rm cm},\bm{r},t) \mapsto S ({\bs x}_{\rm cm}, \bm{r}, t), \quad
  \mathrm{Adj} ({\bs x}_{\rm cm},\bm{r},t) \mapsto e^{ig \theta^a ({\bs x}_{\rm cm},t) T^a} \mathrm{Adj} ({\bs x}_{\rm cm},\bm{r},t) e^{- ig \theta^a ({\bs x}_{\rm cm},t) T^a}\,.
\end{align}
Hence, the effective Lagrangian (\ref{eq:lagr}) is invariant under the ultrasoft gauge transformation.

The equations of motion of the free singlet and adjoint fields are Schr\"odinger equations with the Hamiltonians organized by powers of $\frac{1}{M}$ or equivalently, $v$:
\begin{align}
H_{\rm s} &= \frac{(i\bs \nabla_\ma{cm})^2}{4M} + \frac{(i\bs \nabla_\ma{rel})^2}{M} + V_{s}^{(0)}(r) + \frac{V_{s}^{(1)}(r)}{M} + \frac{V_{s}^{(2)}(r)}{M^2} + \cdots\\
\label{eqn:Ho}
H_{\rm adj} &= \frac{(i\bs D_\ma{cm})^2}{4M} + \frac{(i\bs \nabla_\ma{rel})^2}{M} + V_{\rm adj}^{(0)}(r) + \frac{V_{\rm adj}^{(1)}(r)}{M} + \frac{V_{\rm adj}^{(2)}(r)}{M^2} + \cdots\,.
\end{align}
At leading order in the nonrelativistic expansion, the Hamiltonians can be simplified as
\begin{align}
\label{eqn:hamiltonian}
H_{{\rm s},{\rm adj}}  =  \frac{(i\bs \nabla_\ma{rel})^2}{M} + V_{{\rm s},{\rm adj}}^{(0)}(r)\,.
\end{align}
Here $V_{{\rm s},{\rm adj}}^{(0)}$, $V_A$ and $V_B$ are Wilson coefficients. Perturbatively, at leading order in $\alpha_s(Mv)$ ($\alpha_s\equiv g^2/4\pi$), we have
\begin{align}
\label{eqn:match}
V_{\rm s}^{(0)}(r) = -C_F \frac{\alpha_s}{r}\,,\quad V_{\rm adj}^{(0)}(r) = - \left[ C_F - \frac{N_c}{2} \right]\frac{\alpha_s}{r}\,,\quad V_A=V_B=1\,,
\end{align}
where the quadratic Casimir $C_F$ of the fundamental representation is defined by $T^a T^a = C_F {\bs 1}_{N_c}$, and equals $C_F = T_F (N_c^2 - 1)/N_c$.

The main difference to the Abelian case, i.e., the U($1$) gauge theory that was recently studied in Ref.~\cite{Binder:2020efn}, is that the covariant derivatives $D_0$ and ${\bs D}_\ma{cm}$\footnote{
At leading order in the $v$ expansion, the c.m.\ covariant derivative term $\frac{{\bs D}_\ma{cm}^2}{4M}$ is suppressed in powers of $v$ for ultrasoft modes, compared with the $D_0$ term. However, for Coulomb modes that mediate the Coulomb interaction between the c.m.\ motion of the adjoint field and the gauge field, the $\frac{{\bs D}_\ma{cm}^2}{4M}$ term is at the same leading order in $v$ as the $D_0$ term.
The $\frac{{\bs D}_\ma{cm}^2}{4M}$ term does not affect dynamics at finite time in a non-singular gauge such as the $R_\xi$ gauge. However, it affects dynamics at infinite time and is crucially important for the construction of a gauge invariant electric field correlator in general~\cite{Yao:2020eqy}.
} introduce in the Lagrangian density (\ref{eq:lagr}) extra couplings between the subsystem of the two-body fields and the non-Abelian plasma. Complete accounting of these extra couplings will be crucial in showing the full gauge invariance of our NLO result later.
To take the $D_0$ term into account in an elegant way, we define a new (non-local) field $\widetilde{\mathrm{Adj}}(\bm{x}_{\text{cm}}, \bm{r}, t)$ through
\begin{align}
\label{eq:O_redef}
\mathrm{Adj}(\bm{x}_{\text{cm}}, \bm{r}, t) \to W_{[(\bm{x}_{\text{cm}}, t_0),(\bm{x}_{\text{cm}},t)]}
\widetilde{\mathrm{Adj}}(\bm{x}_{\text{cm}}, \bm{r}, t)
W_{[(\bm{x}_{\text{cm}},t),(\bm{x}_{\text{cm}}, t_0)]}  \,,
\end{align}
where $t_0$ is an arbitrary constant that cancels out in the end. The Wilson line $W$ for a representation ${\bs R}$ which connects $(\bm{x}_{\text{cm}}, t_f)$ and $(\bm{x}_{\text{cm}}, t_i)$ is defined by
\begin{align}
 W_{[(\bm{x}_{\text{cm}}, t_f),(\bm{x}_{\text{cm}}, t_i)]} = \ml{P}\exp \left[ ig\int^{t_f}_{t_i} \diff s A^a_0(\bm{x}_{\text{cm}}, s) T^a_{{\bs R}} \right] \,.
\end{align}
In the case at hand, the Wilson lines are in the adjoint representation, which we denote by $\W$ in the rest of this work. After this field redefinition, the $D_0$ covariant derivative term of the Lagrangian becomes canonical, and the EFT under consideration can be expressed as
\begin{align}
\ml{L}_\ma{pNREFT} \supset \int \diff^3r \Tr \Big[ & \ma{S}^{\dagger}(i\partial_0-H_{\rm s})\ma{S} +\widetilde{\ma{Adj}}^{\dagger}( i\partial_0 - H_\mathrm{adj} )\widetilde{\ma{Adj}} \nn\\
&- g ( \widetilde{\ma{Adj}}^{\dagger} r_i     \widetilde{E}_i  \ma{S} + \ma{S}^\dagger r_i  \widetilde{E}_i \widetilde{\ma{Adj}} ) 
-  \frac{g}{2}\widetilde{\ma{Adj}}^{\dagger}\{ r_i  \widetilde{E}_i, \widetilde{\ma{Adj}}\}
\Big] \,, \label{eq:action}
\end{align}
where
\begin{align}
\widetilde{E}_i(\bm{x}_{\text{cm}},t) = \W_{[(\bm{x}_{\text{cm}}, t_0),(\bm{x}_{\text{cm}},t)]}
{E}_i(\bm{x}_{\text{cm}},t)
\W_{[(\bm{x}_{\text{cm}}, t),(\bm{x}_{\text{cm}},t_0)]}\,.
\end{align}
This field redefinition neatly captures the fact that the time evolution operator for the adjoint configuration includes a Wilson line.

By identifying the subsystem Hamiltonian, $H_S$, as the free Hamiltonian of the singlet and adjoint states and the interaction Hamiltonian, $H_I$, as the electric dipole operator,
one can treat quarkonium as an open quantum system to study bound state formation and dissociation, while carefully taking into account the original $D_0$ (now stored in the Wilson lines $W$) and ${\bs D}_{\text{cm}}$ contributions. Special care has to be taken, especially in regard to the field redefinition, due to nontrivial operator orderings being relevant in the formulation of correlation functions in thermal field theory.
Such a derivation was explicitly done in Ref.~\cite{Yao:2020eqy}. We will explain the open quantum systems setup in the next section, derive explicit formulae for the quarkonium occupancies after a time $t$, and discuss different limits in which evolution equations that are local in time (i.e., Markovian) may be derived.

\section{Quarkonium as an open quantum system}
\label{sect:open}
Our starting point is an interacting quantum system consisting of a subsystem and a thermal environment, where the full Hamiltonian $H$ can be written as 
\begin{align}
H = H_S +H_E + H_I\,,
\end{align}
where $H_S$ is the \emph{subsystem} Hamiltonian (quarkonium), $H_E$ denotes the \emph{environment} Hamiltonian (quark-gluon plasma), and $H_I$ contains the \emph{interactions} between the subsystem and the environment. The time evolution of the density matrix of the full system is given by
\begin{align}
\frac{\diff \rho(t)}{\diff t} = -i[H,\rho(t)] \,.
\end{align}
In the interaction picture $\rho^{(\ma{int})}(t) = e^{i(H_S+H_E)t} \rho(t) e^{-i(H_S+H_E)t}$, the formal solution can be written as
\begin{align}
\rho^{(\ma{int})}(t) = U(t) \rho^{(\ma{int})}(0)  U^\dagger(t) \,,
\end{align}
where the time evolution is given by
\begin{align}
U(t) &= \ml{T}\exp\Big( -i\int_0^t \diff t' H_I^{(\ma{int})}(t')\Big)\\
H_I^{(\ma{int})}(t) &= e^{i(H_S+H_E)t} H_I(t) e^{-i(H_S+H_E)t} \,.
\end{align}
The time evolution of the subsystem can be written as
\begin{align}
\label{eqn:rho_S(t)}
\rho_S^{(\ma{int})}(t) = \Tr_E \big[ \rho^{(\ma{int})}(t)  \big] = \Tr_E \big[  U(t) \rho^{(\ma{int})}(0)  U^\dagger(t)  \big] \,.
\end{align}
When the subsystem and the environment are weakly interacting, we can assume the density matrix of the full system factorizes
\begin{align}
\rho(t) = \rho_S(t) \otimes \rho_E \,,
\end{align}
where the environment density matrix is set to be thermal $\rho_E = \frac{1}{Z} e^{-\beta H_E}$ and thus independent of time. Under the assumption of factorization, it is known that Eq.~(\ref{eqn:rho_S(t)}) can be written as a Lindblad equation in two limits: the quantum Brownian motion limit and the quantum optical limit. These two limits are specified by the hierarchies of time scales. Relevant time scales include the environment correlation time $\tau_E \sim T^{-1}$, the subsystem intrinsic time scale $\tau_S = 1/\Delta E \sim (Mv^2)^{-1}$ given by the inverse of the typical energy gap in the spectrum, and the subsystem relaxation time $\tau_R \sim T/(H_{\rm int})^2 \sim (Mv)^2/T^3$. The quantum Brownian motion limit is valid if $\tau_R \gg \tau_E$ and $\tau_S \gg \tau_E$, that is to say, if $Mv \gg T \gg Mv^2$, while the quantum optical limit is valid when $\tau_R \gg \tau_E$ and $\tau_R \gg \tau_S$, i.e., when $Mv \gg T$ and $M^3 v^4/T^3 \gg 1$ (in particular, the latter is well-justified if $T \lesssim Mv^2$). We discuss these in~\ref{sec:prev}. However, we find it useful to first establish how the dynamics of the quarkonium density matrix depend on the correlations of QGP, regardless of which limit is to be taken.

\subsection{Dynamics of the quarkonium density matrix} \label{sec:rho-dynamics}

Assuming an initial condition of the form $\rho(t) = \rho_{S}(0) \otimes \rho_E$,
the time evolution of the heavy quark pair density matrix, up to second order in perturbation theory, is given by
\begin{align}
\rho_{Q\bar{Q}}(t) &= {\rm Tr}_E\{U(t,t_0) [\rho_{Q\bar{Q}}(t_0) \otimes \rho_E] U(t_0,t) \} \nn\\
&= \rho_{Q\bar{Q}}(t_0) - \int_{t_0}^t dt_2 \int_{t_0}^{t_2} dt_1 {\rm Tr}_E\{[H_I(t_2) ,[ H_I(t_1) , \rho_{Q\bar{Q}}(t_0) \otimes \rho_E] ]\} \, , \label{eq:time-evol-rhoQQ}
\end{align}
where we take $H_I$ to be given by
\begin{equation}
    H_I(t) = S^\dagger(t) r_i \cdot E_i^a(t) {\rm Adj}^a(t) + {\rm Adj}^{a\dagger}(t) r_i \cdot E_i^a(t) S(t) \, ,
\end{equation}
that is to say, we omit adjoint-adjoint transitions. The reason why we do so is because, although they do contribute to the quantum dynamics of the density matrix, they do not contribute to the final quarkonium singlet abundances at leading order in perturbation theory on $T/(Mv)$, as they do not correspond to a recombination or dissociation process. (Rather, they give rise to corrections to the propagator of the adjoint state, suppressed in our power-counting by powers of $T/(Mv)$. We note, nonetheless, that this term is often included in the Quantum Brownian Motion limit analysis, e.g.,~\cite{Brambilla:2020qwo,Brambilla:2022ynh}.) 

To proceed, we assume an initial condition of the form
\begin{equation} \label{eq:rhoQQinit-cond}
    \rho_{Q\bar{Q}}(t_0) =  \rho_{\rm ss}(t_0) \ket{S} \bra{S} + \rho_{\rm oo}(t_0) \W^{ab}_{[t_0 - i\beta ,t_0]} \ket{{\rm Adj}^a} \bra{{\rm Adj}^b}  \, ,
\end{equation}
where $\ket{S} \bra{S}$ and $\ket{{\rm Adj}^a} \bra{{\rm Adj}^b}$ are operators that project the density matrix of the system onto a given color state, and all the rest of the degrees of freedom (position, spin) are encoded in $\rho_{ss}$ and $\rho_{oo}$. The adjoint Wilson line $\W^{ab}_{[t_0-i\beta,t_0]}$ (at the spatial position ${\bs x}_{\rm cm}$, which we omit throughout in what follows) along the imaginary time direction indicates that the color degrees of freedom of the heavy quark pair are assumed to be in local thermal equilibrium with the bath, or, conversely, that the QGP bath is in local thermal equilibrium in the presence of an adjoint color charge.

Strictly speaking, this initial condition does not correspond to a factorized state of the form $\rho = \rho_S \otimes \rho_E$, as there is an explicit dependence on the gauge field through the Wilson line in Eq.~\eqref{eq:rhoQQinit-cond}. However, the joint requirement of local thermal equilibrium plus gauge invariance require that the color state of the heavy quarks be of this form if they are in the octet (adjoint) configuration.

Unravelling the time evolution in Eq.~\eqref{eq:time-evol-rhoQQ}, we find
\begin{align}
    & \rho_{Q\bar{Q}}(t) - \rho_{Q\bar{Q}}(t_0) \\ 
    &=  - \frac{T_F}{N_c} \int_{t_0}^t dt_2 \int_{t_0}^{t_2} \! dt_1 {\rm Tr}_E \bigg\{ \nn \\
    & \quad + \ket{S} U^{({\rm s})}_{[t,t_2]} g E_i^a(t_2) r_i U^{({\rm o}),ab}_{[t_2,t_1]} g E_j^b(t_1) r_j U^{({\rm s})}_{[t_1,t_0]} \rho_E \otimes \rho_{\rm ss}(t_0) U^{({\rm s})}_{[t_0,t]} \bra{S}  \nn \\ 
    & \quad + \ket{{\rm Adj}^a} U^{({\rm o}),ab}_{[t,t_2]} g E_i^b(t_2) r_i U^{({\rm s})}_{[t_2,t_1]} g E_j^c(t_1) r_j U^{({\rm o}),cd}_{[t_1,t_0]} \rho_E \otimes \rho_{\rm oo}(t_0) W^{de}_{[t_0-i\beta,t_0]} U^{({\rm o}),ef}_{[t_0,t]} \bra{{\rm Adj}^f}  \nn \\ 
    & \quad - \ket{{\rm Adj}^a} U^{({\rm o}),ab}_{[t,t_2]} g E_i^b(t_2) r_i U^{({\rm s})}_{[t_2,t_0]} \rho_E \otimes \rho_{\rm ss}(t_0) U^{({\rm s})}_{[t_0,t_1]} g E_j^c(t_1) r_j U^{({\rm o}),cd}_{[t_1,t]} \bra{{\rm Adj}^d} \nn \\ 
    & \quad - \ket{S} U^{({\rm s})}_{[t,t_2]} g E_i^a(t_2) r_i U^{({\rm o}),ab}_{[t_2,t_0]} \rho_E \otimes \rho_{\rm oo}(t_0) W^{bc}_{[t_0-i\beta,t_0]} U^{({\rm o}),cd}_{[t_0,t_1]} g E_j^d(t_1) r_j U^{({\rm s})}_{[t_1,t]} \bra{S}  \nn \\ 
    & \quad  - \ket{S} U^{({\rm s})}_{[t,t_1]} g E_i^a(t_1) r_i U^{({\rm o}),ab}_{[t_1,t_0]} \rho_E \otimes \rho_{\rm oo}(t_0) W^{bc}_{[t_0-i\beta,t_0]} U^{({\rm o}),cd}_{[t_0,t_2]} g E_j^d(t_2) r_j U^{({\rm s})}_{[t_2,t]} \bra{S}  \nn \\ 
    & \quad - \ket{{\rm Adj}^a} U^{({\rm o}),ab}_{[t,t_1]} g E_i^b(t_1) r_i U^{({\rm s})}_{[t_1,t_0]} \rho_E \otimes \rho_{\rm ss}(t_0) U^{({\rm s})}_{[t_0,t_2]} g E_j^c(t_2) r_j U^{({\rm o}),cd}_{[t_2,t]} \bra{{\rm Adj}^d} \nn \\ 
    & \quad + \ket{{\rm Adj}^a} U^{({\rm o}),ab}_{[t,t_0]} \rho_E \otimes \rho_{\rm oo}(t_0) W^{bc}_{[t_0-i\beta,t_0]} U^{({\rm o}),cd}_{[t_0,t_1]} g E_i^d(t_1) r_i U^{({\rm s})}_{[t_1,t_2]} g E_j^e(t_2) r_j U^{({\rm o}),ef}_{[t_2,t]}   \bra{{\rm Adj}^f}  \nn \\ 
    & \quad + \ket{S} U^{({\rm s})}_{[t,t_0]} \rho_E \otimes \rho_{\rm ss}(t_0)  U^{({\rm s})}_{[t_0,t_1]} g E_i^a(t_1) r_i U^{({\rm o}),ab}_{[t_1,t_2]} g E_j^b(t_2) r_j U^{({\rm s})}_{[t_2,t]}  \bra{S} \bigg\} \nn \, ,
\end{align}
where we have denoted
\begin{equation}
    U^{({\rm o}),ab}_{[t,t']} = U^{({\rm o})}_{[t,t']} \W^{ab}_{[t,t']} \, ,
\end{equation}
and introduced the single-particle time evolution operators $U^{({\rm o})}, U^{({\rm s})}$ as the unitary operators that generate the time evolution of the singlet and adjoint configurations as if there were no transitions induced by the medium 
\begin{align}
    \partial_t U^{({\rm o})}_{[t,t']} &= - i H_{\rm adj} U^{({\rm o})}_{[t,t']} \, , \\
    \partial_t U^{({\rm s})}_{[t,t']} &= - i H_{\rm s} U^{({\rm s})}_{[t,t']} \, ,
\end{align}
with the initial condition $U^{({\rm o,s})}_{[t,t]} = \mathbbm{1}$.

Projecting onto singlet and octet components, we get an expression for the corresponding singlet and octet occupancies at time $t$. For the singlet configuration, we have
\begin{align}
    & \rho_{\rm ss}(t) - \rho_{\rm ss}(t_0) \label{eq:evolution-singlet} \\ 
    &=  - \frac{T_F}{N_c}  \int_{t_0}^t dt_2 \int_{t_0}^{t_2} \! dt_1  \bigg\{ \nn \\
    & \quad +  {\rm Tr}_E \left[ g E_i^a(t_2) \W^{ab}_{[t_2,t_1]}  g E_j^b(t_1) \rho_E \right] U^{({\rm s})}_{[t,t_2]} r_i U^{({\rm o})}_{[t_2,t_1]} r_j U^{({\rm s})}_{[t_1,t_0]} \rho_{\rm ss}(t_0) U^{({\rm s})}_{[t_0,t]}  \nn \\ 
    & \quad - {\rm Tr}_E \left[ g E_i^a(t_2) \W^{ab}_{[t_2,t_0]} \rho_E \W^{bc}_{[t_0-i\beta,t_0]} \W^{cd}_{[t_0,t_1]} g E_j^d(t_1) \right] U^{({\rm s})}_{[t,t_2]} r_i U^{({\rm o})}_{[t_2,t_0]} \rho_{\rm oo}(t_0)  U^{({\rm o})}_{[t_0,t_1]} r_j U^{({\rm s})}_{[t_1,t]}  \nn \\ 
    & \quad - {\rm Tr}_E \left[ g E_i^a(t_1) \W^{ab}_{[t_1,t_0]} \rho_E \W^{bc}_{[t_0-i\beta,t_0]} \W^{cd}_{[t_0,t_2]} g E_j^d(t_2) \right]  U^{({\rm s})}_{[t,t_1]} r_i U^{({\rm o})}_{[t_1,t_0]} \rho_{\rm oo}(t_0)  U^{({\rm o})}_{[t_0,t_2]}  r_j U^{({\rm s})}_{[t_2,t]}  \nn \\ 
    & \quad + {\rm Tr}_E \left[ g E_i^a(t_1) \W^{ab}_{[t_1,t_2]} g E_j^b(t_2) \rho_E \right] U^{({\rm s})}_{[t,t_0]} \rho_{\rm ss}(t_0)  U^{({\rm s})}_{[t_0,t_1]}  r_i U^{({\rm o})}_{[t_1,t_2]}  r_j U^{({\rm s})}_{[t_2,t]}  \bigg\} \nn \, ,
\end{align}
and for the octet configuration we sum over final states as $\bra{{\rm Adj}^a} \cdot \ket{{\rm Adj}^a}$, obtaining
\begin{align}
    & \rho_{\rm oo}(t) - \rho_{\rm oo}(t_0) \label{eq:evolution-octet} \\ 
    &=  - \frac{T_F}{N_c}  \frac{1}{ \W^{aa}_{[t_0- i\beta, t_0]} } \int_{t_0}^t dt_2 \int_{t_0}^{t_2} \! dt_1 \bigg\{ \nn \\
    & \quad + {\rm Tr}_E \left[ g E_j^c(t_1) \W^{cd}_{[t_1,t_0]} \rho_E \W^{de}_{[t_0-i\beta,t_0]} \W^{ef}_{[t_0,t_2]} g E_i^f(t_2) \right] U^{({\rm o})}_{[t,t_2]} r_i U^{({\rm s})}_{[t_2,t_1]} r_j U^{({\rm o})}_{[t_1,t_0]} \rho_{\rm oo}(t_0) U^{({\rm o})}_{[t_0,t]} \nn \\ 
    & \quad - {\rm Tr}_E \left[ g E_j^a(t_1) \W^{ab}_{[t_1,t_2]} g E_i^b(t_2) \rho_E \right]
    U^{({\rm o})}_{[t,t_2]} r_i U^{({\rm s})}_{[t_2,t_0]} \rho_{\rm ss}(t_0) U^{({\rm s})}_{[t_0,t_1]} r_j U^{({\rm o})}_{[t_1,t]} \nn \\
    & \quad - {\rm Tr}_E \left[ g E_i^a(t_2) \W^{ab}_{[t_2,t_1]}  g E_j^b(t_1) \rho_E \right] U^{({\rm o})}_{[t,t_1]} r_i U^{({\rm s})}_{[t_1,t_0]} \rho_{\rm ss}(t_0) U^{({\rm s})}_{[t_0,t_2]} r_j U^{({\rm o})}_{[t_2,t]} \nn \\ 
    & \quad + {\rm Tr}_E \left[ g E_j^e(t_2) \W^{eb}_{[t_2,t_0]} \rho_E \W^{bc}_{[t_0-i\beta,t_0]} \W^{cd}_{[t_0,t_1]} g E_i^d(t_1) \right] U^{({\rm o})}_{[t,t_0]} \rho_{\rm oo}(t_0) U^{({\rm o})}_{[t_0,t_1]} r_i U^{({\rm s})}_{[t_1,t_2]} r_j U^{({\rm o})}_{[t_2,t]} \bigg\} \nn \, ,
\end{align}
where we have defined $\rho_{\rm oo}(t) = \bra{{\rm Adj}^a} \rho_{Q\bar{Q}}(t) \ket{{\rm Adj}^a} / \W^{aa}_{[t_0 - i\beta, t_0]}$.

These are self-contained expressions that factorize the time evolution of the subsystem (the heavy quark pair described by an interaction potential) with the time evolution of the QGP medium. Provided we are given $H_{\rm s}$ and $H_{\rm adj}$, all we need to specify to calculate the quarkonium abundances after a time $t$ are the chromoelectric correlators defined by the traces over the environment degrees of freedom in these last two equations.

The correlators we have just introduced are analogous to \textit{Wightman} correlators in thermal field theory~\cite{Laine:2016hma}, as they have their operators ordered explicitly as written. Furthermore, one can show that the correlators do not depend on the time $t_0$ at which the system is initialized, and as such, it will be convenient to send $t_0 \to -\infty$ in practical calculations.

\subsection{Chromoelectric field correlators and Generalized Gluon Distributions} \label{sec:GGDs}

In the preceding discussion, the physical (Wightman) correlation functions that govern quarkonium transport made their first appearance. These are nonperturbative properties of QGP that can be measured or constrained from HIC data, provided knowledge of the quarkonium spectrum, its production mechanisms, and the thermal history of QGP in a heavy-ion collision. We call these \textit{Generalized Gluon Distributions} (GGDs), and we define them as
\begin{align}
[g_{\rm adj}^{++}]^>(t) &= \frac{1}{Z} \frac{g^2 T_F}{3N_c} {\rm Tr}_{\mathcal{H}} \left[ E_i^a(t) \W^{ac}_{[t,+\infty]} \W^{cb}_{[+\infty,0]} E_i^b(0) e^{-\beta H} \right] \label{eq:gE++>} \\
[g_{\rm adj}^{++}]^<(t) &= \frac{1}{Z} \frac{g^2 T_F}{3N_c} {\rm Tr}_{\mathcal{H}} \left[ \W^{cb}_{[+\infty,0]} E_i^b(0) E_i^a(t) \W^{ad}_{[t,+\infty]}  e^{-\beta H} \W^{dc}_{[+\infty -i\beta,+\infty]} \right] \label{eq:gE++<} \\
[g_{\rm adj}^{--}]^>(t) &= \frac{1}{Z} \frac{g^2 T_F}{3N_c} {\rm Tr}_{\mathcal{H}} \left[ \W^{cb}_{[-\infty,t]} E_i^b(t) E_i^a(0) \W^{ad}_{[0,-\infty]}  e^{-\beta H} \W^{dc}_{[-\infty -i\beta,-\infty]} \right] \label{eq:gE-->} \\
[g_{\rm adj}^{--}]^<(t) &= \frac{1}{Z} \frac{g^2 T_F}{3N_c} {\rm Tr}_{\mathcal{H}} \left[ E_i^a(0) \W^{ac}_{[0,-\infty]} \W^{cb}_{[-\infty,t]} E_i^b(t) e^{-\beta H} \right] \, , \label{eq:gE--<}
\end{align}
where $H$ is the environment (QGP) Hamiltonian, ${\rm Tr}_{\mathcal{H}}$ denotes a trace over states in the Hilbert space of the theory, and $Z = {\rm Tr}_{\mathcal{H}} \left[ e^{-\beta H} \right]$ is the partition function of the QGP environment\@. We have written the thermal averages explicitly, with the (Euclidean) adjoint Wilson line at ${\rm Re}\{t\} = -\infty$ in the definition of $[g_{E}^{--}]^>(t)$ accounting for the fact that in the corresponding physical situation, the QGP environment and the color degrees of freedom of the point adjoint color charge have thermalized together. To make explicit that $[g_{E}^{++}]^<(t)$ is the time-reversed version of it, we have set the imaginary time Wilson line at $t = +\infty$, even though, as we mentioned before, this is arbitrary and a choice we can make to simplify calculations. Furthermore, we have contracted the spatial indices of the electric fields, as the thermal state will be, by definition, isotropic and homogeneous, thus guaranteeing that even if we kept the indices, we would have $[g_{\rm adj}^{\pm \pm'}]^{\lessgtr}_{ij}(\omega) = \delta_{ij} [g_{\rm adj}^{\pm \pm'}]^{\lessgtr}(\omega)$ in terms of the definitions we just introduced.

In frequency space, we define the GGDs as
\begin{align}
    [g_{\rm adj}^{\pm \pm'}]^{\lessgtr}(\omega) = \int_{-\infty}^\infty \!\! dt \, e^{i \omega t} [g_{\rm adj}^{\pm \pm'}]^{\lessgtr}(t) \, ,
\end{align}
in terms of which the KMS relations are given by
\begin{align}
[g_{\rm adj}^{++}]^>(\omega) &= e^{\omega/T} [g_{\rm adj}^{++}]^<(\omega) \\
[g_{\rm adj}^{--}]^>(\omega) &= e^{\omega/T} [g_{{\rm adj}}^{--}]^<(\omega) \, ,
\end{align}
and are related by the action of time-reversal
\begin{equation}
\label{eqn:kms}
[g_{{\rm adj}}^{++}]^>(\omega) = [g_{\rm adj}^{--}]^<(-\omega) \,,
\end{equation}
where we have assumed the QGP state is symmetric under these discrete symmetries (as it is for a thermal state). A short proof of the KMS relations and the time reversal property is given in Appendix~\ref{app:kms}\@. From this proof it is also clear that the normalization factor $Z$ of the KMS conjugates $[g_E^{\pm \pm'}]^{\lessgtr}(\omega)$ must be the same. Perturbative calculations~\cite{Binder:2021otw}, to be discussed in Section~\ref{sect:nlo}, verify that the QGP partition function adequately serves this purpose.

One may then define spectral functions that respect the KMS relations:
\begin{align}
\rho_{{\rm adj}}^{++}(\omega) & = [g_{{\rm adj}}^{++}]^>(\omega) - [g_{{\rm adj}}^{++}]^<(\omega) \label{eq:rho-++-def} \\
\rho_{{\rm adj}}^{--}(\omega) & = [g_{{\rm adj}}^{--}]^>(\omega) - [g_{{\rm adj}}^{--}]^<(\omega) \, .
\end{align}
Contrary to typical spectral functions, the ones we have just introduced are not guaranteed to have a definite parity under $\omega \to -\omega$. Rather, because of parity and time-reversal, they are related to each other via:
\begin{align}
    \rho_{{\rm adj}}^{++}(\omega) = - \rho_{{\rm adj}}^{--}(-\omega) \, ,
\end{align}
so it is still true that complete knowledge of one spectral function fully determines all correlation functions introduced above. Conversely, complete knowledge of one of the physical (Wightman) correlation functions also fully determines all the rest of the correlation functions.

One may also introduce a time-ordered version of the correlation function:
\begin{align} \label{eq:g-T-ordered}
    [g_{\rm adj}^{{\mathcal{T}}}](t) = \frac{g^2 T_F}{3 N_c} \langle \hat{\mathcal{T}} E_i^a(t) \mathcal{W}^{ab}_{[t,0]} E_i^b(0) \rangle_T \, ,
\end{align}
which can be written in terms of the GGD $[g_{\rm adj}^{++}]^>$ as
\begin{equation} \label{eqn:g>_to_gT}
    [g_{\rm adj}^{{\mathcal{T}}}](t) = \theta(t) [g_{{\rm adj}}^{++}]^>(t) + \theta(-t) [g_{{\rm adj}}^{++}]^>(-t) \, . 
\end{equation}
Furthermore, with $[g_{\rm adj}^{\mathcal{T}}](t)$ in hand we also have access to the anti-time-ordered correlator through complex conjugation (denoted by a star ${}^*$)
\begin{align}
    [g_{\rm adj}^{{\overline{\mathcal{T}}}}](t) = \frac{g^2 T_F}{3 N_c} \langle \hat{\overline{\mathcal{T}}} E_i^a(t) \mathcal{W}^{ab}_{[t,0]} E_i^b(0) \rangle_T = \left\{ [g_{\rm adj}^{{\mathcal{T}}}](t) \right\}^* \, .
\end{align}
Then, we can also express the GGD $[g_{E}^{++}]^>(t)$ in terms of its (anti)time-ordered counterparts:
\begin{align}
\label{eq:>fromT}
    [g_{{\rm adj}}^{++}]^>(t) = \theta(t) [g_{\rm adj}^{{\mathcal{T}}}](t) + \theta(-t) [g_{\rm adj}^{{\overline{\mathcal{T}}}}](t) \, .
\end{align}
This means that once we obtain the time-ordered correlator, we can evaluate all the other correlation functions, in particular the physical (Wightman) correlators that enter the Boltzmann and rate equations.

We can use all of the above to write Eq.~\eqref{eq:>fromT} in frequency space, thus explicitly obtaining the GGD $[g_{E}^{++}]^>(\omega)$ in terms of the time-ordered correlator:
\begin{equation}
    [g_{{\rm adj}}^{++}]^>(\omega) = {\rm Re}\!\left\{ [g_{\rm adj}^{{\mathcal{T}}}](\omega) \right\} + \frac{1}{\pi} \int_{-\infty}^\infty d p_0 \, \mathcal{P} \left( \frac{1}{p_0} \right) {\rm Im}\! \left\{ [g_{\rm adj}^{{\mathcal{T}}}](\omega + p_0) \right\} \, ,
\end{equation}
where $\ml{P}$ denotes the Cauchy principal value.
The inverse map that determines the time-ordered correlator in terms of the GGD $[g_{E}^{++}]^>(\omega)$ is given by\footnote{To prove that they are the inverse of each other, one needs to use a particular representation of the Dirac delta:
\begin{equation}
    \int_{-\infty}^\infty dx \, \mathcal{P} \left( \frac{1}{(x-1) (x^2 - a^2)} \right) = \frac{\pi^2}{2} \delta(|a| - 1) \, ,
\end{equation} which may be verified by direct action of this distribution on functions whose arguments are the parameter $a$.}
\begin{equation}
    [g_{\rm adj}^{{\mathcal{T}}}](\omega) = \frac12 \left( [g_{{\rm adj}}^{++}]^>(\omega) + [g_{{\rm adj}}^{++}]^>(-\omega) \right) + \frac{1}{2\pi i} \int_{-\infty}^\infty dp_0 \, \mathcal{P} \left( \frac{2 p_0}{p_0^2 - \omega^2} \right) [g_{{\rm adj}}^{++}]^>(p_0) \, .
\end{equation}

All of the above allows one to do the calculation of the correlators or GGDs in whichever form is most convenient, depending on the tools that are available in the regime of interest. For example, the tools that are most effective to do a calculation at weak coupling are different than the ones that are most effective at strong coupling (concretely, Section~\ref{sect:nlo} discusses the weakly coupled limit, for which we calculate the spectral function, and Section~\ref{sec:strong-coupling} discusses a strongly coupled calculation of this family of correlators starting from the time-ordered correlator).

Once the GGDs have been calculated, one can go back to the evolution equations~\eqref{eq:evolution-singlet} and~\eqref{eq:evolution-octet} and write, for the singlet configuration,
\begin{align}
    \rho_{\rm ss}(t) - \rho_{\rm ss}(t_0) 
    &=  - \int_{t_0}^t dt_2 \int_{t_0}^{t_2} \! dt_1  \bigg\{ \nn \\
    & \quad + [g_{\rm adj}^{++}]^>(t_2-t_1) U^{({\rm s})}_{[t,t_2]} r_i U^{({\rm o})}_{[t_2,t_1]} r_i U^{({\rm s})}_{[t_1,t_0]} \rho_{\rm ss}(t_0) U^{({\rm s})}_{[t_0,t]}  \nn \\ 
    & \quad - [g_{\rm adj}^{--}]^>(t_1-t_2) U^{({\rm s})}_{[t,t_2]} r_i U^{({\rm o})}_{[t_2,t_0]} \rho_{\rm oo}(t_0)  U^{({\rm o})}_{[t_0,t_1]} r_i U^{({\rm s})}_{[t_1,t]}  \nn \\ 
    & \quad - [g_{\rm adj}^{--}]^>(t_2-t_1) U^{({\rm s})}_{[t,t_1]} r_i U^{({\rm o})}_{[t_1,t_0]} \rho_{\rm oo}(t_0)  U^{({\rm o})}_{[t_0,t_2]}  r_i U^{({\rm s})}_{[t_2,t]}  \nn \\ 
    & \quad + [g_{\rm adj}^{++}]^>(t_1-t_2) U^{({\rm s})}_{[t,t_0]} \rho_{\rm ss}(t_0)  U^{({\rm s})}_{[t_0,t_1]}  r_i U^{({\rm o})}_{[t_1,t_2]}  r_i U^{({\rm s})}_{[t_2,t]}  \bigg\} \label{eq:singlet-from-GGD} \, ,
\end{align}
and for the octet configuration,
\begin{align}
    \rho_{\rm oo}(t) - \rho_{\rm oo}(t_0)  
    &=  - \frac{1}{ \W^{aa}_{[t_0- i\beta, t_0]} } \int_{t_0}^t dt_2 \int_{t_0}^{t_2} \! dt_1 \bigg\{ \nn \\
    & \quad + [g_{\rm adj}^{--}]^>(t_2-t_1) U^{({\rm o})}_{[t,t_2]} r_i U^{({\rm s})}_{[t_2,t_1]} r_i U^{({\rm o})}_{[t_1,t_0]} \rho_{\rm oo}(t_0) U^{({\rm o})}_{[t_0,t]} \nn \\ 
    & \quad - [g_{\rm adj}^{++}]^>(t_1-t_2)
    U^{({\rm o})}_{[t,t_2]} r_i U^{({\rm s})}_{[t_2,t_0]} \rho_{\rm ss}(t_0) U^{({\rm s})}_{[t_0,t_1]} r_i U^{({\rm o})}_{[t_1,t]} \nn \\
    & \quad - [g_{\rm adj}^{++}]^>(t_2-t_1) U^{({\rm o})}_{[t,t_1]} r_i U^{({\rm s})}_{[t_1,t_0]} \rho_{\rm ss}(t_0) U^{({\rm s})}_{[t_0,t_2]} r_i U^{({\rm o})}_{[t_2,t]} \nn \\ 
    & \quad + [g_{\rm adj}^{--}]^>(t_1-t_2) U^{({\rm o})}_{[t,t_0]} \rho_{\rm oo}(t_0) U^{({\rm o})}_{[t_0,t_1]} r_i U^{({\rm s})}_{[t_1,t_2]} r_i U^{({\rm o})}_{[t_2,t]} \bigg\} \label{eq:octet-from-GGD} \, ,
\end{align}
which are now explicit expressions in terms of the GGDs. As such, given the spectrum and wavefunctions of quarkonia and a QGP temperature, these equations predict the final abundances of singlet and octet states propagating through QGP if one knows the GGDs. Conversely, one can constrain the GGDs by using quarkonium suppression data in HICs.\footnote{This is possible because the temperature in a HIC is a slowly varying quantity relative to all of the energy scales in the problem. Otherwise, it would not be possible to use the GGDs calculated in thermal equilibrium.}

From this starting point, one can derive simpler, more compact descriptions of quarkonium propagation in the so-called Quantum Brownian Motion limit and in the Quantum Optical limit, which coincide with what other authors have previously derived making use of extra approximations. We describe these previously developed transport formalisms in what follows.

\subsection{Previously developed transport formalisms} \label{sec:prev}

When the plasma temperature is very high, which is the case in the early stage of heavy ion collisions, the interaction between a $Q\bar{Q}$ pair can be significantly screened if the separation between the pair is much bigger than the inverse temperature, i.e., if $r T \gg 1$. As a result, the in-medium dynamics of a $Q\bar{Q}$ pair can be described in terms of two independent heavy quarks that diffuse and dissipate in the plasma. This dynamics can be approximately described by a Langevin equation with drag and diffusion and the heavy quark diffusion coefficient has been calculated nonperturbatively via lattice methods~\cite{Banerjee:2011ra,Francis:2015daa,Brambilla:2020siz,Altenkort:2020fgs}\@. One can improve the Langevin equation by adding an attractive potential effect that is screened when the $Q\bar{Q}$ pair is far away in space. This potential effect can only last for a time scale given by the imaginary potential obtained from lattice studies~\cite{Rothkopf:2011db,Larsen:2019zqv,Larsen:2019bwy,Bala:2021fkm,Petreczky:2021zmz}\@.

After the formation of the QGP in heavy ion collisions, the plasma expands quickly and cools down. When the temperature drops to the region where $T\sim\frac{1}{r}$, the interaction between a $Q\bar{Q}$ pair can no longer be neglected and must be included in the Langevin description. When the temperature further drops, $M\gg  \frac{1}{r} \gg T \gg |E_b|$, a different description comes into play\footnote{Albeit for a short period of time, because in reality there is not much space between the scales $Mv$ and $Mv^2$ to add an additional strong inequality between them.}, in which the color correlation between the $Q\bar{Q}$ pair becomes important. The in-medium dynamics of a $Q\bar{Q}$ pair can be described by a Lindblad equation in the quantum Brownian motion limit~\cite{Brambilla:2016wgg,Brambilla:2017zei}
\begin{align}
\frac{d\rho_S(t)}{d t} &= -i\big[ H_S + \Delta H_S,\, \rho_S(t) \big] + \kappa_{\rm adj} \Big( L_{\alpha i} \rho_S(t) L^\dagger_{\alpha i} - \frac{1}{2}\big\{  L^\dagger_{\alpha i}L_{\alpha i},\, \rho_S(t)\big\} \Big)\,,
\end{align}
where the Hamiltonian is given by
\begin{align}
H_S = \frac{{\bs p}_{\text{rel}}^2}{M} + \begin{pmatrix} 
- \frac{C_F\alpha_s}{r} & 0 \\
 0 & \frac{\alpha_s}{2N_cr}
\end{pmatrix} \,, \qquad\quad \Delta H_S = \frac{\gamma_{\rm adj}}{2} r^2 \begin{pmatrix}
1 & 0\\
0 & \frac{N_c^2-2}{2(N_c^2-1)}
\end{pmatrix} \,,
\end{align}
and the density matrix is assumed to be block diagonal in the color singlet and octet basis
\begin{align}
\rho_S(t) = \begin{pmatrix}
\rho_S^{(s)}(t) & 0 \\
0 & \rho_S^{(o)}(t)
\end{pmatrix} \,.
\end{align}
The Lindblad operators are given by
\begin{align}
L_{1i} &= \Big(r_i + \frac{1}{2MT}\nabla_i - \frac{N_c}{8T}\frac{\alpha_s r_i}{r} \Big)\begin{pmatrix}
0 & 0\\
1 & 0
\end{pmatrix}  \\
L_{2i} &= \sqrt{\frac{1}{N_c^2-1}}\Big(r_i + \frac{1}{2MT}\nabla_i + \frac{N_c}{8T}\frac{\alpha_s r_i}{r} \Big)\begin{pmatrix}
0 & 1\\
0 & 0
\end{pmatrix}  \\
L_{3i} &= \sqrt{\frac{N_c^2-4}{2(N_c^2-1)}}
\Big(r_i + \frac{1}{2MT}\nabla_i \Big)\begin{pmatrix}
0 & 0\\
0 & 1
\end{pmatrix} \,,
\end{align}
where $i=x,y,z$.
The transport coefficients $\kappa_{\rm adj}$ and $\gamma_{\rm adj}$ are defined in terms of the chromoelectric field correlators~\cite{Eller:2019spw} as
\begin{align}
\label{eqn:kappa_gamma_adj}
\kappa_{\rm adj} &= \frac{g^2 T_F }{3 N_c} {\rm Re} \int d t\, \big\langle \hat{\ml{T}} E^a_i(t) \ml{W}^{ab}_{[t,0]} E^b_i(0) \big\rangle_T \\
\gamma_{\rm adj} &= \frac{g^2 T_F }{3 N_c} {\rm Im} \int d t\, \big\langle \hat{\ml{T}} E^a_i(t) \ml{W}^{ab}_{[t,0]} E^b_i(0) \big\rangle_T \,,
\end{align}
where $T_F$ is defined by ${\rm Tr}(T^a_FT^b_F)=T_F\delta^{ab}$ with $T_F^a$ being the generator in the fundamental representation and $\ml{W}^{ab}(t,0)$ denotes a time-like Wilson line in the adjoint representation from time $0$ to $t$:
\be
\ml{W}_{[x,y]} = {P} \exp \left( ig \int_y^x \!\! d z^\mu A_\mu^a(z) T^a_A \right) \,,
\ee
in which $x$ and $y$ are Minkowski position 4-vectors connected by a straight path. The expectation value of an operator $O$ is defined as $\langle O \rangle_T \equiv \tr(O e^{-\beta H_E})/\tr(e^{-\beta H_E})$ where $\beta = 1/T$ is the inverse of the plasma temperature and $H_E$ denotes the Hamiltonian of light quarks and gluons in the QGP\@. As can be seen from here, in this high temperature limit where the quantum Brownian motion limit of the open quantum system framework is valid, it is the zero frequency limit of the time-ordered correlator that which is relevant for quarkonium in-medium dynamics. As we showed in~\eqref{eq:>fromT}, this correlator can be written in terms of the GGDs,\footnote{Writing the transport coefficients $\kappa_{\rm adj}$ and $\gamma_{\rm adj}$ in terms of the correct electric field correlator was first done rigorously by the authors of~\cite{Eller:2019spw}, where they emphasized the importance of the Wilson line configuration in the correlator to define the correct physical observable.} but here we have chosen to preserve the form in which they first appeared in the literature.

As the QGP temperature continues dropping and finally becomes of the same order as the binding energy, $M\gg  \frac{1}{r} \gg T \sim |E_b|$, we need another description that is based on a classical Boltzmann equation which can be derived by using the open quantum system framework in the quantum optical limit, pNRQCD, the Wigner transform and semiclassical gradient expansion~\cite{Yao:2020eqy} (a subtlety of using the quantum optical limit can be resolved by working in the semiclassical limit, as explained in Ref.~\cite{Yao:2021lus})\@. If we further integrate over the momentum distribution of the phase space distribution, we will arrive at a rate equation for the density of a quarkonium state $n_b$ with the quantum number $b$,
\begin{align}
\label{eqn:rate}
\frac{d n_b(t,{\bs x})}{d t} = -\Gamma\, n_b(t,{\bs x}) + F(t,{\bs x}) \,,
\end{align}
where $\Gamma$ is the dissociation rate and $F$ denotes the contribution of quarkonium formation (in-medium recombination). They are given by
\begin{align}
\label{eqn:disso}
\Gamma &=  \int \frac{d^3p_{\ma{rel}}}{(2\pi)^3} 
| \langle \psi_b | {\bs r} | \Psi_{{\bs p}_\ma{rel}} \rangle |^2 [g^{++}_{\rm adj}]^{>}\Big(-|E_b| - \frac{p^2_\ma{rel}}{M}\Big) \\
\label{eqn:reco}
F(t,{\bs x}) &=  \int \frac{d^3p_{\ma{cm}}}{(2\pi)^3}  \frac{d^3p_{\ma{rel}}}{(2\pi)^3} 
| \langle \psi_b | {\bs r} | \Psi_{{\bs p}_\ma{rel}} \rangle |^2 
[g^{--}_{\rm adj}]^{>}\Big(\frac{p^2_\ma{rel}}{M}+|E_{b}|\Big) f_{Q\bar{Q}}(t, {\bs x}, {\bs p}_{\ma{cm}}, {\bs x}_{\rm rel}=0, {\bs p}_{\ma{rel}}) \,, 
\end{align}
where $\langle \psi_b | {\bs r} | \Psi_{{\bs p}_\ma{rel}} \rangle$ is the dipole transition amplitude between a bound quarkonium state wavefunction $\psi_b$ and an unbound $Q\bar{Q}$ wavefunction $\Psi_{{\bs p}_\ma{rel}}$ that is a scattering wave with momentum ${\bs p}_{\rm rel}$\@. The two-particle phase space distribution $f_{Q\bar{Q}}(t, {\bs x}, {\bs p}_{\ma{cm}}, {\bs x}_{\rm rel}=0, {\bs p}_{\ma{rel}})$ is for an unbound $Q\bar{Q}$ pair with the center-of-mass (cm) position ${\bs x}_{\rm cm}={\bs x}$, cm momentum ${\bs p}_{\ma{cm}}$, relative position ${\bs x}_{\rm rel}=0$ and relative momentum ${\bs p}_{\ma{rel}}$\@. The relative position is fixed to be $0$ which is a result of a gradient expansion used in taking the semiclassical limit. The $Q\bar{Q}$ phase space distribution does not factorize into the product of two single particle distributions
\be
f_{Q\bar{Q}}(t, {\bs x}, {\bs p}_{\ma{cm}}, {\bs x}_{\rm rel}=0, {\bs p}_{\ma{rel}} ) \neq f_Q(t, {\bs x}, {\bs p}_Q) f_{\bar{Q}}(t, {\bs x}, {\bs p}_{\bar{Q}})\,,
\ee
which means that the formation term $F$ in the rate equation can account for both correlated and uncorrelated recombination~\cite{Yao:2020xzw}\@. The Boltzmann and rate equations have been extensively used in phenomenological studies of quarkonium and exotics production in heavy ion collisions~\cite{Du:2017qkv,Yao:2017fuc,Yao:2018zze,Zhao:2021voa,Du:2022uvj,Wu:2023djn}\@.

\section{Chromoelectric correlator at weak coupling in QCD}

\label{sect:nlo}

In this section, we discuss the NLO calculation of the non-Abelian electric field correlator
\begin{align}\label{eq:elcorrelatorcalc}
[g_E^{++}]^{>}_{ji}(y,x) \equiv \Big\langle {E}_j(y) \ml{W}_{[( y^0, {\bs x}), (+\infty, {\bs x})]} \ml{W}_{[(+\infty, {\bs x}),(x^0, {\bs x})]} {E}_i(x) \Big\rangle_T \,.
\end{align}
We distinguish the $[g_E]$ family of correlators from $[g_{\rm adj}]$ by the lack of the normalization factor $g^2 T_F/(3N_c)$ present in the latter, and keep the original notation introduced in~\cite{Binder:2021otw} throughout this Section. 

We define $\Tr_E(\rho_EO) \equiv \langle O \rangle_T$ as a short-hand notation for the correlation of operators in the thermal environment. Gauge fields written as $A = A^aT_{\bs N}^a$ are in the fundamental representation while those written as $\ml{A} = A^a T_{\bs{adj}}^a$ are in the adjoint representation. The fundamental representation is normalized by $\Tr_c(T_{\bs N}^a T_{\bs N}^b) = C({\bs N}) \delta^{ab}$ where $C({\bs N})=1/2$ and $\Tr_c$ denotes the trace in the color space.
The adjoint representation is given by $(T_{\bs{adj}}^a)^{bc} = -if^{abc}$.
The chromoelectric field is defined as $E_i^a = F_{0i}^a = \partial_0 A_i^a - \partial_i A_0^a + g f^{abc} A_0^b A_i^c$.

Since it is a rather lengthy calculation, we will focus on describing the main aspects of the calculation so that the interested reader can straightforwardly reproduce it. First, we discuss the formulation of the calculation and define the conventions we will use throughout. We then proceed to describe the contributing diagrams, and we explicitly verify that they give a gauge-invariant result by using the $R_\xi$ gauge. After that, we explain the calculation of the electric field correlator, discussing both the vacuum theory results ($T=0$) as well as the intrinsically finite-temperature pieces, and examining some aspects of the infrared and collinear limits of the relevant diagrams. We close this section by adding up all contributions so that we can readily apply this calculation to the computation of the bound state formation and dissociation rates.

\subsection{Formulation and conventions}

In this subsection we outline the formalism we use to perform our calculation, as well as making the appropriate definitions of our conventions regarding: field branches on the Schwinger-Keldysh contour, momentum flows in Feynman diagrams with the corresponding sign conventions in the Feynman rules, and introducing the main objects that are involved in these rules. As the final part of this subsection, we enumerate all Feynman diagrams that contribute to the non-Abelian electric field correlator~\eqref{eq:elcorrelatorcalc}, and therefore set up all of the groundwork needed to carry out the explicit calculation in later sections.

\begin{figure}
    \centering
    \includegraphics[width=0.9\textwidth]{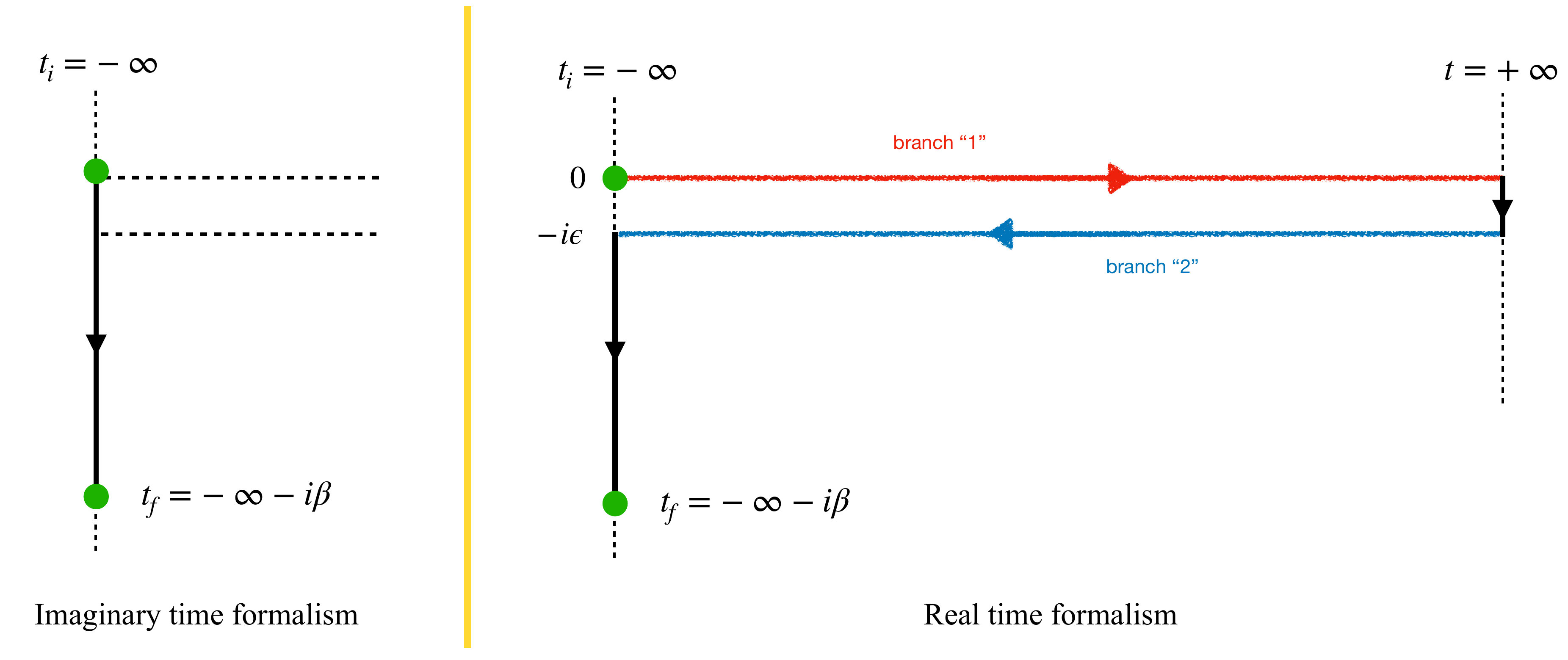}
    \caption{Left: (Imaginary) Time integration contour in the imaginary-time formalism, where the partition function $\int \ml{D} \phi\, e^{-\beta H} $ is calculated by writing $e^{-\beta H}$ using a path integral representation, as one does in quantum mechanics for $e^{-iHt}$, but with time going from $t_i= - \infty $ to $t_f = - \infty -i\beta $ (the reference real time, in this case ${\rm Re} \{t_i\} = {\rm Re} \{t_f\} = -\infty$ is arbitrary for time-independent observables). Right: The conventional Schwinger-Keldysh contour in the real-time formalism, where the imaginary-time path is deformed to allow for insertions of operators at arbitrary real times. To close the path and allow for insertions at an arbitrary real time, it is necessary to go from $t=-\infty$ to $t=+\infty$ and back. From the real time $t$ point of view, there are two copies of the fields: one in a time-ordered branch (branch 1, in red) and the other in an anti-time-ordered branch (branch 2, in blue). Both contours are ``closed'' in the sense that the field configurations at the initial $t_i$ and final time $t_f$ are identified when taking the trace~\eqref{eq:def-expectation}.}
    \label{fig:keldysh-contour}
\end{figure}

\subsubsection{Correlations on the Schwinger-Keldysh contour}

We first review the standard real-time formalism of thermal field theory. In this framework, the number of fields is doubled in order to be able to compute correlation functions at arbitrary (real) time separation. Formally speaking, one would write
\begin{equation} \label{eq:def-expectation}
    \langle \O \rangle = {\rm Tr}\left[ \rho \O \right]\,,
\end{equation}
with $\O$ some operator of which we want to know the expectation value, and in thermal equilibrium $\rho = \frac{1}{{\rm Tr} ( e^{-\beta H} )} e^{-\beta H} $, with $\beta = 1/T$ the inverse temperature. To make progress, one formally evaluates the trace in~\eqref{eq:def-expectation} by inserting a complete set of basis states
\begin{equation}
    {\rm Tr}\left[ \rho \O \right] = \sum_{\phi, \phi'} \bra{\phi'} \rho \ket{ \phi } \bra{\phi} \O \ket{ \phi' }\,,
\end{equation}
where, in field theory, we can think of $\ket{\phi}$ as an eigenstate of a field operator $\hat{\varphi}({\bf x})$, with a function-valued eigenvalue $\phi({\bf x})$, i.e., $\hat{\varphi}({\bf x}) \ket{\phi} = \phi({\bf x}) \ket{\phi}$. Then, if the operator $\ml{O}$ depends on $\hat{\varphi}$, we can replace the $\varphi$ operator by its corresponding eigenvalue $\phi$ when $\ml{O}$ acts on $|\phi\rangle$, i.e. $\ml{O}(\hat{\varphi})|\phi\rangle = \ml{O}(\phi)|\phi\rangle$. However, the set of basis states is complete at any given time slice, meaning that field configurations (i.e., \textit{states}) at different times can be expressed in terms of each other, which in general leads to the possibility that they may have complicated non-local expressions in terms of each other because the time-evolution operator will act non-trivially on $|\phi \rangle$.\footnote{Here we are thinking about having an initial basis of states $|\phi_i(t=0)\rangle$, which in the Schr\"odinger picture is evolved to $|\phi_i(t)\rangle = U(t) |\phi_i(t=0)\rangle$. In general, the overlap $\langle \phi_i(t) | \phi_j(0) \rangle $ is nonzero, meaning that the action of field operators acting on one basis will be different from the other.} This means that if the operator $\O$ depends on $\hat{\varphi}$ at different time slices, it is most convenient to insert a basis of field eigenstates around each field operator $\hat{\varphi}(x)$ at a given time, with the basis eigenstates being at the same time slice, so that the operator can be conveniently replaced by a function corresponding to an eigenvalue of $\hat{\varphi}(x)$ at that time. 

The standard way to insert the field operator eigenstates is the path integral formalism. All we have to do is to insert (infinitely) many bases of states that smoothly connect the starting point of the ``time-evolution'' operator $e^{-\beta H}$, which we can take to be anywhere in the $t$-complex plane, to the final point of the time evolution that is displaced by $-i\beta$ from the starting point. Two standard contours are shown in Figure~\ref{fig:keldysh-contour}, which correspond to the imaginary-time and real-time formalisms of thermal field theory respectively. Both contours are equally valid in the sense of connecting the starting and ending points. But they allow for a different set of possible insertions of operators. In particular, the real-time formalism explicitly allows for operators evaluated at any combination of real times to be inserted in the thermal expectation value. To allow for operator insertions at all real times, we choose this initial time to be $t_i=-\infty$.\footnote{In the sense that we take the starting point $t_i \to -\infty$ in every calculation.} The ordering of the operators is encoded in the path integral by the position along the time contour on which that operator is placed. For instance, operators on branch 2 are always behind operators on branch 1 in a correlation function in the sense of contour ordering. More generally, operators with a time coordinate ``closer'' to $t_f=-\infty -i\beta$ appear later than operators that are ``closer'' to $t_i=-\infty$ along the contour. The path integral representation of the generating functional of correlation functions in terms of the fields that are supported on each segment of the contour can be written as:
\begin{equation}
    Z[J_1,J_2] = \int \ml{D} \phi_E\, \ml{D}\phi_1\, \ml{D} \phi_2\, e^{-S_E[\phi_E] + iS[\phi_1] - iS[\phi_2] - \int_x \left[ J_1(x) \phi_1(x) - J_2(x) \phi_2(x) \right] }\,,
\end{equation}
where the $\phi_E$ field lives on the ``imaginary-time'' part of the contour, and the $\phi_1$ and $\phi_2$ fields live on the branches 1 and 2 respectively. $S_E[\phi]$ is the Euclidean action, and $S[\phi]$ is the real-time action. The symbol $\int_x$ is a short hand notation for $\int\diff^4x$. The negative sign associated with terms on the branch 2 is originated in the Hermitian conjugate of the time evolution operator.

Therefore, to enforce the explicit operator ordering in the correlator~\eqref{eq:elcorrelatorcalc}, we would place the fields at position ${\bs x}$ on the first branch of the Schwinger-Keldysh contour, and the fields at position ${\bs y}$ on the second branch. We note that since the first branch is time-ordered, and the second one is anti time-ordered, the path ordering of the operators in the Wilson lines is naturally implemented in each case. Then, we would expand in powers of the coupling constant and compute the $\ml{O}(g^2)$ correction to the correlator.

Conceptually, there is nothing preventing us from computing~\eqref{eq:elcorrelatorcalc} directly (without any reference to other operator orderings). However, in the light of certain issues (which we will discuss later) that are easier to address if we employ the KMS relations that the correlator satisfies at any definite temperature, we will formulate our calculation by starting from a more general object
\begin{align}\label{eq:elcorrelatorgeneral}
[g_E^{++}]^{da}_{ji,JI}(y,x) \equiv \Big\langle  \ml{T}_C \big[{E}_j(y) \ml{W}_{[( y^0, {\bs x}), (+\infty, {\bs x})]} \big]^d_J
\big[ \ml{W}_{[(+\infty, {\bs x}),(x^0, {\bs x})]} {E}_i(x) \big]^a_I \Big\rangle\,,
\end{align}
where $\ml{T}_C$ denotes the contour ordering (fields further along the contour are placed to the left) and $I,J$ are indices that indicate on which branch of the Schwinger-Keldysh contour the fields are located. This general correlator (\ref{eq:elcorrelatorgeneral}) reduces to the physically relevant correlator (\ref{eq:elcorrelatorcalc}) when we take $I=1$, $J=2$, and it also provides working definitions of other finite-temperature correlators with different contour orderings. For example, we can define the {\it time-ordered} version of this correlator to be the one where we take $I = J = 1$,\footnote{The correlation defined by $I=J=1$ is in general different from the following correlator
\begin{align}
\theta(y^0 - x^0) [g_E^{++}]^>(y,x)  + \theta(x^0 - y^0) [g_E^{++}]^<(y,x) \,, \nonumber 
\end{align}
See the discussions at the end of Section~\ref{sec:add-results}.}
and also construct the {\it retarded} correlator by using that, in general, the time-ordered $G^{\ml{T}}$ and retarded $G^R$ 2-point functions in a thermal background satisfy
\begin{align}
G^{\ml{T}}(p) = G^R(p) + G^<(p) \,.
\end{align}

We stress that, since the physical object that determines the rates is given by the correlator~\eqref{eq:elcorrelatorcalc}, introducing these new correlators is, so far, more of a mathematical tool than an additional point of view revealing of some physical aspects of the calculation. However, we will comment on any physical aspect that becomes apparent from the new correlators as we explain the calculation. For now, given the KMS relations shown in Section~\ref{sec:GGDs}, we know that, mathematically, all we need to know to describe the rates is the following spectral function:
\begin{align}
\big[\rho_E^{++} \big]^{da}_{ji}(y,x) \equiv \big[g_E^{++} \big]^{>,da}_{ji}(y,x) - \big[g_E^{++} \big]^{<,da}_{ji}(y,x) \,,
\end{align}
and as such, we will focus on computing the spectral function. The usefulness of doing the calculation this way will become apparent when we discuss collinear finiteness of the result. 
We want to emphasize that the lesser correlation function $[g_E^{++}]^{<,da}_{ji}(y,x)$ is not equal to the correlator with $I=2,J=1$ in general, due to the different ordering of the gauge fields with respect to the electric field. Calculating $[g_E^{++}]^{<,da}_{ji}(y,x)$ requires going beyond the usual Schwinger-Keldysh contour, while calculating the correlator with $I=2,J=1$ can be done in the Schwinger-Keldysh formalism. In the following, when this subtlety becomes crucial in certain diagrams, we will use the KMS relation to replace  $[g_E^{++}]^{<,da}_{ji}(y,x)$ with $[g_E^{--}]^{>,da}_{ji}(-y,-x)$, which can be calculated in the conventional Schwinger-Keldysh formalism.

\subsubsection{Sign conventions and Feynman rules}
\label{sec:signandfeyn}

Before proceeding to any actual calculation, we must first establish some conventions regarding the flow of momenta through the diagrams, our working definitions of propagators, and the SU$(N_c)$ gauge theory Feynman rules. While these conventions are usually devoted to an appendix, we consider that, albeit somewhat technical, they highlight important aspects of the calculation that we will perform. Nonetheless, because we do not want to overload our development with technical details, in this section we will establish the mathematical machinery that we will use only for the purely gauge boson sector of the theory, leaving the details of the fermion and ghost propagators to Appendix~\ref{app:FeynmanRules}, where we will also repeat the definitions we now present.

As a starting point, let us introduce the \textit{free} gauge boson propagators of the theory in $R_\xi$ gauge. Depending on where the gauge boson fields $A_\mu^a$ are inserted on the Schwinger-Keldysh contour, we can have different types of propagators, with a general structure given by
\begin{align}
D^{Y,ab}_{\mu \nu}(k) = \delta^{ab} P_{\mu \nu}(k) D^Y(k)\,,
\end{align}
where $Y$ can be any of $>,<,\ml{T},\overline{\ml{T}}$, and
\begin{align}
P_{\mu \nu }(k) = - \left[ g_{\mu \nu} - (1 - \xi) \frac{k_\mu k_\nu}{k^2} \right]\,.
\end{align}
with the metric signature $(+,-,-,-)$. The different types of propagators in the free theory ($g=0$) are given by
\begin{align}
D^>(k) &= \left( \Theta(k_0) + n_B(|k_0|) \right) 2\pi \delta(k^2)\,, &  D^<(k) &= \left( \Theta(-k_0) + n_B(|k_0|) \right) 2\pi \delta(k^2)\,, \nonumber \\
D^{\ml{T}}(k) &= \frac{i}{k^2 + i0^+} + n_B(|k_0|) 2\pi \delta(k^2)\,, &  D^{\overline{\ml{T}}}(k) &= \frac{-i}{k^2 - i0^+} + n_B(|k_0|) 2\pi \delta(k^2)\,,
\end{align}
which are called Wightman functions (the two propagators on the first line), time-ordered propagator, and anti time-ordered propagator, respectively. As introduced earlier, $n_B(k_0) = (\exp(k_0/T)-1)^{-1}$ is the Bose-Einstein distribution. It is also useful to define
\begin{align}
    D^R(k) &= \frac{i}{k^2 + i0^+ {\rm sgn}(k^0) }\,, &  D^A(k) &= \frac{i}{k^2 - i0^+ {\rm sgn}(k^0) }\,, \nonumber \\ D^S(k) &= D^>(k) + D^<(k) = (1 + 2n_B(|k_0|)) 2 \pi \delta(k^2)\,, & &
\end{align}
as the \textit{free} retarded, advanced, and symmetric propagators respectively.

\begin{figure}
	\centering
	\begin{tabular}{  c  c  l  }
    \raisebox{-0.26in}{\includegraphics[height=0.6in]{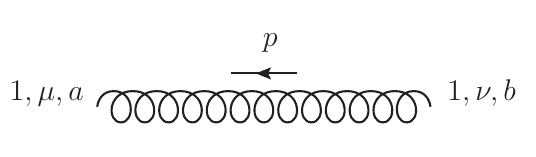}}  &=& $ \delta^{ab} P_{\mu\nu} D^{\ml{T}}(p)  $\\
	\raisebox{-0.26in}{\includegraphics[height=0.6in]{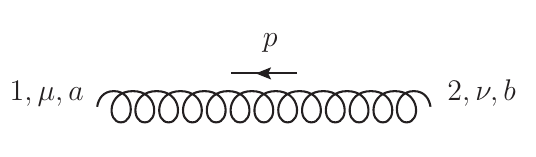}}  &=& $ \delta^{ab} P_{\mu\nu} D^{<}(p) $  \\
	\raisebox{-0.26in}{\includegraphics[height=0.6in]{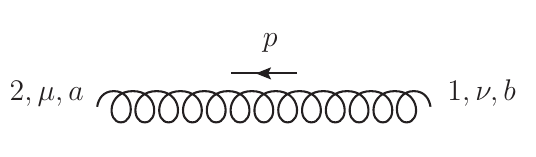}}  &=& $ \delta^{ab} P_{\mu\nu} D^{>}(p)  $\\
	\raisebox{-0.26in}{\includegraphics[height=0.6in]{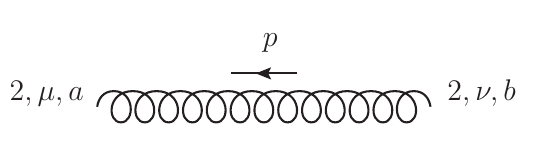}}  &=& $ \delta^{ab} P_{\mu\nu} D^{\overline{\ml{T}}}(p) $ 
	\end{tabular}
\caption{Feynman rules associated with different types of gauge boson propagators.}
\label{fig:rules-props-maintext}
\end{figure}

In terms of the indices of the Schwinger-Keldysh contour, we can compactly write our propagators as
\begin{align}
\mathbb{D}(p)_{JI} =  \begin{bmatrix} D^{\ml{T}}(p)  &  D^<(p) \\
D^>(p) & D^{\overline{\ml{T}}}(p)  \end{bmatrix} _{JI}\,,
\end{align}
for example, $\mathbb{D}(p)_{21} = D^>(p)$, and $\mathbb{D}(p)_{11} = D^{\ml{T}}(p)$. With
these definitions, correlations in position space of two fields in branches $I$ and $J$ of the Schwinger-Keldysh contour are given by
\begin{align}
\langle \phi_J(y) \phi_I(x) \rangle = \mathbb{D}(y-x)_{JI} = \int \frac{\diff^4 k}{(2\pi)^4} e^{-i k \cdot (y-x) } \, \mathbb{D}(p)_{JI}\,,
\end{align}
where $\phi$ can be thought of as a scalar field since we have taken out the non-Abelian indexes and the Lorentz structure.
Crucially, this convention of the Fourier transform defines the signs of the momenta appearing in the Feynman rules, to which we now turn. Pictorially, the momentum ``flow'' in a propagator $\mathbb{D}(p)_{JI}$ should be diagrammatically depicted as going from the field insertion of type $I$ towards the field insertion of type $J$. The Feynman rules that illustrate the relevance of having a consistent definition of momentum flow most clearly are those of the propagators themselves, in a manner consistent with our previous definitions. We list them in Figure~\ref{fig:rules-props-maintext}.

\begin{figure}
	\centering
	\begin{tabular}{  c  c  l  }
	\raisebox{-0.8in}{\includegraphics[height=1.6in]{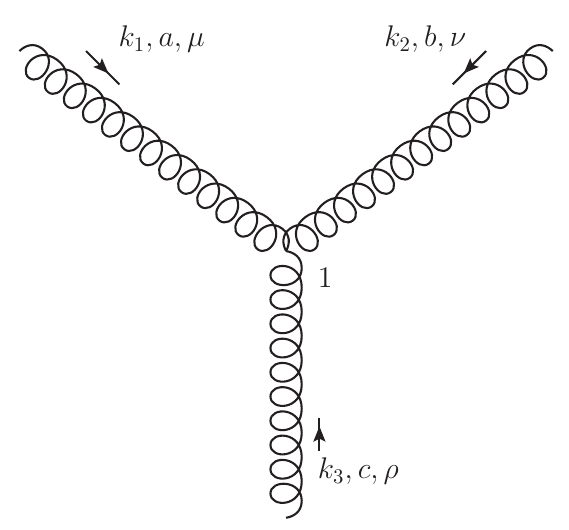}}  &=& $ gf^{abc}\Big( g_{\mu\nu} (k_1-k_2)_\rho + g_{\nu\rho} (k_2-k_3)_\mu + g_{\rho\mu} (k_3-k_1)_\nu \Big) $  \\
	\raisebox{-0.8in}{\includegraphics[height=1.6in]{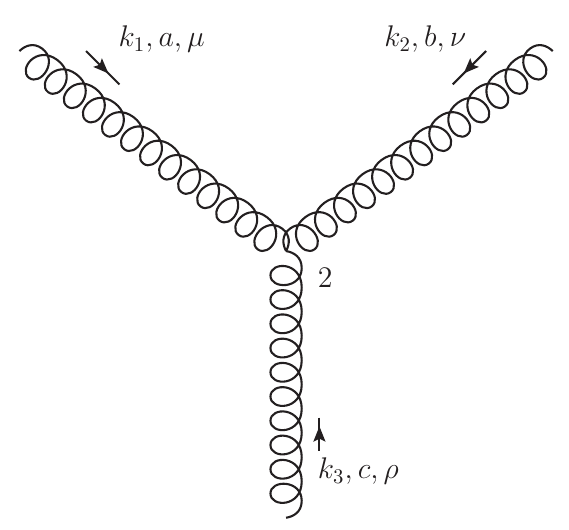}}  &=& $ -gf^{abc}\Big( g_{\mu\nu} (k_1-k_2)_\rho + g_{\nu\rho} (k_2-k_3)_\mu + g_{\rho\mu} (k_3-k_1)_\nu \Big) $
	\end{tabular}
\caption{Feynman rules associated to the 3-gauge boson vertex, for both types of fields on the branches of the Schwinger-Keldysh contour along the real-time directions.}
\label{fig:rules-3gluon-maintext}
\end{figure}

Two other ingredients in our diagrammatic calculations are particularly sensitive to the choice of signs related to the momentum flow. One ingredient is the 3-gauge boson vertex, which involves the incoming/outgoing momentum from each of its external legs explicitly in the corresponding Feynman rule. We show the 3-gauge boson vertex in Figure~\ref{fig:rules-3gluon-maintext}, for both fields of type 1 and 2 on the Schwinger-Keldysh contour, just to emphasize that we have doubled the field content of the theory from the start. From now on, we will write all possible vertex insertions that come from expanding the interacting pieces of the action with a $(-1)^{I+1}$ factor beside the vertex, where $I\in \{1,2\}$ is an index that tells us on which branch the vertex is to be evaluated, so that we can effectively write all possible combinations of indices more compactly. For example, by writing
\begin{equation}
\label{eqn:example_3vertex}
    \D(k)_{I'I} \D(p)_{I'J} \D(q)_{I'K} (-1)^{I'+1} \,,
\end{equation}
we indicate a 3-particle vertex with three incoming particles that have momenta $k,p,q$, and live on branches $I,J,K$, respectively. We implicitly sum over the repeated $I'$ indices in the last expression (\ref{eqn:example_3vertex}) (by taking values $I'\in\{1,2\}$).

\begin{figure}
	\centering
	\begin{tabular}{  c  c  l  }
	\raisebox{-0.8in}{\includegraphics[height=2in]{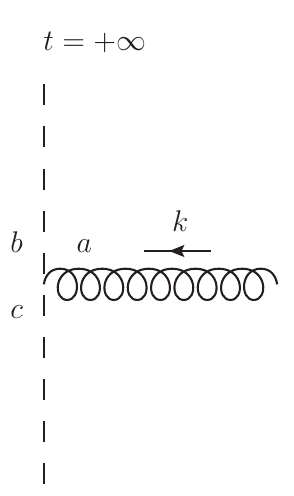}}  &=& $ \dfrac{gf^{abc}}{i k_0 + 0^+}$
	\end{tabular}
\caption{Feynman rule associated with gauge boson insertions of Wilson lines. The color index $b$ is to be contracted with the next operator along the Wilson line towards $t=+\infty$, while the index $c$ is contracted with the next operator in the direction towards the electric field insertion, which is at the lower end of the vertical dashed line and not shown here explicitly.}
    \label{fig:rules-momentum-Wilson}
\end{figure}

The other Feynman rule that is particularly sensitive to the sign convention of the momentum flowing through the diagram is the gauge boson insertion from the Wilson lines. There are two possibilities, depending on the sign convention of the incoming/outgoing momentum:
\begin{enumerate}
    \item Momentum flows away from the gauge boson insertion of the Wilson line at the spacetime point $z_s=(s,{\bs z})$ on the contour branch $K$ with color (index of the adjoint representation) given by $a$, towards the rest of the diagram (with vertex spacetime location, contour branch, color, and Lorentz index $(t',{\bs z}')$, $K'$, $a'$, and $\mu'$, respectively). In this case we have from the expansion of the Wilson line
    \begin{equation}
    \begin{split}
        \int_t^\infty \!\! \diff s \, e^{-\varepsilon s} \, \D(z'-z_s)_{\mu'0,K'K}^{a'a} &=  \delta^{a'a} \int_t^\infty \!\! \diff s \, e^{-\varepsilon s} \int \frac{\diff^d k}{(2\pi)^d} e^{-i k \cdot (z' - z_s)} P_{\mu' 0}(k) \D(k)_{K'K} \\
        &= \delta^{a'a} \int \frac{\diff^d k}{(2\pi)^d} e^{-i k \cdot (z' - z_t)} \frac{1}{-i k_0 + 0^+} P_{\mu' 0}(k) \D(k)_{K'K}\,,
    \end{split}
    \end{equation}
    where $z_t=(t,{\bs z})$ and we have introduced a positive infinitesimal $\varepsilon$ to make the Wilson line a well-defined operator as it approaches infinite time.
    \item Momentum flows towards the gauge boson insertion of the Wilson line at the spacetime point $z_s=(s,{\bs z})$ on the contour branch $K$ with color $a$, from the rest of the diagram (with vertex spacetime location, contour branch, color, and Lorentz index $(t',{\bs z}')$, $K'$, $a'$, and $\mu'$, respectively). In this case we have
    \begin{equation}
    \begin{split}
        \int_t^\infty \!\! \diff s \, e^{-\varepsilon s}\, \D(z_s-z')_{0\mu',KK'}^{aa'} &=  \delta^{aa'} \int_t^\infty \!\! \diff s \, e^{-\varepsilon s} \int \frac{\diff^d k}{(2\pi)^d} e^{-i k \cdot ( z_s-z')} P_{0\mu'}(k) \D(k)_{KK'} \\
        &= \delta^{aa'} \int \frac{\diff^d k}{(2\pi)^d} e^{-i k \cdot (z_t-z')} \frac{1}{i k_0 + 0^+} P_{0\mu'}(k) \D(k)_{KK'} \,.
    \end{split}
    \end{equation}
\end{enumerate}
We summarize these results diagrammatically in Figure~\ref{fig:rules-momentum-Wilson}, where we also include the relevant color factors $f^{abc}$, which come from expanding the Wilson line in the adjoint representation (cases 1 and 2 are related by flipping the sign of $k$). We give the rest of the Feynman rules in Appendix~\ref{app:FeynmanRules}.

We will work in $d$ spacetime dimensions to regulate the potentially ultraviolet (UV) divergence, and take the limit $d\to 4$ at the end of the calculation after renormalization. In practice, we take the spacetime to be $(1,d-1)$ dimensional, i.e., the limit $d\to 4$ is taken by varying the number of spatial dimensions, while always holding the number of time-like dimensions fixed.

\subsubsection{Contributing Feynman diagrams}

Now we proceed to depict all the diagrams that contribute to the generalized thermal electric correlator~\eqref{eq:elcorrelatorgeneral} at next-to-leading order. As we will see momentarily, a natural way to group the terms coming from each Feynman diagram is to take a look at their propagator structures in terms of the different propagator combinations that appear. The first criterion to do this is to separate the different contributions by the number of propagators that appear in each diagram shown in Figure~\ref{fig:diagrams}.

\begin{figure}
    \centering
    \begin{subfigure}[b]{0.33\textwidth}
        \centering
        \includegraphics[height=1.4in]{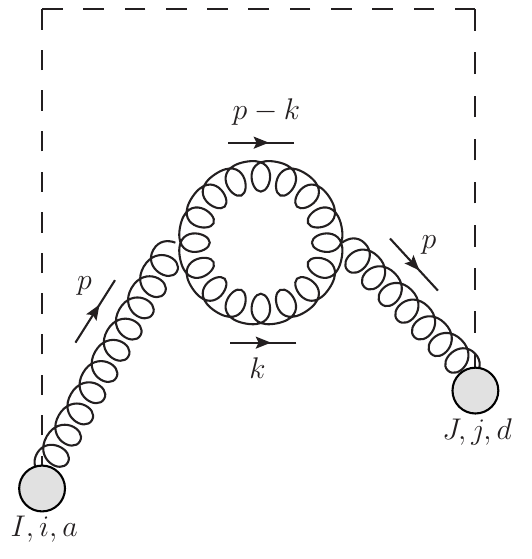}
        \caption*{$(1)$}\label{subfig:1}
    \end{subfigure}%
    ~ 
    \begin{subfigure}[b]{0.33\textwidth}
        \centering
        \includegraphics[height=1.4in]{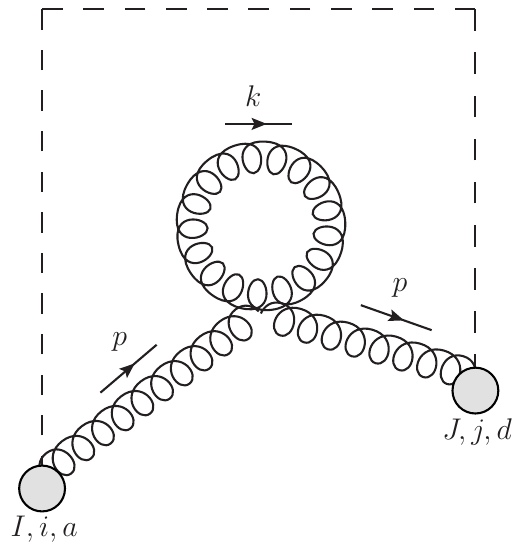}
        \caption*{$(2)$}\label{subfig:2}
    \end{subfigure}%
    ~ 
    \begin{subfigure}[b]{0.33\textwidth}
        \centering
        \includegraphics[height=1.4in]{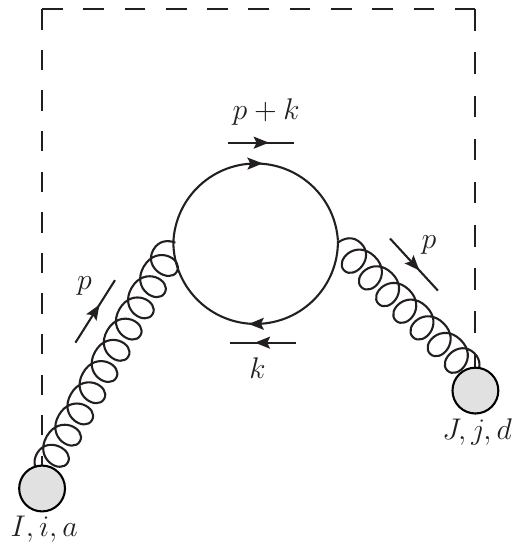}
        \caption*{$(f)$}\label{subfig:f}
    \end{subfigure}%
    \vspace{0.7cm}
    \begin{subfigure}[b]{0.33\textwidth}
        \centering
        \includegraphics[height=1.4in]{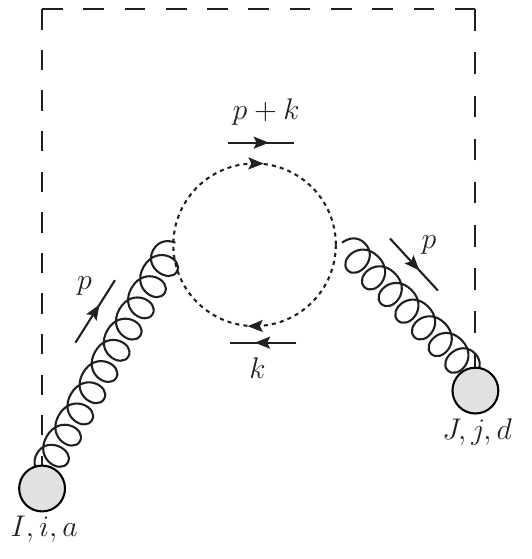}
        \caption*{$(g)$}\label{subfig:g}
    \end{subfigure}%
    ~ 
    \begin{subfigure}[b]{0.33\textwidth}
        \centering
        \includegraphics[height=1.4in]{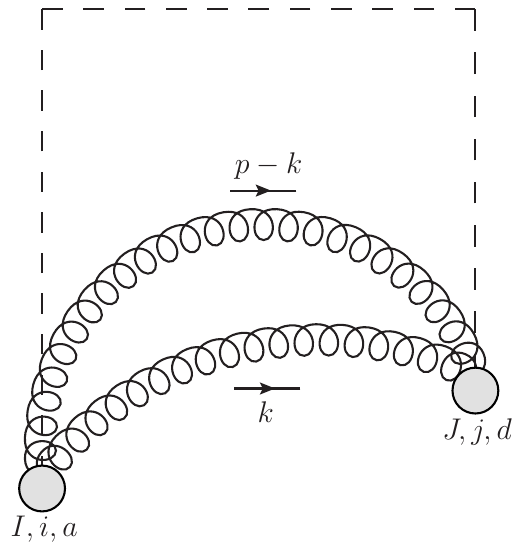}
        \caption*{$(3)$}\label{subfig:3}
    \end{subfigure}%
    ~ 
    \begin{subfigure}[b]{0.33\textwidth}
        \centering
        \includegraphics[height=1.4in]{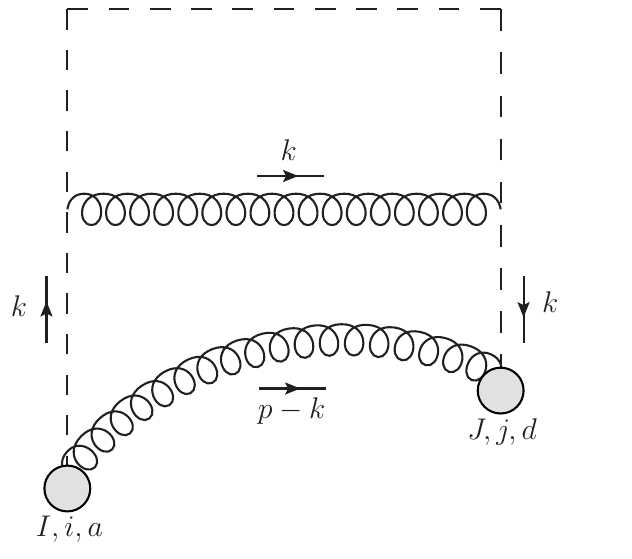}
        \caption*{$(4)$}\label{subfig:4}
    \end{subfigure}%
    \vspace{0.7cm}
   \begin{subfigure}[b]{0.33\textwidth}
        \centering
        \includegraphics[height=1.4in]{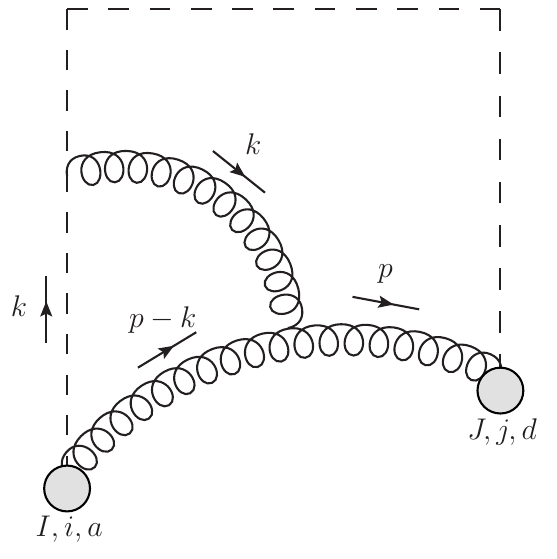}
        \caption*{$(5)$}\label{subfig:5}
    \end{subfigure}%
    ~ 
    \begin{subfigure}[b]{0.33\textwidth}
        \centering
        \includegraphics[height=1.4in]{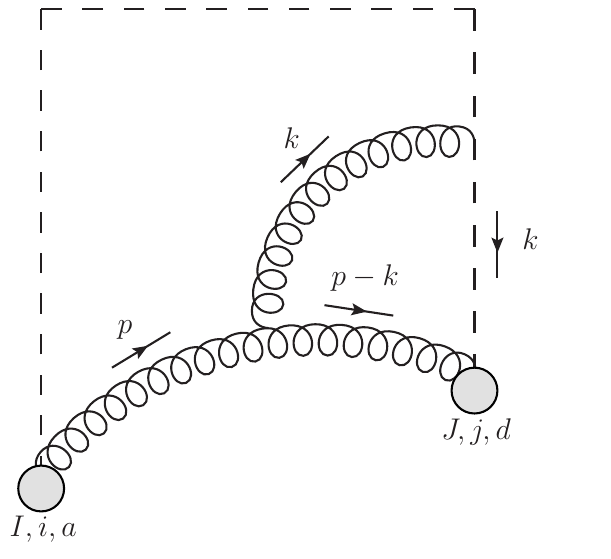}
        \caption*{$(5r)$}\label{subfig:5r}
    \end{subfigure}%
    ~  
    \begin{subfigure}[b]{0.33\textwidth}
        \centering
        \includegraphics[height=1.4in]{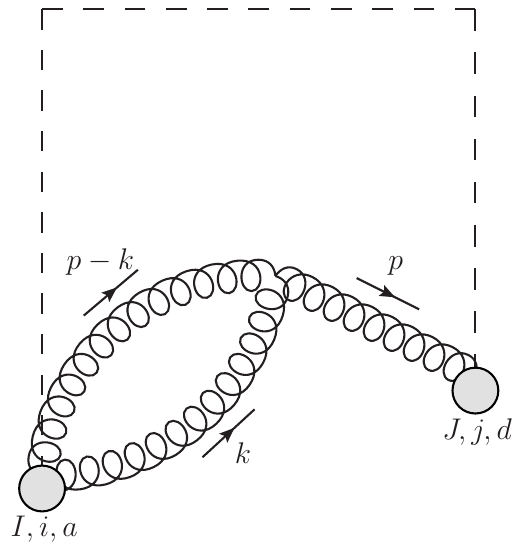}
        \caption*{$(6)$}\label{subfig:6}
    \end{subfigure}%
    \vspace{0.7cm}
    \begin{subfigure}[b]{0.33\textwidth}
        \centering
        \includegraphics[height=1.4in]{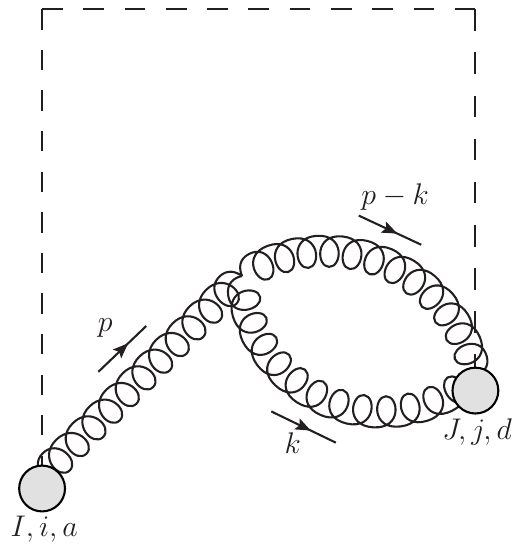}
        \caption*{$(6r)$}\label{subfig:6r}
    \end{subfigure}%
    ~
    \begin{subfigure}[b]{0.33\textwidth}
        \centering
        \includegraphics[height=1.4in]{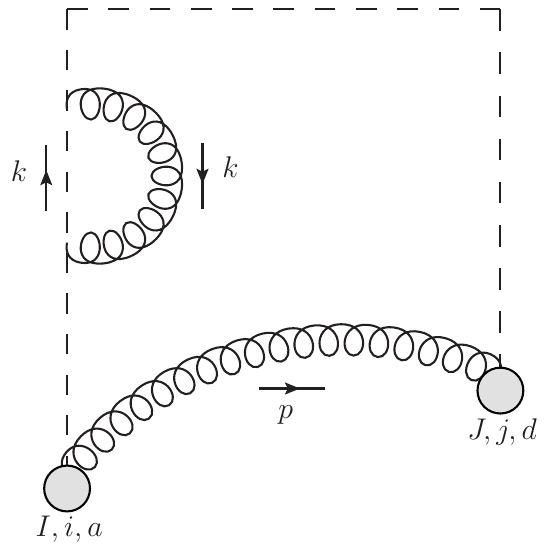}
        \caption*{$(7)$}\label{subfig:7}
    \end{subfigure}%
    ~      
   \begin{subfigure}[b]{0.33\textwidth}
        \centering
        \includegraphics[height=1.4in]{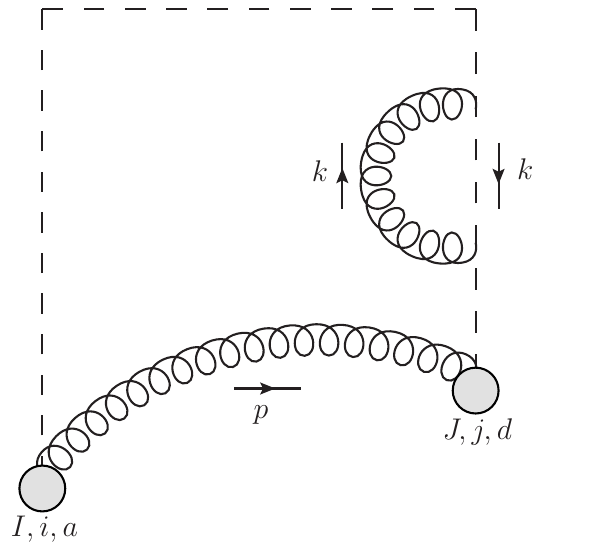}
        \caption*{$(7r)$}\label{subfig:7r}
    \end{subfigure}%
\end{figure}
\begin{figure}\ContinuedFloat
     \begin{subfigure}[b]{0.33\textwidth}
        \centering
        \includegraphics[height=1.4in]{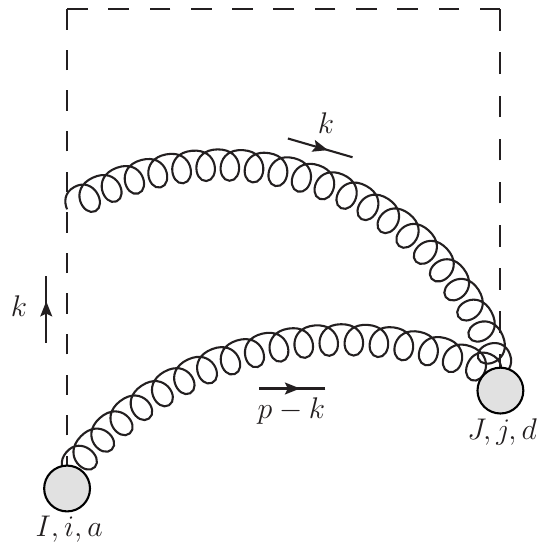}
        \caption*{$(8)$}\label{subfig:8}
    \end{subfigure}%
    ~ 
     \begin{subfigure}[b]{0.33\textwidth}
        \centering
        \includegraphics[height=1.4in]{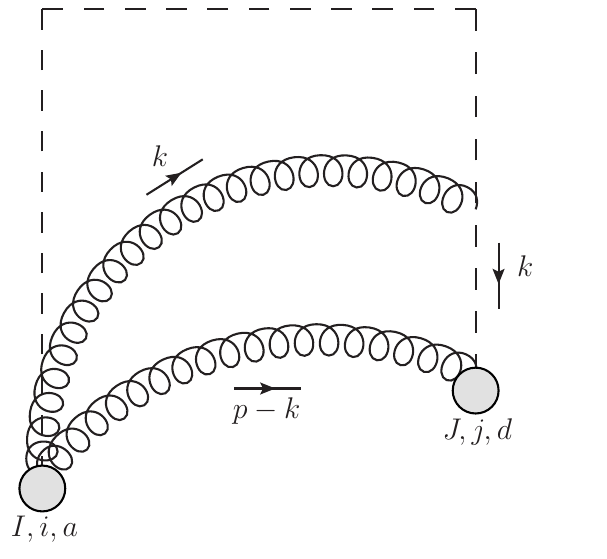}
        \caption*{$(8r)$}\label{subfig:8r}
    \end{subfigure}%
    ~
    \begin{subfigure}[b]{0.33\textwidth}
        \centering
        \includegraphics[height=1.4in]{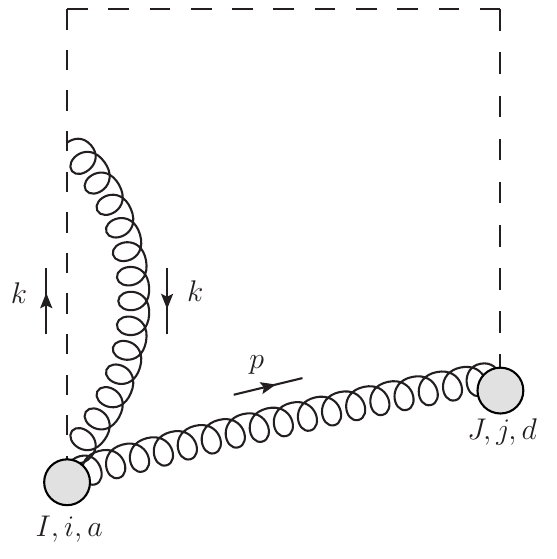}
        \caption*{$(9)$}\label{subfig:9}
    \end{subfigure}%
    \vspace{0.7cm}
    \begin{subfigure}[b]{0.33\textwidth}
        \centering
        \includegraphics[height=1.4in]{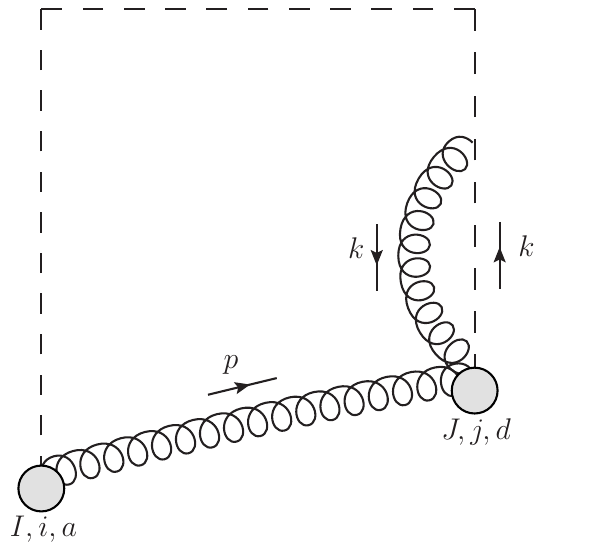}
        \caption*{$(9r)$}\label{subfig:9r}
    \end{subfigure}%
    ~   
    \begin{subfigure}[b]{0.33\textwidth}
        \centering
        \includegraphics[height=1.4in]{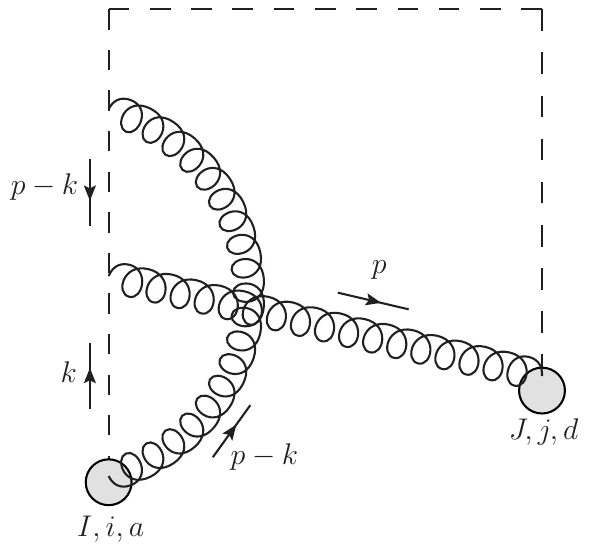}
        \caption*{$(10)$}\label{subfig:10}
    \end{subfigure}%
    ~    
    \begin{subfigure}[b]{0.33\textwidth}
        \centering
        \includegraphics[height=1.4in]{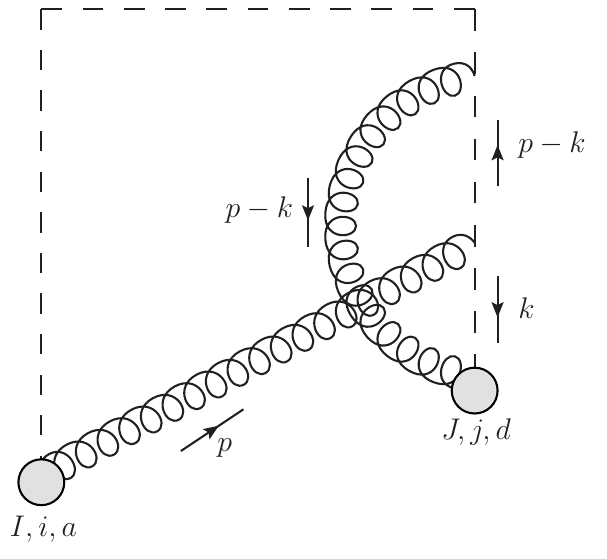}
        \caption*{$(10r)$}\label{subfig:10r}
    \end{subfigure}%
    \vspace{0.7cm}
    \begin{subfigure}[b]{0.33\textwidth}
        \centering
        \includegraphics[height=1.4in]{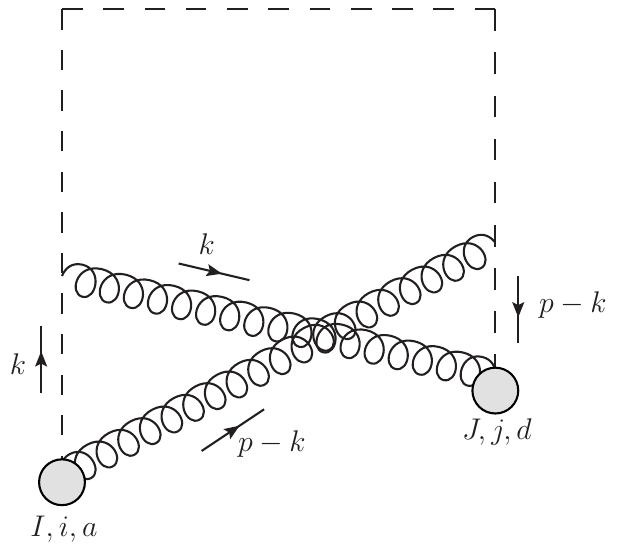}
        \caption*{$(11)$}\label{subfig:11}
    \end{subfigure}%
    \caption{List of all diagrams contributing to the electric field correlator~\eqref{eq:elcorrelatorgeneral}, with free indices $(I,i,a)$ and $(J,j,d)$ for the corresponding electric field insertions. The long dashed lines represent the Wilson lines while the short dashed lines label the ghost field. Solid lines indicate the fermion field. The two grey blobs are the electric fields.}
    \label{fig:diagrams}
\end{figure}

Each diagram in Figure~\ref{fig:diagrams}, labeled by $(X)$, can be written in the following form
\begin{align}
\big(X\big)^{da}_{ji,JI}(p) = \delta^{ad} g^2 \int \frac{\diff^d k}{(2\pi)^d} \ml{Q}^{(X)}(p,k)_{JI} V^{(X)}(p,k)_{ji}\,,
\end{align}
where $\ml{Q}^{(X)}(p,k)_{JI}$ is a sum of products of $\mathbb{D}$ propagators associated with the diagram $(X)$, with momenta flows that depend only on $p$ and $k$, and $V^{(X)}(p,k)_{ji}$ is a rational function of the momenta $p_\mu, k_\nu$ given by the appropriate vertex factors appearing in each diagram.

\begin{table*}
\begin{center}
\begin{tabular}{c||c}
\toprule
    Diagram $(X)$ & Propagator structure $\ml{Q}^{(X)}(p,k)_{JI}$ \\ \midrule
    $ (1), (g) $ & $\D(p)_{I'I}  \D(k)_{J'I'} \D(p-k)_{J'I'} \D(p)_{JJ'} (-1)^{I'+J'}$ \\ \midrule
    $ (f) $ & $ \D(p)_{I'I} {\rm Tr}[ \gamma^{\mu} \mathbb{S}(p-k)_{J'I'} \gamma^{\nu} \mathbb{S}(k)_{J'I'}  ] \D(p)_{JJ'} (-1)^{I'+J'}$ \\
 \bottomrule
\end{tabular}
\end{center}
\caption{Summary of propagator structures of diagrams with 4 propagators contributing to $\big[g_E^{++}\big]_{ji,JI}^{da}$. Summation over repeated $I',J'$ indices is implicit, even when there are three or more instances of such indices. Because fermionic propagators are intrinsically matrix-valued objects, we have made the choice to define the propagator structures with the gamma matrices included from the start. Fermionic propagators $\mathbb{S}_{IJ}$ are defined in Appendix~\ref{app:FeynmanRules}.}
\label{tab:4prop-structure}
\end{table*}

\begin{table*}
\begin{center}
\begin{tabular}{c||c}
\toprule
    Diagram $(X)$ & Vertex factors $V^{(X)}(p,k)_{ji}$  \\ \midrule \midrule
    $ (1) $ & $\dfrac{N_c}{2} (i p_0 g_{i \rho''}  - ip_i g_{0\rho''} ) P_{\mu \mu'}(k)  P_{\nu \nu'}(p-k)  (-i p_0 g_{j\rho'} + i p_j g_{0 \rho'})$ \\ & $\times \left[ g^{\rho'' \mu'} (p+k)^{\nu'} + g^{\mu' \nu'}(p - 2k)^{\rho''} + g^{\nu' \rho''} (k - 2p)^{\mu'} \right]$ \\ & $\times  \left[ g^{\rho' \mu} (-p-k)^{\nu} + g^{\mu \nu} (2k-p)^{\rho'} + g^{\nu \rho'}(2p-k)^{\mu} \right]$ \\ \midrule
    $ (g) $ & $(-1) N_c (i p_0 g_{i \mu}  - ip_i g_{0\mu} ) P^{\mu \mu'}(k) k_{\mu'} (p_{\nu'}-k_{\nu'}) P^{\nu' \nu}(p-k)  (-i p_0 g_{j\nu} + i p_j g_{0 \nu}) $\\ \midrule
    $ (f) $ & $ n_f C({\bs N}) (i p_0 g_{i \mu}  - ip_i g_{0\mu} ) (-i p_0 g_{j\nu} + i p_j g_{0 \nu})  $ \\
 \bottomrule
\end{tabular}
\end{center}
\caption{Summary of vertex factors of diagrams with 4 propagators contributing to $\big[g_E^{++}\big]_{ji,JI}^{da}$. The free $\mu,\nu$ indices in the fermion loop diagram $(f)$ are to be contracted with the free $\mu,\nu$ indices in the corresponding propagator structure.}
\label{tab:4prop-vertex}
\end{table*}

For diagrams with four propagators, 
we list their propagator structures in Table~\ref{tab:4prop-structure}, and their respective vertex factors in Table~\ref{tab:4prop-vertex}. Whenever two diagrams have the same propagator structure, we will list them together. As one might expect, it can be seen that the ghost diagram $(g)$ can be added to the gauge boson loop diagram $(1)$ without altering the propagator structure. This is not only convenient, but also necessary, because the Faddeev-Popov gauge-fixing procedure does not depend on whether the calculation is at finite temperature or not.

Next, we list the propagator structures of all diagrams with three propagators in Table~\ref{tab:3prop-structure}, and their respective vertex factors in Table~\ref{tab:3prop-vertex}. For the three-propagator structure, all contributions come purely from the Yang-Mills sector of the theory, which is,  in some sense, a result of the gauge invariance of the electric field correlator. As we will show in the next section, we will need all of these diagrams (plus those with two propagators) to compensate for the gauge dependence of diagram $(1)$.

\begin{table*}
\begin{center}
\begin{tabular}{c||c}
\toprule
    Diagrams $(X)$ & Propagator structure $\ml{Q}^{(X)}(p,k)_{JI}$ \\ \midrule
    $ (2) $ & $ \D(p)_{I'I}  \D(p)_{JI'}   \D(k)_{I'I'} (-1)^{I'+1} $ \\ \midrule
    $ (5),(6) $ & $ \D(p-k)_{I'I} \D(k)_{I'I} \D(p)_{JI'} (-1)^{I'+1} $ \\ \midrule
    $ (5r),(6r) $ & $\D(p-k)_{JI'} \D(k)_{JI'} \D(p)_{I'I} (-1)^{I'+1} $ \\ 
 \bottomrule
\end{tabular}
\end{center}
\caption{Summary of propagator structures of diagrams with 3 propagators contributing to $\big[g_E^{++}\big]_{ji,JI}^{da}$. Summation over repeated $I'$ indices is implicit, even when there are three or more instances of such indices.}
\label{tab:3prop-structure}
\end{table*}

\begin{table*}
\begin{center}
\begin{tabular}{c||c}
\toprule
    Diagram $(X)$ & Vertex factors $V^{(X)}(p,k)_{ji}$  \\ \midrule \midrule
    $ (2) $ & $ i N_c (i p_0 g_{i\mu} - ip_i g_{0\mu}) P_{\rho \rho'}(p) (-i p_0 g_j^{\rho} + i p_j g_{0}^{\rho}) \left[ g^{\mu \rho'} P_\nu^\nu(k) - P^{\mu \rho'}(k) \right]$ \\ \midrule
    $ (5) $ & $N_c \left[ g^{\rho \mu} (k-2p)^{\nu} + g^{\mu \nu}(p- 2k)^{\rho} + g^{\nu \rho} (k+p)^{\mu} \right]  P_{0\nu}(k) $  \\ 
    & $\quad \times  (p_0 g_{j\rho} - p_j g_{0 \rho}) ( (p-k)_0 g_{i\mu} - (p-k)_i g_{0\mu} )/(-ik_0 + 0^+)$\\ \midrule
    $ (5r) $ & $N_c \left[ g^{\rho \mu} (k-2p)^{\nu} + g^{\mu \nu}(p- 2k)^{\rho} + g^{\nu \rho} (k+p)^{\mu} \right]  P_{0\nu}(k)$ \\ &  $ \quad \times (p_0 g_{i\rho} - p_i g_{0 \rho}) ( (p-k)_0 g_{j\mu} - (p-k)_j g_{0\mu} )/({-ik_0 - 0^+})$ \\ \midrule
    $ (6) $ & $ N_c \left[ g^{\rho \mu} (-k-p)^{\nu} + g^{\mu \nu} ( 2k-p)^{\rho} + g^{\nu \rho} (2p-k)^{\mu} \right] $ \\ & $ \quad \times (i p_0 g_{j\rho} - i p_j g_{0 \rho})  P_{0\mu}(k) P_{i\nu}(p-k)$ \\ \midrule
    $ (6r) $ & $N_c \left[ g^{\rho \mu} (-k-p)^{\nu} + g^{\mu \nu} ( 2k-p)^{\rho} + g^{\nu \rho} (2p-k)^{\mu} \right] $ \\ &  $ \quad \times (i p_0 g_{i\rho} - i p_i g_{0 \rho})   P_{0\mu}(k) P_{j\nu}(p-k)$ \\
 \bottomrule
\end{tabular}
\end{center}
\caption{Summary of vertex factors of diagrams with 3 propagators contributing to $\big[g_E^{++}\big]_{ji,JI}^{da}$}
\label{tab:3prop-vertex}
\end{table*}

Finally, we discuss all remaining diagrams with two gauge boson propagators. In Figure~\ref{fig:diagrams}, we have presented the complete list of non-vanishing diagrams after performing the color indices contraction. From the diagrams with two gauge boson propagators, it is clear that the momenta flows in the propagators are decoupled (one of them carries $p$ and the other can be chosen to carry momentum $k$), and then we only need to analyze the vertex factors. It turns out that diagrams $(9),(9r),(10),(10r)$ do not contribute because of spacetime symmetries. Diagrams $(9)+(9r)$ vanish because one piece of the integrand is odd under $k_0 \to -k_0$, as can be verified explicitly using the Feynman rules, and the other piece of the sum is proportional to the integral of $\k$, which vanishes by rotational invariance of the plasma. It is necessary to take the sum so that the contribution from the pole in the $i0^+$ prescription from the Wilson line propagators cancels unambiguously. Similarly, $(10)$ and $(10r)$ vanish (separately) by rotational invariance because the integrand is proportional to $\k$. Therefore, regarding the diagrams with two gauge boson propagators, we only show the relevant propagator structures of non-vanishing diagrams in Table~\ref{tab:2prop-structure}, and their respective vertex factors in Table~\ref{tab:2prop-vertex}.

\begin{table*}
\begin{center}
\begin{tabular}{c||c}
\toprule
    Diagrams $(X)$ & Propagator structure $\ml{Q}^{(X)}(p,k)_{JI}$ \\ \midrule
    $ (3),(4),(8),(8r),(11) $ & $ \D(p-k)_{JI} \D(k)_{JI} $ \\ \midrule
    $ (7) $ & $ \D(k)_{II} \D(p)_{JI} $ \\ \midrule
    $ (7r) $ & $ \D(k)_{JJ} \D(p)_{JI} $ \\ \midrule
 \bottomrule
\end{tabular}
\end{center}
\caption{Summary of propagator structures of diagrams with 2 propagators contributing to $\big[g_E^{++}\big]_{ji,JI}^{da}$.}
\label{tab:2prop-structure}
\end{table*}

\begin{table*}
\begin{center}
\begin{tabular}{c||c}
\toprule
    Diagrams $(X)$ & Vertex factors $V^{(X)}(p,k)_{ji}$  \\ \midrule \midrule
    $ (3) $ & $ N_c \left( P_{00}(k) P_{ij}(p-k) - P_{0j}(k) P_{i0}(p-k) \right) $ \\ \midrule
    $ (4) $ & $ (-1) N_c \dfrac{(p_0-k_0)^2 g_{ij} + (p-k)_i (p-k)_j g_{00} }{k_0^2} P_{00}(k) $ \\ \midrule
    $ (7), (7r) $ & $ \dfrac{N_c}2 (p_0^2 g_{ij}  + p_i p_j g_{00}) \dfrac{P_{00}(k) }{k_0^2} $ \\ \midrule
    $ (8) $ & $N_c \dfrac{ i(p-k)_0 g_{ij} P_{00}(k) + i(p-k)_i  g_{00} P_{0j}(k) }{-ik_0 +0^+} $ \\ \midrule
    $ (8r) $ & $ N_c \dfrac{ i(p-k)_0 g_{ij} P_{00}(k) + i(p-k)_j  g_{00} P_{0i}(k) }{-ik_0 -0^+} $ \\ \midrule
    $ (11) $ & $ N_c \dfrac{-(p-k)_i k_j}{(-ik_0 + 0^+) (i (p-k)_0 + 0^+) }  $ \\
 \bottomrule
\end{tabular}
\end{center}
\caption{Summary of vertex factors of diagrams with 2 propagators contributing to $\big[g_E^{++}\big]_{ji,JI}^{da}$.}
\label{tab:2prop-vertex}
\end{table*}

Tables~\ref{tab:4prop-structure},~\ref{tab:4prop-vertex},~\ref{tab:3prop-structure},~\ref{tab:3prop-vertex},~\ref{tab:2prop-structure} and~\ref{tab:2prop-vertex}, describe the complete NLO calculations of the electric field correlator $\big[g_E^{++} \big]_{ji,JI}^{da}$ in full generality, for any type of correlation. As advertised, our objective is to compute the correlation with $J=2$ and $I=1$, but we will take a slightly less direct path and first compute the spectral function $\big[\rho_E^{++} \big]_{ji}^{da}$. A non-trivial check of our calculations is to verify that independently of our choice of $I$ and $J$ (i.e., on what branches of the Schwinger-Keldysh contour we evaluate the fields) the result is gauge invariant, since the electric field correlator is defined in a gauge invariant way. We now discuss the gauge invariance of the calculation in $R_\xi$ gauge.

\subsection{Gauge invariance in \texorpdfstring{$R_\xi$}{R_x} gauge} 

As a consistency check of our calculation, we verify our result is gauge invariant. To recapitulate, the purpose of doing this is twofold: i) we can make sure that we have included and accounted for all diagrams contributing to the correlator at NLO, which also partially justifies the neglect of the Wilson lines at infinite time in the more general object we introduced in Eq.~\eqref{eq:elcorrelatorgeneral}, whose contributions vanish in our calculations, and ii) we will verify that the way gauge-dependent parts cancel is independent of what type of correlation function we are calculating. That is to say, our proof will not rely on whether we choose to calculate a Wightman or a time-ordered correlator: we will prove gauge invariance for the full matrix-valued correlation function with the Schwinger-Keldysh contour indices.

We must show all the $\xi$ gauge parameter dependent parts cancel when summing over all diagrams. Depending on the number of gauge boson propagators in a diagram, we may have $\xi^2$, $\xi^3$ and $\xi^4$ terms showing up in the calculations. First, we note that due to the nature of the electric field, a single gauge boson propagator connected with the electric field automatically has its $\xi$ dependent part in the propagator removed. This reduces the naive number of powers of $\xi$ in a diagram by 2. It turns out that at NLO in $R_\xi$ gauge, there are two independent cancellations that must take place: one for the terms proportional to $(1-\xi)^2$, and the other with the pieces proportional to $1-\xi$.

An important issue is how to connect the $\xi$ coefficients of diagrams with different numbers of propagators. For instance, diagram $(1)$ is the only diagram with four propagators that has a term proportional to $\xi^2$, so the $\xi^2$ term must be cancelled by diagram(s) with fewer gauge boson propagators. The crucial property that allows us to prove these cancellations without actually performing the integrals, (which is apparent from the free gauge boson equation of motion) is:
\begin{align}\label{eq:propcancellation}
-i p^2 \D(p)_{JI} (-1)^{I+1} = \mathbbm{1}_{JI} \,,
\end{align}
where no summation is implied by repeated indices, and $I=1$ is assumed to be a time-ordered branch of the Schwinger-Keldysh contour. With this, one can reduce the number of propagators in any diagram if a factor of the corresponding momentum squared appears.

We will verify the cancellation of the terms proportional to $(1-\xi)^2$ explicitly first, and then outline the technically more involved cancellation of the terms proportional to $(1-\xi)$, which is explained in Appendix~\ref{app:gauge-invariance} in detail.

\subsubsection{Cancellation at \texorpdfstring{$O((1-\xi)^2)$}{O_{(1-x)}}}

There are four diagrams that contribute at order $\xi^2$ when we plug
\begin{align}
P_{\mu \nu }(k) = - \left[ g_{\mu \nu} - (1 - \xi) \frac{k_\mu k_\nu}{k^2} \right] \,,
\end{align}
into our vertex factors, which are diagrams $(1)$, $(3)$, $(6)$, and $(6r)$.

\begin{figure}
    \begin{subfigure}[b]{0.49\textwidth}
        \centering
        \includegraphics[height=2.0in]{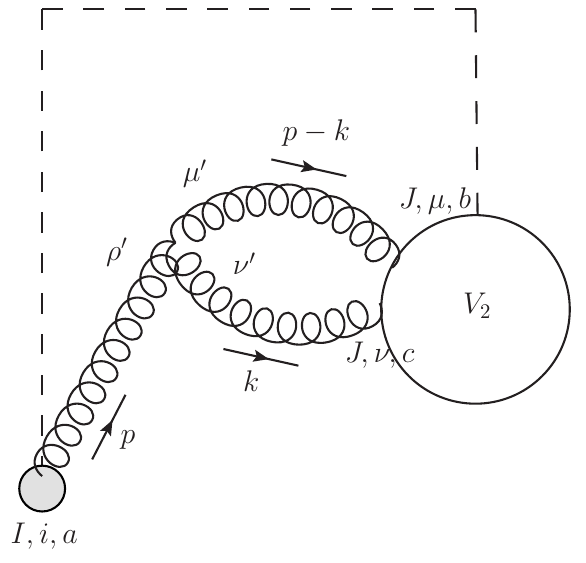}
        \caption{}\label{subfig:V4_1}
    \end{subfigure}%
    ~
    \begin{subfigure}[b]{0.49\textwidth}
        \centering
        \includegraphics[height=2.0in]{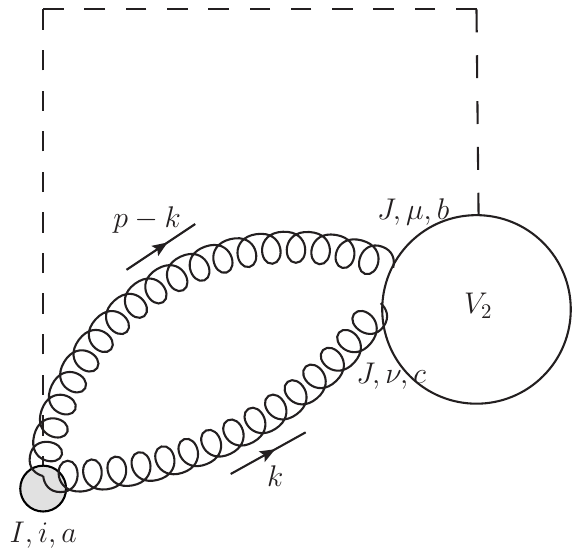}
        \caption{}\label{subfig:V4_2}
    \end{subfigure}%
    \caption{Diagrams with $O((1-\xi)^2)$ gauge dependence, with $V_2$ an arbitrary 2-gauge boson vertex to which the electric field, represented by the grey blob on the left, is connected in either of the two ways (a) and (b) shown.}
    \label{fig:gaugeProofDiagsxi2}
\end{figure}

It turns out that the cancellation can be proven from the common sub-diagram structures of these four. 
In particular, we consider the sub-diagrams of Figure~\ref{fig:gaugeProofDiagsxi2}. As we already emphasized, any diagram containing a term proportional to $(1-\xi)^2$ contributes in this way: the only diagrams that can contain two factors of $(1-\xi)$ are those that have two propagators that are not directly connected to an electric field via a single gauge boson line. 
Moreover, if we group our diagrams as (1)+(6r) and (3)+(6), we find each combination corresponds exactly to one of the two structures shown in  Figure~\ref{fig:gaugeProofDiagsxi2}. If we explicitly add the diagrams in Figure~\ref{fig:gaugeProofDiagsxi2}, we obtain\footnote{For the purposes of our calculation, we only need to consider the situation where the gauge boson propagators that are attached to $V_2$ have the same Schwinger-Keldysh indices $J$. In general, the $J$ indices can be different.}
\be
    && \big(ip_0 \D_{i\rho'}^{aa'}(p)_{I'I} - ip_i \D_{0 \rho'}^{aa'}(p)_{I'I} \big) \D_{ \mu' \mu}^{b' b}(p-k)_{JI'} \D_{\nu' \nu}^{c'c}(k)_{JI'} (-1)^{I'+1} \nonumber \\
&\times& g f^{a'b'c'} \left[ g^{\rho' \mu'} (2p-k)^{\nu'} + g^{\mu' \nu'} (2k-p)^{\rho'} + g^{\nu' \rho'} (-k-p)^{\mu'} \right] \nonumber \\
& +& gf^{ab'c'} \left[ \D_{i\mu}^{c'b}(p-k)_{JI} \D_{0\nu}^{b'c}(k)_{JI} + \D_{i\nu}^{c'c}(k)_{JI} \D_{0\mu}^{b'b}(p-k)_{JI} \right] \,.
\ee
Focusing on the $(1-\xi)^2$ piece, we can replace the propagators $\D_{\mu \nu}^{ab}(p-k)$ and $\D_{\mu \nu}^{ab}(k)$ with
\begin{align}
\D_{\mu \nu}^{ab}(p)_{JI} \to \delta^{ab} (1-\xi) \frac{p_\mu p_\nu}{p^2} \D(p)_{JI} \,,
\end{align}
where the last propagator $\D(p)_{JI}$ is the scalar thermal propagator. The replacement yields
\begin{align}
    & (-ip_0 g_{i \rho'} + ip_i g_{0 \rho'}) (1 - \xi)^2 \frac{(p-k)_{\mu} (p-k)_{\mu'} }{(p-k)^2} \frac{k_\nu k_{\nu'}}{k^2} \D(p)_{II'} \D(p-k)_{I'J} \D(k)_{I'J} (-1)^{I'+1} \nonumber \\
    & \quad \quad \quad \times g f^{abc} \left[ g^{\rho' \mu'} (2p-k)^{\nu'} + g^{\mu' \nu'} (2k-p)^{\rho'} + g^{\nu' \rho'} (-k-p)^{\mu'} \right] \nonumber \\
    & + gf^{abc} (1-\xi)^2 \left[  - \frac{(p-k)_i (p-k)_\mu}{(p-k)^2} \frac{k_0 k_\nu}{k^2} + \frac{k_i k_\nu}{k^2} \frac{(p-k)_0 (p-k)_\mu}{(p-k)^2} \right] \D(p-k)_{IJ} \D(k)_{IJ} \nonumber \\
    &= g f^{abc} (1-\xi)^2 \frac{k_\nu (p-k)_\mu}{k^2 (p-k)^2} \bigg[ \D(p)_{II'} \D(p-k)_{I'J} \D(k)_{I'J} (-1)^{I'+1} i p^2 (p_0 k_i - p_i k_0) \nonumber \\
    & \quad \quad \quad\quad\quad\quad\quad\quad\quad\quad\quad\quad\quad\quad\quad\quad\quad\quad + \D(p-k)_{IJ} \D(k)_{IJ} (p_0 k_i - k_0 p_i) \bigg] \nonumber \\
    &= g f^{abc} (1-\xi)^2 \frac{k_\nu (p-k)_\mu}{k^2 (p-k)^2} (p_0 k_i - p_i k_0) \bigg[ - \mathbbm{1}_{II'} \D(p-k)_{I'J} \D(k)_{I'J} + \D(p-k)_{IJ} \D(k)_{IJ}  \bigg] \nonumber \\
    &= 0 \,.
\end{align}
This proves that all the terms proportional to $(1-\xi)^2$ in Figure~\ref{fig:gaugeProofDiagsxi2} cancel out. Since all the diagrams with the contributions proportional to $(1-\xi)^2$ appear in this way, we have shown that the full NLO result has no $(1-\xi)^2$ terms. Now we have to verify that the remaining $(1-\xi)$ dependence also cancels.

\subsubsection{Cancellation at \texorpdfstring{$O(1-\xi)$}{O_{1-x}}} \label{sec:gauge-inv-1m-xi1}

As illustrated by the previous cancellation, the general strategy we will use to prove gauge invariance relies on identifying the factor of a momentum flowing through the diagram ($p$, $k$, or $p-k$) squared, and using them to ``cancel'' some propagator insertions so that the diagrams with a greater number of propagators, such as $(1)$, can be added seamlessly with other diagrams with fewer gauge boson propagators.

Because of the reflection symmetry relating diagrams $(X)$ in Figure~\ref{fig:diagrams} with diagrams $(Xr)$ (implemented by reversing the flow of momentum and exchanging the external color, spatial, and contour indices), we actually only need to verify that the following cancellation takes place:
\begin{align}\label{eq:gauge-cancellation}
\frac{\partial}{\partial (1-\xi)} \left( \frac{1}{2} \big[ (1) + (2) + (3) + (4) \big] + (5) + (6) + (7) + (8) \right) = 0 \,,
\end{align}
for any value of $\xi$. We have not included diagram $(11)$ in this list because it is automatically $R_\xi$ gauge-invariant: all of its gauge boson propagators are connected to a electric field via a single attachment, which gives a momentum structure that automatically removes all $\xi$ dependence in the propagator:
\begin{align}
\big[ i p_0 g_{i\mu} - i p_i g_{0\mu} \big] P^{\mu \nu}(p) =  (-1) \big[ i p_0 g_{i}^{\nu} - i p_i g_{0}^{\nu} \big] \,.
\end{align}

We show that~\eqref{eq:gauge-cancellation} holds identically in Appendix~\ref{app:gauge-invariance} by explicitly working out each diagram. Specifically, we show that the contributions linear in $(1-\xi)$ cancel. It turns out that all of the gauge-dependent diagrams need to be taken into account: the cancellation does not hold for any separate subset of diagrams. To illustrate schematically how the cancellations take place, we summarize them as follows:
\begin{align}
    (1)_{(1-\xi), \, \not\propto p^2} + (2)_{(1-\xi)} &= 0\,, \\
    \Big(\frac{1}{2} (1)_{(1-\xi), \, \propto p^2} + \big((5)+(6)\big)_{(1-\xi)} \Big)_{\propto (p-k)^2} + (7)_{(1-\xi)} &= 0\,, \\
    \Big(\frac{1}{2} (1)_{(1-\xi), \, \propto p^2} + \big((5)+(6)\big)_{(1-\xi)} \Big)_{\propto p^2} + \frac{1}{2} \left( (3)_{(1-\xi)} + (4)_{(1-\xi)} \right) + (8)_{(1-\xi)} &= 0\,,
\end{align}
where the notation ${(X)}_{(1-\xi)}$ represents the piece of diagram $(X)$ that is linear in $(1-\xi)$. Also, with the subscripts ${}_{\propto p^2}$ we only keep the terms in a given vertex factor that are proportional to the quantity in the subscript while with the subscripts ${}_{\not\propto p^2}$ we keep all the remaining terms that are not proportional to the quantity in the subscript. These subscripts essentially label which propagator in the diagram is cancelled via~\eqref{eq:propcancellation}.\footnote{However, we warn the reader that they depend on the convention taken for the momentum flow in the diagram, and so this structure of cancellations is strictly true only if the calculation is carried out consistently with the conventions we have taken so far and in Appendix~\ref{app:gauge-invariance}.}

In summary, after adding up all the diagrams, the result is independent of $\xi$. This verifies the expectation from the construction of the electric field correlation function, where the Wilson line insertions exactly guarantee the gauge invariance of the correlator. This means that we can confidently choose any (non-singular) gauge to perform the calculation. Throughout the rest of this work, we will choose the Feynman gauge, $\xi=1$.

\subsection{Calculations in Feynman gauge} \label{sec:feyn-calc}

We now proceed to evaluate each diagram that contributes to the chromoelectric correlator. However, to simplify the analysis, we will solely focus on calculating the object that contributes to the inclusive bound state formation/dissociation rates, which is the integrated spectral function
\begin{align}
\label{eqn:rho_p_integrated}
\varrho^{++}_E(p_0) = \frac{1}{2} \int \frac{\diff^{d-1} \p}{(2\pi)^{d-1}} \delta^{ad} \delta_{ij} \big[ \rho_E^{++} \big]_{ji}^{da}(p_0,\p) \,,
\end{align}
directly related to the spectral function~\eqref{eq:rho-++-def} we introduced earlier by an overall prefactor $T_F g^2/(3N_c)$, which in turn determines the dynamics of quarkonium in medium through Eqs.~\eqref{eq:singlet-from-GGD} and~\eqref{eq:octet-from-GGD}. The overall factor of $1/2$ is a choice of normalization, just to cancel the factor of $2$ when we write the spectral function as the real part of the retarded propagator, see Eq.~(\ref{eq:spectral-retarded}) from below. 

As a baseline, we first show, explicitly in $d=4-\epsilon$, how one can obtain the spectral function at LO in a way that is also applicable to higher order contributions. Then, we will consider the diagrams contributing at NLO. Since a naive treatment of the integrals in $d=4$ leads to UV divergences, we use will dimensional regularization with $d = 4 - \epsilon$ and the $\overline{\rm MS}$ renormalization scheme throughout the calculation of these diagrams. We will only take the limit $\epsilon \to 0$ at the very end for diagrams that are UV divergent while for those that are UV finite, we take $d=4$ as early as possible. We will split the calculation into three blocks: First, we start with the calculation of the textbook diagrams that contribute to the gauge boson self-energy in Section~\ref{sec:gluon-self-energy}, quoting the calculation of the $T=0$ quantum field theory result in the $\overline{\rm MS}$ scheme from textbooks as an input to our calculation, and also explicitly calculating the finite-temperature pieces, verifying that the results are consistent. Then, we devote a section to the diagrams with three propagators $(5)$, $(5r)$, $(6)$, $(6r)$ in~\ref{sec:3-propagator-evaluation}, because they share similar properties in the pole structure of their propagators. We calculate these four diagrams for both the vacuum ($T=0$) contribution and the finite temperature contribution. Finally, in section~\ref{sec:2-propagator-evaluation} we discuss the calculation of the remaining diagrams, which are all made up of only two propagators.

In the first two forthcoming subsections, we will make direct use of the relation
\begin{align}\label{eq:spectral-retarded}
\rho(p) = G^>(p) - G^<(p) = 2{\rm Re} \left\{ G^R(p) \right\} = 2 {\rm Re} \left\{ G^T(p) - G^<(p) \right\} \,,
\end{align}
which is a well-known relation between the spectral function and the retarded correlation function~\cite{Laine:2016hma}. The relation between the spectral function and the retarded propagator generally holds for two-point functions of operators that are local in time. This is directly applicable to the LO result, as well as to the NLO contributions from the traditional gauge boson self-energy (diagrams $(1),(2),(f),(g)$), because they only involve correlation functions of local operators. In the case of diagrams involving Wilson lines, it is not obvious that the relation between the spectral function and the retarded propagator is true. Therefore we feel compelled to first present the calculation in a way that does not use this property explicitly, and thus we will resort to using the usual definition $\rho_G(p) = G^>(p) - G^<(p)$ for these cases. 

\subsubsection{Leading order result: a single gauge boson propagator}

As an introductory step, we consider the LO calculation of the momentum-integrated spectral function:
\begin{equation}
    \left. \varrho_E^{++}(p_0) \right|_{\rm{LO}} = \delta_{ij} \delta^{ad}  \int \frac{\diff^{d-1}\p}{(2\pi)^{d-1}} \delta^{ad}\, {\rm Re} \left\{  \frac{(i p_0 g_{i\mu} - i p_i g_{0 \mu} ) i P^{\mu \nu}(p) (-i p_0 g_{j\nu} + i p_j g_{0 \nu} ) }{p_0^2 - \p^2 + i0^+ {\rm sgn}(p_0)}  \right\}\,,
\end{equation}
which we have calculated by taking the real part of the corresponding retarded correlation. The retarded correlation for the electric fields can be constructed by applying derivatives on the retarded correlation function of two gauge fields (viz. the $(i p_0 g_{i\mu} - i p_i g_{0 \mu} )$ factors in the above equation).\footnote{At this point, there is a subtlety that needs to be addressed in the definition of the retarded correlation function, because derivatives do not commute with the time-ordering operator that defines the correlator (in this case, a retarded correlation). The usual choice when defining time-ordered correlation functions in QFT textbooks is to take the derivatives outside the time-ordering symbol (i.e., $\mathcal{T}^\ast$-product), because it is this choice that respects Lorentz covariance. For our present purposes this is inconsequential, because when we take the real part of the retarded correlator the ambiguity is removed as $p_0$ can be replaced by $|\p|$. But in any calculation where retarded objects are involved, one needs to have a consistent definition. We follow standard practice, and define retarded/time-ordered correlations with the time-ordering operator acting first on the fundamental fields, and then calculate the action of derivatives on them so that the time-ordered correlator is compatible with Lorentz covariance.} Evaluating directly in $d$ dimensions, we obtain
\begin{align}
\left. \varrho_E^{++}(p_0) \right|_{\rm{LO}} = (N_c^2-1) \frac{(d-2) \pi \Omega_{d-1}}{2(2\pi)^{d-1}}  |p_0|^{d-1} {\rm sgn}(p_0) \,.
\end{align}
We will keep these $d$-dependent prefactors wherever possible in our calculation, as they will simplify the analysis when we later introduce the coupling constant renormalization to compensate for the UV divergences.

\subsubsection{Traditional gauge boson self-energy: diagrams \texorpdfstring{$(1)$}{(1)}, \texorpdfstring{$(2)$}{(2)}, \texorpdfstring{$(g)$}{(g)}, \texorpdfstring{$(f)$}{(f)}} \label{sec:gluon-self-energy}

Now we turn to the calculation of the the gauge boson self-energy contributions to the integrated spectral function. We start by quoting the textbook result~\cite{Schwartz:2013pla} for the $T=0$ case, where the time-ordered self-energy is given in the $\overline{\rm MS}$ scheme
\begin{equation}
\begin{split}
        i \mathcal{M}_{\ml{T}}^{ad, \mu \nu}(p) = i \delta^{ad} \frac{g^2}{16 \pi^2} (p^2 g^{\mu \nu} - p^\mu p^\nu) & \bigg[ N_c \left( \frac{10}{3\epsilon} + \frac{31}{9} + \frac{5}{3} \ln \left( \frac{\mu^2}{-p^2-i0^+} \right) \right) \\ & - n_f C({\bs N}) \left( \frac{8}{3\epsilon} + \frac{20}{9}  + \frac{4}{3} \ln \left( \frac{\mu^2}{-p^2-i0^+} \right) \right) + \O(\epsilon) \bigg] \,,
\end{split}
\end{equation}
where ${\ml{T}}$ indicates the quantity is time-ordered.
The retarded self-energy can be obtained by replacing $(p^2 + i0^+)$ with $(p^2 + i0^+{\rm sgn}(p_0))$. Since the spectral function is an odd function of $p_0$, we can actually focus only on $p_0>0$ and therefore ignore the distinction between retarded and time-ordered self-energies in this case. Then, as long as we take $p_0>0$, we can compute the vacuum contribution to the momentum-integrated spectral function as
\begin{equation}
\begin{split}
    & \left. \varrho_E^{++}(p_0) \right|_{{\rm NLO}, \, T=0}^{\rm self-energy} \\ &= \delta^{ad} \int \frac{\diff^{d-1} \p}{(2\pi)^{d-1}} {\rm Re} \left\{ \frac{i (p_0 g_i^{\mu'} - p_i g_0^{\mu'} ) (i P_{\mu \mu'}(p)) i \mathcal{M}_{\ml{T}}^{ad, \mu \nu}(p) (i P_{\nu \nu'}(p)) (-i p_0 g_j^{\nu'} + i p_j g_0^{\nu'} ) }{(p^2 + i0^+)^2}  \right\}\,,
\end{split}
\end{equation}
where we have included the two ``external'' propagators that are present in our original formulation in the Schwinger-Keldysh contour. One can directly check from the path integral that the complete retarded correlation has the structure $D^R(p) i \ml{M}_R D^R(p) $ in terms of the retarded self-energy $\ml{M}_R$. It then follows that
\begin{equation}
\begin{split}
    \left. \varrho_E^{++}(p_0) \right|_{{\rm NLO}, \, T=0 }^{\rm self-energy} &= (N_c^2 - 1) \frac{(d-2) \pi \Omega_{d-1}}{2 (2\pi)^{d-1}} \frac{g^2}{4\pi^2} p_0^{d-1} \bigg[ N_c \left( \frac{5}{6\epsilon} + \frac{14}{9} + \frac{5}{12} \ln \left( \frac{\mu^2}{4 p_0^2} \right)  \right)   \\ & \quad \quad \quad \quad \quad \quad \quad \quad \quad \quad \quad \quad \quad  - n_f C({\bs N}) \left( \frac{2}{3\epsilon} + \frac{10}{9} + \frac{1}{3}  \ln \left( \frac{\mu^2}{4 p_0^2} \right) \right) \bigg]\,,
\end{split}
\end{equation}
is the full contribution from the gauge boson self-energy for $p_0>0$. The extension to $p_0<0$ is given by naturally continuing the function with the rule $ \varrho_E^{++}(-p_0) = -\varrho_E^{++}(p_0) $. It is also straightforward to check that the fermion sector matches the result of the Abelian case~\cite{Binder:2020efn}.

Now we explain the calculation of the finite temperature pieces. Since the fermion loop contribution has already been discussed in previous work by some of us, we will simply quote the result at the end of this section, and focus only on the gauge boson and ghost diagrams for the purpose of this discussion. We commence by analyzing the propagator structure of these diagrams in the retarded case, i.e., (looking at the first row of Table~\ref{tab:4prop-structure})
\begin{align}
\ml{Q}^{(1)}_{11} - \ml{Q}^{(1)}_{12} = [D^R(p)]^2 \left[ D^{\ml{T}}(k) D^{\ml{T}}(p-k) - D^<(k) D^<(p-k) \right]\,,
\end{align}
where the equality follows after some algebra. Here we plan to compute the spectral function by taking the real part of the retarded propagator since the diagram (1) only consists of local operators, as discussed below Eq.~(\ref{eq:spectral-retarded}). This also verifies that the correspondence between retarded and time-ordered self-energies we used earlier is consistent: if we set $p_0>0$ in the $T=0$ case, the combination of propagators $D^<(k) D^<(p-k)$ vanishes identically because $D^<(p-k)$ is proportional to $\theta(-p_0+k_0)$ and is nonzero only if $p_0-k_0 < 0$, implying $k_0>0$, but then $D^<(k)$ vanishes due to its own theta function. Then, the retarded self-energy reduces to the structure $D^{\ml{T}}(k) D^{\ml{T}}(p-k)$, which is exactly the time-ordered self-energy, meaning that, as expected, $G^R(p) = G^T(p)$ in vacuum when $p_0>0$.

For the full calculation of both vacuum and finite temperature contributions, what we need to calculate is
\begin{equation}
\begin{split}
    & \left. \varrho_E^{++}(p_0) \right|_{{\rm NLO} }^{\rm self-energy} \\ &= (N_c^2 - 1) g^2 \int_\p \int_k N(p,k) [D^R(p)]^2 \left[ D^{\ml{T}}(k) D^{\ml{T}}(p-k) - D^<(k) D^<(p-k) \right]\,,
\end{split}
\end{equation}
where, in terms of Table~\ref{tab:4prop-vertex}, $N(p,k) = \delta_{ij} \left[V^{(1)}(p,k)_{ji} + V^{(g)}(p,k)_{ji} \right]$, and we have introduced the shorthands $\int_\p = \int \frac{\diff^{d-1}\p}{(2\pi)^{d-1}}$, $\int_k = \int \frac{\diff^d k}{(2\pi)^d}$. One can then manipulate the propagators that depend on $k$ to show that\footnote{The procedure that leads to this equality is only valid if the numerator $N(p,k)$ does not have any poles as a function of $k_0$. This is not strictly true in the case where we have Wilson lines, whose nonlocality in time introduces a pole at $k_0 = 0$, and a more careful treatment would be needed.}
\begin{equation}
\begin{split}
    & \left. \varrho_E^{++}(p_0) \right|_{{\rm NLO} }^{\rm self-energy} \\ &= (N_c^2 - 1) g^2 \int_\p \int_k N(p,-k) [D^R(p)]^2 \left[ D^<(k) D^R(p+k) +  D^<(p+k) D^A(k) \right]\,,
\end{split}
\end{equation}
which has been written in a form that is more appropriate for the finite temperature calculation. To see why this is so, note that we can write the advanced and retarded propagators as
\begin{align}
D^R(k) &= i \int \frac{\diff k_0'}{2\pi} \frac{D^>(k_0',\k) - D^<(k_0',\k)}{k_0 - k_0' + i0^+} \\
D^A(k) &= i \int \frac{\diff k_0'}{2\pi} \frac{D^>(k_0',\k) - D^<(k_0',\k)}{k_0 - k_0' - i0^+}\,,
\end{align}
which, after some algebra, leads to (we define $\q = \p + \k$)
\begin{equation} \label{eq:integrated-gluon-self-energy-before-integration}
\begin{split}
\left. \varrho_E^{++}(p_0) \right|_{{\rm NLO} }^{\rm self-energy} &=  (N_c^2 - 1) g^2 \int_{\k, \q} \frac{1+2n_B(|\k|) }{ 2|\k| 2|\q|} \\ & \quad \times {\rm Re} \left\{ \frac{-i}{( (p_0+ i0^+)^2 - (\q - \k)^2 )^2 } \sum_{\sigma_1,\sigma_2} \frac{\sigma_2 N((p_0,\q-\k),-k)_{k_0 = \sigma_1 |\k|} }{p_0 + \sigma_1 |\k| - \sigma_2 |\q| + i0^+ } \right\} \,.
\end{split}
\end{equation}
This is the expression for the complete contribution (vacuum + finite temperature) in an arbitrary number of dimensions $d$. Since we already evaluated the vacuum contribution, we can drop the $1$ in $1 + 2n_B(k)$ and set $d=4$ right away to evaluate the remaining piece, because the presence of the Bose-Einstein distribution guarantees that the integrals are convergent in the UV limit.

Naively, however, an issue arises in the collinear limit, where $\k$ becomes parallel to $\p$ (or, equivalently, to $\q$). This is because when one takes the real part of this expression, as there is a factor of $i$ in the numerator, one is effectively evaluating the sum of the residues of the poles at the locations where the denominator becomes zero. This is equivalent to what one would do, in a standard QFT textbook, by using ``cutting rules''. The problem appears when one tries to evaluate these ``cuts'' separately, because the integrals over the remaining momentum degrees of freedom (after taking the residue) separately diverge for each pole, and only their sum gives a finite, meaningful result. There are various ways to deal with this, both with and without introducing extra regulators, and, importantly, the result is independent of the choice of the methods we use to deal with this divergence. We discuss two of these methods in detail in Appendix~\ref{app:collinear-integrals}.

Regardless of the choice of integration ordering, one arrives at
\begin{equation}
\begin{split}
&\left. \varrho_E^{++}(p_0) \right|_{{\rm NLO}}^{\rm gauge boson+ghost} - \left. \varrho_E^{++}(p_0) \right|_{{\rm NLO}, \, T=0}^{\rm gauge boson+ghost} \\ &=  \frac{(N_c^2 - 1) g^2}{(2\pi)^3} N_c \int_{0}^\infty \diff k \,2 n_B(k) \left[ -2 k p_0 + (k^2+p_0^2) \ln \left| \frac{k+p_0}{k-p_0} \right| + k p_0 \ln \left| \frac{k^2-p_0^2}{p_0^2} \right| \right] \,,
\end{split}
\end{equation}
which has no known (to us) closed form. The interested reader can note that, asymptotically, the $\O(1/k)$ piece in the series expansion of the term in square brackets is $\frac{5}{3} \frac{p_0^3}{k}$, which is closely related to the fact that the dimensionally-regularized $T=0$ UV divergence in these diagrams goes as $\frac56 \frac{1}{\epsilon}$. To get the factor of $1/2$ correctly, one has to do the computation completely in dimensional regularization, which requires a level of involvement that is unnecessary for our present purposes, since here we focus on the finite temperature piece which has no UV divergence.

Combining self-energy contributions from the fermion loop, the ghost loop and the gauge boson loop, we obtain
\begin{equation}
\begin{split}
&\left. \varrho_E^{++}(p_0) \right|_{{\rm NLO}}^{\rm self-energy} \\
&= g^2 (N_c^2 - 1) \bigg\{ \frac{(d-2) \pi \Omega_{d-1}}{2 (2\pi)^{d-1} (4\pi^2)} p_0^{d-1} \bigg[ N_c \left( \frac{5}{6\epsilon} + \frac{14}{9} + \frac{5}{12} \ln \left( \frac{\mu^2}{4 p_0^2} \right)  \right)   \\ & \quad \quad \quad \quad \quad \quad \quad \quad \quad \quad \quad \quad \quad \quad \quad - n_f C({\bs N}) \left( \frac{2}{3\epsilon} + \frac{10}{9} + \frac{1}{3}  \ln \left( \frac{\mu^2}{4 p_0^2} \right) \right) \bigg] \\ 
& +  \int_{0}^\infty \!\! \diff k \frac{2 N_c n_B(k)}{(2\pi)^3}   \bigg[ -2 k p_0 + (k^2+p_0^2) \ln \left| \frac{k+p_0}{k-p_0} \right|  + k p_0 \ln \left| \frac{k^2-p_0^2}{p_0^2} \right| \bigg] \\
&  + \int_0^\infty \!\! \diff k \frac{2 N_f n_F(k)}{(2\pi)^3} \bigg[ -2k p_0 + (2k^2 + p_0^2) \ln \left| \frac{k+p_0}{k-p_0} \right| + 2 k p_0 \ln \left| \frac{k^2 - p_0^2}{p_0^2} \right| \bigg] \bigg\}\,,
\end{split}
\end{equation}
up to $\O(\epsilon)$ terms that we need not keep track of in the limit $\epsilon \to 0$. We do keep $\O(\epsilon)$ terms in the prefactor of the $1/\epsilon$ pole because they will also appear in front of other UV poles, and thus contribute at $\O(\epsilon^0)$ in the end.

\subsubsection{Diagrams \texorpdfstring{$(5)$}{(5)}, \texorpdfstring{$(5r)$}{(5r)}, \texorpdfstring{$(6)$}{(6)}, and \texorpdfstring{$(6r)$}{(6r)}} \label{sec:3-propagator-evaluation}

We will compute both the vacuum and finite temperature contributions of these diagrams. Given the presence of non-local operators in the correlation, to avoid having to deal with the $1/k_0$ poles coming from the Wilson lines using contour integration, we will not calculate the spectral function by taking the real part of the corresponding retarded correlator, but rather, we will write down the spectral function directly in terms of the Wightman functions. As before, we work with $p_0>0$.

It is instructive to note where the difficulty with the $1/k_0$ poles comes from, and what is necessary to do in order to properly deal with them. Our approach allows us to evaluate the difference between the $(I=1, J=2)$ and $(I=2, J=1)$ correlation functions, but for these diagrams, the $[g_E^{++}]_{12}$ does not agree exactly with the definition of the correlator $[g_E^{++}]^{<}$ that is the KMS conjugate of the physical correlator $[g_E^{++}]^{>}$. This is so because the time-ordering implicit in $[g_E^{++}]_{12}$ does not follow the corresponding matrix product ordering of the Wilson lines.\footnote{Similar features of the time-ordering of operators explain the difference between the heavy-quark diffusion coefficient and the quarkonium transport coefficients at NLO.} To formulate the spectral function of physical interest without extending the Schwinger-Keldysh contour (so as to accommodate non-trivial operator orderings), it is more efficient to use the relation $[g_E^{++}]^{<}(t,{\bs x}) = [g_E^{--}]^{>}(-t,-{\bs x})$. The only new ingredient to do this calculation is the Feynman rule of a gauge boson insertion in a Wilson line going towards $t=-\infty$, which, after some algebra, on can show that only amounts to flipping the sign of the $i0^+$ prescription in our Feynman rule for the Wilson lines going towards $t=+\infty$.

We denote the sum of the vertex factors from diagrams $(5)$ and $(6)$ as
\begin{equation}
\begin{split}
    N^{(5),(6)}(p,k) = \delta_{ij} & \big[ V^{(5)}(p,k)_{ji} + V^{(6)}(p,k)_{ji}  \big]\,,
\end{split}
\end{equation}
where the superscript refers to the corresponding diagrams. After some algebra, using the relations between $[g_E^{--}]^>$ and $[g_E^{++}]^<$, and
using the propagator structures in Table~\ref{tab:3prop-structure}, one finds that
\begin{equation}
\begin{split}
   \frac{1}{2} \delta^{ad} \delta_{ij} \big[ \rho_E^{++} \big]^{da}_{ji} &= g^2 (N_c^2-1) \int_k {\rm Re} \bigg\{ N^{(5),(6)}(p,k) \big[ \big( D^>(p) - D^<(p) \big) D^{\ml{T}}(k) D^{\ml{T}}(p-k)  \\
   & \quad \quad \quad \quad \quad \quad \quad \quad   - D^{\overline{\ml{T}}}(p) \big( D^>(k) D^>(p-k) - D^<(k) D^<(p-k) \big) \big] \bigg\} \, .
\end{split}
\end{equation}
This can be decomposed into the contributions that we can obtain from the $(I=1, J=2)$ and $(I=2, J=1)$ correlation functions, and an extra piece, as we will show in a moment.

It is helpful to denote the sum of all of the vertex factors associated to these diagrams, symmetrized under $k \to p-k$, by
\begin{equation}
\begin{split}
    N_{3p}(p,k) = \frac{\delta_{ij}}{2} & \big[ V^{(5)}(p,k)_{ji} + V^{(5r)}(p,k)_{ji} + V^{(6)}(p,k)_{ji} + V^{(6r)}(p,k)_{ji} \\ & \, +  V^{(5)}(p,p-k)_{ji} + V^{(5r)}(p,p-k)_{ji} + V^{(6)}(p,p-k)_{ji} + V^{(6r)}(p,p-k)_{ji} \big]\,,
\end{split}
\end{equation}
where the subscript $3p$ indicates that the vertex factor originates from the diagrams with three propagators. This quantity, as opposed to $N^{(5),(6)}$ (which has both real and imaginary parts), is manifestly a purely imaginary number. 

Using these definitions, it follows that
\begin{equation}
\begin{split}
 \left. \varrho_E^{++}(p_0) \right|_{\rm NLO}^{5-6} &= g^2 (N_c^2-1) \! \int_{k,{\bs p}} \!\!\! {\rm Re} \bigg\{ i \left[ {\rm Im} \, N^{(5),(6)}(p,k)\right] \big[ \big( D^>(p) - D^<(p) \big) D^{\ml{T}}(k) D^{\ml{T}}(p-k)  \\
   & \quad \quad \quad \quad \quad \quad \quad \quad \quad \quad \quad - D^{\overline{\ml{T}}}(p) \big( D^>(k) D^>(p-k) - D^<(k) D^<(p-k) \big) \big] \bigg\} \\
   & \quad + g^2 (N_c^2-1) \! \int_{k,{\bs p}} \!\!\! {\rm Re} \bigg\{\left[  {\rm Re} \, N^{(5),(6)}(p,k)\right] \big[ \big( D^>(p) - D^<(p) \big) D^{\ml{T}}(k) D^{\ml{T}}(p-k)  \\
   & \quad \quad \quad \quad \quad \quad \quad \quad \quad \quad \quad - D^{\overline{\ml{T}}}(p) \big( D^>(k) D^>(p-k) - D^<(k) D^<(p-k) \big) \big] \bigg\} \\
   &= i g^2  (N_c^2-1) \tilde{\mu}^\epsilon \int_{\k,\p} \frac{1+2n_B(k)}{2k} \sum_{\sigma_1}  N_{3p}((p_0,\p),(\sigma_1 k, \k)) \\
   & \quad \quad \quad \quad \quad \quad \quad \quad \quad \times {\rm Re}\left\{ \frac{i}{((p_0+i\epsilon)^2 -\p^2 )((p_0- \sigma_1 k + i\epsilon)^2 - (\p-\k)^2 )} \right\} \\
   & \quad + g^2  (N_c^2-1) \pi \int_{\k,\p}   \frac{ \left[ k_0 N^{(5),(6)}(p,k) \right]_{k_0=0} }{\k^2} \ml{P} \left( \frac{1}{p_0^2 - (\p - \k)^2} \right) \delta(p^2)   \,,
\end{split}
\end{equation}
where the first set of integrals, containing the Bose-Einstein distribution explicitly, is to be identified with the contribution from the $(I=1, J=2)$ and $(I=2, J=1)$ correlation functions, whereas the last line is a consequence of carefully handling the $k_0=0$ pole using the physical KMS conjugate $[g_E^{++}]^{<}$ of $[g_E^{++}]^{>}$.

We have kept everything in $d$ dimensions explicitly. The last line can be done explicitly, and $d=4$ may be set right away for this piece in the final result. To compute the remaining 6-dimension integral, our strategy is to first carry out the $\p$ integral by introducing Feynman parameters to merge the denominators using the standard formulae for loop integrals in $d=4-\epsilon$ dimensions, and then perform the integrals over the remaining solid angle of $\k$ and the Feynman parameters, leaving only the integral over $k$ to be dealt with (details of the calculation can be found in Appendix~\ref{app:3-prop-detail}). We then get a result of the form
\begin{equation}
 \left. \varrho_E^{++}(p_0) \right|_{\rm NLO}^{5-6} = g^2 p_0^3  N_c (N_c^2-1) \bigg[  \frac{\pi^2/2}{(2\pi)^3} -  \frac{\Omega_{3-\epsilon} \tilde{\mu}^\epsilon}{(4\pi)^{(3-\epsilon)/2}} \Gamma \! \left( \frac{-1+\epsilon}{2} \right) \int_0^\infty \!\!\! \diff k \frac{ 1+2n_B(k)}{(2\pi)^{d-1}} K(k;\epsilon) \bigg]
 \,,
\end{equation}
with only the final integral over $k$ to be evaluated. The expression of $K(k;\varepsilon)$ can be found in Appendix~\ref{app:3-prop-detail}, and is obtained by carrying out the steps we just outlined. An important feature of $K(k;\epsilon)$ is that if one naively takes the limit $\epsilon \to 0$ before integrating over $k$, the integrand becomes UV divergent. However, as we show in Appendix~\ref{app:3-prop-detail}, it turns out that the sum of these diagrams in dimensional regularization is UV finite in vacuum as long as we take $\epsilon \to 0$ after performing the $k$ integration.
After dealing with the potentially divergent pieces in a careful way, i.e., extracting the terms that become UV divergent in the $\epsilon \to 0$ limit and integrating over them before taking $\epsilon \to 0$, one obtains
\begin{equation} \label{eq:to-checkApp52}
\begin{split}
    \left. \varrho_E^{++}(p_0) \right|_{{\rm NLO}, \, T=0 }^{5-6}
    = \frac{g^2 N_c }{(2\pi)^3} (N_c^2-1) p_0^3 \left[  1 + \frac{\pi^2}{3}  \right] + \mathcal{O}(\epsilon)\,.
\end{split}
\end{equation}

The $T>0$ contribution is free of UV divergences because of the presence of the Bose-Einstein distribution, and so one can simply take $\epsilon=0$ to get the physical result. One then obtains the full NLO contribution from these diagrams
\begin{equation}
\begin{split}
    &\left. \varrho_E^{++}(p_0) \right|_{{\rm NLO}}^{5-6} = \frac{g^2 N_c  (N_c^2-1)}{(2\pi)^3} p_0^3 \bigg[  1 + \frac{\pi^2}{3}   \\ 
    & \quad \quad \quad \quad \quad \quad \quad \quad \quad + \int_0^\infty \diff k \, \frac{2 n_B(k)}{p_0} \frac{k^2 p_0 + (k^3 + p_0^3) \ln \big| \frac{k-p_0}{p_0} \big| + (p_0^3 - k^3) \ln \big| \frac{k + p_0}{p_0} \big| }{k p_0^2 - k^3} \bigg] \,,
\end{split}
\end{equation}
plus $\O(\epsilon)$ terms which are irrelevant as the result is already finite. The denominator $k p_0^2 - k^3$ is actually a Cauchy principal value distribution, which gives a finite result after integration.

\subsubsection{Diagrams \texorpdfstring{$(3)$}{(3)}, \texorpdfstring{$(4)$}{(4)}, \texorpdfstring{$(7)$}{(7)}, \texorpdfstring{$(7r)$}{(7)}, \texorpdfstring{$(8)$}{(8)}, \texorpdfstring{$(8r)$}{(8r)}, \texorpdfstring{$(11)$}{(11)} } \label{sec:2-propagator-evaluation}

Finally, we evaluate the diagrams that have no 3-gauge boson vertices, and are purely contractions of fields through propagators. {There are no operator ordering issues here, because only diagrams where the gauge boson insertions at the Wilson lines are contracted with fields on the other side of the correlator contribute. The diagrams in which the gauge bosons from the Wilson lines are contracted with fields on the same side, i.e., diagrams $(9)$, $(9r)$, $(10)$, and $(10r)$ give vanishing contributions.} There are two types of propagator structures here, as can be seen from Table~\ref{tab:2prop-structure}, but it turns out it is necessary to calculate them together in order to cancel possible IR divergences, specifically between diagrams $(4)$, $(7)$, and $(7r)$. Calculating the momentum-integrated spectral function from the difference of the corresponding Wightman functions gives, after performing all integrals that do not involve the temperature dependent pieces,
\begin{align}
\left. \varrho_E^{++}(p_0) \right|_{{\rm NLO}}^{3-11} &= g^2 (N_c^2 - 1) N_c p_0^{-\epsilon} \frac{\pi \Omega_{3-\epsilon}^2}{4(2\pi)^{6-2\epsilon}} {\tilde{\mu}}^\epsilon \int_0^\infty \frac{\diff k }{k^{1+\epsilon}} \nonumber  \\
& \quad \times \bigg( (n_B(k) -n_B(k+p_0) ) (-(3-\epsilon)p_0^2 + (p_0+k)^2 )\Theta(p_0  + k) (p_0 + k) \nonumber \\
& \quad - (n_B(k) -n_B(k-p_0) ) (-(3-\epsilon)p_0^2 + (p_0-k)^2 ) \Theta( k - p_0) (-p_0 + k) \nonumber \\
& \quad + (1 + n_B(k) + n_B(p_0-k)  ) (-(3-\epsilon)p_0^2 + (p_0 - k)^2 ) \Theta(p_0 - k) (p_0 - k) \nonumber \\
& \quad - (1 + n_B(k) + n_B(-p_0-k)  ) (-(3-\epsilon)p_0^2 + (p_0 + k)^2) \Theta(-p_0 - k) (-p_0 - k) \nonumber \\
& \quad- (1 + 2n_B(k)) (-(2-\epsilon) p_0^2) p_0 \bigg) \,,
\end{align}
which contains both the finite-temperature contributions as well as the vacuum parts, in arbitrary $d = 4 - \epsilon$ dimensions.

It is instructive to evaluate the vacuum part separately from the rest. We obtain (assuming $p_0>0$ as before)
\be
\left.\varrho_E^{++}(p_0) \right|_{{\rm NLO},\, T=0}^{3,4,7,7r,8,8r,11} &=& (N_c^2 -1) \frac{(d-2) \pi \Omega_{d-1}}{2 (4\pi^2) (2\pi)^{d-1}} g^2 p_0^{d-1}  N_c \nn\\
&\times& \bigg[ \frac{1}{\epsilon} + \frac{7}{12}  + \frac{1}{2}  \ln \left( \frac{\mu^2}{p_0^2} \right) + \frac{\gamma_E + \psi(3/2)}{2} \bigg] \,,
\ee
where we have kept the same prefactors that appeared in the gauge boson self-energy and in the LO result in arbitrary dimensions, so that the UV divergent pieces can be added straightforwardly.

The contribution at finite temperature can be written in $d=4$, or equivalently $\epsilon=0$, without any issue of possible divergences. Doing so, we get
\be
& & \varrho_E^{++}(p_0) \big|_{{\rm NLO}}^{3,4,7,7r,8,8r,11} \nn\\
&=& g^2 (N_c^2 -1) N_c \bigg\{ \frac{(d-2) \pi \Omega_{d-1}}{2 (4\pi^2) (2\pi)^{d-1}}  p_0^{d-1}   \bigg[ \frac{1}{\epsilon} + \frac{7}{12}  + \frac{1}{2}  \ln \left( \frac{\mu^2}{p_0^2} \right) +  1 - \ln(2) \bigg] \nn\\[4pt]
&+& \frac{1}{2(2\pi)^3} \int_0^\infty \frac{\diff k}{k} \bigg[  6k^2 p_0 n_B(k) 
+ |k-p_0| (-3p_0^2 + (p_0 - k)^2) n_B(|p_0-k|) \nn \\[4pt]
&-& |k+p_0| (-3p_0^2 + (p_0+k)^2 ) n_B(|p_0+k|)  \bigg] \bigg\}\,, \label{eq:2-prop-fin-result}
\ee
which completes the calculation of all the NLO pieces. We will add all results in the subsequent section~\ref{sec:add-results}, where we will further rearrange~\eqref{eq:2-prop-fin-result} such that $n_B(k)$ becomes an overall factor in the integrand.

\subsection{Remarks on infrared and collinear safety}

Up to the UV renormalization that we will discuss in the next subsection, we have obtained explicitly finite results of the spectral function $\varrho_E^{++}(p_0)$ at NLO. That is to say, all potentially infrared and collinear divergent diagrams and subdiagrams have been added up to a physical result. In this subsection, we will discuss the infrared and collinear structures of the diagrams in detail, highlighting the aspects that require particular care when performing these calculations.

\subsubsection{IR aspects}

We first discuss the cancellation of the divergences in the IR limit. Specifically, in bosonic thermal field theory the Bose-Einstein distribution introduces an extra $1/|k_0|$ factor as we let $k_0 \to 0$ in $n_B(|k_0|)$, which means that Feynman diagrams at 1-loop and beyond are potentially more singular than their vacuum counterparts in the low-energy limit $|k_0| \ll T$. Moreover, if we look at an individual Feynman diagram, e.g., diagram $(4)$ (see Table~\ref{tab:2prop-vertex}), one can explicitly see that one encounters an IR divergence because of the extra factors of $1/k_0$ coming from the Wilson line, in addition to the one that originates from the thermal distribution.

Operationally, however, individual Feynman diagrams with a single gauge boson insertion from a Wilson line are not IR divergent because the corresponding $1/k_0$ factor flips sign when approaching zero from either side (positive or negative), while the $1/|k_0|$ factor originated in the thermal distribution appearing in the free thermal propagator does not change sign, meaning that the integrand is an odd function of $k_0$ near the origin $k_0=0$. Since the real parts of the $1/k_0$ factors coming from Wilson lines must be interpreted as principal values, the integral around $k_0=0$ gives a finite contribution for these diagrams.

Diagrams with two gauge boson insertions from Wilson lines, with those gauge bosons connected through a propagator, are nonetheless individually divergent, because now we have a factor of $1/k_0^2$ coming from the Wilson lines that does not cancel with its reflected version $k_0 \to -k_0$ as we take $k_0 \to 0$. There are two such types of diagrams in our calculation: diagram $(4)$, which has a net momentum flow $p$ through the loop, and diagrams $(7)$, $(7r)$, which have their momentum flow $p$ disconnected from the loop momentum $k$ (see Figure~\ref{fig:diagrams}). To be explicit, if we take $(7)$ and $(7r)$ in Feynman gauge at $T=0$, we have
\begin{align}
    (7) &= \frac{N_c}{2} (p_0^2 \delta_{ij} - p_i p_j) \D(p)_{JI} \int_k \frac{1}{k_0^2} \D(k)_{II} \,, \\
    (7r) &= \frac{N_c}{2} (p_0^2 \delta_{ij} - p_i p_j) \D(p)_{JI} \int_k \frac{1}{k_0^2} \D(k)_{JJ} \,.
\end{align}

The problem is then clear: As we have written them, $(7)$ and $(7r)$ are unambiguously IR divergent for any $\p$, as is their sum, so it appears we have an ill-defined intermediate result. Nonetheless, in the case of our spectral function, we can get extra guidance on the IR divergence cancellation because the IR divergence in the contributions of $(7)$ and $(7r)$ to the spectral function becomes localized at $p_0^2 = \p^2$. It turns our that the integrals over $k$ are independent of whether the propagator indices are $I$ or $J$. So when we build the spectral function by taking $(J=2,I=1) - (J=1,I=2)$, we get a distribution with support only on $p_0^2 = \p^2$. Since distributional functions need to be integrated if one is interested in calculating an experimental observable, the cancellation of this IR divergence, if it happens, must be obtained by integrating over $\p$ in the remaining diagrams near $\p^2 = p_0^2$. Indeed, that is what actually happens. To illustrate this point, we point out that the contribution to the spectral function coming from diagrams $(3)$, $(4)$, $(8)$, $(8r)$, $(11)$, can be calculated explicitly in vacuum (for $p_0>0$) by integrating over $k$ in $d=4$:
\begin{equation}
\left. \frac{\delta_{ij} \delta^{ad} [\rho_E^{++}]^{da}_{ji}(p_0,\p)}{g^2 N_c (N_c^2 -1)} \right|^{(3),(4),(8),(8r),(11)}_{{\rm NLO}, \, T=0} = \frac{-1}{4\pi} \Theta (p_0 - |\p|) \frac{4 |\p| p_0^3 + (p_0^4 - \p^4) {\rm arctanh}(|\p|/p_0) }{ p_0 |\p| (p_0^2 - \p^2)}\,,
\end{equation}
which is manifestly divergent as $|\p| \to p_0$. Crucially, if one integrates this over a region including $|\p|=p_0$, then a divergence appears, with opposite sign to the one originated from diagrams $(7)$ and $(7r)$. This is one way to see the cancellation of this IR divergence between diagrams $(7)$, $(7r)$ and diagrams $(3)$, $(4)$, $(8)$, $(8r)$, $(11)$, in which we first integrate over $k$ and then integrate ${\bs p}$ over a region that contains $|{\bs p}|=p_0$. Both ${\bs p}$ integrals (for $(7)$, $(7r)$ and for $(3)$, $(4)$, $(8)$, $(8r)$, $(11)$) give divergent results, but with opposite signs, and these divergent pieces will cancel each other explicitly to give a finite result if we regulate the Wilson line $1/k_0$ contributions in the same way in all diagrams and then remove the regulator after the sum is performed.

On the other hand, if one first integrates over ${\bs p}$ on an arbitrary region, holding off on the IR divergent integrals over $k$, adds the results, and then performs the integration over $k$ for all diagrams ($(3)$, $(4)$, $(7)$, $(7r)$, $(8)$, $(8r)$, $(11)$) simultaneously, one does not encounter any IR divergences at all. This can be interpreted as an explicit verification that the result of performing the loop momentum $k$ integrals defines a distribution as a function of ${\bs p}$ that gives finite results over any integration region.
While at finite temperature it is no longer possible to get explicit expressions in terms of ordinary functions after integrating over $k$, the cancellation between the IR singularities of the different diagrams happens in the same way.

In vacuum, one could have tried to deal with these divergent diagrams by using dimensional regularization to regulate both the UV and IR divergences, and since diagrams $(7)$ and $(7r)$ give scaleless integrals, one can take them to be zero in dimensional regularization. But then, when one calculates diagram $(4)$ and integrates over some range of $\p$ that includes $|\p|=p_0$, one will encounter an uncompensated IR divergence. Since the theory has been already regulated in dimensional regularization with the above choice ($\epsilon_{\rm UV} = \epsilon_{\rm IR} = \epsilon$), one will find that to calculate the correct renormalization group running of the theory (which is in principle a UV effect), we would need to calculate the $1/\epsilon$ pole contribution by inspecting the IR divergence of diagram $(4)$ instead (whereas this $1/\epsilon$ pole would have come from the UV divergence of diagrams $(7)$ and $(7r)$ if we had not taken these diagrams to be zero). Of course, the simplest way to get the renormalization group equations is to just keep track of the UV divergences separately from the IR divergences. This is even more natural when we go to finite temperature, where the finite temperature contributions in the loop integrals of $(7)$ and $(7r)$ are manifestly UV finite, but clearly IR divergent. For that reason, we decided to first compute the integral over $\p$ in our actual calculation, because with that approach we render the IR sector finite from the start and thus we can identify all divergences as UV divergences, up to unphysical collinear singularities that we now discuss.

\subsubsection{The collinear limit}

Another nontrivial aspect of the calculation, at least from the point of view of a naive diagrammatic computation of the formation/dissociation rates, is how the result becomes finite in the limit where the momentum flow through the propagators in our diagrams becomes collinear, i.e., $\k$ becomes parallel to $\p$.

\begin{figure}
    \centering
    \includegraphics[width=\textwidth]{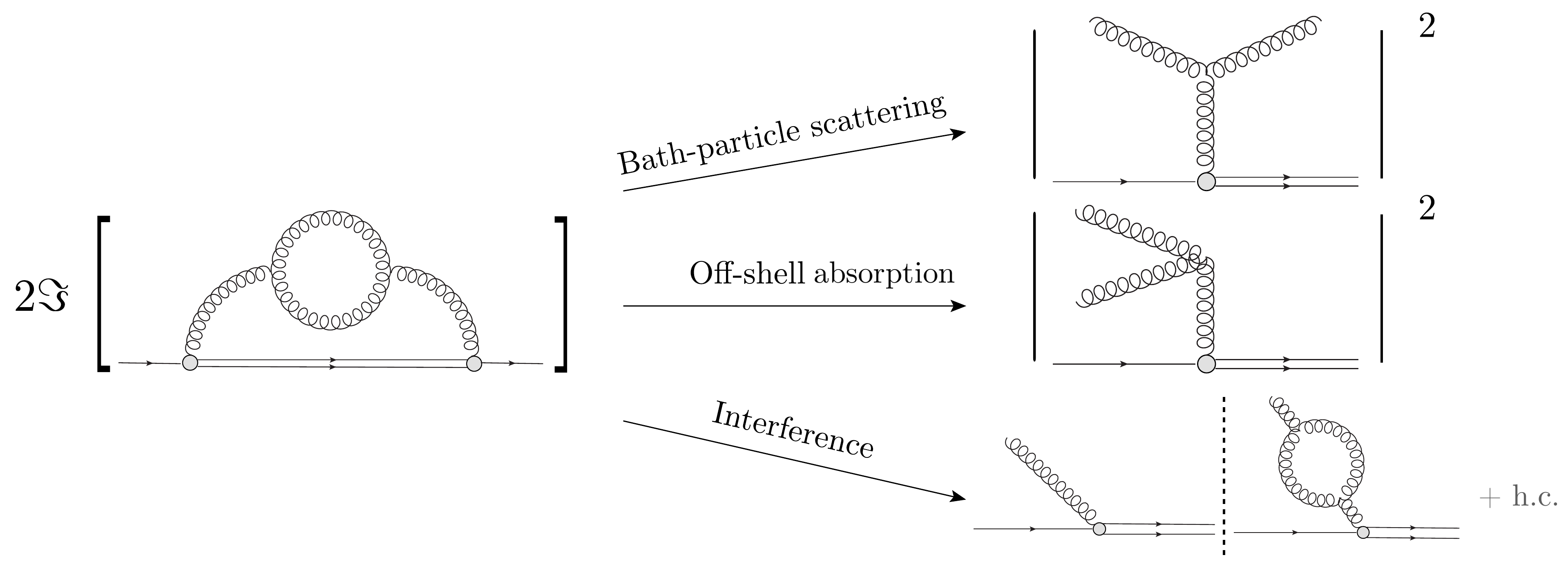}
    \caption{Decomposition of the imaginary part of the forward scattering amplitude of a singlet field with a gauge boson self-energy in terms of ``cut'' diagrams. When taken individually, these diagrams present singularities in the collinear limit, which only cancel after all terms are added.}
    \label{fig:collinear-gluon-self-energy}
\end{figure}

The way in which potential collinear divergences are cancelled is most clearly highlighted in Figure~\ref{fig:collinear-gluon-self-energy}, where the imaginary part of the forward scattering amplitude of a singlet field, associated with the gauge boson self-energy is decomposed in terms of products of tree-level amplitudes by means of the optical theorem.\footnote{Note that the cutting rules at finite temperature can involve loop momentum flow in either direction, i.e., the cuts do not determine that momentum flows in only one direction as in vacuum.} If treated individually, the ``cut'' diagrams are each separately divergent in the collinear limit $\p \parallel \k $, and therefore when one integrates over momenta to obtain a cross-section for each diagram the result appears to be divergent as well. This also means that, in general, one cannot get a reliable estimate of the size of the cross-section by simply looking at one diagram, e.g., scattering by particles in the thermal bath (upper right subdiagram in Figure~\ref{fig:collinear-gluon-self-energy}).

In practice, this means that when we evaluate the imaginary part of an amplitude, we must sum over all ``cuts'', which manifest as poles of propagators before carrying out the integrals. In our approach to the calculation, this is relevant when we take the real part of a retarded correlation, as shown in section~\ref{sec:gluon-self-energy}. When we wrote the contributions to the spectral function from diagrams $(5)$, $(5r)$, $(6)$, $(6r)$ in section~\ref{sec:3-propagator-evaluation}, we cannot evaluate the contributions from each pole of the propagators separately. Broadly speaking, one has to either regulate the divergence, calculate each contribution, add them up, and then remove the regulator, or, calculate the contribution of all poles simultaneously from the start. In obtaining our results, we follow two different integration orders, one that relies on the introduction of a regulator, and the other that does not, both of which are discussed in Appendix~\ref{app:collinear-integrals}. Both approaches agree and give the same final result, even though the resulting algebraic expressions, while equivalent, may have significant differences in the way they look (i.e., in terms of what functions they are expressed) when obtained from one method or the other.

A rigorous justification of why the result is finite is as follows: When integrating over $\p$, the numerator is a holomorphic function of the integration momentum $|\p|$ and the angle between $\p$ and $\k$. For definiteness, let us work under the assumption that we first perform the integral over $|\p|$. Given the propagator structure of the diagrams, all of the integrands we consider can be written in the form
\begin{equation}
    \ml{I}(z) = \frac{H(z)}{(z-z_1)^n (z-z_2)^m} \,,
\end{equation}
where $z$ is the complexified $|\p|$ integration variable, $z_1$ and $z_2$ are the positions of the poles (which are functions of the variables over which we are not integrating at this point, and are infinitesimally displaced off the real line by $\pm i0^+$, as usual in a propagator), and $n,m$ give the order of the poles, assuming that $H$ has no zeros at $z_1$ or $z_2$. Then, because of the concrete expressions we have for the vertex factors, $H(z)$ can be taken to be real on the real line, and then we can obtain the contribution to the retarded correlation functions by taking the imaginary part of the integral of $\ml{I}(z)$. This means that all we have to do is to evaluate the residues at their corresponding poles, because only they can yield an imaginary part. This leads to
\begin{equation}
\begin{split}
    {\rm Im} \left\{ \int \ml{I}(z) \diff z \right\} &=  \frac{\pi}{(n-1)!} \left. \frac{\diff^{n-1}}{\diff z^{n-1}} \left( \frac{H(z)}{(z-z_2)^m} \right) \right|_{z=z_1} + \frac{\pi}{(m-1)!} \left. \frac{\diff^{m-1}}{\diff z^{m-1}} \left( \frac{H(z)}{(z-z_1)^n} \right)  \right|_{z=z_2} \,.
\end{split}
\end{equation}
In this language, the collinear limit is realized when the two poles merge, i.e., $z_2 \to z_1$. If we naively evaluate each pole separately, this will generically lead to divergences because we will have contributions of the form $\lim_{z_2\to z_1} H(z_1)/(z_1-z_2)^k$ for some positive power $k$. However, if we take $z_1=z_2$ from the start, we see that the result actually must be finite and equal to
\begin{equation}
    {\rm Im} \left\{ \int \ml{I}(z) \diff z \right\} =  \frac{\pi}{(m+n-1)!} \left. \frac{\diff^{m+n-1}}{\diff z^{m+n-1}} \big( H(z)\big) \right|_{z=z_1=z_2} \,,
\end{equation}
which is simply a matter of evaluating the residue of a pole of higher order. We can also see that the limit is well defined from the point of view of $z_1 \neq z_2$, for which we show the two cases that are relevant for our purposes:
\begin{enumerate}
    \item $n=m=1$ gives
    \begin{equation}
        {\rm Im} \left\{ \int \ml{I}(z) \diff z \right\} = \pi \frac{H(z_1) - H(z_2)}{z_1-z_2} \xrightarrow{z_2 \to z_1} \pi H'(z_1) \,,
    \end{equation}
    which is the definition of the derivative of a function.
    \item $n=2$, $m=1$ case was already discussed in Ref.~\cite{Binder:2020efn} and gives
    \begin{equation}
    \begin{split}
        {\rm Im} \left\{ \int \ml{I}(z) \diff z \right\} &= \pi \left[ \frac{H'(z_1)}{(z_1-z_2)} - \frac{H(z_1)}{(z_1-z_2)^2} + \frac{H(z_2)}{(z_2-z_1)^2} \right] \\
        &= \pi \frac{H(z_2) - H(z_1) - (z_2 - z_1) H'(z_1) }{(z_2-z_1)^2} \xrightarrow{z_2 \to z_1} \frac{\pi}{2} H''(z_1) \,,
    \end{split}
    \end{equation}
    as can be verified, e.g., by a Taylor expansion.
\end{enumerate}

This conclusively shows that the apparent collinear singularities, which arise because setting one propagator on-shell induces a divergent behavior on other propagators in the collinear limit where $\p \parallel \k$, are all compensated within each uncut diagram, and therefore that any consistent method of calculating the integrals will give the same result.

\subsection{The complete NLO integrated spectral function}

\subsubsection{Coupling constant renormalization} \label{sec:coupling-renorm}

Before proceeding to give the final result at NLO, we discuss including the effects of the renormalization of the coupling constant. This is most easily understood if we recall that the quantity appearing in the transition rates is actually
\begin{equation}
    g^2 [g_E^{++}]^{da}_{ji}(y,x) = g^2 \Big\langle \big[E_j(y) \ml{W}_{[( y^0, {\bs y}), (+\infty, {\bs y})]} \big]^d
\big[ \ml{W}_{[(+\infty, {\bs x}),(x^0, {\bs x})]} E_i(x) \big]^a \Big\rangle,
\end{equation}
i.e., with an extra factor of $g^2$ that will be renormalized due to its own loop effects. Since the above expression comes directly from vertices in the Lagrangian, we can interpret every appearance of $g^2$ as a factor of the bare coupling constant, and then the appropriate substitution to include the corrections due to coupling constant renormalization is to substitute
\begin{align}
g^2 \to Z_{g^2} g^2.
\end{align}
Since $Z_{g^2} = 1 + \O(g^2)$, including modifications due to the renormalization of $g^2$ only affects our NLO calculation by changing the prefactor of the LO calculation, because any substitution of $g^2 \to Z_{g^2} g^2$ in our NLO results would lead to NNLO effects, of which we are not keeping track here.

At NLO, we can write
\be
\big( g^2 \varrho_E^{++}(p_0) \big) \big|_{\rm NLO} =  g^2(Z_{g^2}-1)\varrho_E^{++}(p_0)\big|_{\rm LO} + g^2 \varrho_E^{++}(p_0)\big|_{\rm NLO}\,.
\ee
So we need to combine the NLO result of the spectral function with the LO result multiplied by the factor $(Z_{g^2}-1)$. It is well known that
\begin{equation}
    Z_{g^2} =  1 - \frac{g^2}{8\pi^2 \epsilon} \left( \frac{11 N_c}{3} - \frac{4}{3} n_f C({\bs N}) \right) \,.
\end{equation}
Therefore, we obtain that
\begin{align}
g^2(Z_{g^2}-1)\varrho_E^{++}(p_0)\big|_{\rm LO} =  - g^4 (N_c^2-1) \frac{(d-2) \pi \Omega_{d-1}}{2 (4\pi^2) (2\pi)^{d-1}}  p_0^{d-1} \left( \frac{11 N_c}{6 \epsilon} - \frac{2}{3 \epsilon} n_f C({\bs N}) \right),
\end{align}
which we can now add to our previous results.

\subsubsection{Adding all results} \label{sec:add-results}

We are now in a position to present the final result for the momentum-integrated spectral function. Importantly, the $1/\epsilon$ pole cancels altogether in $g^2\varrho_E^{++}(p_0)$, which means this quantity and $g^2[g_E^{++}]$ are scale independent. Conveniently, we have kept all of the prefactors of the $1/\epsilon$ poles in the same form, and so the cancellation is straightforward. All that remains is to add up the finite pieces, where the limit $d \to 4$ (or $\epsilon \to 0$) is taken explicitly because there is no obstacle to it at this point. Therefore, adding up our results from sections~\ref{sec:gluon-self-energy},~\ref{sec:3-propagator-evaluation},~\ref{sec:2-propagator-evaluation}, and~\ref{sec:coupling-renorm}, we obtain
\begin{equation} \label{eq:spectral-full-result}
\begin{split} 
    &\left. g^2 \varrho_E^{++}(p_0) \right|_{{\rm LO} + {\rm NLO}} \\
    &= \frac{ g^2 ( N_c^2 -1 ) p_0^3}{(2\pi)^3} \bigg\{ 4\pi^2 + g^2  \bigg[ \left( \frac{11}{12} N_c - \frac13 N_f \right) \ln \left( \frac{\mu^2}{ p_0^2}\right) + a_g N_c  - a_F N_f   \bigg] \bigg\} \\
    & \,\, + \frac{g^4 (N_c^2-1)}{(2\pi)^3} \bigg\{ \int_0^\infty \!\! \diff k \, 2 N_f n_F(k) \bigg[ -2k p_0 + (2k^2 + p_0^2) \ln \left| \frac{k+p_0}{k-p_0} \right|  + 2 k p_0 \ln \left| \frac{k^2 - p_0^2}{p_0^2} \right| \bigg] \\
    & \quad \quad \quad \quad \quad \quad \, + \int_{0}^\infty \!\! \diff k \, 2 N_c n_B(k) \bigg[ -2 k p_0 + (k^2+p_0^2) \ln \left| \frac{k+p_0}{k-p_0} \right|  + k p_0 \ln \left| \frac{k^2-p_0^2}{p_0^2} \right| \bigg]  \\ 
    & \quad \quad \quad \quad \quad \quad \, + \int_0^\infty \diff k \, \frac{2 N_c n_B(k)}{k} \ml{P} \left( \frac{p_0^2}{ p_0^2 - k^2} \right) \bigg[ k^2 p_0 + (k^3 + p_0^3) \ln \left| \frac{k-p_0}{p_0} \right| \\ 
    & \quad \quad \quad \quad \quad \quad \quad \quad \quad \quad \quad \quad \quad \quad \quad \quad \quad \quad \quad \quad \quad \quad \quad  + ( -k^3 + p_0^3) \ln \left| \frac{k + p_0}{p_0} \right| \bigg]  \\
    & \quad \quad \quad \quad \quad \quad \, + \int_0^\infty \!\! \diff k \, 2N_c n_B(k) \mathcal{P} \left( \frac{k^3 p_0}{k^2 - p_0^2} \right) \bigg\}\,,
\end{split}
\end{equation}
where we have rearranged the integrand of~\eqref{eq:2-prop-fin-result} so that $n_B(k)$ appears as an overall factor. Also, we have denoted $N_f = n_f C({\bs N})$, and the numbers $a_g$ and $a_F$ are given by
\begin{align}
    a_g &\equiv {\frac{149}{36} + \frac{\pi^2}3 - \frac{11}{12}\ln(4)  \approx 6.157987}\,, \nonumber \\
    a_F &\equiv \frac{10}{9} - \frac{1}{3} \ln(4) \approx 0.649013 \,, \label{eq:consts}
\end{align}
and the coupling constant is given in the $\overline{\rm MS}$ scheme, with all finite pieces accounted for. Since in light of section~\ref{sec:coupling-renorm}, $g^2 \varrho_E^{++}$ does not run after the coupling constant renormalization is included, the RHS of~\eqref{eq:spectral-full-result} multiplied by $g^2$ is independent of $\mu$. So we have
\begin{equation}
\begin{split}
    &0 = 4\pi^2 \frac{\diff g^2}{\diff\ln \mu} \left(1 + \O(g^2) \right) + g^4  \left( \frac{11}{6} N_c - \frac23 N_f \right) + \O(g^6) \\
    &\implies \beta(\alpha) = \frac{\diff \alpha}{\diff\ln \mu} = -\frac{\alpha^2}{2\pi} \left( \frac{11}{3} N_c - \frac43 N_f \right) + \O(\alpha^3)\,,
\end{split}
\end{equation}
where $\alpha = \frac{g^2}{4\pi}$, verifying the expected running of the coupling for non-Abelian gauge theory.

\begin{figure}
    \centering
    \includegraphics[width=0.78\textwidth]{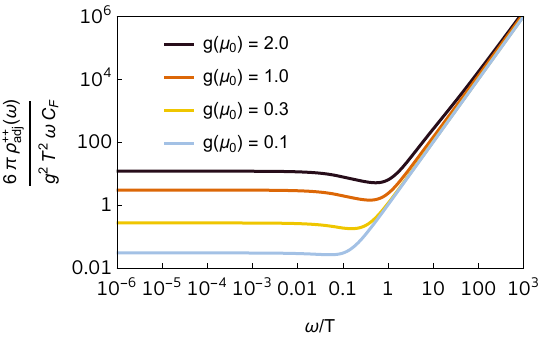}
    \caption{Spectral function for quarkonium transport in weakly coupled QCD, as given by Eq.~\eqref{eq:spectral-full-result}, with 2 light (massless) quarks for different values of the coupling at the reference scale $\mu_0 \approx 8.1 T$. We introduce the prefactor $6\pi/(C_F g^2 \omega T^2)$ such that the asymptotic behavior of all the curves is $\omega^2/T^2$ at large $\omega/T$. The temperature $T$ enters Eq~\eqref{eq:spectral-full-result} through the Fermi-Dirac $n_F$ and Bose-Einstein $n_B$ distributions, and $C_F = T_F (N_c^2 - 1)/N_c$. The coupling constant $g(\mu)$ is evolved to high energies using the 2-loop QCD beta function. Following~\cite{Burnier:2010rp}, and in the same way as we will do later on in this chapter, we choose the reference scale $\mu$ as a function of $\omega$ and $T$:
    $\mu(\omega, T) = \sqrt{T^2 \exp \left[ \ln(4 \pi) - \gamma_E - \frac{N_c - 8 \ln(2) N_f}{2 (11 N_c - 2 N_f)}\right]^2 + 
    \omega^2 \exp \left[\ln(2) + \frac{(6 \pi^2 - 149) N_c + 20 N_f}{6 (11 N_c - 2 N_f)}\right]^2}$.}
    \label{fig:spectral-noHTL}
\end{figure}

Although somewhat lengthy, it is remarkable that we can write an explicit expression (even though there is one integral left to be done, with which we deal numerically) for the full momentum-integrated spectral function at NLO, which involves $\mathcal{O}(20)$ diagrams in non-Abelian gauge theory with Wilson line insertions. See Fig.~\ref{fig:spectral-noHTL} to see the result of Eq.~\eqref{eq:spectral-full-result}, with the normalization convention $\rho_{\rm adj}^{++} = \frac{T_F g^2}{3 N_c} \rho_E^{++}$, as a function of $p_0/T$ (for later convenience, we relabel $p_0 \to \omega$), for various choices of the coupling constant.

{At this point, let us summarize the similarities and differences with the previous results in the literature. The first thing to note is that, up to an overall normalization factor, the contribution of fermions to the spectral function is the same as that obtained in the U$(1)$ case~\cite{Binder:2020efn}, as expected. Secondly, by assembling together all terms that have the same logarithm factors, the finite temperature part of our NLO result of the spectral function can be shown to be equal to that obtained in~\cite{Burnier:2010rp}, which calculates the correlator
\begin{align} \label{eq:correlator-intro-Eller}
\left\langle {\rm Tr}_{\rm color} \left[ U(-\infty,t) E_i(t) U(t,0) E_i(0) U(0,-\infty) \right] \right\rangle_T \,,
\end{align} 
by means of the imaginary time formalism. This correlator is relevant for open heavy quark diffusion. However, we believe this is just a coincidence at NLO. Since the zero temperature piece obtained here differs from that of~\cite{Burnier:2010rp}, we expect that at NNLO in perturbation theory the results of these two correlators will be different in the temperature-dependent terms as well. This is consistent with our initial observation in the introduction that they are, in fact, different correlators. 

At this point, it is worth to expound on the fact that in the quantum theory there is a clear difference between the two correlators
\begin{align}
    T_F \left\langle {E}^a_i(t) \ml{W}^{ab}(t,0)  {E}^b_i(0) \right\rangle_T \neq \left\langle {\rm Tr}_{\rm color} \left[ U(-\infty,t) E_i(t) U(t,0) E_i(0) U(0,-\infty) \right] \right\rangle_T \label{eq:quarkonia-neq-hq}
\end{align}
given by the fact that operators in quantum mechanics at different times, in general, do not commute. In the correlator on the left of~\eqref{eq:quarkonia-neq-hq}, all of the gauge field $A_0$ operator insertions from Wilson lines occur between the two electric fields (relative to the thermal density matrix $e^{-\beta H}$), as opposed to the correlator of the heavy quark diffusion coefficient, which has explicit gauge field insertions that are in between the electric fields as well as adjacent to the thermal density matrix. This means that, even if one could conceivably relate the SU$(N_c)$ color structures of both correlators, going from one correlator to the other in the full quantum theory involves evaluating several non-trivial quantum-mechanical commutators, and there is therefore no reason to expect that the two correlation functions be equal.

Finally, since we computed the zero temperature result explicitly as well, we can compare it with the appropriate limit of the field strength correlator considered in~\cite{Eidemuller:1997bb}:
\begin{align} \label{Eidemuller-Jamin-correlator}
\left\langle 0 \left| \ml{T} \left( F_{\mu \nu}^a(t)  \ml{W}_{[t,0]}^{ab} F_{\rho \sigma}^b(0) \right)  \right| 0 \right\rangle \,.
\end{align}
After rearranging their results, we find that the $\overline{\rm MS}$ finite constant in this time-ordered correlator and the one in our present work are the same:
\begin{align}
a_g^{\text{ref.~\cite{Eidemuller:1997bb}}} = a_g^{\rm this \, work}  \,.
\end{align}

While this is a strong check of our results, we want to stress that depending on how one implements the interplay between time-ordering of the operators acting on the Fock space of the theory and the SU$(N_c)$ matrix products in the Wilson lines, one could be defining mathematically different objects that are relevant to different physics. For instance, due to the non-local nature of the electric fields dressed with Wilson lines, we have in general
\begin{align}
\Big\langle \ml{T} & \big[{E}_i(y) \ml{W}_{[( y^0, {\bs y}), (+\infty, {\bs y})]} \big]^a
\big[ \ml{W}_{[(+\infty, {\bs x}),(x^0, {\bs x})]} {E}_i(x) \big]^a \Big\rangle_T \nonumber \\ 
&\neq \theta(y^0 - x^0) \Big\langle  \big[{E}_i(y) \ml{W}_{[( y^0, {\bs y}), (+\infty, {\bs y})]} \big]^a
\big[ \ml{W}_{[(+\infty, {\bs x}),(x^0, {\bs x})]} {E}_i(x) \big]^a \Big\rangle_T \nonumber \\ 
& \quad + \theta(x^0 - y^0) \Big\langle  
\big[ \ml{W}_{[(+\infty, {\bs x}),(x^0, {\bs x})]} {E}_i(x) \big]^a \big[{E}_i(y) \ml{W}_{[( y^0, {\bs y}), (+\infty, {\bs y})]} \big]^a \Big\rangle_T \,, \label{eq:discrepancy-T-ordered}
\end{align}
where we have omitted the Wilson lines at infinite time since they do not contribute in our calculations. As the notation suggests, the difference between them is due to operator ordering. In our notation, the correlator on the right hand side corresponds to the time-ordered version of the object we are physically interested in, whereas the one on the left hand side corresponds to the choice $I=J=1$ on the Schwinger-Keldysh contour because of the explicit time-ordering symbol. Concretely, the prescription ordering the operators along the Schwinger-Keldysh contour implies that the disposition of the vector potential $A_\mu$ operators in the Wilson lines, while remaining the same in terms of SU$(N_c)$ indices, will be different in terms of the quantum-mechanical operator ordering for the two correlators. However, as we show in Appendix~\ref{app:T-ordered-vacuum-5-5r} by direct computation, this prescription ($I=J=1$) gives the same result at NLO for the vacuum finite piece result after we set ${\bs y}\to {\bs x}$, as the one given in~\cite{Eidemuller:1997bb}, which is equal to that of the physical answer for heavy particles bound state formation/dissociation given in the present work. Therefore, while we have shown that at NLO one cannot distinguish between the correlator in~\eqref{Eidemuller-Jamin-correlator}, and the ones in~\eqref{eq:discrepancy-T-ordered} with ${\bs y}\to{\bs x}$, we do not expect such an equality to hold exactly, nor at higher orders in perturbation theory. Hence, we stress that it is of utmost importance to rigorously define the correlator that is of physical interest to each situation when comparing similar-looking correlation functions.}

\subsection{Understanding the differences with open heavy quark diffusion} \label{sec:axial-gauge}

In this section, we examine the difference we just discussed in a broader gauge theory context. Specifically, we use gauge transformations to highlight the different nature of these two correlators, and show why temporal axial gauge, in which the two would naively be equal, cannot be applied directly.

Gauge theory plays an essential role in the development of modern physics, highlighted in the formulation of the Standard Model of particle physics~\cite{Itzykson:1980rh,peskin1995introduction,Weinberg:1996kr,Burgess:2006hbd,Srednicki:2007qs,Schwartz:2014sze}. Besides high energy and particle physics, gauge theory also has wide applications in studies of condensed matter physics~\cite{kleinert1989gauge,fradkin2013field}.
Formally, a gauge theory is specified by a gauge group under which the matter fields and force carriers (gauge fields) transform, and described by a Lagrangian density that is invariant under local gauge transformations.

The gauge symmetry corresponds to a redundancy in the degrees of freedom of the theory, which causes difficulties in quantizing the theory. The most widely employed method to overcome the problem is the Faddeev-Popov (FP) path integral approach~\cite{Faddeev:1967fc}. In the FP quantization, one chooses a gauge condition to remove the redundancy in the gauge field degrees of freedom, obtaining different Lagrangian densities for each gauge choice. Calculations with different gauge choices lead to the same results for physical observables, since they are experimentally measurable and thus gauge invariant quantities.

A particular gauge choice, called axial gauge, which sets one component of the gauge field to zero $n^\mu A_\mu=0$,\footnote{Here $n^\mu$ is a fixed 4-vector. Our definition of axial gauge is general and includes temporal axial gauge ($n^2>0$), spatial axial gauge ($n^2<0$) and light-cone gauge ($n^2=0$).} has been widely investigated~\cite{Schoenmaker:1981eq,Ball:1981vd,Caracciolo:1982dp,West:1982gg,Landshoff:1985fv,Cheng:1986hv,Leibbrandt:1987qv,James:1990fd,Nyeo:1991if,Joglekar:1999zt}.
However, the use of axial gauge has often led to confusing and seemingly inconsistent results that lack simple physical interpretation due to extra prescriptions required in the calculation. These subtleties become unavoidable in multi-loop calculations for time-ordered quantities~\cite{Leibbrandt:1987qv,Nyeo:1991if}. A famous example where axial gauge causes a subtlety is the transverse momentum dependent parton distribution function (TMD)~\cite{Boer:2011fh,Angeles-Martinez:2015sea,Shanahan:2019zcq,Ebert:2022cku,Ebert:2022fmh}. In light-cone gauge, the gauge links along the light-cone direction in TMDs become trivial and it is essential to include a transverse gauge link at $x^-=\infty$~\cite{Ji:2002aa},\footnote{This corresponds to the infinite light-cone time for a parton moving along the $-z$ direction, such as the struck quark in deep inelastic scattering~\cite{Brodsky:2002cx}.} which has different physical interpretations depending on the boundary conditions of the gauge fields. Different boundary conditions give different prescriptions in the axial gauge gluon propagator but the final result is the same~\cite{Belitsky:2002sm}.
Another example is the transport coefficients of heavy quarks~\cite{Casalderrey-Solana:2006fio,Caron-Huot:2009ncn} and quarkonia~\cite{Brambilla:2016wgg,Brambilla:2017zei,Yao:2020eqy}, which govern their dynamics in the quark-gluon plasma (QGP), a nearly perfect fluid produced in relativistic heavy ion collisions.
A comparison between the perturbative calculations of two correlation functions that define the heavy quark~\cite{Burnier:2010rp} and quarkonium (as discussed earlier in Section~\ref{sect:nlo}) transport coefficients, suggests that axial gauge can raise consistency issues even at next-to-leading order (NLO) for gauge invariant correlation functions that involve Wilson lines of infinite extent. In particular, Feynman gauge calculations show these two correlation functions differ in values, but they look identical in temporal axial gauge (see the discussion at the end of Section~\ref{sec:add-results}). This is the axial gauge puzzle we want to address in this subsection.\footnote{Ref.~\cite{Eller:2019spw} noted axial gauge could be problematic, but did not explicitly address it.}

We will illuminate the origin of the difficulties in applying axial gauge to calculate these quantities (TMDs and QGP transport coefficients), which are defined through correlation functions of the field strength tensors $F_{\mu \nu} \equiv F_{\mu \nu}^a T^a_F$ dressed with Wilson lines. Here $T^a_F$ denote the generators of the SU($N_c$) gauge group in the fundamental representation that satisfy color trace normalization ${\rm Tr}_c(T_F^aT_F^b)=T_F\delta^{ab}$. Proper Wilson lines are necessary for gauge invariance, since in non-Abelian gauge theories, the field strength transforms as $F_{\mu \nu}(x) \to V(x) F_{\mu \nu}(x) V^{\dagger}(x)$ under a local gauge transformation $V(x)$, and is thus not gauge invariant on its own, unlike its counterpart in Abelian gauge theories. As we will show, the difficulty of using axial gauge is deeply connected with the configuration of the Wilson lines. We will also illuminate under what conditions a naive application of axial gauge leads to a correct result. 

\subsubsection{The axial gauge puzzle in QGP}  

We first discuss the puzzle in more detail by taking the example of the QGP transport coefficients for heavy quarks and quarkonia. The heavy quark diffusion coefficient is defined in terms of the zero frequency limit of a chromoelectric field correlator~\cite{Casalderrey-Solana:2006fio,CaronHuot:2007gq}
\begin{align}
g_E^{\rm Q}(t) = g^2 \left\langle {\rm Tr}_c \left( U_{[-\infty,t]} F_{0i}(t) U_{[t,0]} F_{0i}(0) U_{[0,-\infty]} \right) \right\rangle \label{HQ-corr} \,,
\end{align}
where $g$ denotes the strong coupling, angular brackets represent a thermal expectation value $\langle O \rangle \equiv {\rm Tr}(O \rho)$ with $\rho = e^{-\beta H}/{\rm Tr}(e^{-\beta H})$ and all fields are evaluated at the same spatial point, which is dropped here for notational simplicity. The field operators are ordered as shown (similarly below). The Wilson line $U_{[x,y]}$ is defined in the fundamental representation
\be
U_{[x,y]} = {\rm P} \exp \left( ig \int_y^x \!\! \diff z^\mu A_\mu^a(z) T^a_F \right) \,,
\ee
where ${\rm P}$ denotes path ordering and the path is a straight line connecting the two ends.
The plasma property relevant for small-size quarkonium in-medium dynamics is encoded in a different chromoelectric field correlator in both the quantum optical~\cite{Yao:2020eqy} and quantum Brownian motion limits~\cite{Brambilla:2016wgg,Brambilla:2017zei} (see also Ref.~\cite{Yao:2021lus})
\begin{align}
g_E^{\rm Q\bar{Q}}(t) = g^2 T_F \big\langle  {F}_{0i}^a(t) \ml{W}_{[t, 0]}^{ab}
   {F}_{0i}^b(0)  \big\rangle \,, \label{QA-corr}
\end{align} 
where $\ml{W}_{[x,y]}$ is an adjoint straight Wilson line
\be
\ml{W}_{[x,y]} = {\rm P} \exp \left( ig \int_y^x \!\! \diff z^\mu A_\mu^a(z) T^a_A \right) \,,
\ee
with the adjoint generators $[T^a_A]^{bc} = -i f^{abc}$. Different notations for $g_E^{Q\bar{Q}}$ in the literature are discussed in Appendix~\ref{app:diff-literature}. The correlator for quarkonium was constructed by using the effective field theory potential nonrelativistic QCD~\cite{Brambilla:1999xf,Brambilla:2004jw} and the open quantum system framework~\cite{Akamatsu:2011se,Akamatsu:2014qsa,Katz:2015qja,Brambilla:2017zei,Blaizot:2017ypk,Kajimoto:2017rel,Blaizot:2018oev,Yao:2018nmy,Akamatsu:2018xim,Miura:2019ssi,Sharma:2019xum,Rothkopf:2019ipj,Akamatsu:2020ypb,Sharma:2021vvu,Yao:2021lus}.

\begin{figure}
\centering
\includegraphics[width=0.45\textwidth]{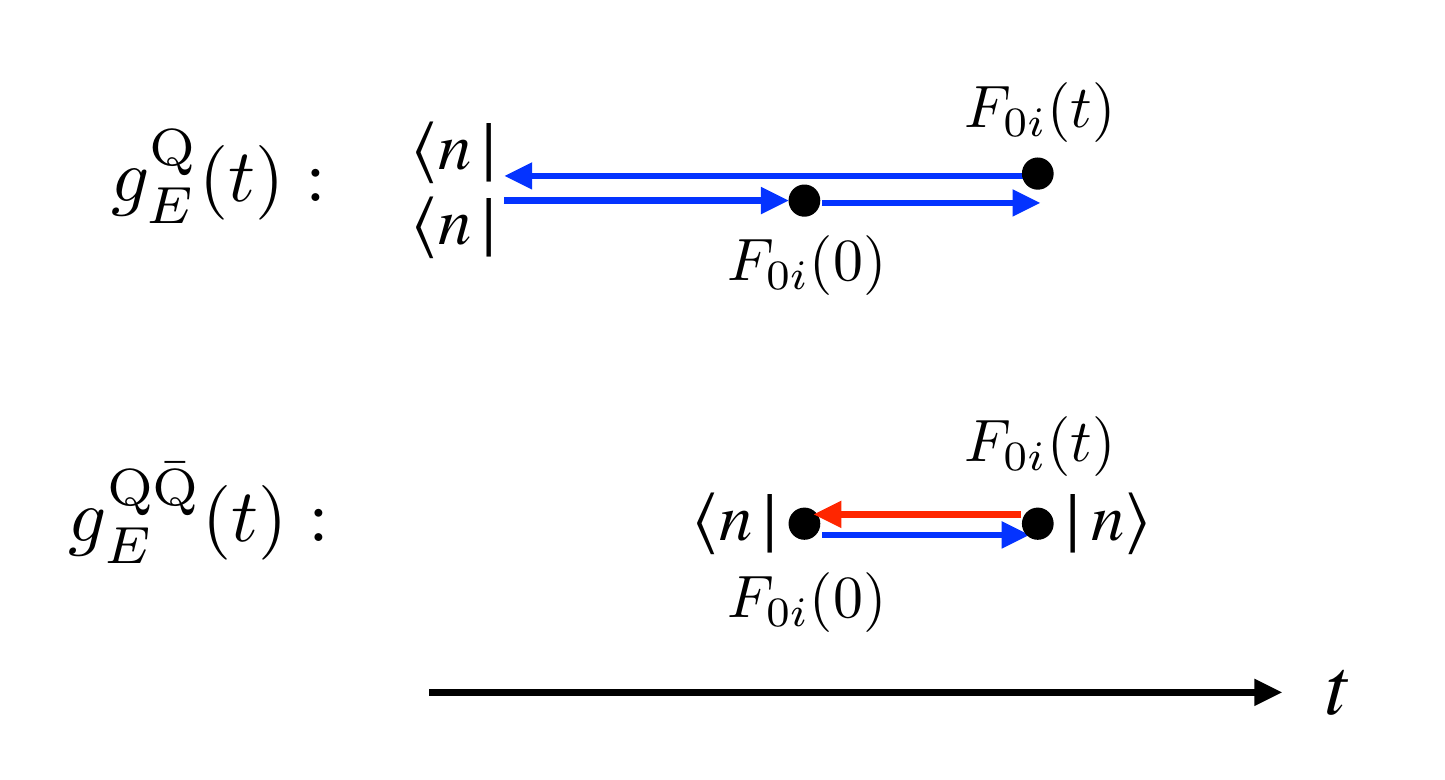}
\caption{Diagrammatic representation of the chromoelectric field correlators for open heavy quarks ($g_E^{\rm Q}(t)$, top row) and quarkonia ($g_E^{\rm Q\bar{Q}}(t)$, bottom row). The dots label the chromoelectric fields. The single and double lines with arrows indicate the Wilson lines in the fundamental and adjoint representations, respectively. The states $|n\rangle$ come from the trace $\Tr(O\rho) \propto \sum_n e^{-\beta E_n}\langle n| O |n \rangle$.}
\label{fig:correlator-rep}
\end{figure}

These two correlators differ in their Wilson line configurations, as shown in Fig.~\ref{fig:correlator-rep}, which contain important physical effects: The open heavy quark carries color through the diffusion process and the Wilson line accounts for both initial and final state interactions. For quarkonium, the Wilson line describes either initial or final state interaction~\cite{Yao:2021lus}. For quarkonium dissociation, the initial state is a heavy quark pair in color singlet which does not interact with the plasma at leading (zeroth) order in the multipole expansion while the final state is a pair in color octet, which does interact with the plasma at leading order, and vice versa for recombination. Explicit NLO calculations for $p_0>0$ showed that these two correlators are already different in vacuum (they also differ by temperature dependent terms, which we will not discuss here)~\cite{Burnier:2010rp,Eidemuller:1997bb,Eller:2019spw,Binder:2021otw}:
\begin{align}
\label{eq:diff-intro}
    \int_{-\infty}^{+\infty} \!\!\!\! \diff t \, e^{i p_0 t} \! \big(  g_E^{\rm Q\bar{Q}}(t) -  g_E^{\rm Q}(t) \big)_{\rm vac} \! = \frac{g^4 N_c (N_c^2 - 1) T_F p_0^3}{(2\pi)^3} \pi^2 \,.
\end{align}
However, these two correlators (\ref{HQ-corr}) and (\ref{QA-corr}) would become identical in temporal axial gauge where $A_0^a=0$ and all the Wilson lines become identities trivially. As a result, their difference is expected to vanish in axial gauge. Now we see the puzzle: The two correlators are defined gauge invariantly and calculations with different gauge choices should give the same result. However, the results in Feynman gauge and axial gauge are explicitly different.

\subsubsection{Resolution of the puzzle} \label{sec:axial-resolution} 

To resolve the puzzle, we first study the correlation functions in temporal axial gauge. For simplicity, we will only consider vacuum correlation functions, as their difference~\eqref{eq:diff-intro} is already apparent in vacuum. The time-ordered gluon propagator can be obtained using the FP procedure, which restricts the path integral over gauge field configurations to be on a ``slice'' determined by the gauge condition:\footnote{One can also choose $G^a[A]= G_A^a[A] - \omega^a(x)$ and then average over $\omega^a$, weighted by $\exp(-\frac{i}{2\xi}\int\diff^4x\, \omega^a\omega^a)$, where $\xi$ is a parameter. \label{fn:gauge-fixing} }
\be
G^a_A[A] = n^\mu A_\mu^a (x) \equiv 0 \,,
\ee
where throughout this section we will use $n^\mu=(1,0,0,0)$, i.e., temporal axial gauge. The time-ordered propagator is given by
\begin{align}
    [D_T(k)]_{\mu \nu}^{ab} = \frac{i \delta^{ab}}{k^2+i\varepsilon } \left[ -g_{\mu \nu} + \frac{n \! \cdot \! k \left( k_\mu n_\nu + n_\mu k_\nu \right) - n^2 k_\mu k_\nu}{(n \! \cdot \! k)^2 + i\varepsilon} \right] \, , 
\end{align}
where $\varepsilon \to 0$ is to be taken at the end of gauge-fixed calculations (we have dropped $\ml{O}(\varepsilon)$ terms in the numerator that do not contribute in this limit). The $\varepsilon$ prescription comes from the time-ordering prescription in the path integral (see SM). In temporal axial gauge, both correlation functions (\ref{HQ-corr}) and (\ref{QA-corr}) simply become $g^2 \langle 0| {\rm Tr}_c(E_i(t)E_i(0))|0\rangle$ for the vacuum part. An explicit NLO calculation for $p_0>0$ gives:\footnote{The Wightman and time-ordered correlators can be related by standard techniques.}
\begin{align}
     \int_{-\infty}^{+\infty} \!\! \diff t\, e^{ip_0 t} \langle 0 |  g^2 \mathcal{T}( {E}^a_{i}(t)  {E}^a_{i}(0) ) | 0 \rangle =  \frac{g^2 (N_c^2 - 1) p_0^3}{(2\pi)^3}  \left\{ 4\pi^2 \! + \! N_c g^2 \! \left[ \frac{11}{12}\ln \! \left( \frac{\mu^2}{4p_0^2} \right) \! + \! \frac{149}{36} \! + \! \frac{\pi^2}{3} \right] \right\}\,,
\end{align}
where $\ml{T}$ denotes time-ordering. This reproduces the Feynman gauge calculation result of Ref.~\cite{Eidemuller:1997bb} and matches the zero temperature limit of the result in Section~\ref{sec:add-results} for Eq.~\eqref{QA-corr}. It also agrees with the corresponding Euclidean correlator in axial gauge; see Appendix~\ref{app:axial-NLO}.

The naive axial gauge calculation does not reproduce the Feynman gauge result for Eq.~\eqref{HQ-corr}, which implies that temporal axial gauge is not smoothly connected with Feynman gauge via a gauge transformation for this observable. To explicitly see the breakdown, we consider a more general gauge-fixing condition
\be
G_M^a[A] = \frac{1}{\lambda} n^\mu A_\mu^a (x) + \partial^\mu A^a_\mu(x) \, ,
\ee
which allows one to smoothly connect Feynman gauge (when $\lambda \to \infty$ for $\xi=1$)\footnote{The definition of $\xi$ is as in footnote~\ref{fn:gauge-fixing}, with $G_A^a$ replaced by $G_M^a$ in the gauge-fixing function.} with axial gauge (when $\lambda \to 0$ for any $\xi$). In this general gauge, the time-ordered gluon propagator with $\xi = 1$ becomes
\begin{align}
\label{eq:mixed-gauge}
    [D_T(k)]_{\mu \nu}^{ab} = \frac{i \delta^{ab}}{k^2+i\varepsilon } \bigg[ -g_{\mu \nu}  + \frac{  k_\mu n_\nu \left( n \! \cdot \! k - i \lambda k^2 \right) + n_\mu k_\nu \left( n \! \cdot \! k + i\lambda k^2 \right) - n^2 k_\mu k_\nu }{(n \! \cdot \! k)^2 + \lambda^2 (k^2)^2 + (1+ 2\lambda^2 k^2) i\varepsilon} \bigg] \,,
\end{align}
where we have dropped $\ml{O}(\varepsilon)$ terms in the numerator. At any finite $\lambda$, one can evaluate the difference between Eqs.~\eqref{HQ-corr} and~\eqref{QA-corr} by carrying out the loop computations using Eq.~\eqref{eq:mixed-gauge}. We find the difference is the same as in Eq.~\eqref{eq:diff-intro} for any $\lambda \neq 0$, as follows.

\paragraph{Difference between heavy quark and quarkonium correlators in mixed axial-Feynman gauge} \hspace{\fill}

In this section we show how to calculate the difference between the two correlators (heavy quark and quarkonium) discussed in the main text in mixed axial-Feynman gauge. To enforce the different operator orderings in the correlation functions, a rigorous calculation of their difference (in particular, the heavy quark correlator) requires introducing the Schwinger-Keldysh contour~\cite{keldysh1965diagram}. As mentioned in the main text, we use a mixed gauge-fixing condition given by
\be
G_M^a[A] = \frac{1}{\lambda} n^\mu A_\mu^a (x) + \partial^\mu A^a_\mu(x) \, ,
\ee
over which we perform the standard average over field configurations $\delta(G_M^a[A] - \omega^a(x))$ weighted by a function $\exp(-\frac{i}{2\xi}\int\diff^4x \omega^a\omega^a)$ with a smearing parameter $\xi$~\cite{Srednicki:2007qs,Schwartz:2014sze}. When $\lambda \to \infty$, one recovers Feynman gauge by setting $\xi = 1$ while axial gauge is recovered for $\lambda \to 0$ for any $\xi$. As such, we shall choose $\xi = 1$ throughout. After performing the gauge-fixing procedure in the path integral, the gluon propagators in this gauge are obtained by inverting the kinetic term in the action for the gauge field: 
\begin{align}
    i S_{\rm kin}[A] = -\frac12 \int_k \begin{pmatrix} \\ A_{(1)}^{a \nu}(-k) \\ \\ A_{(2)}^{a \nu}(-k) \\ {} \end{pmatrix}^{T} \cdot M(k; \xi, \lambda) \cdot \begin{pmatrix} \\ A_{(1)}^{a \mu}(k) \\ \\ A_{(2)}^{a \mu}(k) \\ {} \end{pmatrix} 
\end{align}
where
\begin{equation}
    M(k; \xi, \lambda) = \begin{pmatrix} \begin{matrix} i \Big[ \left(k^2 + i\varepsilon \right) g_{\mu \nu} - \left(1 - \frac{1}{\xi} \right) k_\mu k_\nu \\ \,\,\, + \frac{1}{\xi \lambda^2} n_\mu n_{\nu} + \frac{1}{\xi \lambda} \left( - i k_\mu n_\nu + i k_\nu n_\mu \right)  \Big] \end{matrix}  & 2\varepsilon g_{\mu \nu} \Theta(-k_0) \\ 2\varepsilon g_{\mu \nu} \Theta(k_0) &  \begin{matrix} -i \Big[ \left(k^2 - i\varepsilon \right) g_{\mu \nu} - \left(1 - \frac{1}{\xi} \right) k_\mu k_\nu \\ \,\,\, + \frac{1}{\xi \lambda^2} n_\mu n_{\nu} + \frac{1}{\xi \lambda} \left( - i k_\mu n_\nu + i k_\nu n_\mu \right)  \Big] \end{matrix} \end{pmatrix} \, ,
\end{equation}
and we will set $\xi = 1$ in what follows.
The resulting propagators read
\begin{align}
    [D_T(k)]_{\mu\nu}^{ab} &= \frac{i\delta^{ab}}{k^2+i\varepsilon} \left[ -g_{\mu\nu} + \frac{k_\mu n_\nu f + k_\nu n_\mu f^* - n^2 k_\mu k_\nu }{ |f|^2 + i h \varepsilon } \right] \,, \\
    [D_{\bar{T}}(k)]_{\mu \nu}^{ab} &= \frac{-i\delta^{ab}}{k^2-i\varepsilon} \left[ -g_{\mu\nu} + \frac{k_\mu n_\nu f + k_\nu n_\mu f^* - n^2 k_\mu k_\nu }{ |f|^2 - i h \varepsilon } \right] \,, \\
    [D_>(k)]_{\mu\nu}^{ab} &= \frac{2\varepsilon \Theta(k_0) }{(k^2)^2 + \varepsilon^2} \left[ - g_{\mu \nu} + \frac{k_\mu n_\nu F + k_\nu n_\mu F^* -  k_\mu k_\nu H }{|f|^4 + h^2 \varepsilon^2 } \right] \,, \\
    [D_<(k)]_{\mu\nu}^{ab} &= \frac{2\varepsilon \Theta(-k_0) }{(k^2)^2 + \varepsilon^2} \left[ - g_{\mu \nu} + \frac{k_\mu n_\nu F + k_\nu n_\mu F^* -  k_\mu k_\nu H }{|f|^4 + h^2 \varepsilon^2 } \right] \,,
\end{align}
where we have denoted for brevity
\begin{align}
    f &= (n \! \cdot \! k - i\lambda k^2) \,,  \\
    h &= (n^2+ 2\lambda^2 k^2) \,, \\ 
    F &= k^2 n^2 f + 2 f |f|^2 -  (n \! \cdot \! k) f^2  \,,\\ 
    H &= 3|f|^2 + (n^2)^2 k^2 - (n \! \cdot \! k) (f + f^*) n^2 \, ,
\end{align}
and ${}^*$ denotes complex conjugation. We have also dropped any subleading $\varepsilon$ terms that do not contribute as $\varepsilon \to 0$. 

We then proceed to evaluate the difference between the correlators
\begin{align}
g_E^{\rm Q}(t) &= g^2 \left\langle {\rm Tr}_c \left( U_{[-\infty,t]} F_{0i}(t) U_{[t,0]} F_{0i}(0) U_{[0,-\infty]} \right) \right\rangle \nn \\ &= g^2 \left\langle {\rm Tr}_c \left( U_{[-\infty,t]} F_{0i}(t) U_{[t,+\infty]} U_{[+\infty,0]} F_{0i}(0) U_{[0,-\infty]} \right) \right\rangle \,, \\
g_E^{\rm Q\bar{Q}}(t) &= g^2 T_F \left\langle  \left( {F}_{0i}^a(t) \ml{W}_{[t, 0]}^{ab}
   {F}_{0i}^b(0) \right) \right\rangle = g^2 T_F \left\langle  \left( {F}_{0i}^a(t) \ml{W}_{[t, +\infty]}^{ac} \ml{W}_{[+\infty, 0]}^{cb}
   {F}_{0i}^b(0) \right) \right\rangle \,,
\end{align}
where the last arrangement of operators in each correlator is the one that is better suited for direct evaluation on the Schwinger-Keldysh contour, because in each case we have anti-time ordered operators grouped together on the left and time-ordered operators grouped on the right regardless of whether the time $t$ satisfies $t>0$ or $t<0$. Concretely, in both correlators, we take operators on the left of the Wilson lines extending from $t$ to $t=+\infty$ (including these Wilson lines) to be of type $2$, and operators on the right to be of type $1$.

\begin{figure}
    \begin{subfigure}[t]{0.49\textwidth}
        \centering
        \includegraphics[height=2.2in]{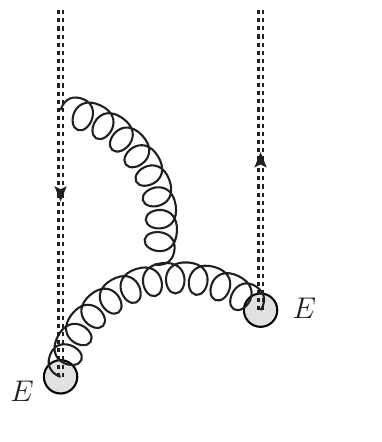}
        \caption{Quarkonium.}
        \label{fig:QQbar_EE}
    \end{subfigure}%
    ~
    \begin{subfigure}[t]{0.49\textwidth}
        \centering
        \includegraphics[height=2.2in]{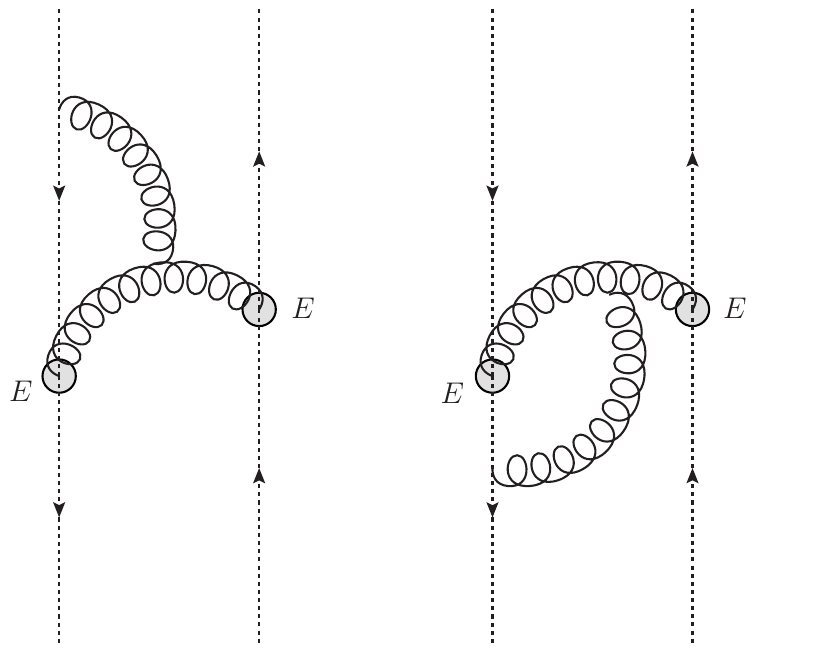}
        \caption{Heavy quark.}
        \label{fig:Q_EE}
    \end{subfigure}%
\caption{Feynman diagrams relevant for the difference between the chromoelectric field correlators for quarkonia (left) and heavy quarks (right). The blobs represent the chromoelectric fields while the double/single-dashed lines indicate the adjoint/fundamental Wilson lines. Similar diagrams where the gluon lines originating from the Wilson lines on the right are also included in the calculation.}
\label{fig:5}
\end{figure}

We perform a perturbative expansion on the coupling constant $g$ to calculate the difference
\be
g_E^{\rm Q\bar{Q}}(p_0) -  g_E^{\rm Q}(p_0) = \int_{-\infty}^{+\infty} \!\! \diff t \, e^{i p_0 t} \! \left(  g_E^{\rm Q\bar{Q}}(t) -  g_E^{\rm Q}(t) \right) \, .
\ee
In the notation of Figure~\ref{fig:diagrams}, the difference comes solely from the diagrams of type $(5)$ and $(5r)$, i.e., diagrams with a triple-gauge boson vertex where only one of the three gluon lines is attached to a Wilson line. In the notation of Refs.~\cite{Eller:2019spw,Burnier:2010rp}, the difference comes from the diagrams labeled as (j). These diagrams are shown in Fig.~\ref{fig:5}. Here we calculate the correlators directly as defined, rather than the time-ordered version as in the naive axial gauge calculation shown in the previous section. Following the calculation details given earlier in Section~\ref{sect:nlo}, we find these diagrams give
\begin{align}
    g_E^{\rm Q\bar{Q}}(p_0) - &  g_E^{\rm Q}(p_0) \nn \\ = \int_{{\bs p}, k} & T_F g^4 N_c (N_c^2 - 1)  \pi \delta(k_0)  \big[ g_{\mu \nu} (p - 2k)_\rho + g_{\nu \rho} (k - 2p)_{\mu} + g_{\rho \mu} (p+k)_\nu \big] \nonumber \\ 
    \times &  (p_0 g_{i\rho'} - p_i g_{0\rho'} ) \big((p_0-k_0) g_{i\nu'} - (p_i - k_i) g_{0\nu} \big) \nonumber \\
    \times &  \Big( [D_{>}(p)]^{\rho' \rho} [D_T(p-k)]^{\nu \nu'} [D_T(k)]^{\mu 0} - [D_{\bar{T}}(p)]^{\rho' \rho} [D_>(p-k)]^{\nu \nu'} [D_>(k)]^{\mu 0} \nonumber \\
    & \,\,\,  - [D_{T}(p)]^{\rho \rho' } [D_>(p-k)]^{\nu' \nu } [D_>(k)]^{ 0 \mu}  + [D_{>}(p)]^{\rho \rho' } [D_{\bar{T}}(p-k)]^{\nu' \nu } [D_{\bar{T}}(k)]^{0 \mu} \Big) \, .
\end{align}
Here we have taken the color contractions out of the propagators and therefore only spacetime indices remain. The Dirac delta $\delta(k_0)$ appears when taking the difference between contributions with different pole prescriptions. These prescriptions are determined by the Wilson line regulator $\eta$ in each of the two correlators. In fact, this is the only difference between the expressions resulting from the Feynman rules for each diagram. It results in contributions of the form
\be
\frac{1}{k_0 + i\eta} - \frac12 \left( \frac{1}{k_0 + i \eta} + \frac{1}{k_0 - i \eta} \right) = \frac12 \left(\frac{1}{k_0 + i \eta} - \frac{1}{k_0 - i \eta} \right) \overset{\eta \to 0}{\longrightarrow} - i \pi \delta(k_0) \nonumber \, ,
\ee
where the first term can be traced back to an adjoint Wilson line attached to $t = +\infty$ (as in the quarkonium correlator), and the second term is the sum of the contributions from a fundamental Wilson line attached to $t=+\infty$ and another one attached to $t=-\infty$ (as in the heavy quark correlator).

The rest of the calculation is tedious, but straightforward. The most sensitive terms come from the propagator that connects the triple-gauge boson vertex with the Wilson line. We list them here explicitly:
\begin{align}
    \left. [D_T(k)]^{\mu 0} \right|_{k_0=0} &= \frac{-i}{{\bs k}^2-i\varepsilon} \left[ -g^{\mu 0} + \frac{i\lambda {\bs k}{}^2 k^{\mu} }{\lambda^2 {({\bs k}{}^2)}{}^2 + (1 - 2\lambda^2 {\bs k}{}^2 ) i \varepsilon } \right] \,, \\
    \left. [D_{\bar{T}}(k)]^{0 \mu} \right|_{k_0=0} &= \frac{i}{{\bs k}^2+i\varepsilon} \left[ -g^{ 0 \mu} + \frac{-i\lambda {\bs k}{}^2 k^{\mu} }{\lambda^2 {({\bs k}{}^2)}{}^2 - (1 - 2\lambda^2 {\bs k}{}^2 ) i \varepsilon } \right]  \,, \\
    \left. [D_>(k)]^{\mu 0} \right|_{k_0=0} &= \frac{\varepsilon}{{({\bs k}{}^2)}{}^2 + \varepsilon^2}  \left[ -g^{\mu 0} + \frac{i \big[ 2\lambda^3 {({\bs k}{}^2)}{}^3 - \lambda {({\bs k}{}^2)}{}^2 \big] k^{\mu} }{\lambda^4 {({\bs k}{}^2)}{}^4 + (1 - 2\lambda^2 {\bs k}{}^2 ){}^2 \varepsilon^2 } \right]  \,,\\
    \left. [D_>(k)]^{0 \mu} \right|_{k_0=0} &= \frac{\varepsilon}{{({\bs k}{}^2)}{}^2 + \varepsilon^2}  \left[ -g^{0 \mu} + \frac{-i \big[2\lambda^3 {({\bs k}{}^2)}{}^3 - \lambda {({\bs k}{}^2)}{}^2 \big] k^{\mu} }{\lambda^4 {({\bs k}{}^2)}{}^4 + (1 - 2\lambda^2 {\bs k}{}^2 ){}^2 \varepsilon^2 } \right] \, .
\end{align}
Here we have used that, as a distribution acting on continuous functions, $\delta(k_0) \Theta(k_0) = \frac12 \delta(k_0)$. Performing the index contractions leads to various integral structures. After isolating the contributions where the ${\bs k}$ momentum flowing in the propagators decouples from ${\bs p}$, using $\int \frac{\diff^d {\bs k} }{(2\pi)^d} {(({\bs k})^{2})}^{n} = 0$  for any integer $n$ in dimensional regularization for $d = 3 -\tilde{\epsilon}$, and using the symmetries of the integrand, one can reduce the expression for the difference to
\begin{align}
    &g_E^{\rm Q\bar{Q}}(p_0) -  g_E^{\rm Q}(p_0)  \\ & =  T_F g^4 N_c (N_c^2-1) \int_{{\bs p},k}  \frac{(2\pi)^2 \delta(k_0) \delta(p^2)}{((p_0^2 - ({\bs p} - {\bs k})^2 )^2 + \varepsilon^2) ( ({\bs k}{}^2)^2 + \varepsilon^2 )}  \big( p_0^2 - ({\bs p} - {\bs k})^2 \big) (d-1) \nn  \\
    & \times \bigg[  - 2  {\bs k}^2 p_0^3  - \frac{ 2  p_0^2 + {\bs k}^2 }{2} \bigg( \frac{\varepsilon \lambda^3 ({\bs k}{}^2)^4 }{\lambda^4 ({\bs k}{}^2)^4 + (1 - 2\lambda^2 {\bs k}{}^2 )^2 \varepsilon^2} - \frac{\varepsilon^3 \lambda {\bs k}{}^2 (1-2\lambda^2 {\bs k}{}^2 ) }{\lambda^4 ({\bs k}{}^2)^4 + (1 - 2\lambda^2 {\bs k}{}^2 )^2 \varepsilon^2} \bigg)  \bigg]\,, \nonumber
\end{align}
where we have assumed $p_0>0$ and taken $\frac{\varepsilon}{{(p^2)}^2 + \varepsilon^2} \to \pi \delta(p^2)$ from the start since this does not cause any singularity in the calculation. 
The first term (i.e., the ${\bs k}^2$ term) in the square bracket gives the Feynman gauge result. The other two terms, together with the overall factor $( ({\bs k}{}^2)^2 + \varepsilon^2 )^{-1}$, give a contribution that is smaller than
$ {\bs k}{}^2 \delta({\bs k}^2)$ in the limit $\varepsilon\to0$, which is automatically vanishing. This can be seen by noting that
\begin{align}
    \frac{1}{({\bs k}{}^2)^2 + \varepsilon^2} \frac{\varepsilon \lambda^2 ({\bs k}{}^2)^4 }{\lambda^4 ({\bs k}{}^2)^4 + (1 - 2\lambda^2 {\bs k}{}^2 )^2 \varepsilon^2} &\leq \frac{\varepsilon \lambda^2 ({\bs k}{}^2)^2 }{\lambda^4 ({\bs k}{}^2)^4 + (1 - 2\lambda^2 {\bs k}{}^2 )^2 \varepsilon^2} \nn \\ &\overset{\varepsilon \to 0}{\longrightarrow} \pi \lambda^2 ({\bs k}{}^2)^2 \delta( \lambda^2 ({\bs k}{}^2)^2 ) = \frac{\pi {\bs k}{}^2 }{2}  \delta({\bs k}{}^2) \nonumber \, , \\
    \frac{1}{({\bs k}{}^2)^2 + \varepsilon^2} \frac{\varepsilon^3  {\bs k}{}^2 |1-2\lambda^2 {\bs k}{}^2 | }{\lambda^4 ({\bs k}{}^2)^4 + (1 - 2\lambda^2 {\bs k}{}^2 )^2 \varepsilon^2} &\leq \frac{\varepsilon}{({\bs k}{}^2)^2 + \varepsilon^2} \frac{ {\bs k}{}^2  }{|1 - 2\lambda^2 {\bs k}{}^2|} \nn \\ &\overset{\varepsilon \to 0}{\longrightarrow} \frac{\pi {\bs k}{}^2  }{|1 - 2\lambda^2 {\bs k}{}^2|} \delta({\bs k}{}^2) = \pi {\bs k}{}^2 \delta({\bs k}{}^2) \nonumber \, ,
\end{align}
in tandem with the fact that the rest of the integral is non-singular at ${\bs k} = 0$ because the possible divergence in the principal value $\mathcal{P} \big[ (p_0^2 - ({\bs p} - {\bs k})^2 )^{-1} \big] $ is tamed by the ${\bs k}{}^2$ factor in the integral measure. Thus, at any finite $\lambda$, only the term that corresponds to the Feynman gauge calculation remains, and the result is independent of $\lambda$. Therefore, by taking $\eta\to0$ and evaluating the integrals first, we find that the limit $\lambda\to0$ gives the same result as the other gauges.

For completeness, we evaluate the remaining (finite) piece that gives the gauge invariant result for the difference. Now we can set $d=3$ because the integrals are strictly convergent and no regularization is required. The result is
\begin{align} 
    &g_E^{\rm Q\bar{Q}}(p_0) -  g_E^{\rm Q}(p_0) \nn \\ &= T_F g^4 N_c (N_c^2-1) \int_{{\bs p},k}  (2\pi)^2 \delta(k_0) \delta(p^2)  \mathcal{P} \left( \frac{ - 2 (d-1) p_0^3 }{(p_0^2 - ({\bs p} - {\bs k})^2 ) {\bs k}{}^2} \right) \nonumber \\
    &= \frac{T_F g^4 N_c (N_c^2-1)}{(2\pi)^3} \left( - 8 p_0^3 \right) \int_0^\infty \diff|{\bs p}| |{\bs p}|^2 \delta(p_0^2 - |{\bs p}|^2) \int_0^{\infty} \diff|{\bs k}| |{\bs k}|^2 \int_{-1}^1 \frac{\diff u}{ {\bs k}{}^2 (2 |{\bs p}| |{\bs k}| u - {\bs k}{}^2 ) } \nonumber \\
    &= \frac{T_F g^4 N_c (N_c^2-1) p_0^3}{(2\pi)^3} (-2) \int_0^\infty \frac{\diff |{\bs k}|}{|{\bs k}|} \ln \left| \frac{1 - |{\bs k}|/(2p_0)}{1+|{\bs k}|/(2p_0) } \right| \nonumber \\
    &= \frac{T_F g^4 N_c (N_c^2-1) p_0^3}{(2\pi)^3} 2 \int_0^\infty \frac{\diff x}{x} \ln \left| \frac{1 + x}{1 - x } \right| \nn \\
    &= \frac{T_F g^4 N_c (N_c^2-1) p_0^3}{(2\pi)^3} \pi^2 \, , \label{eq:result-diff-wightman}
\end{align}
where we used our assumption $p_0>0$. Thus, we reproduced the difference observed in Section~\ref{sec:add-results} by using mixed axial-Feynman gauge.

Therefore, the difference between the two gauge invariant correlators is indeed preserved, even in the limit $\lambda \to 0$, as opposed to the conclusion one would have reached by naively setting $\lambda = 0$ from the start. The problem of naively setting $\lambda = 0$ is caused by a subtlety in the order of taking $\lambda \to 0$ and $\eta \to 0$, where $\eta$ is the regulator that implements how the Wilson line extends to infinity:
\be
U_{[(+\infty) n^\mu, 0]} = {\rm P} \exp \left( i g \int_0^{+\infty} \!\!\! \diff s \, e^{-\eta s} n^{\mu} A_\mu( s n^\mu) \right) \,.\qquad \label{eq:Wilson-line-regulated}
\ee
Specifically, in the calculation of the difference~\eqref{eq:diff-intro} there appear terms of the form (omitting color indices)
\begin{equation} \label{eq:problematic-terms}
\int \frac{\diff^4k}{(2\pi)^4} \frac{\eta}{ (n \! \cdot \! k)^2 + \eta^2}  \left[ D_T(k) \right]_{\nu \mu}  n^{\mu} N(p,k) \, ,
\end{equation}
which are sensitive to the order in which limits are taken. Here $N(p,k)$ is some function of external momentum $p$ and loop momentum $k$ that does not have poles at $ n \cdot k =0 $ or at $k^2=0$. If $\lambda \to 0$ is taken first, then $\lim_{\lambda \to 0} \left[ D_T(k) \right]_{\nu \mu}  n^\mu = 0$, and the result is zero. On the other hand, if one takes $\eta \to 0$ first, Eq.~\eqref{eq:problematic-terms} becomes
\begin{equation}
\int \frac{\diff^4k}{(2\pi)^4} \pi \delta( n \! \cdot \! k ) \left[ D_T(k) \right]_{\nu \mu} n^{\mu} N(p,k) \, ,
\end{equation}
and using the delta function leads to
\begin{align}
\left. \left[ D_T(k) \right]_{\nu \mu} n^{\mu} \right|_{n \cdot k = 0 }  = \frac{(-1)}{k^2 + i\varepsilon} \left[ i n_{\nu} + \frac{  \lambda k^2 k_\nu }{ (\lambda k^2)^2 + (1 + 2\lambda^2 k^2) i\varepsilon } \right] \, , \nn
\end{align}
where we have used $n^2 = 1$.
We find the first term (proportional to $n_{\nu}$) gives the corresponding Feynman gauge result, and the second gives a vanishing contribution for all $\lambda$. Thus, in one order of limits Eq.~\eqref{eq:problematic-terms} is trivially vanishing, whereas in the other it agrees with the Feynman gauge result.
Therefore, we conclude that the gauge invariant result can be reproduced in this general gauge, and naively imposing axial gauge leads to an incorrect result due to an order-of-limit subtlety. 
The correct order for calculating any physical observable that involves Wilson lines extending to infinity is to take $\eta \to 0$ first (as this defines the Wilson line), and only then to vary $\lambda$.\footnote{Our discussion here is about real-time quantities. However, we note that the imaginary time counterparts of Eqs.~\eqref{HQ-corr} and~\eqref{QA-corr} at finite temperature only involve Wilson lines of finite extent. In this case, one cannot gauge-fix the path integral to calculate Eq.~\eqref{HQ-corr} in temporal axial gauge because the Wilson line wraps around the periodic Euclidean time direction, which obtains contributions from gauge field configurations with nontrivial holonomy. In particular, the Polyakov loop would be trivial if such a gauge transformation were possible. However, the Polyakov loop is nontrivial and contains a wealth of information about QCD~\cite{Weiss:1980rj,Gross:1980br}.}

\subsubsection{A nonperturbative viewpoint} 

Another way of seeing the problem with naive axial gauge is to scrutinize the path integral. The FP path integral of a pure gauge theory can be written as
\begin{align}
\int \ml{D}A\, {\rm det}\Big( \frac{\delta G^a(x)}{\delta\theta^b(y)} \Big) \prod_{x,a}\delta \big( G^a(x) \big) e^{iS_{\rm YM}[A^a]} \,,
\end{align} 
where $\theta^b(y)$ denotes the parameter specifying a gauge transformation, $G^a(x)$ is some gauge-fixing condition and $S_{\rm YM}[A^a]$ stands for the action of gauge fields.

We first illustrate the problem in the Abelian case by constructing a gauge transformation that connects Feynman gauge with axial gauge. Under a gauge transformation specified by $\theta(x)$, the Feynman gauge condition transforms as
\begin{align}
G_F(x): \quad \partial_\mu A^\mu(x) \to \partial_\mu A^\mu(x) - \partial^2 \theta(x) \,.
\end{align}
Setting $\partial_\mu A^\mu(x) - \partial^2 \theta(x) = G_A(x) = n_\mu A^\mu(x)$ in momentum space leads to
\begin{align}
& -i k_\mu A^\mu(k) + k^2 \theta(k) = n_\mu A^\mu(k) \,, \nonumber \\
&\implies \theta(k) = \frac{1}{k^2}\big( n_\mu A^\mu(k) + i k_\mu A^\mu(k) \big)\,.
\end{align}
Thus, the gauge transformation needed to transform Feynman gauge to axial gauge is given by
\begin{align}
A^\mu(k) \to M^\mu_{\ \nu} A^\nu(k) \,,
\end{align}
where the transformation matrix is set as
\begin{align}
M^\mu_{\ \nu} = g^\mu_{\ \nu} + \frac{ik^\mu}{k^2}\big( n_\nu + ik_\nu \big) \,.
\end{align}
Inspecting this matrix, we find that $k^\mu$ is an eigenvector of the transformation matrix $M$
\begin{align}
M^\mu_{\ \nu} k^\nu = i\frac{ n \cdot k}{k^2} k^\mu \,,
\end{align}
with eigenvalue $i (n \cdot k)/k^2$. Now we see that the gauge transformation is ill-defined when $n \cdot k = 0$ because the Jacobian of the transformation ${\rm det}(M)$ vanishes, which means the $n\cdot k = 0$ Fourier modes cannot be transformed in this way. Therefore, one cannot transform $n^\mu A_\mu(n \cdot  k = 0)$ to axial gauge from Feynman gauge, that generically has $n^\mu A_\mu(n \cdot  k = 0)\neq0$.

The nonexistence of such a gauge transformation does not always result in the breakdown of axial gauge calculations. To see this more clearly, we consider the gauge field at an arbitrary ``time'' $\bar{n} \cdot x$ ($\bar{n}$ is defined as $(1,-{\bs n})/\sqrt{1+{\bs n}^2}$ for $n=(1,{\bs n})/\sqrt{1+{\bs n}^2}$ so the coordinate $\bar{n} \cdot x$ is the Fourier conjugate of the momentum $n\cdot k$)
\be
A(\bar{n} \cdot x) = \int \diff(n\cdot k)\, e^{i (\bar{n} \cdot x)( n\cdot k) } A(n \cdot k) \,.
\ee
For finite $\bar{n} \cdot x$, the contribution at the point $n\cdot k = 0 $ can typically be neglected, since it has zero measure. However, at infinite ``time'' $\bar{n} \cdot x=\infty$, the dominant contribution to the Fourier transform comes from the region $n \cdot k \approx 0 $. Therefore, when gauge fields at infinite ``time'' are involved in calculations of correlation functions, the breakdown of the gauge transformation at $n \cdot k = 0 $ prevents us from properly gauge-fixing the path integral to axial gauge $n^\mu A_\mu = 0$ globally to perform the calculations. On the other hand, if the correlation function contains no gauge fields at infinite ``time'', the breakdown of the gauge transformation at $n \cdot k = 0$ is irrelevant, since the path integral of fields at $n \cdot k = 0$ only contributes to an overall normalization.
In the latter case, axial gauge calculations work well and give the correct result.

In the non-Abelian case the same breakdown can be seen even more simply by looking at the set of possible local gauge transformations acting on $A_\mu = A_\mu^a T_F^a$:
\begin{align}
A_{\mu}'(x) = V(x) A_\mu(x) V^{-1}(x) - \frac{i}{g} \big(\partial_\mu V(x)\big) V^{-1}(x) \, ,
\end{align}
where $V(x) = \exp(i \alpha^a(x) T_F^a)$. 
For the first term to be well-defined as $\bar{n} \! \cdot \! x \to \infty$, it is necessary that $\lim_{\bar{n}  \cdot  x \to  \infty} \alpha^a(x)$ exists,
which means the $\alpha^a(x)$ specifying the gauge transformation satisfies
\be
\lim_{\bar{n} \cdot x \to \infty} n^\mu \partial_{\mu} \alpha^a(x) = 0 \,.
\ee
Thus, when $\bar{n} \cdot x\! \to \infty$ the projection of the gauge field onto $n^\mu$ only transforms with an SU$(N_c)$ rotation
\begin{equation}
\left. n^{\mu} A_{\mu}' \right|_{\bar{n} \cdot x \to \infty} = \left. V n^\mu A_{\mu} V^{-1} \right|_{\bar{n} \cdot x \to \infty} \, ,
\end{equation}
with no ``shift'' term.
In particular, this means that ${\rm Tr}[ (n^\mu A_\mu (\bar{n} \cdot x = \infty))^2 ]$ cannot be changed by any gauge transformation. Therefore, if we start with a gauge choice in which $n^{\mu} A_\mu^a(\bar{n} \cdot x = \infty) \neq 0 $, we will not be able to set axial gauge $n^{\mu} A_\mu^a(\bar{n} \cdot x = \infty) = 0 $ via gauge transformations.
If the expectation value of an observable ${O}$ has finite contributions from gauge field configurations with $n^{\mu} A_\mu^a(\bar{n} \cdot x = \infty) \neq 0$ in a gauge-fixed path integral, or equivalently from the $n  \cdot  k = 0$ mode of $n^\mu A_\mu$, these contributions cannot be gauge-transformed away. Moreover, the corresponding axial gauge condition $n^\mu A_\mu^a = 0$ becomes inadequate. This is the case for the correlator that defines the heavy quark diffusion coefficient~\eqref{HQ-corr}. On the other hand, when the expectation value of an observable does not contain contributions from the field $n^{\mu} A_\mu^a(\bar{n} \cdot x = \infty)$ in the path integral, it is possible to operationally set $n^{\mu} A_\mu^a = 0$ everywhere and use axial gauge naively. However, we stress that this is possible not because one can effectively set axial gauge for all spacetime points in the path integral, but because the fields at the spacetime points where one cannot do so have no contributions to the expectation value of the operator. This is the case for the quarkonium correlator~\eqref{QA-corr}.

To summarize, we have verified and explained, both perturbatively and nonperturbatively, the origin of the difference between the quarkonium and heavy quark electric field correlators presented in Eq.~\eqref{eq:diff-intro}, namely,
\begin{align}
    \int_{-\infty}^{+\infty} \!\!\!\! \diff t \, e^{i p_0 t} \! \big(  g_E^{\rm Q\bar{Q}}(t) -  g_E^{\rm Q}(t) \big)_{\rm vac} \! = \frac{g^4 N_c (N_c^2 - 1) T_F p_0^3}{(2\pi)^3} \pi^2 \,,
\end{align}
and why temporal axial gauge cannot be used to study it. Our preceding discussion shows that while one may set the Wilson lines to unity in the quarkonium electric field correlator $g_E^{\rm Q\bar{Q}}$ by gauge-fixing, it is impossible to do so for the heavy quark electric field correlator $g_E^{\rm Q}$. Neglecting this point can mislead one into thinking that these two correlators are the same.

\subsubsection{Implications for other physical observables}  

Finally, we discuss the implications of our findings on field strength correlators in other physical contexts. 
In the studies of TMDs, two gluon distributions with different Wilson line configurations exist. The Weizsaecker-Williams (WW) gluon TMD is defined by~\cite{Collins:1981uw,Mulders:2000sh,Ji:2005nu,Meissner:2007rx}
\begin{align}
\frac{1}{xP^+} \int \frac{\diff b^- \diff b^2_\perp}{2(2\pi)^3} e^{-ixb^-P^+ -ib_
\perp \cdot k_\perp }   T_F \big\langle p(P,S) \big|  F^{a+i}(b^-,b_\perp) \ml{W}^{ad} F^{d+j}(0^-,0_\perp) \big| p(P,S) \big\rangle \,, 
\end{align}
where $|p(P,S)\rangle$ denotes the proton state with momentum $P$ and spin $S$. The adjoint Wilson line is $\ml{W}^{ad}=\ml{W}^{ab}_{[(b^-,b_\perp), (+\infty^-,b_\perp)]}\times \ml{W}^{bc}_{[(+\infty^-,b_\perp), (+\infty^-,0_\perp)]}\times \ml{W}^{cd}_{[(+\infty^-,0_\perp), (0^-,0_\perp)]}$.
The dipole gluon TMD is defined as~\cite{Kharzeev:2003wz,Dominguez:2010xd}
\begin{align}
 \frac{1}{xP^+} \int \frac{\diff b^- \diff b^2_\perp}{2(2\pi)^3} e^{-ixb^-P^+ -ib_
\perp \cdot k_\perp }  \big\langle p(P,S) \big| {\rm Tr}_c \big[ U_1F^{+i}(b^-,b_\perp)  U_2
 F^{+j}(0^-,0_\perp) U_3 \big]
\big| p(P,S) \big\rangle  \,, 
\end{align}
where $U_1=U_{[(-\infty^-,0_\perp), (-\infty^-,b_\perp)]}  U_{[(-\infty^-,b_\perp), (b^-,b_\perp)]}$, $U_2=U_{[(b^-,b_\perp), (+\infty^-,b_\perp)]}\times U_{[(+\infty^-,b_\perp), (+\infty^-,0_\perp)]}\times U_{[(+\infty^-,0_\perp), (0^-,0_\perp)]}$, and $U_3=U_{[(0^-,0_\perp), (-\infty^-,0_\perp)]}$ are fundamental Wilson lines.
Their difference is well known~\cite{Kharzeev:2003wz,Dominguez:2010xd,Dominguez:2011wm,Metz:2011wb,Albacete:2012xq,Dumitru:2015gaa,Yao:2018vcg} from small-$x$ studies using the Color Glass Condensate framework~\cite{Iancu:2002xk,Jalilian-Marian:2005ccm,Gelis:2010nm}. They have different $k_\perp$ dependence for small $k_\perp$ while the high $k_\perp$ behavior is the same $\sim1/k_\perp^2$. Therefore, after integrating over the transverse momentum $k_\perp$ and averaging over the spins, we have two different gluon distributions
\begin{align}
\label{eqn:WWpdf}
&\frac{1}{xP^+} \int \frac{\diff b^- }{2(2\pi)} e^{-ixb^-P^+ }   T_F \big \langle p(P) \big| F^{a+i}(b^-) \ml{W}^{ab}_{[b^-,0^-]} F^{b+j}(0^-) \big| p(P) \big\rangle \,, \\
\label{eqn:dipolepdf}
&\frac{1}{xP^+} \int \frac{\diff b^- }{2(2\pi)} e^{-ixb^-P^+ } \big \langle p(P) \big| {\rm Tr}_c \big[ U_{[-\infty^-,b^-]}  F^{+i}(b^-) U_{[b^-,0^-]}  F^{+j}(0^-) U_{[0^-,-\infty^-]} \big] 
\big| p(P) \big\rangle \, .
\end{align}
Naively one would use~\cite{Bomhof:2006dp}
\begin{align}
\label{eqn:FtoA}
U_{[-\infty^-,0^-]} F^{a}_{\mu\nu}(0)T_F^a  U_{[0^-,-\infty^-]} = T^a_F \ml{W}_{[-\infty^-,0^-]}^{ab} F^{b}_{\mu\nu}(0) \,,
\end{align}
to show the two integrated gluon parton distribution functions (PDF) were the same. But Eq.~\eqref{eqn:FtoA} is only valid classically. In quantum theory, Eq.~\eqref{eqn:FtoA} only holds if a path ordering is applied on the left hand side for a path from $0^-$ to $-\infty^-$. Furthermore, the traditional wisdom that inserting a time-ordering operator does not change the physical meaning of the quark PDF~\cite{Jaffe:1983hp} may only apply for Eq.~\eqref{eqn:WWpdf} but not for Eq.~\eqref{eqn:dipolepdf} since the argument given therein relies on using light-cone gauge, which cannot be used naively for Eq.~\eqref{eqn:dipolepdf}.
Thus, these two integrated unpolarized gluon PDFs differ in terms of the operator orderings, similarly to the heavy quark~\eqref{HQ-corr} and quarkonium~\eqref{QA-corr} correlators. 
Therefore, our findings indicate that even though their expressions are identical in naive light-cone gauge, they may have different values. Future work should investigate whether time-ordering can be inserted into Eq.~\eqref{eqn:dipolepdf} without changing its meaning and whether the two gluon PDFs have the same value. If not, gluon PDFs are process dependent and their experimental determination needs systematic reanalysing.

It is also important to discuss in what situations axial gauges still give physically sensible results, even when the observable involves Wilson lines of infinite extent. One ubiquitous such situation is when the observable is constructed as a limit of gauge invariant quantities of finite extent. 
We illustrate this by taking the jet quenching parameter as an example. Albeit not a field strength correlator, it is defined in terms of a Wilson loop~\cite{Wiedemann:2000za,Kovner:2003zj,Liu:2006ug,DEramo:2010wup,DEramo:2012uzl}:
\begin{align}
\hat{q} = \lim_{L^-\to+\infty^-} \frac{\sqrt{2}}{L^-}\int\frac{\diff^2k_\perp}{(2\pi)^2} k_\perp^2 \int\diff^2x_\perp e^{-ix_\perp\cdot k_\perp} \frac{\langle {\rm Tr}_c W^{\ml{R}}_\square \rangle}{d_{\ml{R}}}  \,, \label{eq:jet-quenching}
\end{align}
where $L^-$ is the length of the QGP medium along the $n^\mu=(1,0,0,1)/\sqrt{2}$ direction and $W^{\ml{R}}_\square$ is a rectangular Wilson loop with four corners at $(0^-,0_\perp)$, $(L^-,0_\perp)$, $(L^-,x_\perp)$ and $(0^-,x_\perp)$ in representation $\ml{R}$ of dimension $d_{\ml{R}}$. This is a manifestly gauge invariant object. In light-cone gauge, the transverse gauge links at fixed $0^-$ and $L^-$ become essential, along which $\diff z^\mu A_\mu \neq 0$. Furthermore, the introduction of these transverse gauge links enforces the Wilson loop to lie entirely within a finite spacetime volume in practical calculations, since the transverse gauge links can only be calculated at a finite light-cone distance $L^-$. Once this Wilson loop is calculated, one can take $L^- \to +\infty^-$ and obtain a physically sensible result (even in light-cone gauge). This is also why the light-cone gauge calculation of Ref.~\cite{Belitsky:2002sm} works for TMDs.

\subsubsection{Summary} 

To summarize, in this section we have scrutinized the applicability of axial gauge for computing gauge invariant non-Abelian field strength correlators. We found that attempting to gauge-fix the theory to axial gauge $n^{\mu} A_\mu = 0$ runs into an obstruction for the fields with the Fourier mode $n \cdot k = 0$. As a result, naive axial gauge is not reliable to remove $n^{\mu} A_\mu$ terms when one calculates correlators containing gauge fields at infinite ``time'' $\bar{n} \cdot x$, but it works well for correlators containing only gauge fields at finite $\bar{n} \cdot x$. Our studies further verify the difference between the two correlators defining the heavy quark and quarkonium transport coefficients, which means it is unjustified to use the heavy quark diffusion coefficient calculated via lattice field theory methods~\cite{Banerjee:2011ra,Francis:2015daa,Brambilla:2020siz,Altenkort:2020fgs} in quantum transport equations for small-size quarkonia, as done in Refs.~\cite{Brambilla:2020qwo,Brambilla:2021wkt}. Their difference beyond NLO is unknown. Therefore it is important to study their difference nonperturbatively via lattice field theory methods or the AdS/CFT correspondence~\cite{Maldacena:1997re,Casalderrey-Solana:2011dxg}. (The correlator~\eqref{HQ-corr} has been calculated at strong coupling using the AdS/CFT technique~\cite{Casalderrey-Solana:2006fio}.) Our findings also provide insights into the difference between the unpolarized Weizsaecker-Williams and dipole gluon PDFs, which should be further investigated in the future. These studies will deepen our understanding of QGP transport properties and hadronic structure.

\section{Chromoelectric correlator at strong coupling in $\mathcal{N}=4$ Yang-Mills theory} \label{sec:strong-coupling}

In this section, we calculate this chromoelectric field correlator in strongly coupled $\mathcal{N}=4$ Yang-Mills theory at both zero and nonzero frequency using the AdS/CFT correspondence~\cite{Maldacena:1997re,Casalderrey-Solana:2011dxg}\@. The general strategy of our studies parallels the approaches used in studies of heavy quark diffusion and jet quenching, where effective field theory is applied to describe the strongly coupled physics inside QGP in terms of gauge-invariant objects that require a nonperturbative determination, which can then be subsequently calculated via the AdS/CFT holographic duality. In the case of heavy quark diffusion, heavy quark effective theory was applied to define the heavy quark diffusion coefficient as a chromoelectric field correlator in Ref.~\cite{Casalderrey-Solana:2006fio,Gubser:2006nz}, which was then calculated in the same work via the holographic principle (see Ref.~\cite{Herzog:2006gh,Gubser:2006bz} for a different way of studying heavy quark diffusion in the AdS/CFT approach)\@. In the case of jet quenching, Soft-Collinear Effective Theory~\cite{Bauer:2000ew,Bauer:2000yr,Bauer:2001ct,Bauer:2001yt,Bauer:2002nz} can be used to formulate the jet quenching parameter in terms of a Wilson loop consisting of two light-like Wilson lines~\cite{DEramo:2010wup}, and then be calculated using the AdS/CFT correspondence. While the result of this calculation was first obtained in Refs.~\cite{Liu:2006ug,Liu:2006he}, the more modern formulation in the framework of SCET allows one to organize the calculation in the same way as it is done for heavy quark diffusion, and with the same logic that we will pursue in what follows for quarkonium transport. In this work, we will use the AdS/CFT technique to calculate the chromoelectric field correlator relevant for quarkonium transport.
Our approach will be to first rewrite the chromoelectric field correlator as a path variation of a Wilson loop, which defines a contour on the boundary of an AdS black hole spacetime. Then, using the holographic correspondence, we will calculate the expectation value of the Wilson loop by finding the extremal surface in the bulk of the AdS spacetime that hangs from the contour defined by the Wilson loop. Finally, we will obtain the expectation value of the chromoelectric field correlator by taking the path variation, which amounts to solving linear equations for fluctuations that propagate on the extremal surface.

The result presented here is important for quarkonium phenomenology, because it provides the first nonperturbative picture of the in-medium quarks and gluons that are relevant for quarkonium dynamics in QGP and goes beyond the assumption of a weakly interacting gas. Crucially, the nonperturbative distributions of in-medium quarks and gluons can be process-dependent, as in the case of deep-inelastic scattering (inclusive versus semi-inclusive, polarized versus unpolarized, and so on)\@. In general, it is not expected that the in-medium quarks and gluons relevant for jet quenching would have the same distribution as those affecting heavy quarks.  

This section is organized as follows: The setup of the AdS/CFT calculation will be given in Section~\ref{sect:setup_begin}, followed by details of the calculation in Section~\ref{sec:QQ-setup}, with the final result given at the end of the latter section. We show an extension of the calculation to the case where QGP is flowing in Section~\ref{sec:corr-flow}, and we finalize in Section~\ref{sec:weak-strong} by comparing to the weakly coupled result in QCD.

\subsection{Field strength correlators from Wilson loops}
\label{sect:setup_begin}

We begin by explaining the setup of the calculation of the non-Abelian electric field correlator we wish to obtain. We will start in Section~\ref{sec:setup} by describing how such a correlation function can be obtained by taking variations of a Wilson loop in a purely field-theoretic setup. Then, in Section~\ref{sec:W-loop-AdS} we will proceed to describe how we can evaluate Wilson loops at strong coupling in $\mathcal{N}=4$ supersymmetric Yang–Mills (SYM) theory using the AdS/CFT correspondence. We will discuss the role of the additional parameter $\hat{n} \in S_5$ when constructing the supersymmetric Wilson loop that preserves the features of an adjoint Wilson line in thorough detail in Section~\ref{sec:nhat}. The subsequent Subsection~\ref{sec:EE-setup-AdS} discusses how to take variations of a Wilson loop (i.e., how to introduce field strength insertions) along the contour that defines it from the point of view of the dual gravitational theory. Finally, in Section~\ref{sec:ends-matching}, we will establish the prescriptions necessary to fully define the correlation function for quarkonium in-medium dynamics, and discuss the differences with the correlation function that determines the heavy quark diffusion coefficient. 


\subsubsection{Wilson loops in gauge theory and their variations}
\label{sec:setup}

As we pointed out in the previous section, the transport properties of quarkonia are governed by the correlation functions of chromoelectric fields dressed by Wilson lines, calculated inside a medium. Properties of the medium are encoded in terms of expectation values, which can be determined by a thermal density matrix $Z^{-1} \exp (-\beta H)$\@. However, for the purposes of this subsection we will not need to make reference to the nature of the medium. Rather, we will only discuss how to construct the expectation value we are interested in, by starting from another class of observables that has a well-defined prescription for evaluation at strong coupling using the gauge-gravity duality, which we will discuss explicitly in Subsection~\ref{sec:W-loop-AdS}\@.

Concretely, it is possible to study the correlation functions of gauge theory field strengths $F_{\mu \nu}$ dressed by Wilson lines starting from another class of gauge-invariant operators, namely, Wilson loops $W[\mathcal{C}]$. They are defined as
\begin{align} \label{eq:W-loop}
    W[\mathcal{C}] = \frac{1}{N_c} {\rm Tr}_{\rm color} \! \left[ U_{\ml{C}} \right] = \frac{1}{N_c} {\rm Tr}_{\rm color} \! \left[ {P} \exp \left( ig \oint_{\ml{C}} T^a A^a_\mu dx^\mu  \right) \right] \,,
\end{align}
where $A_\mu^a$ is the SU($N_c$) non-Abelian gauge potential, $T^a$ denotes the generator matrix of the group, $g$ is the coupling constant, and ${P}$ denotes path ordering in the product of group elements $A_\mu = A_\mu^a T^a$\@.
The study of these operators is a cornerstone of much of our understanding of heavy or highly-energetic quarks, and in particular for their dynamics inside a thermal medium. The heavy quark-antiquark interaction potential~\cite{Wilson:PhysRevD.10.2445,Maldacena:1998im,Rey:1998ik}, the jet quenching parameter~\cite{Baier:1996sk,Liu:2006ug}, and the heavy quark diffusion inside a thermal medium~\cite{Casalderrey-Solana:2006fio,Caron-Huot:2009ncn,Burnier:2010rp} have all been formulated and studied through Wilson loops. Crucially, all of them admit a holographic description in $\mathcal{N}=4$ supersymmetric Yang-Mills theory.

As it turns out, one can connect the expectation values of Wilson loops and that of field strengths dressed by Wilson lines by considering functional variations of the path on which the Wilson loop is defined. Concretely, we consider a Wilson loop defined by a path $\mathcal{C}$, and let $\gamma^\mu(s)$ be a parametrization of the path, with $s \in [0, 1]$, and $\gamma^\mu(0) = \gamma^\mu(1)$ for a closed path. Then we consider a deformation of the path, which we denote by $\mathcal{C}_f$ and parametrize by $\gamma_f^\mu(s) = \gamma^\mu(s) + f^\mu(s)$\@. It is then a textbook exercise~\cite{Polyakov:1987ez} to show that
\begin{align} \label{eq:F-insertion-1}
\left.  \frac{\delta}{\delta f^{\mu}(s) } W[\mathcal{C}_f] \right|_{f = 0}  = \frac{(ig)}{N_c} {\rm Tr}_{\rm color} \! \left\{ U_{[1,s]} F_{\mu \rho}[\gamma(s)]\dot{\gamma}^\rho(s) U_{[s,0]}  \right\} \,,
\end{align}
where $U_{[s',s]}$ denotes a Wilson line from $\gamma(s)$ to $\gamma(s')$ in the same representation as $W[\ml{C}_f]$ and $\dot{\gamma}^\rho(s) \equiv d\gamma^\rho(s)/ds$\@. We will abuse the notation a bit to use $U_{[\gamma(s'),\gamma(s)]}$ and $U_{[s',s]}$ interchangeably.
The order of the operators in this expression, and of those in the present discussion before Section~\ref{sec:W-loop-AdS}, only refers to the SU($N_c$) matrix product. Operator ordering in the sense of the order in which they act on a state in the Hilbert space of the theory has yet to be specified (this will be done below, and further developed in Appendix~\ref{sec:App-W-ordering})\@.

Eq.~\eqref{eq:F-insertion-1} provides us with a tool to generate as many field strength insertions as we want along the path of the Wilson loop. By acting on $W[\mathcal{C}]$ with one more derivative and assuming $s_2 > s_1$, 
\begin{align} \label{eq:F-insertion-2}
    & \left.  \frac{\delta}{\delta f^{\mu}(s_2) } \frac{\delta}{\delta f^{\nu}(s_1) } W[\mathcal{C}_f] \right|_{f = 0}  \nonumber \\ 
    & \quad \quad \quad \quad = \frac{(ig)^2}{N_c} {\rm Tr}_{\rm color} \! \left\{ U_{[1,s_2]} F_{\mu \rho}[\gamma(s_2)] \dot{\gamma}^\rho(s_2) U_{[s_2,s_1]} F_{\nu \sigma}[\gamma(s_1)] \dot{\gamma}^\sigma(s_1) U_{[s_1,0]} \right\} \, .
\end{align}
We see explicitly that we can obtain correlation functions of field strength operators $F_{\mu \nu}$ dressed with Wilson lines by taking derivatives with respect to the path on which the Wilson loop~\eqref{eq:W-loop} is defined.\footnote{\label{fn:delta-disc} The only subtlety in this expression is that if $s_1=s_2$, then there is another term (a ``contact term'') on the right hand side of Eq.~\eqref{eq:F-insertion-2}, due to the action of $\delta/\delta f^\mu$ on the $\dot{\gamma}$ present in Eq.~\eqref{eq:F-insertion-1}\@. This term is naturally proportional to derivatives of the delta function $\delta(s_2-s_1)$, and can be easily isolated from the rest of the correlator by looking at positions $s_2 > s_1$ and continuously extending the result to $s_2 = s_1$ (whenever possible)\@. If one looks at the Fourier transform of the left hand side of Eq.~\eqref{eq:F-insertion-2} with respect to $s_2-s_1$, as we will find most natural to do later on, then there will be a contribution from points with $s_2 = s_1$ in the form of a polynomial of positive powers of their Fourier conjugate variable, which we will have to subtract to obtain the correlation function of interest.} It is of course possible to continue beyond two field strength insertions, but for our present purposes it is sufficient to evaluate the two-point deformations.

\begin{figure}
    \centering
    \includegraphics[width=0.8\textwidth]{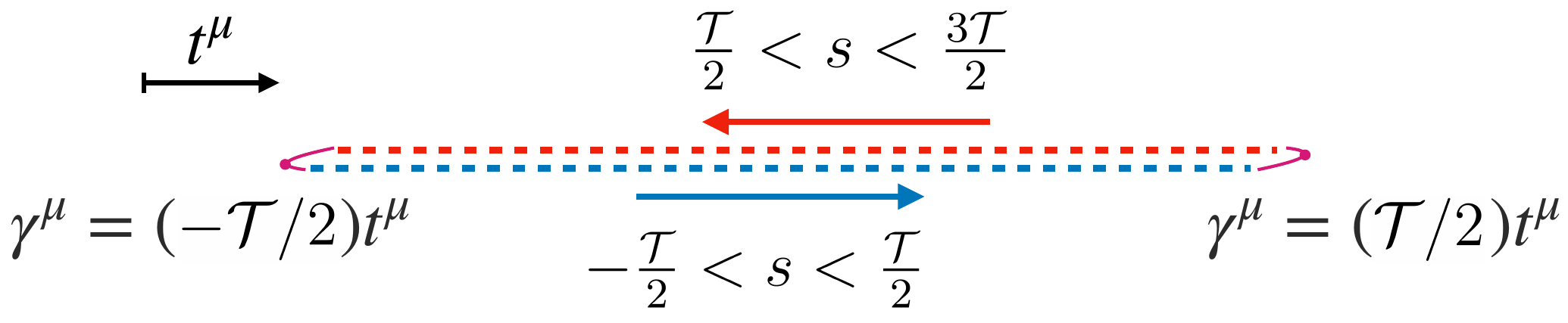}
    \caption{Graphic representation of~\eqref{eq:C-contour-0}. Time increases toward the right of the figure. The contour $\gamma^\mu$ starts at $-(\mathcal{T}/2) t^\mu$, goes in a straight line to $(\mathcal{T}/2) t^\mu$ (in blue; this is the segment where $-\frac{\mathcal{T}}2 t^\mu < s < \frac{\mathcal{T}}2$), and then backtracks over itself (in red; this is the segment where $\frac{\T}2 < s < \frac{3\T}2$).}
    \label{fig:gamma-contour}
\end{figure}

Specifically, our correlator of interest can be obtained by taking $\mathcal{C}$ to be a closed loop parametrized by $\gamma^{\mu}(s)$, where $s \in [-\mathcal{T}/2, 3\mathcal{T}/2]$, as
\begin{align} \label{eq:C-contour-0}
\gamma^\mu(s) = \begin{cases} s t^\mu & -\frac{\T}2 < s < \frac{\T}2 \\ (\T - s) t^\mu &  \frac{\T}2 < s < \frac{3\T}2  \end{cases} \,,
\end{align}
with $t^\mu = (1,0,0,0)$ being a unit vector in the positive time direction. This describes a timelike loop that backtracks upon itself after reaching a maximal value for the time coordinate $t = \T/2$\@. See Figure~\ref{fig:gamma-contour} for a graphic representation. We note that in our setup we must have $W[\mathcal{C}_{f=0}] = 1$, even for time-ordered operators (see Appendix~\ref{sec:App-W-ordering} for details)\@.

Then, taking variations of the path with respect to perturbations in a spatial direction $f^i(s)$ leads to
\begin{align} \label{eq:E-insertion-2}
    &\left.  \frac{\delta}{\delta f^{i}(s_2) } \frac{\delta}{\delta f^{j}(s_1) } W[\mathcal{C}_f] \right|_{f = 0}  =   \nonumber \\
    &\frac{(ig)^2}{N_c} \begin{cases}
    {\rm Tr}_{\rm color} \! \left\{ U_{[3\T/2,\T/2]} U_{[\T/2,s_2]} E_{i}[\gamma(s_2)] U_{[s_2,s_1]} E_j[\gamma(s_1)] U_{[s_1,-\T/2]} \right\} & {\T}/2 > s_2 > s_1 \\ 
    - {\rm Tr}_{\rm color} \! \left\{ U_{[3\T/2, s_2]} E_{i}[\gamma(s_2)] U_{[s_2,\T/2]} U_{[\T/2,s_1]} E_j[\gamma(s_1)] U_{[s_1,-\T/2]} \right\} & s_2 > {\T}/2 > s_1  \\ 
    {\rm Tr}_{\rm color} \! \left\{ U_{[3\T/2, s_2]}  E_{i}[\gamma(s_2)] U_{[s_2,s_1]} E_j[\gamma(s_1)] U_{[s_1,\T/2]} U_{[\T/2,-\T/2]} \right\}  &  s_2 > s_1 >\T/2 \end{cases} \,,
\end{align}
where we have assumed that $s_2 > s_1$\@.
For our purposes, we will take all quantum mechanical operators to be time-ordered, which is exactly what one obtains from the path integral formulation of QFT\@. Some comments on the operator ordering can be found in Appendix~\ref{sec:App-W-ordering}, where we discuss similarities and distinctions with other types of ordering, as well as explain why the time-ordered correlator describes quarkonium dynamics. 
 
We can summarize all of these possibilities as
\begin{align}
\left. \frac{\delta}{\delta f^{i}(s_2) } \frac{\delta}{\delta f^{j}(s_1) } W[\mathcal{C}_f] \right|_{f = 0} = {\rm sgn}[({\T}/2 -s_1)({\T}/2 - s_2)]  \frac{(ig)^2}{N_c} T_F E_i^a(t_2) \mathcal{W}^{ab}_{[t_2,t_1]} E_j^b(t_1) \,,
\end{align}
provided that we choose $s_2 \neq s_1$ such that $\gamma^\mu(s_2) = (t_2,0,0,0)$ and $\gamma^\mu(s_1) = (t_1,0,0,0)$, and $-\frac{\T}2 < t_1<t_2 < \frac{\T}2$\@. In deriving this last expression, we have also used the gauge theory identity
\begin{equation}
\label{eq:adj_fund}
    \W^{ab}_{[t_2,t_1]} = \frac{1}{T_F} {\rm Tr}_{\rm color}  \left[ \hat{\T} T^a U_{[t_2,t_1]} T^b U_{[t_2,t_1]}^\dagger \right] \,,
\end{equation}
where $\hat{\mathcal{T}}$ denotes a time-ordering symbol.

\begin{figure}
    \centering
    \includegraphics[width=.95\textwidth]{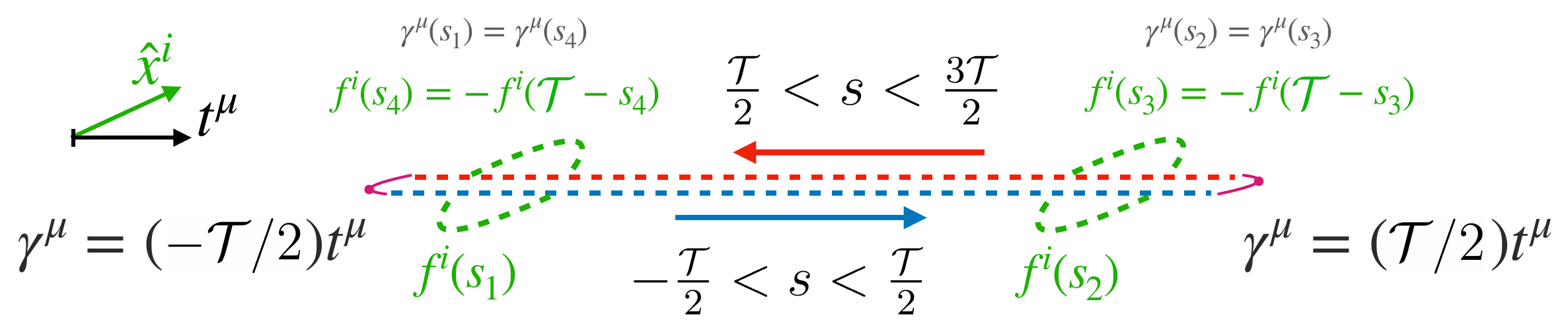}
    \caption{Graphic representation of the antisymmetric deformations~\eqref{eq:f-antisymm} performed on top of the contour defined in Eq.~\eqref{eq:C-contour-0} and depicted in Fig.~\ref{fig:gamma-contour}. The deformations (in green) modify the contour by adding a spatial component to the path.}
    \label{fig:gamma-path-deformed}
\end{figure}

We can further simplify this expression by restricting ourselves to contour deformations of the form
\begin{align} \label{eq:f-antisymm}
f^\mu(s) = \begin{cases} h^\mu(s) & -\frac{\T}2 < s < \frac{\T}2 \\ - h^\mu (\T - s) &  \frac{\T}2 < s < \frac{3\T}2  
\end{cases} \,,
\end{align}
where $h^\mu(t)$ is the independent function with respect to which we will take a variation. See Figure~\ref{fig:gamma-path-deformed} for a graphic representation of thesse deformations. Going through the same arguments given as above, one finds
\begin{align} \label{eq:EE-from-variations}
\left. \frac{\delta}{\delta h^{i}(t_2) } \frac{\delta}{\delta h^{j}(t_1) } W[\mathcal{C}_f] \right|_{h = 0} = 4\frac{(ig)^2}{N_c} T_F E_i^a(t_2) \mathcal{W}^{ab}_{[t_2,t_1]} E_j^b(t_1) \, .
\end{align}
The factor of $4$ is simply a consequence of doubling the size of the deformation to the contour. Intuitively, the reason why the correlator is determined by the antisymmetric deformation is that the symmetric contour deformation, i.e., $f^\mu$ of the form
\begin{align}
f^\mu(s) = \begin{cases} g^\mu(s) & -\frac{\T}2 < s < \frac{\T}2 \\  g^\mu (\T - s) &  \frac{\T}2 < s < \frac{3\T}2  \end{cases} \, ,
\end{align}
with $g^\mu(t)$ as an independent function, gives a vanishing variation
\begin{equation}
    \left. \frac{\delta}{\delta g^{i}(t_2) } \frac{\delta}{\delta g^{j}(t_1) } W[\mathcal{C}_{f}] \right|_{ g = 0} = 0 \, ,
\end{equation}
which is a consequence of the more fundamental statement that $W[\mathcal{C}_{f}] = 1$ for an arbitrary path change given by $g^\mu(t)$ that can be even non-infinitesimal. The reason behind this is that two anti-parallel fundamental Wilson lines $U, V$ that make up the Wilson loop as $W = \frac{1}{N_c} {\rm Tr}_{\rm color} [U V] $ will still satisfy $V = U^{-1}$, even when the path over which they are laid is deformed non-infinitesimally by $f$\@. On the other hand, the antisymmetric deformation described by $h^\mu(t)$ gives a nontrivial result, precisely because in the language we just introduced, we have $V \neq U^{-1}$.

With all of this, we can state that the correlation function we want to calculate is given in terms of path variations of a Wilson loop through the following expression:
\begin{align} \label{eq:EE-corr-from-variations}
\left. \frac{\delta}{\delta h^{i}(t_2) } \frac{\delta}{\delta h^{j}(t_1) } \left\langle W[\mathcal{C}_f] \right\rangle \right|_{h = 0} = 4\frac{(ig)^2}{N_c} T_F \left\langle E_i^a(t_2) \mathcal{W}^{ab}_{[t_2,t_1]} E_j^b(t_1) \right\rangle \, ,
\end{align}
where $\langle\cdots\rangle$ denotes the expectation value.
Having formulated a way to obtain the desired correlation function in terms of operations upon Wilson loops, we now turn to the computational tool that we will use to evaluate this correlation function in $\mathcal{N} = 4$ SYM theory.

\subsubsection{Wilson loops in AdS/CFT} \label{sec:W-loop-AdS}

The AdS/CFT correspondence has proven to be an invaluable tool to gain insight into the strongly coupled regime of non-Abelian gauge theories, by casting a potentially intractable non-perturbative quantum mechanical problem for a conformal field theory (CFT) in terms of a purely classical problem in a concrete gravitational setup of higher dimensionality. In what follows, we will use the real-time formulation of the duality~\cite{Casalderrey-Solana:2011dxg} between $\mathcal{N}=4$ supersymmetric Yang-Mills theory in the Minkowski four dimensional spacetime (Mink${}_4$) in the large $N_c$ limit and type IIB string theory in an (asymptotically) AdS${}_5 \times S_5$ spacetime. The asymptotic boundary of AdS${}_5$ is identified as the Mink${}_4$ on which the supersymmetric Yang-Mills theory lives. At strong coupling in the SYM theory, the dual description on the string theory side reduces to classical string dynamics in a curved spacetime.

The task now is to describe the expectation value of Wilson loops using the duality. This was first done by Maldacena in Ref.~\cite{Maldacena:1998im}\@. However, due to the supersymmetric nature of $\mathcal{N}=4$ SYM, the CFT object that has a simple gravitational dual description in AdS${}_5$ is the locally $1/2$ BPS\footnote{The state that develops this (straight) Wilson line as a phase factor in time evolution is called a $1/2$ Bogomol’nyi-Prasad-Sommerfield (BPS) state, and so the Wilson line in Eq.~\eqref{eq:W-loop-S} is referred to as locally $1/2$ BPS\@. The factor of $1/2$ comes from the fact that the Wilson line commutes with half (eight of the sixteen) supercharges.} Wilson loop
\begin{align} \label{eq:W-loop-S}
W_{\rm BPS}[\mathcal{C};\hat{n}] = \frac{1}{N_c} {\rm Tr}_{\rm color} \! \left[ \mathcal{P} \exp \left( ig \oint_{\ml{C}} ds \, T^a \left[  A^a_\mu \, \dot{x}^\mu + \hat{n}(s) \cdot {\vec{\phi}}^a \sqrt{\dot{x}^2} \right] \right) \right] \, ,
\end{align}
where $\dot{x}^\mu(s) = dx^\mu(s)/ds$ and $\vec{\phi} = (\phi^1, \ldots, \phi^6)$ are the six Lorentz scalar fields in the adjoint representation of SU($N_c$) intrinsic to $\mathcal{N}=4$ SYM. These scalars enter the Wilson loop coupled to a direction $\hat{n}(s) \in S_5$ that specifies the direction along which the string (to be introduced in the next paragraph) ``pulls'' the heavy quark (the string lives in a 10 dimensional space AdS${}_5 \times S_5$ and has a string tension). It can be thought of as an additional property that the heavy quark carries as it propagates through Mink${}_4$, and in general, it depends on the coordinate $s$ along the path $\mathcal{C}$\@. 

Specifically, the AdS/CFT duality gives an explicit prescription to evaluate the expectation value of these generalized Wilson loops. It is given by
\begin{align} \label{eq:duality}
\left\langle W_{\rm BPS}[\mathcal{C};\hat{n}] \right \rangle = \exp \left\{ i  \mathcal{S}_{\rm NG}[\Sigma(\mathcal{C};\hat{n})] - i\mathcal{S}_{0}[\mathcal{C};\hat{n}]  \right\} \, ,
\end{align}
where
\begin{align} \label{eq:NG-action}
\mathcal{S}_{\rm NG}[\Sigma] = - \frac{1}{2\pi \alpha'} \int d\sigma \, d\tau \sqrt{- \det \left( g_{\mu \nu} \partial_\alpha X^\mu \partial_\beta X^\nu \right) } \,,
\end{align}
is the Nambu-Goto action of a string configuration $\Sigma$ described by $X^\mu(\tau,\sigma) \in {\rm AdS}_5 \times S_5$, with $\mu \in \{0,1,\ldots,9\}$\@. $\Sigma(\mathcal{C}; \hat{n})$ is the surface (also referred to as a ``worldsheet'') that extremizes the Nambu-Goto action $\mathcal{S}_{\rm NG}$ with Dirichlet boundary conditions given by $\mathcal{C}$ and $\hat{n}$ at the asymptotic boundary of AdS${}_5$\@. It is in this sense that $\hat{n}$ defines the direction along which the string ``pulls'' the heavy quark in the non-Minkowski directions of the 10-dimensional ${\rm AdS}_5 \times S_5$ space.\footnote{To picture this, it is helpful to note that, asymptotically as $z \to 0$, the part of the metric in Eq.~\eqref{eq:Schwarzschild-AdS} involving $z$ and $\hat{n}$ is proportional to $dz^2 + z^2 d\Omega_5^2$, and as such, $(z, \hat{n})$ may be thought of as a 6-component vector along the direction $\hat{n}$ with length $z$.} The subtraction $\mathcal{S}_{0}[\mathcal{C};\hat{n}]$ is necessary to regularize the result by subtracting the energy associated with the mass of the heavy quark propagating along $\mathcal{C}$, and is also useful because by subtracting the rest mass of the heavy quark it isolates the ``energy'' associated with the interactions described by the Wilson loop at hand. (In QCD, the subtraction also contains the mass renormalization caused by the heavy quark self interaction that is part of the physics contained in the Wilson loop. However, in $\ml{N}=4$ SYM the self interaction diagrams are ultraviolet (UV) finite, and thus the subtraction only contains the bare heavy quark mass~\cite{Drukker:1999zq}.) Only after this subtraction has taken place, may one have an action with a finite value that allows for a comparison of the ``energy'' of different string configurations determined by $\mathcal{S}_{\rm NG}$\@. In the case where one may equate the expectation value of a Wilson loop to that of a time evolution operator over a time period $\T$ for a fixed set of boundary conditions $\mathcal{C}$ and $\hat{n}$, it is the lowest energy $E[\Sigma(\mathcal{C};\hat{n})] =  \left( \mathcal{S}_{0}[\mathcal{C};\hat{n}] - \mathcal{S}_{\rm NG}[\Sigma(\mathcal{C};\hat{n})] \right) /\T$ configuration that determines the expectation value in~\eqref{eq:duality}.

We also need to specify the background metric $g_{\mu \nu}$ that describes the dual ${\rm AdS}_5 \times S_5$ spacetime. Because of our interest in thermal physics, we use the metric for the dual description of a field theory at finite temperature $T$\@. In terms of Poincar\'{e} coordinates, it is given by~\cite{Casalderrey-Solana:2011dxg}
\begin{align} \label{eq:Schwarzschild-AdS}
ds^2 = \frac{R^2}{z^2} \left[ - f dt^2 + d{\bf x}^2 + \frac{dz^2}{f} + z^2 d\Omega_5^2 \right] \, ,
\end{align}
where $f = 1 - (\pi T z)^4$, $R$ is the curvature radius of the AdS metric, $z$ is the radial AdS coordinate, with the asymptotic boundary at $z=0$ and the black hole horizon at $z = (\pi T)^{-1}$\@. The coordinates $t$ and ${\bf x}$ describe Minkowski spacetime at $z=0$, and the $S_5$ coordinates describe a sphere of radius $R$\@. This is the Schwarzschild-AdS metric that is dual to $\mathcal{N}=4$ SYM at finite temperature. The duality also prescribes $R^2/\alpha' = \sqrt{\lambda} = \sqrt{g_{\rm YM}^2 N_c}$, and this single-handedly determines how the coupling constant appears in the results for the strongly coupled limit. In the language of Refs.~\cite{Skenderis:2008dg,Skenderis:2008dh}, this corresponds to a single copy of a Lorentzian manifold on the gravity side of the duality. Therefore, all calculations done in this setup will give time-ordered quantities in terms of the action of operators on the Hilbert space of the quantum theory. If we needed to consider more complicated operator orderings we would be forced to introduce a larger manifold on the gravity side, containing more than one copy of AdS$_{5} \times S_5$\@. This is manifest in the holographic descriptions of heavy quark diffusion and jet quenching~\cite{Casalderrey-Solana:2006fio,DEramo:2010wup}, and is discussed at length in Ref.~\cite{Skenderis:2008dg}\@.

\subsubsection{The role of \texorpdfstring{$\hat{n}$} {} } \label{sec:nhat}

The classic results for Wilson loops in the strong coupling limit that are obtained from the gauge/gravity duality feature a constant value of $\hat{n}$ throughout the quark trajectory~\cite{Maldacena:1998im,Liu:2006ug,Casalderrey-Solana:2006fio}\@. This is a natural choice because, following the discussion of Ref.~\cite{Maldacena:1998im}, the value of $\hat{n}$ is determined by the vacuum expectation value of a Higgs field that has undergone spontaneous symmetry breaking. Given that this is tantamount to a choice of vacuum, the low-energy physics will not modify the value of $\hat{n}$ throughout the trajectory of a heavy W-boson (to be clear, in Ref.~\cite{Maldacena:1998im} Wilson loops are introduced by following the trajectory of a heavy W-boson that is subsequently integrated out. In that discussion, this W-boson is later referred to as ``quark,'' because the Wilson loop also describes the evolution of a heavy quark.)\@. However, inspecting the expression for the $1/2$ BPS Wilson loop~\eqref{eq:W-loop-S}, one realizes that keeping a constant value of $\hat{n} = \hat{n}_0$ throughout both sides of the contour shown in Fig.~\eqref{fig:gamma-contour} violates our expectation for conventional Wilson lines that a closed Wilson loop consisting of two overlapping anti-parallel Wilson lines has $ W[\mathcal{C}] = 1$\@. 
Indeed, the famous result for the heavy quark interaction potential~\cite{Maldacena:1998im}, determined by evaluating a rectangular Wilson loop $\ml{C}_L$ of temporal extent $\T$ and spatial size $L$ (after subtracting the heavy quark mass contribution $\mathcal{S}_{0}[\mathcal{C}_L;\hat{n}=\hat{n}_0]$, and assuming $\T \gg L $) gives
\begin{equation}
    \left\langle W_{\rm BPS}[\mathcal{C}_L;\hat{n}=\hat{n}_0] \right \rangle = \exp \left( i \T \frac{4\pi^2}{\Gamma^4\big(\frac14\big)} \frac{\sqrt{\lambda}}{L} \right) \, ,
\end{equation}
which does \textit{not} satisfy $\lim_{L \to 0} W[\mathcal{C}_L] = 1$\@. The reason behind this is that the contributions from the scalar fields to the locally $1/2$ BPS Wilson loop do not cancel if $\hat{n}$ is held constant~\cite{Maldacena:1998im,Drukker:1999zq,Zarembo:2002an}\@. In other words, this way of approaching a Wilson loop made of coincident anti-parallel fundamental Wilson lines does not have the ``zig-zag'' symmetry~\cite{Polyakov:1997tj,Polyakov:2000ti} that the standard Wilson loop~\eqref{eq:W-loop} respects.\footnote{One may wonder whether there are any mass renormalization effects induced by the self interactions of the $1/2$ BPS Wilson lines at each side of the loop, which would also have to be factored out from the Wilson loop if one wants to isolate the interaction energy between the heavy quarks. However, the supersymmetric nature of $\mathcal{N}=4$ SYM sets these corrections to zero, which may be verified perturbatively~\cite{Drukker:1999zq}, and hence the bare mass subtraction is equivalent to subtracting twice the physical mass of the heavy quark from the energy of the quark-string configuration.}

This is an unavoidable obstacle if one attempts to interpret the variations of the Wilson loop that leads to the heavy quark potential as the dual object of a correlation function of two chromoelectric fields connected by an adjoint Wilson line. Concretely, the standard gauge theory identity for an adjoint representation Wilson line in terms of fundamental representation Wilson lines, i.e., Eq.~\eqref{eq:adj_fund},
is not satisfied if one tries to build such an object by taking variations of a loop at constant $\hat{n}$, because the two anti-parallel fundamental Wilson lines that enter the loop at constant $\hat{n}$ ($U_{[t_2,t_1]}$ and $U_{[t_1,t_2]}$) are not each other's inverse. This had already been noted and discussed previously in Refs.~\cite{Maldacena:1998im,Drukker:1999zq,Zarembo:2002an}\@.

At this point, the appropriate Wilson loop to use in $\mathcal{N} = 4$ SYM is apparent: We have to construct a locally $1/2$ BPS Wilson loop that has two timelike links $U, V$ of temporal extent $\T$ that satisfy $V = U^{-1}$, such that $W_{\rm BPS} = \frac{1}{N_c} {\rm Tr} \left[ U V \right] = 1$\@. This is realized when the $S_5$ coordinates of the Wilson lines are at antipodal points on the $S_5$ on opposite sides of the contour~\eqref{eq:C-contour-0}. In our setup introduced in Section~\ref{sec:setup}, we may choose $\hat{n} = \hat{n}_0$ for $s \in (-\T/2, \T/2)$, and $\hat{n} = -\hat{n}_0$ for $s \in (\T/2, 3\T/2)$. This is a perfectly sensible configuration on the $\mathcal{N}=4$ SYM side, and it actually preserves a maximal number of supersymmetry charges~\cite{Maldacena:1998im,Drukker:1999zq,Zarembo:2002an}\@. Furthermore, it immediately satisfies $W[\mathcal{C}] = 1$, with $\mathcal{C}$ the contour introduced in Eq.~\eqref{eq:C-contour-0}, and it respects the ``zig-zag'' symmetry, in the sense that if we extend the two timelike segments by an arbitrary extent, the contributions from each side of the contour cancel each other. 

Perhaps the most intuitive argument for the time evolution of a heavy quark-antiquark pair to be represented by two Wilson lines with antipodal positions on S${}_5$ comes from inspecting their equations of motion.\footnote{This is simply illustrative, as there are no fermions in the fundamental representation in the $\mathcal{N}=4$ SYM Lagrangian. However, if one were to construct a theory with heavy quarks coupled to the $\mathcal{N}=4$ SYM fields, they should follow Eqs.~\eqref{eq:Q-eom-prop} and~\eqref{eq:Qbar-eom-prop}\@.} Namely, we can define the notion of a heavy antiquark as the object that transforms in the representation conjugate to that of a heavy quark, and therefore follows a evolution equation conjugate to that of the heavy quark. This means that if we take the evolution equation for a heavy quark $Q$ (with its mass already subtracted) to be
\begin{equation}
    (\overrightarrow{\partial}_0 - i A_0 - i \hat{n} \cdot \phi ) Q = 0 \, , \label{eq:Q-eom-prop}
\end{equation}
then the evolution of a heavy antiquark follows
\begin{equation}
    \bar{Q} (\overleftarrow{\partial}_0 + i A_0 + i \hat{n} \cdot \phi ) = 0 \, , \label{eq:Qbar-eom-prop}
\end{equation}
where the arrows on top of $\partial_0$ indicate the directions for it to act.
If one then constructs the supersymmetric Wilson loop~\eqref{eq:W-loop-S} that describes the joint evolution of this heavy quark-antiquark pair, the sign flip in front of $A_0$ is accounted for by flipping the sign of $\dot{x}^\mu$ along the same path, and the sign flip in front of $\phi$ is accounted by inverting the direction $\hat{n}$.

Therefore, based on these considerations, we have arrived at the analogous object to the QCD chromoelectric correlator~\eqref{eq:EE-corr-from-variations} to calculate in the $\mathcal{N} = 4$ SYM theory, namely, the correlation function obtained from a $1/2$ BPS Wilson loop that has two timelike lines at antipodal positions on $S_5$, which has desirable properties distinct from the loop with constant $\hat{n}$\@. This is the setup we use for our calculation and final result in Section~\ref{sec:QQ-setup}. That being said, and even though it will not be part of our final result, we will also evaluate the resulting correlation function from the setup with constant $\hat{n}$ in Appendix~\ref{sec:HQ-setup}, for two reasons that we hope appeal to the expert or interested reader:
\begin{enumerate}
    \item Even if the limit $L \to 0$ of the expectation value of the Wilson loop with constant $\hat{n}$ does not satisfy our expectations, it might still prove instructive to study non-Abelian electric field insertions in a Wilson loop that describes the interaction energy between a pair of heavy quarks;
    \item Ultimately, the motivation behind this calculation is the phenomenology we want to extract for heavy particle pairs in a thermal medium. Since we are using this holographic setup as a model of QCD, we should evaluate all objects that have a reasonable chance to resemble our correlation function of interest, and judge them by their phenomenological implications. 
\end{enumerate}

Before proceeding, we also want to comment on a more recent prescription~\cite{Alday:2007he,Polchinski:2011im} to evaluate the standard Wilson loop~\eqref{eq:W-loop}, which arguably should take precedence in our analysis over the locally $1/2$ BPS Wilson loop because it is constructed from exactly the same fields as in the standard Yang-Mills Wilson loop. This prescription states that the strong coupling limit of Eq.~\eqref{eq:W-loop} is given by extremizing the Nambu-Goto action with Neumann boundary conditions on the $S_5$\@. In practice, this means that most results for the standard Wilson loop at strong coupling are the same as those for the Wilson loop~\eqref{eq:W-loop-S} with constant $\hat{n}$, and they only start to differ at the next order in $1/\sqrt{\lambda}$~\cite{Drukker:2000ep}. This is true because, incidentally, a constant $\hat{n}$ configuration is consistent with Neumann boundary conditions on $S_5$\@.

However, the situation for the limit of our interest, namely, the construction of a standard Wilson loop with two overlapping anti-parallel timelike Wilson lines such that $W[\ml{C}]=1$, using the Neumann boundary condition prescription, might be more subtle. To see this, let us re-examine the arguments that gave rise to this prescription, and to the conclusion that the strong-coupling results of Eqs.~\eqref{eq:W-loop} and~\eqref{eq:W-loop-S} are the same for a general Wilson loop. To facilitate the references to previous works, we will work in Euclidean signature for the remainder of this subsection, unless otherwise noted or if it is explicit from the discussion (e.g., if we refer to lightlike or timelike)\@.

The first reference to Neumann boundary conditions was given by Drukker, Gross and Ooguri in Ref.~\cite{Drukker:1999zq}, where they proposed a boundary condition for an even more general Wilson loop:
\begin{equation}
    W_{\rm DGO}[\mathcal{C};\hat{n}] = \frac{1}{N_c} {\rm Tr}_{\rm color} \! \left[ \mathcal{P} \exp \left( g \oint_{\ml{C}} ds \, T^a \left[  i A^a_\mu \, \dot{x}^\mu + \dot{y}^i \phi_i^a  \right] \right) \right] \, , \label{eq:W-loop-DGO}
\end{equation}
where $y^i = y^i(s) \in \mathbb{R}^6$ is now a general vector. Concretely, the boundary condition they prescribed was
\begin{align}
    X^\mu(\sigma_1,\sigma_2=0) &= x^\mu(\sigma_1) & & {\rm for} \, \mu \in \{0,1,2,3\} \, , \label{eq:bdy-cond-x} \\
    \frac{1}{\sqrt{h}} h_{1b} \epsilon^{bc} \partial_c Y^i(\sigma_1,\sigma_2=0) &= \dot{y}^i(\sigma_1) &  & {\rm for} \, i \in \{1,\ldots,6\} \, , \label{eq:bdy-cond-y}
\end{align}
where $X^\mu$ denotes the usual Mink${}_4$ coordinates  and $Y^i$ is a 6-dimensional vector with the magnitude given by the AdS${}_5$ radial coordinate $z$, and direction specified by $\hat{n} \in S_5$, $h_{ab} = \partial_a X^\mu \partial_b X^\nu g_{\mu \nu}$ is the induced metric on the worldsheet, $\sigma_1$ is the coordinate parametrizing the boundary contour, and $\sigma_2$ is a coordinate that runs into the worldsheet.

An immediate consequence of these boundary conditions, which is discussed explicitly in Ref.~\cite{Drukker:1999zq}, is that the area-minimizing extremal surface reaches the boundary $z = 0$ of AdS${}_5$ if and only if the loop variables obey the constraint $\dot{x}^2 = \dot{y}^2$\@. That is to say, only in this case $\sigma_2 = 0$ corresponds to $z = 0$\@. Once this constraint is incorporated, the inhomogeneous Neumann conditions in Eq.~\eqref{eq:bdy-cond-y} become Dirichlet conditions on the $S_5$ that select $\hat{n}^i(\sigma_1,z=0) = \dot{y}^i/|\dot{y}|$\@.

The boundary condition first proposed in Ref.~\cite{Alday:2007he} for the pure gauge Wilson loop~\eqref{eq:W-loop}, i.e., homogeneous Neumann boundary conditions for the $S_5$ coordinates, is exactly of the same form as in Eq.~\eqref{eq:bdy-cond-y} for the $S_5$ variables with the right hand side vanishing, but prescribes $z=0$ as a Dirichlet condition. This is more explicitly written in Ref.~\cite{Polchinski:2011im}, where the boundary conditions are written as $z = 0$ and $n_a h^{ab} \partial_b U^i = 0$, where, in their notation, $U^i \in S_5$ is the unit vector that we call $\hat{n}$, and $n_a$ is a unit vector normal to the worldsheet boundary.\footnote{Choosing coordinates such that $n_a \leftrightarrow \partial/\partial \sigma_2$, one can show that $n_a h^{ab} \propto h_{1a} \epsilon^{ab}$, and therefore both ways of writing down the Neumann boundary condition are equivalent.}

The motivation behind this prescription of homogeneous boundary conditions for the $S_5$ variables in Ref.~\cite{Alday:2007he} was not disconnected from the preceding discussion of the inhomogeneous Neumann boundary condition shown in Eq.~\eqref{eq:bdy-cond-y} introduced in Ref.~\cite{Drukker:1999zq}\@. Indeed, part of the reasoning that led to this prescription in Ref.~\cite{Alday:2007he} was that when we go to the real time description of Eq.~\eqref{eq:W-loop-DGO} (i.e., to Minkowski signature), provided the condition $\dot{y}^2 = \dot{x}^2$ is met, the coupling to the scalars disappears if we choose $x^\mu$ to be a lightlike path $\dot{x}^2 = 0$, because in this way we force $\dot{y}^i = 0$, and as such, the original Wilson loop~\eqref{eq:W-loop} is recovered.\footnote{See Section 4 of Ref.~\cite{Alday:2007he}\@.} The boundary conditions induced on the $S_5$ variables by this boundary contour are exactly homogeneous Neumann conditions, thus substantiating the proposal in Ref.~\cite{Alday:2007he}\@. It then also follows that, if the path of a supersymmetric Wilson loop at constant $\hat{n}$ can be approximated by a lightlike path in such a way that the extremal worldsheets of both configurations are approximately the same, converging onto each other when the lightlike approximation of the original path becomes better and better (e.g.~by using many small lightlike segments with neighboring segments perpendicular to approximate a straight line), then the expectation values of both Wilson loops in Eqs.~\eqref{eq:W-loop} and~\eqref{eq:W-loop-S} calculated with their respective prescriptions will agree, since a constant $\hat{n}$ fulfills the homogeneous Neumann condition.

However, for the setup where the two anti-parallel Wilson lines are coincident in space, if $\hat{n}$ is held constant, it is not clear whether a lightlike deformation of the (timelike) boundary contour produces a controllable approximation to the worldsheet obtained from the original undeformed contour. Indeed, given that the radial AdS${}_5$ extent of the worldsheet with constant $\hat{n}$ we find in Appendix~\ref{sec:HQ-setup} is proportional to the distance $L$ between the two Wilson lines, which we want to take to zero $L \to 0$, any deformation (lightlike or otherwise) will qualitatively affect the resulting extremal surface, and thus there is no guarantee that the generalized area of the worldsheet with the deformed contour will have the same value as the original undeformed one. Furthermore, for the proper definition of our observable, the first limit we should take is $L \to 0$, because it is only in this limit where we are free to modify the temporal extent of the contour along the time direction without changing the result. This motivates us to look more closely at how the homogeneous Neumann condition can be obtained as a dual description of the standard Wilson loop~\eqref{eq:W-loop} and the role of $\hat{n}$\@.

Another way of arguing for the Neumann prescription for the pure gauge Wilson loop~\eqref{eq:W-loop} has been given in Ref.~\cite{Polchinski:2011im}\@. 
Namely, by considering the value of $\hat{n}$ on the boundary contour as another dynamical variable to be integrated over in the path integral, one can write
\begin{equation}
    \int D \! \left[\hat{n}(s)\right] \langle W_{\rm BPS}[\mathcal{C};\hat{n}(s)] \rangle = \int D \! \left[\hat{n}(s)\right] e^{i\mathcal{S}_{\rm NG} [\Sigma(\mathcal{C};\hat{n})]} \, , \label{eq:Polckinski-proposal}
\end{equation}
where both $\langle W_{\rm BPS}[\mathcal{C};\hat{n}(s)] \rangle$ and $e^{i\mathcal{S}_{\rm NG} [\Sigma(\mathcal{C};\hat{n})]}$ are obtained from a path integral in $\mathcal{N}=4$ SYM and 10-dimensional Supergravity, respectively. On the right hand side of this equality, treating $\hat{n}(s)$ as a dynamical variable gives the Neumann condition as an equation of motion. On the left hand side, one can argue as in Ref.~\cite{Polchinski:2011im} that this path integral gives the pure gauge Wilson loop~\eqref{eq:W-loop}\@. This is achieved by expanding the Wilson loop in a power series in $\hat{n}(s)$, noting that the first term is exactly the Wilson loop~\eqref{eq:W-loop}, and the rest of the terms either vanish by symmetry or are irrelevant operators. 

In fact, in both ways of arguing for the Neumann prescription, no reference to the constant $\hat{n}$ solution is made in its formulation. The connection to the constant $\hat{n}$ solution only appears by noting that, since a constant $\hat{n}$ satisfies the Neumann condition, it follows that the dual descriptions of Eqs.~\eqref{eq:W-loop} and~\eqref{eq:W-loop-S} are governed by the same saddle points, provided that this saddle point is the dominant one, which is usually the case. 

However, direct inspection of the left hand side of Eq.~\eqref{eq:Polckinski-proposal} reveals that there is another saddle point for our contour of interest $\mathcal{C}$, given in Eq.~\eqref{eq:C-contour-0}\@. Namely, the equations of motion that extremize the left hand side of Eq.~\eqref{eq:Polckinski-proposal} have a solution given by $\hat{n}(s) = \hat{n}_0$ for $s \in (-\T/2,\T/2)$ and $\hat{n}(s) = -\hat{n}_0$ for $s \in (\T/2,3\T/2)$, where $\hat{n}_0$ is a fixed direction on $S_5$\@. This is easy to see: any first-order variation of the Wilson loop will give zero because varying it with respect to any field in it will result in an operator insertion along the loop, proportional to an SU$(N_c)$ generator matrix $T^a$\@. For the antipodal $\hat{n}(s)$ configuration that we claim as a solution, in the equation of motion for the $\hat{n}(s)$ variable, the Wilson lines cancel and all that remains is proportional to ${\rm Tr}_{\rm color}[T^a] = 0$\@. The rest of the saddle point configuration is determined by the variations of the $\mathcal{N}=4$ SYM action, which provides its standard equations of motion (i.e., the same equations as in the absence of a Wilson loop)\@.

Furthermore, when we select $\hat{n}$ with antipodal positions on the $S_5$, because the Wilson lines are at separate, antipodal coordinates, it seems plausible that the contour $\mathcal{C}$ as given in Eq.~\eqref{eq:C-contour-0} does admit a lightlike approximation that modifies the extremal surface in such a way that it converges to the unmodified solution as the approximation is made finer. This is so because, as opposed to the case of constant $\hat{n}$ (discussed in detail in Appendix~\ref{sec:HQ-setup}), the size of the deformation here can indeed be taken to be much smaller than the radial extent of the unperturbed worldsheet we will find in Section~\ref{sec:QQ-setup}, which is $(\pi T)^{-1}$\@. Proceeding in this way, we will find a saddle point that should\footnote{We say ``should'' because, as will be made apparent by our discussion in Section~\ref{sec:QQ-setup}, at present we have no explicit extremal surface solution with which to verify the homogeneous Neumann solution at the endpoints of the contour~\eqref{eq:C-contour-0} shown in Fig.~\ref{fig:gamma-contour}, i.e., at $s = \pm \T/2$\@. This is an interesting direction that we leave open to future work.} respect the homogeneous Neumann boundary condition (it is guaranteed to respect it if the proposal of Ref.~\cite{Polchinski:2011im}, expressed through Eq.~\eqref{eq:Polckinski-proposal}, holds), and yields a result $\langle W[\mathcal{C}] \rangle = 1$\@. Therefore, because the pure gauge Wilson loop~\eqref{eq:W-loop} satisfies the unitarity bound $\langle W[\mathcal{C}] \rangle \leq 1$, we would have necessarily found an extremal solution of minimal energy, and consequently, this saddle point would be the one that provides a dual description of the Wilson loop on the contour shown in Eq.~\eqref{eq:C-contour-0}\@. We stop short of claiming a proof of this result because an explicit verification should also provide the extremal worldsheet that is dual to the Wilson loop on the path $\mathcal{C}$ given by Eq.~\eqref{eq:C-contour-0} and explicitly verify all of the boundary conditions discussed above. However, we do conjecture that the pure gauge adjoint Wilson line has the antipodal $\hat{n}$ configuration as its gravitational dual through the gauge/gravity duality.

Another way of saying this is that one may simply look for the dominant contributions to the path integral that defines the Wilson loop expectation value: 
\begin{equation}
\label{eq:nhat-prescription}
    \langle \hat{\mathcal{T}} W[\mathcal{C}_0] \rangle_T = \mathcal{N}_{\mathcal{C}} \int \! D\hat{n} \, \langle W_S[\mathcal{C}, \hat{n}] \rangle \,, = \frac{\mathcal{N}_{\mathcal{C}}}{ \mathcal{Z}} \int \! D\hat{n} \,{\rm Tr}_{\mathcal{H}} \! \left(  e^{-\beta H} \hat{\mathcal{T}} W_S[\mathcal{C}_0,\hat{n}] \right)  \,,
\end{equation}
where $\mathcal{N}_{\mathcal{C}}$ is a path-dependent (re)normalization factor (the need for it is clear when considering the Euclidean calculation of the heavy quark interaction potential~\cite{Maldacena:1998im}, as the LHS of this equation is bounded by 1 and the RHS isn't [in Euclidean signature]). The dominant contribution to the integral over $S_5$ comes from configurations where $\hat{n}$ takes antipodal positions on opposite sides of the contour, such that $\hat{\ml{T}}W_S[\ml{C}_0,\hat{n}]=1$\@.
This follows from the fact that $e^{-\beta H}$ is a positive definite matrix on the Hilbert space and that the time-ordered Wilson loop $\hat{\mathcal{T}}W_S[\mathcal{C}_0,\hat{n}]$ is constructed from a unitary time-evolution operator, and as such, 
\begin{equation} \label{eq:claim-to-show-AppB}
    \left| \frac{1}{ \mathcal{Z}} {\rm Tr}_{\mathcal{H}} \! \left(  e^{-\beta H} \hat{\mathcal{T}} W_S[\mathcal{C}_0,\hat{n}] \right) \right| \leq 1 \, .
\end{equation}
An explicit proof of this bound is given in Appendix~\ref{app:proof-bound}\@. 

To close this section, we note that when there is a nonvanishing spatial separation between the Wilson lines that comprise a Wilson loop, as in the case of the heavy quark interaction potential, one can indeed use the same solution as at constant $\hat{n}$ to describe the extremal worldsheet that gives the expectation value for the pure gauge Wilson loop~\eqref{eq:W-loop}, in accordance to the conjecture laid out in Ref.~\cite{Alday:2007he}\@. One may then worry about the unitarity bound $\langle W[\mathcal{C}_L] \rangle \leq 1$ being violated. However, the situation here is rather similar to that of QCD: only after regularizing and renormalizing does the statement $\log \langle W[\mathcal{C}_L] \rangle \propto 1/L $ make sense. For concreteness, we consider QCD on a 4-dimensional Euclidean lattice, characterized by a lattice spacing $a$\@. If one considers simply the expectation value $\langle W[\mathcal{C}_L] \rangle$ in the limit $L\to0$ at fixed $a$, the value will only converge to $1$ if lattice artifacts are taken into account, which happen when $L\lesssim a$\@. 
This means that to access $\langle W[\mathcal{C}_{L=0}] \rangle$ on the lattice, one has to take $a\to0$ first before taking $L\to0$\@. At finite $L$ the expectation value $\langle W[\mathcal{C}_L] \rangle$ contains both an $L$-dependent term and an $L$-independent diverging term, and they cancel at $L=0$\@.
However, the ratio $\langle W[\mathcal{C}_{L_1}] \rangle/\langle W[\mathcal{C}_{L_2}] \rangle$ will be finite as $a\to 0$ with $L_1, L_2 > 0$ and will exhibit features of a Coulomb potential at distances small compared to the non-perturbative scale $\Lambda_{\rm QCD}^{-1}$\@. While it is true that $\langle W[\mathcal{C}] \rangle = 1$ for the contour shown in Eq.~\eqref{eq:C-contour-0}, its ratio with $\langle W[\mathcal{C}_L] \rangle$ at $L>0$ in the continuum limit is formally infinite. To give meaning to the Wilson loop at a finite spatial separation $L$ and extract energy differences from $\langle W[\mathcal{C}_L] \rangle$, renormalization and regularization is required. Taking ratios achieves this immediately. The need for extra regularization is also clear from calculations in the gravitational description where the boundary contour is modified to be lightlike, see, e.g., Eq.~(3.7) in Ref.~\cite{Alday:2007he}\@.

\subsubsection{Variations of Wilson loops in AdS/CFT} \label{sec:EE-setup-AdS}

Having introduced the necessary concepts to formulate the calculation of the expectation value of a Wilson loop and its path variations (i.e., functional derivatives) that provide the non-Abelian electric field correlator we want to calculate in Yang-Mills theory, we now proceed to describe how the calculation of these path variations takes place in the gravitational description of a Wilson loop in $\mathcal{N}=4$ SYM\@. Concretely, in this section we will lay out the steps one needs to follow to extract correlation functions from the calculation of an extremal worldsheet in a gravitational description and its derivatives. 

Our goal is to insert field strength operators along the boundary contour $\mathcal{C}$\@. As we described in Section~\ref{sec:setup}, this is achieved by taking functional derivatives with respect to deformations of the contour $\mathcal{C}$, parametrized by $h^i(t)$\@. The corresponding operation in the gravitational description is to take functional derivatives with respect to the boundary conditions of the extremal surface. Operationally, we can achieve this by: 
\begin{enumerate}
    \item introducing fluctuation fields $y_i(\tau,\sigma)$ in all directions $i \in \{0,1,2,3,4\}$ on the worldsheet, removing the linear combinations that constitute a coordinate reparametrization, 
    \item expanding the Nambu-Goto action up to a given power $p$ in these perturbations $y_i$,
    \item solving the equations of motion of $y_i$ up to the same order $p$ as a function of arbitrary boundary conditions $h^i(t)$,
    \item and finally, evaluating the Nambu-Goto action expanded up to order $p$, i.e., $\mathcal{S}^{(p)}_{\rm NG}[\Sigma]$ on the worldsheet solution $\Sigma = \Sigma^{(p)}(\ml{C};h)$ obtained up to order $p$ in step 3.
\end{enumerate}
This can be done systematically, starting from the lowest order up to the desired number of powers in the perturbation. The result may be organized as
\begin{align}
    \mathcal{S}_{\rm NG}^{(p)}[\mathcal{C};h] &= \mathcal{S}_{\rm NG}[\mathcal{C};h=0] + \sum_{n=2}^p \frac{1}{n!} \int dt_1 \ldots dt_n \left. \frac{\delta^n \mathcal{S}[\mathcal{C};h]}{\delta h^{i_1}(t_1) \cdots \delta h^{i_n}(t_n)} \right|_{h=0} \, h^{i_1}(t_1) \cdots h^{i_n}(t_n)  \, ,
\end{align}
where the kernel $\left. \frac{\delta^n \mathcal{S}_{\rm NG}[\mathcal{C};h]}{\delta h^{i_1}(t_1) \cdots \delta h^{i_n}(t_n)} \right|_{h=0}$ can be obtained from the four steps listed above. With this definition, $\mathcal{S}_{\rm NG}^{(p)}[\mathcal{C};h]$ is the generating functional for (non-Abelian electric) field strength insertions along the contour $\mathcal{C}$ up to order $p$, which allows one to evaluate correlation functions with up to $p$ insertions of operators.
This object fully characterizes the $n$-point functions of non-Abelian electric field strength insertions along the contour $\mathcal{C}$ at leading order in the large 't Hooft coupling limit, as discussed in the paragraph before Eq.~\eqref{eq:Schwarzschild-AdS}\@.
To marginally ease the notation, let us introduce
\begin{align}
    \Delta_{ij}(t_1,t_2) \equiv - \left. \frac{\delta^2  \mathcal{S}_{\rm NG}[\mathcal{C};h] }{\delta h^i(t_1) \delta h^j(t_2)} \right|_{h=0} \, ,
\end{align}
which we will determine by explicit calculation in the following sections.

From the definition of our operator of interest in terms of Wilson loop variations~\eqref{eq:EE-from-variations}, it should be clear that, \textit{if} the Nambu-Goto action gave a dual description of the pure gauge Wilson loop~\eqref{eq:W-loop}, then its linear response kernel $\Delta_{ij}^{EE}$ would be equal (up to an overall factor) to the chromoelectric field correlation function of our interest dressed by the respective Wilson loop:
\begin{align}
4(ig)^2 T_F \frac{1}{N_c} \frac{\langle \hat{\mathcal{T}} E_i^a(t_2) \mathcal{W}^{ab}_{[t_2,t_1]} E_j^b(t_1) \rangle_T}{\langle \hat{\mathcal{T}} W[\mathcal{C}] \rangle} &=
\left. \frac{1}{\langle \hat{\mathcal{T}} W[\mathcal{C}_h] \rangle}  \frac{\delta}{\delta h^{i}(t_2) } \frac{\delta}{\delta h^{j}(t_1) } \langle \hat{\mathcal{T}} W[\mathcal{C}_h] \rangle \right|_{ h = 0}  \nonumber \\
&= \left. \frac{1}{e^{i \mathcal{S}_{\rm NG}[\mathcal{C};h] }} \frac{\delta}{\delta h^{i}(t_2) } \frac{\delta}{\delta h^{j}(t_1) } e^{i \mathcal{S}_{\rm NG}[\mathcal{C};h] } \right|_{h = 0}  \nonumber \\
&= -i  \Delta^{EE}_{ij}(t_2 - t_1)  \, , \label{eq:EE-true-from-Delta}
\end{align}
where $\ml{C}_h$ denotes the contour $\ml{C}$ perturbed by $h$\@.

We note, however, that the Wilson loop in the duality~\eqref{eq:duality} is \textit{not} the pure gauge loop~\eqref{eq:W-loop}, because it also involves scalar fields~\eqref{eq:W-loop-S}, and so its path variations also give rise to a  contribution from the scalars that modifies the chromoelectric field operators. Namely, assuming that the fluctuations $h^i$ are only in the spatial directions, the non-Abelian field strength insertions are modified to
\begin{equation}
    \dot{x}^\nu F_{i\nu}^a(t) \to \dot{x}^\nu F_{i\nu}^a(t) + \hat{n} \! \cdot \! \big[ D_i \vec{\phi} \, \big]^a \, ,
\end{equation}
where $\hat{n}$ is the direction on the $S_5$ that appears in~\eqref{eq:W-loop-S}, and $\left[ D_i \phi \right]^a = \partial_i \phi^a + g f^{abc} A_i^b \phi^c$ is the gauge covariant derivative of the scalar field. Therefore, after defining 
\begin{equation}
    \tilde{E}_i^a(t) = E_i^a(t) - {\rm sgn}(\dot{x}^0) \, \hat{n} \! \cdot \! \big[ D_i \vec{\phi} \, \big]^a \, ,
\end{equation}
what we can calculate using the duality~\eqref{eq:duality} is
\begin{align}
4(ig)^2 T_F \frac{1}{N_c} \frac{\langle \hat{\mathcal{T}} \tilde{E}_i^a(t_2) \mathcal{W}^{ab}_{[t_2,t_1]} \tilde{E}_j^b(t_1) \rangle_T}{\langle \hat{\mathcal{T}} W_{\rm BPS}[\mathcal{C}] \rangle} &=
\left. \frac{1}{\langle \hat{\mathcal{T}} W_{\rm BPS}[\mathcal{C}_h] \rangle}  \frac{\delta}{\delta h^{i}(t_2) } \frac{\delta}{\delta h^{j}(t_1) } \langle \hat{\mathcal{T}} W_{\rm BPS}[\mathcal{C}_h] \rangle \right|_{ h = 0}  \nonumber \\
&= \left. \frac{1}{e^{i \mathcal{S}_{\rm NG}[\mathcal{C};h] }} \frac{\delta}{\delta h^{i}(t_2) } \frac{\delta}{\delta h^{j}(t_1) } e^{i \mathcal{S}_{\rm NG}[\mathcal{C};h] } \right|_{h = 0}  \nonumber \\
&= -i  \Delta_{ij}(t_2 - t_1)  \,.
\label{eq:EE-from-Delta}
\end{align}
In this expression, $\hat{n}$ takes the sign that it has on the part of the contour where the derivative is taken.\footnote{We note that since the sign of $\dot{x}^\mu$ flips for the two timelike segments of our contour $\mathcal{C}$, for the operator insertions to be equal it is also necessary to flip the sign of $\hat{n}$, as prescribed by our setup in Section~\ref{sec:QQ-setup}\@. In the setup of Appendix~\ref{sec:HQ-setup}, operators inserted on opposite sides of the contour will have a different relative sign between the $E_i^a$ fields and the scalars $\phi^a$\@.} If our conjecture at the end of Section~\ref{sec:nhat} holds true, then we also have $\Delta_{ij}(t) = \Delta^{EE}_{ij}(t)$, and the distinction becomes idle. In this case, according to the prescription presented in Ref.~\cite{Alday:2007he}, the contribution from the scalars disappears.

In these expressions, we have included an extra normalization factor given by the unperturbed Wilson loop. As explained before, we should have $\langle \hat{\mathcal{T}} W[\mathcal{C}] \rangle = 1$ for a Wilson loop in pure Yang-Mills theory (or QCD), and also $\langle \hat{\mathcal{T}} W_{\rm BPS}[\mathcal{C}] \rangle = 1$ for the configuration we discuss in Section~\ref{sec:QQ-setup}, but, as we will remind the reader later, this is not the case for the loop that describes the heavy quark interaction potential, which we discuss in Appendix~\ref{sec:HQ-setup}\@. In this situation it is appropriate to normalize the correlator by the expectation value of the ``background'' Wilson loop.

When we do have $\langle \hat{\mathcal{T}} W[\mathcal{C}] \rangle = 1$, we can summarize the above result as
\begin{align} \label{eq:EE-Delta}
\frac{3}{T_F} [g_{\rm adj}^{\T}]_{ij}(t_2 - t_1) = \frac{g^2}{N_c} \langle \hat{\mathcal{T}} E_i^a(t_2) \mathcal{W}^{ab}_{[t_2,t_1]} E_j^b(t_1) \rangle_T = \frac{i}{2}  \Delta_{ij}(t_2 - t_1) \, ,
\end{align}
where we have used the standard normalization $T_F = 1/2$ for the fundamental representation of SU($N_c$)\@. Therefore, the kernel $\Delta_{ij}$ is exactly the object we are interested in, for each worldsheet configuration of interest. Before proceeding to calculate them, we will take a small digression to discuss aspects of the worldsheet near the turning points at $t = \pm \T/2$\@. In particular, we will discuss how the time-ordering $i\epsilon$ prescription emerges in this setup, and we will also comment on the interplay between the chosen form for the boundary conditions $h^i$ and how the fluctuations behave near $t = \pm \T/2$\@. 

\subsubsection{The Schwinger-Keldysh contour in AdS/CFT} \label{sec:ends-matching}

Because we are interested in the real-time correlation functions of operators in a thermal ensemble, in this section we will discuss the setup of our calculation on the Schwinger-Keldysh contour and how it is realized holographically in the dual gravitational description. Specifically, we will discuss the $i\epsilon$ prescription appropriate for time-ordered quantities, which is exactly the nature of the correlation function we want to calculate. We will also discuss the qualitative differences between the observable we will calculate and the correlation function that defines the heavy quark diffusion coefficient.

The fact that we want to calculate a thermal expectation value requires us to introduce the Schwinger-Keldysh (SK) contour~\cite{keldysh1965diagram} in order to represent the observable of interest through path integrals. The holographic realization of the Schwinger-Keldysh contour in AdS/CFT dates back to the early work of Herzog, Son and Starinets~\cite{Son:2002sd,Herzog:2002pc}, which was more recently expanded and refined by Skenderis and van Rees~\cite{Skenderis:2008dg,Skenderis:2008dh,vanRees:2009rw}\@. In a nutshell, each segment of the Schwinger-Keldysh contour is the boundary of an asymptotically AdS${}_5$ bulk geometry, and these bulk spacetimes are glued together according to appropriate matching conditions, discussed at length in Ref.~\cite{Skenderis:2008dg}\@. This allows one to formulate the Schwinger-Keldysh contour (and, consequently, the bulk geometries) in the complex time plane holographically. As in any quantum-mechanical theory, the thermal nature of the average is dictated by the fact that modes with energy $\omega$ are coupled to themselves across a Euclidean time direction of extent $\beta$, which gives rise to the characteristic thermal statistics through factors of $e^{-\beta \omega}$, and this is realized in the holographic setup by matching the bulk manifolds accordingly. Importantly, the contour in the complex time plane also defines the $i\epsilon$ prescriptions necessary to define the correlation functions in the limit where we take $\T \to \infty$\@. This limit is convenient to do calculations because it restores local translational invariance in the time direction, which is also necessary if one wants to extract the Fourier components of a correlation function at an arbitrary frequency $\omega$, and crucially, to have a continuous limit of it as $\omega \to 0$\@. Because of this, and as we will state explicitly later on, we will indeed take the limit $\T \to \infty$ in all of our calculations. 

We will derive the appropriate $i\epsilon$ prescription for our setup. Before proceeding, it is also important to note that not all observables are equally sensitive to the thermal nature of the contour. In particular, for our correlation function of interest, which is defined through a Wilson loop that ``backtracks'' over the path that it traverses to cover the distance between the two electric field insertions, the extremal surface that defines the expectation value of this Wilson loop will lie only inside the bulk manifold that has the time-ordered part of the Schwinger-Keldysh contour as its boundary. This is so because, crucially, all operators are time-ordered in our setup. The main consequence of this is that the fluctuations that propagate on top of this extremal worldsheet will lose sensitivity to the Euclidean (thermal) part of the Schwinger-Keldysh contour. This is one of the most important qualitative differences between our setup and the heavy quark diffusion calculation in Ref.~\cite{Casalderrey-Solana:2006fio}. We will elaborate further on this at the end of this section, after establishing the prescriptions to select the mode solutions in our setup, in accordance with the boundary conditions that the fluctuations must satisfy.

\paragraph{The $i\epsilon$ prescription for time-ordered correlation functions} \hspace{\fill}

We will first make use of the deformability of the Schwinger-Keldysh contour to derive the $i\epsilon$ prescription that is appropriate for the time-ordered correlation function we want to calculate. Secondly, and as a consistency check, we will verify that this prescription is the same as that one would get by considering the contributions from the turnaround points ($t = \pm \T/2$) of the Wilson loop where the path $\mathcal{C}$ becomes spacelike.

\begin{figure}
    \centering
    \includegraphics[width=0.98\textwidth]{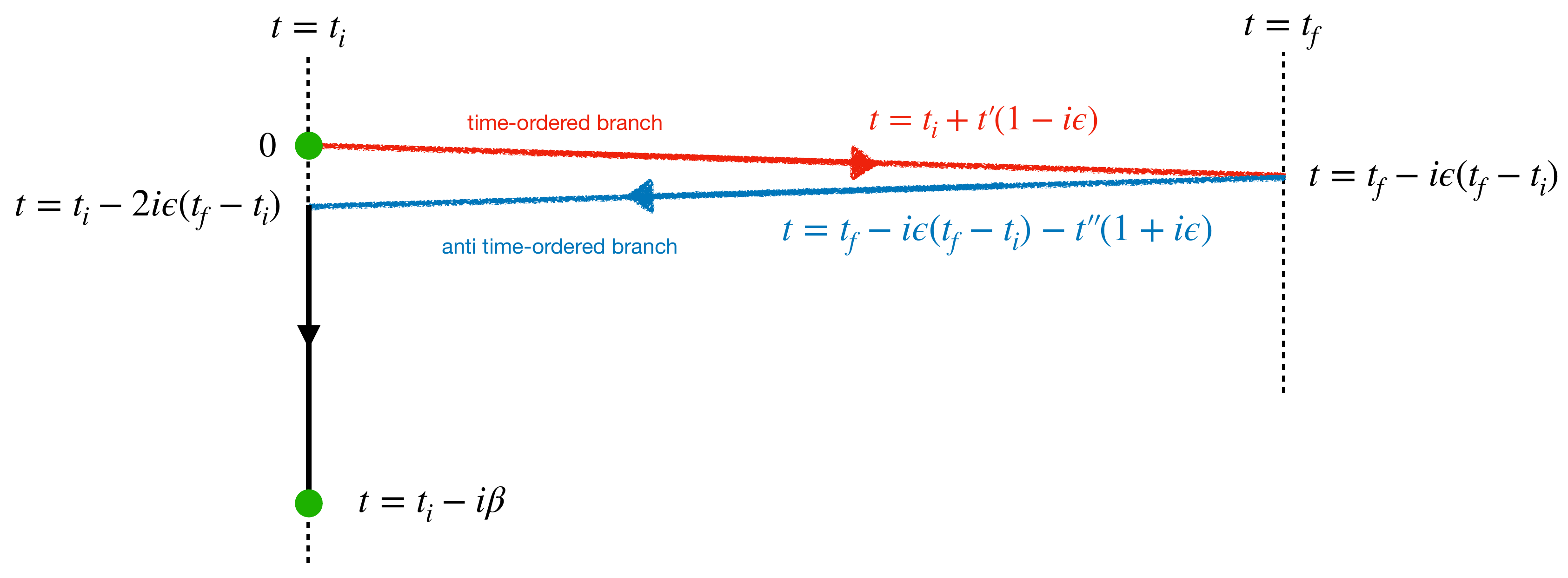}
    \caption{The Schwinger-Keldysh contour, as discussed in this section, including the $i\epsilon$ prescription in full detail. To recover the contour in Figure~\ref{fig:keldysh-contour}, one has to take $t_f - t_i \to \infty$ holding $(t_f - t_i) \epsilon$ fixed, and then relabel $(t_f - t_i) \epsilon \to \epsilon/2$.}
    \label{fig:schwinger-keldysh-deformed}
\end{figure}

To carry out this derivation, it is first necessary to make some remarks about the nature of the Schwinger-Keldysh contour. In principle, any contour starting at $t = t_i$ and ending at $t = t_i -i\beta$ in the complex time plane is adequate in order to evaluate thermal expectation values. The only requisites are that:
\begin{enumerate}
    \item The operators of interest are placed at some point along this contour, and
    \item The contour itself never goes upward in the complex time plane, i.e., its tangent vector always has a non-positive imaginary part. This is necessary for the path integral to be evaluated by a saddle point approximation.
\end{enumerate}
The standard Schwinger-Keldysh contour achieves this by first going from $t=t_i$ along the real axis to some final time $t_f$ (which defines its time-ordered segment), then turning around and going from $t_f$ to $t_i$ along the real time axis (infinitesimally displaced by $-i\epsilon$, which defines the anti-time-ordered segment), and finally going from $t_i$ to $t_i - i\beta$ to close the contour. In fact, the second point of our requisites suggests that we can tilt the time-ordered and the anti-time-ordered parts of the Schwinger-Keldysh contour slightly: Instead of drawing the time-ordered contour exactly along the real axis, parametrized by $t_c = t'$, where $t' \in [0, t_f - t_i]$ is the parameter running along the contour, the natural way to have a well-defined saddle point is to take $t_c = t_i + t'(1 - i\epsilon)$ for the time-ordered branch, with $\epsilon$ a small (infinitesimal) positive number. The anti-time-ordered branch, going back to $t_i$ must then be parametrized by $t_c = t_f - i\epsilon(t_f - t_i) - t''(1 + i\epsilon)$, with $t'' \in (0, t_f - t_i)$\@. The contour is then closed by a straight line along the imaginary time axis at ${\rm Re}(t_c) = t_i$ downward until reaching $t = t_i - i\beta$. For a graphic representation of this, see Figure~\ref{fig:schwinger-keldysh-deformed}. As we will see momentarily, these $i\epsilon$ deformations provide exactly the standard time-ordered and anti-time ordered prescriptions to evaluate correlation functions.

With this setup, we can consider the Nambu-Goto action that describes the worldsheet in the spacetime that has the time-ordered segment of the Schwinger-Keldysh contour as its boundary. In terms of the parameter $t'$, which we write as $t$ in what follows, the metric reads as
\begin{equation}
    ds^2 = \frac{1}{z^2} \left[ - f(z) \, dt^2 (1 - i\epsilon) + d{\bf x}^2 + \frac{1}{f(z)} dz^2 + z^2 d\Omega_5^2 \right] \, .
\end{equation}
The more general case we will consider for our purposes is that of a worldsheet parametrized by
\begin{equation}
    X^\mu = ( t(s,z), x(s,z) , y(s,z) , 0, z, \hat{n}(s,z)) ) \, ,
\end{equation}
where $s$ is a worldsheet coordinate that parametrizes the Wilson loop, which we may define at $z=0$ to be the arc length on the loop. The spacetime coordinates $t, x, z, \hat{n}$ describe the background solution, while $y$ describes the fluctuations we seek to solve for. In our setup, $\hat{n}$ is completely determined by a large circle angle $\phi(s,z)$ due to symmetry considerations. In what follows, we will denote the derivatives of a coordinate $a$ by $\frac{da}{dz} = a'$ and $\frac{da}{ds} = \dot{a}$\@.

Since we will be considering infinitesimal perturbations $y$, whether the worldsheet is spacelike or timelike is wholly determined by the background solution. Expanding up to quadratic order in $y$, one obtains the following action:
\begin{align} \label{eq:quadratic-action-general}
    \mathcal{S}_{\rm NG}[\Sigma] &= - \frac{\sqrt{\lambda}}{2\pi} \left[ S_{\rm NG}^{(0)}[\Sigma_0] + S_{\rm NG}^{(2)}[\Sigma_0;y] + \ldots \right] \, , \\
    S_{\rm NG}^{(0)}[\Sigma_0] &= \int \frac{\diff s \, \diff z}{z^2} \bigg\{ \left[ \dot{t}^2  + f \big( \dot{t} x' - t' \dot{x} \big)^2  + f z^2 \big( \dot{t} \phi' - \dot{\phi} t' \big)^2 \right] (1 - i\epsilon) \nonumber \\ & \quad \quad \quad \quad \quad \quad \quad \quad \quad \quad \quad \quad \quad  - \frac{\dot{x}^2}{f} - \frac{z^2 \dot{\phi}^2 }{f} - z^2 \big( \dot{x} \phi' - x' \dot{\phi} \big)^2 \bigg\}^{1/2} \, , \label{eq:bkg-action-general} 
\end{align}
\begin{align}
    S_{\rm NG}^{(2)}[\Sigma_0;y] &= \int \frac{\diff s \, \diff z}{2 z^2} \left[ f \big(\dot{t} y' - t' \dot{y} \big)^2 (1 - i\epsilon) - \big( \dot{x} y' - x' \dot{y} \big)^2 - z^2 \big( \dot{y} \phi' - y' \dot{\phi} \big)^2 - \frac{\dot{y}^2}{f} \right] \nonumber \\ & \quad \quad \quad \quad \times \bigg\{ \left[ \dot{t}^2  + f \big( \dot{t} x' - t' \dot{x} \big)^2  + f z^2 \big( \dot{t} \phi' - \dot{\phi} t' \big)^2 \right] (1 - i\epsilon) \nonumber \\ & \quad \quad \quad \quad \quad \quad \quad \quad \quad \quad \quad \quad \quad  - \frac{\dot{x}^2}{f} - \frac{z^2 \dot{\phi}^2 }{f} - z^2 \big( \dot{x} \phi' - x' \dot{\phi} \big)^2 \bigg\}^{-1/2} \label{eq:fluct-action-general} \,.
\end{align}
Moreover, Eq.~\eqref{eq:fluct-action-general} provides a prescription that affects the equations of motion of $y$\@. To see this in a concrete setup (which will be relevant in Section~\ref{sec:QQ-setup}), we consider the case of a radially infalling background worldsheet at constant $x$ and $\hat{n}$, with $t = s$\@. The action for the fluctuations simplifies to
\begin{equation}
    S^{(2)}_{\rm NG}[\Sigma_0;y] = \int \frac{\diff s \,\diff z}{2 z^2} \left[ f y'{}^2 (1-i\epsilon)^{1/2} - \frac{1}{f} \dot{y}^2 (1-i\epsilon)^{-1/2} \right] \, ,
\end{equation}
which implies that at the level of the equations of motion, the frequency $\omega$ of the mode solutions will always appear as $\omega^2 (1 + i\epsilon)$. This in turn defines the pole prescription to evaluate the propagator. As we will see later in Section~\ref{sec:EE-calculation-QQ}, this also determines which mode solution should be used when calculating correlation functions.

One may also wonder whether the behavior of the worldsheet around the turnaround times $t = \pm \T/2$ affects this conclusion. Specifically, one can wonder whether one can get extra imaginary terms in the equations of motion by having a transition where the induced metric on the worldsheet goes from having Minkowski signature (i.e., timelike) to having Euclidean signature (i.e., spacelike)\@. We discuss this in Appendix~\ref{app:turnaround-worldsheet}. This analysis cross-checks and verifies the $i\epsilon$ prescription we obtained.

\paragraph{Differences with the heavy quark diffusion coefficient} \hspace{\fill}

Finally, let us comment on how this calculation differs from the one for the heavy quark diffusion coefficient~\cite{Casalderrey-Solana:2006fio}\@. To do this, it is most helpful to use the construction put forth by Skenderis and van Rees~\cite{Skenderis:2008dg,Skenderis:2008dh,vanRees:2009rw}, where the Schwinger-Keldysh contour has a concrete holographic realization. This is done by constructing a bulk manifold made up of several submanifolds satisfying appropriate matching conditions, where the boundary of each of these submanifolds is identified with the lower-dimensional spacetime that corresponds to a given segment of the Schwinger-Keldysh contour in the boundary CFT\@. This can be done in the same way for the correlator we are presently considering and the one that determines the heavy quark diffusion coefficient.

\begin{figure}
    \centering
    \includegraphics[width=\textwidth]{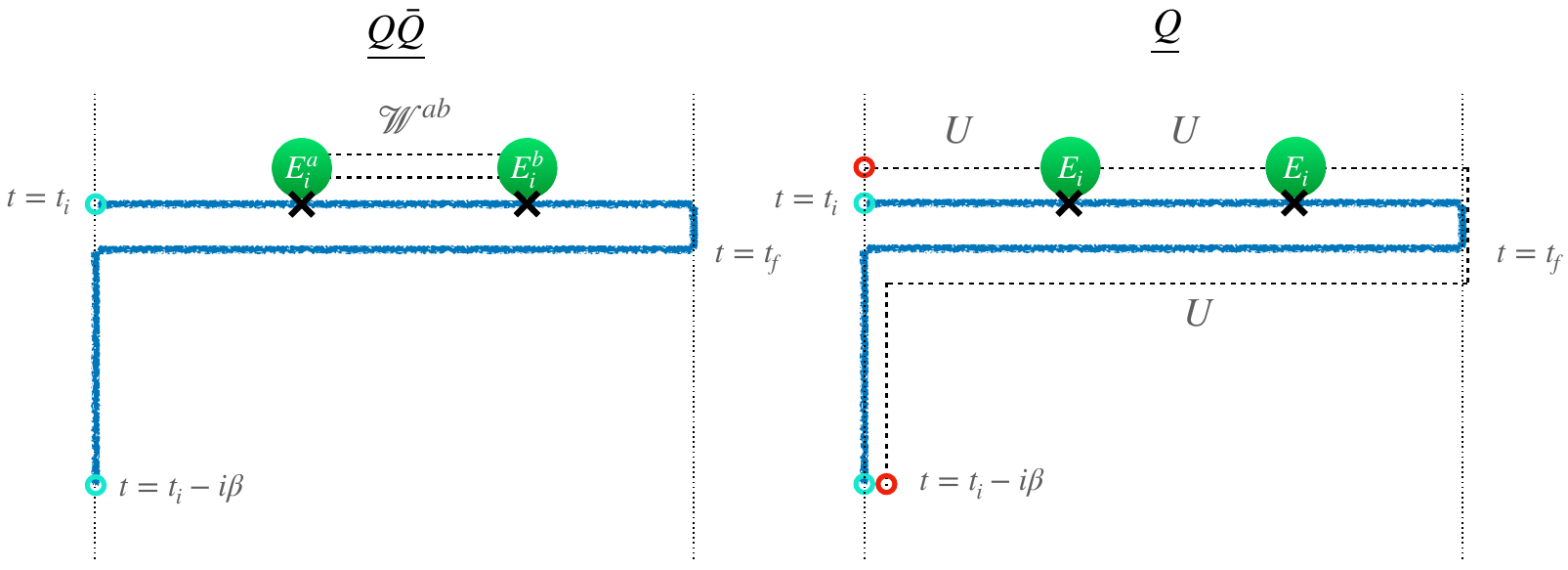}
    \caption{Schematic representation of the chromoelectric field correlators relevant for quarkonium transport (left) and for heavy quark diffusion (right)\@. The blue contour is the Schwinger-Keldysh contour. The open cyan and red circles reflect that the corresponding ends of the contours should be identified (respectively, for the Schwinger-Keldysh contour itself and for the fundamental Wilson line, which we describe in what follows). The adjoint Wilson line is denoted by $\W^{ab}$, and the fundamental lines by $U$. The are represented, respectively, by two dashed lines and a single dashed line. The difference in the Wilson line configuration reflects the different natures of the initial state of the QGP: in the quarkonia case, it is taken to be $\rho = \frac{1}{Z_{\rm QGP}} e^{-\beta H_{\rm QGP} } $, while in the single heavy quark case it is $\rho = \frac{1}{Z_{\rm HQ}} \sum_{Q} \langle Q | e^{-\beta H_{\rm tot}} | Q \rangle$, where $H_{\rm tot} = H_{\rm QGP} + H_Q$ and the sum over $Q$ runs over all states $| Q \rangle$ containing one heavy quark~\cite{McLerran:1981pb,Casalderrey-Solana:2006fio}\@.}
    \label{fig:EE-difference-SK}
\end{figure}

However, the manifold on which the fluctuations that are of interest to us propagate is not the full manifold associated to the Schwinger-Keldysh contour of the full CFT\@. Rather, the fluctuations propagate on a lower-dimensional manifold given by the background solution for the worldsheet configuration. Whether this worldsheet configuration spans every region of the Schwinger-Keldysh contour is solely determined by the shape of the boundary Wilson loop. In the case of the heavy quark diffusion coefficient setup, the corresponding Wilson loop consists of a single Wilson line that winds around the Schwinger-Keldysh contour once. This means that the background worldsheet configuration that defines the manifold on which fluctuations will propagate spans the whole bulk space at a single fixed position coordinate on the boundary ${\bs x}$, and is parametrized by a radial AdS coordinate and a temporal coordinate that goes over both Minkowski and Euclidean regions of the bulk geometry. The topology of the boundary manifold is that of a circle, and the end points of the real-time segments have to be matched with those of the imaginary-time segments, in consistency with the Schwinger-Keldysh contour (see Fig.~\ref{fig:EE-difference-SK})\@. Then, if one seeks mode solutions for the fluctuations on top of this 2-dimensional geometry, the matching conditions discussed in Refs.~\cite{Skenderis:2008dg,Skenderis:2008dh,vanRees:2009rw} imply that Fourier modes $e^{-i\omega t} F_\omega(z)$ on the real-time segments (where $F_\omega(z)$ is the radial AdS profile of a solution with a frequency $\omega$ on the boundary) have to be matched with solutions of the form $e^{\pm \beta \omega} F_\omega(z)$ on the Euclidean segments. Therefore, factors of $e^{\beta \omega}$ naturally appear in the response functions. This is what gives rise to the KMS relations between the different types of correlation functions that can be calculated by introducing fluctuations and evaluating the response functions on different segments of the SK contour.\footnote{We note that for the heavy quark diffusion coefficient setup, the operator orderings of the chromoelectric field correlators that are related via KMS relations refer only to the chromoelectric field insertions, and do not affect the operator orderings of the Wilson lines (see also Appendix~\ref{sec:App-W-ordering})\@. This is a consequence of this correlator being derived from correlation functions of quark currents and subsequently integrating out the massive quark~\cite{Caron-Huot:2009ncn}\@. The fact that the heavy quark is present at all times means that, when one integrates it out, one should actually regard the Wilson line as a modification to the bath Hamiltonian enforcing a modified Gauss' law due to the presence of the static point color charge, which is felt by the bath at all times.}

On the other hand, the Wilson loop that defines the correlation function we are presently interested in does \textit{not} wind around the SK contour. That means that the ``thermal'' contributions $e^{\beta \omega}$ that come from matching the fluctuations around the contour will not be present in this case, and therefore all of the temperature dependence that will appear is going to be due to temperature effects on the bulk geometry of the Minkowski part of the manifold that holographically realizes the path integral associated to the Schwinger-Keldysh contour. Hence, the way in which both observables are defined has manifestly distinct effects in the values that the correlation functions take. After discussing the calculations in detail throughout the following section, we will see how these differences manifest themselves in the final result.

\subsection{AdS/CFT calculation of the chromoelectric field correlator} \label{sec:QQ-setup}

As we discussed in Section~\ref{sec:nhat}, the standard choice to do calculations of Wilson loops using the AdS/CFT correspondence in strongly coupled $\mathcal{N}=4$ SYM is to set a constant value for $\hat{n}$ throughout the Wilson loop. This is indeed the setup used in the celebrated paper by Maldacena~\cite{Maldacena:1998im} to calculate the heavy quark interaction potential at strong coupling. However, as we pointed out in Section~\ref{sec:nhat}, this setup does not satisfy the properties we require of the Wilson loop relevant for quarkonium propagation. To be thorough, we anyways carry out the calculation of the electric field correlator based on a Wilson loop with a constant value of $\hat{n}$ in Appendix~\ref{sec:HQ-setup}, and demonstrate that the result of taking the transverse separation $L$ of the antiparallel Wilson lines to zero $L \to 0$ is inconsistent with the properties of the Wilson loop relevant for quarkonium propagation. Mathematically, this is due to the fact that the result does not have a well-defined limit as $L \to 0$.

Having studied the supersymmetric Wilson loop with constant $\hat{n}$, we now proceed to investigate the relevant configuration, where we have $\hat{n} = \hat{n}_0$ on one of the timelike segments comprising the loop, and $\hat{n} = -\hat{n}_0$ on the other antiparallel segment of the loop.

This configuration has received less attention in the AdS/CFT studies of heavy quarks, mainly because it does not generate a Coulomb interaction potential between a heavy quark pair. However, while it has been usually less emphasized, it has been discussed in many AdS/CFT studies, starting from the same works that discussed the heavy quark interaction potential~\cite{Maldacena:1998im,Rey:1998ik}\@. As we will also verify momentarily, it has the crucial property that $\langle W_{\rm BPS}[\mathcal{C}] \rangle = 1$ for a contour going from one point to another and coming back to the starting point along the same path. Apart from the fact that one can verify this identity by hand in the CFT, this relation is protected by supersymmetry~\cite{Zarembo:2002an}\@. Moreover, the fact that this configuration might be relevant for the dynamics of a quark-antiquark pair was hinted at in Ref.~\cite{Rey:1998ik}, where the first appearance of quark pairs and heavy quark pairs featured antipodal positions on the $S_5$\@.

Given the relative lack of attention that this configuration has received, especially for phenomenological applications, we will try to make our discussions as detailed as possible.

\subsubsection{Background}

We consider the $\T \to \infty$ limit of the contour $\mathcal{C}$ that defines the Wilson loop from which we can extract the correlator relevant for quarkonium transport. In $\mathcal{N}=4$ SYM, we also have to specify the position on the $S_5$ that the Wilson loop goes over for it to have a dual gravitational description in terms of an extremal surface in AdS${}_5 \times S_5$\@. As discussed in Section~\ref{sec:nhat}, the relevant Wilson loop is constructed from two long, antiparallel timelike Wilson lines with antipodal positions on the $S_5$\@. Without loss of generality, we can describe the distance between the $S_5$ coordinates on the two boundary segments by a large circle angular coordinate $\phi \in [0,\pi]$ (the coordinate itself runs over $\phi \in [0,2\pi)$, but the angular separation between two points on a circle is at most $\pi$), and then we may substitute $d\Omega_5^2$ into the metric shown in Eq.~\eqref{eq:Schwarzschild-AdS} by $d\phi^2$\@.

First we study whether there is an extremal surface connecting the two boundary segments that can be described by
\begin{align}
X^\mu(\tau,\phi) = (\tau, 0, 0 , 0, z(\phi), \hat{n}(\phi)) \, ,
\end{align}
for which the Nambu-Goto action reads:
\begin{equation} \label{eq:NG-z-of-phi}
    \mathcal{S}_{\rm NG}[\Sigma] = - \frac{\sqrt{\lambda}}{2\pi} \int d\tau \, d\phi \frac{\sqrt{z'{}^2+ f z^2}}{z^2} \, ,
\end{equation}
where now  we define $z' = \partial z/\partial \phi$\@. The action of the resulting extremal surface should be compared with the action of two disconnected worldsheets falling into the black hole. If we find a positive\footnote{This is due to the overall minus sign in the definition of $\mathcal{S}_{\rm NG}$\@.} regularized action by extremizing the action~\eqref{eq:NG-z-of-phi}, then it will be the preferred configuration, as it will be the one of the lowest energy. On the other hand, if the energy of the configuration we find by extremizing the action~\eqref{eq:NG-z-of-phi} is higher than that of two disconnected worldsheets (i.e., if their regularized action is negative), then the dynamically favored configuration will be the ``trivial'' one, given by the two disconnected worldsheets. We note that no spatial separation between the two timelike Wilson lines is necessary in this configuration, as the angular separation on the $S_5$ already provides the surface $\Sigma$ a non-vanishing coordinate region for it to extend itself over.

Therefore, let us calculate the extremal surfaces that can be derived from Eq.~\eqref{eq:NG-z-of-phi}\@. This action has a conserved quantity due to the explicit independence of the action on $\phi$,
\begin{equation}
    \frac{f(z)}{\sqrt{z'{}^2+ f(z) z^2}} = \sqrt{\frac{f(z_m)}{z_m^2}} \, ,
\end{equation}
which allows us to find an implicit solution for $z(\phi)$ by direct integration
\begin{equation}
\label{eqn:arc_length}
    \int_0^{z(\phi)}  \frac{dz}{\sqrt{z_m^2 f(z) - z^2 f(z_m)}} \sqrt{\frac{f(z_m)}{f(z)}} = \phi \, ,
\end{equation}
where we have chosen one of the timelike Wilson lines to lie at $\phi = 0$\@. The above expression determines the worldsheet configuration up to its maximal radial value $z_m$\@. The value of $z_m$ may then be related to (half of) the total angular distance spanned by the worldsheet $\Delta \phi$ by replacing the upper limit $z(\phi)$ with $z_m$ in Eq.~\eqref{eqn:arc_length}:
\begin{equation} \label{eq:delta-phi-zm}
    \Delta \phi = \int_0^{z_m}  \frac{dz}{\sqrt{z_m^2 f(z) - z^2 f(z_m)}} \sqrt{\frac{f(z_m)}{f(z)}}  \, .
\end{equation}

\begin{figure}
    \centering
    \includegraphics[width=0.4\textwidth]{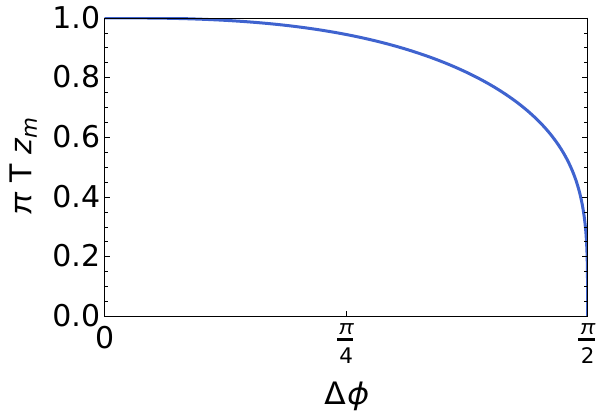}
    \includegraphics[width=0.4\textwidth]{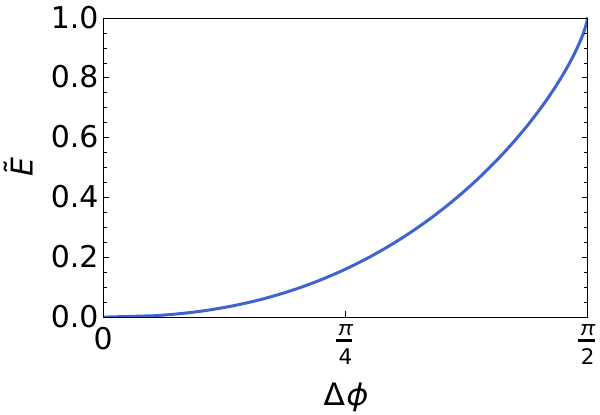}
    \caption{Left: the relation $\Delta \phi ( \pi T z_m )$ that determines the angular distance spanned on the $S_5$ by the connected configuration that reaches a maximal AdS radial coordinate $z = z_m$\@. Right: dimensionless configuration energy $\widetilde{E}$ for the extremal worldsheet that is described by a connected configuration $z = z(\phi)$\@.}
    \label{fig:S5-background}
\end{figure}

One can then evaluate the regularized action $\mathcal{S}_{\rm NG}[\Sigma] - \mathcal{S}_0$ with this solution. A straightforward calculation gives the energy of the configuration as
\begin{align}
    \mathcal{S}_{\rm NG}[\Sigma] - \mathcal{S}_0 &= - \frac{\sqrt{\lambda} \T }{2\pi} \int_0^\pi d\phi \frac{\sqrt{z'{}^2+ f z^2}}{z^2} + \frac{\sqrt{\lambda} \T}{\pi} \int_0^{(\pi T)^{-1}} \frac{dz}{z^2} \nonumber \\ 
    &= - \frac{\sqrt{\lambda} \T }{\pi} \int_0^{z_m} \frac{dz}{z^2} \left( \frac{1}{ \sqrt{1 - \frac{z^2 f(z_m)}{z_m^2 f(z)} }} - 1 \right) + \frac{\sqrt{\lambda} \T}{\pi} \int_{z_m}^{(\pi T)^{-1}} \frac{dz}{z^2} \nonumber \\
    &= - \frac{\sqrt{\lambda} \T}{\pi} (\pi T) \frac{1}{\pi T z_m} \left[ \int_0^1 \frac{du}{u^2} \left( \frac{1}{\sqrt{1 - u^2 \frac{1- (\pi T z_m)^4 }{1 - (\pi T z_m)^4 u^4} }} - 1 \right) - (1 - (\pi T z_m) ) \right] \nonumber \\
    &\equiv - \sqrt{\lambda} \T T \widetilde{E}(\pi T z_m) \, .
\end{align}
where $\widetilde{E}(\pi T z_m)$ is a function of a single variable that characterizes the configuration energy as a function of the $S_5$ angular separation $2\Delta \phi \in [0, \pi]$ of the two boundary timelike Wilson lines, determined by Eq.~\eqref{eq:delta-phi-zm}\@. (That is to say, we define $\Delta \phi$ to satisfy $\Delta \phi \in [0, \pi/2]$.) We plot this quantity in Fig.~\ref{fig:S5-background}, i.e., we plot the energy of the configuration in units of the temperature times $\sqrt{\lambda}$, together with the relation that determines $\Delta\phi = \Delta \phi(\pi T z_m)$ as prescribed by Eq.~\eqref{eq:delta-phi-zm}\@. We see that for worldsheets that can be parametrized by the functions $z(\phi)$, and therefore are connected, the map $\Delta \phi = \Delta \phi (\pi T z_m)$ is one to one, and that at any $\Delta \phi > 0$, their energy is strictly greater than that of two disconnected, radially infalling worldsheets each hanging from their respective boundary toward the bulk of AdS${}_5$\@. 

Crucially, this means that for our present purpose, which is to find the minimal energy configuration for $\Delta \phi = \pi/2$, the relevant extremal surface is that of two disconnected (at least at times $|t| \ll \T/2$) worldsheets hanging radially into the bulk of AdS${}_5$\@. The alternative solution, which is energetically disfavored, is a surface that lies at $z=0$ and can be parametrized by time $t \in (-\T/2,\T/2)$ and the angle $\phi \in (0 , \pi)$\@. It is interesting to note that in the strict limit $T = 0$, all connected configurations have the same energy as the radially infalling solution (i.e., zero), for any value of $z_m$\@. However, since we are interested in the physics in the presence of a thermal plasma, there is no ambiguity in terms of which solution to choose.

Therefore, the solution presently relevant to our purpose is parametrized by two disjoint surfaces
\begin{align}
X_L^\mu(\tau,z) = (\tau, 0, 0 , 0, z, -\hat{n}_0) \, ,
\end{align}
and
\begin{align}
X_R^\mu(\tau,z) = (\tau, 0, 0 , 0, z, \hat{n}_0) \, ,
\end{align}
where $z \in (0, (\pi T)^{-1} )$ is an independent coordinate in this description, and plays the role of one of the worldsheet coordinates.
We note that the solution we found above also applies if the two timelike sides of the contour $\mathcal{C}$ are at nonzero spatial separation $L>0$ (provided they remain at antipodal positions $\Delta \phi = \pi/2$ on the $S_5$), because allowing for a $\phi$-dependent $y_1$ coordinate in the connected background solution can only increase the configuration energy. 

In summary, this configuration features two worldsheets that fall into the black hole, which intuitively represents the propagation of two unbound heavy quarks, with their interactions being screened by the thermal medium~\cite{Friess:2006rk,Liu:2006ug,Casalderrey-Solana:2011dxg}\@. By construction, this configuration has $E^d = 0$ (after subtracting the energy associated to the heavy quark masses), as required to satisfy $\langle W_{\rm BPS}[\mathcal{C}] \rangle = 1$\@.

\subsubsection{Fluctuations}

To calculate the response kernel of interest for this configuration $\Delta_{ij}$, we study the dynamics of fluctuations on top of the background worldsheet we just found. Consistent with the preceding discussion, we work in the limit $\T \to \infty$ (concretely, $|t_1|,|t_2|\ll \T$), which also simplifies the calculations because of the time translational invariance. In this setup, whether $\T$ is finite or infinite
does not affect the final result, as long as the time domain of interest is covered by the timelike Wilson lines, since the parts of the timelike Wilson lines that are out of the time domain of interest cancel due to $U U^{-1} = \mathbbm{1}$\@.\footnote{Up to operator ordering subtleties that do not affect this conclusion. We discuss this in Appendix~\ref{sec:App-W-ordering}\@.} As such, taking $\T \to \infty$ is not an approximation, but rather, it is explicitly equal to the result at any value of $\T > |t_2|, |t_1|$.

As discussed in Section~\ref{sec:setup}, the appropriate boundary deformations of the contour $\mathcal{C}$ with which to evaluate the chromoelectric field correlator in the pure SU($N_c$) gauge theory are the antisymmetric ones, as shown in Eq.~\eqref{eq:f-antisymm}\@. By the same argument as in that Section, these are also the only nontrivial deformations from which we can extract the desired correlation function in our present setup. The reason is that symmetric boundary deformations of the contour $\mathcal{C}$ preserve the value of the (supersymmetric) Wilson loop $W_{\rm BPS}[\mathcal{C}] = 1$, which is also a consequence of having each antiparallel timelike Wilson line with antipodal positions on the $S_5$\@.

Having made these remarks, we now proceed to introduce perturbations on top of the background worldsheet to enable us to evaluate the path functional derivatives on the boundary. Compared to the setup in the previous section, the parametrization of the perturbed worldsheet here is remarkably simpler:
\begin{align}
    X^\mu_L(t,z) &= (t, - y_1(t,z), - y_2(t,z), - y_3(t,z), z, -\hat{n}_0) \, , \\
    X^\mu_R(t,z) &= (t, + y_1(t,z), + y_2(t,z), + y_3(t,z), z, + \hat{n}_0) \, ,
\end{align}
where we have already set to zero the fluctuations corresponding to reparametrization invariance, $y_0$ and $y_4$\@. We have also already incorporated the fact that, because of the antisymmetry of the boundary conditions we will use, the solutions for the fluctuations on either surface will be equal but with opposite signs. 

The action, up to quadratic order for the fluctuations, takes the form
\begin{align}
    \mathcal{S}_{\rm NG}[\Sigma] - \mathcal{S}_0[\mathcal{C};\hat{n}] = - 2 \times \frac{\sqrt{\lambda}}{2\pi} \left[ S_{{\rm NG}}^{(2),\perp}[y_1] + S_{{\rm NG}}^{(2),\perp}[y_2] + S_{{\rm NG}}^{(2),\perp}[y_3] \right] \, ,
\end{align}
where we have subtracted the action corresponding to the energy of two heavy quarks at rest, which is incidentally equal to the background action in this case. The form of the action for the fluctuations is the same for all components and is given by
\begin{align}
    S_{{\rm NG}}^{(2),\perp}[y] = \int_{-\infty}^\infty \!\! dt \int_0^{(\pi T)^{-1}} \!\!\!\!\!\! dz \left[ \frac{f}{2z^2} y'{}^2 - \frac{1}{2z^2 f} \dot{y}{}^2 \right] \, ,
\end{align}
where a prime denotes the derivative with respect to $z$ and a dot represents the derivative with respect to $t$\@.

Integrating by parts, and using the equations of motion, we can evaluate the on-shell action as
\begin{align} \label{eq:S-fluct-thermal-on-shell}
    S_{{\rm NG}}^{(2),\perp}[y]_{\rm on-shell} = { \frac12} \int_{-\infty}^\infty dt \left[ \lim_{z \to (\pi T)^{-1} } \frac{f y'(t, z ) y(t, z )}{z^2} - \lim_{z \to 0} \frac{f y'(t, z ) y(t, z )}{z^2} \right] \, ,
\end{align}
which involves two explicitly nontrivial limits. Let us first focus on the first limit, where $z \to (\pi T)^{-1}$\@. While it is tempting to conclude that it is zero because $f(z=(\pi T)^{-1}) = 0$, we actually have to solve for the mode functions first and verify that its product with $y' y$ indeed goes to zero. In the following Section~\ref{sec:EE-calculation-QQ}, after selecting appropriate boundary conditions at the black hole horizon, we will confirm that this is the case. Therefore, we shall drop this term in the remainder of this section. 

The second limit, where $z \to 0$, is even more subtle, because it can be manifestly divergent if $y'$ does not go to zero fast enough. Nonetheless, after solving for the mode functions and investigating them at small $z$, we shall see that it contains a $1/z$ divergent term of the form discussed around Eq.~\eqref{eq:F-insertion-2} and in the footnote.\footref{fn:delta-disc} That is to say, it is generated by the contact term that comes from evaluating the variational derivatives with respect to the path deformations $f^\mu$ at the same point. 
Moreover, this contribution has exactly the same form and value as that calculated in Appendix~\ref{sec:EE-calculation-HQ} for the Wilson loop with constant $\hat{n}$, meaning that it exactly encodes the contact terms that are not part of the correlation function we are after.
As such, we are justified to simply subtract the second limiting term from our final result. Furthermore, using that we will find the relevant mode functions satisfy $y'(t,z=0) = 0$, and we can conclude by repeated use of L'Hopital's rule that
\begin{align}
    S_{{\rm NG},d}^{(2),\perp}[y]_{\rm on-shell} = - \frac{1}{ 4} \int_{-\infty}^\infty dt \frac{\partial^3 y(t, z = 0)}{\partial z^3}  y(t, z=0) \, .
\end{align}
Therefore, the response function we will need to calculate has a third derivative. The other way of distributing the three derivatives gives terms of the form $ \frac{\partial^2 y}{\partial z^2} \frac{\partial y}{\partial z}$, which vanish because the mode functions satisfy $y'(t,z=0) = 0$\@.

Let us now define the response function we will calculate. We identify the value of the fluctuation on the boundary with that of the contour deformation $y(t,z=0) = h^\perp(t)$\@. As such, we introduce the response kernel
\begin{align} \label{eq:K-perp-d}
    \frac{\partial^3 y}{\partial z^3} (t, z = 0) = -\int_{-\infty}^\infty dt' K_\perp(t-t') h^\perp(t') \, ,
\end{align}
with which we find
\begin{align}
    S_{{\rm NG}}^{(2),\perp}[y]_{\rm on-shell} = { \frac14} \int_{-\infty}^\infty dt \, dt' \, h^\perp(t) K_\perp(t-t') h^\perp(t') \, .
\end{align}
With this, the correlation function we seek is determined by
\begin{equation} \label{eq:Delta-disconnected}
    \Delta_{ij}(t_2 - t_1) =  \frac{\sqrt{\lambda}}{{ 2}\pi} \delta_{ij} K_\perp(t_2-t_1) \, .
\end{equation}
Consequently, all that remains to be done is to evaluate the response function $K_{\perp}$\@.

\subsubsection{Calculation of the time-ordered non-Abelian electric field correlator} \label{sec:EE-calculation-QQ}

To calculate the response function $K_{\perp}$, we proceed by varying the action $S_{{\rm NG}}^{(2),\perp}$ with respect to $y$ to obtain its equations of motion, and then transform to the frequency domain. Then, introducing $\xi = \pi T z$, the equation we want to solve is 
\begin{equation} \label{eq:eom-y-th}
 \frac{\partial^2 y_\omega}{\partial \xi^2} - \frac{2}{\xi}  \frac{1+\xi^4}{1 - \xi^4}   \frac{\partial y_\omega}{\partial \xi} + \frac{\omega^2}{(\pi T)^2} \frac{1}{(1 - \xi^4)^2} y_\omega = 0 \, ,
\end{equation}
which is actually equivalent to the one found in Ref.~\cite{Casalderrey-Solana:2006fio} where it was used to calculate the heavy quark diffusion coefficient in strongly coupled $\mathcal{N}=4$ SYM theory. For the benefit of the reader, we note that in their notation, the independent variable that parametrizes the worldsheet is $u = \xi^2$\@. 

To find the solutions to Eq.~\eqref{eq:eom-y-th}, we proceed as in Ref.~\cite{Casalderrey-Solana:2006fio} to factor out the highly oscillatory piece that is generated close to the black hole event horizon. 
Therefore, we introduce
\begin{equation} \label{eq:mode-functions-yF}
    y^{\pm}_\omega(\xi) = (1-\xi^4)^{\pm \frac{i \Omega}{4} } F^{\pm}_{\omega}(\xi) \, ,
\end{equation}
with $\Omega \equiv \omega/(\pi T)$, and where the prefactor $(1-\xi^4)^{\pm {i \Omega}/{4} }$ follows from defining 
\begin{equation}
    y^{\pm}_\omega(\xi) = \exp \left( \mp i \Omega \int_0^\xi \frac{\xi'^3 d\xi'}{1 - \xi'^4} \right) F^{\pm}_\omega(\xi) \, ,
\end{equation}
to remove the highly oscillatory contribution induced by the $\frac{\omega^2}{(\pi T)^2} \frac{1}{(1 - \xi^4)^2} y_\omega$ term in Eq.~\eqref{eq:eom-y-th}. Basically, we have used the same method as in the celebrated WKB approximation to separate slowly oscillating pieces from highly fluctuating contributions.
To facilitate comparison, we have written $y_\omega^\pm$ in the way of Eq.~\eqref{eq:mode-functions-yF} such that the resulting mode functions are the same as in Ref.~\cite{Casalderrey-Solana:2006fio}\@. With this definition, $F^\pm_\omega$ satisfies
\begin{align} \label{eq:F-thermal}
    \frac{\partial^2 F^\pm_\omega}{\partial \xi^2} - 2 \left[ \frac{1 + \xi^4}{\xi(1-\xi^4)} \pm \frac{i \Omega \xi^3}{1-\xi^4} \right] \frac{\partial F^\pm_\omega}{\partial \xi} + \left[ \mp \frac{i \Omega \xi^2}{1-\xi^4} + \frac{\Omega^2 (1 - \xi^6) }{(1-\xi^4)^2} \right] F^\pm_\omega = 0 \, .
\end{align}

By casting Eq.~\eqref{eq:eom-y-th} in terms of $F^\pm_\omega$, we have isolated the highly oscillatory phase from $y_\omega^\pm(\xi)$ and obtained an equation for $F_\omega^\pm(\xi)$ such that a regular solution can be found at $\xi = 1$\@. This condition must be imposed by hand, because Eq.~\eqref{eq:F-thermal} has two independent solutions: one regular at the horizon, and the other oscillating twice as fast as the solutions for $y_\omega^\pm(\xi)$\@.
Examining the differential equation for $F_\omega^\pm$ and demanding regularity at the horizon, we find this implies
\begin{align}
    & \lim_{\xi \to 1} (1-\xi^4) \left[ - 2 \left[ \frac{1 + \xi^4}{\xi(1-\xi^4)} \pm \frac{i \Omega \xi^3}{1-\xi^4} \right] \frac{\partial F^\pm_\omega}{\partial \xi} + \left[ \mp \frac{i \Omega \xi^2}{1-\xi^4} + \frac{\Omega^2 (1 - \xi^6) }{(1-\xi^4)^2} \right]  F^\pm_\omega \right] = 0 \nonumber \\
    & \implies \frac{1}{F^\pm_\omega(\xi=1)} \frac{\partial F^\pm_\omega(\xi=1)}{\partial \xi} = \mp \frac{i \Omega}{4} \frac{1 \pm \frac{3 i \Omega}{2} }{1 \pm \frac{i \Omega}{2}} \label{eq:F-horizonder} \, ,
\end{align}
where the last condition fully determines the mode solution, up to an overall normalization. This condition allows one to find numerical solutions to Eq.~\eqref{eq:F-thermal} ensuring regularity at the horizon.

The other input required to determine the correlation function is the boundary conditions, i.e., the prescription to select the appropriate linear combination of the mode functions that determines the response kernel. Because we have extended our contour to infinity by taking the limit $\T \to \infty$, the boundary conditions are determined by the time-ordering prescription $\omega \to \omega (1 + i\epsilon)$, which is a consequence of the aspects discussed in Section~\ref{sec:ends-matching}\@. For concreteness, let us focus on the case $\omega > 0$\@. This is without loss of generality, because we are calculating a time-ordered correlation, and so, the full result will be immediately obtained by taking $\omega \to |\omega|$\@.

With these preliminaries, we can now analyze the mode functions~\eqref{eq:mode-functions-yF} under an infinitesimal complex rotation $\omega \to \omega (1 + i\epsilon)$\@. To select or discard a solution, we need to know whether one of the modes generates a divergent limit in the action~\eqref{eq:S-fluct-thermal-on-shell}\@. In particular, whether the limit
\begin{align} \label{eq:limit-onshell-thermal}
     \lim_{z \to (\pi T)^{-1} } \frac{f y'(t, z ) y(t, z )}{z^2}
\end{align}
exists for the mode solutions $y^{\pm}_\omega$ when the $i\epsilon$ prescription is taken into account. The reason why this particular limit is relevant is because of its explicit appearance in the expression for the on-shell action~\eqref{eq:S-fluct-thermal-on-shell}, which must be finite for the action to be at a well-defined extremum.

As discussed before, $F_\omega^\pm$ is regular and finite at the horizon $\xi=1$, and by inspecting the differential equation~\eqref{eq:F-thermal} that defines it, it is also analytic in $\omega$\@. As such, no singularity will appear in $F_\omega^\pm(\xi=1)$ by rotating the frequency $\omega$ by a small amount from the real axis into the complex plane. Therefore, the deformation by $i\epsilon$ affects the result predominantly through the WKB factor $\exp( \pm \frac{i\Omega - \Omega \epsilon}{4} \ln (1 - \xi^4) ) $\@. However, this means that $y^+_\omega$ will grow as $e^{\epsilon \Omega |\ln (1-\xi^4) |/4 }$ close to the horizon for $\omega > 0$, and therefore, substituting the mode function $y_\omega^+$ into Eq.~\eqref{eq:limit-onshell-thermal} leads to a divergent limit. We then conclude that we must keep only $y^-_\omega$ as the allowed mode solution for $\omega > 0$\@. By extension, the mode solution to be kept at arbitrary $\omega$ is $y^-_{|\omega|}$\@.

Now we can evaluate the response function $K_\perp(\omega)$ by means of substituting $y_\omega^-$ into Eq.~\eqref{eq:K-perp-d}\@. Given what we just showed, the WKB factor of $y_\omega^-$ goes like $\exp( - \frac{i\Omega - \Omega \epsilon}{4} \ln (1 - \xi^4) ) $, which goes to zero for $\omega >0$ as $\xi \to 1$\@. Consequently, the first term of the on-shell action~\eqref{eq:S-fluct-thermal-on-shell} vanishes, and we are left with the second term only. By direct inspection of the mode equation~\eqref{eq:F-thermal}, we see that regularity of $F^\pm_\omega$ requires
\begin{align}
    \frac{\partial F^\pm_\omega}{\partial \xi}(\xi=0) = 0 \, , & & \frac{\partial^2 F^\pm_\omega}{\partial \xi^2}(\xi=0) = \Omega^2 F^\pm_\omega(\xi=0) \, , \label{eq:boundary-behavior-F}
\end{align}
which verifies our earlier claim that $y'(t,z=0) = 0$\@.

The other claim of the previous section that we have yet to verify is that the (divergent) contact terms are the same as in the heavy quark interaction potential case. To show this, we may write the unregularized response kernel ${K}_{0\perp}$ from the second term in Eq.~\eqref{eq:S-fluct-thermal-on-shell}, and find
\begin{align}
    {K}_{0\perp}(\omega) &= - \frac{1}{y^-_{|\omega|}(z=0)} \lim_{z\to 0} \frac{2}{z^2} \frac{\partial y^-_{|\omega|}}{\partial z} = - \frac{1}{y^-_{|\omega|}(z=0)} \lim_{z\to 0} \frac{1}{z} \frac{\partial^2 y^-_{|\omega|}}{\partial z^2}\,,
\end{align}
where we have used L'Hopital's rule to obtain the second equality. By virtue of the second condition in Eq.~\eqref{eq:boundary-behavior-F}, $\frac{\partial^2 y^-_\omega}{\partial z^2}(z=0) = \omega^2 y^-_\omega(z=0)$, we can add and subtract the divergent part to get the unregularized response kernel as
\begin{align}
    {K}_{0\perp}(\omega) &=  - \frac{1}{y^-_{|\omega|}(z=0)} \lim_{z\to 0} \frac{\partial^3 y^-_{|\omega|}}{\partial z^3} - \omega^2 \lim_{z\to 0} \frac{1}{z} \, . \label{eq:K-thermal-unreg}
\end{align}
The last term in this equation is a contact term, exactly of the form discussed around Eq.~\eqref{eq:F-insertion-2} and in footnote.\footref{fn:delta-disc} Furthermore,
by comparison with our calculation of the correlator on the setup with constant $\hat{n}$ discussed in Appendix~\ref{sec:HQ-setup}, concretely, with Eq.~\eqref{eq:Delta-yy-LL-c}, and keeping in mind that the $z_m^2/\sqrt{f_m}$ in that expression is cancelled by the relative prefactor in the definition of $\Delta_{ij}^c$, it is clear that the nature of the last term in Eq.~\eqref{eq:K-thermal-unreg} is a local divergence that comes from the contact terms we first discussed in  footnote.\footref{fn:delta-disc} 

Therefore, switching back to $\xi = \pi T z$, we may write the regularized response kernel (which is the one that enters the chromoelectric field correlator) as
\begin{equation} \label{eq:K-perp-d-sol}
    K_\perp(\omega) = -  \frac{(\pi T)^3}{y^-_{|\omega|}(\xi=0)} \frac{\partial^3 y^-_{|\omega|}}{\partial \xi^3}(\xi=0) = -  \frac{(\pi T)^3}{F^-_{|\omega|}(0)} \frac{\partial^3 F^-_{|\omega|}}{\partial \xi^3}(0) \, .
\end{equation}
The last equality follows from direct inspection of the mode functions shown in Eq.~\eqref{eq:mode-functions-yF}, and the fact that the prefactor $(1-\xi^4)^{- i\Omega/4}$ has no effect in the result because all of its first three derivatives with respect to $\xi$ vanish.

In terms of $\Delta_{ij}^d$, the result is
\begin{align}
    \Delta_{ij}(\omega) = - \delta_{ij} \frac{\sqrt{\lambda} \pi^2 T^3}{{ 2} F^-_{|\omega|}(0)} \frac{\partial^3 F^-_{|\omega|}}{\partial \xi^3}(0) \, .
\end{align}
This result may be evaluated numerically after plugging the boundary condition shown in Eq.~\eqref{eq:F-horizonder} to the differential equation~\eqref{eq:F-thermal}, which defines $F_\omega^-$\@.\footnote{Contrary to what one might hope, the third derivative with respect to $\xi$ may not be evaluated directly from the differential equation~\eqref{eq:F-thermal}\@. If one takes another derivative and $\xi \to 0$, one ends up with an identity.} We plot the result for a general temperature in the next section, where we give the final result of our calculation.

\begin{figure}
    \centering
    \includegraphics[width=\textwidth]{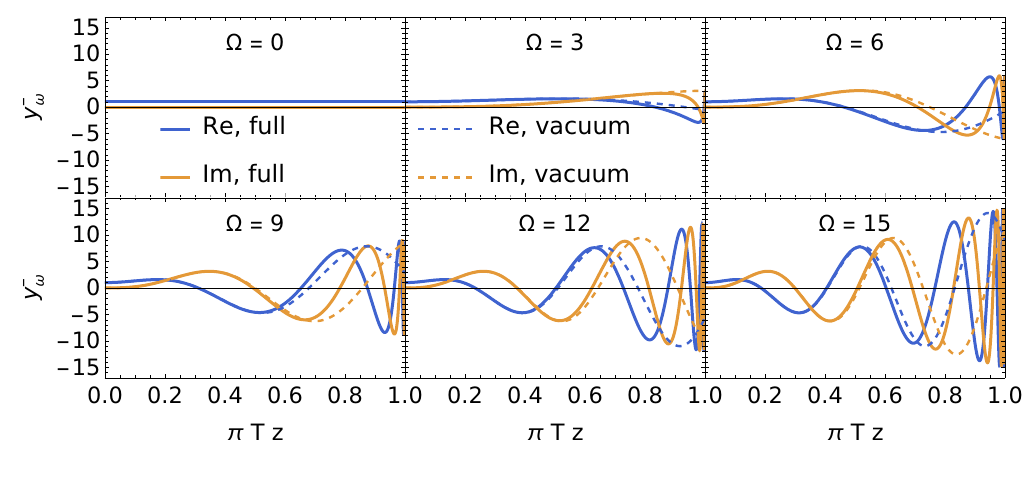}
    \caption{Solid lines: real (blue) and imaginary (orange) parts of the mode solution $y_\omega^{-}$ at selected values of $\Omega = \omega / (\pi T)$\@. Dashed lines: real (blue) and imaginary (orange) parts of the mode solution $y_\omega^{-}$ for $T = 0$. The arguments of the vacuum solutions are rescaled by $(\pi T)^{-1}$, which is the position of the event horizon in black hole AdS, to allow for a clean visual comparison at the same physical value of the AdS${}_5$ radial coordinate $z$\@.}
    \label{fig:mode-functions}
\end{figure}

Before proceeding further, it is also instructive to evaluate the zero temperature limit of our expression. To do so, it is most convenient to go back to the original AdS coordinate $z$ in the mode equation for $y_\omega$\@. Then, setting $T=0$, Eq.~\eqref{eq:eom-y-th} becomes
\begin{align}
    \frac{\partial^2 y_\omega}{\partial z^2} - \frac{2}{z} \frac{\partial y_\omega}{\partial z} + \omega^2 y_\omega = 0
\end{align}
and the solutions are relatively simpler:
\begin{align}
    y_\omega^\pm(z) = \left( 1 \pm i \omega z \right) e^{ \mp i \omega z} \, .
\end{align}
We choose the sign labelling so that the solutions match their finite-temperature counterparts in the limit $T\to 0$\@. For visual clarity, we show the mode solutions as a function of $z$ for a range of frequencies rescaled by the temperature in Fig.~\ref{fig:mode-functions}\@. The fact that we have an explicit expression then allows us to evaluate Eq.~\eqref{eq:K-perp-d-sol} explicitly, obtaining
\begin{equation}
    K_\perp(\omega)_{T=0} = - 2 i |\omega|^3 \, .
\end{equation}
We note that the cubic power in the frequency is exactly what one expects from dimensional analysis of the correlation function we are interested in from the field theory perspective. In terms of $\Delta_{ij}$, we have
\begin{equation}
    \Delta_{ij}(\omega)_{T = 0} = - \delta_{ij} \frac{i \sqrt{\lambda}} {\pi} |\omega|^3 \, .
\end{equation}

In summary, we have obtained the response kernel $K_\perp$ that determines the on-shell Nambu-Goto action up to second order in the contour deformations of the Wilson loop expectation value that is dual to it. Contrary to the results in Appendix~\ref{sec:HQ-setup} for a constant value of $\hat{n} \in S_5$, these are well-behaved, have a well-defined result for a Wilson loop made up of two antiparallel Wilson lines at the same position, and so provide a quantitative description of the dynamics of in-medium quarkonium. Furthermore, they are not sensitive to modes outside the domain of a low-energy description, in contrast to what happens in Appendix~\ref{sec:HQ-setup}, where the contribution from UV modes causes the result to diverge as $1/L^3$ as the limit $L \to 0$ (by which the antiparallel Wilson lines are drawn to the same spatial position) is taken. From our discussion in Section~\ref{sec:nhat}, the background configuration where $\hat{n}$ takes antipodal positions on the $S_5$ is also well-founded. Therefore, we conclude that this is the $\mathcal{N}=4$ observable that most closely resembles the analogous QCD correlation function, and will use it as the $\mathcal{N}=4$ result for the quarkonium transport coefficients.\footnote{Insofar as $\mathcal{N}=4$ SYM is a different theory than QCD, analogous results are, in the end, all we can get.} We give the expressions and plots for the chromoelectric field correlator that we extract from this holographic calculation in the next section, where we also discuss its implications as a baseline for phenomenological applications.

\subsubsection{Final result} \label{sec:ads-cft-final-result}




Putting everything together, we find from Eq.~\eqref{eq:EE-Delta} that the electric field correlator relevant to quarkonium transport is given by
\begin{align}
     \frac{g^2}{N_c} [g_E^{\T}]^{\mathcal{N}=4}_{ij}(\omega) =  \delta_{ij}  \frac{ (\pi T)^3 \sqrt{\lambda}}{{ 4}\pi} \left( \frac{-i}{F^-_{|\omega|}(0)} \frac{\partial^3 F^-_{|\omega|}}{\partial \xi^3}(0) \right) \, ,
\end{align}
where $F_{|\omega|}^{-}$ is determined by the regular solution to Eq.~\eqref{eq:F-thermal} when the sign choice corresponding to the superscript in $F_{|\omega|}^{-}$ is made.

\begin{figure}
    \centering
    \includegraphics[width=0.49\textwidth]{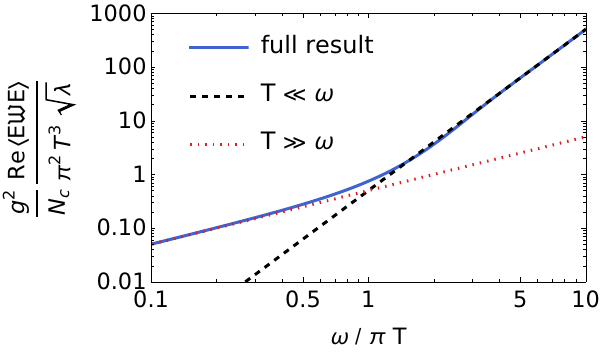}
    \includegraphics[width=0.49\textwidth]{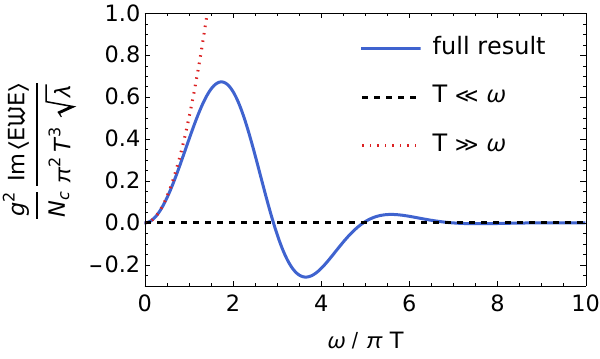}
    \caption{Real part (left) and imaginary part (right) of the non-Abelian electric field correlator of interest. The finite temperature result is shown in solid lines, and the zero temperature limit is shown in black dashed lines. The leading low-frequency limit is shown in red dotted lines. As before, the arguments of the functions at zero temperature have been rescaled by $\pi T$ to have a clean visual comparison.}
    \label{fig:correlator}
\end{figure}

This is the main result of this section. We plot this in Fig.~\ref{fig:correlator}\@. Furthermore, its zero-temperature limit is given by
\begin{align} \label{eq:EE-SYM-zero-T-limit}
     \frac{g^2}{N_c} [g_E^{\T}]^{\mathcal{N}=4}_{ij}(\omega)_{T = 0} = \delta_{ij}\frac{ \sqrt{\lambda}}{{ 2}\pi} |\omega|^3  \, .
\end{align}

The other limit of interest is the low-frequency limit, which can be extracted analytically by solving the mode equation~\eqref{eq:F-thermal} up to linear order in $\Omega$. The algebraic steps necessary to do this are the same as those in the heavy quark diffusion coefficient calculation~\cite{Casalderrey-Solana:2006fio}\@. The result is
\begin{align}
     \frac{g^2}{N_c} [g_E^{\T}]^{\mathcal{N}=4}_{ij}(\omega) = \delta_{ij} \frac{ \sqrt{\lambda} (\pi T)^3 }{{ 2}\pi} \left[ \frac{|\omega|}{\pi T}  + i \frac{\omega^2}{(\pi T)^2} + O \! \left( \left(\frac{\omega}{\pi T}\right)^3 \right) \right] \, , \label{eq:small-omega-expansion}
\end{align}
where we have kept one higher order in $\omega/T$ than that explicitly shown in Ref.~\cite{Casalderrey-Solana:2006fio}\@. The details of how this expansion was carried out can be found in Appendix~\ref{sec:App-omega-expansion}\@.

With the expression for the correlation function in hand, we can now use it to describe how a heavy quark-antiquark pair will propagate through the thermal $\mathcal{N}=4$ SYM plasma. Specifically, we can calculate the spectral function that determines the transition rates of in-medium quarkonium within a potential non-relativistic EFT description as shown in Sections~\ref{sec:rho-dynamics} and~\ref{sec:prev}, and draw the phenomenological implications thereof for strongly coupled plasmas. 

\subsubsection{Evaluation of chromoelectric field spectral function} \label{sec:ads-cft-spectral-eval}

\begin{figure}
    \centering
    \includegraphics[width=0.8\textwidth]{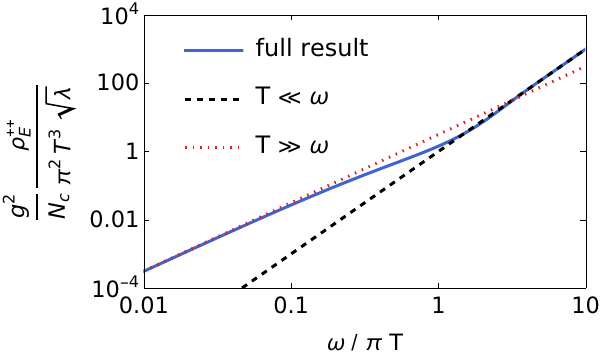}
    \caption{Spectral function for quarkonium transport calculated in $\mathcal{N} = 4$ SYM. Only the positive frequency domain is shown, as $\rho_E^{++}$ vanishes for $\omega < 0$\@. In Section~\ref{sec:weak-strong} we will compare this result with the weakly coupled limit of QCD, given by Eq.~\eqref{eq:spectral-full-result} and displayed in Fig.~\ref{fig:spectral-noHTL}.}
    \label{fig:Spectral}
\end{figure}

We are now ready to evaluate the spectral function that encodes the effect of the plasma on quarkonium transport, by using the relations introduced in Section~\ref{sec:GGDs}, as dictated by our $\mathcal{N}=4$ SYM result. Specifically, we use
\begin{equation} \label{eq:gE++-from-T-ordered-sec}
    [g_{E}^{++}]^>(\omega) = {\rm Re} \left\{ [g_E^{{\mathcal{T}}}]^{\mathcal{N}=4}(\omega) \right\} + \frac{1}{\pi} \int_{-\infty}^\infty d p_0 \, \mathcal{P} \left( \frac{1}{p_0} \right) {\rm Im} \left\{ [g_E^{{\mathcal{T}}}]^{\mathcal{N}=4}(\omega + p_0) \right\} \, .
\end{equation}
In general, the above equation would be an integral expression that can only be evaluated numerically. However, in $\mathcal{N}=4$ Yang-Mills theory, the fact that one obtains $[g_E^{{\mathcal{T}}}]^{\mathcal{N}=4}(\omega)$ by selecting modes using the time-ordering prescription $\omega \to \omega(1 + i\epsilon)$ gives us more analytic control. As we show in Appendix~\ref{sec:App-analytic-rot}, this property allows us to prove that
\begin{equation} \label{eq:analytic-prop-gE}
    \frac{1}{\pi} \int_{-\infty}^\infty d p_0 \, \mathcal{P} \left( \frac{1}{p_0} \right) {\rm Im} \left\{ [g_E^{{\mathcal{T}}}]^{\mathcal{N}=4}(\omega + p_0) \right\} = {\rm sgn}(\omega) {\rm Re} \left\{ [g_E^{{\mathcal{T}}}]^{\mathcal{N}=4}(\omega) \right\} \, ,
\end{equation}
and consequently, the generalized gluon distribution (GGD) that enters the quantum and classical quarkonium time evolution equations is given by
\begin{equation}
\label{eqn:gE++}
    [g_{E}^{++}]^>(\omega) = 2 \theta(\omega) {\rm Re} \left\{ [g_E^{{\mathcal{T}}}]^{\mathcal{N}=4}(\omega) \right\} \, .
\end{equation}
Some additional care is required to treat the UV contributions on the LHS of Eq.~\eqref{eq:analytic-prop-gE}, which we discuss in Appendix~\ref{sec:App-UV-div-Wick}. One more step gives us an explicit expression for the spectral function
\begin{equation} \label{eq:rho-strong}
    \rho_E^{++}(\omega) = 2 \theta(\omega) \big(1 - e^{- \omega/T} \big) {\rm Re} \left\{ [g_E^{{\mathcal{T}}}]^{\mathcal{N}=4}(\omega) \right\} \, ,
\end{equation}
which we plot in Fig.~\ref{fig:Spectral}\@. This function is manifestly neither even nor odd, as expected from the evidence coming from the perturbative calculations in Section~\ref{sect:nlo}\@.

One immediate implication of our results, which may already be seen from Fig.~\ref{fig:correlator} is that the transport coefficients introduced in the Quantum Brownian motion limit~\cite{Brambilla:2016wgg,Brambilla:2017zei,Eller:2019spw,Brambilla:2020qwo,Brambilla:2021wkt} of the open quantum system approach to in-medium quarkonium, namely, the analogs to
\begin{align}
\kappa_{\rm adj} &= \frac{T_F g^2}{3 N_c} {\rm Re} \int dt\, \big\langle \hat{\ml{T} }E^a_i(t) \mathcal{W}^{ab}(t,0) E^b_i(0) \big\rangle_T \\
\gamma_{\rm adj} &= \frac{T_F g^2}{3N_c} {\rm Im} \int dt\, \big\langle \hat{\ml{T}} E^a_i(t) \mathcal{W}^{ab}(t,0) E^b_i(0) \big\rangle_T \,,
\end{align}
vanish in strongly coupled $\mathcal{N}=4$ supersymmetric Yang-Mills theory. Explicitly,
\begin{equation}
    \kappa_{\rm adj}^{\mathcal{N}=4 } = \gamma_{\rm adj}^{\mathcal{N}=4 } = 0 \, .
\end{equation}
The quantity $\gamma_{\rm adj}$ represents the mass shift of quarkonium states inside a plasma. The result $\gamma_{\rm adj}^{\mathcal{N}=4 } = 0$ is consistent with a recent lattice QCD study~\cite{Bala:2021fkm}\@.

In the quantum optical limit where the quarkonium time evolution can be effectively described by a Boltzmann equation as shown in Eq.~\eqref{eqn:rate}, it is the finite frequency part of the chromoelectric field correlator that enters the quarkonium dissociation and recombination rates. Because the argument of $[g_E^{++}]^>$ is negative in Eq.~\eqref{eqn:disso}, our result in Eq.~\eqref{eqn:gE++} indicates that the dissociation rate of a small-size quarkonium state in a strongly coupled QGP vanishes. Using Eq.~\eqref{eqn:kms}, we also see that the recombination rate in Eq.~\eqref{eqn:reco} also vanishes in the same limit.

It is worth emphasizing again that from the field theory perspective, the correlation functions that characterize quarkonium and open heavy quark in-medium dynamics are fundamentally different~\footnote{Comparing to the results of~\cite{Casalderrey-Solana:2006fio}, we find a simple relation between the spectral functions for open heavy quark $\rho_{\rm fund}$ and quarkonium $\rho_{\rm adj}^{++}$ in $\mathcal{N}=4$ SYM:
\begin{equation*}
    \rho_{\rm adj}^{++}(\omega) =  \frac12 \theta(\omega) \big(1 - e^{-\omega/T} \big) \rho_{\rm fund}(\omega) \, .
\end{equation*}
}.
The spectral function for quarkonium $\rho_{\rm adj}^{++}(\omega) = (1 - e^{- \omega/T} ) 
[g_{\rm adj}^{++}]^>(\omega)$ is non-odd in $\omega$ and vanishes at negative frequencies, whereas that for heavy quark diffusion is odd in $\omega$. 

\subsection{Chromoelectric correlator in a flowing medium} \label{sec:corr-flow}

Phenomenologically, the situation where the rest frame of the plasma is moving with velocity $v$ relative to that of the $Q\bar{Q}$ pair is extremely relevant because heavy quarks need not be produced at rest relative to the QGP medium. Rather, the converse situation is the most common in HICs, i.e., that the $Q\bar{Q}$ pair is produced with a nonzero relative velocity to the QGP environment they will propagate through.

This can be complicated to calculate in the quantum field theory, because in general this amounts to writing the grand canonical density matrix of the plasma $\propto \exp \left( \int d\Sigma_\mu T^{\mu \nu} \beta_\nu \right)$ in a boosted frame instead of in its rest frame. 
However, the holographic duality provides us with a tool to calculate the correlator even when the medium is not at rest. All that needs to be done is to perform an appropriate change of coordinates. 
The dual gravitational description of this system is obtained by boosting the metric dual to a static plasma via a Lorentz transformation. This is straightforward to obtain, as the dual description of a static plasma on which thermal expectation values may be computed is an $AdS_5$-Schwarzschild spacetime ($\times S_5$). Such a description has been explored and discussed before in~\cite{Liu:2006he,Liu:2006nn}.

The calculation of the correlator is analogous to that in a medium at rest, described in Section~\ref{sec:QQ-setup}\@, except that the string configuration that hangs from each side of $\mathcal{C}_0$ is given by the trailing string of~\cite{Herzog:2006gh,Gubser:2006bz}\@, instead of a string hanging straight into the bulk. This is the lowest energy configuration because $\hat{n}$ takes antipodal positions on opposite sides of the contour. 

\subsubsection{Calculation of the correlator in a flowing medium}

Here we describe the calculation of the chromoelectric correlator that determines the in-medium dynamics of quarkonium for the case when the QGP environment is moving with respect to the heavy quark pair. We first discuss the setup of the background worldsheet calculation, which has been studied in the past~\cite{Herzog:2006gh,Gubser:2006bz}, and then proceed to discuss the dynamics of fluctuations on this surface. Some degree of parallel with~\cite{Casalderrey-Solana:2007ahi} will be explicit, but, in the same way as the calculation of Section~\ref{sec:QQ-setup} differs from that of~\cite{Casalderrey-Solana:2006fio}, there are important conceptual differences to be highlighted, which we will come to in due time.

The reason why the relevant background worldsheet is the string trailing a single heavy quark trajectory, and not the worldsheet configuration discussed in~\cite{Liu:2006he,Liu:2006nn} for purposes of studying the modification to the heavy quark interaction potential in a ``hot wind'' of QGP, is because each side of the contour is located at opposite points on the $S_5$. In the limit $\mathcal{T} \to \infty$ (essentially, $\mathcal{T} \pi T \gg 1$), locally, the lowest energy configuration for the worldsheet hanging from each side of the contour is to fall inwards as if there were only a single Wilson line. Physically, this is consistent with the fact that two heavy quarks in the octet representation cannot form a singlet bound state, and so propagate independently through the same trajectory. This is also consistent with the expectation that the heavy quark potential in the octet channel vanishes in the large $N_c$ limit. Furthermore, this configuration satisfies the expectation $\hat{\mathcal{T}} W[\mathcal{C}_0] = 1$ after subtracting the divergence due to the heavy quark masses. Conversely, the worldsheet configuration discussed in~\cite{Liu:2006he,Liu:2006nn} describes a $Q\bar{Q}$ bound state, and the Wilson loop need not satisfy $\hat{\mathcal{T}} W[\mathcal{C}_0] = 1$ because of the finite transverse separation between the antiparallel (timelike) Wilson lines. This also means that the antipodal positions on the $S_5$ play no special role, and the string configuration of least action selects a single point on the $S_5$ (see the discussion in Section~\ref{sec:nhat}).

To characterize the trailing worldsheet, we go to the rest frame of the heavy quarks, where the Wilson lines extend purely along the time direction, and the metric dual to the $\mathcal{N} = 4$ SYM plasma boosted along the $x_1$ direction is
\begin{align}
    {\rm d}s^2 = \frac{R^2}{z^2} \bigg\{ & - {\rm d}t^2 + {\rm d}x_1^2 + {\rm d}x_2^2 + {\rm d}x_3^2 + \frac{{\rm d}z^2}{f(z)} + z^2 {\rm d}\Omega_5^2 \nonumber \\ & + [1 - f(z)] \left( \cosh^2 \! \eta \, {\rm d}t^2 + \sinh 2\eta \, {\rm d}x_1 {\rm d}t + \sinh^2 \! \eta \, {\rm d}x_1^2 \right) \bigg\} \, ,
\end{align}
where $f = 1 - (\pi T z)^4$. By symmetry considerations, the background worldsheet may be locally described (at times $|t| \ll \mathcal{T}/2$) by
\begin{equation}
    X^{\mu} \to \left( t, \chi(z), 0, 0, z, \hat{n}_0 \right) \, ,
\end{equation}
and the Nambu-Goto action is therefore given by
\begin{equation}
    S_{\rm NG} = - 2 \times \frac{\sqrt{\lambda} \mathcal{T}}{2\pi} \int \frac{{\rm d}z}{z^2} \sqrt{\chi'{}^2 f + \cosh^2 \! \eta - \frac{\sinh^2 \! \eta}{f} } \, .
\end{equation}
(The factor of 2 is due to having two copies of the worldsheet at $\pm \hat{n}_0$.) The extremal surface that solves the equations of motion is determined by
\begin{equation}
    \chi'(z) = - \sinh \eta \frac{\sqrt{1 - f}}{f} = - \sinh \eta \frac{(\pi T z)^2}{1 - (\pi T z)^4} \, ,
\end{equation}
together with $\chi(0) = 0$. One may immediately verify that
\begin{equation}
    \sqrt{\chi'{}^2 f + \cosh^2 \! \eta - \frac{\sinh^2 \! \eta}{f} } = 1 \, ,
\end{equation}
as expected for the mass term that has to be subtracted in order to isolate the expectation value of the Wilson loop. This completely determines the background solution.

Following~\cite{Casalderrey-Solana:2007ahi}, to study the fluctuations on top of this solution it is convenient to introduce a shift in the time coordinate, namely, $\bar{t} = t + F(z)$.
Equivalently, the parametrization of the time coordinate on the worldsheet is now $t = \bar{t} - F(z)$. For obvious reasons, we will drop the bar in what follows. Also, as discussed in Section~\ref{sec:QQ-setup}, the $i\epsilon$ prescription that enforces time ordering on the Schwinger-Keldysh contour can be accounted for by taking $t$ to be slightly tilted into the negative imaginary direction of the complex time plane. On top of all of these ingredients, we introduce fluctuations parallel and perpendicular to the worldsheet, denoted, respectively, by $\Delta(t,z), \delta(t,z)$, and $y(t,z)$.\footnote{Because the worldsheet has a nontrivial profile along $z$, it is mathematically convenient to keep explicit fluctuations on both the $x_1$ and $z$ coordinates, denoted respectively by $\delta$ and $\Delta$. As will become apparent later, only one linear combination of them represents the physical longitudinal perturbation, and the other linear combination does not appear in the equations of motion as it encodes the reparametrization invariance of the worldsheet.} As such, the worldsheet parametrization is now
\begin{equation}
    X^{\mu} \to \left( t(1 - i\epsilon) - F(z), \chi(z) + \delta(t,z), y(t,z), 0, z + \Delta(t,z), \hat{n}_0 \right) \, .
\end{equation}
Choosing $F(z)$ to remove cross-terms in the differential equations for the fluctuations lead to choosing it to satisfy
\begin{equation} \label{eq:F-coord-redef}
    F'(z) = \frac{\sinh^2 \! \eta \cosh \eta}{f \cosh^2 \! \eta - \sinh^2 \! \eta} \frac{(1 - f)^{3/2}}{f} \, .
\end{equation}
Concretely, this sets to zero the coefficients of the terms proportional to $y' \dot{y}$ in the quadratic part of the action in the next paragraph (we denote ${\rm d}/{\rm d}z = ()'$, ${\rm d}/{\rm d}t = \dot{()}/(1 - i\epsilon)$). 

From now on, we choose units such that $\pi^2 T^2 \cosh \eta = 1$. (This is allowed because of conformal symmetry.) With this, the Nambu-Goto action, expanded up to quadratic order on $\Delta(t,z), \delta(t,z)$, and $y(t,z)$, reads
\begin{align}
    & S_{\rm NG}^{(0-2)} = - \frac{\sqrt{\lambda} (1 - i\epsilon) }{\pi}  \int  \frac{{\rm d}t \, {\rm d}z}{z^2} \nonumber \\ \times & \bigg[ 1  + \frac{z^4 \tanh \eta}{1-z^4} \dot{\delta} + z^2 \tanh \eta \, \delta' + \frac{4 \cosh^2 \! \eta \left( 1 - 5z^4 + 2z^8 + (1 + z^4) \cosh 2\eta \right)}{z (1 - 2z^4 + \cosh 2\eta)^2} \Delta \nonumber \\  
    &   - \frac{2 (1 - z^4) \cosh^2 \! \eta }{1 - 2z^4 + \cosh 2\eta } \Delta'  + \frac{\dot{y}^2}{2 (1 - z^4)}  - \frac{(1 - z^4)}{2} y'{}^2 + \frac{\dot{\delta}^2}{2 (1 - z^4)}  - \frac{(1 - z^4)}{2} \delta'{}^2 \nonumber \\ 
    &  + \frac{z^2 \sinh 2\eta}{1 - 2z^4 + \cosh 2\eta} \left( \frac{1}{ 1 - z^4 } \dot{\delta} \dot{\Delta} + z^2 (\dot{\delta} \Delta' - \delta' \dot{\Delta} ) -  (1 - z^4) \delta' \Delta' \right) \nonumber \\
    &  + \frac{2(1 + 2z^4 + \cosh 2\eta) \sinh 2\eta}{(1 - 2z^4 + \cosh 2\eta)^2} \left( z^3 \dot{\delta} \Delta - z(1-z^4) \delta' \Delta \right) \nonumber \\ 
    &  + \frac{2 z^4 \cosh^2 \! \eta \, \sinh^2 \! \eta}{(1 - 2z^4 + \cosh 2\eta)^2} \left( \frac{1}{1 - z^4} \dot{\Delta}^2 - (1 -z^4) \Delta'{}^2 \right) + \frac{4 z^5 \sinh^2 \! 2\eta }{(1-z^4) (1-2z^4 + \cosh 2\eta)^2} \dot{\Delta} \Delta \nonumber \\
    & + \frac{2 \cosh^2 \! \eta \left( P^{\Delta' \Delta}_{0}(z) + P^{\Delta' \Delta}_{2}(z) \cosh 2\eta + P^{\Delta' \Delta}_{4}(z) \cosh 4\eta \right) }{z(1-2z^4 + \cosh 2\eta)^3} \Delta' \Delta \nonumber \\
    & - \frac{\cosh^2 \! \eta \left( P^{\Delta \Delta}_{0}(z) + P^{\Delta \Delta}_{2}(z) \cosh 2\eta + P^{\Delta \Delta}_{4}(z) \cosh 4\eta + P^{\Delta \Delta}_{6}(z) \cosh 6\eta \right) }{2z^2 (1 - 2z^4 + \cosh 2\eta)^4 } \Delta^2 \bigg] \, , \!
\end{align}
where we have denoted, for brevity,
\begin{align}
    P^{\Delta' \Delta}_{0}(z) &= 3(1-4z^4 + 9z^8 - 4z^{12}) \, , \\ \nn
    P^{\Delta' \Delta}_{2}(z) &= 4(1-3z^4 -z^8 +z^{12}) \, , \\ \nn
    P^{\Delta' \Delta}_{4}(z) &= 1+z^8 \, , \\ \nn
    P^{\Delta \Delta}_{0}(z) &= 30 - 146z^4 + 32z^8 (10 - 17z^4 + 4z^8) \, , \\ \nn
    P^{\Delta \Delta}_{2}(z) &= (45 - 193 z^4 + 290 z^8 + 176 z^{12} - 32z^{16}) \, , \\ \nn
    P^{\Delta \Delta}_{4}(z) &= - 2(-9 + 23z^4 + 8z^8(2 + z^4)) \, , \\ \nn
    P^{\Delta \Delta}_{6}(z) &= (3 + z^4 - 2z^8) \, .
\end{align}

The first thing to note is the presence of linear terms in $\delta$, $\Delta$ in the action. These terms are, naturally, total derivatives, and do not contribute to the equations of motion. However, they could, as written, contribute to the on-shell value of the action. This is not expected nor acceptable on physical grounds, as a non-vanishing contribution at linear order in the perturbations would mean that, firstly, the action was not at an extremum to begin with, and secondly, it would generate a non-vanishing 1-point function of the chromoelectric field on the field theory side of the duality (which is unacceptable because ${\rm Tr} E_i = 0$ where the trace also includes summation over colors). Such considerations imply that consistent solutions for the mode functions of $\delta, \Delta$ will cancel these contributions.

Nonetheless, there is a simpler approach to deal with this potential issue. Geometrically, one can interpret the linear terms in the action for the fluctuations as deformations that are non-orthogonal to the background surface (if they were orthogonal, the action would start at quadratic order). Moreover, the physical perturbations, i.e., those that correspond to a genuine deformation of the surface, are exactly the ones that are orthogonal to the extremal surface. Consequently, the linear terms are associated with the reparametrization invariance of the string worldsheet. Consistently with reparametrization invariance, one can check that the Euler-Lagrange equations derived from extremizing $S_{\rm NG}^{(0-2)}$ with respect to $\delta$ and $\Delta$ are equivalent. Therefore, we can isolate the physical perturbations by setting
\begin{equation} \label{eq:physical-constraint}
    \Delta' = \frac{2}{z} \frac{\cosh^2 \! \eta  - (3 - \cosh^2 \! \eta)z^4 + z^8}{(1 - z^4)(1 - z^4 \, {\rm sech}^2 \eta ) \cosh^2 \! \eta } \Delta + z^2 \tanh \eta \, \frac{1 - z^4 \, {\rm sech}^2 \eta}{1-z^4} \delta' \, .
\end{equation}
This makes the perturbations orthogonal to the worldsheet along $z$. The $y$ perturbations are already orthogonal. As a side note, one may wonder what happens with the $\dot{\delta}$ term, which we have not cancelled by this choice. As it turns out, this can be dealt with in the same way if we had included perturbations for the time component of the worldsheet, i.e., $t(1-i\epsilon) - F(z) \to t(1-i\epsilon) - F(z) + \tau(t,z)$. Including the temporal perturbations $\tau(t,z)$ generates a linear term in the action, which can be chosen to compensate the $\dot{\delta}$ term, thus maintaining the perturbations orthogonal to the worldsheet. One can also verify that the equations of motion for $\tau(t,z)$ are trivial (i.e., all terms in the action that involve this perturbation are total derivatives).

It turns out one can integrate~\eqref{eq:physical-constraint} analytically. Because of local time translation invariance, from here on we Fourier transform $\delta$ and $\Delta$ from functions of time $t$ to functions of frequency $\omega$ (also, whenever we write $\omega$, we actually mean $\omega(1 + i\epsilon)$ due to the slight tilt of the Schwinger-Keldysh contour). The result is
\begin{equation}
    \Delta_\omega(z) = z^2 \tanh \eta \, \frac{1 - z^4 {\rm sech}^2 \eta}{1 - z^4} \left[ \delta_\omega(z) + a_\omega \right] \, ,
\end{equation}
where $a_\omega$ is an integration constant. Then, replacing this constraint in the equation of motion for $\delta$ (or $\Delta$, they are equivalent) to eliminate $\delta$ in favor of $\Delta$, one obtains
\begin{equation}
    \Delta_\omega''(z) - \frac{2}{z} \frac{3 - z^4}{1 - z^4} \Delta_\omega'(z) + \frac{2}{z^2} \frac{5 - z^4}{1 - z^4} \Delta_\omega(z) + \frac{\omega^2}{(1-z^4)^2} \Delta_\omega = \frac{a_\omega z^2 \omega^2 \tanh \eta}{(1 - z^4)^2} \, .
\end{equation}
One may directly verify that the particular solution to this equation is simply $a_\omega z^2 \tanh \eta$. It follows that we can write
\begin{equation}
    \Delta_\omega(z) = a_\omega \tanh \eta \, z^2 + A z^2 \tilde{\Delta}_\omega(z) \, ,
\end{equation}
where $A$ is a normalization constant and $\tilde{\Delta}_\omega(z)$ obeys
\begin{equation}
    \tilde{\Delta}_\omega''(z) - \frac{2}{z} \frac{1 + z^4}{1 - z^4} \tilde{\Delta}_\omega'(z) + \frac{\omega^2}{(1 - z^4)^2} \tilde{\Delta}_\omega(z) = 0 \, .
\end{equation}
Remarkably, this is the same equation that the perturbations satisfy in the case where the direction of the Wilson lines coincide with the rest frame of the medium. The only qualitative difference is the position of the (worldsheet) horizon, which here is at $z = (\pi T \sqrt{\cosh \eta})^{-1}$. The solutions to this equation at arbitrary $\omega$ have been studied in Section~\ref{sec:QQ-setup}. Due to the $i\epsilon$ prescription, the physical, regular solution to the equations of motion is given by only one of the mode functions that solve the homogeneous equation above, which, in the notation of Section~\ref{sec:QQ-setup}, corresponds to $\tilde{\Delta}_\omega(z) \propto (1 - z^4)^{-i|\omega|/4} F_{|\omega|}^{-}(z)$. Consequently, we have fully determined the mode functions for the fluctuations $\Delta$, $\delta$. Including the normalization constant $A$ for the above mode solutions, we find
\begin{align}
    \delta_\omega(z) &= - \frac{a_\omega \tanh^2 \! \eta \, z^4 }{1 - z^4 \, {\rm sech}^2 \eta} + A \frac{(1 - z^4)^{1 - i |\omega|/4}}{1 - z^4 \, {\rm sech}^2 \eta} F^{-}_{|\omega|}(z) \, , \\
    \Delta_\omega(z) &= a_\omega \tanh \eta \, z^2 + A z^2 (1 - z^4)^{-i |\omega|/4} F_{|\omega|}^{-}(z) \, .
\end{align}
Similarly, the mode functions for the transverse fluctuations $y$ are given by
\begin{equation}
    y_{\omega}(z) = B (1 - z^4)^{-i |\omega|/4} F_{|\omega|}^{-}(z) \, ,
\end{equation}
where $B$ is a normalization constant.

Finally, as discussed in Section~\ref{sec:QQ-setup}, the time-ordered correlator as a function of $\omega$ is obtained by evaluating the action on the solution with boundary conditions specified by Fourier mode deformations $y(t,z=0), \delta(t,z=0) = e^{- i \omega t}$. Specifically, in position space the correlator is obtained by extracting the quadratic part of the action
\begin{equation}
    \frac{g^2}{N_c} [g_{\rm adj}^{\mathcal{T}}]_{ij}(t_2-t_1) = \frac{g^2 }{ N_c} \langle \hat{\mathcal{T}} E_i^a(t_2) \mathcal{W}^{ab}_{[t_2,t_1]} E_j^b(t_1) \rangle_T = - \frac{i}{2}  \left. \frac{\delta^2  S_{\rm NG}[\mathcal{C};h] }{ \delta h^i(t_2) \delta h^j(t_1)} \right|_{h=0} \, .
\end{equation}
Integrating by parts and using the equations of motion, the on-shell boundary action in the presence of non-vanishing deformations at $z=0$ is given by
\begin{align}
    S_{\rm NG}^{(0-2)} - S_{0} &= \frac{\sqrt{\lambda} (1 - i\epsilon) }{\pi}  \int {\rm d}t  \nonumber \\ \lim_{z \to 0} \bigg[ & - \frac{(1 - z^4)}{2 z^2} y y'  - \frac{(1 - z^4)}{2 z^2} \delta \delta' - \frac{\sinh 2\eta \, (1-z^4)}{2(1 - 2z^4 + \cosh 2\eta)} \left( \delta \Delta' + \delta' \Delta \right) \nonumber \\ 
    &  - \frac{ (1-z^4) (1 + 2z^4 + \cosh 2\eta) \sinh 2\eta}{z (1 - 2z^4 + \cosh 2\eta)^2} \delta \Delta  - \frac{2 z^2 (1 -z^4) \cosh^2 \! \eta \, \sinh^2 \! \eta}{(1 - 2z^4 + \cosh 2\eta)^2} \Delta \Delta' \nonumber \\
    & + \frac{\cosh^2 \! \eta \left( P^{\Delta' \Delta}_{0}(z) + P^{\Delta' \Delta}_{2}(z) \cosh 2\eta + P^{\Delta' \Delta}_{4}(z) \cosh 4\eta \right) }{z^3 (1-2z^4 + \cosh 2\eta)^3} \Delta^2 \bigg]  \, ,
\end{align}
where the upper limit of integration $z=1$ gives a vanishing contribution, provided we set $a_\omega = 0$. The reason why the upper limit of integration for the fluctuations is $z=1$ and not $z = \sqrt{\cosh \eta}$ is the following: in the parametrization we have chosen for this calculation, $z=1$ lies on the past infinity hypersurface in the Poincar\'e patch, because the $z$-dependent shift $-F(z)$ in the time coordinate (determined by Eq.~\eqref{eq:F-coord-redef}) goes to $-\infty$ as $z \to 1^-$. This means that the propagation of the perturbations we introduced at the AdS boundary will go outside the Poincar\'e patch when $z>1$, and thus the action for the fluctuations will not receive contributions from $z>1$. (It is important to stress at this point that the $z=1$ contribution to the on-shell value of the action only vanishes if the mode solution is chosen as in~\cite{Nijs:2023dks}, i.e., with the $i\epsilon$ prescription that singles out $F_{|\omega|}^{-}(z)$. The other solutions are discarded because they would give a divergent contribution to the action.)

It would be interesting to study deformations on a Wilson loop of finite extent $\mathcal{T}$, where the way in which the worldsheet is closed at the temporal endpoints must be accounted for explicitly, and see how our current considerations change.

Finally, we are at the point where we can give our result.
Because the mode functions for $\Delta$ go as $z^2$ near $z=0$, the only non-vanishing contributions to $S_{\rm NG}^{(0-2)} - S_0$ come from the $yy'$ and $\delta \delta'$ terms. By analogy with the previous section, it follows that
\begin{align}
    \frac{g^2}{N_c} [g_{E}^{\mathcal{T}}]^{\mathcal{N}=4}_{ij}(\omega) =   \frac{\sqrt{\lambda} \delta_{ij}}{ 4\pi} \left( \frac{-i}{F^-_{|\omega|}(0)} \frac{\partial^3 F^-_{|\omega|}}{\partial z^3}(0) \right) \, .
\end{align}
The final step is to restore units by inserting $\pi T \sqrt{\cosh \eta}$ whenever a mass dimension 1 quantity is appropriate.

\subsubsection{Result}

Remarkably, we find that the result for the time-ordered correlation function in a moving plasma is equal to that in the static case, but with the substitution $T \to T \sqrt{\cosh{\eta}} = T \sqrt{\gamma}$, where $\gamma = (1 - v^2)^{-1/2}$ is the Lorentz boost factor, in the same way that previous AdS/CFT studies of the heavy quark potential~\cite{Liu:2006he,Liu:2006nn} have shown. This is also true for the heavy quark diffusion coefficient~\cite{Gubser:2006nz,Casalderrey-Solana:2007ahi}: while the results in those works are often quoted as
\begin{align}
    \kappa_T &= \pi \sqrt{\lambda} \gamma^{1/2} T^3 \, , \\
    \kappa_L &= \pi \sqrt{\lambda} \gamma^{5/2} T^3 \, ,
\end{align}
these results are derived in the rest frame of QGP. In fact, a Lorentz transformation of these results to the rest frame of the heavy quark generates additional factors of $\gamma$ that make $\kappa_{T}^{\rm HQ \, rest} = \kappa_{L}^{\rm HQ \, rest}$. In the simplest possible terms, both diffusion coefficients receive a factor of $\gamma$ due to time dilation, and $\kappa_L$ receives a second additional factor of $\gamma^{-2}$ due to length contraction. With this,
\begin{align}
    \kappa_{T}^{\rm HQ \, at \, rest} &= \pi \sqrt{\lambda} \gamma^{3/2} T^3 \, , \\
    \kappa_{L}^{\rm HQ \, at \, rest} &= \pi \sqrt{\lambda} \gamma^{3/2} T^3 \, ,
\end{align}
in perfect consistency with the $T \to T \sqrt{\gamma}$ identification as the effective temperature felt by heavy quarks in a flowing QGP.

In the same way, in the rest frame of the heavy quark pair the longitudinal and transverse components of the chromoelectric field correlator relative to the velocity of the medium are the same. Explicitly, the result is:
\begin{align}
\label{eqn:results}
     \frac{g^2}{N_c} [g_{E}^{\mathcal{T}}]^{\mathcal{N}=4}_{ij}(\omega) = \sqrt{\lambda} \, \delta_{ij} \frac{ (\pi T \sqrt{\gamma})^3 }{4\pi} \left( \frac{-i}{F^-_{|\Omega|}(0)} \frac{\partial^3 F^-_{|\Omega|}}{\partial \xi^3}(0) \right) \, .
\end{align}
where $F^-_\Omega$ is defined as the regular solution of
\begin{align} \label{eq:F-thermal-2}
    \frac{\partial^2 F^-_\Omega}{\partial \xi^2} - 2 \left[ \frac{1 + \xi^4}{\xi(1-\xi^4)} - \frac{i \Omega \xi^3}{1-\xi^4} \right] \frac{\partial F^-_\Omega}{\partial \xi}  + \left[ \frac{i \Omega \xi^2}{1-\xi^4} + \frac{\Omega^2 (1 - \xi^6) }{(1-\xi^4)^2} \right] F^-_\Omega = 0  \, .
\end{align}
In the above, $\Omega = \omega/(\pi T \sqrt{\gamma})$, and $\lambda = g^2 N_c$ is the 't Hooft coupling of the $\mathcal{N}=4$ SYM theory. An immediate consequence of Eq.~\eqref{eqn:results} is that the moving medium effect on quarkonium dynamics is that the temperature it experiences gets increased by a factor of $\sqrt{\gamma}$ (and therefore, that the frequency dependence of the correlators is the same as in Fig.~\ref{fig:correlator} with the substitution $T \to T \sqrt{\gamma}$)\@. Qualitatively, when the medium is boosted, the light quarks and gluons interacting with quarkonium are more energetic and thus the corresponding quarkonium dynamics occurs faster. Following our results in Section~\ref{sec:QQ-setup}, we find that the generalized gluon distribution (GGD) for quarkonium in-medium dynamics in the strong coupling limit is given by
\begin{align}
\label{eqn:gE++-2}
    [g_{E}^{++}]^>(\omega) &= 2 \theta(\omega) {\rm Re} \left\{ [g_{E}^{{\mathcal{T}}}]_{ii}(\omega) \right\} \,.
\end{align}

In summary, we have calculated the GGD that characterizes the in-medium dynamics of quarkonium and determines its dissociation and recombination rates in a strongly coupled $\mathcal{N}=4$ SYM plasma moving at velocity $v$ relative to the $Q\bar{Q}$ pair. 
The velocity dependence is a rescaling of the temperature $T$ to $\sqrt{\gamma} \, T$, in consistency with the effect of a ``hot wind'' on quarkonium screening in AdS/CFT~\cite{Liu:2006he,Liu:2006nn}. This effect can be interpreted as the result of a Lorentz transformation of the energy density under a boost $\mathcal{E} \to \gamma^2 \mathcal{E}$, and thus the effective temperature the $Q\bar{Q}$ pair experiences due to this energy density will be rescaled according to $\mathcal{E} \propto T^4$. This effect is not small when the quarkonium momentum is larger than its mass, which is highly relevant for quarkonium production measured in current HIC experiments, and will generically make dissociation/recombination processes faster (as long as the multipole expansion $ \sqrt{\gamma} \, T \ll M v_{\rm rel}$ is under control). This effect will compete with the fact that a $Q\bar{Q}$ pair of higher-$p_T$ generally has less time to interact with the medium.

\subsection{From weak to strong coupling} \label{sec:weak-strong}

A natural question we can ask is how the calculation result in the strong coupling limit compares with that in the weak coupling limit. We also want to understand if they allow for an interpolation at intermediate couplings. To do this, we first compare the time-ordered chromoelectric correlators at weak and strong coupling in $\mathcal{N} = 4$ SYM, and then we proceed to compare the spectral functions in weakly coupled QCD and in strongly coupled $\mathcal{N} = 4$ SYM so that we can lay out potential phenomenological implications at intermediate couplings. We do this comparison at zero relative velocity $v=0$ between the $Q\bar{Q}$ pair and the QGP environment.

\subsubsection{Comparison of the time-ordered chromoelectric correlators in $\mathcal{N}=4$ SYM}

At weak coupling, the non-Abelian electric field correlation function we have set out to calculate can be evaluated directly using the standard real-time perturbation theory in the Schwinger-Keldysh formalism. In the large $N_c$ limit of $\mathcal{N}=4$ SYM, it reads
\begin{equation} \label{eq:LO-SYM}
    \frac{g^2}{N_c} [g_E^{\T}]_{ij}(\omega) = \delta_{ij} \frac{\lambda}{6\pi} |\omega|^3 \coth \! \left( \frac{|\omega|}{2T} \right) \, .
\end{equation}
This result may be read off directly from the LO term in Eq.~\eqref{eq:spectral-full-result}, as the only contribution comes from the gluon propagator and is thus equal for QCD and $\mathcal{N}=4$ SYM. It is apparent that the small $\omega/T$ limit of the strong coupling and weak coupling results is different: in the weakly coupled case it goes as $|\omega|^2$, while for the strongly coupled limit it is linear in $|\omega|$\@. That is to say, for the range of frequencies that is sensitive to thermal effects, the physics at weak and strong coupling is different. Note that the $|\omega|^2$ behavior displayed above implies that the transport coefficients $\kappa_{\rm adj}^{\ml{N}=4}$ and $\gamma_{\rm adj}^{\ml{N}=4}$ vanish at leading order in perturbation theory. At NLO, however, $\kappa_{\rm adj}^{\mathcal{N}=4}$ is known to be nonzero~\cite{Caron-Huot:2008dyw}, and is actually equal to $\kappa_{\rm fund}^{\ml{N}=4}$ up to this order in perturbation theory. Therefore, it seems that the most interesting thermal physics lies in the intermediate coupling regime. However, a first approximation to this regime via interpolation between weakly and strongly coupled results would require to calculate the first nonvanishing contributions from both sides, which is a challenging computation we do not undertake in this work.

\begin{figure}
    \centering
    \includegraphics[width=0.8\textwidth]{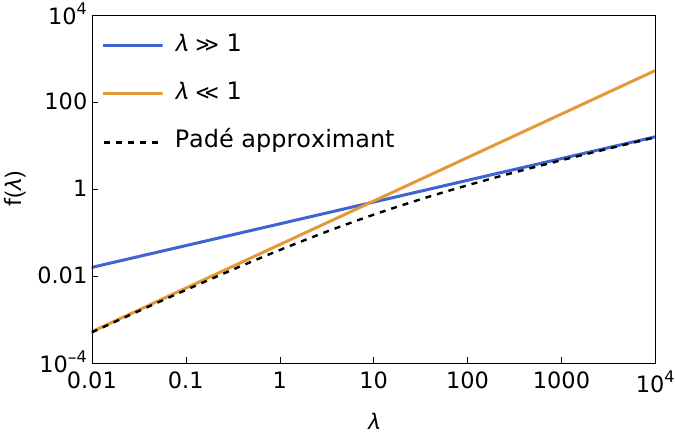}
    \caption{Coupling dependence of the time-ordered chromoelectric correlator in vacuum $T=0$. The solid lines depict the information currently available at weak and strong coupling, and the dashed line is the lowest order Pad\'e approximant consistent with both asymptotic behaviors.}
    \label{fig:interpolation-SYM}
\end{figure}

On the other hand, in the $T=0$ limit, the frequency dependence of both results agrees: both are proportional to $|\omega|^3$\@. This is unsurprising given that $\mathcal{N}=4$ supersymmetric Yang-Mills theory is a conformal field theory, and there is thus no other scale available to give rise to a different behavior (note that because of this the limits $\omega \to \infty$ at fixed $T$ and $T \to 0$ at fixed $\omega$ are equivalent). As such, in vacuum we have 
\begin{equation}
    \frac{g^2}{N_c} [g_E^{\T}]_{ij}(\omega) = \delta_{ij} f(\lambda) |\omega|^3 \, ,
\end{equation}
where (from taking $T \to 0$ in~\eqref{eq:LO-SYM} and from Eq.~\eqref{eq:EE-SYM-zero-T-limit})
\begin{equation}
    f(\lambda) \approx \begin{cases} \dfrac{\lambda}{6\pi} & \lambda \ll 1 \\ & \\  \dfrac{\sqrt{\lambda}}{{ 2}\pi} & \lambda \gg 1 \end{cases} \,\,\, .
\end{equation}
We plot both limits in Fig.~\ref{fig:interpolation-SYM}, together with the Pad\'e approximant of order $[2/1]$ in $\sqrt{\lambda}$ that interpolates between the two limits. Such an interpolation constitutes, at most, an educated guess of the result for the chromoelectric correlation function in the intermediate coupling regime around $\lambda \sim \ml{O}(10)$\@. As it should be clear from the comparison, we only expect the result to be valid asymptotically. 

Nonetheless, such a comparison may provide valuable insight into what the behavior of the correlator is at intermediate couplings. In fact, Fig.~\ref{fig:interpolation-SYM} is just the first step towards a more complete understanding of the intermediate coupling regime, as the tools to make progress on either limiting case are already available. At weak coupling, what is required is a next-leading order calculation analogous to what has already been done for QCD in Section~\ref{sect:nlo}, but this time for $\mathcal{N}=4$ SYM\@. At strong coupling, one would have to evaluate the quantum corrections to the string worldsheet action in the so-called semiclassical expansion, which is tantamount to a 1-loop calculation of the fluctuation fields on the worldsheet~\cite{Drukker:2000ep}\@. Both are necessary steps towards a more complete understanding of the correlator, which are along the path that we want to follow in the future (in the hope that the convergence of the series is comparable to that of other observables in $\mathcal{N}=4$ SYM, e.g., the thermodynamic pressure~\cite{Du:2021jai})\@.

\subsubsection{Comparison of the spectral functions: weakly coupled QCD and strongly coupled $\mathcal{N}=4$ SYM}

Finally, we compare the strong coupling and weak coupling results at the level of the spectral functions, in order to assess their phenomenological implications and shed light on the physics at intermediate coupling. Note that in what follows, we use the notation convention $[g_{\rm adj}^{\pm\pm'}]^{\lessgtr}(\omega)$ for the family of correlators relevant for quarkonia, as opposed to the $[g_E^{\pm\pm'}]^{\lessgtr}(\omega)$ convention we used so far in Section~\ref{sec:strong-coupling}. These conventions are related by $[g_{\rm adj}^{\pm\pm'}]^{\lessgtr}(\omega) = \frac{g^2 T_F}{3N_c} [g_E^{\pm\pm'}]^{\lessgtr}(\omega)$.

The spectral function for quarkonium transport in weakly coupled QCD at positive frequencies was calculated in Section~\ref{sect:nlo}, and its negative frequency part will be discussed in detail in Section~\ref{sect:non-odd}. Up to order $g^4$, it reads
\begin{align}
\label{eqn:rho_UV}
    & \rho^{++}_{\rm adj}(\omega)  =   \frac{g^2 T_F (N_c^2-1) \omega^3 }{ 3  \pi N_c } \times \nonumber \\ &  \bigg\{  1  + \frac{g^2}{(2\pi)^2} \bigg[ \left( \frac{11 N_c}{12} - \frac{N_f}{6} \right) \ln \left( \frac{\mu^2}{4 \omega^2} \right)   + N_c \left( \frac{149}{36} - \frac{\pi^2}{6} + \frac{\pi^2}{2} {\rm sgn}(\omega) \right) - \frac{5 N_f}{9}  \bigg]  \nonumber  \\
    &  + \frac{g^2}{(2\pi)^2} \bigg[ \int_0^\infty \!\! \diff k \,  N_f n_F(k) \bigg( -2k \omega + (2k^2 + \omega^2) \ln \left| \frac{k+\omega}{k-\omega} \right|  + 2 k \omega \ln \left| \frac{k^2 - \omega^2}{\omega^2} \right| \bigg) \nonumber \\
    &  + \int_{0}^\infty \!\! \diff k \, 2 N_c n_B(k) \bigg( -2 k \omega + (k^2+\omega^2) \ln \left| \frac{k+\omega}{k-\omega} \right|  + k \omega \ln \left| \frac{k^2-\omega^2}{\omega^2} \right| +\mathcal{P} \left( \frac{k^3 \omega}{k^2 - \omega^2} \right)\bigg) \nonumber \\ 
    &  + \int_0^\infty \diff k \, \frac{2 N_c n_B(k)}{k} \ml{P} \left( \frac{\omega^2}{ \omega^2 - k^2} \right) \bigg( k^2 \omega + (k^3 + \omega^3) \ln \left| \frac{k-\omega}{\omega} \right| + ( -k^3 + \omega^3) \ln \left| \frac{k + \omega}{\omega} \right| \bigg) \bigg] \bigg\} \nonumber \\
    & + \rho_{\rm HTL}(\omega)
\end{align}
where the hard thermal loop contribution (HTL) can be read off from the heavy quark transport spectral function, as the HTL-resummed diagrams that contribute to them up to $\mathcal{O}(g^4)$ in perturbation theory are the same. Explicitly, it is given by
\begin{align}
    \rho_{\rm HTL}(\omega) &=  \frac{g^2 T_F (N_c^2 -1 ) m_D^2 \,\omega}{3\pi N_c} \nonumber \\ &  \times
 \bigg\{ \int_{\hat\omega}^{\infty} \!  \frac{{\rm d}\hat k \, \hat k}{2} \, 
 \frac{\hat\omega^2
 \Big( 1 - \frac{\hat\omega^2}{\hat k^2}\Big)}
 {
  \Big( 
    \hat k^2 - \hat\omega^2 + \frac12 
     \Big[ 
       \frac{\hat\omega^2}{\hat k^2} + 
       \frac{\hat\omega}{2\hat k} 
       \Big( 1 - \frac{\hat\omega^2}{\hat k^2}\Big) 
       \ln\frac{\hat k + \hat\omega}{\hat k - \hat\omega}
     \Big]
  \Big)^2
 + \Big( 
     \frac{\hat\omega\pi}{4\hat k}
  \Big)^2
       \Bigl( 1 - \frac{\hat\omega^2}{\hat k^2}\Big)^2 
 } \nonumber \\ 
& \quad +  
 \int_{0}^{\infty} \! \frac{ {\rm d}\hat k \, \hat k^3}{2} 
 \bigg[ 
 \frac{
 \theta(\hat k - \hat\omega) 
 }
 {
  \Big( 
    \hat k^2 
     + 1 - 
       \frac{\hat\omega}{2\hat k} 
       \ln\frac{\hat k + \hat\omega}{\hat k - \hat\omega}
       \Big)^2
 + \Big( 
     \frac{\hat\omega\pi}{2\hat k}
  \Big)^2
 } 
 - \frac{1}{(\hat k^2 + 1)^2}
 \bigg]
 \nonumber 
\\ 
& \quad +  
 \left. 
   \frac{ {2 \hat\omega} \hat k_T^3 (\hat\omega^2 - \hat k_T^2)}
   {|3(\hat k_T^2 - \hat\omega^2)^2 -\hat\omega^2|}
 \right|_{\hat k_T^2 - \hat\omega^2 + \frac12
      [\frac{\hat\omega^2}{\hat k_T^2}+
        \frac{\hat\omega}{2\hat k_T} 
       ( 1 - \frac{\hat\omega^2}{\hat k_T^2} ) 
       \ln\frac{\hat\omega + \hat k_T }{\hat\omega - \hat k_T}
       ] \, = \, 0 }
\nonumber 
\\ 
& \quad +  
 \left. 
   \frac{\hat k_E^3 (\hat\omega^2 - \hat k_E^2)}
   { \hat\omega |3(\hat k_E^2 - \hat\omega^2) + 1|}
 \right|_{\hat k_E^2 + 1 -
        \frac{\hat\omega}{2\hat k_E} 
       \ln\frac{\hat\omega + \hat k_E }{\hat\omega - \hat k_E} \, = \, 0 }
  - \frac{\omega^2}{m_D^2} + 
     \frac12  
     \bigg(\ln\frac{2\omega}{m_D} - 1 \bigg)
\bigg\} \,
\end{align}
where, following~\cite{Caron-Huot:2009ncn}, we have denoted $\hat{\omega} = \omega/m_D$, and we have written both the ``naive'' and the ``resummed'' corrections (c.f.~\cite{Burnier:2010rp}) in a single function.

The final step to evaluate this expression is to choose the renormalization scheme, i.e., how to define $\mu$. We choose it following the notion that the best choice of $\mu$ is the one that makes the result the least sensitive to higher order corrections on $g$. In the UV regime, $|\omega| \gg T$, we choose it to compensate for the NLO correction to the $\omega^3$ term of the spectral function. While mathematically we could also choose $\mu$ to compensate the $|\omega|^3$ term, it seems unphysical to let the renormalization group scale depend on the sign of the energy transferred in a physical process (only its magnitude should set the scale). In the IR, we follow~\cite{Burnier:2010rp} and use the EQCD result of~\cite{Kajantie:1997tt} to set the scale. Putting these together, we choose to interpolate and set the scale for each background temperature $T$ with the following formula:
\begin{equation}
    \mu(\omega, T) = \sqrt{T^2 \exp \left[ \ln(4 \pi) - \gamma_E - \frac{N_c - 8 \ln(2) N_f}{
    2 (11 N_c - 2 N_f)}\right]^2 + 
 \omega^2 \exp \left[\ln(2) + \frac{(6 \pi^2 - 149) N_c + 20 N_f}{6 (11 N_c - 2 N_f)}\right]^2} \, .
\end{equation}
We then choose the value of the coupling constant at the scale $\mu_0$ determined by $\omega = 0$, which means $\mu_0 \approx 8.1 T$, and evolve the coupling constant to higher scales (i.e., $|\omega|>0$) using the 2-loop QCD beta function:
\begin{equation}
    \frac{{\rm d} \alpha_s}{{\rm d} \ln \mu} = -2 \alpha_s \left[ \left(\frac{11 N_c}{3} -\frac{2 N_f}{3} \right) \left(\frac{\alpha_s}{4 \pi}\right) + \left( \frac{34 N_c^2}{3} - 
      \frac{10 N_c N_f}{3} - \frac{(N_c^2 - 1) N_f}{N_c}\right)\left(\frac{\alpha_s}{4 \pi}\right)^2 \right] \, .
\end{equation}

\begin{figure}[t]
    \centering
    \includegraphics[width=0.78\textwidth]{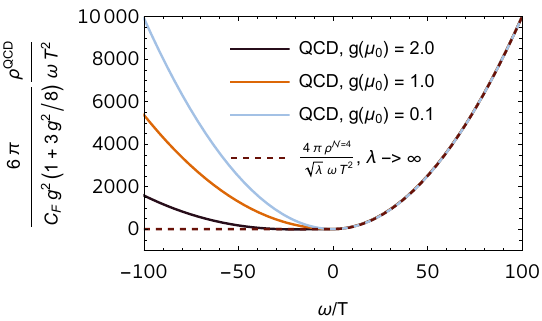}
    \caption{Spectral function for quarkonium transport in weakly coupled QCD with 2 light (massless) quarks for different values of the coupling at the reference scale $\mu_0 \approx 8.1 T$\@. The coupling constant is evolved to high energies using the 2-loop QCD beta function. The scale of the plot differs from that of Fig.~\ref{fig:spectral-noHTL}, as it is chosen to highlight the UV behavior of the spectral function at negative frequencies. For this reason, we have normalized each curve in this plot by a factor that depends on the coupling constant, in such a way that the asymptotic behavior of every curve agrees at large positive frequencies.}
    \label{fig:spectral-2sided}
\end{figure}

Qualitatively, one of the striking features of the result in strongly coupled $\mathcal{N}=4$ SYM is that 
$[g_{\rm adj}^{++}]^>(\omega<0)=0$\@. 
As we can see from Fig.~\ref{fig:spectral-2sided}, increasing the coupling in the perturbative calculation leads to the same feature: if we normalize the spectral function such that its behavior in the ultraviolet (UV) $\omega/T \gg 1$ is fixed, then the spectral function at negative $\omega$ becomes smaller as the coupling increases\@. As such, the trend at weak coupling is compatible with the (supersymmetric) strongly coupled result.

\begin{figure}[t]
    \centering
    \includegraphics[width=0.78\textwidth]{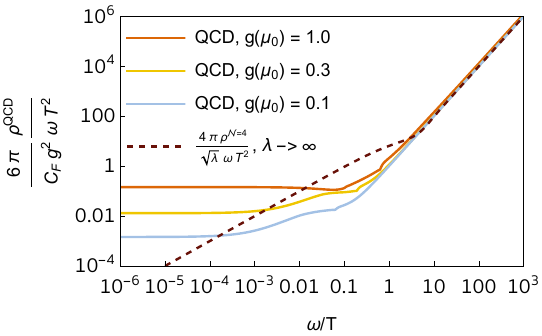}
    \caption{Same as Fig.~\ref{fig:spectral-2sided}, but focusing on the transition between IR and UV physics at positive frequencies, including resummed contributions from Hard Thermal Loop effective theory. The scale of the ordinate axis is chosen in the same way for the QCD results here as we chose it in Fig.~\ref{fig:spectral-noHTL}. Because the perturbative difference between $\omega > 0$ and $\omega < 0$ is a temperature-independent term, the IR features of the weakly coupled result at negative frequencies are qualitatively the same as those at positive frequencies. The only qualitative difference occurs at $\omega \sim -T$, where the weakly coupled QCD curves cross, in accordance with the $\omega < 0$ behavior of Fig~\ref{fig:spectral-2sided}.}
    \label{fig:spectral-IR}
\end{figure}

We then focus on the infrared (IR) regime $|\omega|/T \lesssim 1$ in Fig.~\ref{fig:spectral-IR}\@.
We have chosen the normalization such that the leading contribution to each curve goes as $\omega^2$ at $\omega/T \gg 1$.
On the one hand, the asymptotic IR behavior of $\rho(\omega)/\omega$ is constant at weak coupling, and linear in $\omega$ at strong coupling. 
On the other hand, as before, there is a consistent trend between weak and strong coupling, in the sense that the transition between IR and UV regimes takes place gets pushed to higher values of $\omega/T$ with increasing coupling. This means that, except for the regime $|\omega| \ll T$ (where the convergence of perturbation theory in QCD is generally poor), the perturbative result moves toward the strongly coupled one as the coupling is increased in a consistent trend, both at positive and negative frequencies.

Future phenomenological studies using the GGDs at different couplings have the potential to tell which value of the coupling provides the best description of the experimental data for each quarkonium species. However, our findings imply that this will not be straightforward. 
Previous phenomenological studies solved Markovian transport equations such as Boltzmann equations~\cite{Yao:2020eqy} or Lindblad equations~\cite{Brambilla:2022ynh}, in which either the $\omega = -\Delta E < 0$ (if $T \sim \Delta E$) or $\omega = 0$ (if $T \gg \Delta E$) part of $\rho_{\rm adj}^{++}(\omega)$ gives the dominant contribution. This is certainly the case at weak coupling (i.e., the dominant contribution comes from $\omega \sim - |\Delta E|$).
Our results imply that no such contribution exists in the strongly coupled limit, and thus the leading mechanisms driving quarkonium dynamics, coming from the positive $\omega \sim T$ part of $\rho_{\rm adj}^{++}(\omega)$, must be non-Markovian. That is to say, QGP memory effects are not negligible for quarkonium transport in a strongly coupled QGP. Physically, there is no quasi-gluon that a bound $Q\bar{Q}$ pair can absorb resonantly and incoherently in a strongly coupled plasma. Rather, the plasma responds coherently through strong correlations between different points in time, as opposed to behaving as independent, point-like sources.

Our findings motivate formulating the in-medium dynamics of $Q\bar{Q}$ pairs in a general non-Markovian setup, without which it may be impossible to provide reliable phenomenological predictions for quarkonium transport in strongly coupled plasmas. It is also worth exploring at which finite coupling the non-Markovian contribution becomes more important than the Markovian one. In the future, by following this direction, we expect to deepen our knowledge of QGP's microscopic structure.

\section{On the calculation of the chromoelectric correlators in lattice QCD} \label{sec:latticec}

Various properties of QGP are encoded in terms of gauge invariant correlation functions of field operators that often define transport coefficients showing up in the time evolution equations of the probes in the medium. Well-known examples include the shear viscosity (defined as a correlator of stress-energy tensors), the jet quenching parameter (a correlator of light-like Wilson lines) and the heavy quark diffusion coefficient (a correlator of two chromoelectric fields dressed with Wilson lines). Since QGP is a strongly coupled fluid, nonperturbative determinations of these transport coefficients are crucial in our understanding of QGP and QCD at finite temperature. As it turns out, the only systematic, nonperturbative method currently available in QCD itself is lattice QCD. However, this is limited to Euclidean calculations, which means that one has to formulate correlation functions in Euclidean signature and then relate them to real-time correlations. 

The correlator for quarkonium transport is similar to but different from the correlator defining the heavy quark diffusion coefficient~\cite{Casalderrey-Solana:2006fio,Caron-Huot:2009ncn} in terms of the ordering of the fields contained in the Wilson lines. The perturbative calculations in $R_\xi$ gauge in Section~\ref{sect:nlo} showed that the spectral function of the correlator for quarkonium transport differs from that for heavy quark transport~\cite{Burnier:2010rp} by a temperature independent constant at next-to-leading order (NLO). However, if both calculations had been performed in temporal axial gauge ($A_0=0$), one would, at first sight, have concluded that the two correlators were identical. This resulted in a puzzle: Since both correlators are defined in a gauge invariant way, calculations with different gauge choices must give the same result. This puzzle was resolved in Section~\ref{sec:axial-gauge}, establishing the difference between the two correlators on a more solid ground in QCD. Beyond NLO, the heavy quark diffusion coefficient has been studied by using hard-thermal-loop resummation~\cite{Caron-Huot:2007rwy}, as well as nonperturbatively via the lattice QCD method~\cite{Banerjee:2011ra,Ding:2012sp,Francis:2015daa,Brambilla:2020siz,Altenkort:2020fgs} and the AdS/CFT correspondence~\cite{Herzog:2006gh,Gubser:2006qh,Casalderrey-Solana:2006fio}. On the other hand, our AdS/CFT calculation in Section~\ref{sec:strong-coupling} showed that the analog quarkonium transport coefficients in $\mathcal{N}=4$ supersymmetric Yang-Mills (SYM) theory are zero, in stark contrast to the heavy quark diffusion coefficient value of $ \sqrt{\lambda} \pi T^3$ at large coupling $\lambda = g^2 N_c \gg 1$. This difference is surprising because the heavy quark and quarkonium transport coefficients are defined by similar chromoelectric field correlators. Therefore, it is well motivated to study the quarkonium transport properties nonperturbatively in QCD. It is also crucial and urgent, since quarkonium production serves as an important probe of strongly coupled QGP that is produced in current heavy ion collision experiments.

In this section, we discuss how to extract the quarkonium transport coefficients from lattice QCD calculations of a specific Euclidean chromoelectric correlator. It is organized as follows: in Section~\ref{sec:euclidean} we will discuss the Euclidean version of the correlator and how to relate it to its real time counterpart. Next, in Section~\ref{sec:lattice} the setup of a lattice QCD calculation of this Euclidean correlator will be discussed, with a focus on how to renormalize it. Finally, we will conclude and present our outlook for a lattice QCD determination of quarkonium transport properties in Section~\ref{sec:lattice-summary}.

\subsection{Euclidean Correlators and Transport Coefficients}
\label{sec:euclidean}

As is well known, lattice QCD methods can only calculate correlation functions in Euclidean space and thus cannot be applied directly to be used to study the real-time correlators defined in Eqs.~\eqref{eq:gE++>}-~\eqref{eq:gE--<}.
In this section, we will introduce a Euclidean version of the correlator for quarkonium transport and discuss how to extract the quarkonium transport coefficients from the evaluation of this Euclidean correlator. As we will show, both the Euclidean correlator itself and the method to extract the quarkonium transport coefficients are different from the case of heavy quark diffusion in subtle and important aspects. To make the comparison more explicit, and also to take advantage of the apparent similarities between them, we will first review the extraction of the heavy quark diffusion coefficient from the corresponding Euclidean correlator. 

\subsubsection{Heavy Quark Diffusion}
The Euclidean correlator relevant for the heavy quark diffusion case is given by~\cite{Caron-Huot:2009ncn}
\begin{align}
\label{eqn:Gfund}
G_{\rm fund}(\tau) = - \frac{1}{3} \frac{\big\langle {\rm Re} {\rm Tr}_c[ U(\beta,\tau) gE_i(\tau) U(\tau,0) gE_i(0) ] \big\rangle_T}{\big\langle {\rm Re} {\rm Tr}_c[U(\beta,0)] \big\rangle_T} \,,
\end{align}
where $\beta=1/T$ is the inverse of the QGP temperature and $\langle\cdot\rangle_T = \Tr(\cdot e^{-\beta H})/\Tr(e^{-\beta H})$, with $H$ the Hamiltonian of the hot QCD matter that comprises QGP in the absence of any external color source. It has been shown that the heavy quark transport coefficients can be obtained from $G_{\rm fund}(\tau)$ via~\cite{Caron-Huot:2009ncn,Eller:2019spw}
\begin{align}
\label{eqn:fund_from_rho}
\kappa_{\rm fund} &= \lim_{\omega\to0} \frac{T}{\omega} \rho_{\rm fund}(\omega) \, , \\
\gamma_{\rm fund} &= -\int_0^\beta \diff\tau \, G_{\rm fund}(\tau) \, ,\nn
\end{align}
where the spectral function $\rho_{\rm fund}(\omega)$ is related to the Euclidean correlator through\footnote{Our convention for the Fourier transform is $O(\omega) = \int \diff t e^{i\omega t}O(t)$.}
\begin{align}
\label{eqn:Gfund_rho}
G_{\rm fund}(\tau) = \int_0^{+\infty} \frac{\diff\omega}{2\pi}  \frac{\cosh\big(\omega(\tau-\frac{1}{2T})\big)}{\sinh\big( \frac{\omega}{2T} \big)} \rho_{\rm fund}(\omega) \,.
\end{align}
This correlator is constructed such that the standard Kubo-Martin-Schwinger (KMS) and analytic continuation relations hold as in textbook thermal field theory. Given an analytic expression for {$\tilde{G}_{\rm fund}(\omega_n)$}, with $\omega_n = 2\pi T n$, $n \in \mathbb{Z}$ the Matsubara frequencies, one can extract the spectral function by taking the real part\footnote{Many studies define correlation functions with an imaginary unit prefactor, and there the spectral function corresponds to the imaginary part of the retarded correlator, which has a factor of $1/2$ compared with the spectral function defined by the difference between the $>$ and $<$ Wightman correlators in frequency space.} of the retarded correlator obtained by analytic continuation $\omega_n \to -i (\omega + i\epsilon)$ of this Euclidean correlator. This has been done both at weak~\cite{Burnier:2010rp} (QCD) and strong~\cite{Casalderrey-Solana:2006fio} ($\mathcal{N}=4$ SYM) coupling. However, at physical values of the coupling in QCD, the only tool available at the moment is lattice gauge theory, and as such, the reconstruction of the spectral function $\rho_{\rm fund}$ through the relation~\eqref{eqn:Gfund_rho} has received much attention in recent years~\cite{Altenkort:2020fgs,Altenkort:2023oms,Brambilla:2022xbd}.

Comparatively, the theoretical treatment of quarkonium transport coefficients has received less attention. We now aim to fill in this gap, and subsequently, to provide a recipe to determine these transport coefficients from lattice QCD calculations. To this end, we need to first construct a Euclidean version of the correlator for quarkonium transport that can be calculated via lattice QCD methods, and then explain how to extract the quarkonium transport coefficients from the evaluation of such an Euclidean correlator. We will answer these two questions in the following two subsections. Details of the lattice calculation of the Euclidean correlator will be discussed in the next section.

\subsubsection{Euclidean Correlator for Quarkonium Transport} 
To construct the Euclidean correlator for quarkonium transport, we first note that because of the operator ordering in the definitions~\eqref{eq:gE++>}-~\eqref{eq:gE--<}, we can equivalently write
\begin{equation}
    [g_{\rm adj}^{++}]^>(t) = \frac{g^2 T_F }{3 N_c} \big\langle E_i^a(t)W^{ab}(t,0) E_i^b(0) \big\rangle_T \, .
\end{equation}
To perform the analytic continuation, it is best to explicitly isolate the $t$ dependence from the field operators and write it purely in terms of time evolution factors. We let $H$ be the Hamiltonian of the thermal bath QGP in the absence of any external color charge. When an external adjoint color charge is present, the Hamiltonian of the thermal bath is given by $[H \mathbbm{1} - g A_0^c(0) T^c_{\rm adj} ]^{ab}$.
The reason for the appearance of this modified Hamiltonian can be seen from converting the adjoint Wilson line back to the Schr\"odinger picture from the interaction picture
\be
e^{-i H t} W^{ab}(t,0) =  \left[e^{- i (H - g A_0^c(0) T_{\rm adj}^c ) t}\right]^{ab} \,. \label{eq:W-as-eiHt}
\ee
Eq.~\eqref{eq:W-as-eiHt} has the following physical interpretation: during the time interval between $0$ and $t$ the QGP environment evolves in the presence of an adjoint color charge, which is manifest in the modification of the Hamiltonian by $- g A_0$. It is essentially a local modification to Gauss's law\footnote{An interesting question one can ask of this expression is whether we still have explicit gauge invariance. The answer is, naturally, affirmative. However, this is not as easy to see when considering time-dependent gauge transformations as it is for time-independent gauge transformations. This is because the Hamiltonian also changes if one considers time-dependent gauge transformations, which is something to keep in mind when quantizing the theory. We will not pursue this further here, and we shall assume that $H$ is already determined. For a thorough discussion on the quantization of gauge theories, we refer the reader to Ref.~\cite{Henneaux:1992ig}.}, revealing the presence of a color octet $Q\bar{Q}$ pair. Outside this time interval the QGP environment evolves in the absence of external color sources.

Using Eq.~\eqref{eq:W-as-eiHt}, one can write:
\begin{align}
    \frac{3 N_c}{g^2 T_F } [g_{\rm adj}^{++}]^>(t)  = \frac{{\rm Tr}_{\mathcal{H}} \! \left[ e^{ i H t} E^a_i(0) \! \left[e^{- i (H - g A_0^c(0) T_{\rm adj}^c ) t}\right]^{ab} \!  E^b_i(0) e^{-\beta H} \right]}{{\rm Tr}_{\mathcal{H}} \left[  e^{-\beta H} \right]} \, , 
\end{align}
where the trace ${\rm Tr}_{\mathcal{H}}$ runs over physical states of the QGP environment. The analytic continuation is now direct, because all of the time dependence is in the exponentials. We just set $t \to - i\tau$, and identify the Euclidean gauge field $A_4$ with the Minkowski one by $A_0(0) = i A_4(0)$ {(which in turn means that the electric field picks up a factor of $i$)}, to find
\begin{align}
    & [g_{\rm adj}^{++}]^>(-i\tau) \nn \\ 
    &= { -} \frac{g^2 T_F}{3 N_c}  \frac{{\rm Tr}_{\mathcal{H}} \! \left[ e^{ H \tau} E^a_i(0) \! \left[e^{-  (H - g A_0^c(0) T_{\rm adj}^c ) \tau }\right]^{ab} \! E^b_i(0) e^{-\beta H} \right]}{ {\rm Tr}_{\mathcal{H}} \left[  e^{-\beta H} \right] } \, \nn \\
    &= { -} \frac{g^2 T_F}{3 N_c}  \big\langle E_i^a(\tau) \! \left[ {\rm P} \exp \left( ig \int_0^\tau \diff \tau' \, A_4^c(\tau') T_{\rm adj}^c \right) \right]^{ab} \! E_i^b(0) \big\rangle_T \nonumber \\
    & = { -} \frac{g^2 T_F }{3 N_c} \big \langle E_i^a(\tau) W^{ab}(\tau,0) E_i^b(0) \big\rangle_T \nn\\
    &\equiv  G_{\rm adj}(\tau) \, , \label{eq:G-analytic-cont}
\end{align}
where ${\rm P}$ denotes path-ordering.
That is to say, we have proven that one of the real-time correlations we want to evaluate is related to an Euclidean correlation function by $ [g_{\rm adj}^{++}]^{>}(-i\tau) = G_{\rm adj}(\tau)$. We note that the absence of the denominator term as in Eq.~\eqref{eqn:Gfund} is a result of the absence of a Wilson line along the imaginary time direction at $t=-\infty$ in the definition of $[g_{\rm adj}^{++}]^{>}$. In quarkonium dissociation, the initial state is a color singlet, whereas in heavy quark diffusion, the initial state is in a color triplet representation, whose effect appears explicitly in the initial thermal state.

\subsubsection{Extraction of Quarkonium Transport Coefficients from Euclidean QCD}
Now we discuss how to extract the quarkonium transport coefficients from $G_{\rm adj}(\tau)$. Even though this correlation function has been studied in the past~\cite{Eidemuller:1997bb,Eidemuller:1999mx,DElia:2002hkf}, its precise connection to quarkonium transport has remained unexplored, until now. 
It turns out that neither Eq.~\eqref{eqn:fund_from_rho} nor Eq.~\eqref{eqn:Gfund_rho} is valid for the quarkonium case. This is so because Eq.~\eqref{eqn:fund_from_rho} is a result of the standard KMS relation, which, as we will show momentarily, is more complicated for the quarkonium correlator. Furthermore, Eq.~\eqref{eqn:Gfund_rho} relies on the spectral function being odd in $\omega$, which is crucially not true for the quarkonium correlator, as we will discuss in what follows. 

\paragraph{KMS Relation and Non-odd Spectral Function} \hspace{\fill}

\label{sect:non-odd}
To explain the non-oddness of the spectral function for quarkonium transport, we follow Appendix~\ref{app:kms} to use the usual proof of the KMS relation, plus the time-reversal operation and find
\be \label{eqn:standard_kms}
[g_{\rm adj}^{++}]^>(\omega) = e^{\omega/T} [g_{\rm adj}^{--}]^>(-\omega) \,,
\ee
which is the necessary KMS relation for proper thermalization of the internal degrees of freedom of the heavy quark pair (their relative motion and internal quantum numbers~\cite{Yao:2017fuc}). We then introduce the spectral function that governs quarkonium transport as
\be \label{eqn:rho_adj}
\rho_{\rm adj}^{++}(\omega) = [g_{\rm adj}^{++}]^>(\omega) - [g_{\rm adj}^{--}]^>(-\omega) \, ,
\ee
which, by definition satisfies $[g_{\rm adj}^{++}]^>(\omega) = (1 + n_B(\omega)) \rho_{\rm adj}^{++}(\omega)$, with $n_B(\omega) = (e^{\beta \omega}-1)^{-1}$. We have kept the superscripts ``$++$'' in the label of this spectral function because we can  also define
\be
\rho_{\rm adj}^{--}(\omega) = [g_{\rm adj}^{--}]^>(\omega) - [g_{\rm adj}^{++}]^>(-\omega) \, ,
\ee
which contains the same information, and satisfies $\rho_{\rm adj}^{--}(\omega) = - \rho_{\rm adj}^{++}(-\omega)$.

Here comes the most important part: The spectral function~\eqref{eqn:rho_adj} is not odd in $\omega$. In the standard thermal field theory setup, we define $\rho(\omega) = g^>(\omega) - g^<(\omega)$ where
$g^>(t) = \langle \phi(t) \phi(0) \rangle$ and
$g^<(t) = \langle \phi(0) \phi(t)\rangle$, which are related via $g^>(\omega) = g^<(-\omega)$ in frequency space by time translational invariance. This immediately leads to $\rho(\omega) = -\rho(-\omega)$. However, the relation $g^>(\omega) = g^<(-\omega)$ is not true for $[g_{\rm adj}^{++}]^>(\omega)$ and $[g_{\rm adj}^{--}]^>(\omega)$ due to the path ordering of field operators and the additional Wilson line along the imaginary time in $[g_{\rm adj}^{--}]^>$. That is to say, $[g_{\rm adj}^{--}]^>(t) \neq [g_{\rm adj}^{++}]^>(t)$. Therefore, we do not know how $\rho_{\rm adj}^{++}(\omega)$ transforms under $\omega\to-\omega$ a priori.

To see this more formally, one may also write the spectral function as a spectral decomposition in terms of the eigenvalues/eigenstates of $H$, denoted by $\{E_n, |n\rangle\}$, and those of $[H \mathbbm{1} - g A_0^c(0) T^c_{\rm adj} ]^{ab}$, denoted by $\{\tilde{E}_n , |\tilde{n}^a \rangle \}$, where $a$ is interpreted as a component of the state, rather than a label. With these definitions, it follows that
\begin{align}
    \rho_{\rm adj}^{++}(\omega) = \frac{g^2 T_F}{3 N_c} \sum_{n, \tilde{n}}  (2\pi) & \delta(\omega + E_n - \tilde{E}_{\tilde{n}})  | \langle n | E_i^a(0) | \tilde{n}^a \rangle |^2 \nn \\ & \quad   \times \left[ e^{-\beta E_n} - e^{-\beta \tilde{E}_{\tilde{n}} } \right] \, .
\end{align}
There is no reason why this expression would be odd under $\omega \to -\omega$, because the energies $E_n$ and $\tilde{E}_n$ can (and will) be different in general.

Indeed, explicit perturbative calculations at NLO show that $\rho_{\rm adj}^{++}(\omega)$ contains both $\omega$-odd, which is the usual case, and $\omega$-even parts (see Appendix~\ref{app:spectral}). The final result Eq.~\eqref{eq:spectral-full-result} shown in Section~\ref{sec:add-results} is only for $\omega>0$, as mentioned there. We have performed a similar calculation for $\omega<0$ and found an $\omega$-even part, which originates from the diagrams (5, 5r) of Fig.~\ref{fig:diagrams}, or diagrams (j) of Refs.~\cite{Eller:2019spw,Burnier:2010rp}
\begin{align} \label{eq:rho-diff}
\Delta \rho(\omega) \equiv \big(\rho^{++}_{\rm adj}(\omega) - \rho_{\rm fund}(\omega) \big) = \frac{g^4 T_F (N_c^2-1) \pi^2}{3 (2\pi)^3}  |\omega|^3 \,,
\end{align}
where we have also added a factor of $2$ since the definition of the spectral function given in Eq.~\eqref{eq:spectral-full-result} differs from Eq.~\eqref{eqn:rho_adj} by a factor of $2$ (see Eq.~\eqref{eqn:rho_p_integrated}).

To demonstrate the importance of the $\omega$-even part, we use it to calculate the difference between $\gamma_{\rm fund}$ and $\gamma_{\rm adj}$ at the order of $\alpha_s^2$
\be
\label{eqn:delta_gamma}
\Delta \gamma \equiv \gamma_{\rm adj} - \gamma_{\rm fund} = - \frac{16 \zeta(3)}3 T_F C_F N_c  \alpha_s^2 T^3   \,,
\ee
where $C_F = \frac{N_c^2 - 1}{2N_c}$. This difference was first calculated in Ref.~\cite{Eller:2019spw}.  
Some algebra and use of the definitions for $[g_{\rm adj}^{\pm \pm}]^{>}$ leads to
\begin{align}
\gamma_{\rm adj} &= {\rm Im} \int_{-\infty}^{+\infty} \diff t \big( \theta(t) [g^{++}_{\rm adj}]^>(t) + \theta(-t) [g^{++}_{\rm adj}]^>(-t) \big) \nn\\
\Delta \gamma &= -\frac1{\pi}  \int_{-\infty}^{+\infty} \! \frac{\diff \omega }{|\omega|} \big(\theta(\omega) + n_B(|\omega|) \big) \Delta \rho(\omega) \,, 
\end{align}
where we have used $[g_{\rm adj}^{\pm \pm}]^{>}(\omega) = (1+n_B(\omega))\rho_{\rm adj}^{\pm \pm}(\omega)$ and used that they are translationally invariant in time.
The piece proportional to $\theta(\omega)$ is a pure vacuum contribution that vanishes in dimensional regularization. The second term inside the integral, however, gives a thermal contribution:
\begin{align} \label{eq:Delta-gamma-proof}
    \Delta \gamma &= - \frac{4 g^4 T_F}{3 (2\pi)^4} \pi^2 (N_c^2 - 1)  \int_0^{+\infty} \frac{\omega^2 \diff \omega}{e^{\omega/T} - 1} \\
    &= -\frac{16 \zeta(3)}{3} T_F C_F N_c \alpha_s^2  T^3 \, , \nn
\end{align}
which is exactly the difference given in Eq.~\eqref{eqn:delta_gamma}. This settles a long-standing issue noted in~\cite{Eller:2019spw} regarding the consistency of the gauge-invariant chromoelectric correlators in the adjoint and fundamental representation, and verifies explicitly that the spectral function relevant for quarkonium transport is qualitatively different from that for heavy quark diffusion. The above discrepancy $\Delta \gamma$ is explained precisely because $\rho^{++}_{\rm adj}(\omega)$ is not odd in frequency.

With these theoretical foundations in hand, we can now proceed to write down the formula analogous to Eq.~\eqref{eqn:Gfund_rho}, which will allow for the extraction of $\kappa_{\rm adj}$ and $\gamma_{\rm adj}$ from the evaluation of the Euclidean correlator $G_{\rm adj}(\tau)$.

\paragraph{Extraction Formulas} \hspace{\fill}

Using the fact that $G_{\rm adj}(\tau)$ is the analytic continuation of $[g_{\rm adj}^{++}]^>(t)$ to Euclidean signature, we can write
\begin{align}
    G_{\rm adj}(\tau) &= \int_{-\infty}^{+\infty} \frac{\diff \omega}{2\pi} e^{-\omega \tau} [g_{\rm adj}^{++}]^{>}(\omega) \label{eq:G-adj-inversion} \\
    &=  \int_{-\infty}^{+\infty} \frac{\diff \omega}{2\pi} \frac{\exp \big( \omega ( \frac{1}{2T} - \tau ) \big)}{2\sinh \big( \frac{\omega}{2T} \big) } \rho_{\rm adj}^{++}(\omega) \, .\nn
\end{align}
However, in contrast to Eq.~\eqref{eqn:Gfund_rho}, the integrand may not be symmetrized with respect to $\omega$ because $\rho_{\rm adj}^{++}(\omega)$ is neither even nor odd. We note that, as one might suspect from Eq.~\eqref{eq:G-analytic-cont} and is apparent from Eq.~\eqref{eq:G-adj-inversion}, the analytic continuation holds provided that $0 < \tau < \beta$. This is precisely the range where we discuss the calculation of $G_{\rm adj}$ in the next section.
A direct calculation using Eqs.~\eqref{eqn:Gfund_rho},~\eqref{eq:rho-diff}, and~\eqref{eq:G-adj-inversion} shows that the diagrams (5, 5r) of Figure~\ref{fig:diagrams} lead to
\begin{align}
    \Delta G(\tau) &\equiv G_{\rm adj}(\tau) - G_{\rm fund}(\tau) \\ 
    &=  \int_{-\infty}^{+\infty} \frac{\diff \omega}{2\pi} \frac{\exp \big( \omega ( \frac{1}{2T} - \tau ) \big)}{2 \sinh \big( \frac{\omega}{2T} \big) } \Delta \rho (\omega)  \nonumber \\
    &= \frac{g^4 T_F (N_c^2 - 1)}{(2\pi)^3} \pi T^4 \big[ \zeta ( 4, \tau T ) - \zeta ( 4, 1 - \tau T ) \big] \nonumber \\ & \quad + \mathcal{O}(g^6) \, , \nonumber
\end{align}
where $\zeta(s,a) = \sum_{k=0}^\infty (k+ a)^{-s}$ is the Hurwitz zeta function.

After extracting $\rho_{\rm adj}^{++}(\omega)$ from the lattice QCD calculated $G_{\rm adj}(\tau)$, which will be discussed in the next section, we can obtain $\kappa_{\rm adj}$ and $\gamma_{\rm adj}$ as
\begin{align}
\label{eqn:extraction}
\kappa_{\rm adj} &= \lim_{\omega\to0} \frac{T}{2\omega} \left [\rho_{\rm adj}^{++}(\omega) - \rho_{\rm adj}^{++}(-\omega) \right] \\
\gamma_{\rm adj} &= - \int_0^\beta \diff \tau \, G_{\rm adj}(\tau)  \nn \\
& \quad - \frac{1}{2\pi} \int_{-\infty}^{+\infty} \!\!\! \diff \omega  \frac{1 + 2n_B(|\omega|)}{|\omega|} \rho_{\rm adj}^{++}(\omega) \,, \nn
\end{align}
where the expression we have written for $\kappa_{\rm adj}$ makes it manifest that only the $\omega$-odd part of $\rho_{\rm adj}^{++}(\omega)$ contributes to it. (One can show this by using Eqs.~\eqref{eqn:kappa_gamma_adj} and~\eqref{eqn:g>_to_gT}.)
We note that $\gamma_{\rm adj}$ may be substantially more difficult to extract than in the fundamental representation case. While the first term is indeed the same as in the fundamental case by virtue of $ \int_{-\infty}^{+\infty} \frac{\diff \omega}{2\pi} \frac{\rho_{\rm adj}^{++}(\omega)}{\omega} = \int_0^\beta \diff \tau G_{\rm adj}(\tau)$, the fact that $\rho_{\rm adj}^{++}$ is not necessarily odd under $\omega \to - \omega$ means that the last term can contribute. Indeed, it does so in perturbation theory, as demonstrated by our calculation of $\Delta \gamma$ in Eq.~\eqref{eq:Delta-gamma-proof}. There is even an additional complication in that the $1$ in $1+2n_B$ of the second line will usually generate ultraviolet divergences that have to be regulated analytically (e.g., by dimensional regularization).
Furthermore, the first term may also require regularization for the integration regions where $\tau \approx 0, \beta$.

\begin{figure*}[t]
    \centering
    \includegraphics[width=0.89\textwidth]{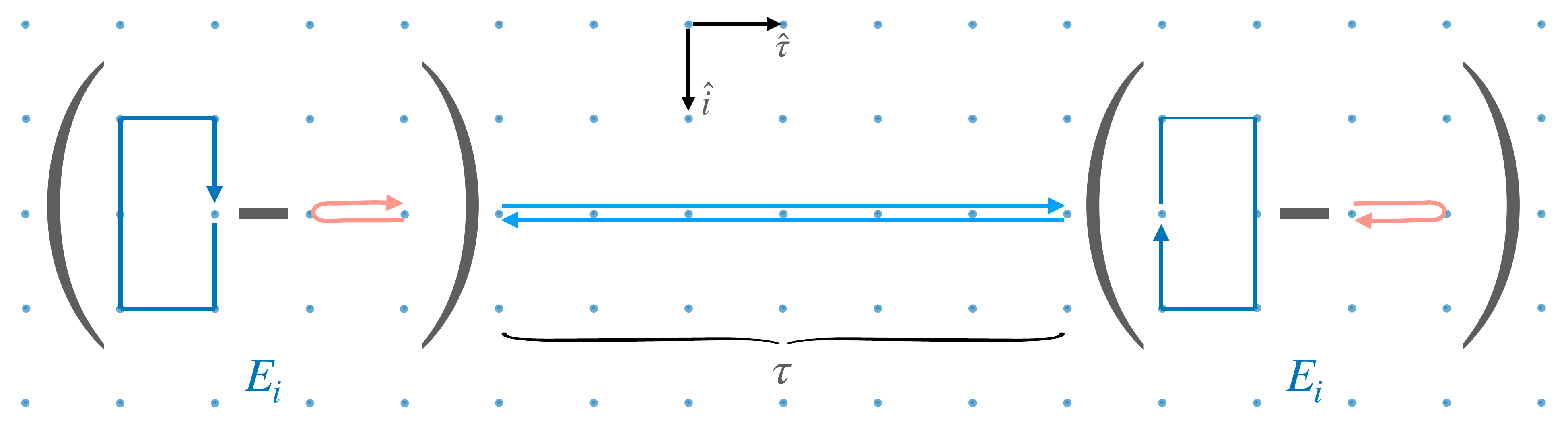}
    \caption{Lattice discretization of the chromoelectric field correlator. The electric field insertions are constructed by taking the difference between the products of gauge links over the blue and red contours at the ends of the light blue contours, which represents an adjoint Wilson line. In this setup, the adjoint Wilson line is equivalent to two antiparallel fundamental Wilson lines.}
    \label{fig:EE-picture}
\end{figure*}

\subsection{Lattice QCD Determination of $G_{\rm adj}(\tau)$ and Renormalization}
\label{sec:lattice}
In this section, we discuss how to perform a lattice QCD calculation of $G_{\rm adj}$ and extract $\rho_{\rm adj}^{++}$. We will first show a discretized version of $G_{\rm adj}$ and then discuss how to renormalize the lattice QCD result when taking the continuum limit. Finally we will give a fitting ansatz to extract $\rho_{\rm adj}^{++}$ from the calculated $G_{\rm adj}$, which can then be plugged into Eq.~\eqref{eqn:extraction} to obtain the quarkonium transport coefficients.

\subsubsection{Lattice Discretization}
The main ingredient we require in order to construct a lattice formulation of the correlator that determines quarkonium transition rates is a discretized formula for the gauge field strength $F_{\mu \nu} = \partial_\mu A_v - \partial_\nu A_\mu - i g [A_\mu, A_\nu ] $ in terms of link variables $U_\mu(n) = \exp (i a g A_\mu(n) )$ :
\begin{align}
    [\Delta U]_{\mu \nu}(n) &= U_{-\nu}(n + \hat{\nu} ) U_{-\mu}(n + \hat{\mu}+\hat{\nu}) U_\nu(n+\hat{\mu}) \nonumber \\ & \quad \times U_{\nu}(n+\hat{\mu}-\hat{\nu}) U_\mu(n - \hat{\nu}) U_{-\nu}(n) - 1 \nonumber \\ &= 2 i g a^2 F_{\mu \nu}(n) + \mathcal{O}(a^3) \, .
\end{align}
This discretization is different from the standard square plaquette. We chose this one because it makes the operator symmetric around the Wilson line direction, as shown in Fig.~\ref{fig:EE-picture}.
One can then write an expression purely in terms of link variables for the correlator:
\begin{align}
    G_{\rm adj}(\tau; a) &= \frac{(-1)}{12 a^4 N_c} \left\langle \! {\rm Tr}_{c} \! \left\{ \! \left( \prod_{n= n_\tau-1}^{0} U^\dagger_{0}(n) \right) \!  [\Delta U]_{\tau i}( n_\tau ) \right. \right. \nonumber \\ &  \quad  \left. \left. \times \! \left( \prod_{n=0}^{n_\tau - 1} U_{0}(n) \right) \!  [\Delta U]_{(-\tau) (-i)}( 0 ) \! \right\} \! \right\rangle_E   ,
\end{align}
where $\tau = a n_\tau$, and the products are ordered in such a way that the lower limit of the index labels corresponds to the operator that is most to the right in the product, and the upper limit to the one that is most to the left. A graphic representation of the correlator can be found in Fig.~\ref{fig:EE-picture}. The average $\langle \cdot \rangle_E$ represents the expectation value under the measure defined by the Euclidean lattice path integral, i.e., $\langle O \rangle_E = \frac{1}{\ml{Z}_E} \int DU \exp(- S_E[U]) O[U] $ where $\ml{Z}_E = \int DU \exp(- S_E[U])$.

\subsubsection{Renormalization and Infrared Renormalon}
The bare chromoelectric correlator $G_{\rm adj}(\tau; a)$ can be evaluated by the lattice method explained above. For physical quantities, the lattice calculation result needs proper renormalization. Since the operator involves a Wilson line, it is expected that $G_{\rm adj}(\tau; a)$ contains a linear divergence (which has not been explicitly checked and should be done so in the future via, e.g., a calculation in lattice perturbation theory), in addition to the usual logarithmic divergence. Therefore, we renormalize the bare correlator via
\begin{align}
\label{eqn:GR_adj}
G^R_{\rm adj}(\tau,\mu) = Z(\mu,a) e^{\delta m\cdot\tau} G_{\rm adj}(\tau; a) \,,
\end{align}
where $Z$ stands for the renormalization factor for the logarithmic divergence of the composite operator, with $\mu$ the renormalization scale and $\delta m$ the mass renormalization associated with the self energy of the Wilson line.

It has been shown that this form of the renormalization factor for the nonlocal operator is consistent with the fact that when the nonlocal operator is expressed as a weighted sum of local lattice operators, they mix in the renormalization group flow~\cite{Musch:2010ka}. In this work, we will not address the potential mixing between similar correlators with different Wilson line paths connecting the two chromoelectric fields.

Our NLO calculation of the real-time partner of $G_{\rm adj}$, i.e., $[g_{\rm adj}^{++}]^>$ in Section~\ref{sect:nlo} has shown that
\begin{align}
Z' = 1 + \frac{0}{\epsilon} + {\rm finite\ terms\ at\ }g^2 +\ml{O}(g^4) \,,
\end{align}
where we used $Z'$ to distinguish the renormalization factor for $[g_{\rm adj}^{++}]^>$ from the $Z$ for $G_{\rm adj}$. The ``$0$'' coefficient of the $1/\epsilon$ term emphasizes that $[g_{\rm adj}^{++}]^>$ has no logarithmic divergence at NLO.
The calculation was performed in the continuum by using dimensional regularization. The divergent term should be the same in the dimensionally regularized and lattice regularized perturbative calculations. Only the finite terms can be different. If we want to obtain the renormalized result in the $\overline{\rm MS}$ scheme, the finite difference between the lattice scheme result and the $\overline{\rm MS}$ result should still be accounted for. In the case of $G_{\rm fund}$, the difference is known at NLO~\cite{Christensen:2016wdo}. We leave the calculations of $Z$ for the Euclidean $G_{\rm adj}$ in both schemes to future studies. (As can be seen by comparing to Ref.~\cite{Christensen:2016wdo}, such calculations are research projects on their own.)

Since the $\delta m$ term is associated with the self energy of the Wilson line, one can use lattice perturbative calculations to determine it, but the uncertainties are expected to be large due to infrared renormalons. In particular, $\delta m$ is expected to be of the form
\begin{align}
\delta m = \frac{m_{-1}(a \Lambda_{\rm QCD})}{a} + m_0(\Lambda_{\rm QCD})\,,
\end{align}
where $m_{-1}$ is constant at leading order in lattice perturbation theory, but it has a residual dependence on $a$ at higher orders via, e.g., $a\Lambda_{\rm QCD}$ due to renormalization effects. On the other hand, $m_0$ is independent of the lattice spacing $a$, but it is scheme dependent as well. (Both $m_{-1}$ and $m_0$ also depend on the other mass scales of the theory, if there are any.) The infrared renormalon ambiguity leads to an uncertainty in summing the perturbative series for $m_{-1}$, which is compensated by the same uncertainty in determining $m_0$. The fact that both $m_{-1}$ and $m_0$ are scheme dependent is reflected in the systematic uncertainty of fitting the $a$ dependence from lattice calculations at small $a$, as shown in the recent study on renormalizing the quasi parton distribution function (quasi-PDF)~\cite{LatticePartonCollaborationLPC:2021xdx}.

Here we discuss a strategy to reduce the uncertainty caused by the infrared renormalons in determining the renormalization factor $\delta m$ by using lattice QCD calculation results, which is motivated by the recent work on self renormalization of the quark quasi-PDF~\cite{LatticePartonCollaborationLPC:2021xdx,Zhang:2023bxs}. The first step is to fit $m_{-1}$ from the $a$ dependence of $Z(\mu, a)G_{\rm adj}(\tau;a)$ when $a$ is small for some $\tau$. Different choices of $\tau$ are expected to give the same fitting result, as long as we maintain $\tau\gg a$ to have negligible lattice artifacts). Due to the unknown nonperturbative dependence of $m_{-1}$ on $a$, different parametrizations may be used in the fitting and they do not lead to the same result necessarily, which reflects the scheme dependence of $m_{-1}$. Then we define $G^{R'}_{\rm adj}(\tau,\mu) \equiv Z(\mu, a) e^{m_{-1} \tau/a} G_{\rm adj}(\tau;a)$, i.e., we only absorb the extracted $a$-dependent linear divergence and the logarithmic divergence into the renormalization factor and perform an operator production expansion (OPE) at small $\tau$ (i.e., $\beta \gg \tau$ but we still require $\tau \gg a$)
\begin{align}
\label{eqn:remove_IRR}
& G^{R'}_{\rm adj}(\tau,\mu) = e^{-m_0\tau} \sum_n C_n(\alpha_s(\mu), \mu\tau) \tau^n \langle O_n \rangle_T^R(\mu)  \\
& \xrightarrow{\tau\to 0} (1-m_0\tau) \sum_{n=0,1} C_n \tau^n \langle O_n \rangle_T^R  + \ml{O}(\tau^2) \,, \nn 
\end{align}
where $O_n$ denotes the local operators in the OPE and $\langle O_n \rangle_T^R(\mu)$ represents their renormalized expectation values at the same temperature $T$. The expectation values of $O_n$ can be calculated by standard lattice QCD methods and renormalized perturbatively by calculating the corresponding logarithmic renormalization factors via lattice perturbative calculations, in the same way as it is done for the logarithmic renormalization factor $Z$ for $G_{\rm adj}$. These local operators do not involve Wilson lines and thus do not have linear divergence, so it is expected that their renormalization is insensitive to the effects from infrared renormalons. The local OPE operators that may contribute include
\begin{align}
O_0&:\quad \mathbbm{1}\,,\ \Tr_{\rm c}(F_{0i}F_{0i})\,,\ \Tr_{\rm c}(F_{ij}F_{ij})\,,\ m_q\bar{q}q \\
O_1&:\quad e_\rho \Tr_{\rm c}(F_{0i} D^\rho F_{0i})\,,\ e_\rho \Tr_{\rm c}(F_{ij} D^\rho F_{ij})\,,\ e_\rho m_q\bar{q}D^\rho q \,, \nn
\end{align}
where $e_\rho$ is a unit vector along the spacetime direction $\rho$. The short-distance Wilson coefficients $C_n$ can be calculated in perturbation theory at the scale $\mu=1/\tau$. The calculation of these coefficients is an active area of research~\cite{Braun:2020ymy,Braun:2021cqe}. In practice, we can determine $m_0$ via Eq.~\eqref{eqn:remove_IRR} by calculating the lattice renormalized $G^{R'}_{\rm adj}(\tau,\mu)$ and $\langle O_n \rangle_T^R(\mu)$. With $m_0$ determined, we can obtain $G_{\rm adj}^R(\tau,\mu)$ from $G^{R'}_{\rm adj}(\tau,\mu)$ by including the renormalization factor associated with $m_0$. As suggested in Ref.~\cite{Zhang:2023bxs}, to reduce the uncertainty caused by the infrared renormalons, one resums the leading infrared renormalons in $C_n$ by regulating the renormalon poles in the Borel space and applying the inverse Borel transformation. As shown therein, this strategy removes a large uncertainty in the determination of the quark PDF. We expect a similar uncertainty reduction to happen for the determination of $G_{\rm adj}^R$ by using this strategy.

After determining the renormalized $G_{\rm adj}^R$ in the lattice regularization, we can convert it into the $\overline{\rm MS}$ scheme if we know the difference between the perturbative results of the logarithmic divergence in these two schemes. As part of the conversion process, one has to take care of the fact that in dimensional regularization with $d=4-\epsilon$ and $\epsilon\to0$, the linear divergence is absent. Any residual finite terms from this linear divergence are accounted for through $m_0$ in the OPE matching.

\subsubsection{Fitting Ansatz for $\rho_{\rm adj}^{++}$} \label{sec:fitting-ansatz-rho}
Once we obtain the renormalized $G_{\rm adj}^R$, we can use Eq.~\eqref{eq:G-adj-inversion} to fit the spectral function $\rho_{\rm adj}^{++}$. Since we only have a limited number of data points in $\tau$, we need a fitting ansatz. One ansatz that has been used in the lattice studies of the heavy quark diffusion coefficient is of the form~\cite{Altenkort:2023oms}
\begin{align}
\label{eqn:ansatz}
\rho_{\rm adj}^{++}(\omega) = \sqrt{\rho_{\rm IR}^2(\omega) + \rho_{\rm UV}^2(\omega)} \,,
\end{align}
where $\rho_{\rm IR}$ and $\rho_{\rm UV}$ represent ansatzes in the small and large $\omega$ regions, respectively. We will construct ansatzes motivated from perturbative studies.

The {$\omega$-even part of the} large frequency behavior of $\rho_{\rm adj}^{++}(\omega)$ is determined by Eq.~\eqref{eq:rho-diff}. {The remaining ($\omega$-odd) terms can be read off directly from~\eqref{eq:spectral-full-result}, where $\rho_{\rm adj}^{++}(\omega)$ was calculated at $\omega > 0$.}
Explicitly,
\begin{align}
\label{eqn:rho_UV-2}
    \rho^{++}_{\rm adj}(\omega)  \overset{\omega \gg T}{=} \, &  \frac{g^2 T_F (N_c^2-1) \omega^3 }{ 3  \pi N_c } \times   \\  \bigg\{ 1 & + \frac{g^2}{(2\pi)^2} \bigg[ \left( \frac{11 N_c}{12} - \frac{N_f}{6} \right) \ln \left( \frac{\mu^2}{4 \omega^2} \right)  + N_c \left( \frac{149}{36} - \frac{\pi^2}{6} + \frac{\pi^2}{2} {\rm sgn}(\omega) \right) - \frac{5 N_f}{9}  \bigg] \bigg\} \nonumber \\ &+ \mathcal{O}(g^6) \, , \nonumber
\end{align}
where $N_f$ is the number of light (massless) quark flavors in the theory. It was shown in Ref.~\cite{Burnier:2010rp} that up to $\mathcal{O}(g^4)$, the leading temperature-dependent contributions {(which are the same for $\rho_{\rm adj}^{++}$ and $\rho_{\rm fund}$, cf. Eq.~\eqref{eq:spectral-full-result})} at large frequency go as $T^4/\omega$, which are omitted in Eq.~\eqref{eqn:rho_UV-2} since they are subleading.

On the infrared side, one needs to use the hard thermal loop effective theory to capture the behavior of correlation functions when $|\omega| \lesssim g T \propto m_D$, where $m_D$ is the so-called Debye mass of the QGP environment, given (perturbatively) by $m_D^2 = g^2 T^2 \left( \frac{N_c}{3} + \frac{N_f}{6} \right) $, which quantifies color-electric screening in a thermal plasma. To see the difference between the $\rho_{\rm fund}$ and $\rho_{\rm adj}$ in the small $\omega$ region, one needs to consider the same type of diagrams that led to the difference shown in Eq.~\eqref{eq:rho-diff}, which has a prefactor of $g^4$, meaning that the dominant corrections in the regime $|\omega| \lesssim m_D$ will be of order $g^4 m_D^2 |\omega| \propto g^6 T^2 |\omega|$. This means that we cannot make quantitative statements by considering only the 1-loop diagram that leads to Eq.~\eqref{eq:rho-diff} (replacing the propagators with their HTL-resummed counterparts), as we can get competing effects from 2-loop diagrams in QCD, which contribute at order $g^6$. In practice, one would also need to calculate these 2-loop diagrams to be able to match the HTL result to full QCD. We will leave such calculations to future studies. Here we only list the leading contribution in the infrared regime, which can be written in terms of the well-known heavy quark diffusion coefficient $\kappa_{\rm fund}$ at NLO:
\begin{align}
    \rho^{++}_{\rm adj}(\omega)  \overset{\omega \ll gT}{=} \rho_{\rm fund}(\omega)  \overset{\omega \ll gT}{=} \frac{\kappa_{\rm fund} \omega}{ T} + \mathcal{O}(g^6) \, , 
\end{align}
where $\kappa_{\rm fund}$ is given by~\cite{Caron-Huot:2007rwy,Caron-Huot:2008dyw}:
\begin{align}
    \kappa_{\rm fund} &= \frac{g^4 T_F (N_c^2-1) T^3 }{9 (2\pi) N_c} \times \bigg[ \left( N_c + \frac{N_f}{2} \right) \left( \ln \frac{2 T}{m_D}  + \frac12 - \gamma_E + \frac{\zeta'(2)}{\zeta(2)} \right) + \frac{N_f}{2} \ln 2 + \frac{N_c m_D}{T} C \bigg] \nonumber \\ &\quad + \mathcal{O}(g^6)  \, ,
\end{align}
with $C \approx 2.3302$, as given in Ref.~\cite{Caron-Huot:2008dyw}. The fact that the low-frequency limit of the adjoint and fundamental correlators do not differ up to this order had already been noticed in Ref.~\cite{Caron-Huot:2008dyw}.

Motivated by the above perturbative analyses, we suggest to use Eq.~\eqref{eqn:rho_UV-2} as $\rho_{\rm UV}$ in the fitting ansatz~\eqref{eqn:ansatz} and use $\kappa_{\rm adj} \omega + c |\omega|$ to parametrize $\rho_{\rm IR}$ with $c$ some constant that does not contribute to $\kappa_{\rm adj}$. The appearance of the $c |\omega|$ term in $\rho_{\rm IR}$ is a crucial difference from the case of the heavy quark diffusion coefficient and is motivated by perturbative calculations shown in Section~\ref{sect:non-odd}. The fitting of $\rho_{\rm adj}^{++}$ will not only provide the quarkonium transport coefficient $\kappa_{\rm adj}$, but also the frequency dependence of $\rho_{\rm adj}^{++}$, which is important to evaluate $\gamma_{\rm adj}$, as well as the frequency-dependent correlators $g^{\pm \pm}_{\rm adj}(\omega)$ that determine the quarkonium dissociation and recombination rates.

\subsection{Summary and outlook for a Lattice QCD calculation}
\label{sec:lattice-summary}
In this section, we explained how to determine the real time quarkonium transport properties from a Euclidean chromoelectric field correlator. This determination requires to reconstruct a spectral function in a way that is different from more intensively studied spectral function reconstruction problems, such as the one required for the extraction of the heavy quark diffusion coefficient. The key results are shown in Eq.~\eqref{eqn:extraction}. We then discussed the lattice determination of the Euclidean correlator, and in particular, a method to reduce the uncertainty caused by infrared renormalons in obtaining the renormalization factor for the linear divergence of the correlator. This method is quite involved and several perturbative calculations needed to implement the method are left to future studies, such as the lattice-regularized perturbative calculation of the logarithmic renormalization factor $Z$ in Eq.~\eqref{eqn:GR_adj} and the Borel-resummed calculation of the Wilson coefficients in the OPE~\eqref{eqn:remove_IRR}. Our work paves the way towards a nonperturbative determination of the quarkonium transport properties in the QCD hot medium, which generalizes the use of a weakly interacting gas of quarks and gluons as a microscopic model of QGP in Boltzmann (rate) equations~\cite{Rapp:2017chc,Yao:2018sgn,Du:2019tjf,Yao:2020xzw} for quarkonium to the strongly coupled case.
This not only deepens our understanding of QGP and quarkonium production in heavy ion collisions, but may also provide insights for studies of exotic heavy flavor production~\cite{Yao:2018zze,Wu:2022blx,Wu:2023djn} and dark matter bound state formation in the early universe~\cite{Binder:2020efn,Binder:2021otw,Biondini:2023yxt,Biondini:2023zcz}.

\section{Phenomenology outlook: quarkonium suppression at strong coupling} \label{sec:pheno}

To understand the dynamics of quarkonium in a strongly coupled plasma, in the light of our results in Section~\ref{sec:strong-coupling}, it is necessary to revisit the derivation of the transport formalisms where the Generalized Gluon Distributions appear. To do this, we recall that the full quantum dynamics of the system, without any approximations, determines the (reduced) $Q\bar{Q}$ density matrix by evolving the full density matrix of the system and then tracing out the environment degrees of freedom (QGP), i.e.,
\begin{align}
    \rho_{Q\bar{Q}}(t) = {\rm Tr}_{\rm QGP} \left[ U(t) \rho_{\rm tot}(t=0) U^\dagger(t) \right] \, , \label{eq:oqs-general}
\end{align}
where $\rho_{\rm tot}(t=0)$ is the initial density matrix of the whole system.

The small parameter $T/(Mv) \sim r T$ allows one to expand the time evolution operators $U(t)$ in Eq.~\eqref{eq:oqs-general} in this power counting parameter and keep only terms at the first nontrivial order. From this point forward, derivations of evolution equations for the reduced density matrix $\rho_{Q\bar{Q}}$ usually rely on additional assumptions beyond those required to set up the pNRQCD description of the heavy quark pair. For instance, in the quantum Brownian Motion limit~\cite{Brambilla:2016wgg,Brambilla:2017zei} it is assumed that $T \gg Mv^2 \sim \Delta E$, so that the typical time scale of the medium is much shorter than the characteristic time associated to the energy level splittings $\Delta E$ of quarkonium. This scale separation means that the dynamics will be determined by the zero frequency limit of $[g_{\rm adj}^{++}]^>(\omega)$, which vanishes in the strongly coupled $\mathcal{N}=4$ plasma. 
Another example is the quantum optical limit~\cite{Yao:2018nmy}, which is a semi-classical description applicable when $T \sim \Delta E$, where the off-diagonal elements of the density matrix are {grouped into the Wigner function}. As a consequence of the hierarchy of energy scales and the semi-classical limit, the contributions to in-medium quarkonium dynamics come only from the negative frequency part of $[g_{\rm adj}^{++}]^>(\omega)$, which also vanishes in the strongly coupled $\mathcal{N}=4$ plasma.

However, without these extra assumptions, a direct perturbative expansion of~\eqref{eq:oqs-general} gives a nonvanishing result, signalling a clear departure from the regime of validity of the quantum Brownian Motion and quantum optical limits. For definiteness, we consider an initial $Q\bar{Q}$ state in the color octet, as a particular case of Eq.~\eqref{eq:singlet-from-GGD}. One can then show that the probability that the $Q\bar{Q}$ pair is in a singlet state with quantum numbers $n,l$, after being in contact with a thermal bath between times $\tau_i$ and $\tau_f$, is given by
\begin{align} \label{eq:pert-Upsilon-prob}
    &\langle nl | \, \rho_{Q \bar{Q}}(\tau_f) \, | nl \rangle \\ &= \int_{\tau_i}^{\tau_f} \!\!\! d\tau_1 \int_{\tau_i}^{\tau_f} \!\!\! d\tau_2 \, [g_{\rm adj}^{--}]^{>}({\tau_2,\tau_1}) \, \langle nl | U^{\rm singlet}_{[\tau_f,\tau_1]} r_i U^{\rm octet}_{[\tau_1, \tau_i]} | \psi_0 \rangle \big( \langle nl | U^{\rm singlet}_{[\tau_f,\tau_2]} r_i U^{\rm octet}_{[\tau_2, \tau_i]} | \psi_0 \rangle \big)^\dagger \, , \nonumber
\end{align}
where $| \psi_0 \rangle$ is the initial wavefunction for the relative position coordinate of the $Q\bar{Q}$ pair in the octet state, and $U^{\rm singlet}_{[t,t']}$, $U^{\rm octet}_{[t,t']}$ are the one-body time evolution operators from time $t'$ to time $t$ for states in the singlet and octet representations, respectively, acting only on the wavefunction for the relative position coordinate between the heavy quark pair in each case. For more details on these objects, see Section~\ref{sec:rho-dynamics}.

\begin{figure}
    \centering
    \includegraphics[width=0.85\textwidth]{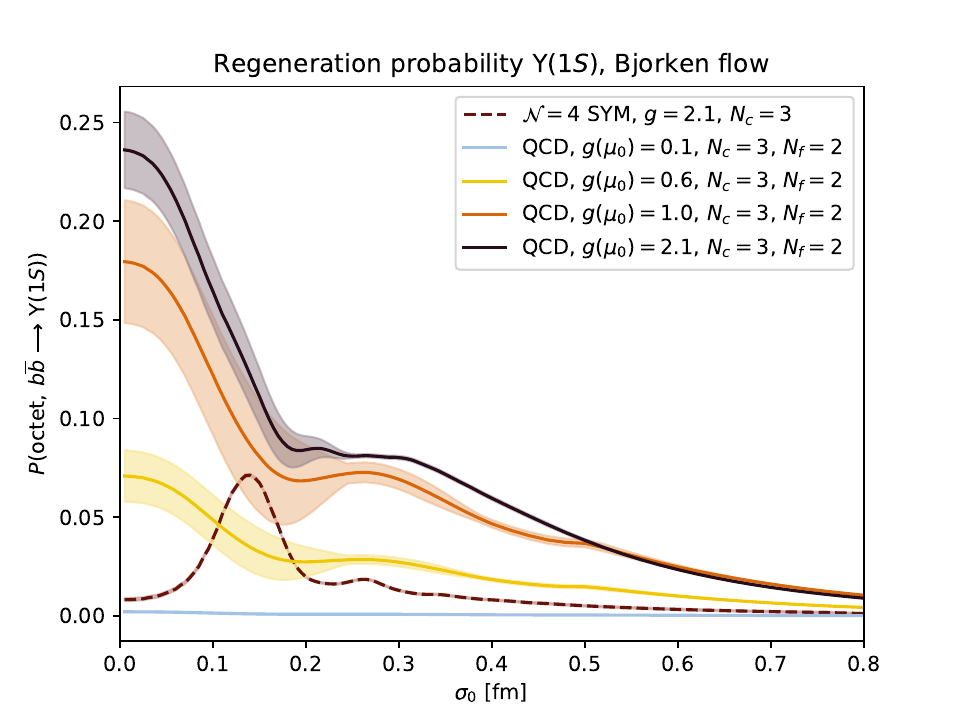}
    \caption{Regeneration/formation probability for an $\Upsilon(1S)$ state as a function of the initial separation $\sigma_0$ between the two heavy quarks, calculated with the weakly coupled QCD result~\eqref{eqn:rho_UV} we obtained in Section~\ref{sec:weak-strong} after including HTL corrections (solid lines), and the strongly coupled result~\eqref{eq:rho-strong} in $\mathcal{N}=4$ SYM at large $N_c$ (dashed line). The error bands indicate the uncertainty in the result due to truncation effects when solving for the evolution of the wavefunction. These errors are propagated into the final result in different ways for different correlators, because they can be more or less sensitive to different kinematic regions of the $Q\bar{Q}$ pair wavefunction (see discussion in the footnote\footref{fn:errorbands} in the main text). The temperature profile experienced by the heavy quark pair was set to be given by Bjorken flow scaling, $T(\tau) = (\tau_f/\tau)^{1/3} T_f $, with $T_f = 155 \, {\rm MeV}$, $\tau_i = 0.6 \, {\rm fm/c}$, and $\tau_f = 10 \, {\rm fm/c}$. The initial condition for the wavefunction in the radial component of the relative coordinate was given by $\psi_0(r) \propto r Y_{1m}(\theta, \varphi) \exp(- r^2/(2\sigma_0^2)) $, where $Y_{1m}$ is a spherical harmonic. The reason to choose $\ell=1$ as the initial state is that the transition to the $1S$ state is allowed by the dipole interaction of pNRQCD at the order we work in the EFT. The wavefunction is appropriately normalized to have unit probability, and the final result is averaged over $m$. The weakly coupled curves can have qualitatively different behavior for different values of the coupling, as their frequency and temperature dependencies are different for the $g^2$ and $g^4$ terms. Nonetheless, at weak coupling $g(\mu_0) < 1$ the result is approximately proportional to $g^2(\mu_0)$, and the different curves can be obtained by an overall rescaling (the orange, yellow, and light blue curves can be related in this way). On the other hand, the strongly coupled result only depends on $g$ through an overall factor of $\sqrt{\lambda}$, so rescaling $g$ will simply rescale the result of the calculation with the $\mathcal{N}=4$ SYM GGD by the same overall factor.}
    \label{fig:prob}
\end{figure}

To illustrate this formula, we plot it as a function of different initial conditions in Figure~\ref{fig:prob}, using the GGD $[g_{\rm adj}^{--}]^>$ calculated in a weakly coupled approximation in QCD (in solid lines) for $g(\mu_0) \in \{ 0.1, 1, 2.1\}$ with two flavors of light fermions, and the GGD $[g_{\rm adj}^{--}]^>$ calculated in a strongly coupled approximation in $\mathcal{N} = 4$ SYM (dashed line) setting $\lambda = g^2 N_c = 13.23$, corresponding to $g = 2.1$, $N_c = 3$. Strictly speaking, neither represents the (unknown) exact QCD result at realistic couplings, but we expect that by comparing these results we will gain intuition on what the features of the QCD result may be. To do this, we need to specify an interaction potential model to construct the time evolution operators, which we take to be a Karsch-Mehr-Satz potential~\cite{Karsch:1987pv} for the singlet, and no potential for the octet. 
After specifying the potential model, the only remaining parameter that needs to be specified is the temperature of the system as a function of time, which we take to be given by Bjorken flow $T(\tau) = (\tau_f/\tau)^{1/3} T_f $, with $T_f = 155 \, {\rm MeV}$, $\tau_i = 0.6 \, {\rm fm/c}$, and $\tau_f = 10 \, {\rm fm/c}$. 

In practice, we have to make an additional approximation: because $T = T(\tau)$ is time-dependent, we need to incorporate this into how we evaluate $[g_{\rm adj}^{--}]^>$, for which we only have expressions in equilibrium. We use the following (uncontrolled) approximation to relate the (slightly) out-of-equilibrium GGD $[g_{\rm adj}^{--}]^>(\tau_2,\tau_1)$ to the one we have calculated in thermal equilibrium $[g_{\rm adj}^{--}]^>(\omega,T)$:
\begin{equation}
    [g_{\rm adj}^{--}]^>(\tau_2,\tau_1) \approx \int \frac{\diff \omega}{2\pi} e^{-i \omega (\tau_2 - \tau_1) } \sqrt{[g_{\rm adj}^{--}]^>(\omega, T(\tau_1) ) [g_{\rm adj}^{--}]^>(\omega, T(\tau_2) ) } \, .
\end{equation}
This approximation becomes a strict equality if the temperature is time-independent (and deviations from the equality are small if $|\partial_\tau T| \ll T^2$, which is certainly satisfied for hydrodynamic QGP in a HIC). Furthermore, we set a cutoff for the frequency integral $|\omega| \leq 800 \, {\rm MeV}$, in consistency with the range of applicability of the effective theory (the mass scale that was integrated out is $Mv \sim 1 \, {\rm GeV}$).

The result in Figure~\ref{fig:prob} shows many interesting features. First and foremost, the regime in which quarkonium can form starting from an unbound octet state is when the initial separation between the heavy quarks is well below $1$ fm, always within the regime of applicability of pNRQCD $r T < 1$. Second, the formation/regeneration probability is suppressed in magnitude if one uses the correlator calculated in strongly coupled $\mathcal{N}=4$ SYM relative to its weakly coupled counterpart in QCD if the coupling is set by hand to be equal to a realistic value $g = 2.1$ for both cases. Even at a weaker coupling of $g = 1$, the perturbative QCD calculation yields a larger regeneration probability than the strongly coupled one at $g = 2.1$. Our physical interpretation of this is as follows: because there are no quasiparticles in strongly coupled QGP, it is not possible for an unbound $Q\bar{Q}$ pair to transition to a bound state efficiently by emitting a gluon that is absorbed by the QGP environment. This is certainly possible in the weakly coupled limit, where gluons are quasiparticles, and furthermore, their emission rate at finite temperature is Bose-enhanced due to the bosonic nature of their quantum statistics.

This interpretation also explains why the formation probability is enhanced at small $\sigma_0$ for the GGD calculated in the weakly coupled approximation, relative to the strongly coupled case: taking the strongly coupled case as a baseline, where the $\sigma_0$ dependence of the regeneration probability~\eqref{eq:pert-Upsilon-prob} is mostly determined by the overlap of wavefunctions at the initial and final times (as evidenced by the clear maximum of the probability at $\sigma_0 \sim 0.14 \, {\rm fm}$ for the dashed curve in Fig.~\ref{fig:prob}), the formation probability gets enhanced, rather than suppressed, as $\sigma_0$ is decreased. What happens is that as $\sigma_0$ is decreased, the $Q\bar{Q}$ wavefunction gets components of larger and larger momenta, meaning that the state is of higher energy, and therefore there is more phase space available to emit a gluon (with a Bose-enhanced probability) and induce a transition to a singlet state. This enhancement is absent in the strongly coupled case, as there are no quasiparticles in QGP that can give rise to this effect.\footnote{\label{fn:errorbands} This also explains why the error bands are larger at small initial separation: this sector of the plots is more sensitive to higher energy states, which are more sensitive to truncation effects when numerically solving a Schr\"odinger equation. The error band is larger for $g = 1$ than for $g = 2.1$ because, in proportion to the total emission at each value of the coupling, the emission of quasiparticles (which are controlled by the negative frequency behavior of $\rho^{++}_{\rm adj}$) is smaller due to the the fact that $\rho_{\rm adj}^{++}$ in QCD approaches the shape of $\rho_{\rm adj}^{++}$ in $\mathcal{N}=4$ SYM as the coupling is increased, as demonstrated by Fig.~\ref{fig:spectral-2sided}.} 
In contrast, at strong coupling the processes of quarkonium dissociation and recombination are driven by fluctuations of the QGP chromoelectric field, which, in absence of a quasiparticle picture, can be thought of as intrinsic properties of the fluid. Given the successes of the hydrodynamic description of QGP, it would be interesting to explore whether such fluctuations can be described within the framework of hydrodynamics, or an extension thereof, and whether the GGDs can be thought of as encoding (hydrodynamic) medium response. We leave this as an open question ripe for exploration.


We stress that if either of the transport formalisms implied by the quantum Brownian Motion limit and the quantum optical limit, respectively, had been used to calculate the regeneration probability~\eqref{eq:pert-Upsilon-prob} with the correlator calculated in $\mathcal{N}=4$ SYM at strong coupling, the result would have been zero. However, the physical effect of forming/regenerating bound quarkonium is still there, and the same statement can be made for dissociation, with different  values for the regeneration probability and qualitatively different physical mechanisms mediating this process. Therefore, in order to be able to interpret quarkonium suppression data in terms of an underlying quantum field theory that can be strongly coupled, it will be necessary to have a transport formalism that can account for both Markovian and non-Markovian effects. Constructing such a transport formalism is a brave new challenge calling for new theory developments. Doing so is especially pressing because we do not know which kind of effect dominates quarkonium dynamics in QCD at realistic values of the coupling. By using it, we hope to constrain the GGDs from data in a way that can inform us directly about the microscopic structure of QGP. Once a first-principles QCD calculation of the GGDs is available (most likely via lattice QCD methods), comparing them with the constraints from HIC data in the way we just described will become a test of our understanding of QCD.

\section{Conclusions}

In this chapter we carried out a systematic study of the correlation functions/Generalized Gluon Distributions (GGDs) that govern the dynamics of quarkonium inside QGP. We discussed how these correlation functions appear from a first-principles description of quarkonium using effective field theory techniques, rigorously defining them for the first time. Then we calculated them up to next-to-leading order in the coupling constant at weak coupling in QCD, and in the strongly coupled limit of large $N_c$ $\mathcal{N} = 4$ SYM. We compared these results and thoroughly examined the qualitative features in each limit, highlighting the differences and similarities between the chromoelectric field correlators we calculated and other chromoelectric field correlators that encode different aspects of heavy quark physics as they propagate through QGP. After doing so, we discussed the necessary ingredients and setup to extract the GGDs from a lattice QCD calculation in Euclidean (imaginary) time. To conclude, we presented the prospects, and the challenges that will have to be met, to systematically use these correlators in phenomenological studies of quarkonium propagating through QGP, aimed at describing the suppression ratios that have been so precisely measured in collider experiments (see, e.g., Fig~\ref{fig:Quarkonium-Suppression-CMS}).

Our calculation in weakly coupled QCD revealed an unusual, but crucial, property of the spectral function $\rho^{++}_{\rm adj}$ that describes the physical processes of quarkonium dissociation and recombination. Namely, that this spectral function does not satisfy the usual parity property under $\omega \to - \omega$. That is to say, we explicitly verified that $\rho^{++}_{\rm adj}(\omega) \neq - \rho^{++}_{\rm adj}(-\omega)$. As we highlighted in Section~\ref{sect:non-odd}, this property encodes the difference between the transport coefficients $\gamma_{\rm adj}$ and $\gamma_{\rm fund}$ that had been found in previous studies using perturbation theory in QCD. These studies had limited themselves to calculating the zero-frequency limit of the chromoelectric field correlation functions. Our studies in Sections~\ref{sect:nlo} and~\ref{sec:latticec} clarified that this difference is encoded in the whole frequency dependence of $\rho^{++}_{\rm adj}$, and not only in its low-frequency limit. This happens in a nontrivial way precisely because $\rho^{++}_{\rm adj}$ does not have definite parity under $\omega \to - \omega$. This property might cause large effects when the coupling constant is larger, thus precluding a perturbative expansion, which motivated us to do an analog calculation in strongly coupled $\mathcal{N}=4$ SYM.

The results of such a calculation, discussed in Section~\ref{sec:strong-coupling}, were striking, in the sense that we found that in the strong coupling limit the spectral function $\rho^{++}_{\rm adj}(\omega)$ is maximally away from being either even or odd in $\omega$, as it vanishes for $\omega < 0$ and is nonzero for $\omega > 0$. Because of this, the results of our calculation in strongly coupled, large $N_c$ $\mathcal{N}=4$ SYM indicate that the two transport coefficients in the Lindblad equation for quarkonium in-medium dynamics in the quantum Brownian motion limit, as defined through the chromoelectric field correlator we studied, vanish in a strongly coupled $\ml{N}=4$ SYM plasma. Furthermore, the correlation functions $[g_E^{\pm\pm}]^>$ that determine quarkonium dissociation and recombination in the Boltzmann equation that is valid in the quantum optical limit also vanishes in a strongly coupled $\ml{N}=4$ SYM plasma. This means the in-medium dynamics of small-size quarkonium states is trivial at leading order in both the quantum Brownian Motion and quantum optical limits which are often used to simplify the evolution equations of quarkonium in QGP.
In hindsight, this is not too big of a surprise, because both of the above limits rely on some form of weakly coupled physics or the existence of quasiparticles: the quantum Brownian Motion limit assumes that scatterings off the medium are decorrelated in time, and the quantum optical limit assumes that quasiparticles of definite energy can be absorbed/emitted by the $Q\bar{Q}$ pair. Neither is likely to be true for strongly coupled QGP formed in HICs where there are no quasiparticles.

However, we showed in Section~\ref{sec:pheno} that quarkonium dynamics is nontrivial even when the GGD that governs its dynamics is given by the result of the strongly coupled $\mathcal{N}=4$ SYM calculation. For this specific instance, this signals a breakdown of the quantum Brownian Motion and quantum optical limits, as the leading contribution to the dynamics is given by terms that are omitted in both of these limits. The degree to which these limits approximate the real-time dynamics of quarkonium produced in HICs remains as an open question to be answered by future studies.
Therefore, our results demonstrate the need for a non-perturbative determination of the chromoelectric correlator that determines in-medium quarkonium dynamics in QCD. At present, the only option to do this in the foreseeable future is to do a calculation of the GGDs using lattice QCD techniques, along the lines of our discussion in Section~\ref{sec:latticec}. This has the potential to make a significant contribution to phenomenological studies of in-medium quarkonium, as we will then be able to pin down the precise physical mechanism that drives quarkonium dissociation and recombination in QGP.


Finally, from a data-driven perspective, one should try to extract this correlation from phenomenological studies and experimental measurements by applying Bayesian analysis techniques, where one uses some ansatz for the correlation that is well-motivated from both weak-coupling and strong-coupling studies, varies the parameters in the ansatz and compares the calculation results of quarkonium nuclear modification factors and elliptic flow coefficients with experimental data. The Bayesian analysis is then applied to systematically find the best set of parameters in describing the data and estimate the parameters' uncertainties. 
For example, it will be particularly instructive to see how data from 200 GeV Au-Au collisions, particularly from the sPHENIX program, can be used to better constrain the finite frequency dependence of the correlator via a Bayesian analysis.
We expect data coming from these collision energies will be sensitive to its finite frequency dependence, because the prevailing temperature regime in this experiment is of low temperature comparable to
the energy level splittings of quarkonia states.
All these studies will deepen our understanding of quarkonium in-medium dynamics and the relevant transport properties of the QGP\@. 

%% file: hydrodynamization.tex

\chapter{Dynamics of Hydrodynamization and Emergence of Hydrodynamics in Heavy-Ion Collisions} \label{ch:hydrodynamization-in-HIC}

The discovery that quark-gluon plasma (QGP) is a strongly coupled fluid~\cite{PHENIX:2004vcz,BRAHMS:2004adc,PHOBOS:2004zne,STAR:2005gfr,Gyulassy:2004zy} has opened a window to study the many-body physics of hot liquid QCD matter under reproducible conditions using particle colliders, particularly via heavy-ion collisions (HICs). While strongly correlated many-body systems are ubiquitous across physics, QGP is special because it is the most perfect liquid ever discovered, its constituents are the elementary quarks and gluons of the standard model of particle physics and --- because the theory that describes these constituents, QCD, is asymptotically free --- we know that the dynamics at the earliest moments of the very high energy collisions in which QGP is later formed is weakly coupled. Therefore, by studying how strongly coupled liquid QGP forms (hydrodynamizes) from initially weakly coupled quarks and gluons, how a droplet of this liquid evolves, expands and cools in HICs, we deepen our understanding of the fundamental building blocks of matter
in conditions as extreme as those present in the first microseconds after the Big Bang, as well as learning how a complex strongly correlated phase of matter can emerge from the fundamental laws that govern matter at the shortest distance scales.

Among the plethora of QGP-related phenomena that have been studied in the past two decades via heavy-ion collisions (for reviews see Refs.~\cite{Muller:2006ee,Jacak:2012dx,Muller:2012zq,Heinz:2013th,Shuryak:2014zxa,Akiba:2015jwa,Romatschke:2017ejr,Busza:2018rrf,Nagle:2018nvi,Schenke:2021mxx,Harris:2023tti}), the discovery that the time it takes for the highly excited, far-from-equilibrium, initially weakly coupled quarks and gluons liberated just after the collision to become near-hydrodynamic is around 1~${\rm fm}/c$ was one of the first and one of the most striking~\cite{Heinz:2001xi,Heinz:2002un,Kolb:2003dz,Heinz:2004pj}. This discovery emerged via comparison of early RHIC measurements of the anisotropic flow of off-center heavy ion collisions with early hydrodynamic calculations and the realization that if the onset of hydrodynamic behavior were much later than 1~fm$/c$, anisotropies as large as those observed were not possible. Later estimates~\cite{Shen:2010uy,Shen:2012vn} applying the same logic with more fully developed calculations and analyses of data indicated that the hydrodynamization time in RHIC collisions could be as short as 0.4~fm$/c$ to 0.6~fm$/c$, with the hydrodynamic fluid formed at that time having a temperature $\sim 500$~MeV to $\sim 350-380$~MeV.
These estimates of the hydrodynamization timescale were seen as surprisingly rapid, simply because they are comparable to the time it takes light to cross a proton and also because early attempts to use the theory of how thermalization proceeds at weak coupling in QCD kinetic theory~\cite{Baier:2000sb} seemed to yield longer timescales.
This became less puzzling when
calculations done in strongly coupled theories with holographic duals~\cite{Chesler:2008hg,Chesler:2009cy,Chesler:2010bi,Heller:2011ju,Heller:2012je,Heller:2012km,vanderSchee:2012qj,Heller:2013oxa,Casalderrey-Solana:2011dxg,Chesler:2015lsa,Chesler:2015fpa,Heller:2016gbp} showed that in the strong coupling limit hydrodynamization should be expected within $\sim 1/T_{\rm hyd}$ after the collision, where $T_{\rm hyd}$ is the temperature of the hydrodynamic fluid that forms. This implicit criterion corresponds to a hydrodynamization time in HICs that is comparable to the fastest estimates that had previously been inferred from RHIC data.  
That said, all of these developments posed a pressing challenge: can we find
a microscopic understanding of this process in terms of the kinetic theory of the fundamental degrees of freedom of QCD itself, noting that at the earliest moments of a high energy collision their dynamics must be weakly coupled.
This
has been a subject of intense study over the past decade; for reviews, see Refs.~\cite{Schlichting:2019abc,Berges:2020fwq}. The successes of hydrodynamic modeling in describing HIC data only reinforce the need for a qualitative and quantitative understanding of the processes that connects the initial state (two highly Lorentz-contracted atomic nuclei) with a hydrodynamic droplet of strongly coupled QGP in local thermodynamic equilibrium.

In this Chapter we present and develop a new avenue in our understanding of the hydrodynamization process of QGP in the framework of kinetic theory, by enriching and extending the pioneering work of Brewer, Yan and Yin~\cite{Brewer:2019oha}, where they formulated the Adiabatic Hydrodynamization (AH) scenario. The framework it entails should also provide valuable insight for the thermalization process of more general many-body systems. Section~\ref{sec:intro-AH-BSY} discusses how self-similarity in the evolution of the distribution function is related to the presence of an ``adiabatic frame,'' in which the distribution function is fully described by the adiabatically evolving lowest energy eigenstate of the time evolution operator of the theory in that frame. Our study in the initial Section of this Chapter revolves around the so-called first stage of the `bottom-up' thermalization scenario~\cite{Baier:2000sb}, where we verify that the AH scenario is indeed realized. Section~\ref{sec:adiab-beyond-scaling} further develops the AH framework and proposes a method for finding the adiabatic frame by optimizing the degree to which the evolution of the system is adiabatic, and shows how this frame can be defined even when the distribution function does not exhibit self-similar phenomena. Our main result is that, in a simplified QCD kinetic theory with only small angle gluon scatterings, we find that the AH scenario is realized sequentially, with two separate steps of memory loss of the initial condition. This process of memory loss is first explained by the dominance of a group of low-energy instantaneous eigenstates of the time evolution operator of the kinetic theory. As the system approaches hydrodynamics, only one of these states remains of low energy and then later dominates the evolution, signalling an almost complete memory loss of the initial condition, encoded in the fact that only the hydrodynamic degrees of freedom remain. In this way, we have achieved a unified picture of the process of hydrodynamization in this theory. Section~\ref{sec:outlook} summarizes our findings.

\section{Scaling and adiabaticity: the early stages of a weakly coupled plasma in a Heavy Ion collision} \label{sec:intro-AH-BSY}

Studies of far-from-equilibrium QCD have revealed a surprising self-similar ``scaling'' behavior of the quark and gluon distribution functions.
A distribution function is said to exhibit scaling behavior if the shape of the (rescaled) distribution function remains stationary when expressed in terms of a rescaled momentum variable.
This self-similar evolution has been observed in classical field simulations~\cite{Berges:2013eia,Berges:2013fga}, for small-angle scatterings in kinetic theory~\cite{Tanji:2017suk},
and later in simulations of QCD EKT~\cite{Mazeliauskas:2018yef}. 
In addition to showing that the distribution function reaches the self-similar scaling form, the study of Ref.~\cite{Mazeliauskas:2018yef} further demonstrated that
the distribution function can take the scaling form with time-dependent scaling exponents much before the scaling exponents attain their fixed-point (time-independent) values.\footnote{
In Ref.~\cite{Mazeliauskas:2018yef}, this time-dependent scaling is called ``prescaling".} 
This interesting and important finding suggests that the early time evolution of quark-gluon matter created in heavy-ion collisions might be simply characterized by a scaling function together with the evolution of a handful of scaling exponents. Time-dependent scaling in Bose gases has also been studied in Ref.~\cite{Schmied_2019}. However, it remains unclear what causes the emergence of time-dependent scaling and how general the resulting exponents are.

In this section we concentrate on the early non-equilibrium stage of a heavy-ion collision and aim at gaining qualitative lessons about the emergence of time-dependent scaling and the evolution of the scaling exponents.
For this purpose, 
we shall consider a Bjorken-expanding gluon plasma and study the kinetic Boltzmann equation with a highly occupied initial condition. 
We will employ the small angle approximation to the Boltzmann equation, which then takes the form of a Fokker-Planck (FP) equation. We will refer to this equation as ``FP equation'' throughout. 
This equation has been studied previously in Ref.~\cite{Tanji:2017suk}, where they showed that it featured solutions with time-independent scaling behavior. 
Our results further demonstrate that this FP equation exhibits \textit{time-dependent} scaling for hard gluons, and that its solutions capture key qualitative features of the scaling seen from QCD EKT results reported in Ref.~\cite{Mazeliauskas:2018yef}. 
This supports our view that the relatively simple FP equation can be utilized as a qualitatively accurate effective description of time-dependent scaling behavior of hard gluons.

\begin{figure}[t]
\centering
\includegraphics[width=0.9\textwidth]{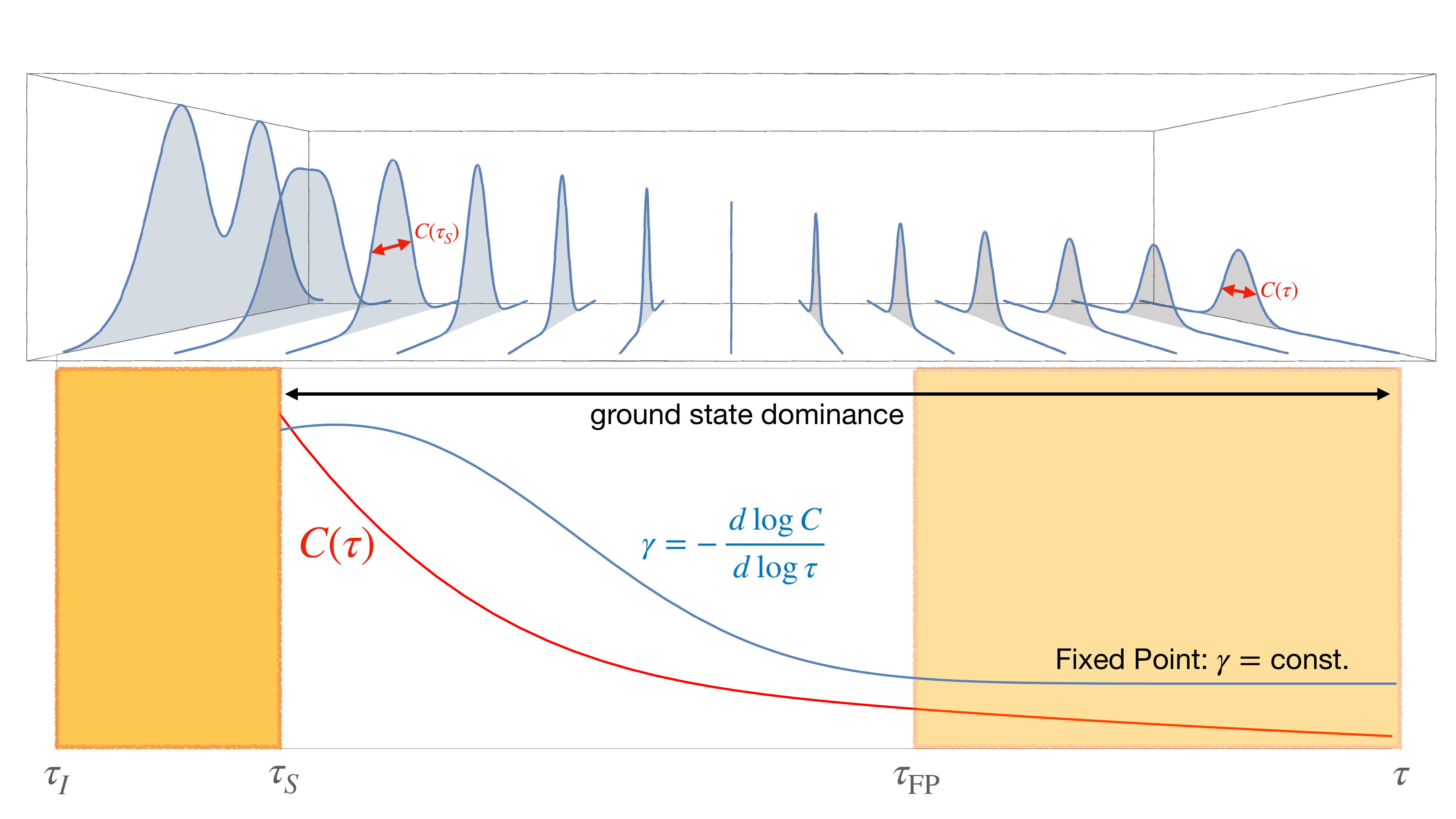}
  \caption{
  \label{fig:Evo-Ill}
We illustrate time-dependent scaling behavior and its connection to adiabaticity. On the top, we show a typical evolution of a distribution function in the present work.
Below, we show the temporal evolution of the characteristic scale $C$ and its associated scaling exponent $\gamma$ in red and blue solid curves, respectively. Though the evolution of the distribution function begins at $\tau_{I}$, the scale and exponents are only well-defined after the time $\tau_S$ when the distribution function reaches its self-similar scaling form.
Although the scaling exponent $\gamma$ will eventually approach its fixed point value at 
$\tau_{\rm FP}$, 
the distribution function may take the scaling form at $\tau_{S}<\tau_{\rm FP}$.
Within the present set-up, we find the emergence of scaling behavior around $\tau_{S}$ is associated with the decay of excited modes, as will be explained throughout the main text. 
The ground state mode can then be associated with the scaling form of the distribution, giving the dominant contribution to the state of the system during the scaling stage, and hence the distribution function's self-similar evolution becomes equivalent to adiabatic evolution. We note that, in this scenario, all of this happens well before the system becomes hydrodynamic: $\tau_{\rm FP} \ll \tau_{\rm Hydro}$.}
\end{figure}

One of the novel results in this work is that 
for the case of the FP equation, 
we show explicitly and analytically the equivalence between scaling in the evolution of the distribution function and the adiabatic evolution of the distribution function by extending the adiabatic scenario for rapidly expanding gluon plasmas first proposed in Ref.~\cite{Brewer:2019oha}. 
In our adiabatic picture, the scaling function can be identified as the instantaneous ground state of a non-Hermitian operator describing the evolution of a rescaled distribution function. 
In this framework, the emergence of self-similarity is due to the decay of instantaneous excited states. Excited states naturally decay over time because the non-Hermitian nature of the time evolution operator considered herein implies that their time evolution factors decay exponentially over time, like the evolution of states in quantum mechanics under Euclidean time evolution. The time scale over which this decay happens is determined by the inverse of the energy gap $\Delta E$ between the ground and lowest excited state. If this energy gap is larger than the rate $\Gamma_{0 \to e}$ at which transitions induced by the time-dependence of the time evolution operator move the system away from the ground state, then one says that the evolution is adiabatic, and furthermore, it is a good approximation to describe the evolution of the whole system by that of its ground state.
This naturally explains why a wide range of initial distributions would approach the scaling function, and showcases the generality of time-dependent scaling (see Fig.~\ref{fig:Evo-Ill} for an illustration). In general, whenever such a description can be set up, this provides a simple and straightforward way to understand the emergence of pre-hydrodynamic attractors.

To describe the evolution of the scaling exponents,
we derive a set of closed-form equations by imposing the adiabaticity condition $ \Gamma_{0 \to e} / \Delta E \ll 1$ for the rescaled distribution function, which in the case of the FP equation we study can be made exact by demanding $\Gamma_{0 \to e} = 0$, ensuring that the state cannot transition away from the ground state.   
We verify numerically that those equations not only give a reasonable description for scaling exponents extracted from the FP equation, but also for those from QCD EKT~\cite{Mazeliauskas:2018yef}.
From those equations, 
we obtain non-universal corrections to the scaling exponents near the fixed point, analogous to anomalous dimension corrections in quantum field theory.

Section~\ref{sec:intro-AH-BSY} is organized as follows: 
we review the pertinent ingredients of time-dependent scaling behavior and specify the FP equation we solve throughout this work in Sec.~\ref{sec:scaling-setup}. 
Then, in Sec.~\ref{sec:scaling-num}, we demonstrate the scaling behavior of the FP equation numerically, and in Sec.~\ref{sec:analytic} we present analytical results for the scaling solution for a simplified case in order to gain some intuition. Next, in Sec.~\ref{sec:Adiabatic} we establish the connection between adiabaticity and scaling, demonstrating our claim that describing the distribution function in terms of an adiabatic evolution of the system makes manifest the underlying phenomena that lead to attractor behavior. In Sec.~\ref{sec:evo}, we formulate and study the evolution equations for the scaling exponents that define the frame in which adiabaticity is optimized, and compare with available numerical results to test our formalism.
We give our concluding remarks in Sec.~\ref{sec:sum-adiab-early}.

\subsection{Set-up} 
\label{sec:scaling-setup}

In this work, 
we consider the early-time, far-from-equilibrium evolution of gluonic matter created in a heavy-ion collision undergoing Bjorken expansion.
We shall assume that the initial gluon distribution is given by the saturation scenario (see Ref.~\cite{Gelis:2010nm} for a review), i.e., the typical gluon momentum is the saturation scale $Q_{s}$ and the occupation number of hard gluons is much larger than $1$.
The gluon distribution will subsequently evolve because of the longitudinal expansion and interactions among gluons. 
Within the above picture,
we will investigate how a self-similar evolution of the gluon distribution function $f(\pz,\pT;\tau)$ (which depends on transverse and longitudinal momentum $p_{\perp}, p_{z}$ and Bjorken time $\tau$) can emerge.

In this section, we will establish the concepts we will need in our subsequent analysis. Specifically, we review pertinent ingredients of time-dependent scaling in subsection~\ref{sec:scaling-review} and specify the collision integral we use in subsection~\ref{sec:kin-setup}.

\subsubsection{Time-dependent scaling
\label{sec:scaling-review}
}

Let us begin by writing an arbitrary distribution function $f(\pz,\pT;\tau)$ as 
\begin{align}
\label{f-w}
    f(\pT,\pz; \tau)= \tA(\tau)\, w(\zeta,\xi; \tau)\, , 
\end{align}
where we have introduced the rescaled variables
\begin{align}
\label{zeta-xi-def}
  \zeta \equiv \frac{p_{\perp}}{B(\tau)}\, , 
  \qquad
  \xi \equiv \frac{p_{z}}{C(\tau)}\, .
\end{align}
Given that the function $w$ is time-dependent at this point, there is no loss of generality as any function $f$ can be written in this way. For simplicity in the notation, we shall henceforth keep the time-dependence of $A,B,C$ implicit.

The choice of $A,B,C$ can be viewed as a choice of frame. For a given distribution function $f$, there is a family of frames resulting in a family of rescaled distribution functions $w$. Though $A,B,C$ at this point are arbitrary, an appropriate frame choice may illuminate the underlying physics.\footnote{See also Refs.~\cite{2001nlin.....11055A,kevrekidis2017revisiting} for examples in the study of self-similar solutions for partial differential equations.} For convenience, 
we shall take $A,B,C$ to be of the order of the characteristic occupancy number, transverse, and longitudinal momentum respectively, so that $w$ is order one for $\zeta,\xi\sim 1$. Furthermore, if one is able to find a frame such that these properties are preserved under time evolution, then a great reduction in complexity is achieved because the characteristic scales of the problem are immediately apparent. Finding such a frame is one of the main tasks that we will undertake throughout the rest of this work.



The evolution of $A,B,C$ can be characterized by their percentage rate of change,
 \begin{align}
   \label{exp-def}
   \dot{A} \equiv \frac{\tau \pd_{\tau}A}{A} = \a(\tau)\, , 
   \qquad
   \dot{B}= -\beta(\tau)\, , 
   \qquad
   \dot{C}= -\g(\tau)\, . 
 \end{align}
Throughout this work, we will use the ``dot'' to denote the logarithmic derivative with respect to $\log\tau$, e.g., $\dot{X} \equiv \partial_{\log \tau} \log X$, and keep the time-dependence of $\a,\b,\g$ implicit unless otherwise specified.

To gain intuition for these changes of frames, and what to expect for the values of the scaling exponents throughout the system's evolution, we note that for a plasma undergoing rapid longitudinal expansion, the characteristic longitudinal momentum $C$ should drop as $1/\tau$ in the free-streaming limit, corresponding to $\g=1$. 
Once interactions become relevant, one expects that the momentum exchange among gluons would slow the decay of $C$, so we expect $0< \g < 1$. 
On the other hand, the change of the characteristic transverse momentum $B$ is solely due to interactions and hence is slower than that of $C$. 
This implies that generically during the early stages of the evolution we will have
\begin{align}
\label{BC-ratio}
    r\equiv\frac{C}{B}\ll 1\, , 
    \qquad
    \frac{|\beta|}{|\g|}\ll 1\, ,
\end{align}
(see also Ref.~\cite{Baier:2000sb}). 
When the collision integral is dominated by momentum diffusion, the width of the transverse momentum distribution broadens and we expect that $\beta \leq 0$.


A distribution function is said to exhibit \textit{scaling} if there exists a special (time-dependent) frame $\As,\Bs,\Cs$ in which $w$ becomes time-independent, i.e.,
\begin{align}
w(\zeta,\xi;\tau)=w_{S}(\zeta,\xi) \, ,
\end{align}
and the distribution function takes the scaling form
\begin{align}
   \label{prescaling-1}
  f(p_\perp,p_z;\tau) = \As(\tau) \, w_{S}\left( \frac{p_\perp}{\Bs(\tau)},\frac{p_z}{\Cs(\tau)} \right)\, .
  \end{align}
The distribution function $f$ generally changes rapidly in a fast-expanding gluon plasma. Scaling is the special property that this time-dependence can be absorbed into that of $\As,\Bs,\Cs$ so that the shape of the gluon distribution in rescaled coordinates may evolve slowly or become stationary (as in \eqref{prescaling-1}).

Fixed points of the evolution are characterized by the special case that $\als,\betas,\gas$ in Eq.~\eqref{exp-def} are time-independent, and therefore
%
\begin{align}
\label{ABC-scale}
    \As\sim \tau^{\als}\, , 
    \qquad
    \Bs\sim \tau^{-\betas}\, , 
    \qquad
    \Cs\sim \tau^{-\gas}\, . 
\end{align}
Because of Eq.~\eqref{ABC-scale}, $\als,\betas, \gas$ are commonly referred to as the \textit{scaling exponents}. 
Different values of $\als,\betas, \gas$ specify different fixed points. 
For example, in the bottom-up thermalization scenario, 
the gluon plasma transits from the free-streaming fixed point $(\als,\betas,\gas)=(0,0,1)$ to the Baier-Mueller-Schiff-Son (BMSS) fixed point $(\als,\betas,\gas)=(-2/3,0,1/3)$ first found in~\cite{Baier:2000sb}.

It is conceivable that a distribution function could take the scaling form before it evolves to the fixed point. 
In this case, $w$ approaches the time-dependent scaling function $w_{S}$ while the scaling exponents $\als,\betas,\gas$ still change in time. As a matter of fact,
Ref.~\cite{Mazeliauskas:2018yef} demonstrated that the gluon and quark distribution functions exhibit this time-dependent scaling (also called ``prescaling'') in numerical simulations of QCD effective kinetic theory (EKT). This observation suggests a surprising simplification in the far-from-equilibrium evolution of the distribution function. The goal of the present work is to gain qualitative insight into this behavior.\footnote{
The time-dependent scaling of a distribution function bears a certain similarity to the crossover phenomenon of a critical Ising system.
In this case, the critical exponents evolve as a function of temperature $T$ (and/or magnetic field) from the mean-field values to those of Wilson-Fisher fixed point as $T$ approaches the critical temperature~\cite{cha95}.
}

\subsubsection{Kinetic equation and the small angle scattering approximation
\label{sec:kin-setup}
}

We will work in the weak coupling regime $ g^{2}_{s}f\ll 1$, with $g_{s}$ the coupling constant. 
In this regime the evolution of the distribution function can be described by 
the Boltzmann equation~\cite{Mueller:2002gd}
\begin{equation}
\label{kin}
  \pd_{\tau}\,f - \frac{p_z}{\tau} \partial_{p_{z}} f = - \C[f]\, ,
\end{equation}
where $\C$ is the collision integral.
As we mentioned at the beginning of Section~\ref{sec:intro-AH-BSY}, 
we shall employ the small angle scattering approximation to the collision integral, which, as the name suggests, assumes that gluons interact exclusively through small-angle elastic scatterings. 
Then, the collision integral is reduced to a Fokker-Planck-like diffusive kernel~\cite{Mueller:1999pi,Blaizot:2013lga}
\begin{equation}
	\label{eq:small-angle-kernel}
  \C_{{\rm FP}}[f] = -\lambda_0 \lcb [f] \left[ I_a[f] \nabla_{\p}^2 f + I_b[f] \nabla_{\p} \cdot \left( \frac{\p}{p} (1+f) f \right) \right]\, ,
\end{equation}
where $\lambda_0 = \frac{g^{4}_{s}}{4\pi} N_c^2 = \frac{\lambda_{'t \, Hooft}^2}{4\pi}$ (recall $\lambda_{'t \, Hooft} \equiv g^2 N_c$).
Throughout this work we refer to the Boltzmann equation~\eqref{kin} with the collision integral~\eqref{eq:small-angle-kernel} as the Fokker-Planck (FP) equation.  
The functionals $I_a$, $I_b$ are given by
\begin{align}
	\label{eq:IaIb}
  I_a[f] = \int_{\vp} f (1+f), \, 
  \qquad
  I_b[f] = \int_{\vp}\, \frac{2}{p} f\, , 
\end{align}
where here and throughout we use the shorthand notation $\int_{\vp} \equiv \int \frac{d^3 p}{(2\pi)^3}$. 
The integrand of $I_{a}$ is proportional to the density of possible scatterers and hence will be enhanced by the Bose factor when $f>1$.
$I_{b}$ is related to the Debye mass $m_{D}$, the typical momentum exchange per collision, by (see for example Ref.~\cite{Arnold:2008zu})
\begin{equation}
\label{mD}
  m_D^2 = 2 N_c g^2_{s} I_b\, .
\end{equation}
The Coulomb logarithm $\lcb$ represents a (perturbatively divergent) integral over the small scattering angle~\cite{Mueller:1999pi} 
\begin{equation}
\label{eq:lcb-def}
  \lcb [f] =\ln \left( \frac{p_{\rm UV}}{p_{\rm IR}} \right)\, ,
\end{equation}
where $p_{\rm UV}$ and $p_{\rm IR}$ are UV and IR cutoffs, respectively.
This IR divergence originates from the long range nature of the color force and is regularized by
the thermal medium-induced mass, so we take $p_{\rm IR}$ to be $m_D$. 
Since the distribution function has finite support in momentum space, 
we take $p_{\rm UV}$ to be the characteristic hard momentum of gluons above which the occupation number starts to decrease.
When the typical transverse momentum scale is much greater than the longitudinal momentum scale, which is the case when the medium is undergoing rapid longitudinal expansion (see Eq.~\eqref{BC-ratio}),  
we use
\begin{align}
p_{{\rm} UV}=\sqrt{\langle p^{2}_{\perp}\rangle}\, ,
\end{align}
where the average over the distribution function is defined in a standard way 
\begin{align}
\label{average}
  \langle\ldots \rangle\equiv \frac{\int_{\vp}\, (\ldots)\,f}{\int_{\vp}\, f}\, . 
\end{align}
As a result, for our present purposes the expression for the Coulomb logarithm can be explicitly written as~\cite{Mueller:1999fp} 
\begin{equation}
\label{lcb-0}
  \lcb [f] = \ln \left( \frac{\sqrt{\langle p^{2}_{\perp}\rangle}}{m_{D}} \right)\, .
\end{equation}
Since both $p_{\rm IR}$ and $p_{\rm UV}$ are functionals of the distribution function, they themselves are time-dependent, and therefore so is $\lcb$.
We will later demonstrate in sec.~\ref{sec:a-dim} that
the temporal dependence of $\lcb$ plays an interesting role in determining the precise behavior of the scaling exponents near the fixed points.

In the coming section, we will first establish the emergence of time-dependent scaling in the FP equation in the hard transverse momentum regime $\zeta \geq 1$ for all $\xi$. 
Gluons in this regime have typical longitudinal momentum much smaller than their typical transverse momentum, and therefore $r=C/B$ is small (see Eq.~\eqref{BC-ratio}). 
This allows us to analyze the scaling behavior order by order in $r$. To the zeroth order in the small $r$ limit,
it is sufficient to consider only longitudinal momentum diffusion in the collision integral
\begin{equation}
\label{CIa-0}
    C[f] = -\lambda_0 \lcb[f] I_a[f]\, \partial^2_{p_{z}} f \, .
\end{equation}
At finite $r$ we find that setting $I_b=0$ in $C_{{\rm FP}}[f]$, i.e.
\begin{equation}
\label{CIa-1}
    C[f] = -\lambda_0 \lcb[f] \, I_a[f] \, \nabla_{\bf p}^2 f \, ,
\end{equation}
accurately describes sufficiently hard gluons as long as $A>1$ (see Appendix~\ref{app:Ib}).
This anticipation will be corroborated by the numerical calculations in sec.~\ref{sec:scaling-num}.
We therefore use eqs.~\eqref{CIa-0} and \eqref{CIa-1} for the analytic part of our study of self-similarity and the  scaling behavior of the distribution function.

Before closing this Section, we note that conservation laws can impose important constraints on the possible values of scaling exponents.
For example, a Bjorken-expanding medium with a collision integral that conserves particle number (such as \eqref{eq:small-angle-kernel}) satisfies
\begin{align}
\label{n-evo}
  \dot{n}= - 1\,
\end{align}
where the gluon number density is given by $n=\int_{\vp}\, f$. 
In this case, it is easy to show that $\als,\betas,\gas$ must satisfy the relation
\begin{equation}
\label{scaling-relation}
  \als-2\betas-\gas=-1\,  
\end{equation}
even if the exponents themselves are time-dependent. Their fixed-point values and dynamics are determined by the precise form of the collision integral.

\subsection{Scaling in the Fokker-Planck equation}
\label{sec:scaling-num}

Scaling around the BMSS fixed point has been observed before in the FP equation~\cite{Tanji:2017suk}. In this section we will establish that this equation also exhibits scaling with a well-defined set of time-dependent scaling exponents prior to approaching the fixed point.

To do this, we numerically solve the Boltzmann equation~\eqref{kin} with the collision kernel~\eqref{eq:small-angle-kernel} (see Appendix~\ref{app:numerics} for details on the numerical implementation). 
Following Ref.~\cite{Mazeliauskas:2018yef}, we initialize the gluon distribution at the initial time $\tau_I Q_s = 70$ as
\begin{align}
\label{fI}
 f(\pT,\pz;\tau_I) = \frac{\sigma_0}{g_s^2} \exp \left( - \frac{p_\perp^2 + \xi^2_{0} p_z^2}{Q_s^2} \right)\, , 
\end{align}
where $\xi_{0}$ in Eq.~\eqref{fI} characterizes the initial anisotropy and we take $\xi_0 = 2$, and $Q_s$ is the characteristic momentum scale of the initial condition. 
The parameter $\sigma_{0}$ specifies the overoccupancy of hard gluons at the initial time, i.e. $g^{2}_{s} f(p=Q_{s};\tau_{I})\sim \sigma_{0}$.
For the kinetic description to be valid, we require $\sigma_{0}<1$.

To explore the scaling behavior, we first follow the proposal of Ref.~\cite{Mazeliauskas:2018yef} and study the moments
\begin{equation}
\label{moment-def}
    \text{n}_{m,n}(\tau) \equiv \int_{\vp}  p_\perp^m |p_z|^n f(p_\perp,p_z;\tau)\, , 
\end{equation}
for non-negative integers $m,n$. 
For $m,n<0$, the integration \eqref{moment-def} can potentially be IR divergent.
If the distribution function takes the scaling form~\eqref{prescaling-1}, 
one can substitute this distribution into the definition of moments~\eqref{moment-def}, and find that the percentage change rate of the moments is expressible in terms of scaling exponents as 
\begin{equation}
\label{moment-eq}
    \dot{\text{n}}_{m,n} = \als - (m+2)\betas - (n+1)\gas\, .
\end{equation}
From numerical solutions to the FP equation we can compute the change rate of any three different moments, and estimate $\alpha(\tau)$, $\beta(\tau)$, and $\gamma(\tau)$ from Eq.~\eqref{moment-eq}. 
Throughout we use $\als,\betas,\gas$ to refer to exponents derived by assuming that the distribution has the scaling form, while we use $\alpha,\beta,\g$ for exponents extracted from a general distribution function (as in our numerical results).
In the scaling regime, the exponents extracted from any set of three moments $(m,n)$ via Eq.~\eqref{moment-eq} will agree with each other. Conversely, if the system is not in the scaling regime, the exponents extracted from two different sets of moments will generally not agree.
In practice, we obtain the scaling exponents in Eq.~\eqref{moment-eq} from three sets of moments 
$(m,n) \in \{(0,0),(1,0), (0,1)\}$;
$(m,n)\in\{(0,0),(2,0), (0,2)\}$; and $(m,n)\in\{(0,1),(2,0), (1,1)\}$.
We take the agreement of exponents extracted from these three sets of three moments as an approximate criterion for the emergence of scaling.

\begin{figure}[t]
\includegraphics[width=\textwidth]{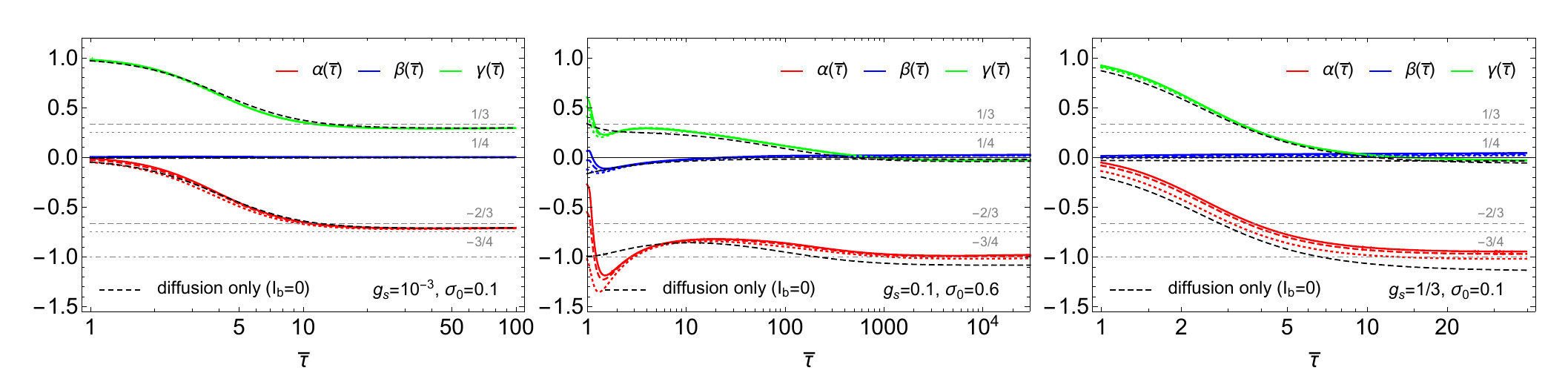}
  \caption{
  \label{fig:FP-exp}
Evolution of time-dependent scaling exponents $\alpha,\beta,\gamma$ for $(g_s,\sigma_{0})=(10^{-3},0.1)$ (left), $(0.1,0.6)$ (middle) and $(1/3,0.1)$ (right) as a function of the rescaled time coordinate $\bar{\tau}=\tau/\tau_{I}$. 
Colored curves show exponents extracted from numerical solutions to the FP equation, with different dashing styles indicating exponents extracted from different combinations of moments in Eq.~\eqref{moment-eq}. 
For comparison, black dashed curves show the exponents extracted from solutions with $I_b=0$ in the collision integral~\eqref{eq:small-angle-kernel} (see text in the three final paragraphs of this section). We include dashed horizontal lines at the values of BMSS and dilute fixed points, and additionally at $1/4$ and $-3/4$ for visual clarity.
}
\end{figure}

In Fig.~\ref{fig:FP-exp} we present the evolution of the extracted exponents for different combinations of the coupling and initial occupancy, $(g_s,\sigma_0) = (10^{-3},0.1)$, $(0.1,0.6)$, and $(1/3,0.1)$, as a function of the dimensionless time coordinate $\bar{\tau} \equiv \tau/\tau_I$.
The scaling exponents extracted from different sets of moments agree well from $\bar{\tau} \gtrsim 3$ for all $(g_s,\sigma_0)$ combinations, indicating emergence of time-dependent scaling at early times.\footnote{
We note that for $(g_s,\sigma_0)=(0.1,0.6)$ we observe that the relation \eqref{scaling-relation} between the exponents is violated by up to $20\%$ while it is satisfied within a few percent for other $(g_s,\sigma_0)$ combinations shown here.
This effect has also been observed in several previous works on the FP equation \cite{Blaizot:2014jna,Tanji:2017suk} that have suggested it may be related to the gluon condensate at $p=0$. 
}
The late-time behavior of the scaling exponents depends on the combination $(g_{s},\sigma_{0})$. 
For $(g_s,\sigma_{0})=(10^{-3},0.1)$, the late-time values of the exponents are close to (but with visible difference from) the BMSS values $(\als,\betas,\gas)=(-2/3,0,1/3)$. 
This result is also in good agreement with the late-time values of the exponents in QCD effective kinetic theory calculated for the same values of $(g_s,\sigma_0)$ \cite{Mazeliauskas:2018yef}. 
We will discuss in sec.~\ref{sec:a-dim} how this deviation from the exact BMSS scaling exponents can be attributed to an ``anomalous dimension" correction.

Remarkably, in addition to the BMSS fixed point
we also observe a new late-time fixed point in the middle and right panel of Fig.~\eqref{fig:FP-exp}, with (up to small corrections)
\begin{align}
\label{dilute-0}
\textrm{dilute:}\, \qquad
    (\als,\betas,\gas)=(-1,0,0)
\end{align}
We refer to it as the ``dilute" fixed point, since we find that the system evolves to it when the typical occupancy becomes small, $A\ll 1$. 
 For $(g_s,\sigma_0)=(0.1,0.6)$, the exponents tend toward the BMSS values before finally transiting to the dilute fixed point~\eqref{dilute-0}. 
For $(g_s,\sigma_0)=(1/3,0.1)$ the exponents go directly to this new fixed point~\eqref{dilute-0}.
We will further elaborate on its physical origin in sec.~\ref{sec:fixed-point}.

\begin{figure}[t]
  \includegraphics[width=\textwidth]{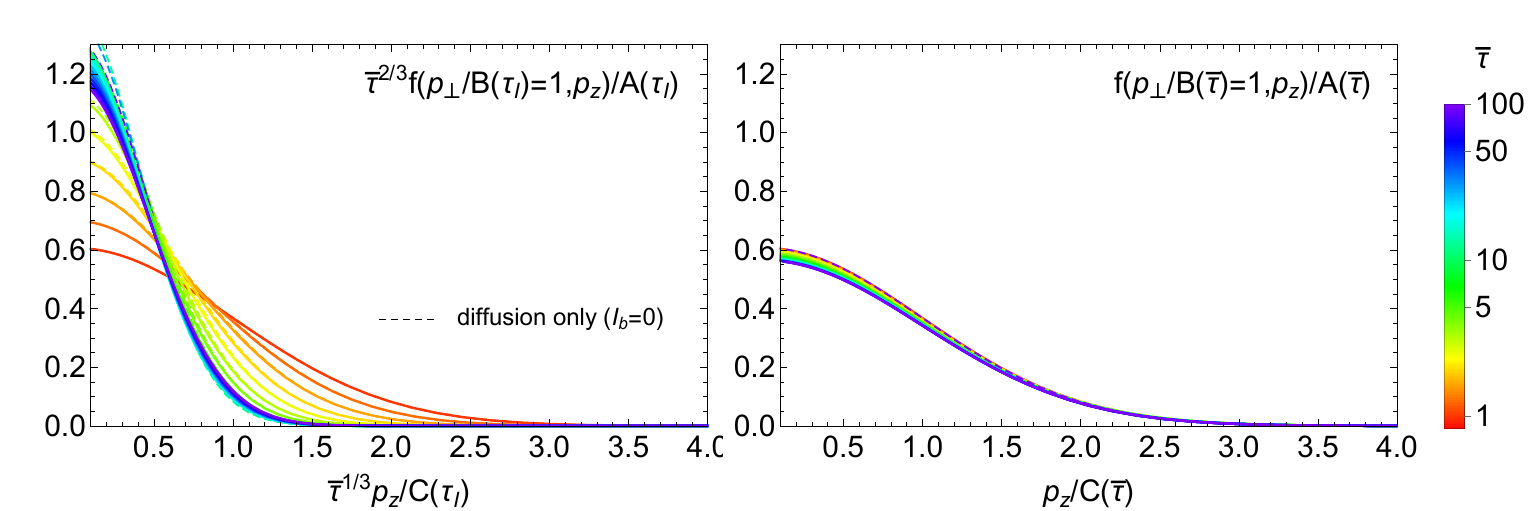}
\includegraphics[width=\textwidth]{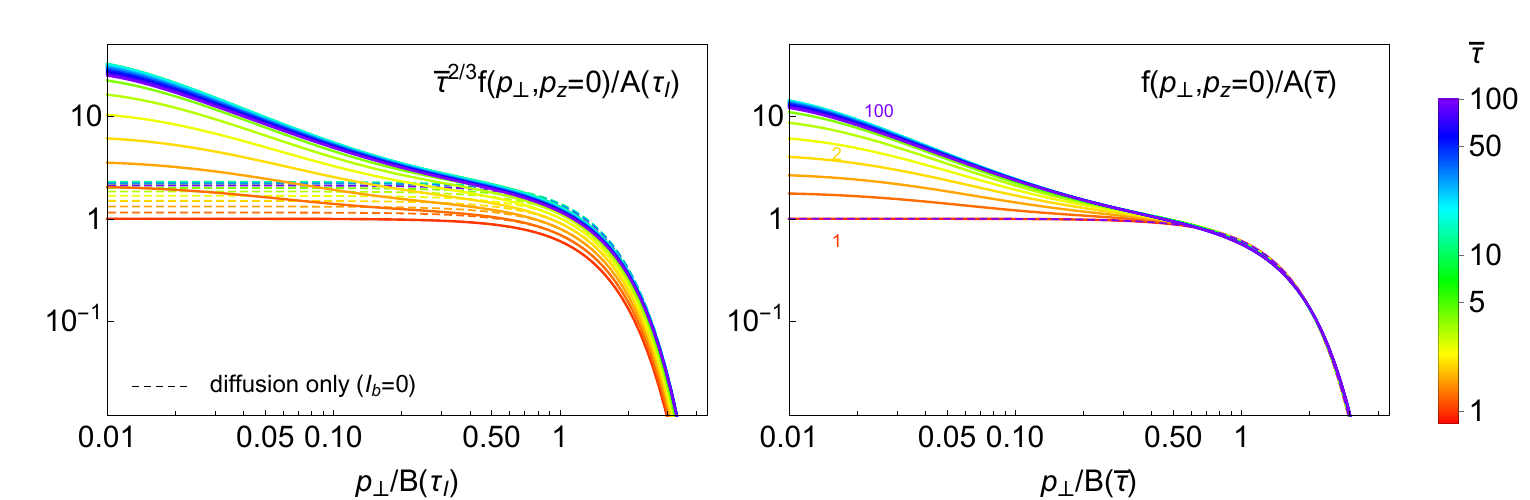}
  \caption{
   \label{fig:FP_case1}
  The rescaled distribution function $w$~\eqref{f-w} for the numerical solutions of the Fokker-Planck equation with $(g_s,\sigma_0)=(10^{-3},0.1)$. 
  The left panel shows the results with $A,B,C$ determined by the BMSS exponent while the right panel shows the same but with time-dependent scaling exponents extracted from Fig.~\ref{fig:FP-exp}. 
  Colors show the evolution in the rescaled time coordinate $\bar{\tau}$. 
  Dashed curves show the analytic scaling solution obtained for $I_b=0$, i.e., Eq.~\eqref{w-sol1}, that has no dependence on $\tau$, as shown later in this work.
}
\end{figure}

\begin{figure}[t]
    \includegraphics[width=\textwidth]{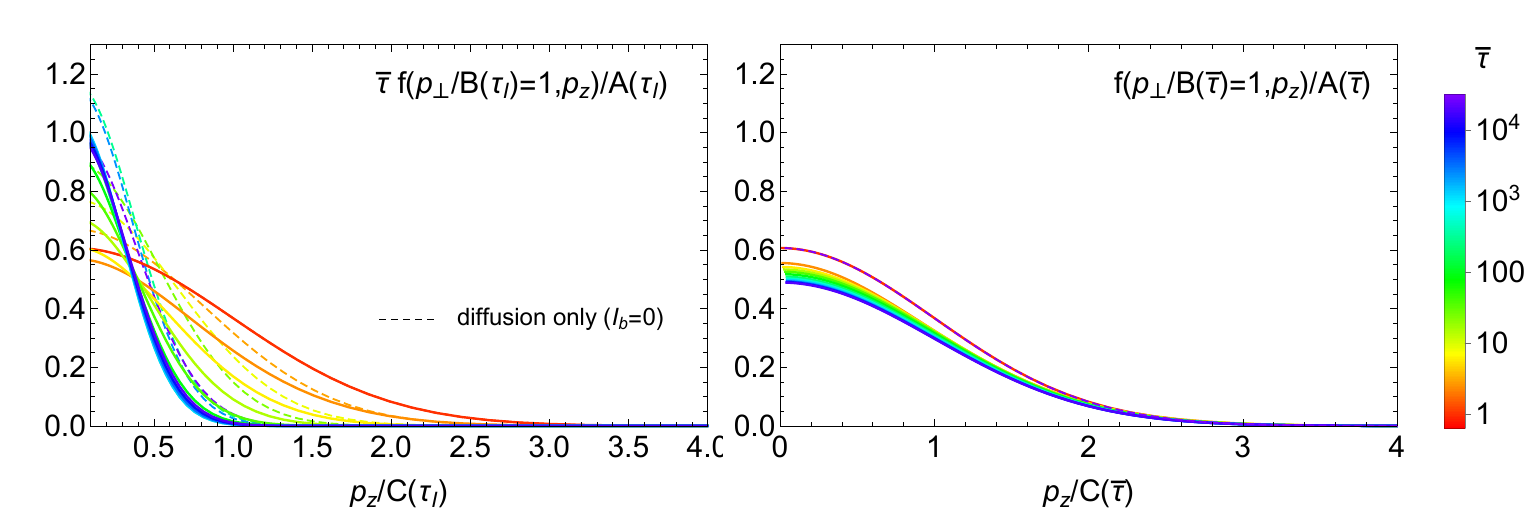}
    \includegraphics[width=\textwidth]{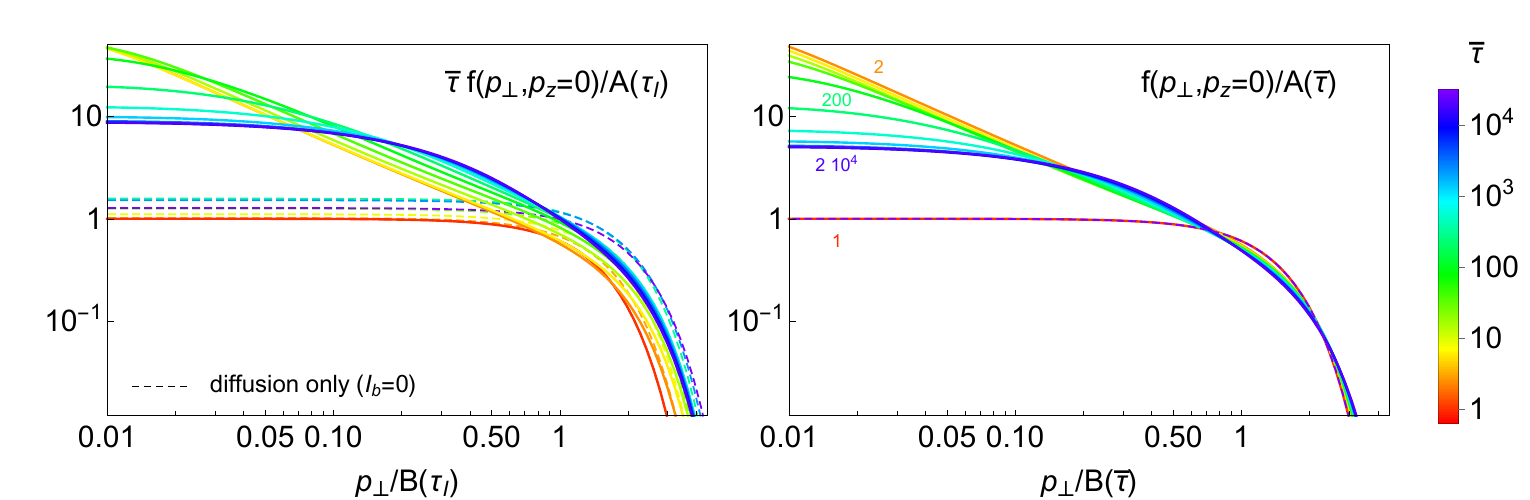}
  \caption{
  \label{fig:FP_case2}
  The same as Fig.~\ref{fig:FP_case1} but for $(g_s,\s_{0})=(0.1,0.6)$. 
  Note in the left figures we use $A,B,C$ determined not by BMSS but dilute fixed point exponents~\eqref{dilute}. 
}
\end{figure}

Though the analysis based on the moment equation~\eqref{moment-eq} shows clearly the evolution of the scaling exponents, the scaling of the distribution is seen more clearly from the full distribution function.
Figs.~\ref{fig:FP_case1} and \ref{fig:FP_case2} show the $\xi$-dependence of the rescaled distribution function $w$ at fixed $\zeta=1$ (top panels) and the $\zeta$-dependence at $\xi=0$ (lower panels) for $(g_s,\sigma_0)=(10^{-3},0.1)$ (Fig.~\eqref{fig:FP_case1}) and $(g_s,\sigma_0)=(0.1,0.6)$ (Fig.~\ref{fig:FP_case2}). We compare the scaling of the distribution function around the fixed point (left panels) to the scaling with time-dependent exponents (right panels). In all panels we take the initial values of $A,B,C$ to be the characteristic occupation number, transverse, and longitudinal momentum of the initial distribution~\eqref{fI}, which gives
$A_{I}=\sigma_{0}/g^{2}_{s}, B_{I}=Q_{s}/\sqrt{2}, C_{I}=Q_{s}/(2\sqrt{2})$. For the left panels we fix (time-independent) exponents $\als,\betas,\gas$ according the late-time fixed point (BMSS in Fig.~\eqref{fig:FP_case1} and dilute in Fig.~\eqref{fig:FP_case2}). In the right panels, we instead estimate the time-dependent scaling exponents $\alpha(\tau)$, $\beta(\tau)$, $\gamma(\tau)$ by averaging the extracted scaling exponents from three sets of moments shown in Fig.~\ref{fig:FP-exp}.

We observe in Fig.~\ref{fig:FP_case1} that, after a short time, the distribution function scales to an excellent degree with time-dependent exponents (right panel) even though the exponents have not yet reached the fixed point.\footnote{To see this from this figure, note that the left and right panels would be equal if the scaling exponents had reached their fixed point values.}
In Fig.~\ref{fig:FP_case2}, scaling appears in the hard regime $\zeta\geq 1$.
Importantly, we see that in both cases scaling occurs before the system reaches the late-time fixed point.  
This agrees with the results shown in Fig.~\ref{fig:FP-exp}. 
We note that the absence of scaling at early times in the soft regime in Figs.~\ref{fig:FP_case1} and \ref{fig:FP_case2} does not contradict our preceding analysis based on the evolution of moments.  
This is because moments with $(m,n)>0$ are mainly determined by the behavior of the distribution in the hard regime $\zeta,\xi>1$, but are less sensitive to that in the soft regime.

%
%

We note that the $\xi$-dependence of the scaling distribution is Gaussian, as is the $\zeta$-dependence in the hard regime.
We shall provide analytic insight into this Gaussianity in sections~\ref{sec:analytic} and~\ref{sec:Adiabatic}.
Collisions among gluons with typical momentum transfer of the order of the Debye mass $m_{D}$~\eqref{mD} will rapidly cascade gluons from the hard regime to the soft regime. 
The growth of the occupancy in the soft regime will in turn expedite the thermalization of soft gluons. 
Therefore, we observe $1/\zeta$ behavior in Fig.~\ref{fig:FP_case1}, corresponding to the small momentum limit of Bose-Einstein distribution. 
For Fig.~\ref{fig:FP_case2}, the system transits from the BMSS fixed point to the dilute fixed point (see Fig.~\ref{fig:FP-exp} (middle)), meaning the typical occupancy decreases from $A\gg 1$ to $A\ll 1$. 
Accordingly, the distribution at small $\zeta$ first shows $1/\zeta$ behavior and then becomes more similar to a Boltzmann distribution at later times.

As we demonstrate in Appendix~\ref{app:Ib}, the $I_{b}$ term is less important than the $I_{a}$ term
in the hard regime and when $A\geq 1$. 
To verify this in our numerical approach, we also compute the scaling exponents for a purely diffusive kernel by setting $I_b=0$ in Eq.~\eqref{eq:small-angle-kernel}. 
The resulting exponents are shown in dashed black lines in Fig.~\ref{fig:FP-exp}. We observe good agreement between the exponents obtained from solving the full FP equation and those obtained with setting $I_{b}=0$. We also show the scaling function for $I_{b}=0$ in dashed lines in Figs.~\ref{fig:FP_case1} and ~\ref{fig:FP_case2}. 
The scaling distribution with $I_{b}=0$ describes that of the hard gluons very well, in particular when $\As$ is not too small. We will show in sec.~\ref{sec:analytic} that 
the scaling function $w_{S}$ can also be computed analytically for $I_{b}=0$, see Eq.~\eqref{w-sol1}.

In summary, we have observed time-dependent scaling behavior in the FP equation for hard gluons $\zeta=p_{\perp} \Bs\geq 1$.
We find that the FP equation captures the key qualitative behavior of time-dependent scaling in EKT in this hard regime for $(g_s,\sigma_0)=(10^{-3},0.1)$, as was first shown in Ref.~\cite{Mazeliauskas:2018yef}.
In contrast, we do not observe early time scaling for soft gluons, indicating the importance of inelastic scattering in the soft regime (as was already noticed in Ref.~\cite{Mazeliauskas:2018yef}).
Nevertheless, 
our finding suggests that scaling of hard gluons is mainly driven by the longitudinal expansion and  $2\leftrightarrow 2$ small angle scatterings that are present in the FP equation. 
When the system is not too dilute, 
the solution to the FP equation in the hard regime is well-described by considering only the diffusive term (proportional to $I_{a}$) in the collision integral. Of course, the $I_{b}$ term is important in the soft regime where gluons are in equilibrium, since the equilibrium distribution in this regime is crucially determined by the balance between $I_{a}$ and $I_{b}$ terms.

In the coming sections we will understand the emergence of scaling in an analytically transparent way for hard gluons, by studying the FP equation with $I_{b}=0$. We will do so incrementally. First, in sec.~\ref{sec:analytic} we will study the solutions that exhibit scaling considering only the longitudinal part of the collision kernel~\eqref{CIa-0}. Second, in sec.~\ref{sec:Adiabatic} we will demonstrate the relevance of adiabatic evolution in this problem, and explain why the self-similar solutions are dynamically preferred. Finally, in sec.~\ref{sec:evo} we will derive the evolution equations for the time-dependent scaling exponents and compare with the numerical solutions to the FP equation as well as the results from EKT simulations in Ref.~\cite{Mazeliauskas:2018yef}.

\subsection{Analytic scaling solution for longitudinal diffusion
\label{sec:analytic}
}

In this section, 
we shall derive the scaling solution to the FP equation analytically in the limit that the typical longitudinal momentum is much smaller than the typical transverse momentum, i.e. $C/B \ll 1$. 
To leading order in small $C/B$,
the collision integral~\eqref{eq:small-angle-kernel} is reduced to Eq.~\eqref{CIa-0}, and we can write the FP equation as
\begin{align}
\label{f-simple}
\pd_{y}f=\le(p_z \pd_{p_z} +q\, \pd_{p_z}^2\ri)\, f\, .
\end{align}
Here, we have defined the effective momentum diffusion coefficient
\begin{equation}
\label{q-def}
  q \equiv \lambda_0 \lcb\, I_a[f]\, \tau \, .
\end{equation}
For later convenience, we will follow the evolution of the system in terms of a (dimensionless) logarithmic temporal variable
\begin{align}
y\equiv \log(\tau/\tau_{I})\, . 
\end{align} 
Though $q$ is a functional of $f$, for notational brevity we leave this dependence implicit.
Since the simplified collision integral~\eqref{CIa-0} does not change transverse momentum, in this section
we shall suppress the $\pT$-dependence in the distribution function and set $\betas=0$.

To look for a self-similar solution, 
we substitute Eq.~\eqref{prescaling-1} into Eq.~\eqref{f-simple} to obtain an equation for $w_{S}$:
\begin{align}
\label{w-eq-full}
  \pd_{y}w_{S}&=-\als w_{S} + (1-\gas)\xi\,\pd_{\xi}w_{S}+\frac{\qs}{C^{2}_{S}}\pd^{2}_{\xi}w_{S}
\nonumber  \\
  &=
  - (1-\gas)\, 
  \le[
  \frac{\qs}{(1-\gas)\Cs^{2}}\pd^{2}_{\xi}+\xi\,\pd_{\xi}-\frac{\als}{(1-\gas)}
  \ri] w_S
\end{align}
where we have used the definitions of $\als,\gas$ from~\eqref{exp-def}. 
Here the scaling variable $\xi$~\eqref{zeta-xi-def} is evaluated with $C=\Cs$ and the subscript ``$S$" in $\qs$ reminds us that $q$ is evaluated with the scaling distribution function as its argument. 

Then, by definition, a distribution undergoing scaling satisfies $\partial_y w_S = 0$, and consequently Eq.~\eqref{w-eq-full} becomes
\begin{align}
\label{ws}
 w_{S}+\xi\,\pd_{\xi}\,w_{S}+ \frac{\qs}{(1-\gas)\Cs^{2}}\pd^{2}_{\xi}\,w_{S}=0\, , 
\end{align}
where we have used the relation among scaling exponents~\eqref{scaling-relation} with $\betas=0$, namely $\als=-1+\gas$.
In the analysis above, we have assumed $\gas\neq 1$.
In the special case $\gas=1$, the condition on the scaling solution can be read from the first line of Eq.~\eqref{w-eq-full}: $\qs\,\pd^{2}_{\xi}w_{S}=0$, which has no bounded solution unless $\qs=0$. 
The latter corresponds to the free-streaming (collisionless) limit with scaling exponents given by $(\als,\betas,\gas)=(0,0,1)$.

For $w_{S}$ determined by Eq.~\eqref{ws} to be time-independent, we must have
\begin{align}
  \frac{(1-\gas)}{\qs/\Cs^{2}}=\textrm{const}\, . 
\end{align}
Without loss of generality, we choose the normalization of $\Cs$ such that
\begin{align}
\label{C-cond}
  \frac{(1-\gas)}{\qs/\Cs^{2}}=1\, ,
\end{align}
and with this choice, we can write the equation for $w_{S}$~\eqref{ws} as
\begin{align}
\label{ws-0}
   &\, \pd^{2}_{\xi}\,w_{S}+\xi\,\pd_{\xi} w_{S} + w_{S} = 0\, . 
\end{align}
Note that Eq.~\eqref{C-cond} imposes a non-trivial self-consistency condition for $\Cs$ since $\qs$ itself is a functional of the distribution function. 
Moreover, because $\gamma_S = -\dot{C}_S$, this equation is also implicitly a differential equation for $\Cs$.
Therefore, the evolution of $\Cs$, and consequently $\gas$, can be determined by solving Eq.~\eqref{C-cond} (see sec.~\ref{sec:evo} for more details). 
Up to a normalization constant, the solution to Eq.~\eqref{ws-0} is
\begin{align}
\label{w-sol}
  \lim_{\Cs/\Bs\to 0}\, w_{S}(\zeta,\xi)=  e^{-\frac{\xi^2}{2}}\, .
\end{align} 
The other linearly independent solution to the differential equation~\eqref{ws-0} does not give a finite number density when integrated over the momentum domain $\xi$, and hence has to be discarded.

As we have noted in sec.~\ref{sec:kin-setup}, when the typical occupancy is large
the first order corrections from $\Cs/\Bs$ can be accounted for by setting $I_{b}=0$ in the FP collision integral~\eqref{eq:small-angle-kernel}, but keeping the derivatives with respect to $p_{\perp}$ in the $I_{a}$ term. 
In this case, the collision integral is reduced to Eq.~\eqref{CIa-1}. 
By a straightforward generalization of the analysis presented in this section, we find that the scaling solution is given (up to normalization) by
\begin{align}
\label{w-sol1}
    w_{S}(\zeta,\xi)=  e^{-\frac{\zeta^{2}+\xi^2}{2}}\, . 
\end{align}
Eq.~\eqref{w-sol} and its generalization Eq.~\eqref{w-sol1} are the main results of this section. 
They tell us that the momentum dependence is Gaussian in the scaling regime, which is also what we observed numerically in the previous section. 
In the next section, we shall explain why the distribution function is attracted to this scaling form, using the adiabatic theorem of quantum mechanics as our main guiding principle.

\subsection{Scaling and adiabaticity} 
\label{sec:Adiabatic}

Here we set out to demonstrate the close connection between the emergence of scaling behavior and adiabaticity in the temporal evolution of the distribution function. 
Let us first recall that in a time-dependent quantum mechanical problem where the Hamiltonian changes with time, a system prepared in its ground state will remain in the instantaneous ground state as long as the transition rates between the ground and excited states are small compared to the energy gap between them. This is referred to as adiabatic evolution and characterizes many real-time dynamical problems in quantum mechanics~\cite{APT}.

In Ref.~\cite{Brewer:2019oha},
the idea of adiabatic evolution has been employed to describe the far-from-equilibrium evolution of the Boltzmann equation for a Bjorken-expanding plasma under the relaxation time approximation 
(see also Ref.~\cite{Blaizot:2021cdv}). With a natural, yet non-trivial, extension, 
we shall see that the scaling evolution obtained in the previous section can be viewed as an example of adiabatic evolution. 
In particular, we will show that adiabaticity naturally explains why the rescaled distribution function $w$ will generically be attracted to and stay in a time-dependent scaling function  $w_{S}$.

\subsubsection{Adiabatic frame
\label{sec:adi-pz}
}

For definiteness, we shall begin with the simplified collision integral~\eqref{CIa-0} and suppress the $\pT$ dependence of the distribution function.
In the next subsection we will extend our analysis to the collision integral defined by~\eqref{CIa-1}.

To make contact with quantum mechanics, we recast the evolution equation~\eqref{w-eq-full} for the rescaled distribution function $w$ into the form
\begin{equation} 
\label{w-eq-H}
    \partial_y w = - {\cal H}\, w\, ,
\end{equation}
where the ``Hamiltonian" operator reads
\begin{equation}
    \label{H}
    {\cal H} = - (1-\g)\, 
  \le[
  \tq\, \pd^{2}_{\xi}+\xi\,\pd_{\xi}-\frac{\alpha}{(1-\gamma)}
  \ri]\, ,
\end{equation}
and we have defined
\begin{equation}
\label{tq-def}
  \tq = \frac{q}{C^{2}(1-\g)}\, .
\end{equation}
Note that $\tq$ is a functional of $A,C$ since $q$ (defined in Eq.~\eqref{q-def}) depends on the distribution function $f$ and hence in general is evolving in time. 
Eq.~\eqref{w-eq-H} is analogous to the Schr\"odinger equation except that the operator ${\cal H}$ is non-Hermitian because the system under study is expanding and involves dissipative processes due to collisions.

Since $\mathcal{H}$ is a non-Hermitian operator, its left and right eigenvectors are not necessarily related to each other by complex conjugate. We have
\bes
\begin{align}
  {\cal H}(y) \phi_{n}^R(\xi;y)&= \sE_{n}(y)\phi_{n}^R(\xi;y)\, 
\\
  {\cal H}^\dagger(y) \phi_{n}^L(\xi;y)&= \sE_{n}(y)\phi_{n}^L(\xi;y)\, 
\end{align}
\ees
where the conjugate of ${\cal H}$ is given by
\begin{equation}
    {\cal H}^\dagger w = -(1-\gamma) \left[\tq \partial_{\xi}^2 w - \partial_{\xi} ( \xi w)  - \frac{\alpha}{1-\gamma} w \right]\, .
\end{equation}
The eigenfunctions of ${\cal H}$ represent a specific form of the distribution function in phase space, and as such, must have finite support in $\xi$ space.
Furthermore, assuming inversion symmetry about the longitudinal axis $p_z \to -p_z$, the eigenfunctions should be even in $\xi$.
Taking these constraints into account, we find 
\begin{align}
\label{phi}
&\, \phi_{n}^L = 
{\rm He}_{2n} \! \left(\frac{\xi}{\sqrt{\tq}} \right) \, ,
\qquad
  \phi_{n}^R= 
\frac{1}{\sqrt{2\pi \tq}(2n)!}\, {\rm He}_{2n}\! \left(\frac{\xi}{\sqrt{\tq}} \right)\, e^{-\frac{\xi^{2}}{2\tq}}\, ,
\\
\label{En}
&\,\sE_{n} = 2n(1-\g) +(\a-\g+1)\, , 
\end{align}
for $n=0,1,\ldots$.
Here ${\rm He}_{2n}$ denote probabilist's Hermite polynomials, 
and we have chosen the normalization of eigenstates by requiring $\int_{-\infty}^\infty d\xi \, \phi_m^L(\xi) \, \phi_n^R(\xi) = \delta_{mn}$. For obvious reasons, we refer to the $n=0$ mode as the instantaneous ground state
\begin{align}
\label{gs}
  \phi_{0}^R(\xi;y)=\frac{1}{\sqrt{2\pi \tq}}\,e^{-\frac{\xi^2}{2\tq}}\, 
\end{align}
and to modes with $n>0$ as instantaneous excited states.

The defining property of a system undergoing adiabatic evolution is that the contribution from excited states through transitions is suppressed.
To quantify the weight of excited states in the rescaled distribution $w$, we write
\begin{align}
w(\xi;y)=\sum_{n=0}\, \ta_{n}(y)\phi_{n}^R(\xi;y)\, .
\end{align}
From eqs.~\eqref{w-eq-H},~\eqref{H}, and the orthogonality of the eigenbasis, it can be shown that the coefficients $\ta_n(y)$ follow the evolution equation~(see also Ref.~\cite{Brewer:2019oha})
\begin{equation}
\label{an}
    \partial_y \ta_n + \sum_{m \neq n}  V_{nm}(y) \ta_m = -\sE_n(y) \ta_n\, ,
\end{equation}
with
\begin{equation}
\label{Vmn}
    V_{mn} = \int^{\infty}_{-\infty} d\xi \, \phi_m^L(\xi;y) \, \partial_y \phi_n^R(\xi;y)=  \dot{\tq}\,  m(2m-1) \delta_{m-1,n}\, .
\end{equation}
Transitions between different eigenstates occur only through $V_{mn}$ in Eq.~\eqref{an}. Since the eigenstates~\eqref{phi} depend on time through the time-dependence of $\tq$, 
the transition rate~\eqref{Vmn} is proportional to $\pd_{y}\tq$. When this transition rate is not small, the ground state can mix with excited states and adiabaticity will break down.

However, at this point we have the freedom to choose $A,C$ at will, so we will look for $A,C$ that minimize the transition rate $V_{mn}$. 
This goal can be achieved by imposing the condition
\begin{align}
\label{q-choice}
\tq = \frac{q}{C^{2}(1-\g)}=1\,  ,
\end{align}
so that $V_{mn}$~\eqref{Vmn} vanishes. 
As in sec.~\ref{sec:analytic}, 
we shall assume $\g<1$, so that the ground state is gapped from the excited states $\phi_{n>0}$ by $2n(1-\g)$. With $V_{nm}=0$ and an energy gap between the ground and excited states, the conditions for adiabaticity are satisfied.
With~\eqref{q-choice}, the eigenfunctions~\eqref{phi} do not depend on time explicitly and become
\begin{align}
\label{phi-2}
   \phi_{n}^R(\xi)&= \frac{1}{\sqrt{2\pi}(2n)!}\, \text{He}_{2n}(\xi)\, e^{-\frac{\xi^{2}}{2}}\, , 
\end{align}
Up to now, we still have the freedom to specify $\a$. 
A natural choice is to take
\begin{align}
\label{alpha-choice}
\alpha = \gamma -1\,      
\end{align}
such that the ground state energy $\sE_{0}$ in Eq.~\eqref{En} vanishes,
\begin{align}
    \sE_{0}=0\, . 
\end{align}
We will define the frame satisfying conditions~\eqref{q-choice},~\eqref{alpha-choice} as the ``adiabatic frame,'' and denote the associated rescalings by $A_{\ad},C_{\ad}$.

Since we have seen that the conditions for adiabaticity are satisfied in the adiabatic frame, we expect that $w$ will approach the ground state following the decay of excited states,
\begin{align}
\label{gs-co}
   w\to \phi_{0}^R&= \frac{1}{\sqrt{2\pi}}\, e^{-\frac{\xi^{2}}{2}}\, ,
\end{align}
and then will remain in the ground state because the evolution is adiabatic. 
In this case, $\pd_{y}w=-\sE_{0}w=0$ by construction. 
Therefore, we can identify $\phi_{0}$ with the scaling distribution $w_{S}$. This identification can also be confirmed by looking at the explicit expression for $w_{S}$~\eqref{w-sol}. 
We also note that the conditions~\eqref{C-cond},~\eqref{scaling-relation} determining  $\As,\Cs$ are the same as those specifying the adiabatic frame~\eqref{q-choice},~\eqref{alpha-choice}.
We therefore conclude that during the adiabatic evolution $A_{\ad},C_{\ad}$ and $\As,\Cs$ coincide and that scaling behavior for the collision integral~\eqref{CIa-0} is an example of adiabatic evolution.

We wish to emphasize the similarities and differences between $\As, \Cs$ and $A_{\ad},C_{\ad}$. 
Only in the scaling regime is it possible to identify $\As,\Cs$ such that the rescaled distribution $w$ becomes time-independent.
On the other hand, the adiabatic frame $A_{\ad},C_{\ad}$ is defined by requiring the evolution of the instantaneous eigenstates of ${\cal H}$ to be as slow as possible. 
Such a frame exists even if $w$ is different from $w_S$. 
Indeed, for a given $q$, the corresponding $C_{\ad}$ can be obtained by solving \eqref{q-choice}, without having to require that the system is in the ground state.

We finally note that one could study the dynamics of the distribution function in different frames, and still solve the same physical problem. 
The advantage of using the adiabatic frame is that this frame reveals the adiabatic nature of the scaling evolution.  
Moreover, in this frame we can conveniently describe how a self-similar evolution for the distribution function arises from a generic initial condition: the ground state, i.e., the scaling distribution, becomes the dominant contribution to the state of the system through the decay of excited states. Explicitly, since $V_{nm}=0$ in the adiabatic frame, the evolution equation for $\ta_n$~\eqref{an} reads
\begin{align}
\label{a-evo}
  \pd_{y} a_{n}=
  -\sE_{n}\, a_{n}=
  - 2n (1-\g) \,a_{n}\, ,
\end{align}
from which it is clear that the excited modes decay as $\sim e^{-2(1-\gamma)ny}$, and only the ground state survives after a transient time. 
This is why the scaling distribution is an attractor of the evolution.

\subsubsection{Time scales for approaching the scaling function and approaching the fixed point} 
\label{sec:time-compare}

Time-dependent scaling, such as that observed in Ref.~\cite{Mazeliauskas:2018yef}, occurs when the time scale for the system to approach the scaling distribution $\tau_S$ is much shorter than that for the exponents to reach their fixed point values $\tau_\text{FP}$.
In this section we wish to understand the condition under which $\tau_S \ll \tau_\text{FP}$.
In this situation, 
the evolution of the distribution function is captured by the evolution of the scaling exponents from $\tau_S$ to $\tau_\text{FP}$.

The analysis in the previous section tells us that distribution will approach the scaling form (ground state) after the damping of excited states. 
Therefore, $\tau_{S}$ is set by the inverse of the energy gap between the ground and excited states in the adiabaticity frame. 
For illustrative purposes we can estimate $\tau_{S}$ from Eq.~\eqref{a-evo}, assuming that we are not very close to the free-streaming limit so that $\gamma$ in the adiabatic frame is not very close to $1$. 
Under this condition, 
the time scale for the decay of the $n^\text{th}$ excited state is $\tau_{I} \exp(\frac{3}{4n})$.
On the other hand, 
Eq.~\eqref{g-analytic} (derived in sec.~\eqref{sec:evo}) 
tells us that deviations from the BMSS fixed point value $\g_{{\rm BMSS}}=1/3$ will decay as $e^{-2y}$, meaning that $y_{\rm FP} \sim 1/2$ or $\tau_{{\rm FP}} \sim \tau_{I}\, \sqrt{e}$.
Therefore, the excited states with sufficiently large $n$ decay at a scale much shorter than $\tau_{\rm FP}$. 
However, low lying excited states decay on a similar time scale to the approach to the fixed point and therefore may defer the emergence of scaling and shorten, or even remove altogether the time-dependent scaling regime.  
A long duration of time-dependent scaling requires a clean separation $\tau_{S}\ll \tau_{{\rm FP}}$, which requires that the contribution from the low-lying excited states in the initial distribution be sufficiently small. 
This is consistent with our numerical observation in sec.~\ref{sec:scaling-num} that with a Gaussian initial condition, scaling starts at early times. Gaussian initial conditions were also used for solving QCD EKT in Ref.~\cite{Mazeliauskas:2018yef}.

\paragraph{Contribution from the first excited state} \hspace{\fill}

Our previous observation notwithstanding, even when the contribution from the first excited state is significant, one can still show that a time-dependent scaling phase exists prior to reaching the fixed point values. Consider a distribution function given by
\begin{equation}
    w = a_0 \phi^{R}_{0} + a_1 \phi^{R}_{1} = \frac{a_0}{\sqrt{2\pi}} e^{-\frac{\xi^2}{2}} \left[ 1 + \delta \frac{\xi^2 - 1}{2} \right] \, ,
\end{equation}
where we have introduced $\delta \equiv a_1/a_0$, representing the relative contribution of the first excited state to the full distribution function.

When $\delta > 1$, there is no reason to expect that any kind of self-similarity will appear in the distribution function. However, if $\delta < 1$, one can manipulate the previous expression into
\begin{equation}
    w = \frac{a_0}{\sqrt{2\pi (1+\delta)}} e^{-\frac{\xi^2}{2(1+\delta)}} + O(\delta^2) \, ,
\end{equation}
from which it is apparent that the full distribution function $f(p_z; \tau) = A\, w(p_z/C ; \tau )$ has a scaling form (at least when $\delta$ is perturbatively small), albeit with a different set of rescaling functions:
\begin{align}
    A &\to A_{\delta} = \frac{A}{\sqrt{1+\delta}} \, , \\
    C &\to C_{\delta} = C \sqrt{1+\delta} \, .
\end{align}
Then, by using that $\partial_y \delta = \partial_y \ta_1 / \ta_0 = - 2(1-\gamma) \delta$, one immediately infers that the distribution function $f$ will exhibit scaling, with exponents given by
\begin{align}
    \alpha_\delta &= \alpha + \delta (1 - \gamma) + O(\delta^2) \, , \\
    \gamma_\delta &= \gamma + \delta (1 - \gamma) + O(\delta^2) \, .
\end{align}
What is perhaps most remarkable about this is that a precise notion of scaling survives up to linear order in $\delta$, which expands the domain of time-dependent scaling phenomena even further. This result guarantees that, at the very least, there will always be a short time-dependent scaling phase once $\delta$ becomes sufficiently small before reaching the attractor.

\subsubsection{
Generalization to isotropic diffusive kernel
\label{sec:with-pT}
}

To complete our discussion on adiabaticity for the FP equation, in this section we will extend the adiabatic analysis in sec.~\eqref{sec:adi-pz} to the collision integral~\eqref{CIa-1}, which includes transverse momentum diffusion. 
The evolution equation for $f$ is now
\begin{equation}
\label{FP_diffusive}
    \pd_{y}f=\le(p_z \pd_{p_z} +q \nabla^{2}_{\bf p}\ri) f\, . 
\end{equation}
Since this equation describes diffusion in transverse momentum, we shall reinstate the $p_T$-dependence of the distribution function.
 From the definition of the scaled distribution function $w$~\eqref{f-w}, Eq.~\eqref{FP_diffusive} can be rewritten as $\partial_y w = - {\cal H} w$, with
\begin{equation}
    \label{H-iso}
    {\cal H} = \alpha - (1-\g) 
  \le[
  \tq\, \pd^{2}_{\xi}+\xi\,\pd_{\xi}
  \ri] + \b  
  \le[
   - \frac{q}{B^2 \b}\,( \pd^{2}_{\zeta} +  \frac{1}{\zeta} \pd_{\zeta}) + \zeta\,\pd_{\zeta}
  \ri]\, ,
\end{equation}
where $\tq$ is defined in Eq.~\eqref{tq-def}.

Analogously to our analysis in sec.~\ref{sec:adi-pz}, 
we choose $A,B,C$ to ensure the adiabatic evolution of the states in this system. 
It is straightforward to show that the appropriate choice is 
\begin{align}
\label{q-general}
   \tq = \frac{q}{C^2 (1-\gamma)}=1\, ,
   \qquad
\tqB \equiv - \frac{q}{B^2 \b}=1 \, .
\end{align}
Furthermore, imposing the condition
\begin{align}
\label{alpha-general}
\alpha = \gamma + 2\beta - 1 \, ,
\end{align}
we can make the ground state energy zero (note that when the distribution is in the scaling regime, this is implied by number conservation).
In this adiabatic frame the eigenvalues of ${\cal H}$ are
\begin{equation}
\label{eigen-iso}
    \sE_{n,m} = 2n (1-\gamma) - 2m \beta \quad \quad n,m = 0,1,2,\ldots \, ,
\end{equation}
which can be verified explicitly by solving for the eigenfunctions of each operator in the square brackets of~\eqref{H-iso}. For the longitudinal part, they are Hermite functions as before, whereas for the transverse part they are given by confluent Hypergeometric functions: 
\begin{align}
    &\, \phi_{n,m}^L = 
{\rm He}_{2n} \! \left(\frac{\xi}{\sqrt{\tq}} \right) {}_1 F_1 \! \left(-2m,1,\frac{\zeta^2}{2 \tqB }\right) \, , \\
  & \, \phi_{n,m}^R= 
\frac{1}{\sqrt{2\pi \tq} (2n)!} \frac{1}{\tqB} \, {\rm He}_{2n}\! \left(\frac{\xi}{\sqrt{\tq}} \right)  {}_1 F_1 \! \left(-2m,1,\frac{\zeta^2}{2 \tqB }\right) e^{-\frac{\xi^{2}}{2\tq}-\frac{\zeta^{2}}{2\tqB}} \, \, .
\end{align}
One can verify that these Hypergeometric functions are actually polynomials and that the states are normalized under the inner product
\begin{equation}
    \int_{-\infty}^\infty \!\!\!\! d\xi \int_0^\infty \!\!\!\! d\zeta \, \zeta \, \phi^L_{n_L,m_L}(\xi,\zeta) \phi^{R}_{n_R,m_R}(\xi,\zeta) = \delta_{n_L, n_R} \delta_{m_L,m_R} \, .
\end{equation} 
For the reasons listed below Eq.~\eqref{BC-ratio}, and in consistency with~\eqref{q-general}, 
we have assumed $\beta\leq 0$. 
With the choice $\tq = \tqB = 1$, the ground state $(n,m)=(0,0)$ is given exactly by
\begin{equation}
\phi_{0,0}^R = \frac{1}{\sqrt{2\pi}} e^{-\frac{\xi^2+\zeta^2}{2}} \, ,
\end{equation}
which coincides with the scaling solution~\eqref{w-sol1} of the same collision integral. 
This again illustrates the connection between adiabaticity and scaling evolution.

Since $\beta$ is assumed to be small, we note that 
the energy gap  $-2 m \beta$ between the ground state $\phi^{R}_{0,0}$ and ``transverse'' excited states $\phi^{R}_{0,m}$ is not particularly large. 
This means that in general, the longitudinal profile approaches the Gaussian form much earlier than the transverse profile does. 
Physically, this difference means that the longitudinal expansion changes the longitudinal momentum distribution rather rapidly. 
Applying the argument presented in sec.~\ref{sec:time-compare}, we conclude that for the transverse profile to exhibit scaling with a universal distribution form $w_S$ prior to approaching the fixed point, the initial transverse distribution should be close to a Gaussian, because deviations from Gaussianity (i.e., from excited states) would typically be long-lived.

\subsection{The evolution of scaling exponents
\label{sec:evo}
}

An important implication of scaling is that it simplifies the description of the gluon plasma evolution far from equilibrium. 
Once the scaling function is given, the evolution in the scaling regime can be described by that of time-dependent scaling exponents $\als,\betas,\gas$. In this section we derive evolution equations for scaling exponents from the same conditions that ensure adiabaticity for the collision kernel~\eqref{CIa-1}.
As shown below, the resulting equations lead to various fixed points, and provide a precise description of the flow between those fixed points.

\subsubsection{Deriving evolution equations
\label{sec:evo-eq}
}

In the previous section, we have demonstrated that one can choose $A,B,C$ (and consequently $\a,\b,\g$) such that the scaling distribution $w_{S}$ is the ground state of the Hamiltonian ${\cal H}$ that describes the evolution of $w$ with zero eigenvalue. 
This leads to the self-consistency conditions~\eqref{q-general}, \eqref{alpha-general}, which in the scaling regime become
\begin{align}
\label{alpha-relation}
\als &=-2\betas-\gas-1
\\
  \label{beta-q}
  -\betas &= \frac{\qs}{\Bs^{2}}, 
  \\
\label{gamma-q}
  -\gas &= -1 + \frac{\qs}{\Cs^{2}}\, , 
\end{align}
where we have denoted $q$ evaluated on the scaling distribution by $q_S$. Alternatively, these consistency conditions can be obtained by inspecting~\eqref{H-iso} for the requirements on $\beta,\gamma$ such that $w$ can be time-independent.

Before continuing, let us pause to develop some physical intuition for eqs.~\eqref{beta-q} and \eqref{gamma-q}. 
Following Ref.~\cite{Mueller:1999pi}, we consider the following phenomenological equation describing the temporal evolution of the average longitudinal momentum for a system undergoing Bjorken expansion 
\begin{align}
\label{pz2-evo}
  \pd_{y}\langle p^{2}_{z}\rangle = - 2 \langle p^{2}_{z}\rangle + 2 D\, .
\end{align}
The average over the phase space weighted by the distribution $\langle\ldots\rangle$ is defined in Eq.~\eqref{average}. 
The first and second terms on the right hand side of Eq.~\eqref{pz2-evo} account for the effects of the longitudinal expansion and diffusion in momentum space with diffusive constant $D$, respectively. 
In the scaling regime, we further have $\langle p_z^2 \rangle = c_0 \Cs^2$ where $c_{0}$ is a constant of order one. Using the definition of $\g$ in Eq.~\eqref{exp-def}, Eq.~\eqref{pz2-evo} becomes
\begin{align}
\label{gamma-D}
  -\gas = -1 + \frac{D}{c_{0}\Cs^{2}}\, , 
\end{align}
which is equivalent to Eq.~\eqref{gamma-q} with $D\propto q_S$.

The physical interpretation of Eq.~\eqref{gamma-q} is now quite clear. 
Recalling that $\g$ is the rate of change of the characteristic longitudinal momentum $C$,
Eq.~\eqref{gamma-D} indicates that it is given by the combined effects of longitudinal expansion and momentum diffusion. 
Eq.~\eqref{beta-q} can be interpreted similarly in term of transverse momentum diffusion.

We can write down evolution equations for the scaling exponents by differentiating eqs.~\eqref{beta-q} and~\eqref{gamma-q} with respect to $y$:
\bes
 \label{evo}
\begin{align}
   \pd_{y}\betas&= (\dot{q}_{S}+2\betas)\betas\, ,
   \\
   \pd_{y}\gas&= -(\dot{q}_{S}+2\gas)(1-\gas)\, . 
\end{align}
\ees
The evolution of $\als$ is determined from that of $\betas,\gas$ by Eq.~\eqref{alpha-relation}.

To close the system of equations~\eqref{evo}, we need to express $\dot{q}_{S}$ in terms of $\betas,\gas$. Substituting the scaling form~\eqref{prescaling-1} for the distribution function into the definition of $q$~\eqref{q-def} yields
\begin{align}
\label{q1}
  \qs= \lambda_0\, \lcb \le(
  c_{a}\tau \As^{2}\Bs^{2}\Cs + \tau n \ri)\, , 
\end{align}
where the time-independent constant $c_{a}$ is given by
\begin{align}
  c_{a}&=\int^{\infty}_{0}\, \frac{d\zeta}{2\pi}\,\zeta \int^{\infty}_{-\infty}\frac{d\xi}{2\pi}\, w^2_{S}(\xi,\zeta)\, .
\end{align}
Using the fact that $\tau n$ is constant from Eq.~\eqref{n-evo} along with Eq.~\eqref{alpha-relation}, we find
\begin{align}
  \frac{\pd_{y}\le(\tau n + \tau c_{a}\As^{2}\Bs^{2}\Cs\ri)}{\le(\tau n + \tau c_{a}\As^{2}\Bs^{2}\Cs\ri)}
=  \frac{(-1+2\betas+\gas)\tau c_{a}\As^{2}\Bs^{2}\Cs}{\le(\tau_{I} n_{I} + \tau c_{a}\As^{2}\Bs^{2}\Cs\ri)} \, .
\end{align}
As a result, the rate of change of $\qs$ from Eq.~\eqref{q1} can be written as
\begin{align}
 \label{q-evo}
  \dot{q}_{S}= \frac{(-1+2\betas+\gas)\tau c_{a}\As^{2}\Bs^{2}\Cs}{\le(\tau_{I} n_{I} + \tau c_{a}\As^{2}\Bs^{2}\Cs\ri)}+\lcbdot\, .
 \end{align}
In sec.~\ref{sec:a-dim}, we shall derive an explicit expression for $\lcbdot$ (see Eq.~\eqref{lcb-evo}). 
Since both $\dot{q}_S$ and $\lcbdot$ depend explicitly on $y$ and $\As,\Bs,\Cs$, 
eqs.~\eqref{evo},~\eqref{q-evo}, and ~\eqref{lcb-evo} (together with the relation Eq.~\eqref{exp-def}) form a set of closed equations that can be solved for the evolution of the scaling exponents. The solution to these equations (shown in sec.~\ref{sec:exponent-evo}) is the main result of this section.

However, we find it instructive to first consider two limiting cases where the evolution equations~\eqref{evo} are simplified. 
In the first limit, 
we shall assume the distribution is highly-occupied, $\As\gg 1$. 
Since $n\propto \As\Bs^{2}\Cs$ we can neglect the second term in Eq.~\eqref{q1} to obtain
\begin{align}
\label{q2}
  \qs\approx \lambda_0\, \lcb\, c_{a} \tau \As^{2}\Bs^{2}\Cs\, . 
\end{align}
Then $\dot{q}_S$ in Eq.~\eqref{q-evo} reduces to
\begin{align}
\label{D-evo-1}
  \dot{q}_{S}= -1+2\betas +\gas +\lcbdot\, 
\end{align}
and the evolution equation~\eqref{evo} takes the form
\bes
\label{evo-dense}
\begin{align}
   \textrm{over-occupied ($A_S\gg1$): }\qquad\, \pd_{y}\betas &= \le(\gas+4\betas-1+\lcbdot\ri)\betas \, ,
   \\
   \pd_{y}\gas &= \le(3\gas + 2 \betas -1+\lcbdot\ri)(\gas-1) \, .
 \end{align}
 \ees
In the opposite regime,  we consider a very dilute distribution $\As\ll 1$. 
In this case, the dominant contribution to $\qs$ is from the second term in Eq.~\eqref{q1},
\begin{align}
  \qs\approx \lambda_0 \lcb \tau n\, , 
\end{align}
meaning $\dot{q}_{S} = \lcbdot$ since $\tau n$ is time-independent.
Then, we can write Eq.~\eqref{evo} as
\bes
\label{evo-dilute}
\begin{align}
   \textrm{dilute ($A_S\ll1$): }\qquad\,\pd_{y}\betas &= \le(2\betas+\lcbdot\ri)\betas \, .
   \\
   \pd_{y}\gas &= (2\gas+\lcbdot)(\gas-1) \, .
 \end{align}
 \ees
These simplified evolution equations~\eqref{evo-dense} and~\eqref{evo-dilute} will be used in the next section to discuss the fixed points of the scaling evolution.

Finally, we emphasize that the evolution equations are derived by assuming the simplified collision integral~\eqref{CIa-1}. 
As we have argued in Appendix~\ref{app:Ib}, 
this simplification describes well the scaling evolution of hard gluons with $A \geq 1$. 
In this sense, we should be cautious when applying those equations to a dilute system with $A\leq 1$. 
Nevertheless, we notice numerically in sec.~\ref{sec:scaling-num} that scaling exponents extracted using Eq.~\eqref{CIa-1} agree well with those from solving the full FP equation even near the dilute fixed point.
We therefore expect that the evolution equations shown here be able to describe scaling in the dilute regime, at least qualitatively.

\subsubsection{Fixed points
\label{sec:fixed-point}
}

Before solving the evolution equations~\eqref{evo}, let us first identify the possible (non-thermal) fixed points, which correspond to the values of exponents $\betas,\gas$ such that the right hand side of Eq.~\eqref{evo} vanishes. 
These fixed points play an important role in characterizing the scaling evolution. 
We will first assume that $\lcbdot =0$, and later in this subsection illustrate the qualitative implications of a non-zero but constant $\lcbdot$. At the end of this subsection we will derive self-consistent equations for $\lcbdot$, which we will later put to use in subsection~\ref{sec:exponent-evo}.

We begin our discussion by considering perhaps the simplest possibility
\begin{align}
\label{FS-FP}
  &\textrm{Free streaming:}\qquad\, 
  (\als,\betas,\gas)=(0,0,1)\, . 
\end{align}
These exponents automatically make the right hand side of Eq.~\eqref{evo} vanish and characterize the free streaming fixed point.
Indeed, 
 \begin{align}
 \label{FS-sol}
    f_{{\rm F.S.}}(\pT,\pz;\tau)= f_{I}(\pT,\left(\frac{\tau}{\tau_{I}} \right)\pz)
\end{align}
solves the Boltzmann equation~\eqref{kin} in the collisionless limit for a generic initial condition
$f(\pT,\pz;\tau=\tau_{I})= f_{I}(\pT,\pz)$.
From the free-streaming solution~\eqref{FS-sol}, 
we can read the corresponding exponents~\eqref{FS-FP} directly.

Next, 
we consider the case with $\gas< 1, \betas =0$.
We note from \eqref{beta-q} that since $\qs$ is finite, 
when we say $\betas=0$ we mean $\qs\ll \Bs^{2}$.
When the typical occupancy is large $\As \gg 1$, 
we can use Eq.~\eqref{evo-dense}, which reproduces the BMSS fixed point~\cite{Baier:2000sb} in the absence of $\lcbdot$
\begin{align}
\label{BMSS}
  &\,\textrm{BMSS:}\qquad\, 
  (\als,\betas,\gas)=(-2/3,0,1/3)\, . 
\end{align}
In fact,
for $\gas(y=0)=\g_{I}$ and $\betas=0,\lcbdot=0$, 
we can solve Eq.~\eqref{evo-dense} analytically
\begin{align}
\label{g-analytic}
    \gas = \frac{(\g_{I}-1)+e^{-2y}(1-3\g_{I})}{3(\g_{I}-1)+e^{-2y}(1-3\g_{I})}\, . 
\end{align}
For sufficiently large $y$, 
$\gas$ will flow from $\g_{I}$ to the BMSS fixed point value $1/3$. 
The only exception to this would be if $\gas$ starts at the unstable fixed point $\gamma_I = 1$, in which case the solution would stay there forever. Dynamically, however, the original evolution equation for $\gas$~\eqref{C-cond} sets $\gas <1$ always, and therefore the system always flows to the BMSS fixed point in the regime $f \gg 1$.

Finally, we turn to the situation where the system becomes dilute during its expansion, $\As\ll 1$. 
We then read the third fixed point from Eq.~\eqref{evo-dilute}: 
\begin{align}
\label{dilute}
&\,\textrm{Dilute:}\qquad \gas=\betas=0\, 
. 
\end{align}
In this limit, the solution to Eq.~\eqref{evo-dilute} reads
\begin{align}
    \betas = -\frac{1}{2y - \beta_I^{-1} } \, , & & \gas = \frac{1}{ 1 + \left( \gamma_I^{-1} - 1 \right) e^{2y} }\, ,
\end{align}
which approaches $(\betas, \gas)=(0,0)$ at late times.

The careful reader might ask if imposing the condition $2\betas = -{\dot q}_{S}$ leads to additional fixed points, but it does not, provided $\lcbdot = 0$.
In the limit $\As \gg 1$ with constant $\lcb$, both $-1+\gas$ and $\betas$ are negative by virtue of the consistency conditions~\eqref{beta-q},~\eqref{gamma-q} while ~Eq.~\eqref{D-evo-1} implies that $\dot{q}_S < 0$, meaning there is no solution to $2\betas=-\dot{q}_S$. 
In the dilute regime 
$\dot{q}_S\sim 0$, and so $2\betas + \dot{q}_S = 0$ reduces to $\betas = 0$.

From the possible fixed points discussed above, we anticipate three possible scenarios during the far-from-equilibrium stage of the evolution:  
\begin{enumerate}
\item Scenario~I: The expanding plasma evolves from the free streaming fixed point to the BMSS fixed point. 
After that, thermalization occurs. 
This scenario has been discussed extensively in the literature.  

However, because of the presence of the dilute fixed point~\eqref{dilute}, there are two additional possibilities. 
 
\item Scenario~II: Scaling exponents first approach the BMSS fixed point, and then move to the dilute fixed point.

 \item Scenario~III: The exponents are not attracted to the BMSS fixed point, but transit directly to the dilute fixed point. 
\end{enumerate}

These scenarios are consistent with what we observed in the numerical solutions to the FP equation in sec.~\ref{sec:scaling-num}.

The key new finding in this section is the identification of the dilute fixed point~\eqref{dilute}. 
The presence of this new fixed point leads to two additional scenarios in the far-from-equilibrium evolution, namely Scenarios II and III described above. 
To appreciate the underlying physics, we inspect the relation between $\betas,\gas$ and momentum diffusion rate $\qs$~\eqref{beta-q}, \eqref{gamma-q}. 
The vanishing of $\gas$ around this fixed point means the characteristic longitudinal momentum $\Cs$ approaches a constant value, implying that the change of the typical longitudinal momentum due to the expansion is balanced by the momentum diffusion $\qs$. 
On the other hand, the diffusion of transverse momentum is still small compared with its typical value $\Bs$ so that $\betas\to 0$.

For the dilute fixed point to be realized, the typical occupancy number should become small before thermalization. 
Since the occupancy number is characterized by $\As$, 
we estimate the time scale at which the system becomes dilute by $\As(\tau_{{\rm di}}) \sim 1$. Using the relation between $\As$ and $\als$ in Eq.~\eqref{exp-def} and estimating $\als \sim -1$ gives
 \begin{equation}
 \label{y-di}
   \tau_{{\rm di}}\sim\tau_{I} A_{I}\, ,
 \end{equation} 
indicating that $\tau_{{\rm di}}$ becomes shorter with smaller occupancy.
 Parametrically, we can take $\tau_I Q_s$ to be of order one and consequently $Q_{s}\tau_{{\rm di}}\sim A_I =\sigma_0/g_s^{2}$.
The thermalization time in the FP equation is parametrically $Q_{s}\tau_{{\rm th}}\sim \exp(1/g_s^2)$~\cite{Bjoraker:2000cf}. 
Comparing the two, we anticipate that there is a range of $g_s$ for which $\tau_{{\rm di}}< \tau_{{\rm th}}$ so that Scenario~II and III would occur. 
This expectation has been confirmed numerically in Fig.~\ref{fig:FP-exp}. 
For $(g_s,\sigma_0)=(0.1,0.6)$ as in the middle panel of Fig.~\ref{fig:FP-exp}, $\tau_\text{di}\sim 60\tau_I$ is in good correspondence to the time scale when the exponents turn toward the dilute fixed point. 
For $(g_s,\sigma_0)=(1/3,0.1)$, as in the right panel of Fig.~\ref{fig:FP-exp}, $\tau_\text{di} \sim \tau_I$ and there is no approach to the BMSS fixed point.
We note, however, that the dominant thermalization processes in the FP equation and QCD EKT are different, and the thermalization time in the latter theory is parametrically shorter. 
Therefore, it would be interesting to examine if Scenario~II and III are relevant for QCD EKT. We leave this as an open question for future investigation.

We now turn to discussing the effect of $\lcbdot$ on the fixed points. 
We shall first discuss the modifications to the BMSS fixed point. 
For this discussion, it is sufficient to set $\betas=0$ and use Eq.~\eqref{evo-dense} to find that
\begin{align}
\label{g-lcb}
  \gas = \frac{1}{3}\le(1-\lcbdot\ri)\, ,
\end{align}
which clearly indicates that the scaling exponent $\gas$ differs from the BMSS value $1/3$ due to $\lcbdot$. We interpret the contribution from $\lcbdot$ as an ``anomalous dimension" correction to BMSS scaling exponents, in analogy with the fact that the renormalization group flow in field theories can generate an ``anomalous'' correction to the scaling exponents of correlation functions. 
Remarkably, we will see in the coming section (see Eq.~\eqref{delta-gamma}) that this anomalous dimension does depend on the initial values of $A, B, C$, in contrast to the BMSS fixed point exponents which are independent of the initial conditions.\footnote{
According to the general theory of self-similar evolution developed by Barenblatt, 
the situation that scaling exponents are not fully fixed by dimensional analysis but depend on initial conditions is referred to as self-similarity of the second kind~\cite{barenblatt1996scaling}.
The anomalous dimension correction observed in this work fits into this classification.
See also Ref.~\cite{PhysRevLett.64.1361} for an example of the emergence of an anomalous dimension in non-linear diffusive processes.}

Following similar steps, we obtain the effects of $\lcbdot$ on scaling exponents near the dilute fixed point. In this case,
we see from \eqref{evo-dilute} that the presence of $\lcbdot$ also introduces an anomalous dimension correction to the dilute fixed point
\begin{equation}
\label{delta-gamma-dilute}
    \gas = \betas = - \frac12 \lcbdot.
\end{equation}
The fixed point with $(\betas, \gas) = (0, - \lcbdot/2)$ is also possible, but is unstable under time evolution.

\begin{figure}
    \centering
    \includegraphics[width=0.45\textwidth]{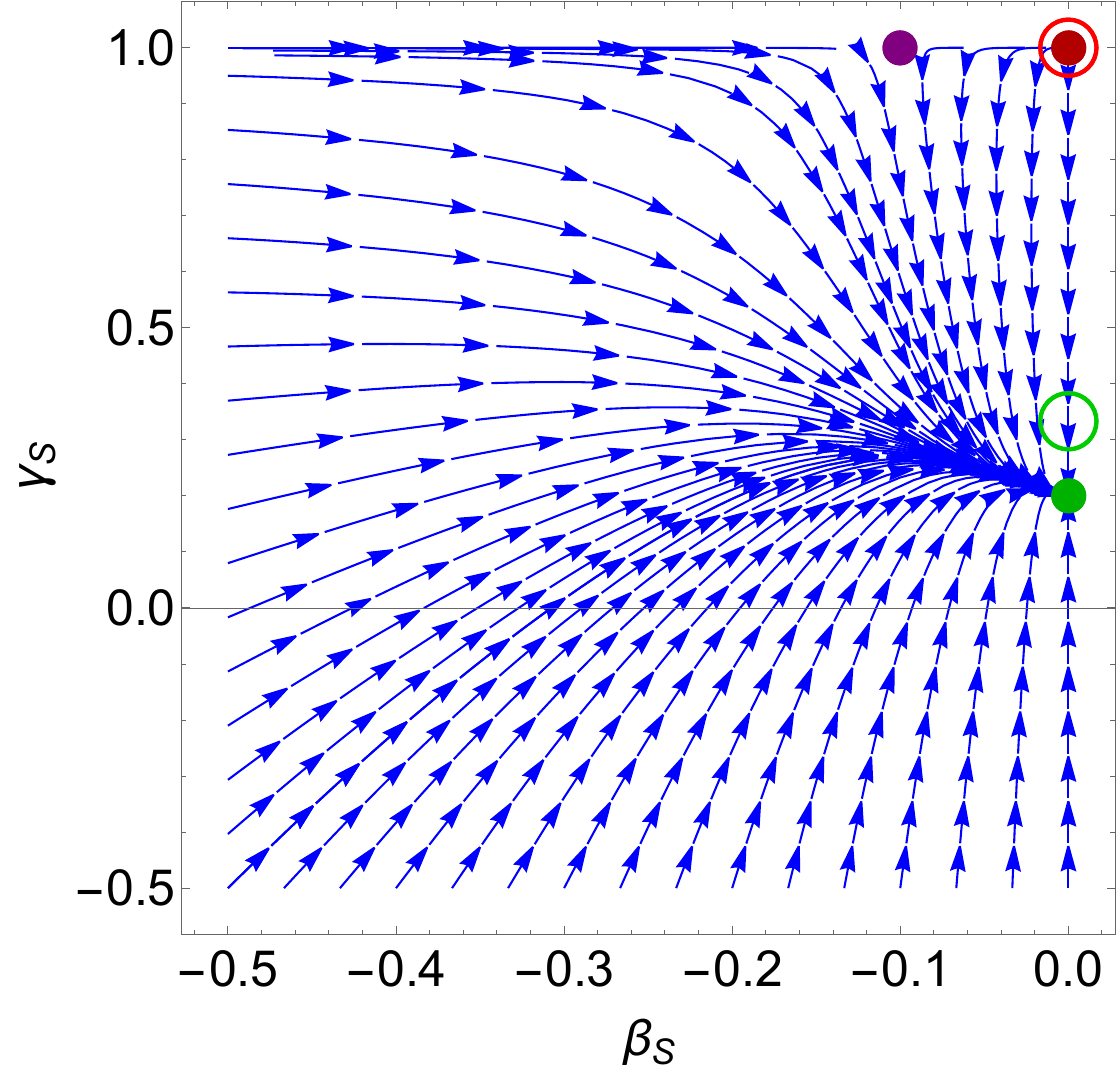} \hspace{0.03\textwidth}
    \includegraphics[width=0.45\textwidth]{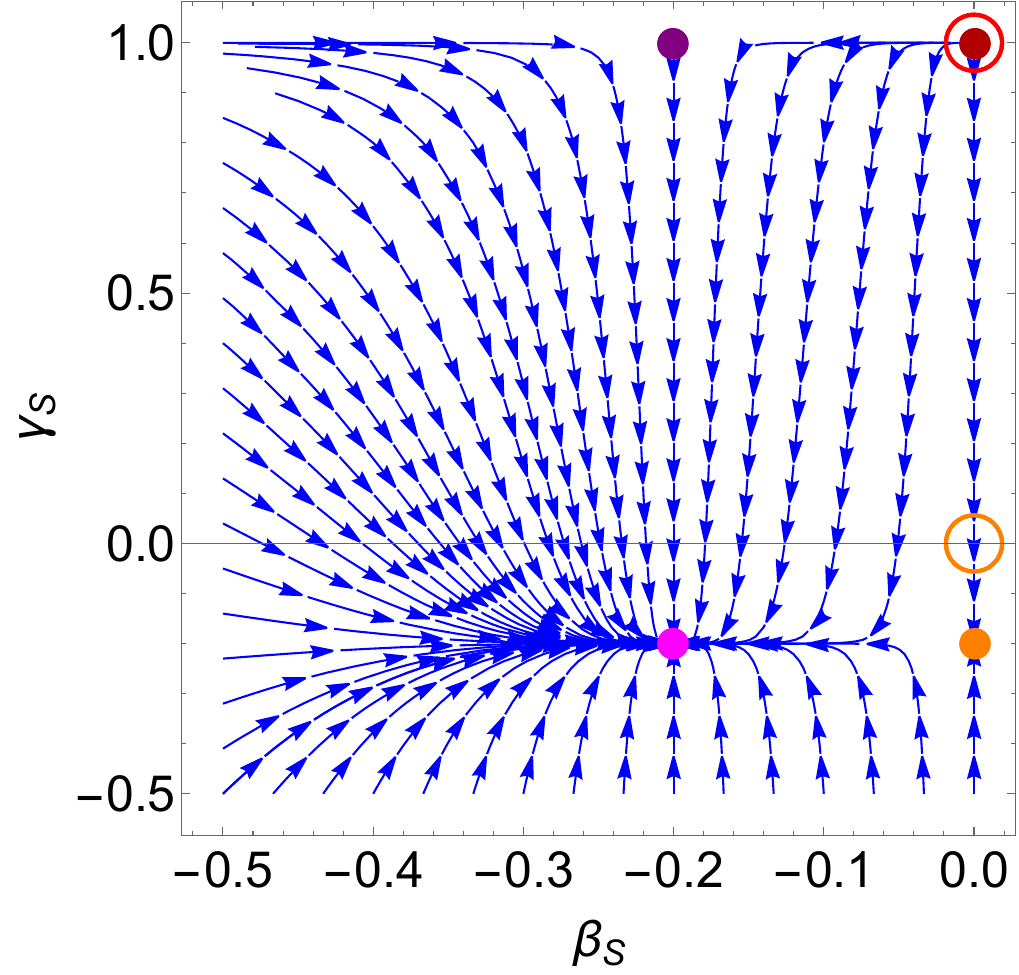}
    \caption{Stream flow of the scaling exponents. Blue arrows represent the flow of the scaling exponents $\betas,\gas$ under time evolution. (left) $f\gg 1$, (right) $f\ll 1$. For illustrative purposes, we set $\lcbdot = 0.4$ and show the corresponding fixed points in filled circles. Fixed points of the evolution equations with $\lcbdot = 0$ are shown as open circles.  Red and purple markers show the free-streaming fixed point with the ``anomalous'' correction and the one without the ``anomalous'' correction, respectively. Green markers show the BMSS fixed point. The orange and pink markers show the dilute fixed point with the anomalous correction in both $\betas$ and $\gas$ and the one with only the ``anomalous'' correction in $\gas$, respectively. 
    }
    \label{fig:flow}
\end{figure}

To summarize this section, we show in Fig.~\ref{fig:flow} the fixed points and flow of exponents in the $(\betas, \gas)$ plane. Though $\lcbdot$ is generally time-dependent, for illustrative purposes we take it to be constant, here fixed to $\lcbdot = 0.4$ for visual clarity. For comparison, we show the fixed points with $\lcbdot = 0$ in open circles.
The left panel shows the overoccupied case, where $f \gg 1$. Here, at early times (earlier in the time evolution flow), the free-streaming fixed point with the ``anomalous'' correction is preferred over the ``non-anomalous'' one (which has no $\lcbdot$-dependent corrections), in the sense that flow lines between them go from the ``non-anomalous'' fixed point towards the ``anomalous'' fixed point. At late times, the exponents flow to the BMSS fixed point, which also includes an anomalous correction due to $\lcbdot$ (albeit that this fixed point has no ``non-anomalous'' counterpart). 
On the other hand, the right panel shows the dilute case, with $f \ll 1$. In this situation, at early times, the free-streaming fixed point with the ``anomalous'' correction is again dynamically preferred over the ``non-anomalous'' one. At late times, the exponents flow to the dilute fixed point that includes the anomalous correction due to $\lcbdot$ for both $\beta$ and $\gamma$.

Therefore, we see that the effects of $\lcbdot \neq 0$ are qualitatively relevant to properly understand the exponents near the stable, attractive fixed points. Hence, a more detailed investigation into the consequences of having a nonzero $\lcbdot$ is warranted.

\paragraph{The Coulomb logarithm} \hspace{\fill}
\label{sec:a-dim}

We will now obtain an explicit expression for $\lcbdot$. We assume the scaling function $w_{S}$ takes the Gaussian form~\eqref{w-sol1} and find
 \begin{align}
 \label{mD-in-ABC}
    m_D^2 &= 4 N_{c} g^2 c_{b}(r_{S}) \As \Bs \Cs\, ,
 \qquad
 \langle\pT^{2}\rangle=2\Bs^{2}
  \, ,
 \end{align}
 where $r_{S}=\frac{\Cs}{\Bs}$, and 
 \begin{align}
 \label{cb-vs-r}
   c_{b}(r_{S})&=
   \int^{\infty}_{0}\, \frac{d\zeta}{2\pi}\,\zeta \int^{\infty}_{-\infty}\,\frac{d\xi}{2\pi} \frac{1}{\sqrt{\zeta^2+r^{2}_{S}\xi^2}}
  w_{S}(\zeta,\xi)\, 
  =  \frac{1}{2\pi^{2}}\, \frac{\arccos(r_{S})}{\sqrt{1-r^{2}_{S}}}\,. 
 \end{align}
Using the definition~\eqref{lcb-0}, we now have
 \begin{align}
 \label{lcb-1}
   \lcb=
   \log \left( \frac{\sqrt{\langle p^{2}_{\perp}\rangle}}{m_{D}} \right)
   =
   \frac{1}{2}\log \le[
   \frac{\Bs}{2 N_{c}g^2_{s} c_{b}(r_{S}) \As\Cs}\, 
   \ri]\, 
 \end{align}
  which in turn gives
 \begin{align}
 \label{lcb-evo}
\lcbdot=\frac{1}{2 \lcb}\le[1-3\betas - \frac{c'_{b}(r_{S})r_{S}}{c_{b}(r_{S})}(\betas-\gas)\ri]\, .
 \end{align}

To obtain a more explicit expression for Eq.~\eqref{g-lcb}, 
we use that $\tau \As \Bs^2 \Cs = \tau_I A_I B_I^2 C_I$ is time-independent due to Eq.~\eqref{scaling-relation}, and we take $B \approx B_I$ to be approximately constant. This is arbitrarily accurate near the BMSS fixed point, since $\betas=0$ there, and is a reasonable approximation near the dilute fixed point, up to $\lcb$-dependent corrections (because there we have $\betas = -\lcbdot/2$). We get
\begin{align}
  \frac{\Bs}{\As\Cs}=
  \frac{e^{y}}{A_{I}}\, \frac{B_{I}}{C_{I}}\, .
\end{align}
The argument of the $\log$ in Eq.~\eqref{lcb-1} now becomes 
\begin{align}
  \frac{\Bs}{2 N_{c}g^2c_{b}(r_{S}) \As\Cs}
  \approx \frac{1}{ 2 N_{c} g^2_{s} c_{b}(r_{S}) }\frac{e^{y}}{A_{I}}\, \frac{B_{I}}{C_{I}}\, ,
\end{align}
so that~Eq.~\eqref{lcb-1} gives
\begin{align}
\label{lcb-2}
  \lcb \approx \frac{1}{2}\le[
  y+ \log \le(\frac{B_{I}}{2 N_{c} g^{2}_{s} c_b(r_{S}) A_{I} C_{I}}\ri) 
  \ri]
 \approx
  \frac{1}{2}y + \lcb^I\, ,
\end{align}
where $\lcb^I$ is the value of $\lcb$ at $y=0$,
\begin{align}
\label{lcb-I}
\lcb^I \approx \frac{1}{2}\, \log \le(\frac{ 2\pi   }{g^{2}_{s} N_{c}A_{I}}\, \frac{B_{I}}{C_{I}}\ri)\, . 
\end{align}
We have assumed $r_{S}\to 0$ so that $c_{b}\approx c_{b}(0)=1/(4\pi)$ does not evolve in time. 
Therefore
\begin{equation}
    \lcbdot = \frac{1}{2 \lcb}
\end{equation}
and the correction to the BMSS value now reads
\begin{align}
\label{delta-gamma}
  \gas- \frac{1}{3}=-\frac{1}{3}\lcbdot\approx
  -\frac{1}{3\le(y+2 \lcb^{I}\ri)}\, . 
\end{align}
As noted above, it is remarkable that, unlike the BMSS fixed point exponents, the anomalous dimension in Eq.~\eqref{delta-gamma} depends on the initial values of $A, B, C$ through its dependence on $\lcb^{I}$.

\subsubsection{Solutions 
\label{sec:exponent-evo}
}

In this section, we shall showcase the solutions to Eq.~\eqref{evo}, with $\dot{q}_S$ and $\lcbdot$ given by eqs.~\eqref{q-evo} and \eqref{lcb-evo}, respectively.
Our goal is to illustrate the three different scenarios for the temporal behavior of the scaling exponents described in sec.~\ref{sec:fixed-point} and the impact of the time evolution of $\lcb$ on the fixed points.

To solve Eq.~\eqref{evo}, we specify initial conditions by matching the scaling form of the distribution Eq.~\eqref{f-w} with the initial condition \eqref{fI} for $\xi_{0}=2$ by choosing $A_{I}=\sigma_{0}/g^{2}_{s}, B_{I}=Q_{s}/\sqrt{2}, C_{I}=Q_{s}/(2\sqrt{2})$. 
The initial values of the exponents $\gamma_I, \beta_I$ are fixed by the consistency conditions~\eqref{gamma-q} and~\eqref{beta-q}, with $q_{S}$ evaluated using~\eqref{q1}.
With $\sigma_0$ fixed, the typical occupation number is controlled entirely by the coupling constant $g_s$. 
Therefore, we anticipate that the transition from Scenario~I to Scenario~II and then to Scenario~III occurs by increasing $g_{s}$.

\begin{figure}[t]
    \centering
        \includegraphics[width=.49\textwidth]{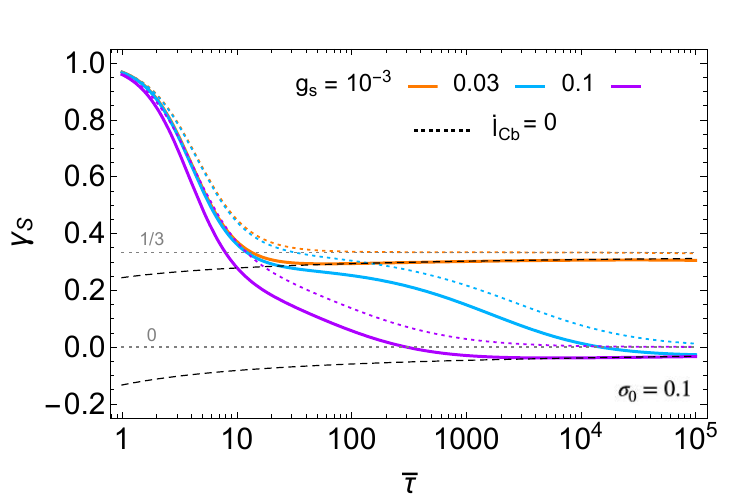}
        \includegraphics[width=.49\textwidth]{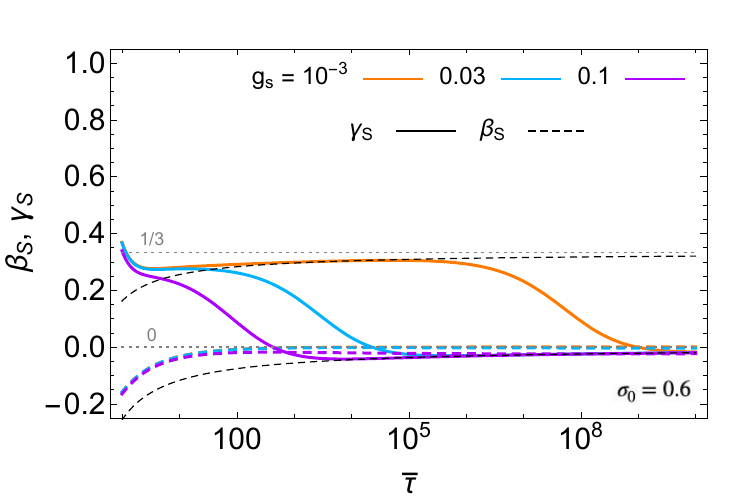}
    \caption{
    \label{fig:scenario}
Evolution of scaling exponents for solutions to Eq.~\eqref{evo} for $\gas$ with representative values of the coupling constant $g_s=10^{-3}$ (orange), $0.03$ (blue), and $0.1$ (purple) are shown in solid lines, for $\sigma_0=0.1$ (left) and $\sigma_0=0.6$ (right). 
The evolution of $\betas$ is shown by colored dashed lines in the right panel ($\betas=0$ in the left panel).
In the left panel, colored dotted lines show solutions with $\lcbdot=0$ for the same set of $g_s$. 
Thin dashed black lines show results for the fixed points including anomalous dimension corrections from eqs.~\eqref{delta-gamma-dilute} and~\eqref{delta-gamma}. 
}
\end{figure}

In Fig.~\ref{fig:scenario}, we show the evolution of the scaling exponents as a function of time for $\sigma_0=0.1$ (left) and $\sigma_0=0.6$ (right), for a range of couplings $g_s$ (indicated by solid colored curves). 
In the left panel we show the evolution of $\gas$ from Eq.~\eqref{evo} with $\lcbdot$ given by Eq.~\eqref{lcb-evo} and $\sigma_0=0.1$. 
We show only $\gas$ since $|\betas| \lesssim 10^{-3}$ and $\als$ is given by~Eq.~\eqref{scaling-relation}. 
For this very small coupling $g_s=10^{-3}$, the scaling exponents approach the BMSS fixed point as in Scenario~I. 
For an intermediate value of the coupling $g_s=0.03$, $\gas$ spends a short time near the BMSS fixed point before transiting to the dilute fixed point as in Scenario~II. 
For larger couplings $g_s=0.1$, $\gas$ goes directly to the dilute fixed point as expected from Scenario~III. 
Therefore we confirm the three scenarios anticipated in the previous qualitative analysis.

It is noteworthy that the late-time values of the exponents at the fixed points are visibly different from the values anticipated in eqs.~\eqref{BMSS} and ~\eqref{dilute}, which are derived by assuming a constant Coulomb logarithm. 
To understand the origin of this deviation, we also show solutions to~Eq.~\eqref{evo} with $\lcbdot=0$ in dotted colored curves.
When $\lcbdot=0$ we see that the asymptotic values of the exponents agree exactly with eqs.~\eqref{BMSS} and ~\eqref{dilute}, thus confirming that the deviation arises from the time evolution of $\lcb$.
Indeed, the modification of the asymptotic values of $\gamma_S$ is quantitatively well-described by Eq.~\eqref{delta-gamma}, which is shown in thin dashed lines.

%
%
%
%

In the right panel of Fig.~\ref{fig:scenario} we show the evolution of $\betas$ and $\gas$ for $\sigma_0=0.6$. 
The evolution of $\gas$ is again shown in solid colored lines and the evolution of $\betas$ is shown in dashed colored lines. 
In this case, $\betas$ can be a few percent, but this non-zero value of $\betas$ has a small impact on the evolution of $\gas$. 
We note that we show a larger time interval in the right panel than we did in the left. 
On this longer timescale, we see that $g_s=10^{-3}$ eventually transits from the BMSS fixed point to the dilute fixed point, as expected since $\tau_\text{di}\sim (\sigma_{0}/g_s^{2})\tau_{I}\sim 10^{5} \tau_{I}$.
In addition to the modification of the fixed point for $\gas$ discussed in the previous paragraph, we also see that the fixed point for $\betas$ is modified from $0$.
The fixed points for $\gas$ are quantitatively described by Eq.~\eqref{delta-gamma} in both the left and right panels of Fig.~\ref{fig:scenario}. For $g_s=0.1$, the late-time fixed point for $\betas$ in the right panel agrees quantitatively with ~Eq.~\eqref{delta-gamma-dilute}. 
Since $\betas$ is close to zero, we note that it can take a long time for the fixed point to be reached.
We anticipate that at later times, $\gas=\betas$ would also be realized for the smaller couplings in the right panel of Fig.~\ref{fig:scenario}.

\subsubsection{Comparison to solutions of kinetic theory
\label{sec:kinetic}
}

\begin{figure}[t]
  \centering
  \includegraphics[width=.49\textwidth]{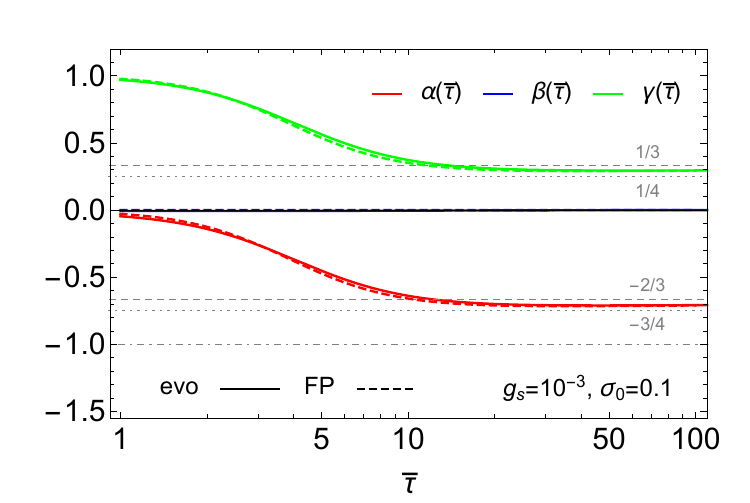}
    \includegraphics[width=.49\textwidth]{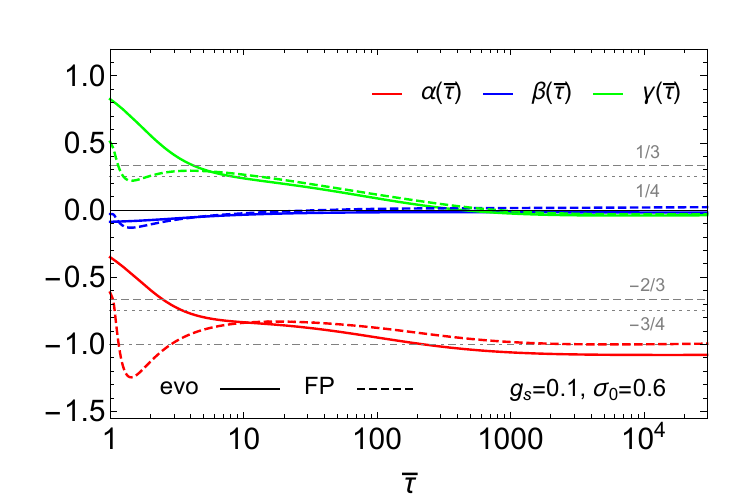}
  \caption{
  \label{fig:FP}
  We compare the evolution of the scaling exponents from Eq.~\eqref{evo} (solid curves) with results from 
  the FP equation (dashed).
  In the left panel we take the same initial distribution function for both the evolution equations and the FP equation
  at $\tau_I$. In the right panel we specify initial conditions for~Eq.~\eqref{evo} at $\bar{\tau}_S=3.1$ (see text for details), corresponding to the approximate time for scaling (see the middle panel of Fig.~\ref{fig:FP-exp}). For clarity of presentation, in both panels the dashed curves are the average of exponents computed from different sets of moments of the distribution function.
  }
\end{figure}

Finally, we compare the evolution of scaling exponents obtained from Eq.~\eqref{evo} to those extracted from full solutions to kinetic theory.
In Fig.~\ref{fig:FP} we compare to solutions of the FP equations with two different combinations of $\le(\s_{0},g_{s}\ri)=
\le(10^{-3},0.1\ri),\le(0.1,0.6\ri)$. 
These FP results have already been presented in Fig.~\ref{fig:FP-exp} (left) and (middle), and are reproduced in Fig.~\eqref{fig:FP}.
We first note that the solutions to Eq.~\eqref{evo} are indistinguishable from the curves for $I_b=0$ in Fig.~\ref{fig:FP-exp} with the same initial conditions for the distribution function, so these are not shown. 
We emphasize that the evolution equations~\eqref{evo} only apply to the evolution in the scaling regime.

For $(g_s,\sigma_0)=(10^{-3},0.1)$, we see from Fig.~\ref{fig:FP-exp} (left) that the distribution function is approximately scaling from $\tau_I$. In this case we can compute initial conditions for Eq.~\eqref{evo} at $\tau_I$ directly from the initial distribution~\eqref{fI}. The results are shown in Fig.~\ref{fig:FP} (left). We observe remarkable agreement between the results from solving Eq.~\eqref{evo} and from numerically solving the FP equation. However, for a distribution function that is not initially scaling, in general we should specify initial conditions for Eq.~\eqref{evo} after the distribution function has taken the scaling form. This is the case for $(g_s,\sigma_0)=(0.1,0.6)$ where we see substantial deviations from scaling at early times in Fig.~\ref{fig:FP-exp} (middle). We estimate the time to reach the scaling form to be $\tau_{S}/\tau_I \approx 3.1$. 
Then we can estimate $\As,\Bs,\Cs$ from the distribution function at $\tau_{S}$ using $n = \As \Bs^2 \Cs/(2\pi)^{3/2}$, $\langle p_T^2 \rangle = 2 \Bs^2$, and $\langle p_z^2 \rangle = \Cs^2$, and calculate $\gas, \betas$ at $\tau_{S}$ from the consistency conditions~\eqref{beta-q}, \eqref{gamma-q}.
These results are shown in Fig.~\ref{fig:FP} (right) and show good agreement with numerical solutions to the FP equation in the scaling regime.
These results illustrate that in the scaling regime, 
the evolution of the gluon plasma can be reduced to describing the evolution of scaling exponents, in the manner we have done here.

\begin{figure}[t]
  \centering
  \includegraphics[width=.49\textwidth]{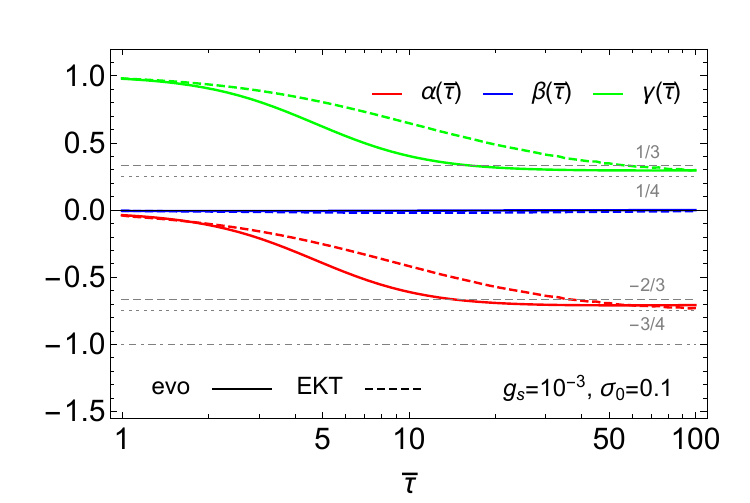}
  \caption{
  \label{fig:EKT}
  The comparison between the evolution of the scaling exponents from Eq.~\eqref{evo} (solid curves) with results from QCD effective kinetic theory (EKT) from Ref.~\cite{Mazeliauskas:2018yef} (dashed curves). We take the same initial distribution function as the EKT results at $\tau_I$. For clarity of presentation, EKT results are the average of exponents computed from different sets of moments of the distribution function.
  }
\end{figure}

As we explained earlier,  
we expect that the small-angle scatterings included in the FP equation play the dominant role for the evolution of hard gluons in QCD EKT. We have shown that the collision integral~\eqref{CIa-1} captures the main features of scaling evolution in the FP equation. We therefore wish to compare the evolution equations~\eqref{evo} we have derived based on collision integral~\eqref{CIa-1} to the evolution of scaling exponents in QCD EKT as presented in Ref.~\cite{Mazeliauskas:2018yef}. 
In Fig.~\ref{fig:EKT} we show this comparison for the same initial distribution function~\eqref{fI} and  $(g_s,\sigma_0)=(10^{-3},0.1)$, and observe not only qualitative but also semi-quantitative agreement, with the largest deviations taking place during the transition between the free-streaming and BMSS fixed points.

Perhaps the most striking observation that one can draw from Fig.~\ref{fig:EKT} is that the values of $\gas$ and $\betas$ from Eq.~\eqref{evo} agree even quantitatively with the exponents from EKT around the BMSS fixed point.
This is highly non-trivial since those asymptotic values are different from their BMSS values.
In Ref.~\cite{Mazeliauskas:2018yef}, 
the authors obtain $\le(\als,\betas,\gas\ri)\approx \le( 0.73, -0.01, 0.29\ri)$ for $(g_s,\sigma_{0})=(10^{-3},0.1)$.\footnote{
In the classical field simulation of Ref.~\cite{Berges:2013eia,Berges:2013fga}, the authors found $\gas= 0.335\pm 0.035$
}
The underlying reason for this deviation from the BMSS value has been the subject of some speculation~\cite{Mazeliauskas:2018yef}.
As we explained in detail in the previous section, the time evolution of $\lcb$ gives rise to an anomalous dimension correction to the scaling exponents (c.f.~Eq.~\eqref{delta-gamma}) in the FP equation. 
We therefore propose that the deviation from the BMSS value in QCD EKT may also arise from the time evolution of the ratio between the typical hard scale and typical momentum exchange per collision.

To further test our speculation, we substitute $A_{I}=\sigma_{0}/g^{2}_{s}$ and $B_{I}/C_{I}=2$ into Eq.~\eqref{delta-gamma} to estimate the deviation of $\gas$ from BMSS expectation
\begin{align}
\label{delta-gamma-1}
  \delta \gamma \equiv \gas-\frac{1}{3}= - \frac{1}{3 \left(y+ \log \le(\frac{ 4 \pi}{N_{c}\sigma_{0}}\ri) \right)}\, . 
\end{align}
In Ref.~\cite{Mazeliauskas:2018yef}, the evolution of kinetic theory starts at $Q_{s}\tau_I=70$ and ends at $Q_{s}\tau=7000$, meaning we should replace $y$ in Eq.~\eqref{delta-gamma-1} with $\log(100)\approx 4.6$. 
For $\sigma_{0}=0.1$ , 
we obtain the correction from $\lcb$ to $\gas$ at the BMSS fixed point: 
\begin{align}
\delta\g = -0.040 \, ,
\qquad
\text{or}\,\,\,\,\,\,\, \gas \approx 0.29
\end{align}
which is in remarkable agreement with the asymptotic value of $\gas$ in EKT.

\subsection{Summary of Adiabatic Hydrodynamization in the earliest stages} \label{sec:sum-adiab-early}

In this work we have studied scaling in a Bjorken expanding gluon plasma described by the Boltzmann transport equation under the small-angle approximation, which takes the form of a  Fokker-Planck (FP) equation. 
For hard gluons, we showed that the FP equation features time-dependent scaling behavior that is qualitatively similar to that observed by solving QCD EKT~\cite{Mazeliauskas:2018yef}. 

We then showed that scaling can be interpreted as arising from adiabatic evolution.
With the simplified collision integral~\eqref{CIa-1}, the kinetic equation can be recast into the form of a Schr\"odinger-like equation. 
Adiabaticity, understood as the property that the eigenstates of the corresponding Hamiltonian do not transition into each other, may be attained to a lesser or greater degree depending on the choice of frame. For the particular case we study here, we find that one can choose a rescaling of the momentum coordinates (frame) such that there are no transitions between eigenstates of the corresponding Hamiltonian.
This means that after some transient time, the excited states have decayed and the distribution function follows the evolution of the instantaneous ground state. 
It is only in this frame that the scaling distribution we observed in the numerical solutions is the ground state.
Without identifying the adiabatic frame, one can still observe scaling phenomena, but the adiabatic nature of the scaling evolution would be obscured. In this sense, we have generalized the notion of the abiabaticity with respect to a fixed set of coordinates $(\tau;p_z,p_\perp)$ to the situation where there exists a ``frame" $(\tau;p_z/C(\tau), p_\perp/B(\tau))$ in which the transition rate from the instantaneous ground state to excited states is suppressed (in this case, zero). We believe that this generalization of adiabaticity may find applications in a broader context.\footnote{
For example, 
consider a time-dependent Hamiltonian in quantum mechanics, and suppose there exists a unitary transformation under which the transformed Hamiltonian evolves slowly. In that case, we can still say that the system described by the original Hamiltonian evolves adiabatically even though this Hamiltonian may change rapidly in time. See Ref.~\cite{scaling-QM} for a similar discussion.
}


From the condition for adiabaticity, we further derived evolution equations for the time dependence of the scaling exponents. 
Our equations can be used to estimate the evolution of typical occupancy and momentum of far-from-equilibrium QGP during the early stages of heavy-ion collisions. In addition to the well-known free-streaming and BMSS fixed points, we found a new ``dilute'' fixed point~\eqref{dilute} that occurs when the typical gluon occupation number becomes small before thermalization.
We also find that the fixed point scaling exponents receive ``anomalous dimension" corrections, arising from the temporal evolution of the Coulomb logarithm, which is determined by the ratio of the hard and soft momentum scales. We compared our results with QCD EKT simulations from Ref.~\cite{Mazeliauskas:2018yef}, and found striking quantitative agreement on the correction to the BMSS exponent in the two theories. In our analysis, this is precisely due to the time evolution of the Coulomb logarithm. In the view that the FP equation we solve here gives an effective description of time-dependent scaling, in qualitative and semi-quantitative agreement with more sophisticated first-principles QCD EKT simulations, our findings suggest that understanding time evolution in terms of an adiabatic evolution may be a valuable approach for describing far-from-equilibrium QCD plasmas.

{The relation between adiabaticity and the non-equilibrium attractor had previously been tested in the simpler single relaxation time approximation~\cite{Brewer:2019oha}. 
Together with the results of Ref.~\cite{Brewer:2019oha}, our finding that the non-thermal scaling evolution of a far-from-equilibrium gluon plasma can be characterized by adiabatic evolution gives compelling support for the claim that the reduction of relevant degrees of freedom in a class of expanding QCD plasmas is due to adiabaticity.} Since we have here shown that this class is larger than it was previously known, we anticipate that a similar study of more general kinetic equations will reveal more connections to adiabaticity. More general collision kernels as well as more realistic heavy-ion collision scenarios including radial expansion in the kinetic description of the plasma should be fertile ground for such an investigation.

We hope some of our qualitative lessons, such as the relation between adiabaticity and scaling, and the emergence of anomalous dimension corrections to scaling exponents, might be instructive when studying other dynamical problems. Examples could include the evolution near a critical point based on the Kibble-Zurek framework~\cite{Kibble_1976,Zurek:1985qw,Zurek:1996sj,KZ2012,Mukherjee:2016kyu,Akamatsu:2018vjr}, and turbulent cascades~\cite{frisch1995turbulence} driven by quantum anomalies~\cite{Hirono:2015rla,Mace:2019cqo}. We defer the investigation of these interesting topics to future work.

\section{Adiabaticity beyond scaling: a complete picture of hydrodynamization} \label{sec:adiab-beyond-scaling}

It is not obvious that one will be able to identify unique underlying principles behind the process of QCD hydrodynamization across all of the energy scales through which QCD matter transits in a HIC, despite the fact that the final state (local thermal equilibrium) is universal in the sense that it has lost all memory of which initial state it originated from. Furthermore, one may worry that the early stages of hydrodynamization could have maximal sensitivity to what the initial state is, and therefore that a separate calculation would need to be carried out in full for each posited initial state in order to have predictive power for observables that are sensitive to the pre-hydrodynamic stages of a HIC.
As it turns out, the situation is not as grim. Indeed, much progress in this direction has been enabled by studies of far-from-equilibrium phenomena that are themselves universal in the sense referred to above and, in particular, of so-called ``attractor'' solutions in either a kinetic theory or a holographic description of pre-hydrodynamic physics. These are (families of) solutions to which generic initial conditions converge in a finite time, meaning that they are dynamically selected in the phase space of the theory. The presence of attractor solutions makes it natural for the system to lose memory of (and have little sensitivity to) its initial state before hydrodynamization. Such attractor solutions have been sought and found in most effective descriptions of out-of-equilibrium QCD matter; see, e.g., Ref.~\cite{Kurkela:2019set}.

What attractor solutions crucially encode is \textit{how} the sensitivity to the initial conditions of a HIC is lost, \textit{what} information is carried through hydrodynamization, and perhaps an answer to \textit{why} the hydrodynamization time of QGP is $\sim 1 \, {\rm fm}/c$ for generic initial conditions, while also preserving quantitative information that may give predictive power over HIC observables that are particularly sensitive to the pre-hydrodynamic stage. Even quantities that are usually thought of as independent from the initial hydrodynamization stage in a HIC, such as heavy quark diffusion 
and the quenching of, and transverse momentum broadening of, high energy partons in a jet shower 
may receive modifications in this initial period~\cite{Avramescu:2023qvv,Boguslavski:2023fdm,Boguslavski:2023alu,Boguslavski:2023waw,Boguslavski:2023jvg,Pandey:2023dzz} that affect how HIC data is to be interpreted -- modifications that may be described by studying the corresponding pre-hydrodynamic attractor~\cite{Boguslavski:2023jvg}. As such, it is desirable to have a framework in which to organize and describe the emergence of attractor solutions systematically. This is the task that concerns us at present. Concretely, we will discuss how the Adiabatic Hydrodynamization (AH) picture~\cite{Brewer:2019oha} provides such a framework using the kinetic theory description of QCD in the small-angle scattering approximation~\cite{Mueller:1999pi,Blaizot:2013lga} as a proof of concept.

A key observation was made in Section~\ref{sec:intro-AH-BSY} regarding the role of time-dependent coordinate redefinitions (in particular, rescalings) in identifying the attractor solution and explaining its rapid emergence relative to the other time scales in the system. Inspired by the fact that the same self-similar scaling had been observed in classical-statistical simulations~\cite{Berges:2013eia,Berges:2013fga}, in small-angle scattering kinetic theory~\cite{Tanji:2017suk} and in QCD EKT~\cite{Mazeliauskas:2018yef}, we proposed that the reduction of dynamically relevant degrees of freedom was most naturally understood in the (time-dependent) frame in which the typical momentum scales of the particle distribution functions are approximately constant. We then demonstrated this explicitly by analytically solving for the instantaneous eigenstates and eigenvalues of the generator of time evolution of the theory (which, out of familiarity with the quantum mechanics nomenclature, was referred to as a Hamiltonian) in a simplified version of the small-angle scattering collision kernel, applicable in the earliest stages of the hydrodynamization process of a weakly coupled, boost-invariant gluon gas. 

In this work, we extend the results of Section~\ref{sec:intro-AH-BSY} in two distinct ways that demonstrate the effectiveness of the AH framework to identify and describe attractors in kinetic theories:
\begin{enumerate}
    \item We show that the close connection between time-dependent scalings, adiabaticity, and universality persists even in situations where one does not have explicit analytic control over the eigenvalues and eigenstates of the time evolution operator.
    \item We demonstrate, within the small-angle scattering approximation and with fewer additional approximations than in Section~\ref{sec:intro-AH-BSY}, that the AH scenario describes the evolution of the gluon distribution function all the way from the times described in Section~\ref{sec:intro-AH-BSY} until local thermal equilibrium is reached and the system hydrodynamizes. The process of relaxation to a thermal distribution takes place in stages, firstly along the longitudinal direction, driven by expansion along the boost-invariant direction, where the distribution relaxes to a set of slow modes characterized by a unique profile in this direction, and secondly from this set of slow modes to a thermal distribution where a unique slow mode is singled out. We show that AH provides a unified description of both stages.
\end{enumerate}

More generally, our work provides a systematic method to study the emergence of universal behavior in kinetic theories and identify attractor solutions. Building on the previous papers on this subject~\cite{Brewer:2019oha,Brewer:2022vkq}, we posit that adiabaticity, in the sense we will define in the next Section, provides a robust underlying principle that allows one to identify the slow (low-energy) degrees of freedom of the theory. While we do not aim to prove that it is always possible to find a description where the evolution of the system is adiabatic, the power of the AH framework is that when an attractor solution exists it provides a natural method by which to single it out from the rest.

For simplicity's sake, the collision kernel we work with omits qualitatively and quantitatively important terms of the full QCD EKT collision kernel; this simplification makes the hydrodynamization time of the system unrealistically long. Our goal in this work is therefore not the explanation of the (rapid) timescale for hydrodynamization in QCD. Rather, we seek (and have found) a formalism
that provides a common physical description of, and intuition for, the processes occuring through all the stages of hydrodynamization in the kinetic theory with a simplified collision kernel that we employ, with the goal of employing this formalism and applying this intuition in QCD EKT in future work.
We also omit spatial gradients, which means that we are neglecting the transverse expansion of the 
droplet of QGP throughout the hydrodynamization process. Relaxing this assumption is also a worthy goal for future work, but we do not anticipate that doing so will make a qualitiative difference to the processes of hydrodynamization in the collisions of nuclei whose transverse extent is much larger than the hydrodynamization timescale.
There is no fundamental barrier to generalizing the tools developed here to a kinetic theory which includes the full QCD EKT collision kernel and transverse expansion. We expect that by applying the AH framework to QCD EKT using the systematic procedure we develop here, we will be able to employ the intuitive understanding of the rapid reduction of dynamically relevant degrees of freedom (i.e. hydrodynamization) that we find in this work in a context where the rapid hydrodynamization expected in a more complete description of QCD is realized. In this way, our work paves the way for a satisfying physical description of and intuition for how and why hydrodynamization occurs in HICs.

This work is organized as follows: In Section~\ref{sec:AH}, we review the AH framework, explaining its usefulness and purpose; we also describe the concrete kinetic theory setup that we shall work with here and the role of conserved quantities in this picture. In Section~\ref{sec:scaling}, we show that the introduction of time-dependent rescalings of the momentum coordinates allows one to find adiabatic descriptions of the hydrodynamization process of a weakly coupled gluon gas. We consider the examples of a static, non-expanding case as well as of a boost-invariant longitudinally expanding gluon gas that exhibits two separate scaling regimes: one at early times (discussed in Section~\ref{sec:intro-AH-BSY}) and one at late times, which is identified with reaching the hydrodynamic regime. While these rescalings are sufficient to describe each regime in terms of an adiabatically evolving state, they do not provide a way to smoothly connect early (pre-hydrodynamic) and late (hydrodynamic) times and attractors. Finally, Section~\ref{sec:non-scaling} provides the tools needed to make this connection, generalizing the relation between scaling and adiabaticity to more general time-dependent variables, and demonstrating point 2 in the preceding discussion via the explicit construction of a unified description of the pre-hydrodynamic and hydrodynamizing attractors. We present our concluding remarks and outlook in Section~\ref{sec:outlook}.

\subsection{Adiabatic Hydrodynamization} \label{sec:AH} 

In this Section, we discuss how the Adiabatic 
Hydrodynamization (AH) framework constitutes a systematic approach to characterize aspects of the out-of-equilibrium dynamics of many-body theories that are common to many systems. Even though beginning in Section~\ref{sec:kin} and then throughout what follows we will employ kinetic theory to describe the dynamics of interest, we expect that the AH framework has applications that go beyond kinetic theory and we therefore introduce its logic with considerable generality.

As laid out in the original work on this approach~\cite{Brewer:2019oha}, the AH scenario can be attained if the dynamics of the system is described by an evolution equation with the form
\begin{equation} \label{eq:general-evol}
    \frac{\partial \ket{\psi}}{\partial t} = - H[\psi; t] \ket{\psi} \, ,
\end{equation}
where $\ket{\psi}$ is a state vector containing all of the many variables needed to describe the state of the many-body system at time $t$. In the case of a strongly coupled many-body theory with a holographic description, the state vector would be specified by the metric and other fields along some hypersurface in a spacetime with one additional dimension. 
More relevant for us in this work, in a kinetic theory the state vector can encode a distribution function via
$f({\bs x},{\bs p},t) = \braket{{\bs x},{\bs p} | \psi(t)}$. In our discussion in the rest of this work, we will often refer to $f$ and $\ket{\psi}$ as the state of the system interchangeably. 
The form of Eq.~(\ref{eq:general-evol})  will enable us to carry over some of the intuition developed for analog problems in quantum mechanics, and in particular will enable us to use the adiabatic approximation as an organizing principle.
Nevertheless, there are important conceptual differences between Eq.~(\ref{eq:general-evol}) and the standard formulation of quantum mechanics itself, all of which are related to the time evolution operator $H$:
\begin{enumerate} 
    \item The prefactor in front of the $H$ on the RHS of~\eqref{eq:general-evol} is real not imaginary, and by convention is chosen to be $(-1)$. 
    \item In general, $H$ will be a non-Hermitian operator.
    \item $H$ can depend on the state of the system $\ket{\psi}$, and therefore the evolution of the system is, in general, nonlinear.
\end{enumerate}
The sign convention in the first point is so that if the real part of the eigenvalue spectrum of $H$ is bounded from below, as will be the case in examples of physical interest, we may be able to organize the directions in the vector space of states by the ``speed'' at which they evolve. This will later allow us to single out ``slow modes,'' i.e., solutions that are long-lived compared to all the others. 
While the first point just described is only a convention that aids us in organizing our description of the dynamics, the second and third points are necessary ingredients to describe interacting many-body theories through a kinetic equation with collisions. We shall discuss these two points in turn.

Non-hermiticity of the time-evolution operator means that the left and right eigenstates of $H$ at time $t$ (henceforth the instantaneous eigenstates) will not be naively related by adjoint conjugation. That is to say, if $\ket{n}_R$ is a right eigenstate of $H$, i.e. $H \ket{n}_R = \epsilon_n \ket{n}_R$, it does not follow that $(\ket{n}_R)^\dagger H = \bra{n}_R H = \bra{n}_R \epsilon_n$. However, there does exist a set of states $\{\bra{n}_L\}_n$, which together with $\{\ket{n}_R\}_n$ and the eigenvalues $\epsilon_n$ satisfy 
\begin{align}
    H \ket{n}_R = \epsilon_n \ket{n}_R \, , & & \bra{n}_L H = \bra{n}_L \epsilon_n \, , & & \bra{m}_L \ket{n}_R = \delta_{mn} \, , \label{eq:mutual-orthogonality}
\end{align}
where we take ${\rm Re}\{\epsilon_m\} \leq {\rm Re}\{\epsilon_n\}$ if $m < n$, and $n,m \in \{0,1,2,\ldots\}$.
We can then decompose the state of the system as a linear superposition of instantaneous eigenstates
\begin{equation} \label{eq:state-decomposition}
    \ket{\psi(t)} = \sum_n a_n(t) \ket{n(t)}_R = \sum_n \tilde{a}_n(t) e^{- \int^t dt' \epsilon_n(t') } \ket{n(t)}_R \, 
\end{equation}
where the coefficients $a_n(t)$ that specify the superposition by telling us the occupation of each of the instantaneous eigenstates at time $t$ evolve as
\begin{equation}
    \partial_t \ln \frac{a_n}{a_m} = \partial_t \ln \frac{\tilde{a}_n}{\tilde{a}_m} - (\epsilon_n(t)-\epsilon_m(t)).
    \label{eq:2.4}
\end{equation}
This means that if there is a ``low-energy'' state $m$ (or a set thereof) such that ${\rm Re} \{\epsilon_n\} > {\rm Re} \{\epsilon_m\} $ for all $n > m$ (that is, ${\rm Re} \{\epsilon_n\} > {\rm Re} \{\epsilon_m\} $ for all $m$ in said set of low energy states and all $n$ not in said set), and if the condition  
\begin{equation} \label{eq:adiabatic-approx}
    \partial_t \ln \left| \frac{\tilde{a}_n}{\tilde{a}_m} \right|   <  {\rm Re} \{\epsilon_n(t)\}-{\rm Re} \{\epsilon_m(t)\},
\end{equation}
is satisfied for all low energy $m$'s and all high energy $n$'s, then via (\ref{eq:2.4}) we have $\partial_t \left| a_n/a_m \right|<0$ and can therefore conclude that the occupation of states with larger relative values of ${\rm Re}\{\epsilon_n\}$ (``high-energy" states) will decay faster than that of ``low-energy'' states. 
In such a system, the long-lived solutions necessarily correspond to states in which only these low-energy states of the ``effective Hamiltonian'' $H$ are occupied. After an early transient period during which the occupation of the high-energy states decays away, the subsequent evolution of the system follows that of the evolving low-energy states, and is referred to as adiabatic.  
The condition (\ref{eq:adiabatic-approx}) is thus the condition that must be satisfied in order for the evolution to become adiabatic, provided the inequality never comes arbitrarily close to being violated.  Furthermore, if all the $\tilde{a}_n$'s and $\tilde{a}_m$'s are constant in time, 
then the adiabatic condition (\ref{eq:adiabatic-approx}) is strictly satisfied as its left-hand side vanishes and
the system is perfectly (and in a sense trivially) described by adiabatic evolution. Section~\ref{sec:intro-AH-BSY} provides us with an example of this circumstance. In such a case, the
so-called ``attractor'' solutions of the theory are  described described exactly by the ground state(s) of $H$. In the situation considered in Section~\ref{sec:intro-AH-BSY} this was achieved by finding a set of time-dependent coordinates in which the instantaneous eigenstates of $H$ were time-independent which, as per Eq.~\eqref{eq:evol-coeffs} below, fulfills the condition (\ref{eq:adiabatic-approx}). 
If instead the system is such that 
$\partial_t \tilde{a}_n \neq 0$ but the
adiabatic approximation~\eqref{eq:adiabatic-approx} is nevertheless satisfied, 
then the ground state(s) of $H$ will describe the attractor(s) up to corrections controlled by the severity of the deviation from $\partial_t \tilde{a}_n = 0$. In practical situations, the attractor will be realized up to transients stemming from a general initial condition, which means that the adiabatic approximation will be useful to identify the long-lived solutions.

By substituting the eigenstate decomposition \eqref{eq:state-decomposition} into the evolution equation \eqref{eq:general-evol}, one finds that 
\begin{equation} \label{eq:evol-coeffs}
    \frac{\partial_t a_n}{a_n} = - \epsilon_n - \sum_{n'} \frac{a_{n'}}{a_n} \bra{n}_L\partial_t \ket{n'}_R \,.
\end{equation}
Let us therefore introduce an adiabaticity criterion that we will find more useful in practice:
\begin{equation}
    \delta_A^{(n,m)} \equiv \left| \frac{\bra{ n}_L\partial_t \ket{m}_R}{\epsilon_n-\epsilon_m} \right| \ll 1 \, , \quad \forall n, m \ \ {\rm s.t.} \ \epsilon_n \neq \epsilon_m \, .
\label{eq:adiabaticitynm}
\end{equation} 
We shall show below that if this criterion is satisfied then the adiabatic criterion that we first introduced in the form \eqref{eq:adiabatic-approx} 
is satisfied.
For our purposes, though, with the goal of identifying attractors in mind we first note that since we expect that the excited states decay faster than the ground state we can conclude immediately that if the condition \eqref{eq:adiabaticitynm} holds for all $m$ in the set of low energy states and all $n$ outside that set then this will ensure that the high-energy states are not driven to large occupation numbers $a_n$ on account of their coupling to the set of instantaneous ground states in the evolution equation~\eqref{eq:evol-coeffs}. 

In fact, we can reproduce~\eqref{eq:adiabatic-approx} from~\eqref{eq:adiabaticitynm} by writing
\begin{align}
    \partial_t \ln \frac{a_n}{a_m}  &= \frac{\partial_t a_n}{a_n} - \frac{\partial_t a_m}{a_m} \nonumber \\
    &= -\epsilon_n + \epsilon_m - \sum_{n'} \frac{a_{n'}}{a_n} \bra{n}_L\partial_t \ket{n'}_R + \sum_{n'} \frac{a_{n'}}{a_m} \bra{m}_L\partial_t \ket{n'}_R  \, , \label{eq:relating-adiab-criteria}
\end{align}
from which it is clear that if the coefficients $a_n$ are all ${\cal O}(1)$ numbers then if the criterion~\eqref{eq:adiabaticitynm} holds for all $n,m$ this implies that $\partial_t \ln |a_n/a_m| \approx - {\rm Re} \{ \epsilon_n - \epsilon_m \}$ (with both of these quantities  $< 0$ if $n > m$, as hypothesized). 
Using \eqref{eq:2.4}, this in turn implies that the adiabatic criterion in the form \eqref{eq:adiabatic-approx} that we first stated is satisfied.
In order to convince ourselves that a strict inequality in~\eqref{eq:adiabatic-approx} is preserved under time evolution, we must check the case in which $|a_n| \ll |a_m|$ if $n$ labels a high-energy state and $m$ labels a low-energy state. In this case, the last sum in~\eqref{eq:relating-adiab-criteria} is either small or ${\cal O}(1)$, 
meaning that the sign of the whole expression will not depend on it provided the gap $\epsilon_n - \epsilon_m$ is sufficiently large. However, $a_{n'}/a_{n}$ in the first sum in \eqref{eq:relating-adiab-criteria} can be large when $n'$ labels a low energy state because $n$ is a high-energy state. In this case it becomes imperative that~\eqref{eq:adiabaticitynm} holds when $m$ is a low-energy state and $n$ is a high-energy state, so that the fact that $a_{n'}/a_{n}$ is large does not prevent the excited states from decaying. This is exactly the statement that long-lived attractor solutions are completely captured by the set of low-energy states. If~\eqref{eq:adiabaticitynm} didn't hold when $m$ is a low-energy state and $n$ is a high-energy state, then excited states would be sourced by the low energy $n'$ states in the first sum of~\eqref{eq:relating-adiab-criteria} and the late time solution would not be dominated by low-energy states alone. 

In keeping with the particular importance of slow variation of low-energy states, we can focus on the case in which the system begins in a state that is close to its ground state (or close to a superposition of its low energy states) which allows us to sharpen \eqref{eq:adiabaticitynm}, because the only aspect of this criterion that then matters is  
\begin{equation} \label{eq:adiabaticityn0}
    \delta_A^{(n)} \equiv \left| \frac{\bra{n}_L\partial_y\ket{0_m}_R}{\epsilon_n-\epsilon_0} \right| \ll 1 
\end{equation}
where $\ket{0_m}_R$ is any one of the low-energy states and where $n$ labels any one of the higher energy states.
If the criterion \eqref{eq:adiabaticityn0} is satisfied, 
the subsequent evolution will rapidly converge to the adiabatically evolving ground state(s).
We see that, in addition to rendering it in a form with practical utility, phrasing the adiabatic criterion 
in the form \eqref{eq:adiabaticityn0} naturally encapsulates the essence of the adiabatic approach that is at the core of all the analyses that follow later in this work.

Returning to the three listed differences between our adiabatic framework and adiabaticity in quantum mechanics, in practice, the main obstacle to making progress is point 3. This is because the instantaneous eigenstates and eigenvalues of $H$ will also depend on the state of the system, and so solving for them in order to write down the decomposition~\eqref{eq:state-decomposition} is, in general, highly nontrivial for an arbitrary state $\ket{\psi}$.
Nonetheless, this does not obstruct the previous reasoning: given a state $\ket{\psi}$ at time $t$, one can calculate the instantaneous eigenstates $\ket{n[\psi,t]}_R$ and decompose the state on this basis. If after doing this one finds that $\ket{\psi}$ is approximately equal to a superposition of the low-energy instantaneous eigenstates, then the system will evolve slowly, remaining dominated by these low-energy states. If the adiabatic criterion is met, then even if the system is initialized with a non-zero occupancy for the excited states, these will decay and only leave the slow modes driving the system.

If one is able to overcome the practical barriers to framing the system's evolution in these terms, as we will outline later in this Section, one gains both intuitive understanding of the physics as well as predictive power. Specifically, one will have achieved:
\begin{itemize}
    \item A systematic organization of the theory in terms of long- and short-lived modes.
    \item A characterization of the dynamically preferred solutions of the theory, i.e., of out-of-equilibrium attractors.
    \item Predictive power for complex systems in terms of a small number of degrees of freedom. Concretely, once the ``relevant'' degrees of freedom have been identified, one can truncate the evolution equation~\eqref{eq:general-evol} to the relevant subspace and solve it only for that small set of degrees of freedom.
\end{itemize}

The key ingredient, which is not guaranteed to be present in a physical system, is that we are able to describe its evolution in a framework where the adiabatic criterion is met. To achieve this, the rest of this Section deals with the problem of how to set up a vector space describing a distribution function where the instantaneous eigenstates of the time-evolution operator stemming from the underlying kinetic theory evolve adiabatically. As we will see, one ingredient that can solve this problem almost entirely is to first characterize the evolution of the relevant dimensionful scales of the system, treat those as ``background'' quantities for the evolution of the system, and then find the slow modes of the remaining degrees of freedom. This was first realized in our work in Section~\ref{sec:intro-AH-BSY}, where the fact that the distribution function had a self-similar evolution was exploited to achieve exact adiabaticity after a time-dependent scaling of the momenta was performed. In Section~\ref{sec:scaling}, we review and extend their results towards a complete treatment of QCD kinetic theory in the small-angle scattering approximation by looking at the different limits where scaling phenomena appear. Later, in Section~\ref{sec:non-scaling} we will see how the reduction of degrees of freedom appears even in solutions where there is no self-similar scaling solution valid for all times, thus explicitly showing how Adiabatic Hydrodynamization can take place in kinetic theories derived directly from QCD.

In the remainder of this subsection, we will introduce the concrete kinetic theory setup that we shall employ in the rest of this Section, and discuss how conserved quantities provide another extra ingredient that will aid us in organizing the theory.

\subsubsection{Kinetic theory setup: an overview of our approximations} \label{sec:kin}

To study the hydrodynamization process of a Yang-Mills plasma, we consider a kinetic theory description of the gluon phase space density encoded in a distribution function $f({\bf x}, {\bf p},t)$, where $(t, {\bf x})$ are Minkowski coordinates and ${\bf p}$ labels the 3-momentum of the gluons. The gluon distribution function evolves according to a kinetic equation, also referred to as a Boltzmann equation:
\begin{equation} \label{eq:Boltzmann}
    \frac{\partial f}{\partial t} + \frac{\bf p}{p} \cdot \nabla_{\bf x} f = - \mathcal{C}[f] \, ,
\end{equation}
which is specified by the collision kernel $\mathcal{C}[f]$. For weakly coupled QCD, the collision kernel was obtained in~\cite{Arnold:2002zm} and the resulting theory is called QCD Effective Kinetic Theory (EKT).

In the context of heavy-ion collisions, it is appropriate to work in coordinates that incorporate some information from the geometry of the collisions. Taking $z$ to be the coordinate along the beam axis, it is convenient to go to Milne coordinates $(\tau,{\bf x}_\perp, \eta)$, specified by
\begin{align}
    t = \tau \cosh \eta \, , & & z = \tau \sinh \eta \, , & & {\bf x}_\perp = {\bf x}_\perp \, ,
\end{align}
where the natural momentum variables at each point $(\tau,{\bf x}_\perp, \eta)$ are
\begin{align}
    p_\tau = p \cosh \eta - p_z \sinh \eta \, , & & p_\eta = p_z \cosh \eta - p \sinh \eta \, , & & {\bf p}_\perp = {\bf p}_\perp \, .
\end{align}
The collision is assumed to take place at $\tau = 0 \implies (t, z) = (0, 0)$.

In terms of these new coordinates, the kinetic equation is given by
\begin{equation} \label{eq:Boltzmann-Milne}
    \frac{\partial f}{\partial \tau} + \frac{{\bf p}_{\perp}}{p_\perp} \cdot \nabla_{{\bf x}_\perp} f + \frac{p_\eta}{\tau p_\perp} \frac{\partial f}{\partial \eta} - \frac{p_\eta}{\tau} \frac{\partial f}{\partial p_\eta} = - \mathcal{C}[f] \, .
\end{equation}
In a realistic description of a heavy-ion collision, one should explicitly take into account the dependence of $f$ on all of these variables. In practice, however, we will use two simplifying assumptions:
\begin{enumerate}
    \item We will assume boost invariance of $f$, i.e., $\partial f/\partial \eta = 0$. In the CM frame of an $AA$ collision, this will be true at mid-rapidity $\eta = 0$, and approximately true for some range of $\eta$ around $\eta=0$ that becomes larger in collisions with higher energy. Therefore, for observables that focus on the mid-rapidity region in collisions at top RHIC energies and at the LHC, this assumption is not too costly. Furthermore, because at $\eta = 0$ we have $p_\eta = p_z$, we will use $p_z$ throughout in the place of $p_\eta$.
    \item We will assume translation invariance in the transverse plane, $\nabla_{{\bf x}_\perp} f = 0$. This is certainly a major simplification that blatantly ignores the fact that the droplets of QGP produced in a HIC have finite transverse size. However, by simple inspection of the kinetic equation, it can be a good approximation at early times $\tau$ because of the $1/\tau$ prefactor in the $\partial f/\partial p_\eta$ term, provided the transverse gradients are not large. In particular, as long as the transverse extent of the droplet is much larger than the hydrodynamization time, neglecting the build up of radial flow during the pre-hydrodynamic epoch is a reasonable starting point for its analysis. 
\end{enumerate}
Clearly, the one of these two assumptions that would be most interesting to relax for HIC phenomenology would be translation invariance in the transverse plane; neglecting the lumpiness (in the transverse plane) of the matter produced initially in a HIC -- for example that arising because nuclei are made of nucleons -- cannot be justified by the argument above. However, we leave the investigation of its consequences to future work, as they are beyond the scope of this work. 
We note, also, that both of these assumptions have been widely used in studies of QCD EKT applied to HICs~\cite{Keegan:2016cpi,Kurkela:2018vqr,Kurkela:2018wud,Mazeliauskas:2018yef}. In fact, the pioneering paper on the ``bottom-up'' thermalization scenario~\cite{Baier:2000sb}, which describes the stages via which thermalization would occur in a HIC within the framework of perturbative QCD and upon assuming weak coupling employs precisely these assumptions.

The only ingredient we have yet to discuss is the collision kernel $\mathcal{C}[f]$. Ideally, we would simply use the full collision kernel provided by QCD EKT. However, since our present purpose is to establish how hydrodynamization can be understood in terms of the instantaneous ground state of an effective Hamiltonian, as prescribed by the AH scenario~\cite{Brewer:2019oha}, we will make further simplifications compared to the full collision kernel. Concretely, we will study the small-angle scattering approximation~\cite{Mueller:1999pi,Serreau:2001xq,Blaizot:2013lga}, which we already introduced in~\eqref{eq:small-angle-kernel}.

Throughout our exposition we will consider this collision kernel in various regimes, starting from the simplest cases and building up to a nearly complete treatment of it. Numerical details of the implementation are available in Appendix~\ref{app:numerics}. 
Our treatment is ``nearly complete'' in that we also make one further simplifying assumption throughout this work, dropping
the term proportional to $I_b f^2$ in the collision kernel~\eqref{eq:small-angle-kernel}. We do so for simplicity, and because we have checked that this is a good approximation in the early pre-hydrodynamic attractor and because it is manifestly a good approximation as hydrodynamization is achieved in our analysis, since $f$ is then small. We provide further discussion of this approximation in Appendix~\ref{app:f2}.  

As we have already discussed at the beginning of Section~\ref{sec:adiab-beyond-scaling}, our goal in this work is to find a common physical description of, and intuition for, the processes occuring during all the stages of hydrodynamization in the kinetic theory with the simplified collision kernel that we have specified here. Extending our analysis to apply it to the computationally more challenging QCD EKT collision kernel is a priority for future work, but we anticipate that the physical intuition that we shall glean from the analyses to come below, in particular from Sect.~\ref{sec:connectstages} where we put all the pieces together, will carry over.

\subsubsection{Conserved quantities in Adiabatic Hydrodynamization}
\label{sec:ConservationLaws}

The fact that hydrodynamics is an effective theory for the dynamical evolution of densities of the conserved quantities of a given system suggests that in the process of hydrodynamization such quantities should also play a central role. Indeed, as we will see momentarily, they can provide crucial information to organize the analysis of the pre-hydrodynamic evolution more efficiently. In the present context, if a quantity $X[f]$ is conserved by the kinetic equation, and the AH scenario is realized, i.e.~if the system approaches hydrodynamics via first being driven into and then following an evolving instantaneous ground state (or set of slow modes which can collectively be considered as ``ground states''), then the conserved quantity must be encoded in such ground state(s).

In the AH framework, the conserved quantities that are most useful are those that can be written in terms projections of the physical state $\ket{\psi}$, i.e., those quantities $K$ that can be written as
\begin{equation}
	K = q_K(\tau) \braket{P_K | \psi(\tau) } \, ,
\end{equation}
where $\bra{P_K(\tau)}$ is a projector that, once applied on the state yields the conserved quantity $K$, and $q_K(\tau)$ is a time-dependent prefactor that does not depend on the state (i.e., its time dependence is explicit).

For a spatially homogeneous, non-expanding gluon gas (i.e., without assuming boost invariance in Milne coordinates), in Minkowski coordinates as in~\eqref{eq:Boltzmann}, and described by the collision kernel~\eqref{eq:small-angle-kernel}, there are two conserved quantities
\begin{align}
	K_N = \braket{P_N | \psi} \equiv \int_{\bf p} f({\bf p}, t) \, , & & K_E = \braket{P_E | \psi} = \int_{\bf p} p \, f({\bf p}, t) \, ,
\end{align}
which are proportional to the number density and energy density of the gluon gas, respectively. 

In the case of a boost-invariant, transversely homogeneous, expanding gluon gas, neither of the previous quantities is conserved. However, the rescaled number density $K_{\tilde{N}} = \tau K_N$ is conserved. Explicitly,
\begin{equation}
	K_{\tilde{N}} = \tau \braket{P_{N} | \psi} \equiv \tau \int_{\bf p} f({\bf p}, \tau) \, ,
\end{equation}
is conserved. Note that the fact that there can be a time-dependent prefactor $q_{\tilde{N}} = \tau$ is crucial for this definition.

As we shall now explain, identifying such a conserved quantity is extremely helpful for implementing the construction that we have described abstractly above,  in particular when choosing a set of left and right basis states to solve the Boltzmann equation~\eqref{eq:Boltzmann-Milne} by writing it in the effective Hamiltonian form~\eqref{eq:general-evol}.
Because of the crucial role that rescalings take in this step, the precise way in which this rewriting is done will be one of the central matters of the following Sections.  Even without an explicit version of~\eqref{eq:Boltzmann-Milne} in the form of~\eqref{eq:general-evol} in hand, one can formally see right away how taking account of conserved quantities will be useful in this regard. If one chooses the left basis $\{\bra{\psi_n}_L\}_{n=1}^{N_{\rm basis}}$ such that the first basis state $\bra{\psi_1}_L$ is exactly the projector $\langle P_N |$, then because acting with $\partial/\partial \tau$ on the conserved quantity $\tau \braket{P_{N} | \psi(\tau) } $ yields zero, from (\ref{eq:general-evol}) we have 
\begin{align}
	\bra{P_{N}} H = \frac{1}{\tau} \bra{P_N} \, ,
\end{align}
i.e., by construction, the first basis state will be a left eigenvector of the Hamiltonian,\footnote{In this discussion, $H$ is the Hamiltonian that generates time evolution with respect to the coordinate $\tau$. It will later turn out to be convenient to introduce a new time coordinate $y = \ln (\tau/\tau_I)$ such that the eigenvalue of the corresponding Hamiltonian acting on this left state is constant.} which we will denote by $\bra{\phi_1}_L$.

Furthermore, this means that there will be a right eigenvector $\ket{\phi_1}_R$ of the Hamiltonian with the same eigenvalue, and, moreover, that all other right eigenvectors $\ket{\phi_n}_R$ (with $n \neq 1$) of the Hamiltonian will be orthogonal to this projector, i.e., $\bra{\phi_1}_L \ket{\phi_n}_R = 0$ $\forall n \neq 1$. As such, all the number density of the system is exclusively contained within the corresponding right eigenvector $\ket{\phi_1}_R$, and the value of the number density is determined by the expansion coefficient $a_1$ in the decomposition~\eqref{eq:state-decomposition}. It follows that if a system described by a positive-definite distribution function $f$ (i.e.~a system with nonzero number density) has an attractor solution described by the AH scenario, then $\ket{\phi_1}_R$ must be (one of) the ground state(s) determining the attractor  and driving the evolution of the system. 

In addition to this, because we know beforehand that the systems we are studying can (and will eventually) hydrodynamize, we also know what the late-time ground state of the system should be: $\ket{\phi_1}_R$ should describe either a Boltzmann distribution if the Bose-enhancement terms are dropped, such that $f \propto \exp(-p/T)$, or a Bose-Einstein distribution $f \propto (\exp(p/T)-1)^{-1}$ if the effects of bosonic quantum statistics are not neglected. It is thus appropriate to choose the first basis state (at least at late times) such that it describes a distribution in local thermal equilibrium, because if the system does hydrodynamize, this will guarantee a quick convergence of the sum in the basis state decomposition of the ground state, and consequently a good quantitative description. Having fixed the first left and right basis states, and provided with an inner product (which we take to be simply determined by $\int_{\bf p}$ or a rescaling thereof, as we discuss in the next Section) and a family of functions to generate the basis, the Gram-Schmidt method uniquely determines the form of all the other (mutually orthogonal) left and right basis states.\footnote{One may wonder what would change in this discussion if we had two conserved quantities, as in the case of a spatially homogeneous, non-expanding gluon gas. In practice, as it turns out, it is impossible to choose a basis where the left basis contains the projectors associated with both number and energy density, and the right basis contains a state describing a positive definite distribution function while satisfying the mutual orthogonality conditions~\eqref{eq:mutual-orthogonality}. This means that in practice when we apply the construction described here we must do so upon choosing only one of the two conserved quantities, and we shall use number conservation.}

Having laid out the AH framework, in the next two Sections we shall make everything we have described above fully explicit (including in particular fully specifying $H$) in several kinetic theory calculations with increasing completeness. First, though, in the next Section we shall discuss the one remaining ingredient that one needs in order to achieve an adiabatic description of out-of-equilibrium kinetic theories. Namely, how should we address
the fact that 
the variables that describe the state of the system (the eigenstates/values of $H$ and the
coefficients that specify the occupation of these eigenstates) and consequently
the typical scales that characterize the state of the system (such as the typical longitudinal momentum of gluons $\sqrt{\langle p_z^2 \rangle}$ or the typical transverse momentum $\sqrt{\langle p_\perp^2 \rangle}$) 
may be rapidly evolving even if the dynamics have driven the system to an attractor solution. Note that the projections of the state $\langle \psi^{(L)}_n | \psi \rangle$ onto a given basis $\{|\psi^{(R)}_n \rangle \}_n$ (satisfying  orthogonality conditions analogous to those for the eigenstate basis) are in practice moments of the distribution function, of which $\langle p_z^2 \rangle$, $\langle p_\perp^2 \rangle$ are examples. This means that in order to make it possible to describe said attractor solutions as adiabatically evolving (sets of) ground state(s), one needs to choose a set of basis states and/or rescale the momentum and time coordinates  in such a way as to recast the description of the evolution of the attractor solution in a way that makes it adiabatic.
There is a large class of collision kernels and kinematic regimes, including all of those described in Section~\ref{sec:scaling}, for which this issue is most appropriately dealt with by introducing ``scaling variables'' that map the physical momentum coordinates to rescaled momentum coordinates, thus giving one the freedom to choose different frames (this use of the term ``frame'' originates from Section~\ref{sec:intro-AH-BSY}, in a loose analogy with the reference frame concept in relativity where one is free to choose coordinates at will) to study the system. We will refer to a frame that makes the evolution adiabatic, in the sense described above and made explicit in the next Section, as an ``adiabatic frame.''

\subsection{Scaling and adiabaticity}
\label{sec:scaling}

Scaling phenomena are a hallmark of universality in physical systems.
They also play a prominent role in the behavior of out-of-equilibrium systems across a whole range of phenomena including HICs~\cite{Baier:2000sb,Berges:2013eia,Kurkela:2014tea,Mukherjee:2016kyu,Mukherjee:2017kxv,Mazeliauskas:2018yef} and cold atoms~\cite{Mikheev:2018adp,Prufer:2018hto,Erne:2018gmz}, as well as more general theories~\cite{PineiroOrioli:2015cpb}, the Kibble-Zurek phenomenon of defect formation after out-of-equilibrium phase transitions in cosmology or condensed matter phyics~\cite{Kibble:1976sj,Zurek:1985qw,Chuang:1991zz,Zurek:1996sj,Chandran:2012cjk}, and even turbulence~\cite{frisch1995turbulence}. In this Section, we will detail how identifying time-dependent scaling in a theory allows us to analyze its underlying adiabatic evolution.

A lot can be gained quantitatively by treating the evolution of the typical scales of the system on a different footing than the rest of the system's evolution. Specifically, if one knows that the evolution of the expectation value of a quantity $x$ is given by some calculable function of time $\langle x\rangle (t)$, then the rest of the features of the distribution of $x$ are easier to analyze as a function of $x/\langle x \rangle$, where the dominant time dependence of the expectation values have been scaled out. Provided that the system can undergo scaling, this is the natural way in which to compare long- and short-lived modes.

To put this on a concrete footing, we now specialize to scaling phenomena in kinetic theory, and demonstrate how the intuitive picture we just described can be realized in terms of the AH framework. For the purposes of this discussion, we will consider a distribution function $f$ that is homogeneous 
in the spatial Milne coordinates $(\eta, {\bs x}_\perp)$ and symmetric under rotations in the ${\bs p}_\perp$ plane.
Such a distribution function can be said to be ``scaling" if it takes the form
\begin{equation} \label{eq:scalingform}
    f(\bm{p},\tau) = A(\tau) w\left(\frac{p_\perp}{B(\tau)},\frac{p_z}{C(\tau)}\right)
\end{equation}
for some time-dependent $A,B,$ and $C$, such that the rescaled distribution function $w = w(\zeta,\xi)$ is independent of time. If the dynamics of the system causes $f$ to generically fall into such a scaling form, $w$ can be viewed as a universal attractor. This is true in, for example, the bottom-up thermalization picture~\cite{Baier:2000sb}, in which the early-time dynamics of the system is described by the scaling form (\ref{eq:scalingform}) with
\begin{equation}
    A_\text{BMSS}(\tau) \propto \tau^{-\frac{2}{3}}, \;\;\; B_\text{BMSS}(\tau)  \propto \tau^0, \;\;\; C_\text{BMSS}(\tau)  \propto \tau^{-\frac{1}{3}} \, , \label{eq:BMSS-scalings}
\end{equation}
where BMSS refers to the authors of~\cite{Baier:2000sb}. In this specific case, the ``scaling exponents"
\begin{equation}
    \alpha \equiv \frac{\tau}{A} \frac{\partial A}{\partial \tau}, \;\;\; \beta \equiv -\frac{\tau}{B} \frac{\partial B}{\partial \tau}, \;\;\; \gamma \equiv -\frac{\tau}{C} \frac{\partial C}{\partial \tau}
    \label{eq:scaling-exponents}
\end{equation}
are time-independent with
\begin{align}
    \alpha_\text{BMSS}= -\frac23 \, , \;\;\; \beta_\text{BMSS}=0 \, , \;\;\;  \gamma_\text{BMSS}=\frac13 \, .
    \label{eq:BMSS-exponents}
\end{align} 
We see from \eqref{eq:scalingform} that the physical interpretation of the scaling exponents defined via \eqref{eq:scaling-exponents} is that they describe the instantaneous logarithmic rate of change of the typical occupancy and momentum scales in the distribution function: 
\begin{equation}
    \alpha =\frac{\partial \ln \langle f\rangle}{\partial \ln \tau}, \;\;\; \beta = -\frac{1}{2}\frac{\partial\ln\langle p_\perp^2\rangle}{\partial\ln\tau}, \;\;\; \gamma =-\frac{1}{2}\frac{\partial\ln\langle p_z^2\rangle}{\partial\ln\tau}\,.
    \label{eq:scaling-exponents-2}
\end{equation}
The scaling phenomenon introduced by BMSS has been verified in numerous numerical simulations, both in classical-statistical Yang-Mills simulations~\cite{Berges:2013eia,Berges:2013fga}, as well as more recently in QCD EKT~\cite{Mazeliauskas:2018yef}, where it was observed that these exponents could be well-defined even away from their fixed point values. That is to say, the distribution function collapsed onto the form in Eq.~\eqref{eq:scalingform} even before Eq.~\eqref{eq:BMSS-exponents} came to be fulfilled.

The values of the scaling exponents $\alpha$, $\beta$ and $\gamma$ in early-time BMSS dynamics were correctly predicted via analytical arguments~\cite{Baier:2000sb} in early work, but only recently was the analytical description refined to include the evolution of the exponents as the fixed point is approached~\cite{Brewer:2022vkq}. (See also Ref.~\cite{Mikheev:2022fdl} for a stability analysis around the fixed point.) Our discussion in earlier Sections motivates interpreting the fact that the scaling exponents can be extended to times before the fixed point is attained by postulating that the scaling form~\eqref{eq:scalingform} is the \textit{effective} ground state of the system, undergoing adiabatic time evolution, and that this form (with the same functional form $w$ as for the fixed point) is approached because of the emergence of an energy gap between this ground state and
all the other shapes that the distribution function can attain. This is exactly what was conjectured and proven in Section~\ref{sec:intro-AH-BSY}.

The BMSS scaling regime, as described above, is not the only scaling regime to (or through) which a weakly coupled gluon plasma can evolve. We observed in Section~\ref{sec:intro-AH-BSY} that the small angle scattering collision kernel admits another fixed point for the scaling exponents when the system undergoes longitudinal expansion, with a distribution function of the same form as~\eqref{eq:scalingform}. At this fixed point, called the \textit{dilute} fixed point, the scaling exponents are time-independent with
\begin{align}
    \alpha_\text{dilute}= -1 \, , \;\;\; \beta_\text{dilute} = 0 \, , \;\;\;  \gamma_\text{dilute} = 0 \, .
\label{eq:dilute-exponents}
\end{align}
Corrections to the values of the exponents as they approach the fixed point, both in the dilute and BMSS cases, can be calculated explicitly as shown in Section~\ref{sec:intro-AH-BSY}, and were shown to evolve logarithmically in proper time $\tau$. A similar correction has been observed in the approach to fixed points in non-expanding QCD EKT~\cite{Heller:2023mah}.

Another example that is important to keep in mind is that in the absence of collisions, i.e., if one sets $C[f] = 0$, the particles in the kinetic theory stream freely. In this case, the evolution of the distribution function (assuming boost invariant longitudinal expansion and homogeneity in the transverse plane) \textit{always} takes the scaling form as in~\eqref{eq:scalingform}, with $w$ determined by the initial condition. The scaling exponents in this case are given by
\begin{align}
    \alpha_\text{fs}= 0 \, , \;\;\; \beta_\text{fs} = 0 \, , \;\;\;  \gamma_\text{fs} = 1 \, ,
    \label{eq:fs-exponents}
\end{align}
where ``fs'' denotes free streaming. 
We note that the dilute scaling regime we just described, with scaling exponents~\eqref{eq:dilute-exponents}, is distinct from free streaming,  in the sense that dilute scaling evolution only arises in the presence of collisions as it involves a balance between longitudinal expansion and collisions. In the analytic calculations of Section~\ref{sec:intro-AH-BSY}, the term that balances the longitudinal expansion is the $I_a \nabla^2 f$ term in the collision kernel.

All the scaling regimes described above are pre-hydrodynamic. Hydrodynamization must end with the system evolving to a
hydrodynamic state in which the kinetic theory is in local thermal equilibrium, with the distribution function taking the form of a thermal distribution (i.e., Fermi-Dirac, Boltzmann, or Bose-Einstein). 
It has been known since Bjorken~\cite{Bjorken:1982qr} that the hydrodynamic evolution of a boost-invariant longitudinally expanding fluid is itself 
a scaling regime. The kinetic theory description of this regime takes the form
\begin{equation} \label{eq:scalingform-late}
    f(\bm{p},\tau) = w\left(\frac{p}{D(\tau)}\right) \, ,
\end{equation}
where $w(\chi)$ is the appropriate thermal distribution, and $D(\tau)$ plays the role of the local temperature, which evolves as $D(\tau) \propto \tau^{-1/3}$ due to the boost-invariant longitudinal expansion of the system. This distribution function can equivalently be written in the form of~\eqref{eq:scalingform}, with the scaling exponents \eqref{eq:scaling-exponents} then taking values
\begin{align}
    \alpha_\text{thermal}= 0 \, , \;\;\; \beta_\text{thermal} = \frac13 \, , \;\;\;  \gamma_\text{thermal} = \frac13 \, .
    \label{eq:hydro-exponents}
\end{align}

We will encounter all of these scaling regimes --- free-streaming, BMSS, dilute, and hydrodynamic --- later on in this Section.
As we found in Sect.~\ref{sec:intro-AH-BSY}, the AH formalism successfully describes the pre-hydrodynamic evolution of a longitudinally expanding gluon gas with a small-angle scattering collision kernel that begins with free-streaming, is rapidly attracted toward the BMSS fixed point, and subsequently evolves toward the dilute fixed point. We reproduce and further elucidate these findings Sect.~\ref{sec:pre-hydro}. In Sect.~\ref{sec:hydro} we introduce a distinct variation of the AH formalism that is designed to describe the approach to hydrodynamic scaling and see explicitly how AH provides an intuitive understanding of hydrodynamization itself.  Finally, in Sect.~\ref{sec:non-scaling} we shall set up the AH framework that provides a unified description of the early evolution governed by a pre-hydrodynamic attractor (that 
takes the kinetic theory distribution from free-streaming to BMSS to dilute) followed by the evolution governed by a hydrodynamizing attractor that takes the distribution from the dilute scaling regime to the hydrodynamic regime. AH provides a unified intuition for this complex dynamics via elucidating how the approach to each new attractor involves the opening up of new gaps in the instaneous spectrum of an evolving effective Hamiltonian, followed by the adiabatic evolution of the remaining ground state(s). We shall see in Sects.~\ref{sec:hydro} and \ref{sec:non-scaling} that hydrodynamization itself occurs when only one isolated instantaneous  
ground state remains, with that state evolving adiabatically toward the hydrodynamic regime where the scaling exponents take the values \eqref{eq:hydro-exponents}.

First, though, in Sect.~\ref{sec:scaling-H} we describe the conceptual underpinnings of the findings of Section~\ref{sec:intro-AH-BSY} and then, in Sect.~\ref{sec:isotropic},  demonstrate their applicability in a setting that is on the one hand simple but that nevertheless goes beyond the regime described in Section~\ref{sec:intro-AH-BSY}, in so doing setting the stage for a complete description of the hydrodynamization process of gluons in the small-angle scattering approximation using the AH framework.

\subsubsection{Effective Hamiltonians for scaling distributions} \label{sec:scaling-H}

Consider a time before the distribution function $f$ takes a scaling form, which, for definiteness, we take to be of the form (\ref{eq:scalingform}) we just discussed. It is always possible to write $f$ in the form
\begin{equation}
    f(\bm{p},\tau) = A(\tau) w\left(\frac{p_\perp}{B(\tau)},\frac{p_z}{C(\tau)},\tau\right),
\end{equation}
and any choice of $A,B,$ and $C$ would comprise a valid choice of ``frame" for $f$, as all of the time dependence could be moved into the time evolution of $w(\cdot,\cdot, \tau)$. However, as $f$ approaches its scaling form, we expect that there is an optimal ``frame", namely an optimal choice of $A,B,$ and $C$ necessary to match Eq.~\eqref{eq:scalingform} as $f$ enters the scaling regime, which is to say necessary to ensure that $w(\zeta,\xi, \tau)$ approaches the functional form of $w(\zeta,\xi)$ in Eq.~\eqref{eq:scalingform}. Equivalently, we could change from one frame to another (and in so doing seek the optimal frame) by making 
time-dependent rescalings of $p_\perp$, $p_z$ and $\tau$.

Let us examine the behavior of a general rescaled distribution function $w$ to try to understand why it should be attracted to the scaling form $w(\zeta,\xi)$ in Eq.~\eqref{eq:scalingform}. Recalling Eq.~\eqref{eq:general-evol}, we shall do so by recasting the Boltzmann equation describing $f$ into an evolution equation for $w$ of the form
\begin{equation} \label{eq:hdef}
    \partial_y w = -H_{\rm eff} w
\end{equation}
where $y \equiv \ln \left( \frac{\tau}{\tau_0} \right)$, and where we shall refer to the operator $H_{\rm eff}$ as the ``effective Hamiltonian''. The effective Hamiltonian will in general be non-Hermitian, non-linear, and depend on our choice of rescalings $A,B,$ and $C$. Furthermore, the variables that define the space of states will be $\zeta$ and $\xi$, with $\zeta \equiv p_\perp/B$ and $\xi \equiv p_z/C$. 

This effective Hamiltonian is determined explicitly by the form of the collision kernel $C[f]$ and the choice of scaling variables. For a kinetic theory undergoing Bjorken flow, in the coordinate basis it reads
\begin{equation}
    H_{\rm eff} = \frac{\partial_y A}{A} - \frac{\partial_y B}{B} \zeta \partial_\zeta - \left( 1 + \frac{\partial_y C}{C} \right) \xi \partial_\xi - \tau \tilde{\mathcal{C}}[f] = A w(\zeta, \xi, \tau)]_{\substack{p_\perp \, = \, \zeta B \\ p_z \, = \, \xi C }} \,\, ,
\end{equation}
where $\tilde{\mathcal{C}}[f]$ is a linear operator defined such that $\mathcal{C}[f] = \tilde{\mathcal{C}}[f] f$, i.e., such that its action on $f$ reproduces the collision kernel. In the case of the small-angle scattering approximation where $C[f]$ is given by Eq.~\eqref{eq:small-angle-kernel}, the effective Hamiltonian takes the explicit form
\begin{align}
    H_{\rm eff} =\alpha + \beta \zeta \partial_\zeta + (\gamma-1) \xi \partial_\xi & - \tau \lambda_0 \ell_{\rm Cb} I_a \left[ \frac{1}{B^2} \left( \frac{1}{\zeta} \partial_\zeta + \partial_\zeta^2 \right) + \frac{1}{C^2} \partial_\xi^2 \right] \label{eq:Heff-smallanglescatt-BC}  \\
& - \tau \lambda_0 \ell_{\rm Cb} I_b \left[ \frac{2(1+Aw)}{p} + \frac{(1+2Aw)}{p} \left( \zeta \partial_\zeta +  \xi \partial_\xi \right) \right] \, , \nonumber
\end{align}
where here $w = w(\zeta, \xi, y)$ because we have not yet shown explicitly that the adiabatic approximation is satisfied and the system rapidly collapses onto an adiabatic evolution in which $w$ depends only on its first two arguments.  The quantities $\ell_{\rm Cb}$, $I_a$, $I_b$ that appear in Eq.~(\ref{eq:Heff-smallanglescatt-BC}) are given by the expressions~\eqref{eq:IaIb} and~\eqref{eq:lcb-def} from Section~\ref{sec:kin}. Furthermore, we have introduced the scaling exponents  $\beta = - \partial_y B/B$, $\gamma = - \partial_y C/C$, $\alpha = \partial_y A/A$ as in~\eqref{eq:scaling-exponents}. In practice, we will choose $\alpha$ so that the real part of the lowest eigenvalue of $H_{\rm eff}$ is 0.

Finding the eigenstates and eigenvalues of $H_{\rm eff}$ can be challenging. Nonetheless, once a frame has been specified by choosing $A,$ $B$ and $C$ we can in principle write down the set of right eigenstates $\ket{n}_R$ of $H_{\rm eff}$ and decompose any state -- i.e.~any distribution function $f$ specified by the function $w$ -- at a given time $y$ in this basis:
\begin{equation}
    w(\zeta,\xi,y) = \sum_n a_n(y) \braket{\zeta,\xi|n(y)}_R \, ,
\label{eq:w-decomposition}
\end{equation}
where the dependence of 
$\ket{n}_R$ on $y$ contains both explicit and implicit dependencies, in the sense that it depends on $y$ through $B(y), C(y)$, and $w(\zeta,\xi,y)$. Furthermore, constructing the basis states $\ket{n}_R$ used in the decomposition \eqref{eq:w-decomposition} requires knowing the state, meaning that \eqref{eq:w-decomposition} is an implicit equation even at one time $y$.
However, this does not impede the applicability or implementation of the AH picture.

Now consider the notion of adiabaticity quantified by Eq.~\eqref{eq:adiabaticityn0} which was discussed in Sec.~\ref{sec:AH}. Inspecting this adiabaticity condition, we can see a clear connection between scaling solutions and adiabatic ground states: if we are able to choose $A, B$ and $C$ such that the ground state $\ket{0}_R$ takes the scaling form, then  $\delta_A^{(n)} = 0$ is automatically satisfied exactly because $\partial_y \ket{0}_R = 0$. This means that if the system is in the ground state of $H_{\rm eff}$ then the evolution will be adiabatic. 
However, it is not guaranteed that any given scaling form for $w$ will be the ground state, but if this scaling form behaves like an attractor, it is reasonable to hypothesize that this interpretation will hold, and this hypothesis was confirmed analytically in Section~\ref{sec:intro-AH-BSY} in the kinetic theory that was analyzed therein. Concretely, we found that if the term in the collision kernel proportional to $I_b$ is neglected, then if $B$ and $C$ are chosen according to
\begin{align}
    \beta = - \frac{\partial_y B}{B} = - \frac{\tau \lambda_0 \ell_{\rm Cb} I_a }{B^2} \, , & & \gamma =-\frac{\partial_y C}{C} = 1 - \frac{\tau \lambda_0 \ell_{\rm Cb} I_a }{C^2} \, ,
\label{eq:bsyscalings} \end{align}
and $A(y)$ is fixed such that $\alpha = \gamma + 2\beta - 1$, the ground state of $H_{\rm eff}$ is time-independent and the spectrum of states is discrete. Consequently, there is an energy gap between the ground state and all the other states, which means that the occupation of all states other than the ground state decays away exponentially as we have discussed and the ground state quickly comes to dominate.  This is the key physical intuition needed to understand how the sensitivity to the initial conditions in a HIC can be lost at very early times, long before hydrodynamization.

In the remainder of this Section, we shall demonstrate the robustness of the adiabatic description of pre-hydrodynamic evolution and subsequent hydrodynamization in kinetic theory via a set of examples in all of which (unlike in the example treated in Section~\ref{sec:intro-AH-BSY}) the eigenvalue solutions are not known analytically and in which (also unlike in Section~\ref{sec:intro-AH-BSY}) the LHS of the adiabatic criterion \eqref{eq:adiabatic-approx} is not identically zero. We will begin in Sec.~\ref{sec:isotropic} by discussing the example of a dilute, non-expanding, gluon gas. This is a case where the appropriate rescaling will be different from that we have just discussed because there is no reason to treat $p_z$ differently from $p_\perp$. In our analysis of this example, we shall discuss how to select the rescaling so that the evolution is as adiabatic as possible. After working out this comparatively simple example, in Sec.~\ref{sec:expanding} we will return to the longitudinally expanding gluon gas and show that the AH framework works in this case even without the simplifying assumptions made in Section~\ref{sec:intro-AH-BSY}. In Sec.~\ref{sec:non-scaling}, we extend our AH description further, all the way until the distribution becomes hydrodynamic. This will require us to extend the AH framework to describe attractors that are not scaling solutions.

\subsubsection{Example 1: Dilute, non-expanding, gas of weakly coupled gluons} \label{sec:isotropic}

As a simple initial demonstration, we will consider an isotropic, non-expanding, dilute gas of gluons in the weak coupling regime. 
With no expansion terms and in the dilute limit $f \ll 1$, the Boltzmann equation~\eqref{eq:Boltzmann-Milne} with the small-angle scattering collision kernel~\eqref{eq:small-angle-kernel} reduces to
\begin{equation} \label{eq:isotropickineq}
    \frac{\partial f}{\partial \tau} = \lambda_0 \ell_{\rm Cb}[f] \big( I_a[f] \nabla_p^2 f + I_b[f] \nabla_p \cdot (\hat{p} f ) \big) \, .
\end{equation}
A consequence of this simplification is that the equilibrium distribution of this kinetic theory is a Boltzmann distribution because 
$f \propto \exp(-p/T_{\rm eff})$ with $T_{\rm eff} = I_a / I_b$ makes the RHS of Eq.~\eqref{eq:isotropickineq} vanish.

We can further assume that the gluon distribution function $f$ is isotropic, and write the rescaled distribution function
\begin{equation}
    f(p,\tau) = A(\tau) w\left( \frac{p}{D(\tau)}, \tau \right) = A(\tau) w(\chi,\tau)
\end{equation}
as a function of the rescaled momentum $\chi = p/D(t)$. We then write the evolution equation in the form
\begin{equation}
    \partial_\tau w = - H w,
\end{equation}
(a slight variation of Eq.~\eqref{eq:hdef} for this simplified case), and find that the effective Hamiltonian operator $H$ is 
\begin{equation} \label{eq:isotropich}
    H = \alpha + \delta \chi \partial_\chi - \lambda_0 \ell_{\rm Cb}[f] \frac{I_a[f]}{ D^2} \left(\frac{2}{\chi} \partial_\chi + \partial_\chi^2\right) -  \lambda_0 \ell_{\rm Cb}[f] \frac{I_b[f]}{D} \left( \frac{2}{\chi} + \partial_\chi \right) \, ,
\end{equation}
where we have introduced $\delta \equiv - \partial_\tau D / D$.

To study the dynamics of this system numerically, it is necessary to describe the full distribution function in terms of a finite number of dynamical variables. In the AH framework, the natural variables are the basis state coefficients (that specify the occupation of each basis state) determined by the series expansion of the distribution function in a given basis. In principle, any basis of integrable functions suffices for this purpose because the integral of the distribution function over all of momentum space is finite (and corresponds to the spatial number density of gluons). However, in an actual calculation we keep only some finite number of basis states and, in order to be least sensitive to the effects of this truncation, it is best to choose a basis that is well adapted to describe the physical phenomena of interest.  We shall see how to do this in the present simple case here, and then subsequently in each of the more complete examples we introduce in later Sections.

As we have seen in Sec.~\ref{sec:AH}, because of the fact that the time evolution operator of the theory is non-Hermitian, a better description of the eigenstates is achieved if we choose different left and right bases. And, as discussed in Sec.~\ref{sec:ConservationLaws}, because the collision kernel conserves particle number the constant function will always be a left eigenstate in all of our examples, and so it behooves us to include it as one of the left basis states.
Furthermore, because in the particular example that we are considering here the late-time solution of Eq.~\eqref{eq:isotropickineq} is a Boltzmann distribution,\footnote{Not a Bose-Einstein distribution because (motivated by the more realistic examples to come, not by this simple example) we have dropped the $I_b f^2$ term in the collision kernel \eqref{eq:small-angle-kernel}. See Sec.~\ref{sec:AH} and Appendix~\ref{app:f2} for further discussion.} we will use the basis
\begin{equation} \label{eq:boltzmannbasis}
    \psi_i^{(R)} = N_i \,{\rm L}_i^2(\chi) e^{-\chi}, \;\; \psi_i^{(L)} = N_i \,{\rm L}_i^2(\chi)
\end{equation}
where ${\rm L}_i^2(\chi)$ are associated Laguerre polynomials, and $N_i$ is a normalization constant chosen such that the basis satisfies
\begin{equation}
    \int d^3\chi \psi_i^{(L)}  \psi_j^{(R)} = \delta_{ij} \, .
\end{equation}
As we will see below, provided the scaling parameter $D$ is chosen appropriately, this basis provides a quickly convergent expansion of the dynamics near thermalization.

Once the basis is set up, we can calculate matrix elements
\begin{equation} \label{eq:hprojection}
    H_{ij} = \int d^3\chi \psi_i^{(L)}  H \psi_j^{(R)}.
\end{equation}
If we truncate at some finite number of basis states, it is straightforward to calculate an approximate instantaneous ground state by solving the eigenvalue problem of the truncated Hamiltonian, for a given state of the system. Furthermore, for the effective Hamiltonian given in 
Eq.~\eqref{eq:isotropich} and the basis in 
Eq.~\eqref{eq:boltzmannbasis}, we notice that the matrix element associated with the first left basis state (which, as discussed in Sect.~\ref{sec:AH}, projects onto the particle number) is given by
\begin{equation} \label{eq:hfirstrow}
    H_{1j} = \delta_{1j} (\alpha - 3\delta) \, ,
\end{equation}
confirming that the constant function is indeed a left eigenstate of $H$. Moreover, this shows that 
we can guarantee that one eigenvalue of $H$ will be zero by choosing $\alpha = 3\delta$. We anticipate that the right eigenvector associated with this eigenvalue will be the ground state, and that this eigenvector will represent the thermal Boltzmann distribution at late times.

\begin{figure}
    \centering
    \includegraphics[width=\textwidth]{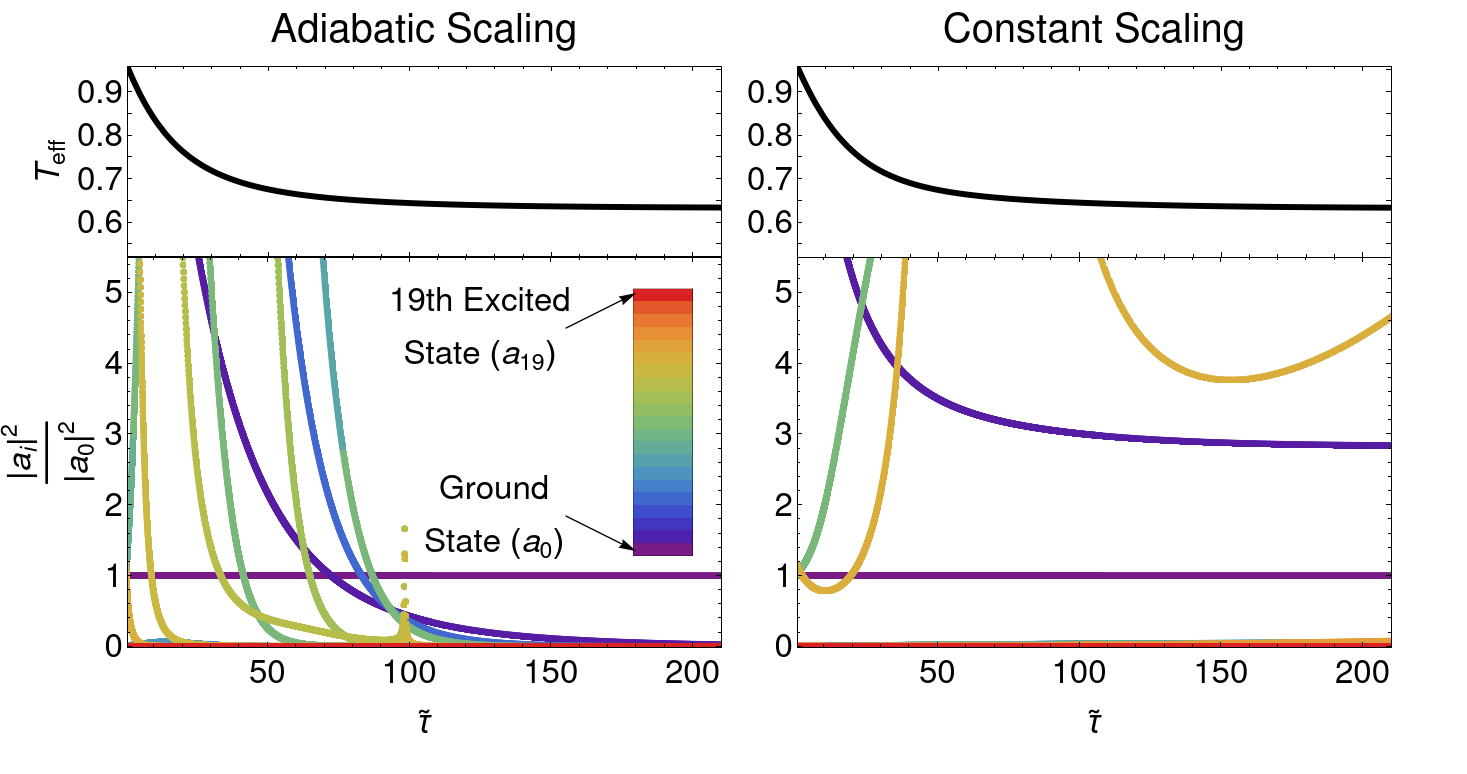}
    \caption{Eigenstate coefficients $a_i$ (bottom panels) ordered by color, and effective temperature $T_{\rm eff} = I_a/I_b$ (top panels) as a function of rescaled time $\tilde{\tau} = \lambda_0 \ell_{\rm Cb} \tau$. In the left panels, the adiabaticity-maximizing choice of scaling $D(\tau)$ described in the text is used, while in the right panels, the scaling is chosen to be constant. 
    Note that the coefficients $a_i$ are normalized relative to the ground state coefficient $a_0$. Therefore, by definition, the ground state occupation appears as a straight line at 1 in both lower panels.
    Both choices for $D(\tau)$ reproduce the same physical dynamics, as exemplified by the identical effective temperatures  in the two top panels. We can see in the lower-left panel that in the adiabatic frame, the ground state of the rescaled distribution function becomes dominant on roughly the same time-scale as the $T_{\rm eff}$ levels off and the system thermalizes, showing that adiabatic evolution provides a reasonable physical interpretation for the rescaled system, even though in this case the initial condition is very far from the instantaneous ground state. Furthermore, the decay of excited state coefficients is somewhat ordered, with the longest-lived excited mode being the first excited state. In the time-independent frame, evolution is clearly non-adiabatic, and the ground state does not become dominant, showing the importance of the choice of frame for understanding the evolution of the system. } \label{fig:moneyplot}
\end{figure}

With these choices, the instantaneous ground state $\ket{0}_R = \ket{0(A,D,\delta,I_a,I_b)}_R$, and we can write
\begin{equation}
    \partial_\tau \ket{0}_R = \left(A\alpha \frac{\partial }{\partial A} - D\delta \frac{\partial }{\partial D} + \frac{\partial \delta}{\partial \tau} \frac{\partial }{\partial \delta} + \frac{\partial I_a}{\partial \tau} \frac{\partial }{\partial I_a} + \frac{\partial I_b}{\partial \tau} \frac{\partial }{\partial I_b} \right) \ket{0}_R\,,
    \label{eq:partial-tau-phi0}
\end{equation}
where because of the dilute limit we in fact have $\frac{\partial I_a}{\partial \tau} = 0$. Then we can seek to minimize the LHS of the adiabaticity condition \eqref{eq:adiabaticityn0}, namely $\delta_A^{(0)}$, by using our ability to choose the one remaining free rescaling parameter $D(\tau)$. We do so by choosing $\frac{\partial \delta}{\partial \tau}$ to minimize $||\partial_\tau \ket{0_R}||^2\equiv\Bra{0}_R\overleftarrow{\partial_\tau}\partial_\tau\Ket{0}_R$. From \eqref{eq:partial-tau-phi0}, this means that we choose
\begin{equation}
    \frac{\partial \delta}{\partial \tau} = -\frac{\text{Re} \Bra{0}_R \overleftarrow{\frac{\partial }{\partial \delta}} \left( A\alpha \frac{\partial }{\partial A} - D\delta \frac{\partial }{\partial D} + \frac{\partial I_b}{\partial \tau} \frac{\partial }{\partial I_b} \right) \Ket{0}_R }{\Bra{0}_R\overleftarrow{\frac{\partial }{\partial \delta}}\frac{\partial }{\partial \delta}\Ket{0}_R}.
\end{equation}
Additionally, for any state
\begin{equation}
    w(\chi,\tau) = \sum_i w_i(\tau) \psi_i(\chi),
\end{equation}
the coefficients evolve according to
\begin{equation}
    \partial_\tau w_i = -\sum_j H_{ij} w_j.
\end{equation}
Using this framework, we can numerically solve for the evolution of the system (through its basis coefficients) and the maximally adiabatic frame $A(\tau),D(\tau)$ by explicitly writing the numerical values of the truncated effective Hamiltonian matrix at each time. Some additional specifics of the numerical implementation of the system evolution are detailed in Appendix \ref{sec:earlynumerical}. In this manner we can solve for an optimally adiabatic choice of time-dependent rescalings $A(\tau)$ and $D(\tau)$ and simultaneously solve for the evolution of the distribution function. An example result of this procedure is shown in Figure \ref{fig:moneyplot}; without dynamical rescaling it is not clear that the AH interpretation holds, but using the adiabaticity-maximizing rescaling, it becomes clear that the ground state dominates on approximately the same time scale as the thermalization of the system in the rescaled picture. The instantaneous ground state that we can identify from the state at the initial time using the maximally adiabatic scaling frame (which we identify without using prior knowledge of the equilibrium solution beyond the choice of basis) proves to be the attractor solution for this system, as we hypothesized in the introduction to this Section. Importantly, unlike in the case considered in Section~\ref{sec:intro-AH-BSY}, the rescaled eigenstates are still time-dependent and although we have found an adiabatic rescaling meaning that the adiabaticity condition {\it is} satisfied it is not satisfied trivially via its LHS vanishing as in Section~\ref{sec:intro-AH-BSY}. 
This is a promising initial demonstration of the broad applicability of the adiabatic framework.

\subsubsection{Example 2: Longitudinally expanding gluon gas} \label{sec:expanding}

We now turn to considering an anisotropic gas of gluons which is expanding only longitudinally, for which we will make no assumptions about whether $f$ is over- or under-occupied. 
We shall assume that the collision kernel takes
the small-angle scattering form~\eqref{eq:small-angle-kernel}.
With these assumptions, Eq.~\eqref{eq:Boltzmann-Milne} 
takes the form
\begin{equation}
    \frac{\partial f}{\partial y} - p_z \frac{\partial f}{\partial p_z} = \tau \lambda_0 \ell_{\rm Cb} \left[ I_a \nabla_{\bm{p}}^2 f + I_b \nabla_{\bm{p}} \cdot \left( \frac{\bm{p}}{p} f (1+f) \right) \right] \,.
\end{equation}
We will assume that the gluon distribution function has azimuthal symmetry. In the next two subsections, we 
shall first describe the early, pre-hydrodynamic, attractor for this kinetic theory and then describe the late-time attractor that can describe hydrodynamization. In both cases we shall use the AH framework, but we shall find it convenient to use different bases 
for the early and late time attractors.

\paragraph{Early-time, pre-hydrodynamic, attractor} \hspace{\fill} \label{sec:pre-hydro}

At early times, we cast $f$ into the rescaled form
\begin{equation}
    f(p_\perp,p_z,y) = A(y) w \left(\frac{p_\perp}{B(y)},\frac{p_z}{C(y)},y\right).
\end{equation}
From here, we can use the effective Hamiltonian in~\eqref{eq:Heff-smallanglescatt-BC} to generate the time evolution of $w$ according to $\partial_y w = - H w$. Furthermore, we make the approximation $p \approx B \zeta$ in the both the denominator of the terms in \eqref{eq:Heff-smallanglescatt-BC} proportional to $I_b$ and the evolution of $I_b$ itself, which is consistent with the fact that at early times the hierarchy $p_z^2 \ll p_\perp^2$ holds. To simplify the analysis, we will also drop 
the $I_b f^2$ term in the collision kernel which corresponds to dropping
the terms in the effective Hamiltonian \eqref{eq:Heff-smallanglescatt-BC} that are explicitly dependent on $w$. We discuss the limitations of this approximation in Appendix~\ref{app:f2}. With these approximations, Eq.~\eqref{eq:Heff-smallanglescatt-BC} becomes
\begin{align} \label{eq:expandingh}
    H =&\alpha + \beta \zeta \partial_\zeta + (\gamma-1) \xi \partial_\xi - \tau \lambda_0 \ell_{\rm Cb} I_a \left[ \frac{1}{B^2} \left( \frac{1}{\zeta} \partial_\zeta + \partial_\zeta^2 \right) + \frac{1}{C^2} \partial_\xi^2 \right] \\ \nonumber
&- \frac{\tau \lambda_0 \ell_{\rm Cb} I_b}{B \zeta} \left[ 2 + \zeta \partial_\zeta + \xi \partial_\xi \right] \,. 
\end{align}
We will use the basis
\begin{equation}
\psi_{ij}^{(R)} = N_{ij} {\rm L}_i^1(\zeta) {\rm He}_{2j}(\xi) \exp{\lbrace-\left(\xi^2/2+\zeta\right)\rbrace}\,, \;\; \psi_{ij}^{(L)} = N_{ij} {\rm L}_i^1(\zeta) {\rm He}_{2j}(\xi)\,,
\end{equation}
where the ${\rm L}_i^1(\zeta)$ are associated Laguerre polynomials, ${\rm He}_{2j}(\xi)$ are probabilist's Hermite polynomials~\cite{abramowitz1965handbook}, and $N_{ij}$ is a normalization constant chosen such that the basis satisfies the orthonormality condition
\begin{align}
    \frac{1}{(2\pi)^2} \int_{-\infty}^\infty d\xi \int_0^\infty d\zeta \; \zeta \; \psi_{ij}^{(L)} \psi_{kl}^{(R)} &= \delta_{ik} \delta_{jl} \,.
\end{align}

\begin{figure}
\centering
\begin{subfigure}{0.65\textwidth}
    \includegraphics[width=\textwidth]{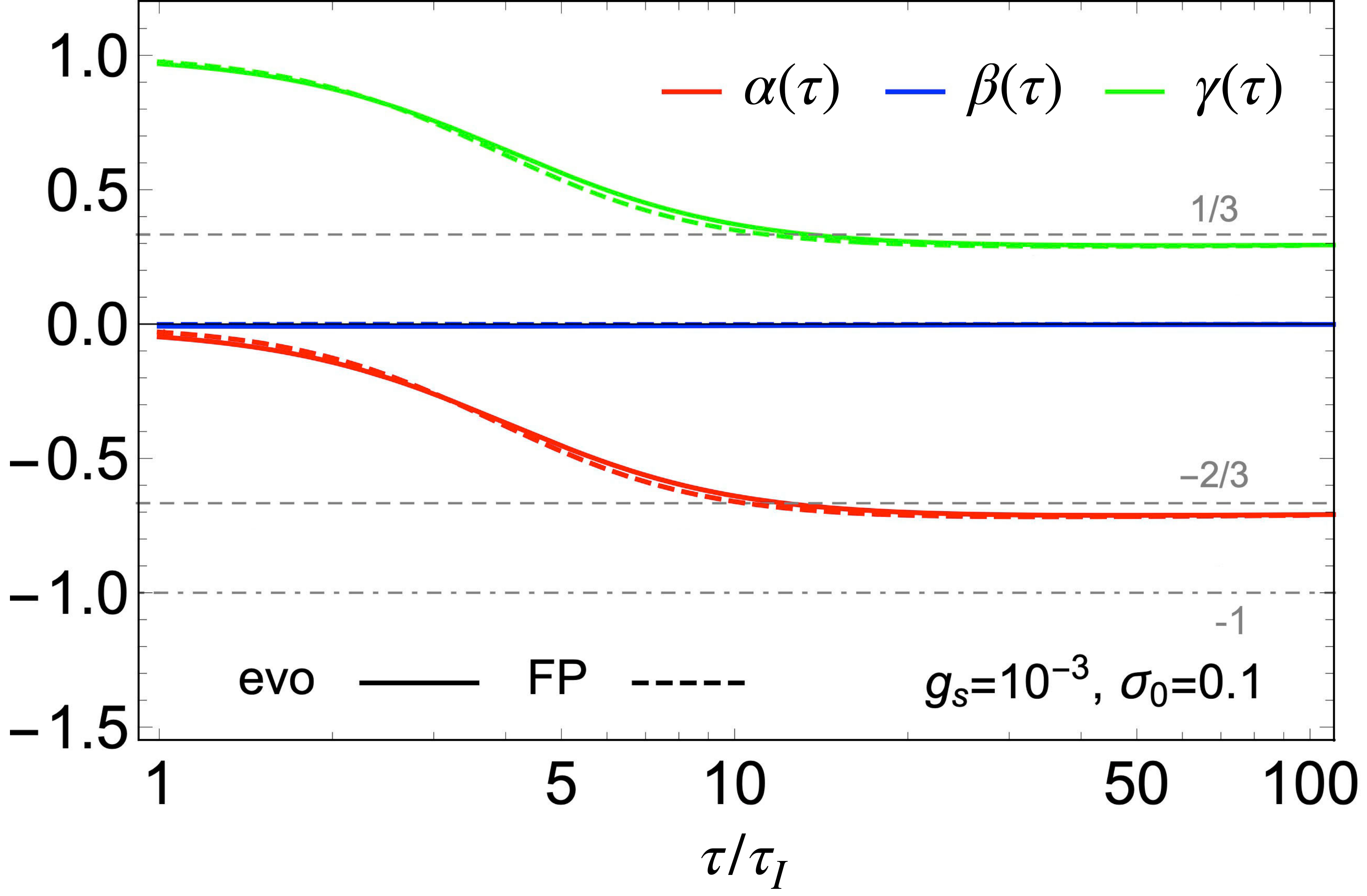}
\end{subfigure}
\hfill
\begin{subfigure}{0.7\textwidth}
    \includegraphics[width=\textwidth]{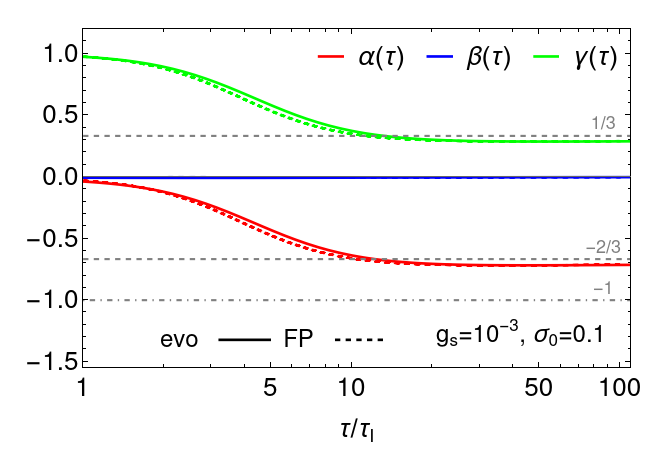}
\end{subfigure}
\caption{A comparison of the time evolution of the extracted scaling exponents $\alpha$, $\beta$ and $\gamma$ in kinetic theory with the weak coupling $g=10^{-3}$ for an initial condition 
 \eqref{eq:BSY-initial-conditions} with $\sigma_0=0.1$ which approaches the BMSS fixed point with $\alpha=-2/3$, $\beta=0$ and $\gamma=1/3$ (indicated by the gray dashed lines;  for later reference, we also include a dot-dashed line describing the dilute fixed point $\alpha = -1$). The top panel is the left panel of Fig.~\ref{fig:FP}; the solid curves represents scaling exponents found analytically in Section~\ref{sec:intro-AH-BSY} by applying the adiabatic hydrodynamization framework to the kinetic theory with a simpler collision kernel than the one we employ.
The colored dashed curves depict the scaling exponents extracted
from a numerical analysis of
the early-time dynamics of the kinetic theory with the full small-angle scattering collision kernel. 
In the bottom panel, the dotted curves are the same numerical calculation as in the top panel, while the solid curves depict the scaling exponents that we have found in this work using the full small-angle scattering collision kernel and adiabaticity-maximizing scalings chosen at each time. We see a satisfying agreement across the three methods.}
\label{fig:bsycomparison1}
\end{figure}

\begin{figure} 
\centering
\begin{subfigure}{0.65\textwidth}
    \includegraphics[width=\textwidth]{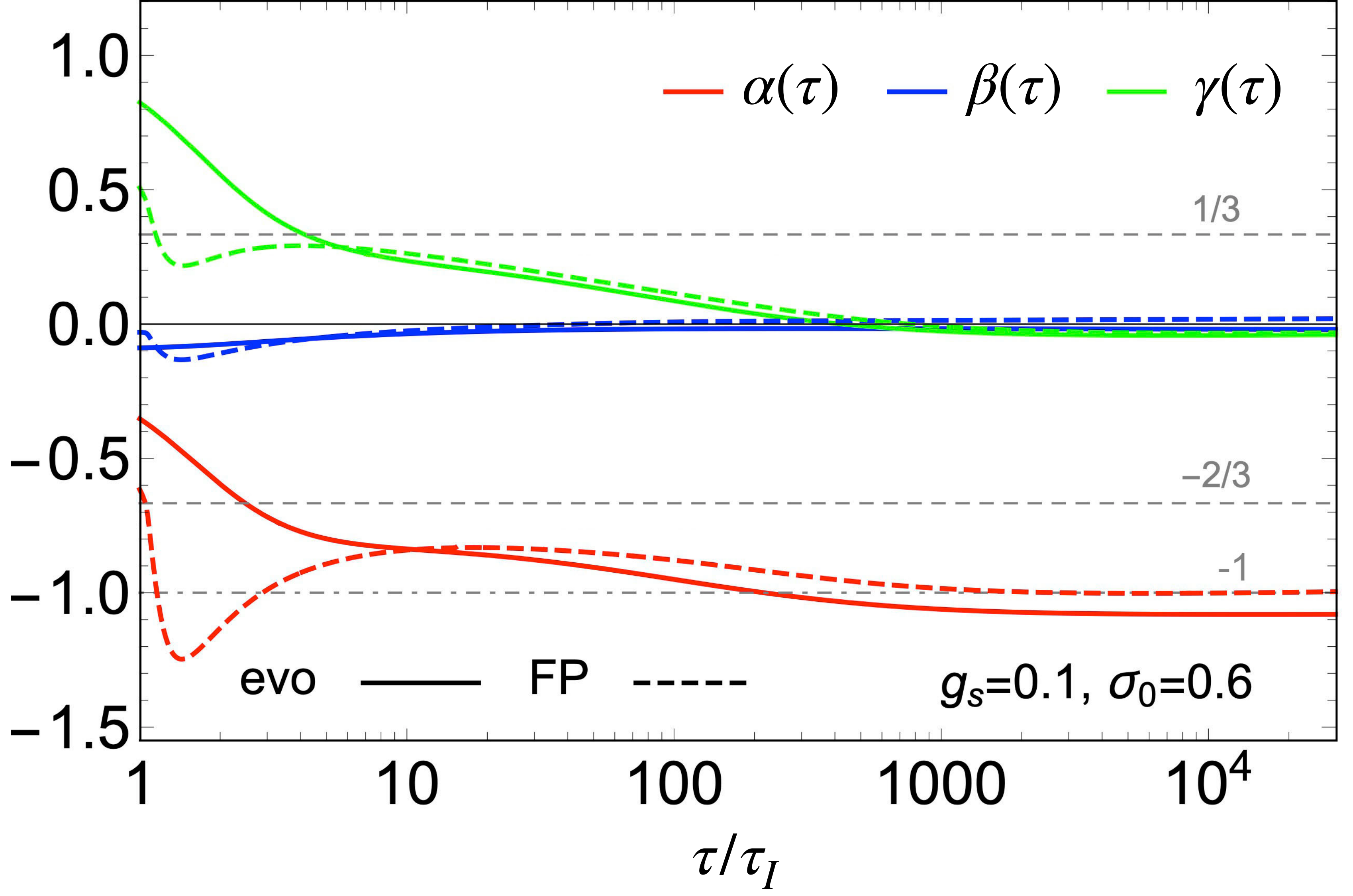}
\end{subfigure}
\hfill
\begin{subfigure}{0.7\textwidth}
    \includegraphics[width=\textwidth]{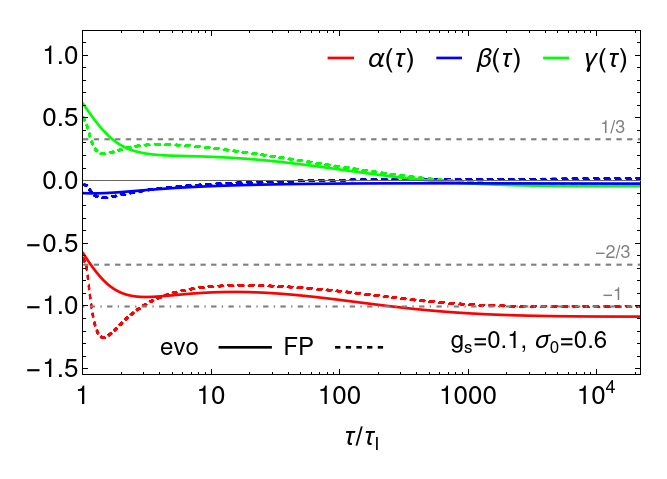}
\end{subfigure}
\caption{A comparison of scaling exponents as in Fig.~\ref{fig:bsycomparison1}, but for a more strongly coupled, highly occupied initial condition {{with $g_s=0.1$ and $\sigma_0=0.6$}} which initially approaches the BMSS fixed point but then evolves to  the ``dilute fixed point'' with $\alpha=-1$, $\beta=\gamma=0$ (indicated with gray dash-dotted lines). As in Fig.~\ref{fig:bsycomparison1}, the colored dashed curves in both panels represent numerically extracted scaling exponents, while the solid curves in the top panel (reproduced from the right panel in Figure~\ref{fig:FP}) were found using the simplified analytic solution of Section~\ref{sec:intro-AH-BSY}. The solid curves in the bottom panel are from this work and were found using the adiabaticity-maximizing method described in the text. All three are in good agreement. At early and intermediate times, our new adiabatic hydrodynamization results perhaps agree somewhat better with the scaling exponents extracted numerically 
than the results in the top panel (coming from Section~\ref{sec:intro-AH-BSY}) do.
}
\label{fig:bsycomparison2}
\end{figure}

\begin{figure}
\centering
\begin{subfigure}{0.7\textwidth}
    \includegraphics[width=\textwidth]{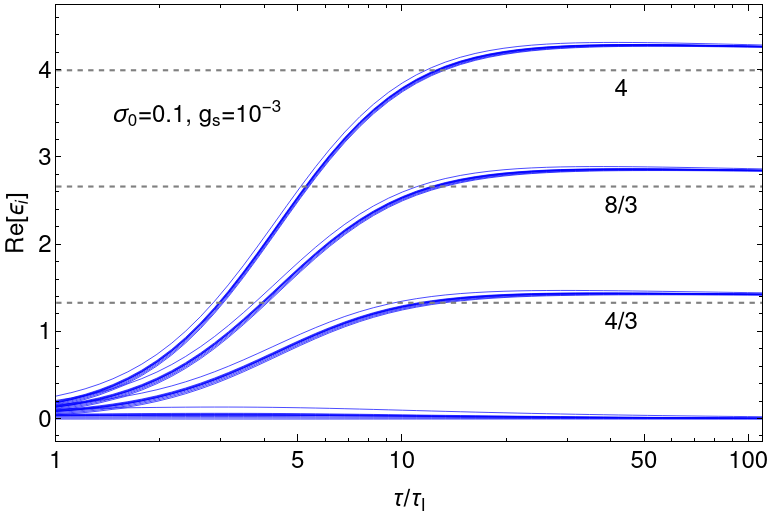}
    \label{fig:bsy1energy}
\end{subfigure}
\hfill
\begin{subfigure}{0.7\textwidth}
    \includegraphics[width=\textwidth]{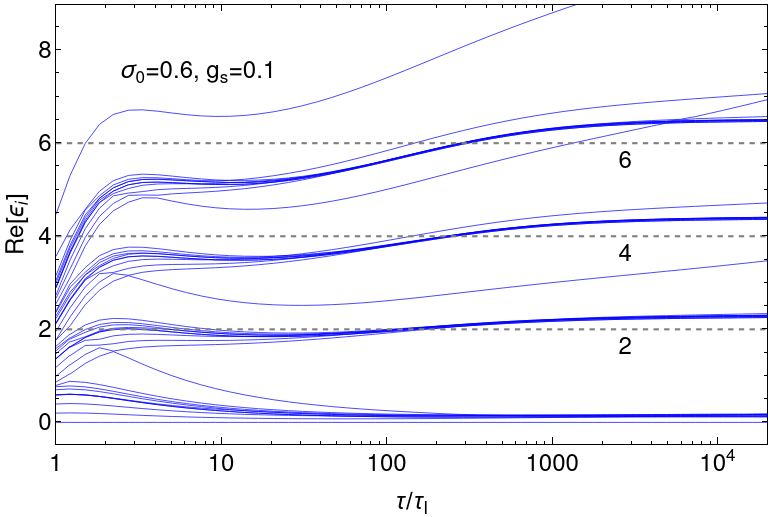}
    \label{fig:bsy2energy}
\end{subfigure}
\caption{Eigenvalues corresponding to the time-dependent eigenstates of the effective Hamiltonian $H$ for the two initial conditions presented in Figs.~\ref{fig:bsycomparison1} and \ref{fig:bsycomparison2}. For both initial conditions, many of these effective energy levels are 
clustered into 
degenerate or near-degenerate groups for much of the system's evolution, with large energy gaps between these groups at late times. We can associate each cluster of eigenvalues with a longitudinal mode, and any splitting within a cluster comes from the small effects of differing transverse modes, as discussed in the text.  Dotted lines show the energy levels expected from Section~\ref{sec:intro-AH-BSY} (see Eq.~\eqref{eq:BSYenergies}) using 
$\alpha=-2/3$, $\beta=0$, $\gamma=1/3$ for the top panel 
and
$\alpha=-1$, $\beta=\gamma=0$ for the bottom panel 
to reflect the approximate late-time values of these scaling exponents, as seen in Figs. \ref{fig:bsycomparison1} and \ref{fig:bsycomparison2} respectively.}
\label{fig:energies}
\end{figure}

\begin{figure} 
\centering
\begin{subfigure}{.7\textwidth}
    \includegraphics[width=\textwidth]{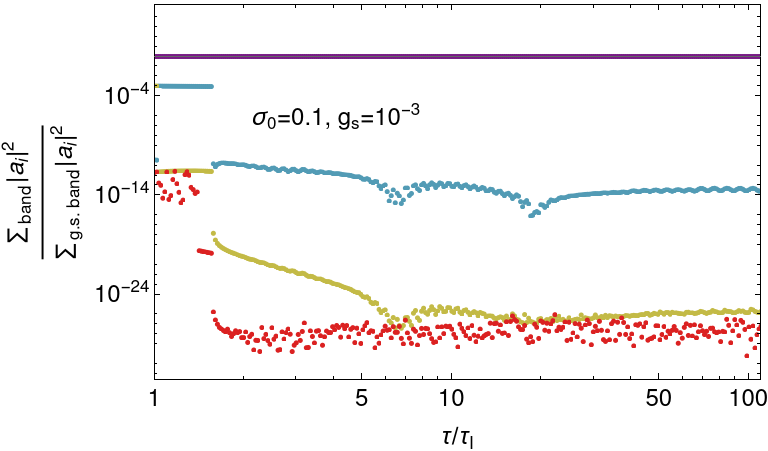}
\end{subfigure}
\hfill
\begin{subfigure}{.7\textwidth}
    \includegraphics[width=\textwidth]{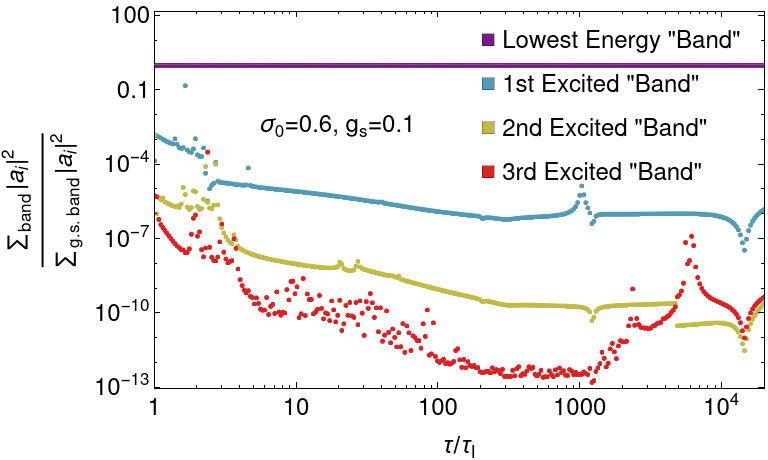}
\end{subfigure}
\caption{Sums over each energy ``band'' of eigenstate coefficients $a_i$ that tell us about the occupation of each band of eigenstates in the distribution function as a function of time for the two initial conditions presented in Figs.~\ref{fig:bsycomparison1} and \ref{fig:bsycomparison2}. $N_\perp=10$ and $N_z=4$ are the number of transverse and longitudinal basis states, respectively. Then each of the $N_z$ ``bands'' contains $N_\perp$ states, ordered by energy. That is, the first ``band" contains the states with the lowest $N_\perp$ energies, the second contains the states with the next-lowest $N_\perp$ energies, etc. The large downward jump in the excited bands at an early time ($\tau/\tau_I\approx 1.6$) in the more weakly coupled example (top panel) is due to a level crossing in which one state moves from the second band to the first band. We report the coefficients in this way because (as seen in Fig. \ref{fig:energies}) the energies of the eigenstates are clustered into well-separated groups at late times. Therefore rather than comparing eigenstate occupation to occupation of a single ground state, it is more meaningful to compare occupation of the higher-energy bands of eigenstates to the lowest-energy band. We can see that for both of these initial conditions, the state of the system is initially very close to its instantaneous ground state, and then remains very close to its ground state as the ground state evolves. This is consistent with the notion of adiabatic evolution.}
\label{fig:coefficients}
\end{figure}

As in the previous example, we will choose $\alpha$ in such a way as to guarantee a zero eigenvalue. Calculating matrix elements according to
\begin{align}
    H_{ijkl} \equiv \frac{1}{(2\pi)^2} \int_{-\infty}^\infty d\xi \int_0^\infty d\zeta \; \zeta \; \psi_{ij}^{(L)} H \psi_{kl}^{(R)} \,,
\end{align}
we note that the first row of the effective Hamiltonian with our choice of basis is
\begin{equation}
    H_{1j1l}=\delta_{1j}\delta_{1l}(\alpha-2\beta-\gamma+1)\ .
\end{equation}
We therefore choose
\begin{equation}
    \alpha = \gamma + 2\beta - 1 
\end{equation}
to ensure that the left basis state which projects onto the particle number has zero eigenvalue.
We will then choose $\gamma(y)$ and $\beta(y)$ according to the adiabatic solutions for these scaling exponents found previously in Section~\ref{sec:intro-AH-BSY} upon analyzing the kinetic theory with a simpler version of the collision kernel (omitting the term proportional to $I_b$) than we are considering here. That is, we choose $\gamma(y)$ and $\beta(y)$ according to Eq.~\eqref{eq:bsyscalings}. In the case of Section~\ref{sec:intro-AH-BSY}, we found that it is possible to choose scalings $\gamma(y)$ and $\beta(y)$ such that their simplified effective Hamiltonian is completely time independent, and therefore perfectly adiabatic. We now adopt the same scalings as before in the hope that the evolution of the rescaled system will still be adiabatic, even though the inclusion of $I_b$ terms in the collision kernel means that our effective Hamiltonian will have some time dependence.
For the sake of comparison with Section~\ref{sec:intro-AH-BSY}, we use the same initial conditions: 
\begin{equation}
    f(p_\perp,p_z;\tau_I) = \frac{\sigma_0}{g_s^2}\exp \left( - \frac{p_\perp^2+\xi_0^2 p_z^2}{Q_s^2}\right)
    \label{eq:BSY-initial-conditions}
\end{equation}
with initial anisotropy $\xi_0=2$ and initial time $\tau_I Q_s=70$, and the values of the coupling $g_s$ and overoccupancy of hard gluons $\sigma_0$ in Figs.~\ref{fig:bsycomparison1} and \ref{fig:bsycomparison2} chosen as in Section~\ref{sec:intro-AH-BSY}. The scaling exponents found for these initial conditions in the way we have described above are plotted in 
Figs.~\ref{fig:bsycomparison1} and~\ref{fig:bsycomparison2} and, as expected, the scaling exponents we extract match both the results from Section~\ref{sec:intro-AH-BSY} and the numerically calculated scaling exponents very well.  
Although this prescription for the scaling exponents 
does not yield a time-independent effective Hamiltonian with perfectly adiabatic time evolution, 
we can see that the eigenvalues of the extracted time-dependent rescaled eigenstates, shown in Fig.~\ref{fig:energies}, are similar to the eigenvalues calculated in Section~\ref{sec:intro-AH-BSY}:
\begin{equation} \label{eq:BSYenergies}
    \epsilon_{nm} = 2n(\gamma-1)-2m\beta \;\;\;\;\; n,m=0,1,2,\dots\,.
\end{equation}
As can be seen both in Fig.~\ref{fig:energies} and in this analytic expression (with the understanding that $\beta$ is small relative to $\gamma$), there is a large separation between the energies of states which have differing longitudinal modes, while within each longitudinal mode the transverse modes are nearly degenerate. That is, unlike in the simple model presented in Sec.~\ref{sec:isotropic} where there was a single ground state separated from all other states by a gap, here we have a subset of states which act collectively as a ``ground state" in our adiabaticity picture, with these states nearly degenerate with each other and separated from all other states by a gap. 
As such, in Fig.~\ref{fig:coefficients}, we compare the relative occupation of the clustered subsets of states rather than occupation of individual eigenstates. We see that for the initial conditions considered, which begin with the system primarily occupying the lowest energy set of states, the system remains very near this ``ground" set of states as the system evolves. In that sense, we might call the system's evolution ``quasi-adiabatic'', since despite the lack of a single adiabatic ground state, we still have a picture in which we have a restricted set of preferred degrees of freedom for the system.
We shall see in Sect.~\ref{sec:non-scaling}  that at later times as the kinetic theory hydrodynamizes, the energies of all but one of this set of ground states rises and the system evolves into a single isolated adiabatically evolving state.

Unfortunately, the choice of basis that we have used here to describe the early pre-hydrodynamic attractor is limited
in its utility later, during hydrodynamization, because
we have employed the approximation $p \approx p_\perp$ in the effective Hamiltonian and in $I_b[f]$. Once the system begins to isotropize, a different basis will be necessary in order to continue to evolve $f$ to hydrodynamization.

\paragraph{Late-time, hydrodynamizing attractor} \hspace{\fill} \label{sec:hydro}

The approximation $p \approx p_\perp$ made in the previous 
Section to calculate the matrix elements was motivated by the fact that at very early times the longitudinal expansion drives $p_z$ downward, making the momentum distribution anisotropic with $p_z\ll p_\perp$.
This motivated us to choose basis states that are products of functions of $p_z$ and $p_\perp$. 
In this Section, we seek to describe hydrodynamization within the AH framework.
In the late stages of hydrodynamization, the distribution function must approach local thermal equilibrium, meaning that the momentum distribution must approach isotropy.
To describe hydrodynamiztion, therefore, it behooves us to choose a basis in which an isotropic
thermal distribution 
is well described by the set of states that the (truncated) basis spans. A natural choice of coordinates via which to accomplish this is $p = \sqrt{p_\perp^2 + p_z^2}$ and $u = p_z/p$. 
As before, we introduce a rescaling in the $p$ coordinate such that $p = D(y) \chi$ to find an adiabatic frame. However, because $u$ is a coordinate that takes values in a bounded interval $-1 \leq u \leq 1$,  rescaling $u$ to a different variable $\tilde{u} = r(y) u $ would induce a more complicated inner product in terms of $\tilde{u}$ because the limits of integration would become time-dependent. For this reason, we shall not introduce any rescaling of $u$ in the present discussion. (However, in the next Section we shall need to introduce the parameter $r(y)$ as a way of recovering adiabaticity beyond the regime where scaling solutions exist.)

With the motivations above, in this Section we cast $f$ in the form
\begin{equation}
    f(p,u,y) = A(y) w \left( \frac{p}{D(y)} , u, y \right) \, ,
\end{equation}
where $w = w(\chi, u , y)$. As in the previous Section and as discussed in Appendix~\ref{app:f2}, we will drop the $I_b f^2$ terms in the collision kernel, which corresponds to dropping the explicitly $w$-dependent terms in the effective Hamiltonian. With this, the effective Hamiltonian that generates the time evolution of $w$ is given by
\begin{align} 
    H &= \alpha + \delta \chi \partial_\chi - u^2 \chi \partial_\chi - u(1-u^2) \partial_u - \tau \lambda_0 \ell_{\rm Cb} \frac{I_b}{D} \left( \frac{2}{\chi} + \partial_\chi \right) \nonumber \\
    & \quad   - \tau \lambda_0 \ell_{\rm Cb} \frac{I_a}{D^2} \left[ \frac{2}{\chi} \partial_\chi + \partial_\chi^2 + \frac{1}{\chi^2} \partial_u \left( (1 - u^2) \partial_u f \right) \right] \, . \label{eq:Heff-thermal-noBose}
\end{align}

Next, we choose the basis 
\begin{align}
    \psi_{nl}^{(R)} = N_{nl} e^{-\chi} L_{n-1}^{(2)}(\chi) P_l(u) \, , & & \psi_{nl}^{(L)} = N_{nl} L_{n-1}^{(2)}(\chi) P_l(u) \, ,
\end{align}
where $L_{n-1}^{(2)}(\chi)$ are associated Laguerre polynomials, $P_l(u)$ are Legendre polynomials, and the normalization coefficients $N_{nl}$ are such that
\begin{align}
    \frac{1}{4\pi^2} \int_{-1}^1 du \int_0^\infty d\chi \, \chi^2 \psi_{mk}^{(L)} \psi_{nl}^{(R)} = \delta_{kl} \delta_{mn} \, .
\end{align}
Since the thermal state of the system described by Eq~\eqref{eq:Heff-thermal-noBose} is a Boltzmann distribution (and not a Bose-Einstein distribution because we dropped the $I_b f^2$ term in the collision kernel), this basis is well-equipped to describe it.

As before, the eigenvalue associated with the mode that carries the particle number can be calculated explicitly. In this case, it is $\alpha - 3\delta + 1$ (where here $\delta\equiv -\partial_y D/D$), and so we set $\alpha=3\delta -1$ so as to fix this eigenvalue to zero. 
Since the ground state is the state that will carry the particle number at late times, we are thus guaranteeing that the late-time thermal distribution has eigenvalue zero.

We then have to choose how we evolve $D$ or, equivalently, $\delta$. In this case, we find that a good description is achieved by taking $D$ to follow the scale that determines the effective temperature at late times, namely
\begin{equation} \label{eq:D-evol}
    \delta=-\frac{\partial_y D}{D} =  -\rho \left( 1 - D \left\langle \frac{2}{p} \right\rangle  \right) \, ,
\end{equation}
where $\rho$ is a dimensionless parameter which we can choose so as to control the speed with which $D(y)$ follows the typical momentum scale of gluons $\langle \frac{2}{p}\rangle^{-1}$. In practice, we choose $\rho = 10$.
When thermalization is approached, $D \to 1/\langle \frac{2}{p}\rangle \sim I_a/I_b = T_{\rm eff}$ where $\langle\frac{2}{p}\rangle \sim I_b/I_a$. Note that $\langle\frac{2}{p}\rangle = I_b/I_a$
only for a dilute system with $f \ll 1$.
Nonetheless, because the final state of the system in the examples we consider is in fact dilute, in practice we will often ignore the distinction.

We choose the initial condition to be
\begin{equation} \label{eq:init-cond-iso}
    f({\bf p}, \tau = \tau_I) = \frac{\sigma_0}{2 g_s^2} e^{- \sqrt{2} p/Q_s}
\end{equation}
and initialize the system at $\ln (\tau_I Q_s ) = 4$, with $\sigma_0 = 10^{-2}$, and scan over couplings $g_s \in \{1, 1.5, 2, 2.6\}$. 
{{In order to ensure that we initialize the evolution in this Section close enough to the hydrodynamization stage of the bottom-up scenario of 
BMSS~\cite{Baier:2000sb}, which occurs after the pre-hydrodynamic stages that we described via AH in Sect.~\ref{sec:pre-hydro},
we could initialize the distribution function at a later time $\tau_I$ with a lower initial occupancy $\sigma_0$ than we employed in that section.
What we have done instead is to choose a smaller $\sigma_0$ with a comparable $\tau_I$, relative to the choices we made in Sect.~\ref{sec:pre-hydro},
but to employ substantially
larger values of the coupling $g_s$.
The stronger coupling ensures that the dynamics of the system will have been more rapid in the previous stages, making these choices appropriate.}} Our goal in Sect.~\ref{sec:non-scaling} will be to find a single formulation that describes the early time dynamics that we 
analyzed in Sect.~\ref{sec:pre-hydro} followed smoothly and adiabatically by the hydrodynamization that we describe here.  Since here we are treating the two regimes separately, we must initialize our analysis of the late-time, hydrodynamizing, attractor here in a way that resembles the state of the system at the end of our AH analysis of the pre-hydrodynamic scaling regime in Sect.~\ref{sec:pre-hydro}.

\begin{figure}
    \centering
    \includegraphics[width=0.49\textwidth]{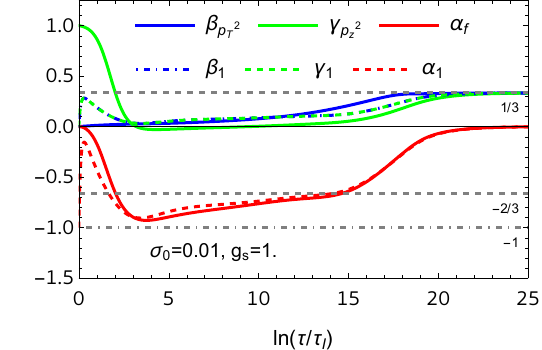}
    \includegraphics[width=0.49\textwidth]{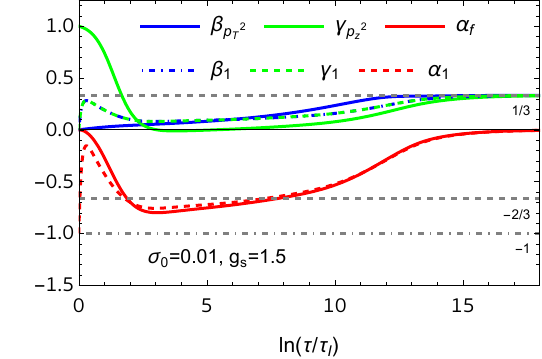}
    \includegraphics[width=0.49\textwidth]{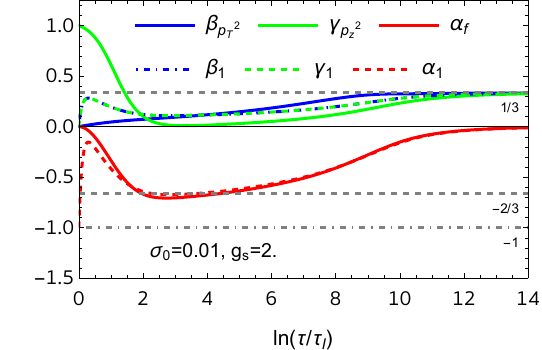}
    \includegraphics[width=0.49\textwidth]{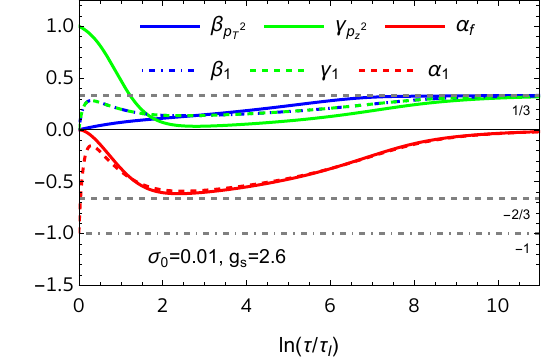}
    \caption{Evolution of the typical momentum scales encoded in the scaling exponents $\alpha, \beta, \gamma$ for the initial condition specified in Eq.~\eqref{eq:init-cond-iso}. From left to right and top to bottom, $g_s = 1, 1.5, 2, 2.6$. In order to test how well-adapted the basis is to the dynamics of the longitudinally expanding gluon gas as it hydrodynamizes, we plot two sets of scaling exponents: the solid lines describe the scaling exponents as calculated from the moments $\langle p_z^2 \rangle$ and $\langle p_\perp^2 \rangle$ computed using the full distribution, and the dashed lines represent the evolution of those scales as would be prescribed only by a single basis state, the one which carries the particle number and which, we shall see below, becomes the lowest energy eigenstate of the effective Hamiltonian at late times.
    For all four values of the coupling, we see the system hydrodynamize: it follows an attractor solution that brings it to $\alpha=0$, $\beta=\gamma=1/3$ which, as described at the beginning of Section~\ref{sec:scaling}, are the values of the exponents that characterize the kinetic theory of a boost-invariant longitudinally expanding hydrodynamic fluid in local thermal equilibrium. Furthermore, the agreement between the dashed and solid curves indicates that the evolution of the typical momentum scales $\langle p_z^2 \rangle_1$ and $\langle p_\perp^2\rangle_1$ described only by the single basis state that carries the particle number are quite similar to those computed from the full distribution function. This agreement at intermediate as well as at late times suggests that, as we shall indeed confirm below, the evolution of the state of the system as it hydrodynamizes is described well by the adiabatic evolution of a single, lowest energy, state of the effective Hamiltonian. 
    }
    \label{fig:scalings-D}
\end{figure}

To test how well our evolution equation~\eqref{eq:D-evol} performs, Figure~\ref{fig:scalings-D} shows the evolution of $\alpha, \beta, \gamma$ {introduced at the beginning of this Section~\ref{sec:scaling}, evaluated as in \eqref{eq:scaling-exponents-2} via the characteristic momenta and occupancies of the distribution function computed via the full evolution equation, namely}
\begin{align}
    \beta_{\langle p_T^2\rangle} &= - \frac{1}{2} \partial_y \ln \langle p_\perp^2 \rangle \, ,  \\ 
    \gamma_{\langle p_z^2\rangle} &= -\frac{1}{2} \partial_y \ln \langle p_z^2 \rangle \, , \\
    \alpha_{\langle f \rangle} &= \partial_y \ln \langle f \rangle 
\label{eq:scaling-exponents-3}
\end{align}
and as calculated from the single basis state $\psi_{10}^{(R)}$ (which carries the particle number) alone, namely 
\begin{align}
    \beta_1 &\equiv -\frac12 \partial_y \ln \langle p_\perp^2 \rangle_1 = - \partial_y \ln D  \, , \\
    \gamma_1 &\equiv -\frac12 \partial_y \ln \langle p_z^2 \rangle_1 = - \partial_y \ln D \, , \\
    \alpha_1 &\equiv \partial_y \ln \langle f_1 \rangle_1 = - 1 - 3 \partial_y \ln D \, ,
\end{align}
where the average $\langle \cdot \rangle_1$ denotes
\begin{equation}
    \langle X \rangle_1 \equiv \frac{\int_{\bf p} X f_1 }{\int_{\bf p} f_1} \, ,
    \label{eq:single-state-average}
\end{equation}
with $f_1$ being the 
particle-number-carrying basis state
\begin{equation}
    f_1({\bf p}, y) = A(y) \psi_{10}^{(R)}(\chi = p/D(y), u = p_z/p ) \, ,
\end{equation}
and where $A$ is chosen to be proportional to $e^{-y} D^{-3}$, such that $\alpha = 3\delta - 1$. As we can see via the comparison between dashed and solid curves in Fig.~\ref{fig:scalings-D}, the information about how the typical longitudinal and transverse scales evolve in time is described well at late times and reasonably well at intermediate times, provided the separation between the rate of change of the longitudinal and transverse scales is not large.

\begin{figure}
    \centering
    \includegraphics[width=0.7\textwidth]{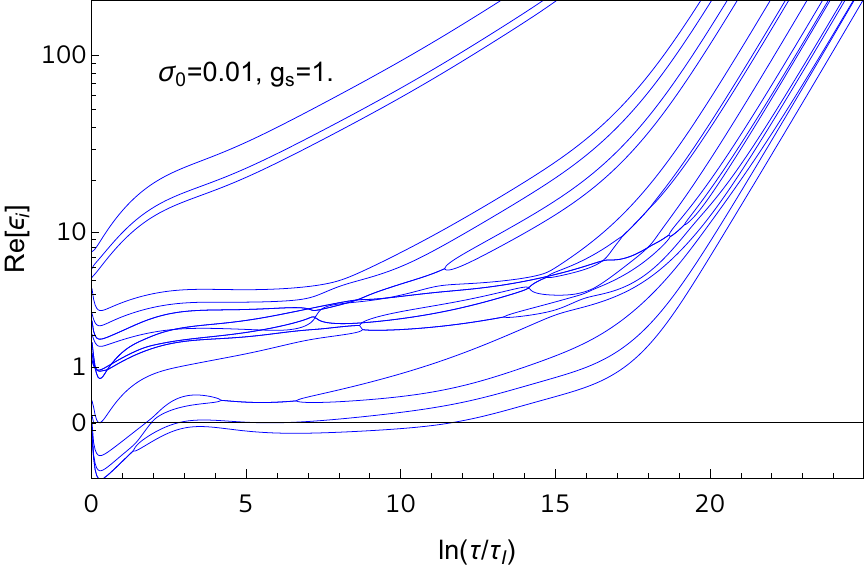}
    \caption{Plot of the eigenvalues of the effective Hamiltonian~\eqref{eq:Heff-thermal-noBose} for $g_s = 1$ and the initial condition given in Eq.~\eqref{eq:init-cond-iso}. We see a cluster of low energy eigenstates (and clusters of higher energy eigenstates also) at early times. Before $\ln(\tau/\tau_I)=15$, all states except one separate from the single ground state of the effective Hamiltonian $H$ that has eigenvalue zero. Comparing to the top-left ($g_s=1$) panel of Fig.~\ref{fig:scalings-D}, we see that the hydrodynamization phenomenon in that Figure is described by the adiabatic evolution of the isolated ground state of the effective Hamiltonian seen here.}
    \label{fig:eigenvals-lateTime}
\end{figure}

\begin{figure}
    \centering
    \includegraphics[width=0.8\textwidth]{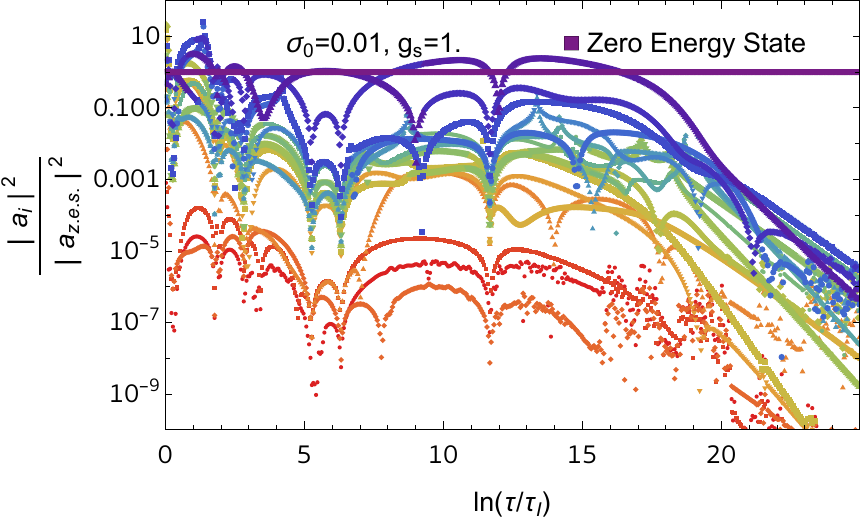}
    \caption{Plot of the coefficients in the $H$ eigenstate decomposition of the distribution function for $g_s = 1$ and the initial condition given in Eq.~\eqref{eq:init-cond-iso}. The coefficients are normalized relative to the occupation of the zero-energy-eigenvalue state that carries the particle number. At late times, the distribution function is composed almost entirely of this state whose adiabatic evolution describes the hydrodynamization of this kinetic theory.}
    \label{fig:coefficients-lateTime}
\end{figure}

\begin{figure}
    \centering
    \includegraphics[width=0.7\textwidth]{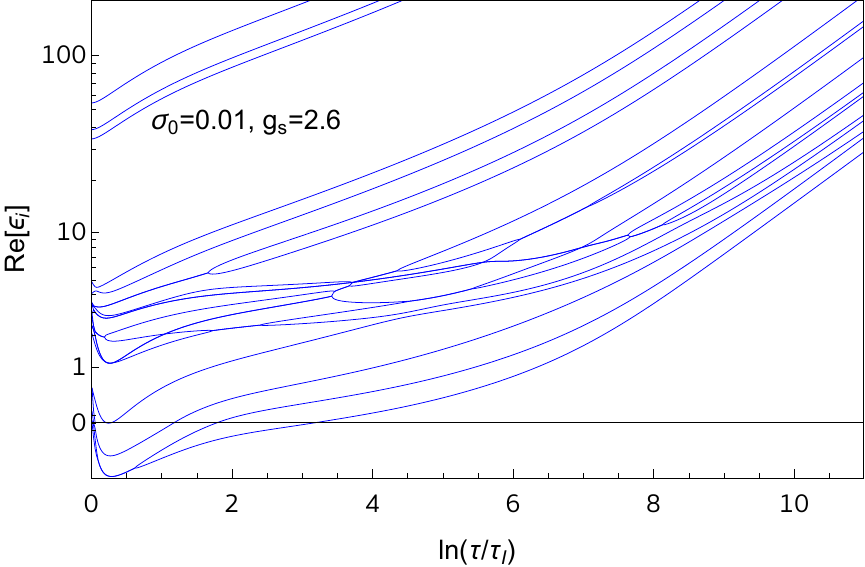}
    \caption{Plot of the eigenvalues of the effective Hamiltonian~\eqref{eq:Heff-thermal-noBose} for $g_s = 2.6$ and the initial condition given in Eq.~\eqref{eq:init-cond-iso}. Before $\ln(\tau/\tau_I)=5$, all states except one separate from the single ground state of the effective Hamiltonian $H$ that has eigenvalue zero. Comparing to the bottom-right ($g_s=2.6$) panel of Fig.~\ref{fig:scalings-D}, we see that the hydrodynamization phenomenon in that Figure is described by the adiabatic evolution of the isolated ground state of the effective Hamiltonian seen here.}
    \label{fig:eigenvals-lateTime-2}
\end{figure}

\begin{figure}
    \centering
    \includegraphics[width=0.8\textwidth]{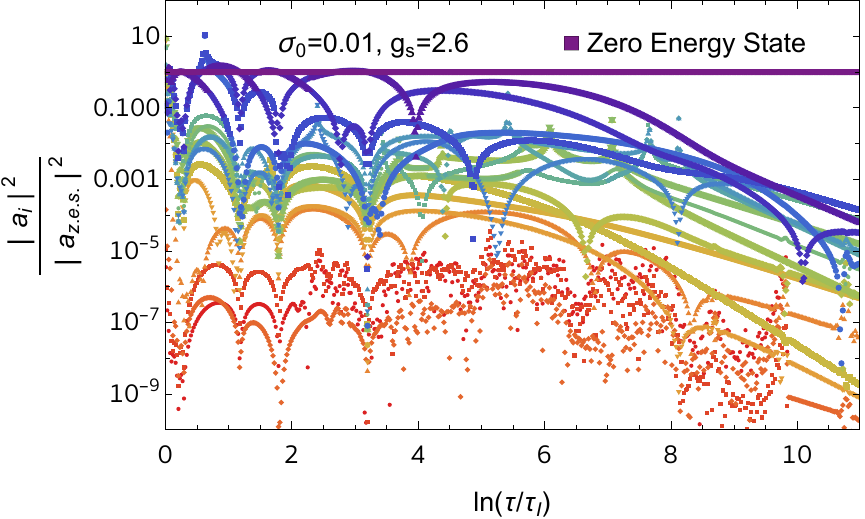}
    \caption{Plot of the coefficients in the $H$ eigenstate decomposition of the distribution function for $g_s = 2.6$ and the initial condition given in Eq.~\eqref{eq:init-cond-iso}. The coefficients are normalized relative to the occupation of the 
    zero-energy-eigenvalue state that carries the particle number. At late times, the distribution function is composed almost entirely of this state whose adiabatic evolution describes the hydrodynamization of this kinetic theory.}
    \label{fig:coefficients-lateTime-2}
\end{figure}

We see {from the final values of the scaling exponents in Fig.~\ref{fig:scalings-D} that all of the solutions we consider here hydrodynamize, reaching the scaling form with the values of the scaling exponents that describe a boost-invariant longitudinally expanding kinetic theory fluid in} local thermal equilibrium at late times. It was not possible to see this with the choice of basis and scalings employed in the previous Section. 
Moreover, as we can see from Figures~\ref{fig:eigenvals-lateTime} and~\ref{fig:eigenvals-lateTime-2}, hydrodynamization follows after an energy gap opens up between a single isolated ground state and all the excited states, with the dynamics from then on being governed by the adiabatic evolution of the ground state of the effective Hamiltonian, with the evolution of this state bringing the system to a thermal distribution at late times.
The dominance of this state, which by construction has energy eigenvalue ${\rm Re}\, \epsilon_n = 0$, is apparent from the late time behavior of 
Figs.~\ref{fig:coefficients-lateTime} and~\ref{fig:coefficients-lateTime-2}.  At late times, during hydrodynamization, the distribution function is composed almost entirely of the adiabatically evolving ground state. The occupation of any of the  higher energy states is small and declining.
Therefore, with the basis and scaling we have introduced here, AH describes an attractor that hydrodynamizes.

In addition to seeing how
pre-hydrodynamic scaling and its attractor (Sect.~\ref{sec:pre-hydro}) and the attractor that brings this kinetic theory through the hydrodynamization process (this Section) are each described naturally via AH, the other important conclusion that we can draw from these results is simply that a longitudinally expanding gas of gluons can reach thermal equilibrium even when the collision kernel only includes small-angle scattering, provided that the coupling is large enough. We shall see in the next Section that if the coupling is too weak (in this kinetic theory with only small-angle scattering)  the system dilutes away before it can hydrodynamize.

\subsection{Adiabaticity beyond scaling}
\label{sec:non-scaling}

In the previous Section, we described the hydrodynamization of a longitudinally expanding gas of gluons in two stages: first, a pre-hydrodynamic attractor; then, an attractor that describes hydrodynamization. For each stage separately, we found a basis and a scaling such that the dynamics are described naturally via the adiabatic evolution of a few (prehydrodynamic attractor) or one (hydrodynamization) lowest energy eigenstate(s) of the (different) effective Hamiltonians that we were able to construct explicitly in each case.  This is a pleasing validation of the power of the AH framework, a confirmation that in two quite different contexts it provides an intuitive description of the loss of memory of earlier stages and the scaling evolution of the attractor.
What would be even better, though, would be a single seamless AH description of the entire dynamics, including the prehydrodynamic attractor and the subsequent hydrodynamization. 
What we have done so far does not tell us how to connect the early time dynamics with the hydrodynamic regime. If we are to capture both of these different scaling regimes that within a single description, that description cannot be described by a single scaling form throughout. This means that to achieve this goal we must
generalize the AH framework beyond the scaling assumption. This is what we do in this Section.

So far, we have discussed how to seek and find adiabatically evolving ground states of theories where the distribution function acquires a scaling form, and all of the time dependence of this state can be encoded in a few time-dependent variables. The main simplification in the dynamics is that by a suitable choice of these time-dependent variables the ground state of the effective Hamiltonian becomes as slowly varying as possible, and therefore evolves adiabatically. From the point of view of the decomposition of the distribution function in basis states, this is implemented by having a basis that is ``co-moving'' with the typical width of the distribution as a function of momentum.

However, a more general picture is possible following the same logic. The essential ingredient of the above is that the basis is well-adapted to the physical dynamics of the problem. When the system undergoes dynamics that drive it to a universal scaling form, what we have discussed so far in the previous sections is ideal. However, when the system transitions between two such phases, the time-dependent parameters of the basis should be adapted in such a way that they can adequately describe the intermediate dynamics.  If we can find such parameters so that the energy levels of the Hamiltonian are gapped and the ground state is slowly evolving in time as before, then we can succeed in connecting the different phases with a simple, unified, adiabatic description.

This is precisely the situation in the kinetic theory for a longitudinally expanding gluon gas with only small-angle scattering in the collision kernel that we have investigated in Sec.~\ref{sec:expanding}. There, we had one scaling form long before hydrodynamics and another as the system hydrodynamizizes. Our goal now is to smoothly connect them. We anticipate that the construction that follows can be extended to more general kinetic theories as well. With this in mind, in Sect.~\ref{sec:non-scaling-H} we will first describe the differences between the approach of this Section and that of Sect.~\ref{sec:scaling-H} in general terms. Then, in 
Sect.~\ref{sec:connectstages} we proceed to describe the boost-invariant kinetic theory of gluons undergoing small-angle scatterings from early to late times, quantitatively.

\subsubsection{Effective Hamiltonians outside the scaling regime} \label{sec:non-scaling-H}

Because of the preceding discussion, we are led to consider distribution functions that are not scaling, but nonetheless have explicit time dependence through a parameter that we introduce by hand. For definiteness, consider the case where we write
\begin{equation}
    f({\bf p},y) = A(y) w \left( \frac{p}{D(y)}, u, r(y), y \right) \, ,
\end{equation}
{{where $u=p_z/p$ as in Sect.~\ref{sec:hydro} and where here we have introduced}} $r(y)$, a parameter that will be encoded \textit{in the definition} of the basis we will expand $w$ on {and that we will use to important ends below. Note now that when we write $\partial_y w$ in what follows, we shall always understand this derivative to be defined as acting only on the last argument of $w$, not on the $y$-dependence inside $r$ and $D$. The idea of this definition is to absorb the ``fast'' evolution of the scales in the distribution function into the parameter $r(y)$ and the scaling function $D(y)$ and, by doing so, make the evolution of the basis state coefficients we choose to set up the description ``slow,'' which in turn enables the eigenstates to be slowly-varying functions of time, as required for AH to take place.}

As before, we assume we have a boost-invariant kinetic theory with a specified collision kernel. The effective Hamiltonian that implements $\partial_y w = - {H}_{\rm eff} w$ is given by
\begin{equation}
    H_{\rm eff} = \frac{\partial_y A}{A} + (\partial_y r) \partial_r - \frac{\partial_y D}{D} \chi \partial_\chi - u^2 \chi \partial_\chi - u(1-u^2) \partial_u  - \tau \tilde{\mathcal{C}}[f = A w(\chi, u, r, y)]_{\chi = p/D(y)} \,\, ,
\end{equation}
where $r(y)$ is a function we have to specify based on the same considerations as before, from which it will follow that the evolution of the distribution function is dominated by the ground state(s) of this effective Hamiltonian.

Upon making the small-angle scattering approximation, this effective Hamiltonian becomes
\begin{align}
    H_{\rm eff} &= \alpha + (\partial_y r) \partial_r + \delta \chi \partial_\chi - u^2 \chi \partial_\chi - u(1-u^2) \partial_u \nonumber \\  
    & \quad - \tau \lambda_0 \ell_{\rm Cb} \frac{I_a}{D^2}  \left[ \frac{2}{\chi} \partial_\chi + \partial_\chi^2 + \frac{1}{\chi^2} \frac{\partial}{\partial u} \left( (1 - u^2) \frac{\partial f}{\partial u} \right) \right] \nonumber \\ 
    & \quad - \tau \lambda_0 \ell_{\rm Cb} \frac{I_b}{D} \left[ \frac{2(1+Aw)}{\chi} + (1+2Aw) \partial_\chi \right] \, , \label{eq:Heff-smallanglescatt-Dr}
\end{align}
where we have reintroduced $\delta \equiv - \partial_y D/D$ and $\alpha \equiv \partial_y A/A$. As before, in what follows we shall drop the $I_b f^2$ terms in the collision kernel, which corresponds to dropping the explicit $w$ terms in the last line of $H_{\rm eff}$.

Other than the fact that there is a new operator in $H_{\rm eff}$ due to the new time-dependent parameter $r(y)$, the next logical steps follow those of Section~\ref{sec:scaling-H} almost identically. If the criterion $\delta_A^{(n)} \ll 1$ is satisfied in tandem with an energy gap between the lowest energy state(s) and all higher energy states, then the ground state will rapidly come to dominate the evolution of the system. When the evolution of the system comes to be governed by the adiabatic evolution of the instantaneous ground state of $H_{\rm eff}$, this provides an intuitive path to understanding how the kinetic theory loses almost all memory of its initial state  as an attractor emerges and to 
identifying what degrees of freedom do get transported from the initial to the final state. Furthermore, here we shall be able to do all of this in a way that follows the evolution of an instantaneous ground state that connects the early-time pre-hydrodynamic evolution in a heavy ion collision smoothly through to
hydrodynamization, and explain the emergence of the small set of low-energy degrees of freedom describing hydrodynamics.

\subsubsection{Example: Longitudinally expanding gluon gas from free-streaming until hydrodynamics} \label{sec:connectstages}

Having discussed in general terms how to find an adiabatically evolving ground state that captures the dynamics of a weakly coupled gluon gas without assuming a single scaling form throughout the evolution, we now discuss the concrete setup that we will use to describe the longitudinally expanding gluon gas from free-streaming through prehydrodynamic attractor to hydrodynamization and hydrodynamics, explicitly and quantitatively.

The first ingredient for this is the choice of basis on which to expand the rescaled distribution function $w(\chi,u,r,y)$. The main requirement that we need in our basis is that it must be able to accurately describe both of the self-similar solutions we have observed in the previous Section: the early-time prehydrodynamic attractor described in Section~\ref{sec:pre-hydro} and the late-time hydrodynamizing attractor described in Section~\ref{sec:hydro}. To this end, we choose
\begin{align}
    \psi_{nl}^{(R)} = N_{nl} e^{-\chi} e^{-u^2 r^2/2} L_{n-1}^{(2)}(\chi) Q_l^{(R)}(u;r) \, , & & \psi_{nl}^{(L)} = N_{nl} L_{n-1}^{(2)}(\chi) Q_l^{(L)}(u;r) \, ,
\end{align}
where the polynomials $Q_l^{(R)}(u;r)$, $Q_l^{(L)}(u;r)$ are polynomials on $u$ of degree $l$, constructed such that
\begin{equation}
    \int_{-1}^1 du \, e^{- u^2 r^2/2} Q_l^{(L)}(u;r) Q_k^{(R)}(u;r) = 2 \delta_{lk} \, .
\end{equation}
That is to say, they are constructed such that they are orthogonal with respect to the measure $e^{- u^2 r^2/2}$ on the $(-1,1)$ interval, and are normalized by setting $Q_0^{(L)} = 1$, $Q_0^{(R)} = 2/J_0(r) $, and $Q_k^{(L)} = J_0(r) Q_k^{(R)}/2$ for $k \geq 1$, where $J_0(r) = \int_{-1}^1 du \, e^{-u^2 r^2/2}$, consistent with the definition~\eqref{eq:Jn-moments} we will introduce later. In fact, in the $r\to 0$ limit, they are equivalent to Legendre polynomials, whereas in the $r \to \infty$ limit they approach Hermite polynomials in shape (after an appropriate rescaling of the $u$ coordinate by $r$).

As before, the normalization coefficients $N_{nl}$ are chosen such that
\begin{align}
    \frac{1}{4\pi^2} \int_{-1}^1 du \int_0^\infty d\chi \, \chi^2 \psi_{mk}^{(L)} \psi_{nl}^{(R)} = \delta_{kl} \, .
\end{align}
In essence, what this basis does is that it reproduces the late-time basis of 
Sect.~\ref{sec:hydro} if $r = 0$, and resembles the early-time basis of Sect.~\ref{sec:pre-hydro} when $r \gg 1$, with the identification $\xi \sim r u$, $\zeta \sim \chi$. In particular, we expect that at early times, when $\langle p_z^2 \rangle \ll \langle p_\perp^2 \rangle$, $r$ will assume the role of $C$ and encode the longitudinal expansion of the system. In this way, it provides us with a parameter that makes the basis flexible enough so that the physical state of the kinetic theory, i.e.~its distribution function, can always be well described by a small set of basis states including both at early and at late times.  This flexibility that this choice of basis incorporates means that the dynamics we describe need not, and will not, be characterized by a single scaling form throughout. We shall see, though, that at sufficiently weak coupling the time evolution dictated by the effective Hamiltonian ensures that at early times the system follows the prehydrodynamic scaling of Sect.~\ref{sec:pre-hydro} and at later times it hydrodynamizes as in Sect.~\ref{sec:hydro}, with the basis chosen so as to yield an efficient description throughout.

In principle, we could choose $r$ by maximizing the degree to which the evolution of the system is adiabatic. However, we know from the previous sections that, at early times, the maximally adiabatic basis choice implies that there is a group of quasi-degenerate ``ground states'' that will drive the evolution. In contrast to this, at later times the ground state is unique and evolves to become the thermal distribution. At early times, even though the gaps among these low-energy states are extremely small, the time derivative of the ground states themselves is also small. As the system approaches thermalization, a gap must open up between an isolated lowest energy state and the other state(s) that were previously almost degenerate, and all these eigenstates of the effective Hamiltonian have to rearrange themselves into different functional forms. Hence, maximizing adiabaticity during this stage of the evolution might be too strict of a condition for what is actually needed, which is that the physical state of the system is dominated by a (set of) ground state(s) throughout. 

Instead of maximizing adiabaticity, we shall choose $r$ here based upon what we already know about the physical behavior of the gluon distribution at early and late times from Sects.~\ref{sec:pre-hydro} and \ref{sec:hydro}, by matching each to an approximate evolution of the distribution function. We motivate our choice of $r$ in the following way: if the distribution function were dependent on $p$ and $u$ in a factorized form as
\begin{equation}
    w \sim e^{-y} \frac{D_0^3}{D^3} g(\chi) \frac{e^{-u^2r^2/2}}{\int_{-1}^1 dv \, e^{-v^2 r^2/2} } \, , \label{eq:w-r-motivation}
\end{equation}
then we can choose $r$ such that the evolution equation for the $\langle u^2 \rangle$ moment of the distribution function is exactly satisfied (for this specific functional form of $w$). That is to say, such that $\int_{\bf p} u^2 \partial_y f = \int_{\bf p} u^2 \big( p_z \partial_{p_z} f  - C[f] \big) $. The idea is to maximize the degree to which the evolution can be approximately described by the first basis state of our construction.
We then introduce the following integral moments
\begin{equation} \label{eq:Jn-moments}
    J_n(r) = \int_{-1}^1 du \, u^n e^{-u^2 r^2/2} \, ,
\end{equation}
which allow us to derive the following equation:
\begin{equation} \label{eq:r-evol}
    \partial_y r = - \frac{1}{r} \frac{J_0}{J_4 J_0 - J_2^2} \left[ -2 ( J_2 - J_4) + \frac{\tau \lambda_0 \ell_{\rm Cb} I_a }{D^2} (J_0 - 3 J_2) \right] \, ,
\end{equation}
which we use to fix the evolution of $r$ in the analysis that follows. {It can be seen from Eq.~\eqref{eq:w-r-motivation} that the parameter $r$ describes how anisotropic the basis states with which we describe the distribution function are. The case when $r = 0$ corresponds to a fully isotropic basis. And indeed, the time evolution described by Eq.~\eqref{eq:r-evol} approaches $r\rightarrow 0$ (isotropizes) at late times in a carefully balanced way: as $\tau$ grows large, the value of $r$ will be driven to a configuration where $J_0 = 3 J_2$
which (by inspection of the expressions for these integral moments)
is only satisfied with $r=0$. (Note that at $r=0$, $J_0=2$, $J_2=2/3$, and $J_4=2/5$.) 
In more detail, the approach to isotropy $r \to 0$ happens in such a way that
the quantity in square brackets on the right-hand side of \eqref{eq:r-evol} evolves toward zero (meaning that $r$ becomes constant) even while
the time-dependent quantity $\tau \lambda_0 \ell_{\rm Cb} I_a/D^2$ grows with time without bound.
That is, $J_0-3J_2$ evolves toward zero inversely proportional to this quantity
with the proportionality constant being
$2(J_2-J_4)$.  This means that 
$r$ evolves toward 0, as that is where $J_0=3J_2$, and this corresponds to isotropization.
With this, we see that the time at which the basis states become isotropic is controlled by the dynamical quantity $\tau \lambda_0 \ell_{\rm Cb} I_a/D^2$.  }

As in \eqref{eq:D-evol}, we let the $D$ scaling evolve as
\begin{equation} \label{eq:Dr-evol}
    \frac{\partial_y D}{D} =  10 \left( 1 - D \left\langle \frac{2}{p} \right\rangle  \right) \, ,
\end{equation}
with the simple idea that $D$ should follow the typical momentum scale of the temperature, and match the (inverse) effective temperature of the system at late times. As before, we choose $A$ such that the eigenstate that carries particle number has a vanishing eigenvalue, i.e., $\alpha = 3\delta - 1$, with $\delta = - \partial_y D / D$.

\begin{figure}
    \centering
    \includegraphics[width=0.49\textwidth]{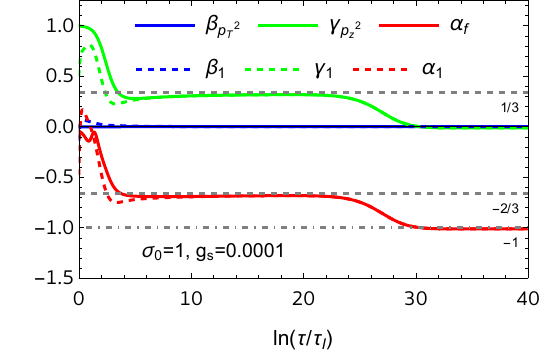}
    \includegraphics[width=0.49\textwidth]{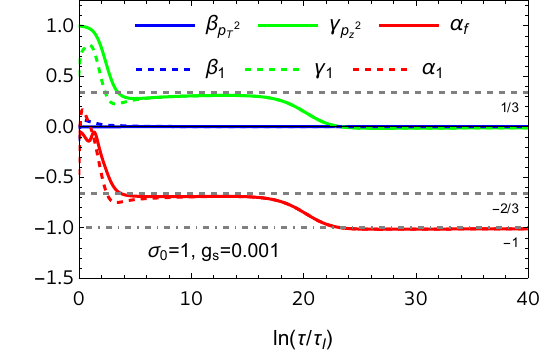}
    \includegraphics[width=0.49\textwidth]{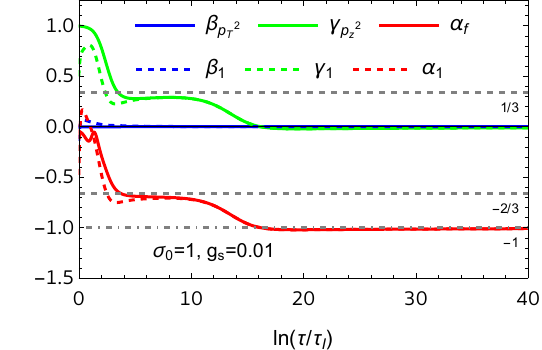}
    \includegraphics[width=0.49\textwidth]{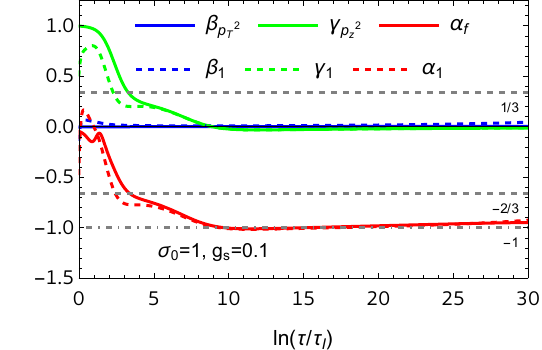}
    \caption{Evolution of the typical momentum scales encoded in the scaling exponents $\alpha, \beta, \gamma$ for weakly coupled kinetic theories with the initial condition specified in Eq.~\eqref{eq:init-cond}. From left to right and top to bottom, $g_s = 10^{-4}, 10^{-3}, 10^{-2}, 10^{-1}$. In order to test how well-adapted the basis is to the dynamics of the gluon gas, we plot two sets of scaling exponents: the solid lines describe the scaling exponents as calculated from the moments $\langle p_z^2 \rangle$ and $\langle p_\perp^2 \rangle$, and the dashed lines represent the evolution of those scaling exponents as described only by the basis state $\psi^{(R)}_{10}$ that carries the particle number.}
    \label{fig:scalings-Dr-weak}
\end{figure}

\begin{figure}
    \centering
    \includegraphics[width=0.49\textwidth]{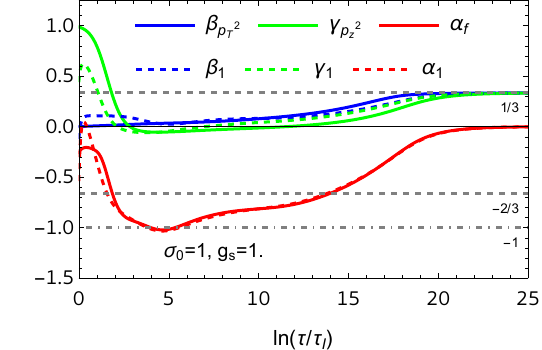}
    \includegraphics[width=0.49\textwidth]{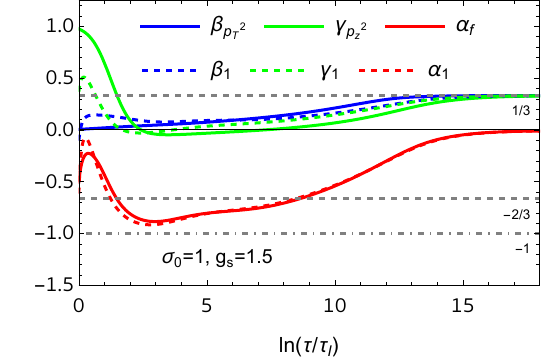}
    \includegraphics[width=0.49\textwidth]{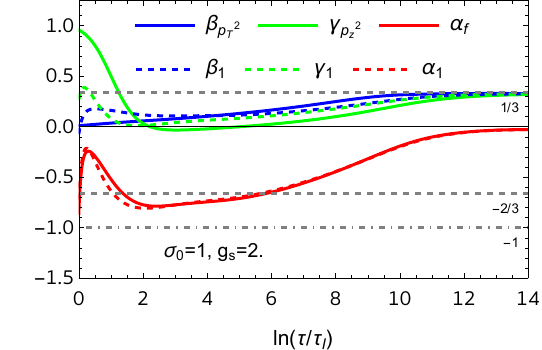}
    \includegraphics[width=0.49\textwidth]{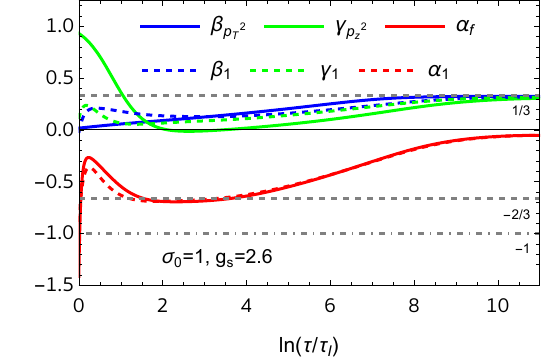}
    \caption{Evolution of the typical momentum scales encoded in the scaling exponents $\alpha, \beta, \gamma$ for more strongly coupled kinetic theories with the initial condition specified by Eq.~\eqref{eq:init-cond} as in Fig.~\ref{fig:scalings-Dr-weak}. From left to right and top to bottom, $g_s = 1, 1.5, 2, 2.6$. As before, the solid lines describe the scaling exponents as calculated from the moments $\langle p_z^2 \rangle$ and $\langle p_\perp^2 \rangle$, and the dashed lines represent the evolution of those scaling exponents as described only by the basis state $\psi^{(R)}_{10}$ that carries the particle number.}
    \label{fig:scalings-Dr-strong}
\end{figure}

At the very early stages, $\tau \to 0$ suppresses the effect of collisions and $r$ becomes large as $r \sim e^y$. When $r$ is large, its evolution equation reduces to that of $C$ in Section~\ref{sec:intro-AH-BSY} with the identification $C = D/r$, where $D$ stays essentially constant because $\langle 1/p \rangle$ is dominated by the $p_\perp$ dependence of the distribution during the BMSS ($\gamma \approx 1/3$ and $\alpha \approx -2/3$) and dilute ($\gamma \approx 0$ and $\alpha \approx -1$) stages 
that we introduced at the beginning of Section~\ref{sec:scaling}, and described in Sect.~\ref{sec:pre-hydro} (see Fig.~\ref{fig:bsycomparison2}). 
Throughout this second stage, $r$ slowly decreases (in a non-exponential way with $y$). As we explained just before, as time progresses the term explicitly proportional to $\tau$ in the evolution equation \eqref{eq:r-evol} for $r$ dominates and the system is driven to a state where $J_2/J_0 = 1/3$,  which corresponds to $ r = 0$, where the basis functions $Q_l$ are Legendre polynomials by construction and thus correspond to spherical harmonics. By itself, this does not mean that the distribution function is isotropic, because this would require that all of the excited states with nontrivial profiles along $u$ decay away, which is something we will verify by showing that an energy gap opens up dynamically as the system hydrodynamizes. Nonetheless, because the hydrodynamic state corresponds to an isotropic distribution, and this state will be dynamically approached from a highly anisotropic state, we expect that having a basis that becomes isotropic at the same time as the system hydrodynamizes will provide a simpler description of the dynamics than an isotropic basis. In this way, this setup is well-equipped to describe isotropization of the gluon distribution and hydrodynamization, as we will demonstrate explicitly in what follows.

In Figs.~\ref{fig:scalings-Dr-weak} and~\ref{fig:scalings-Dr-strong} we see a comparison of the evolution of the typical scales of the problem in terms of the longitudinal and transverse scaling exponents $\gamma$ and $\beta$, respectively, for eight choices of the coupling $g_s$. For these solutions, we have chosen the initial condition to be
\begin{equation} \label{eq:init-cond}
    f({\bf p}, \tau = \tau_I) = \frac{\sigma_0}{g_s^2} e^{- \sqrt{2} p/Q_s} e^{- r_I^2 u^2 /2 } Q_0^{(R)}(u;r_I)
\end{equation}
with $r_I = \sqrt{3}$, $\tau_I Q_s = 1$, and $\sigma_0 = 1$.
Our initial conditions are inspired by previous works~\cite{Kurkela:2015qoa,Mazeliauskas:2018yef,Boguslavski:2023jvg} so as to match them in qualitative terms. Furthermore, they are suitably chosen to study the bottom-up thermalization scenario~\cite{Baier:2000sb} with $f \sim 1/g_s^2$ at the earliest times $\tau Q_s \sim 1$ of a heavy-ion collision when the typical momentum scale characterizing $f$ is $Q_s$. In practice, we choose $\langle p \rangle = 3\, Q_s / \sqrt{2} \approx 2.12 \,Q_s$, which is close to the initial typical momentum in the simulations of Ref.~\cite{Boguslavski:2023jvg}. With all of 
these choices in hand, the only parameter we vary in Figs.~\ref{fig:scalings-Dr-weak} and~\ref{fig:scalings-Dr-strong} is the coupling constant $g_s$.

We stress that our initial condition resembles that of Ref.~\cite{Mazeliauskas:2018yef} more than that of Refs.~\cite{Kurkela:2015qoa,Boguslavski:2023jvg}, which suffices for our purposes as we will not attempt to precisely describe the IR physics of the distribution function. 
  Indeed, such a goal would require us to keep, at the very least, the $I_b f^2$ terms we have neglected in the kinetic equation, and ultimately the $1 \leftrightarrow 2$ processes we have omitted throughout.

To test how well our choice of evolution equations~\eqref{eq:r-evol} and~\eqref{eq:Dr-evol} perform, Figures~\ref{fig:scalings-Dr-weak} and~\ref{fig:scalings-Dr-strong} show the evolution of $\alpha, \beta, \gamma$ as calculated from the characteristic occupancy and momenta of the full distribution via \eqref{eq:scaling-exponents-2}, namely \eqref{eq:scaling-exponents-3}, 
and as calculated from only the evolution of the single basis state that carries the particle number, which is governed by the evolution of $D$ and $r$ {{in such a way that here the resulting scaling exponents are}
\begin{align}
    \beta_1 &\equiv -\frac12 \partial_y \ln \langle p_\perp^2 \rangle_1 = -\frac{1}{2} \partial_y \ln  \left( D^2 - \frac{D^2 J_2(r)}{J_0(r)} \right) \, , \\
    \gamma_1 &\equiv -\frac12 \partial_y \ln \langle p_z^2 \rangle_1 = -\frac{1}{2} \partial_y \ln  \frac{D^2 J_2(r)}{J_0(r)} \, , \\
    \alpha_1 &\equiv \partial_y \ln \langle f_1 \rangle_1 = - 1 - \partial_y \ln \frac{D^3 J_0(r)^2}{J_0(\sqrt{2} r)} \, .
\end{align}
where the average $\langle \cdot \rangle_1$ is as defined previously in \eqref{eq:single-state-average},
where $f_1$ is given by the particle number-carrying state in the basis that we use throughout this Section
\begin{equation}
    f_1({\bf p}, y) = A(y) \psi_{10}^{(R)}(\chi = p/D(y), u = p_z/p ; r(y)) \, ,
\end{equation}
and, as previously mentioned, $A$ is chosen to be proportional to $e^{-y} D^{-3}$, such that $\alpha = 3\delta - 1$.}

As we can see via the (successful) comparison between the dashed and solid curves in Figs.~\ref{fig:scalings-Dr-weak} and~\ref{fig:scalings-Dr-strong}, the information about how the typical longitudinal and transverse scales evolve in time (solid curves) is already well-captured by 
our choices of $D(y)$ and $r(y)$
that determine the dashed curves
for all values of the coupling constant that we have considered. We describe the physical regimes corresponding to the stages seen in the evolution of the scaling exponents in Figs.~\ref{fig:scalings-Dr-weak} and~\ref{fig:scalings-Dr-strong} below, after looking at the behavior of the pressure anisotropy during the same dynamical evolution.

\begin{figure}
    \centering
    \includegraphics[width=0.48\textwidth]{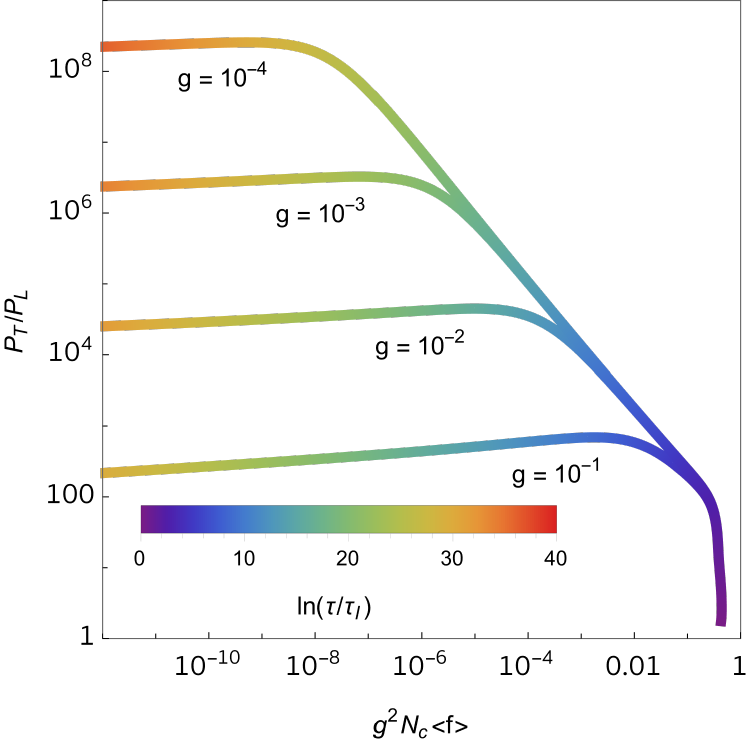}
    \includegraphics[width=0.48\textwidth]{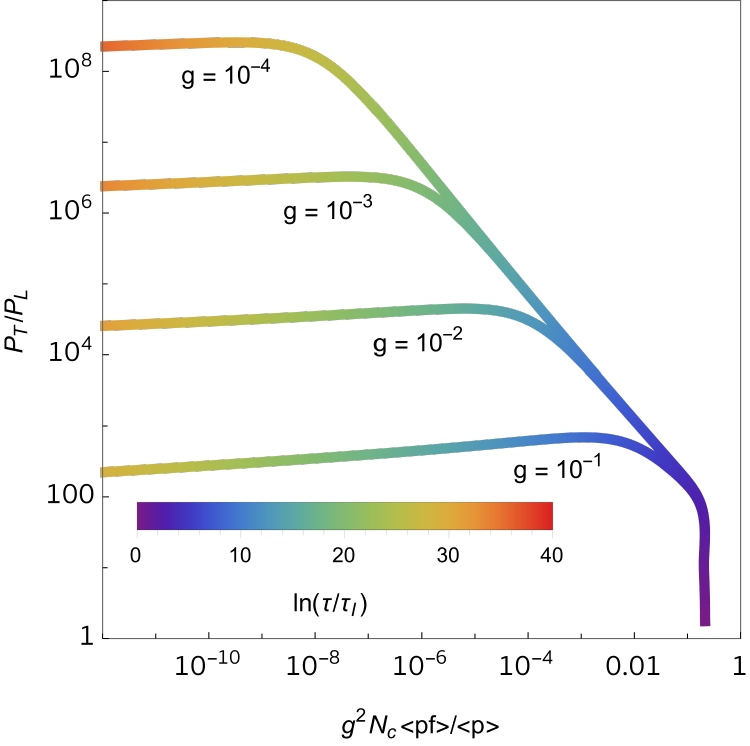}
    \caption{Left: evolution of the pressure anisotropy as a function of the occupancy $g_s^2 N_c \langle f \rangle$ of the distribution function, with the evolution time along each curve depicted by the coloring. Right: evolution of the pressure anisotropy as a function of the energy-weighted occupancy $g_s^2 N_c \langle p f \rangle/\langle p \rangle$ of the distribution function. Both plots were obtained from weakly coupled kinetic theories with $g_s = 10^{-1}, 10^{-2}, 10^{-3}, 10^{-4}$, with initial conditions specified by Eq.~\eqref{eq:init-cond}.}
    \label{fig:occupancies-anisotropies-weak}
\end{figure}

\begin{figure}
    \centering
    \includegraphics[width=0.48\textwidth]{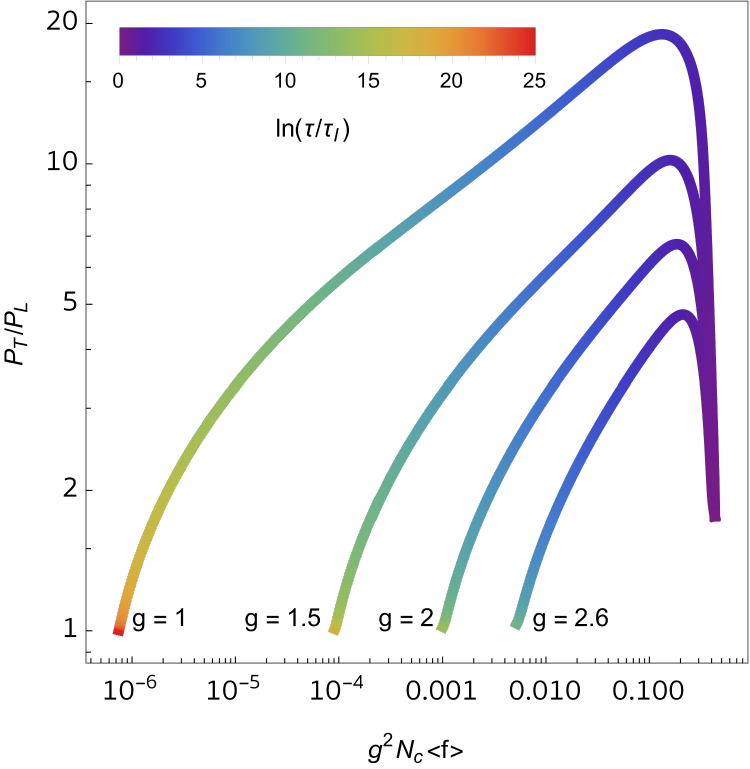}
    \includegraphics[width=0.48\textwidth]{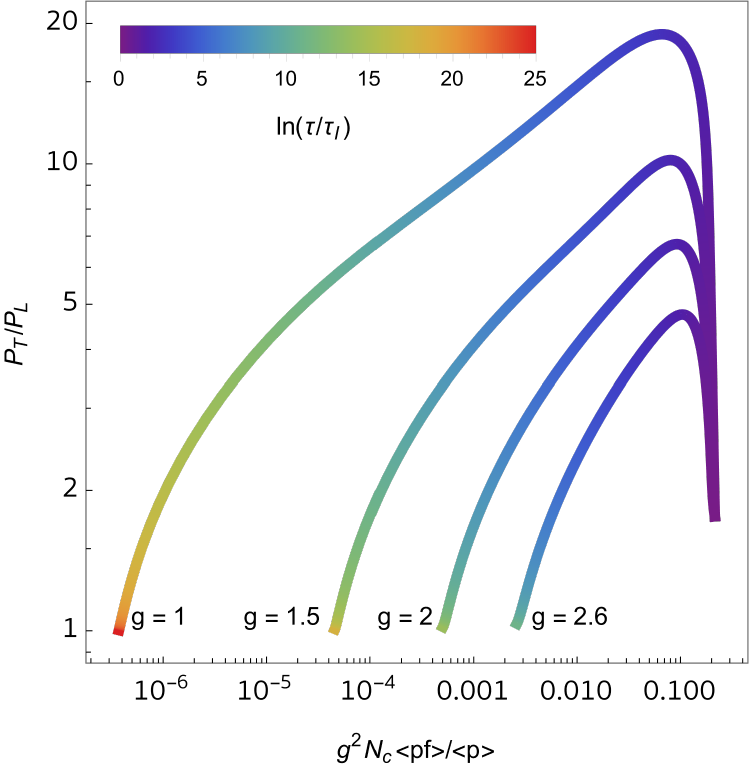}
    \caption{Left: evolution of the pressure anisotropy as a function of the occupancy $g_s^2 N_c \langle f \rangle$ of the distribution function, with the evolution time represented by color. Right: evolution of the pressure anisotropy as a function of the energy-weighted occupancy $g_s^2 N_c \langle p f \rangle/\langle p \rangle$ of the distribution function. Both plots were obtained from more strongly coupled kinetic theories with $g_s = 1, 1.5, 2, 2.6$, with initial conditions specified by Eq.~\eqref{eq:init-cond}.}
    \label{fig:occupancies-anisotropies-strong}
\end{figure}

Let us investigate how the pressure anisotropy and the degree of occupation of the distribution function evolve together as a function of time. We quantify the occupation by calculating $g_s^2 N_c\langle f \rangle$, for the same choice of initial condition and the same values of the coupling constant as in 
Figs.~\ref{fig:scalings-Dr-weak} and~\ref{fig:scalings-Dr-strong}. We follow this quantity together with the pressure anisotropy in the left panel and together with $g_s^2 N_c \langle pf \rangle / \langle p \rangle$ (as introduced in~\cite{Kurkela:2015qoa}) in the right panel of 
Figs.~\ref{fig:occupancies-anisotropies-weak} and~\ref{fig:occupancies-anisotropies-strong}, which show, respectively, weakly coupled kinetic theories ($g_s \in \{10^{-1}, 10^{-2}, 10^{-3}, 10^{-4}\}$) and more strongly coupled kinetic theories ($g_s \in \{1, 1.5, 2, 2.6\}$).

At weak coupling, in Figs.~\ref{fig:scalings-Dr-weak} and \ref{fig:occupancies-anisotropies-weak} one can see three distinct stages in the evolution of the gluon gas:
\begin{enumerate}
    \item There is first a stage of free streaming, where the occupancy stays fixed and the pressure anisotropy grows, driven by a depletion of the distribution at large $p_z$ due to the boost-invariant longitudinal expansion. During this very early free-streaming stage, $\alpha$ and $\beta$ begin near zero and $\gamma$ begins near 1.
    \item The BMSS fixed point~\cite{Baier:2000sb} dominates starting at $y \sim 4$ (that is, $\tau/\tau_I \sim 50$), independently of the coupling constant, and drives the dynamics with its characteristic scalings ($\gamma \approx 1/3, \beta \approx 0, \alpha \approx -2/3$).
    \item The maximal value of the anisotropy scales approximately with $g_s^{-2}$. After this maximal value is reached, which takes longer and longer times at weaker and weaker couplings, the system enters the dilute regime with $\alpha=-1$ and $\beta=\gamma=0$ found in Ref.~\cite{Brewer:2022vkq}, and (very) slowly becomes more isotropic as the occupancy drops rapidly.
\end{enumerate} 
However, at such weak couplings, the system does not reach local thermal equilibrium within a time of at least $\tau/\tau_I \sim 10^{13}$. We do not evolve the system to even larger (arbitrarily large) times because the system will be too dilute for any possible hydrodynamization to be of physical interest to us.

At stronger couplings, we see in Figs.~\ref{fig:scalings-Dr-strong} and \ref{fig:occupancies-anisotropies-strong} that the second stage of the weakly coupled scenario disappears essentially completely, with the system evolving directly from the free-streaming regime to a regime where the scaling exponents are close to the values that, at weak coupling, correspond to the dilute fixed point. There is then, however, a last stage of the evolution, namely hydrodynamization, whose analysis is our goal in this Section. 
The three stages in the evolution at stronger couplings are:
\begin{enumerate}
    \item Growth of the pressure anisotropy without changing the typical occupancy until $y \sim 3$. As at weak coupling, during this very early free-streaming stage, $\alpha$ and $\beta$ begin near zero and $\gamma$ begins near 1.
    \item 
    The system evolves directly from the free-streaming stage to a regime in which $\beta$ and $\gamma$ are close to zero and $\alpha$ is close to -1. At these stronger values of the coupling, the scaling exponents are not constant during this epoch, {{but at least for $g_s$ around 1 to 1.5  we see a regime that resembles the dilute scaling found at weak coupling.}} During this epoch, both the pressure anisotropy and the occupation drop together. During both the free-streaming stage and this stage, the dynamical evolution is approximately described via a scaling function with a pre-hydrodynamic form as in Sect.~\ref{sec:pre-hydro}, which is to say by a pre-hydrodynamic attractor. 
    {{Once we take $g_s$ as large as 2.6, though, it is no longer clear whether there is a distinct dilute-like regime as the system quickly evolves onward to$\ldots$}}
    \item Hydrodynamization. Isotropization and approach to local thermal equilibrium, with a thermalization time that  ranges between $\tau_{\rm th}/\tau_I \sim 10^{4-5}$ at $g_s = 2.6$ and $\tau_{\rm th}/\tau_I \sim 10^{8-10}$ at $g_s = 1$, suggesting that $\tau_{\rm th}/\tau_I$ scales exponentially with an inverse power of the coupling constant, as expected in kinetic theories with only small angle scatterings, 
    whose thermalization times have been estimated to scale parametically as $\tau_{\rm th}/\tau_I 
    \sim \exp(-1/\alpha_s^{1/2})$~\cite{Mueller:1999fp,Bjoraker:2000cf,Tanji:2017suk}. Due to the absence of gluon splittings, the thermalization time of this kinetic theory is much longer than that of QCD EKT~\cite{Kurkela:2015qoa,Kurkela:2014tea}. Note that hydrodynamization, namely the approach to isotropization and thermalization which is signified in Fig.~\ref{fig:scalings-Dr-strong} by the three scaling exponents rising away from their dilute values toward their hydrodynamic values $\alpha=0$, $\beta=\gamma=1/3$, begins at $\tau/\tau_I\sim 10^{5-6}$ for $g_s=1$, {{and at earlier and earlier times for larger values of the coupling.}} Furthermore, at the largest value of the coupling that we have investigated, namely $g_s=2.6$ which corresponds to $\alpha_s\sim0.5$ and $\lambda_{\rm 't~Hooft}\sim 20$, as noted above we see that the evolution proceeds essentially directly from the early free-streaming phase to hydrodynamization, with the intermediate phase 2.~so brief as to be not distinguishable. During hydrodynamization and during the subsequent hydrodynamic evolution, the dynamical evolution is described via a scaling function as in Sect.~\ref{sec:hydro}, which is to say by a hydrodynamizing attractor.
\end{enumerate}

It is important to note that the distribution function takes on different scaling forms in the pre-hydrodynamic stage (stage 2 above; scaling as in Sect.~\ref{sec:pre-hydro}) and during hydrodynamization (stage 3 above; scaling as in Sect.~\ref{sec:hydro}). We have achieved a unified and continuous description of both stages and the transition from one to the other even though the distribution function does not take on a scaling form during that transition.

We see that for values of $g_s$ ranging from 1 (which is much weaker than appropriate for the description of hydrodynamization in a heavy ion collision) to 2 (reasonable) to 2.6 (a little on the large side) we have been able to obtain a complete description of hydrodynamization, beginning from free-streaming, with the form of the scaling function then changing smoothly from its pre-hydrodynamic form (as in Sect.~\ref{sec:pre-hydro}) to the form needed to describe hydrodynamization and ultimately isotropization and thermalization that we first introduced in Sect.~\ref{sec:hydro}. Our description is adiabatic throughout, but the dynamics governed by the adiabatic evolution of the ground state(s) of the effective Hamiltonian changes from that of the pre-hydrodynamic attractor to that of the hydrodynamizing attractor. We describe the transition from one attractor to the next further below.

In the remainder of this Section, we confirm that AH is working as described via careful inspection of how the eigenvalues of the effective Hamiltonian, and the occupation of the associated eigenstates, evolve with time during the dynamics illustrated in the Figs.~\ref{fig:scalings-Dr-weak}-\ref{fig:occupancies-anisotropies-strong} above. Can we confirm that the system is indeed dominated by an adiabatically evolving band of instantaneous ground states and ultimately by an adiabatically evolving isolated instantaneous ground state? If so, these state(s) encode the information about the initial state that survives each stage in the process of hydrodynamization.

\begin{figure}
    \centering
    \includegraphics[width=0.7\textwidth]{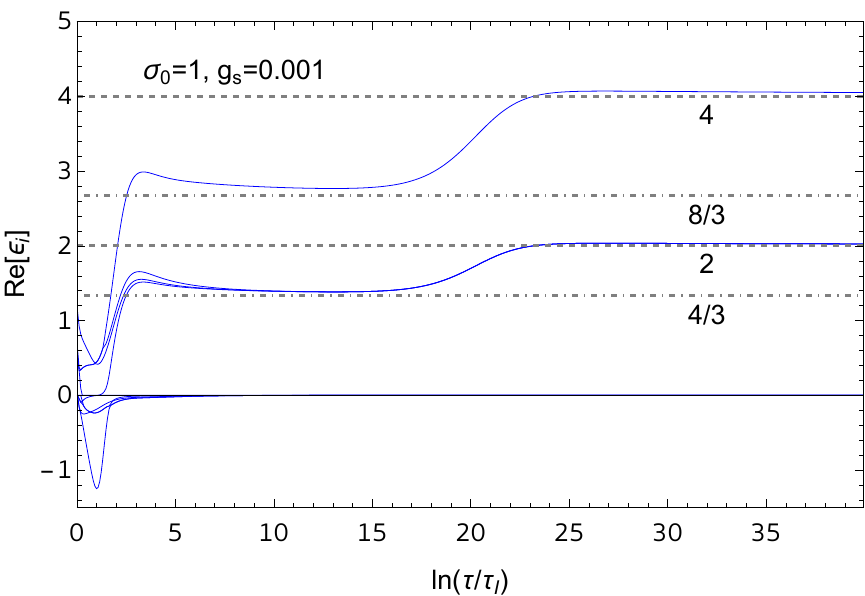}
    \caption{Plot of the instantaenous eigenvalues of the effective Hamiltonian~\eqref{eq:Heff-smallanglescatt-Dr} for $g_s = 10^{-3}$ and the initial condition given in Eq.~\eqref{eq:init-cond}. As first analyzed in Section~\ref{sec:intro-AH-BSY}, the system rapidly reaches the pre-hydrodynamic attractor which first describes evolution to and at the BMSS fixed point and subsequently describes the approach to and evolution at the dilute fixed point.}
    \label{fig:eigenvals-Dr-gs10-3}
\end{figure}

\begin{figure}
    \centering
    \includegraphics[width=0.8\textwidth]{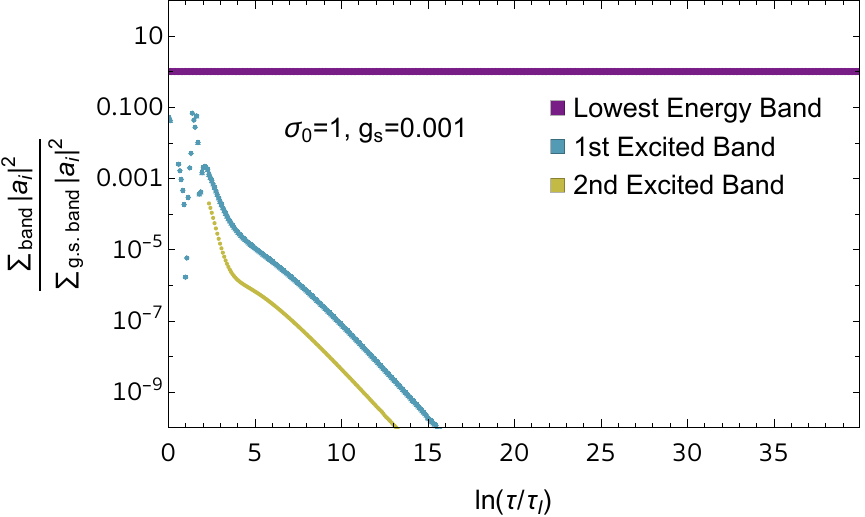}
    \caption{Plot of the coefficients in the $H_{\rm eff}$~\eqref{eq:Heff-smallanglescatt-Dr} eigenstate decomposition of the distribution function for $g_s = 10^{-3}$ and the initial condition given in Eq.~\eqref{eq:init-cond}, grouped by bands of nearly degenerate eigenvalues (see Fig.~\ref{fig:eigenvals-Dr-gs10-3}). The coefficients are normalized relative to the occupation of the lowest energy band that carries the particle number. We see that the occupation of each of the higher bands decay steeply and exponentially starting already during the approach to the BMSS regime. Only the lowest energy band is relevant to describing the evolution of this kinetic theory from then on.}
    \label{fig:coefficients-Dr-gs10-3}
\end{figure}

We start at very weak coupling with $g_s=10^{-3}$, where the dynamics is that which we first explored in Sect.~\ref{sec:pre-hydro}. As we have seen in Figs.~\ref{fig:scalings-Dr-weak} and \ref{fig:occupancies-anisotropies-weak} and discussed above,
the system does not reach hydrodynamization. As we can see from Figures~\ref{fig:eigenvals-Dr-gs10-3} and~\ref{fig:coefficients-Dr-gs10-3}, a band of lowest energy eigenstates of the effective Hamiltonian
quickly becomes dominant, starting from the time at which the BMSS fixed point is reached and continuing through the approach to, and evolution at, the dilute fixed point. In the BMSS regime, the occupation of the excited states decays exponentially, and these states stay irrelevant at later times, even as the instantaneous energy levels change their values during the transition from the BMSS regime to the dilute regime. This is in precise agreement with the picture first introduced in Section~\ref{sec:intro-AH-BSY}, where the adiabatic evolution of a set of states with the same, lowest, instantaneous energy eigenvalue dominates the pre-hydrodynamic dynamics.

\begin{figure}
    \centering
    \includegraphics[width=0.7\textwidth]{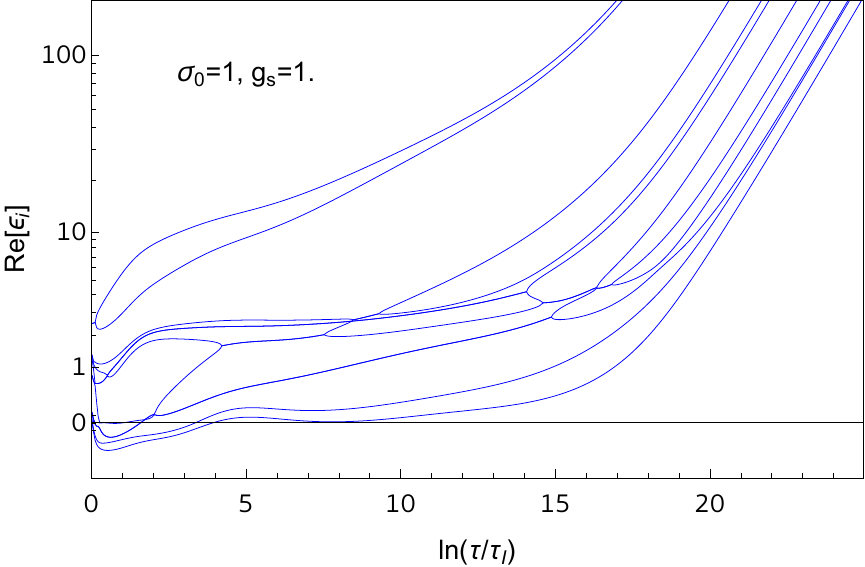}
    \caption{Plot of the instantaneous eigenvalues of the effective Hamiltonian~\eqref{eq:Heff-smallanglescatt-Dr} for an intermediate coupling $g_s = 1$ and the initial condition given in Eq.~\eqref{eq:init-cond}. We see that a gap opens up within the band of nearly degenerate low-lying eigenvalues at $\ln(tau/\tau_I)\sim 12$, which corresponds to the time in the top-left panel of Fig.~\ref{fig:scalings-Dr-strong} when the scaling exponents begin to evolve away from the dilute regime in the direction of their hydrodynamic values. After this gap opens, the evolution is governed by an isolated instantaneous ground state, corresponding to the isolated  lowest eigenvalue. The adiabatic evolution of this state describes hydrodynamization.}  
    \label{fig:eigenvals-Dr-gs1}
\end{figure}

\begin{figure}
    \centering
    \includegraphics[width=0.8\textwidth]{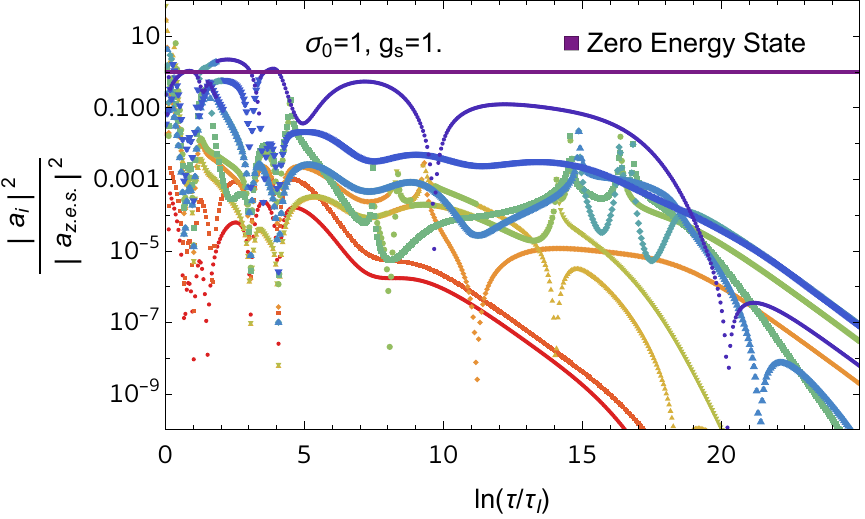}
    \caption{Plot of the coefficients in the $H_{\rm eff}$~\eqref{eq:Heff-smallanglescatt-Dr} eigenstate decomposition of the distribution function for $g_s = 1$ and the initial condition given in Eq.~\eqref{eq:init-cond}. The coefficients are normalized relative to the occupation of the zero-energy-eigenvalue state that carries the particle number, which at late times is the instantaneous ground state with the lowest eigenvalue. We see that after the gap between this eigenvalue and the others opens up (see Fig.~\ref{fig:eigenvals-Dr-gs1}), the occupation of each of the higher energy states falls away.}
    \label{fig:coefficients-Dr-gs1}
\end{figure}

This picture changes qualitatively at late times when we consider the intermediate value of the coupling $g_s = 1$. From Figs.~\ref{fig:eigenvals-Dr-gs1} and~\ref{fig:coefficients-Dr-gs1}, we see that at early times there are bands of energy eigenvalues with the lowest such band remaining close to degenerate until $y \equiv \ln(\tau/\tau_I) \sim 12$, at which time a gap starts opening up, consistent with the beginning of hydrodynamization, i.e.~the beginning of the departure from the dilute regime and the approach to isotropization and local thermal equilibrium seen in the top-left panel of 
Fig.~\ref{fig:scalings-Dr-strong} and discussed above. Meanwhile, we see in Fig.~\ref{fig:coefficients-Dr-gs1} that between $y \sim 2$ and $y \sim 10$ the occupation of a small group of low-lying states is significant while above $y\sim 12$ and certainly for $y > 15$ the occupation of all states other than the one with the isolated lowest eigenvalue drop away and only the instantaneous ground state of the effective Hamiltonian is needed to describe hydrodynamization in this kinetic theory.
This confirms that the AH scenario is realized sequentially in this kinetic theory, with different attractors (first pre-hydrodynamic; later hydrodynamizing) describing different out-of-equilibrium regimes corresponding to different stages in the loss of memory of the initial condition, with each collapsing the state onto a smaller set of degrees of freedom than in the previous stage, until local thermal equilibrium is reached and only the hydrodynamic evolution of the state with a thermal distribution remains.

With these results in hand, we have a unified and continuous description of the transition from one attractor to the next.
The pre-hydrodynamic attractor is described via the adiabatic evolution of a band of lowest-energy eigenstates, with the system having previously been attracted into some superposition of the eigenstates in this band. The mix within this superposition may evolve as the  eigenstates in this band evolve adiabatically but there is  no mixing with higher energy states. 
These few lowest-energy eigenstates, and their time-evolution, encode whatever information about its initial conditions the system ``remembers'' during the epoch when its evolution is described via the pre-hydrodynamic attractor.
The transition from this pre-hydrodynamic attractor to hydrodynamization is precipitated by the energy of all the basis states but one in this band rising, leaving a single isolated ground state of the evolving effective Hamiltonian. As this happens, the occupancy of all 
the other states from what used to be the low-lying band decay rapidly to zero.  That is to say, the system ``forgets'' all aspects of its initial state except those encoded in a single instantaneous eigenstate and its time-evolution.
From then on, during
hydrodynamization the system is described by the adiabatic evolution of a single isolated lowest energy eigenstate of the evolving effective Hamiltonian.

\begin{figure}
    \centering
    \includegraphics[width=0.7\textwidth]{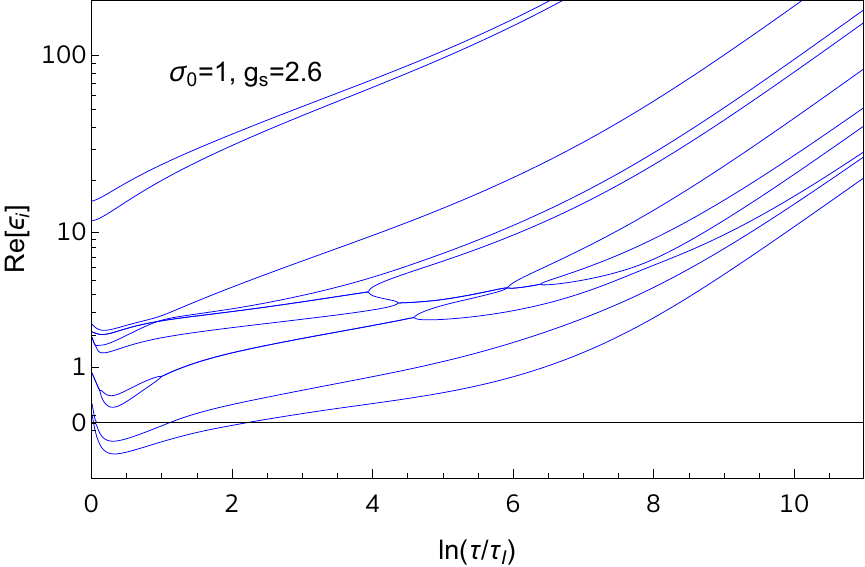}
    \caption{Plot of the eigenvalues of the effective Hamiltonian~\eqref{eq:Heff-smallanglescatt-Dr} for $g_s = 2.6$ and the initial condition given in Eq.~\eqref{eq:init-cond}. At this large coupling, it is hard to say whether there is a regime at early times where a band of low eigenvalues dominates, since a gap between a single lowest eigenvalue and all the others opens up so early. This corresponds to the rapid onset of hydrodynamization that we have seen in the bottom-right panel of Fig.~\ref{fig:scalings-Dr-strong}.}
    \label{fig:eigenvals-Dr-gs26}
\end{figure}

\begin{figure}
    \centering
    \includegraphics[width=0.8\textwidth]{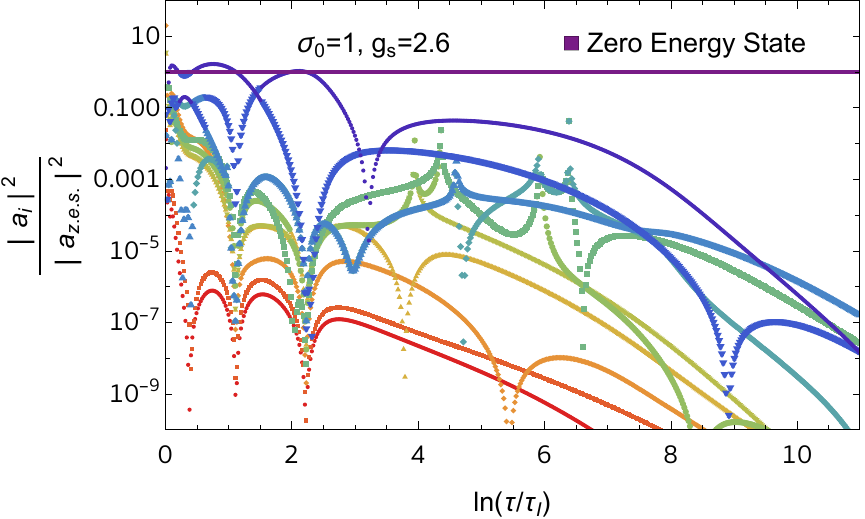}
    \caption{Plot of the coefficients in the $H_{\rm eff}$~\eqref{eq:Heff-smallanglescatt-Dr} eigenstate decomposition of the distribution function for $g_s = 2.6$ and the initial condition given in Eq.~\eqref{eq:init-cond}. The coefficients are normalized relative to the occupation of the zero-energy-eigenvalue state that carries the particle number, which at late time is the instantaneous ground state of $H_{\rm eff}$. We see that after the (here quite early time at which the) gap between this eigenvalue and the others opens up (see Fig.~\ref{fig:eigenvals-Dr-gs1}), the occupation of each of the higher energy states falls away.}
    \label{fig:coefficients-Dr-gs26}
\end{figure}

When we analyze the eigenvalues and eigenstate occupations at the still larger value of the coupling $g_s=2.6$ in Figs.~\ref{fig:eigenvals-Dr-gs26} and \ref{fig:coefficients-Dr-gs26}, the middle stages of the sequential process that we have analyzed have become so short as to be indistinguishable: the system proceeds relatively directly from the earliest pre-hydrodynamic stage to hydrodynamization, which brings it to isotropization and local thermal equilibrium in a shorter time. This evolution is described just as well via the AH scenario as the dynamics with more distinct epochs found at intermediate and weak coupling, above.
We see that
a gap opens up between the instantaneous ground state of the effective Hamiltonian with an isolated lowest eigenvalue and all higher energy states, with this happening 
before $\tau=20\,\tau_I$, which is to say at a much earlier time here with $g_s=2.6$ than what we found with $g_s=1$ in Figs.~\ref{fig:eigenvals-Dr-gs1} and \ref{fig:coefficients-Dr-gs1}.
After this happens, hydrodynamization and the approach to isotropization and equilibrium is duly realized by the rapid decay of excited states, with the subsequent evolution of the full distribution function being described by the adiabatic evolution of the instantaneous ground state. 

In our analyses of this kinetic theory at $g_s=1$ and $g_s=2.6$, we have thus achieved all of the goals that we set out to achieve, seeing in concrete terms how all the physical processes and regimes involved in the early pre-hydrodynamic dynamics and the subsequent hydrodynamization and hydrodynamic evolution in this kinetic theory are described via Adiabatic Hydrodynamization in a fashion that yields understanding and intuition for what is happening when, and why and how it happens.

\section{Conclusions and Outlook} \label{sec:outlook}

We have demonstrated via concrete examples that attractors in kinetic theory can be described by the low-energy instantaneous eigenstates of the operator that generates the time evolution of the system (i.e., the effective Hamiltonian) in an appropriate ``adiabatic'' frame, defined by the requirement that the instantaneous ground state of the effective Hamiltonian evolves in time as slowly as possible. In this way, we have demonstrated that the AH framework is ideally equipped to describe the process of hydrodynamization of kinetic theory, and in particular should greatly simplify the analysis of the QCD EKT description of the initial stages of heavy-ion collisions.
Furthermore, we have formulated a prescription to find the optimal adiabatic frame, even though in practice we showed that the AH scenario can be realized without finding the strictly optimal solution for this frame.

We showed explicitly via our analysis of the kinetic theory for a longitudinally expanding gluon gas obtained by making the small-angle elastic scattering approximation to the QCD EKT collision kernel that the 
processes by which the system loses memory of its initial state proceed in stages -- stages that are distinct if the kinetic theory is sufficiently weakly coupled.  Each out-of-equilibrium attractor stage encodes only a  subset of the information in the initial condition, characterized precisely by the adiabtically evolving low-energy instantaneous eigenstate(s) of the effective Hamiltonian in the adiabatic 
frame. The adiabatic description of the earliest prehydrodynamic era is governed by a band consisting of  degenerate (e.g.~in the analysis of Sect.~\ref{sec:intro-AH-BSY} or Sect.~\ref{sec:pre-hydro}) or almost degenerate (e.g.~in the analysis of Sect.~\ref{sec:non-scaling}) lowest-energy instantaneous eigenstates. Subsequently, further gaps open up and hydrodynamization begins as the system evolves to have a single isolated lowest-energy eigenvalue, with the corresponding adiabatically evolving instantaneous ground state of the effective Hamiltonian describing hydrodynamization and the subsequent hydrodynamic evolution of a distribution in local thermal equilibrium.
As the number of ground states in the adiabatic description of the dynamics
becomes progressively smaller as the system passes from one attractor stage to the next, more information about the initial state of the kinetic theory is forgotten, until only a single state remains. The adiabatic evolution of this state describes hydrodynamization and at late times, as hydrodynamization concludes, this state corresponds to the thermal distribution.

At this point, we comment on our expectations for what will happen when this framework is applied to QCD EKT in full. While it is very likely that for realistic values of the coupling in full QCD EKT 
it may be hard to identify distinct attractor regimes, we anticipate that at sufficiently weak coupling we will see a variation on the themes that we have explored concretely in Sect.~\ref{sec:non-scaling}, where the evolution is described adiabatically throughout but not via a single scaling form, with the system initially following pre-hydrodynamic attractor(s) and subsequently hydrodynamizing, as described by a different attractor. For realistic values of the coupling these processes may become less distinct. But we expect that at larger values of the coupling, as in our analysis of the kinetic theory we have used in Sect.~\ref{sec:non-scaling},  
the entire evolution will be described via the adiabatic evolution of a small number of instantaneous ground states, with that number decreasing to one as gaps in the low-lying spectrum of the effective Hamiltonian open up.
Most of the loss of memory of the initial conditions for the kinetic theory will occur at the very beginning as all eigenstates of the effective Hamiltonian except those in a low-lying band 
rapidly become irrelevant. Further loss of memory of what came before will then correspond to the opening up of further gaps
in the instantaneous spectrum of the effective Hamiltonian, and the last-remaining, lowest energy, instantaneous ground state will adiabatically evolve so as to become the state that describes a thermal distribution function as the system hydrodynamizes. We leave the explicit implementation and verification of this picture in QCD EKT to future work.

There are at least two substantial advances that need to be realized in order to realize the vision above. The first is to restore the full collision kernel of QCD EKT in the kinetic theory and to add quarks and antiquarks to the gluons that we have focused on throughout this work. Likely the most important aspect of this will be to add number-nonconserving processes in the collision kernel, starting with $1 \leftrightarrow 2$ processes, as these are not included in the small-angle scattering collision kernel that we have employed throughout this work. 
Restoring $1 \leftrightarrow 2$ processes will result in hydrodynamization happening more quickly at weak (and likely at any given value of the) coupling.
It will be very interesting to watch the Adiabatic Hydrodynamization scenario 
in action fully explicitly in that case, as we have done here.
We anticipate that the massive simplification of the 
problem we have achieved here by demonstrating that the system quickly becomes dominated by the low-energy eigenstates of the 
time-evolution operator 
will, when realized in QCD EKT, be of practical importance in addition to being of value for the understanding and physical intuition that it yields. A full numerical simulation of QCD EKT carries along a lot of unnecessary information corresponding to the description of the evolution of all the higher energy states of the time-evolution operator.  Realizing its description in the language of adiabatic hydrodynamization should therefore make it more practicable to make advances in the second important direction, namely introducing initial, prehydrodynamic, hydrodynamizing and hydrodynamic states with geometries that are more similar to those of heavy ion collisions (HICs) than the longitudinally-boost invariant transversely-infinite geometry that we have employed throughout this work.
Adding transverse expansion into the kinetic equation is of foremost importance, so as to be able to describe how a (finite) droplet of QCD matter produced in a HIC expands and eventually falls apart, especially in situations where the QCD matter does not fully hydrodynamize. 
Understanding the pre-hydrodynamic dynamics in small collision systems in which hydrodynamization may not be complete is of considerable interest as it is one of the few ways in which experimental measurements may shed light on QCD matter in the act of hydrodynamizing. Applying Adiabatic Hydrodynamization in such a setting will require applying it in geometries where the transverse extent of the matter described by kinetic theory is finite and spatial gradients and transverse expansion are important.  This presents considerable challenges in general; the simplification introduced via understanding and employing the adiabatic hydrodynamization approach could help to overcome these challenges.
Another feature to include in future work is spacetime rapidity dependence in the kinetic equation, which would in principle allow one to make more quantitative statements about observables as one moves away from the mid-rapidity region in a HIC.

Our results pave the way for an intuitive understanding of the dynamics of hydrodynamization in QCD EKT, and more generally of the emergence of scaling 
phenomena during the approach to equilibrium in kinetic theory in other contexts. Presumably after some modification, the approach that we have presented in explicit detail in this work could also be suitable to provide the same kind of understanding and intuition for the physics of thermalization during and after the reheating epoch in the early Universe, which has already been described via kinetic theory 
(see, e.g.,~\cite{Harigaya:2013vwa,Mukaida:2022bbo,Mukaida:2024jiz}). It may also be of interest in the context of the preheating epoch of some inflationary models. And, it almost goes without saying that we expect that the understanding of the dynamics of out-of-equilibrium QCD matter via the approach we have laid out here could be particularly valuable in Bayesian analyses of HICs (e.g., JETSCAPE~\cite{Kauder:2018cdt} or \textit{Trajectum}~\cite{Nijs:2020roc}) due to the simplicity gained by only needing to describe the pre-hydrodynamic and hydrodynamizing distribution function in terms of the instantaneous low-energy eigenstates of the time-evolution operator. Bayesian analyses like these, simplified and empowered via employing adiabatic hydrodynamization in QCD EKT, offer a path to 
making direct connections between experimental data from HICs and the QCD description of their initial stages.

%% file: outlook.tex

\chapter{Outlook} \label{ch:outlook}

The continued exploration of the many-body physics of QCD marches on, with theory and experiment making progress hand-in-hand, and there is still much to learn. As we mentioned in the introduction, much of the QCD phase diagram, especially away from the zero temperature or zero baryon chemical potential lines, remains unexplored experimentally, let alone the out-of-equilibrium many-body physics of QCD approaching such extreme conditions. In this thesis, we have made substantial progress towards understanding two concrete directions in this vast landscape. We have, on the one hand, precisely and rigorously formulated the correlation functions that need to be calculated to describe quarkonium transport in QGP in the heavy quark mass limit, with which we expect that a much closer connection between theory and experiment will be attained regarding quarkonium suppression data in HICs. We carried out their calculation at weak coupling in QCD and at strong coupling in $\mathcal{N}=4$ Yang-Mills theory. Because these correlation functions are attributes of QGP, measuring the final states of quarkonia in HICs deepens our knowledge of QGP properties, and can establish, from the point of view of quarkonium, whether these features of QGP resemble more closely those of a weakly or a strongly coupled plasma by comparing the extracted properties with our theoretical calculations. On the other hand, we have developed tools to understand the process of hydrodynamization in QCD kinetic theory and applied them to a simplified description where only a subset of the QCD scattering mechanisms are included. By doing this, we learned that the process of hydrodynamization in this theory, and specifically, how memory of the initial condition is lost, is a sequential process in which a monotonously shrinking set of low-energy states dominate the dynamics, where the opening of an energy gap relative to the ground state(s) signals the start of each stage of this process. The hydrodynamic attractor is reached when only one low-energy state remains as the ground state, and the system approaches local thermal equilibrium following the adiabatic evolution of this low-energy state. Therefore, our findings provide strong support for the conjecture~\cite{Brewer:2019oha} that the hydrodynamization process of QCD matter into QGP may be understood through the Adiabatic Hydrodynamization scenario, as we just described it.

We hope that these developments will provide new tools and open new directions to study and analyze HICs, both theoretically and from a data-driven perspective (as in a Bayesian analysis). In the case of quarkonium, once the transport formalisms are upgraded to be able to account for memory effects of the QGP environment, we expect that our results will be immediately applicable to $\Upsilon$ suppression data, but further developments will be needed for them to have the same effectiveness for the $J/\psi$ states, simply because the non-relativistic limit required to construct the EFT description we used is not consistent with the kinematics of charmonium states (there is no significant separation between the mass of the charm quark and the inverse size of the bound state it forms, meaning that the typical relative velocity of the bound $c$ and $\bar{c}$ quarks will not be as small as needed to be considered non-relativistic). We also expect that the techniques that we used to set up the description of quarkonium in medium will likely be helpful for us to sharpen the description of momentum broadening of hard probes in QGP, both of heavy quarks as well as of jets made of lighter particles, in several analog directions. The first one would be to calculate field strength correlators at finite lightcone time separation that encode the physics of momentum broadening of high energy, ultra-relativistic particles while preserving the correlations of the QGP environment with itself. This would correspond, in spirit, to a resummation of $T/E$ corrections to the jet quenching parameter $\hat{q}$. Another direction is to revisit the AdS/CFT calculation of $\hat{q}$ itself, as there is a long-standing mismatch between the result at nonzero $v < 1$ heavy quark velocity~\cite{Casalderrey-Solana:2007ahi} and the strict lightlike case $v=1$ calculated in Ref.~\cite{Liu:2006ug} (even though each calculation is derived from a different kinematic regime). We expect that the intuition and technical expertise that the work in this thesis has provided will be an important asset to sort out this puzzle. Finally, it is also conceivable that our results will serve as a stepping stone for the development of an analog description of jet propagation in hot QGP via the open quantum systems formalism, along the same lines of Ref.~\cite{Vaidya:2020cyi}, but with the goal of setting up the description in such a way that it is also possible to follow jets propagating through strongly coupled QGP. Optimistically, such a description will automatically include the creation and the effects of the hydrodynamic wake of a jet through the correlation functions of QGP that enter the envisioned jet transport equations.

Regarding the equilibration process of QCD, the deeper understanding of hydrodynamization we gained for kinetic theories should find direct applicability in studies of full QCD Effective Kinetic Theory~\cite{Arnold:2002zm} (EKT). It is our expectation that at sufficiently weak coupling one will be able to find an analog ``adiabatic frame'' that clearly features two distinct regimes where memory of the initial condition is lost by means of the opening of an energy gap between the ground state(s) and the rest of the states in the eigenstate decomposition of the time evolution operator of the theory. At stronger couplings we expect, in complete analogy with what we saw in Section~\ref{sec:connectstages}, that the two stages of memory loss coalesce into a single process, and a gap only opens as the system begins to approach hydrodynamics with the unique low-energy state defining the hydrodynamic attractor. A verification of this would be a definitive step towards an intuitive understanding of the fast hydrodynamization time of QCD kinetic theory in kinematic regimes that model HICs. Furthermore, we expect that this framework will be useful, to begin with, in any context where self-similar scaling phenomena appear, provided the evolution equations are presented as first-order in time derivatives. For example, it is completely conceivable that hydrodynamization in holographic descriptions of strongly coupled gauge theories can be described in the same way. However, finding the adiabatic frame in terms of a state vector in this theory might require the introduction of new ingredients, in the sense that it is not clear \textit{a priori} what are the optimal time-dependent variables to extract from the state and move into the definition of the ``adiabatic frame.''

In the bigger picture of QCD under extreme conditions, new data will continue to be gathered and analyzed both in the near and not so near future. As we mentioned in the Introduction, we now eagerly await the publication of the analysis of the data taken during the Beam Energy Scan II program at STAR. In the very near future, the data coming from the sPHENIX program~\cite{PHENIX:2015siv,Arslandok:2023utm,Achenbach:2023pba} will provide unprecedented high-statistics measurements of jets and quarkonia production that will give us further quantitative insight into the microscopic nature of QGP. Further into the future, the Electron-Ion Collider (EIC), to be built at Brookhaven National Laboratory~\cite{Accardi:2012qut,Achenbach:2023pba} is projected to give us cutting-edge precision measurements of the structure of atomic nuclei, thus further constraining our theoretical descriptions and modelling to describe matter all around us. Another frontier in the not too distant future is the continuing study of neutron stars, which provide information about the equation of state of QCD at zero temperature and finite baryon chemical potential, via the detection of electromagnetic and gravitational waves. The NICER telescope has already provided constraints on the mass-radius relation of neutron stars~\cite{Miller:2021qha,Vinciguerra:2023qxq} through X-ray timing, and may yet provide us with more data. Also, LIGO~\cite{LIGOScientific:2007fwp,LIGOScientific:2014pky}, VIRGO~\cite{VIRGO:2012dcp} and KAGRA~\cite{KAGRA:2018plz,KAGRA:2020tym} have already detected gravitational waves from inspiral events involving neutron stars~\cite{LIGOScientific:2017vwq,LIGOScientific:2021qlt}, and LISA~\cite{LISA:2017pwj,LISA:2022yao} has been forecast to continue doing so. In tandem with perturbative QCD results, these measurements provide sharp constraints on the QCD equation of state at finite density~\cite{Komoltsev:2021jzg,Gorda:2022jvk,Komoltsev:2023zor,Kurkela:2024xfh}. Last but not least, Run 3 of the LHC is scheduled to continue until the end of 2025, after which upgrades will begin. Operations will restart in 2029 with higher luminosity runs, which will hopefully continue providing us with new data on nuclear and particle physics until the 2040s and give further input to our understanding of QCD, with jets and quarkonia observables being of prime interest. With all of these prospects to continue exploring the physics of QCD, we envision a thriving nuclear and particle physics community with high potential to discover new physics and further understand the microscopic structure of matter all around us.

%% file: appendixa.tex

\chapter{Appendix: Calculation of the Chromoelectric Field Correlator at Weak Coupling in QCD}

\section{KMS relation for electric field correlator}
\label{app:kms}

In this Appendix we will verify the KMS relation between $[g_{\rm adj}^{++}]^>(t)$ and $[g_{\rm adj}^{++}]^<(t)$, as introduced in the main text in Eqs.~\eqref{eq:gE++>} and~\eqref{eq:gE++<}. For brevity, here we write the correlators as $g_E$, which are related to $g_{\rm adj}$ by $g_{\rm adj} = \frac{g^2 T_F}{3N_c} g_E$ for all types of correlators (Wightman, (anti-)time-ordered, spectral functions)\@. We also discuss how time-reversal relates $[g_E^{++}]^>$ and $[g_E^{--}]^<$. 


We begin by noting that an adjoint Wilson line in the interaction picture can be written in terms of time-evolution operators as
\begin{equation}
    \mathcal{W}^{ab}_{[t_f,t_i]} = e^{i H t_f} \left[e^{- i (H - g A_0^c(0) [T_{\rm Adj}^c] ) (t_f - t_i) }\right]^{ab} e^{-i H t_i} \, , \label{eq:W-as-T-evol}
\end{equation}
where $H$ is the QGP Hamiltonian, and one may interpret $ H \delta^{ab} - g A_0^c(0) [T_{\rm Adj}^c]^{ab} $ as the total Hamiltonian when there is a point color charge in the adjoint representation at the position ${\bs x} = 0$\@.

With this, the Wightman correlator $[g_{E}^{++}]^>(t)$ can be written as
\begin{align}
    [g_{E}^{++}]^>(t) &= \frac{1}{Z} {\rm Tr}_{\mathcal{H}} \left[ E_i^a(t) \W^{ac}(t,+\infty) \W^{cb}(+\infty,0) E_i^b(0) e^{-\beta H} \right] \nonumber \\
    &= \frac{1}{Z} {\rm Tr}_{\mathcal{H}} \left[ e^{i H t} E_i^a(0) \left[e^{- i (H - g A_0^c(0) [T_{\rm Adj}^c] ) t}\right]^{ab} E_i^b(0) e^{-\beta H} \right] \, .
\end{align}
The KMS conjugate of this correlator is obtained by shifting $t \to t -i\beta$\@. We therefore obtain
\begin{align}
    [g_{E}^{++}]^<(t) &= [g_{E}^{++}]^>(t - i\beta) \nonumber \\
    &= \frac{1}{Z} {\rm Tr}_{\mathcal{H}} \left[ e^{i H t}  E_i^a(0) \left[e^{- i (H - g A_0^c(0) [T_{\rm Adj}^c] ) (t - i\beta) }\right]^{ab} E_i^b(0) \right] \, . \label{eq:KMS-proof-step}
\end{align}
We can explicitly see in this expression that the thermal ensemble is now determined by the total Hamiltonian $ H \delta^{ab} - g A_0^c(0) [T_{\rm Adj}^c]^{ab} $ instead of just $H$\@.

The equivalence with Eq.~\eqref{eq:gE++<} is then verified by using that
\begin{equation}
    \left[e^{- i (H - g A_0^c(0) [T_{\rm Adj}^c] ) (t - i\beta) }\right]^{ab} = e^{- i H t} \mathcal{W}^{ac}_{[t, t_f]} e^{-\beta H} \mathcal{W}^{cd}_{[t_f - i\beta, t_f]} \mathcal{W}^{db}_{[t_f, 0]} \, ,
\end{equation}
which follows from using Eq.~\eqref{eq:W-as-T-evol} repeatedly. It is interesting to note that this equation holds for any value of $t_f$\@. Plugging this expression into Eq.~\eqref{eq:KMS-proof-step} one obtains
\begin{equation}
    [g_{E}^{++}]^<(t) = \frac{1}{Z} {\rm Tr}_{\mathcal{H}} \left[   E_i^a(t) \mathcal{W}^{ac}_{[t, t_f]} e^{-\beta H} \mathcal{W}^{cd}_{[t_f - i\beta, t_f]} \mathcal{W}^{db}_{[t_f, 0]} E_i^b(0) \right] \, ,
\end{equation}
which is equivalent to Eq.~\eqref{eq:gE++<} as displayed in the main text when $t_f \to +\infty$\@.

The KMS relation between $[g_{E}^{--}]^>(t)$ and $[g_{E}^{--}]^<(t)$ then follows from using the one we just verified above and their relation to $[g_{E}^{++}]^>(t)$ and $[g_{E}^{++}]^<(t)$ through time-reversal respectively.

There is also a relation between $[g_E^{++}]^{>}$ and $[g_E^{--}]^{<}$. To see this, we need to apply a time reversal transformation. The time reversal transformation is given by
\begin{align}
\ml{T}_r A^\mu(t, {\bs x}_{\rm cm}) \ml{T}_r^{-1} &= A_\mu(-t, {\bs x}_{\rm cm}) \\
\ml{T}_r F^{\mu\nu}(t, {\bs x}_{\rm cm}) \ml{T}_r^{-1} &= - F_{\mu\nu}(-t, {\bs x}_{\rm cm}) \,.
\end{align}
Under the time reversal transformation, the Wilson line
\begin{align}
\ml{W}_{[t_f,t_i]} = {\rm P} \exp\Big( ig \int_{t_i}^{t_0} \diff t \, A_0(t, {\bs x}_{\rm cm}) \Big) \,,
\end{align}
where ${\rm P}$ is path ordering, changes according to
\begin{align}
\ml{T}_r  \ml{W}_{[a,b]}  \ml{T}_r^{-1} &= \overline{{\rm P}} \exp\Big( -ig \int_{t_f}^{t_i} \diff t \, A_0(-t, {\bs x}_{\rm cm}) \Big) =
\ml{W}_{[-t_f,-t_i]} \,.
\end{align}
Also, the time reversal operator is anti-unitary, which means under the time reversal operation
\begin{equation}
\langle n | O(t) | m \rangle = \langle n | \ml{T}_r^{-1} \ml{T}_r O(t)   | m \rangle =  \langle  O(-t) m | n\rangle =  \langle m |  O^\dagger(-t)  | n \rangle \,,
\end{equation}
where $O$ is an arbitrary operator.
This implies under a time reversal operation
\be
&&\Tr(O_1(t_1) O_2(t_2) e^{-\beta H}) = \sum_{n,m} e^{-\beta E_n} \langle n| O_1(t_1) |m\rangle \langle m | O_2(t_2) |n\rangle \nn \\
&\xrightarrow{\ml{T}_r} &= \sum_{n,m} e^{-\beta E_n} \langle m | O_1^\dagger (-t_1)| n \rangle \langle n | O_2^\dagger(-t_2) | m \rangle = \Tr(O_2^\dagger(-t_2) O_1^\dagger(-t_1) e^{-\beta H}) \,.
\ee
Thus, applying the time reversal transformations leads to
\be
[g_E^{++}]^>(t)
&=& \Tr_E \Big( E_i^a(t) \ml{W}^{ab}_{[t,+\infty]} 
\ml{W}^{bc}_{[+\infty,0]} E_i^c(0)
\frac{e^{-\beta H_E}}{Z}\Big) \nn \\
\xrightarrow{\ml{T}_r} &=& \Tr_E\Big(  E_i^c(0)
\ml{W}^{cb}_{[0, -\infty]}
\ml{W}^{ba}_{[-\infty, -t]} E_i^a(-t)
\frac{e^{-\beta H_E}}{Z}
\Big) \nn \\
&=& [g_E^{--}]^<(-t) \,.
\ee

\section{Feynman rules} \label{app:FeynmanRules}

In this Appendix we give the remaining definitions that are referred to in Section~\ref{sec:signandfeyn}.

Two-point functions on the Schwinger-Keldysh contour are defined as:
\begin{align}
S(x-y)&= \langle \ml{T}_C \psi(x) \bar{\psi}(y) \rangle,\\
C(x-y)&= \langle \ml{T}_C c(x)\bar{c}(y) \rangle,\\
D_{\mu \nu}(x-y)&= \langle \ml{T}_C A_{\mu}(x)A_{\nu}(y) \rangle,
\end{align}
where $\psi$ is a fermion, $c$ an anti-commuting scalar (ghost) and $A$ a gauge boson field. $\ml{T}_C$ is the time-ordering operator along the contour, placing fields inserted at later sections of the contour (closer to $t=-i\beta$) to the left of the field insertions closer to $t=0$ in the correlation function. For an arbitrary covariant gauge, the propagators are given by
\begin{align}
D^{Y,ab}_{\mu \nu}(k) = \delta^{ab} P_{\mu \nu}(k) D^Y(k),
\end{align}
where $Y$ can be any of $>,<,\ml{T},\overline{\ml{T}}$, and
\begin{align}
P_{\mu \nu }(k) = - \left[ g_{\mu \nu} - (1 - \xi) \frac{k_\mu k_\nu}{k^2} \right].
\end{align}
with metric signature $(+,-,-,-)$. The free propagators in Fourier space are given by
\begin{align}
D^>(k) &= \left( \Theta(k_0) + n_B(|k_0|) \right) 2\pi \delta(k^2)\,, \quad\quad\ \,  D^<(k) = \left( \Theta(-k_0) + n_B(|k_0|) \right) 2\pi \delta(k^2) \nonumber \\
D^\ml{T}(k) &= \frac{i}{k^2 + i0^+} + n_B(|k_0|) 2\pi \delta(k^2) \,, \quad\quad D^{\overline{\ml{T}}}(k) = \frac{-i}{k^2 - i0^+} + n_B(|k_0|) 2\pi \delta(k^2) \nonumber \\ D^S(k) &= D^>(k) + D^<(k) = (1 + 2n_B(|k_0|)) 2 \pi \delta(k^2) \,.
\end{align}
We also need to state what are the Fermionic propagators $\mathbb{S}_{IJ}$. 
We will only explicitly use the Wightman functions $\mathbb{S}_{21}$ and $\mathbb{S}_{12}$, which are given by
\begin{align}
\mathbb{S}_{12} = S^{<}(k)&= -\slashed{k} (2\pi) \delta(k^2)\left[-\Theta(-k_0) + n_{F}(|k_0|) \right],\\
\mathbb{S}_{21} = S^{>}(k)&= -\slashed{k} (2\pi) \delta(k^2)\left[-\Theta(+k_0) + n_{F}(|k_0|) \right],
\end{align}
where $n_F(k_0) = (e^{k_0/T} + 1)^{-1}$.

\begin{figure}
	\centering
	\begin{tabular}{  c  c  l  }
	\raisebox{-0.8in}{\includegraphics[height=1.6in]{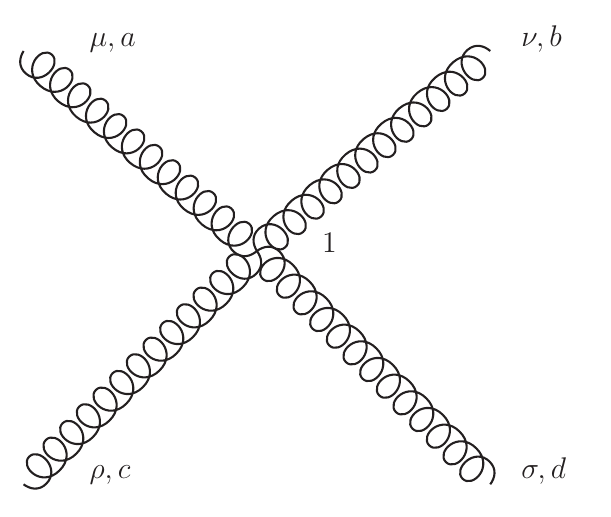}}  & & $ \begin{matrix} = & -ig^2 \big[ f^{abe} f^{cde} (g^{\mu \rho} g^{\nu \sigma} - g^{\mu \sigma} g^{\nu \rho} ) \\ &  + f^{ace} f^{bde} (g^{\mu \nu} g^{\rho \sigma} - g^{\mu \sigma} g^{\nu \rho}) \\ & + f^{ade} f^{bce}(g^{\mu \nu} g^{\rho \sigma} - g^{\mu \rho} g^{\nu \sigma}) \big] \end{matrix} $  \\
	\raisebox{-0.8in}{\includegraphics[height=1.6in]{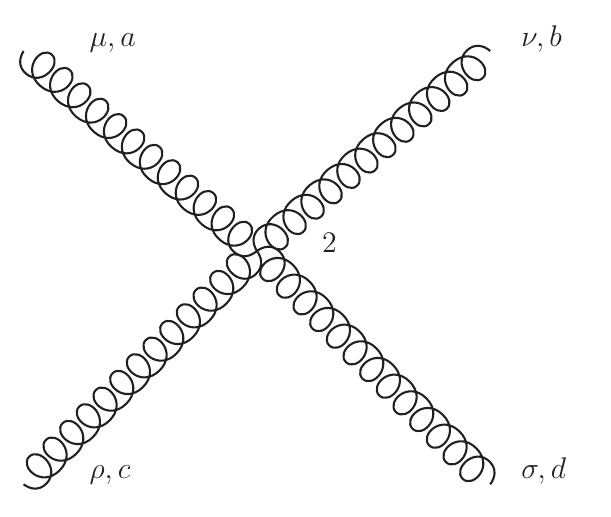}}  & & $ \begin{matrix} = & ig^2 \big[ f^{abe} f^{cde} (g^{\mu \rho} g^{\nu \sigma} - g^{\mu \sigma} g^{\nu \rho} ) \\ &  + f^{ace} f^{bde} (g^{\mu \nu} g^{\rho \sigma} - g^{\mu \sigma} g^{\nu \rho}) \\ & + f^{ade} f^{bce}(g^{\mu \nu} g^{\rho \sigma} - g^{\mu \rho} g^{\nu \sigma}) \big] \end{matrix} $
	\end{tabular}
\caption{Feynman rules associated to the 4-gauge boson vertex, given here for the time-ordered and anti-time ordered branches of the Schwinger-Keldysh contour. The anti-time ordered branch multiplies the vertex factors by $(-1)$, which is naturally included in the notation we adopt in the main text.}
\label{fig:rules-4g-appendix}
\end{figure}

\begin{figure}
	\centering
	\begin{tabular}{  c  c  l  }
	\raisebox{-0.8in}{\includegraphics[height=1.6in]{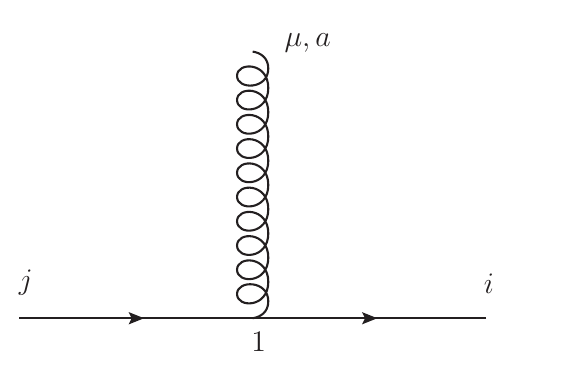}}  &=& $ i g \gamma^\mu \left[T_{\bs{R}}^a\right]_{ij}$  \\
	\raisebox{-0.8in}{\includegraphics[height=1.6in]{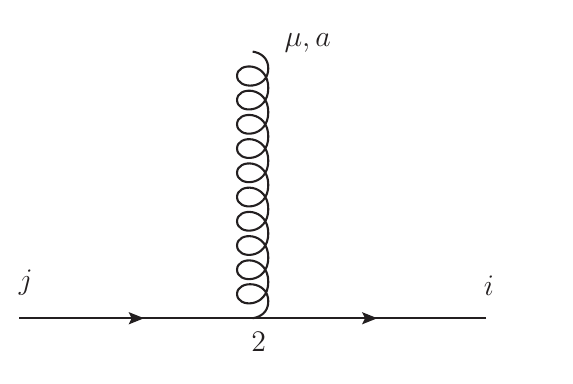}}  &=& $ -i g \gamma^\mu \left[T_{\bs{R}}^a\right]_{ij} $
	\end{tabular}
\caption{Feynman rules associated to the gauge boson-fermion-fermion vertex, given here for the time-ordered and anti-time ordered branches of the Schwinger-Keldysh contour, with the fermions in a representation $\bs{R}$. The anti-time ordered branch multiplies the vertex factors by $(-1)$, which is naturally included in the notation we adopt in the main text.}
\label{fig:rules-g-fermion-appendix}
\end{figure}

\begin{figure}
	\centering
	\begin{tabular}{  c  c  l  }
	\raisebox{-0.8in}{\includegraphics[height=1.6in]{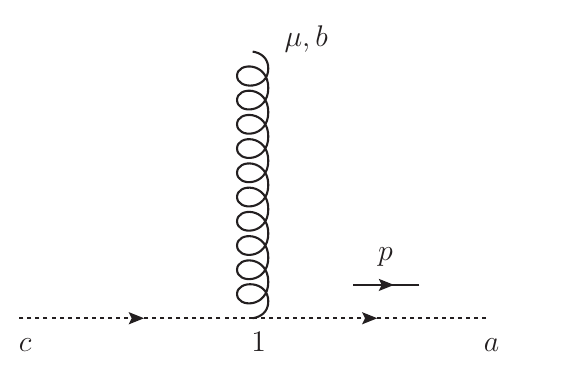}}  &=& $ -g f^{abc} p^\mu $ \\
	\raisebox{-0.8in}{\includegraphics[height=1.6in]{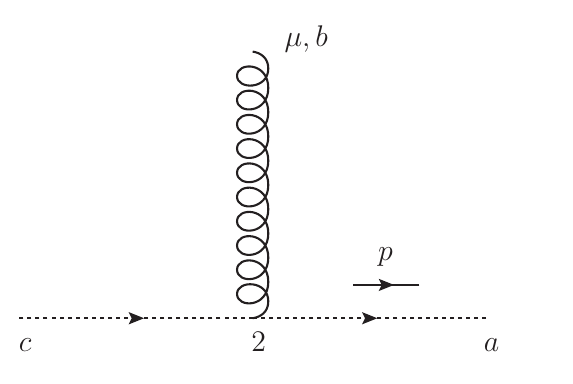}}  &=& $ +g f^{abc} p^\mu $
	\end{tabular}
\caption{Feynman rules associated to the gauge boson-ghost-ghost vertex, given here for the time-ordered and anti-time ordered branches of the Schwinger-Keldysh contour. The anti-time ordered branch multiplies the vertex factors by $(-1)$, which is naturally included in the notation we adopt in the main text.}
\label{fig:rules-g-ghost-appendix}
\end{figure}

For completeness, we also list the anti-commuting scalar field Wightman functions (see e.g., \cite{Hata:1980yr}), 
\begin{align}
C^{<}(k)&= (2\pi) \delta(k^2)\left[\Theta(-k_0) + n_{B}(|k_0|) \right],\\
C^{>}(k)&= (2\pi) \delta(k^2)\left[\Theta(+k_0) + n_{B}(|k_0|) \right].
\end{align}
Finally, we list in Figures~\ref{fig:rules-4g-appendix},~\ref{fig:rules-g-fermion-appendix}, and~\ref{fig:rules-g-ghost-appendix} the remaining Feynman rules that are relevant for the calculations shown in the main text.


\section{Proof of gauge-dependence cancellation at \texorpdfstring{$O(1-\xi)$}{O(1-x)}} \label{app:gauge-invariance}

In this Appendix we demonstrate the cancellation of the gauge-dependent terms that are proportional to $(1 - \xi)$, outlined in Section~\ref{sec:gauge-inv-1m-xi1}.

We start from the diagram $(1)$. Evaluating it, and keeping the linear terms on $(1-\xi)$ only, we can work through the index contractions to find
\begin{align}
&(1)_{(1-\xi)} \nn\\
&= -\frac{i}{2} g^2 N_c \delta^{ad} \D(p)_{I'I} \int_k \D(k)_{J'I'} \D(p-k)_{J'I'} (-1)^{I'+J'} \D(p)_{JJ'} P_{\rho \rho'}(-p) (-i p_0 g_j^{\rho} + i p_j g_{0}^{\rho}) \nonumber \\
& \quad \times (-2)\frac{(1-\xi)}{k^2} \bigg[ (p_0 g_i^{\rho'} - p_i g_0^{\rho'}) (- (p^2)^2 + 2p^2 (p-k)^2 - ((p-k)^2)^2 ) \nonumber \\ & \quad\quad\quad\quad\quad\quad\quad\quad + (p_0 k_i - p_i k_0) \left[  k^{\rho'} ( (p-k)^2  -2 p^2 ) - p^{\rho'} (k^2 - k \cdot p) \right] \bigg].
\end{align}
Because there is a chromoelectric field being contracted through the $\rho,\rho'$ indices, the contribution proportional to $p^{\rho'}$ vanishes, and we can therefore drop it, obtaining
\begin{align}
&(1)_{(1-\xi)} \nn\\
&= -\frac{i}{2} g^2 N_c \delta^{ad} \D(p)_{I'I} \int_k \D(k)_{J'I'} \D(p-k)_{J'I'} (-1)^{I'+J'} \D(p)_{JJ'} P_{\rho \rho'}(-p) (-i p_0 g_j^{\rho} + i p_j g_{0}^{\rho}) \nonumber \\
& \quad \times (-2)\frac{(1-\xi)}{k^2} \bigg[ (p_0 g_i^{\rho'} - p_i g_0^{\rho'}) (- (p^2)^2 + 2p^2 (p-k)^2 - (p-k)^2 (p^2 - 2k \cdot p + k^2) ) \nonumber \\ & \quad\quad\quad\quad\quad\quad\quad\quad + (p_0 k_i - p_i k_0) \left[  k^{\rho'} ( (p-k)^2  -2 p^2 ) \right] \bigg].
\end{align}
Now all terms in the integrand are explicitly proportional to a propagator momentum squared. The ones proportional to $p^2$ will cancel one of the $\D(p)$ propagators and allow this diagram to be put on equal footing as diagrams with only one $\D(p)$ propagator. However, the ones that have no factor of $p^2$ keep their two ``external'' propagators, and therefore any gauge-dependent contribution must be cancelled by a diagram with the same structure. The only other diagram with two $\D(p)$ propagators is (2), the 4-point vertex tadpole.

So we proceed to evaluate the tadpole diagram $(2)$, where if we only keep the $(1-\xi)$-dependent terms, we have
\begin{align}
    (2)_{(1-\xi)} &= (1-\xi)  \D(p)_{I'I}  \D(p)_{JI'} P_{\rho \rho'}(-p) (-i p_0 g_j^{\rho} + i p_j g_{0}^{\rho}) (-ig^2) N_c  \delta^{ad} \nonumber \\
    & \quad \times \int_k (-1)^{I'+1} \D(k)_{I'I'} (-i p_0 g_{i\mu} + ip_i g_{0\mu})\left[ g^{\mu \rho'} - \frac{k^\mu k^{\rho'}}{k^2} \right] \nonumber \\
    &= (1-\xi) g^2 N_c \delta^{ad} \D(p)_{I'I}  \D(p)_{JI'} P_{\rho \rho'}(-p) (-i p_0 g_j^{\rho} + i p_j g_{0}^{\rho}) (-1)^{I'+1} \nonumber \\
    & \quad \times \int_k \D(k)_{I'I'} \left[ (p_0 k_i - p_i k_0) \frac{k^{\rho'}}{k^2} - ( p_0 g_{i}^{\rho'} - p_i g_{0}^{\rho'}) \right] \,. \label{4-pt-gl-bubble}
\end{align}
Comparing this to the contributions from (1) that are not proportional to $p^2$, we find
\begin{align}
(1)_{(1-\xi), \, \not\propto p^2 } &= -\frac{i}{2} g^2 N_c \delta^{ad} \D(p)_{II'} \D(p)_{JJ'} P_{\rho \rho'}(-p) (-i p_0 g_j^{\rho} + i p_j g_{0}^{\rho}) \nonumber \\
& \quad  \int_k \D(k)_{J'I'} \D(p-k)_{J'I'} (-1)^{I'+J'} (-2)\frac{(1-\xi)}{k^2} (p-k)^2  \nonumber \\ & \quad\quad\quad\quad\quad\quad\quad\quad \times  \left[  k^{\rho'} (p_0 k_i - p_i k_0) - (p_0 g_i^{\rho'} - p_i g_0^{\rho'}) (k^2 - 2k\cdot p) \right] \nonumber \\
&= - (1-\xi) g^2 N_c \delta^{ad} \D(p)_{I'I} \D(p)_{JJ'} P_{\rho \rho'}(-p) (-i p_0 g_j^{\rho} + i p_j g_{0}^{\rho}) (-1)^{J'+1}  \nonumber \\
& \quad  \int_k \frac{1}{k^2} \D(k)_{J'I'} \mathbbm{1}_{J'I'}  \left[  k^{\rho'} (p_0 k_i - p_i k_0) - (p_0 g_i^{\rho'} - p_i g_0^{\rho'}) (k^2 - 2k\cdot p) \right] \nonumber \\
&= - (1-\xi) g^2 N_c \delta^{ad} \D(p)_{I'I} \D(p)_{JI'} P_{\rho \rho'}(-p) (-i p_0 g_j^{\rho} + i p_j g_{0}^{\rho}) (-1)^{I'+1}  \nonumber \\
& \quad  \int_k  \D(k)_{I'I'} \left[   (p_0 k_i - p_i k_0) \frac{k^{\rho'}}{k^2} - (p_0 g_i^{\rho'} - p_i g_0^{\rho'}) \right] \,,\label{3-pt-gl-bubble}
\end{align}
where we have dropped the term proportional to $k \cdot p$ in the last equality because it integrates to zero by inversion symmetry $k \to -k$.  The cancellation of \eqref{4-pt-gl-bubble} with \eqref{3-pt-gl-bubble} is obvious at this point: one contribution is just the additive inverse of the other.

Now we have to deal with the remaining pieces, which are all proportional to $p^2$. Concretely, the pieces of interest are given by
\begin{align}
(1)_{(1-\xi), \, \propto p^2} &= i p^2 g^2 N_c \delta^{ad} \D(p)_{I'I} \int_k \D(k)_{J'I'} \D(p-k)_{J'I'}  \D(p)_{JJ'} P_{\rho \rho'}(-p) (-i p_0 g_j^{\rho} + i p_j g_{0}^{\rho}) \nonumber \\
& \quad \times (-1)^{I'+J'} \frac{(1-\xi)}{k^2} \bigg[ (p_0 g_i^{\rho'} - p_i g_0^{\rho'} ) (- p^2 +  (p-k)^2  ) -2 k^{\rho'} (p_0 k_i - p_i k_0) \bigg] \nonumber \\
&= -g^2 N_c \delta^{ad} \int_k \D(k)_{I'I} \D(p-k)_{I'I} (-1)^{I'+1} \D(p)_{JI'} P_{\rho \rho'}(-p) (-i p_0 g_j^{\rho} + i p_j g_{0}^{\rho}) \nonumber \\
& \quad \times \frac{(1-\xi)}{k^2} \bigg[ (p_0 g_i^{\rho'} - p_i g_0^{\rho'} ) (  (p-k)^2 -p^2 ) -2 k^{\rho'} (p_0 k_i - p_i k_0) \bigg] \,.
\end{align}
This now has the same structure as diagrams (5) and (6), so we must evaluate those to compare with this contribution. We start with the $\xi$-dependent part of (5):
\begin{align}
    &(5)_{(1-\xi)} \nn\\
    &= (1-\xi) g^2 N_c \delta^{ad} \D(p)_{JJ'} P_{\rho \rho'}(p) (-i p_0 g_j^{\rho} + i p_j g_{0}^{\rho}) \nonumber \\
    &\quad \times \int_k (-i(p-k)_0 g_{i \mu} + i(p-k)_i g_{0\mu} )\D(p-k)_{I'I} \frac{1}{-ik_0+\epsilon}  \frac{k_0 k_\nu}{k^2} \D(k)_{I'I}  \nonumber \\
    & \quad \times (-1)^{I'+1} \left[ g^{\rho' \mu} (k-2p)^{\nu} + g^{\mu \nu}(p- 2k)^{\rho'} + g^{\nu \rho'} (k+p)^{\mu} \right] \nonumber \\
    &= (1-\xi) g^2 N_c \delta^{ad} \D(p)_{JJ'} P_{\rho \rho'}(p) (-i p_0 g_j^{\rho} + i p_j g_{0}^{\rho}) (-1)^{I'+1} \nonumber \\
    &\quad \times \int_k \D(p-k)_{I'I}\D(k)_{I'I} \frac{1}{k^2} \left[ ((p-k)_0 g_i^{\rho'} - (p-k)_i g_0^{\rho'}) (k^2 - 2 k \cdot p) + (p_0k_i  - p_i k_0)p^{\rho'} \right] \nonumber \\
    &= g^2 N_c \delta^{ad} \int_k \D(p-k)_{I'I}\D(k)_{I'I} (-1)^{I'+1} \D(p)_{JJ'} P_{\rho \rho'}(p) (-i p_0 g_j^{\rho} + i p_j g_{0}^{\rho})  \nonumber \\
    &\quad \times  \frac{(1-\xi)}{k^2}  \left[(p-k)_0 g_i^{\rho'} - (p-k)_i g_0^{\rho'}\right] ( (p-k)^2 - p^2 ) \,,
\end{align}
where in the last line we have dropped the term proportional to $p^{\rho'}$ because it vanishes when contracted with the momentum structure coming from the chromoelectric field.

Similarly for the diagram $(6)$, we find the $\xi$-dependent part is given by
\begin{align}
    (6)_{(1-\xi)} &= - g^2 N_c \delta^{ad} \int_k \D(k)_{I'I} \D(p-k)_{I'I} (-1)^{I'+1}  \D(p)_{JI'} P_{\rho \rho'}(p)  (-i p_0 g_j^{\rho} + i p_j g_{0}^{\rho}) \nonumber \\
    & \quad \times \frac{(1-\xi)}{k^2} \left[ ( k_0 g_i^{\rho'} - k_i g_0^{\rho'}) (2k \cdot p - k^2) + k^{\rho'} (p_0 k_i -k_0 p_i) \right]
\end{align}

These are all the contributions with three propagators, so at the very least we expect a cancellation of the $k^{\rho'}$terms. Adding things up, we have (the prefactor $1/2$ is the symmetry factor).
\begin{align}
    &\frac{1}{2} (1)_{(1-\xi), \, \propto p^2} + \big((5)+(6)\big)_{(1-\xi)} \nn\\
    &= g^2 N_c \delta^{ad}  \D(p)_{JJ'} P_{\rho \rho'}(-p) (-i p_0 g_j^{\rho} + i p_j g_{0}^{\rho})  \nonumber \\
    & \quad \times \int_k \D(p-k)_{I'I}\D(k)_{I'I} (-1)^{I'+1} \nonumber \\
    & \quad \times \frac{(1-\xi)}{k^2} \bigg[ - \frac{1}{2} \left( (p_0 g_i^{\rho'} - p_i g_0^{\rho'} ) (  (p-k)^2 -p^2 ) -2 k^{\rho'} (p_0 k_i - p_i k_0) \right) \nonumber \\
    & \quad \quad \quad \quad \quad \quad + \left[(p-k)_0 g_i^{\rho'} - (p-k)_i g_0^{\rho'}\right] ( (p-k)^2 - p^2 )  \nonumber \\
    & \quad \quad \quad \quad \quad \quad - \left[ ( k_0 g_i^{\rho'} - k_i g_0^{\rho'}) (p^2 - (p-k)^2 ) + k^{\rho'} (p_0 k_i -k_0 p_i) \right] \bigg] \nonumber \\
    &= \frac{1}{2} g^2 N_c \delta^{ad}  \D(p)_{JJ'} P_{\rho \rho'}(-p) (-i p_0 g_j^{\rho} + i p_j g_{0}^{\rho})  \nonumber \\
    & \quad \times \int_k \D(p-k)_{I'I}\D(k)_{I'I} (-1)^{I'+1}  \times \frac{(1-\xi)}{k^2} \left[p_0 g_i^{\rho'} - p_i g_0^{\rho'} \right] \left( (p-k)^2 - p^2 \right) \,.
\end{align}
Due to the last term, the integrand of the above expression has two pieces: one proportional to $p^2$, and another proportional to $(p-k)^2$. The term proportional to $p^2$ has two propagators in the integrand, one with momentum $(p-k)$ and the other with momentum $k$, which thus must be cancelled by the sum of diagrams (3), (4), and (8). The term proportional to $(p-k)^2$ has a loop integral over $k$ that is disconnected from the $p$ momentum flow, so it must be cancelled by contributions from (7) (diagram (9) vanishes).

We focus on the term proportional to $(p-k)^2$ first. A little algebra leads to
\begin{align}
    (7)_{(1-\xi)} = \frac{g^2 N_c}{2} \delta^{ad} \left[p_0^2 g_{ij} + p_i p_j g_{00} \right] \D(p)_{JI} \int_k \frac{(1-\xi)}{k^2} \D(k)_{II} \,.
\end{align}
Now, going back to our result from the three- and four-propagator diagrams, and looking at the piece proportional to $(p-k)^2$, we have
\begin{align}
    &\left(\frac{1}{2} (1)_{(1-\xi), \, \propto p^2} + \big((5)+(6) \big)_{(1-\xi)} \right)_{\propto (p-k)^2} \nn\\
    &=  \frac{1}{2} g^2 N_c \delta^{ad}  \D(p)_{JI'} P_{\rho \rho'}(p) (-i p_0 g_j^{\rho} + i p_j g_{0}^{\rho})  \nonumber \\
    & \quad \times \int_k (p-k)^2\, \D(p-k)_{I'I}\D(k)_{I'I} (-1)^{I'+1}  \frac{(1-\xi)}{k^2} \left[p_0 g_i^{\rho'} - p_i g_0^{\rho'} \right]  \nonumber \\
    &= \frac{g^2 N_c}{2} \delta^{ad} \D(p)_{JI} (- g_{\rho\rho'}) ( p_0 g_j^{\rho} -  p_j g_{0}^{\rho}) \left[p_0 g_i^{\rho'} - p_i g_0^{\rho'} \right] \int_k \frac{(1-\xi)}{k^2} \D(k)_{II} \nonumber \\
    &= -\frac{g^2 N_c}{2} \delta^{ad} \D(p)_{JI} \left( p_0^2 g_{ij} + p_i p_j g_{00} \right) \int_k \frac{(1-\xi)}{k^2} \D(k)_{II} \,,
\end{align}
and thus
\begin{align}
 \left(\frac{1}{2} (1)_{(1-\xi), \, \propto p^2} + \big((5)+(6)\big)_{(1-\xi)} \right)_{\propto (p-k)^2} + (7)_{(1-\xi)} = 0 \,.
\end{align}

Then we consider the remaining contribution from the term proportional to $p^2$
\begin{align}
    &\left(\frac{1}{2} (1)_{(1-\xi), \, \propto p^2} + \big((5)+(6)\big)_{(1-\xi)} \right)_{\propto p^2} \nn\\
    &= \frac{1}{2} g^2 N_c \delta^{ad}  (-p^2) \D(p)_{JI'} P_{\rho \rho'}(p) (-i p_0 g_j^{\rho} + i p_j g_{0}^{\rho})  \nonumber \\
    & \quad \times \int_k \D(p-k)_{I'I}\D(k)_{I'I} (-1)^{I'+1} \frac{(1-\xi)}{k^2} \left[p_0 g_i^{\rho'} - p_i g_0^{\rho'} \right] \nonumber \\
    &= \frac{-ig^2}{2} N_c \delta^{ad} (ip_0 g_{j \rho'} - i p_j g_{0  \rho'}) \left[p_0 g_i^{\rho'} - p_i g_0^{\rho'} \right] \int_k \D(p-k)_{JI}\D(k)_{JI} \frac{(1-\xi)}{k^2} \nonumber \\
    &= \frac{g^2 N_c}{2} \delta^{ad} \left(p_0^2 g_{ij} + p_i p_j g_{00} \right) \int_k \D(p-k)_{JI}\D(k)_{JI} \frac{(1-\xi)}{k^2}\,,
\end{align}
which is expected to be cancelled by diagrams with two propagators. To show this, we need to evaluate diagrams (3), (4), and (8). We start with $(4)$, whose gauge-dependent part is given by
\begin{align}
    (4)_{(1-\xi)} &= - g^2 N_c \delta^{ad} \int_k \big( (p-k)_0^2 g_{ij} +  (p-k)_i (p-k)_j g_{00} \big) \D(p-k)_{JI} \D(k)_{JI} \frac{(1-\xi)}{k^2} \,.
\end{align}
Then we evaluate the linear $(1-\xi)$ gauge-dependent part of the diagram $(3)$:
\begin{align}
(3)_{(1-\xi)} &= g^2 N_c \delta^{ad} \int_k \D(k)_{JI} \D(p-k)_{JI} (1-\xi) \left[ -g_{00} \frac{(p-k)_i (p-k)_j}{(p-k)^2} -g_{ij} \frac{k_0 k_0}{k^2} \right] \nonumber \\
&= -g^2 N_c \delta^{ad} \int_k \D(k)_{JI} \D(p-k)_{JI} \frac{(1-\xi)}{k^2} \left[ k_0^2 g_{ij} + k_i k_j g_{00} \right] \,.
\end{align}
At last, we evaluate the $(1-\xi)$-dependent part of diagram (8), obtaining
\begin{align}
(8)_{(1-\xi)} = -g^2 N_c \delta^{ad} \int_k \D(k)_{JI} \D(p-k)_{JI} \frac{(1-\xi)}{k^2} \left[ (p_0 k_0 - k_0^2) g_{ij} + (p_i k_j - k_i k_j) g_{00} \right] \,.
\end{align}
Finally, adding things up, one finds
\begin{align}
    & \left(\frac{1}{2} (1)_{(1-\xi), \, \propto p^2} + \big((5)+(6)\big)_{(1-\xi)} \right)_{\propto p^2} + \frac{1}{2} \left( (3)_{(1-\xi)} + (4)_{(1-\xi)} \right) + (8)_{(1-\xi)} \nonumber \\
    &= \frac{g^2 N_c}{2} \delta^{ad} \int_k \D(p-k)_{JI} \D(k)_{JI} \frac{(1-\xi)}{k^2} \nonumber \\
    & \quad \quad \quad \bigg[ \left(p_0^2 g_{ij} + p_i p_j g_{00} \right) -  \left[ k_0^2 g_{ij} + k_i k_j g_{00} \right] \nonumber \\
    & \quad \quad \quad \quad - \left( (p-k)_0^2 g_{ij} +  (p-k)_i (p-k)_j g_{00}\right) - 2\left[ (p_0 k_0 - k_0^2) g_{ij} + (p_i k_j - k_i k_j) g_{00} \right] \bigg] \nonumber \\
    &= \frac{g^2 N_c}{2} \delta^{ad} \int_k \D(p-k)_{JI} \D(k)_{JI} \frac{(1-\xi)}{k^2} \nonumber \\
     & \quad \quad \quad \bigg[ \left(p_0^2 - k_0^2 - (p_0^2 - 2p_0 k_0 + k_0^2) - 2p_0 k_0 + 2k_0^2 \right) g_{ij} \nonumber \\
     & \quad \quad \quad \quad + \left( p_i p_j - k_i k_j - (p_i p_j - p_i k_j - p_j k_i + k_i k_j ) - 2(p_i k_j -k_i k_j) \right) g_{00} \bigg] \nonumber \\
     &= 0 \,,
\end{align}
where we have made the symmetry factors of diagrams $(3)$ and $(4)$ explicit, which are $1/2$ for both diagrams.

Therefore, after taking into account all contributions, the result is $R_\xi$ gauge-invariant.
The proof of gauge invariance is now complete.

\section{NLO evaluation} \label{app:collinear-integrals}
Here, we briefly outline two different integration orders that we adopted to check independently the results of our calculations in Section~\ref{sec:feyn-calc}. This provides a verification that the collinear limit of the integrals was handled correctly in all cases, and also provides two different calculation strategies that could be adopted in the future to carry out this type of integrals.

\subsection{Integration order without regulator}

In this Appendix, we carry out the momentum integrals making use of the analytic structure of thermal correlation functions.

After some algebra, one can show that diagrams $(1)$, $(f)$, $(g)$, $(3)$, $(4)$, $(5)$, $(5r)$, $(6)$, $(6r)$, $(8)$, $(8r)$ and $(11)$ in their retarded forms, involve the structure:
\begin{align}
\Pi^R(p)\equiv\int_k N(p,k) \left[ D^{\ml{T}}(k) D^{\ml{T}}(p-k) - D^{<}(k) D^{<}(p-k) \right] \,. \label{eq:a1}
\end{align}
From here, we outline a procedure to arrange the propagator structure in a generic form that makes manifest how the different pole structures are handled, such that we can generically use it for the expressions given in Section~\ref{sect:nlo} to obtain the relevant part of the retarded component. 

Using the Feynman rules, we can write the finite temperature part of Eq.~(\ref{eq:a1}), after some algebra and relabelling, as
\begin{align}
&\Pi^R(p)= i \int \frac{\text{d}^3 k}{(2\pi)^3 2  |\mathbf k |} \frac{\text{d}^3 q}{(2\pi)^3 2  |\mathbf q |} n_{B}(|{\bs k}|) \times \nonumber \\
&\bigg\{ \left[ N(p,k) + N(p,p-k) \right] \left[ \frac{ (2\pi)^3 {\delta}^3({\bs p} - {\bs k} - {\bs q})}{p_0 - |{\bs k}| - |{\bs q}| + i 0^+}- \frac{ (2\pi)^3 {\delta}^3({\bs p} - {\bs k} + {\bs q})}{p_0 - |{\bs k}| + |{\bs q}| + i 0^+}  \right] \nonumber \\
&+ \left[ N(p,-k) + N(p,p+k)\right] \left[\frac{(2\pi)^3 {\delta}^3({\bs p} + {\bs k} - {\bs q})}{p_0 + |{\bs k}| - |{\bs q}| + i 0^+} - \frac{(2\pi)^3 {\delta}^3({\bs p} + {\bs k} + {\bs q})}{p_0+|{\bs k}|+|{\bs q}| + i 0^+}\right] \bigg\} \,. \label{eq:trans}
\end{align}
For example, the numerator factor in the case of diagram (1) (gauge boson self-energy, gauge boson loop) is:
\begin{align}
N(p,k)=- \frac{1}{2} \left[ g^{\mu \nu} (5 p^2 - 2 p \cdot k ) + 10 k^{\mu} k^{\nu} \right] \,,
\end{align}
which leads to the retarded object:
\begin{align}
\Pi^{R,(1)}(p)&= i \int \frac{\text{d}^3 k}{(2\pi)^3 2  |\mathbf k |} \frac{\text{d}^3 q}{(2\pi)^3 2  |\mathbf q |} 2 n_{B}(|{\bs k}|) \bigg\{- \frac{1}{2} \left[ g^{\mu \nu} 4 p^2 + 5 k^{\mu} k^{\nu} \right] \bigg\} \times \nonumber \\
&\bigg\{  \left[ \frac{ (2\pi)^3 {\delta}^3({\bs p} - {\bs k} - {\bs q})}{p_0 - |{\bs k}| - |{\bs q}| + i 0^+}- \frac{ (2\pi)^3 {\delta}^3({\bs p} - {\bs k} + {\bs q})}{p_0 - |{\bs k}| + |{\bs q}| + i 0^+}  \right] \nonumber \\
&+ \left[\frac{(2\pi)^3 {\delta}^3({\bs p} + {\bs k} - {\bs q})}{p_0 + |{\bs k}| - |{\bs q}| + i 0^+} - \frac{(2\pi)^3 {\delta}^3({\bs p} + {\bs k} + {\bs q})}{p_0+|{\bs k}|+|{\bs q}| + i 0^+}\right] \bigg\} \,.\label{eq:selfgeneric}
\end{align}
We can write down similar expressions for all the other diagrams. 

\begin{table*}
\begin{center}
\begin{tabular}{c||c|c|c}
\toprule
   $\sigma_1\sigma_2$ & $z_0$ & $z_p$& $z_m$   \\ \midrule
    $++$ & $\Delta E$ & $-|{\bs k}|\tau + \sqrt{{\bs k}^2\tau^2 + \Delta E^2 + 2 \Delta E |{\bs k}|}$ & $-$ \\ \midrule
    $+-$ & $\Delta E$ & $-$ & $-$ \\  \midrule
    $-+$ & $\Delta E$ & $|{\bs k}|\tau + \sqrt{{\bs k}^2\tau^2 + \Delta E^2 - 2 \Delta E |{\bs k}|}$\,, & $|{\bs k}|\tau - \sqrt{{\bs k}^2\tau^2 + \Delta E^2 - 2 \Delta E |{\bs k}|}$ \,, \\
& & if $  \bigg\{ \Delta E/2 \leq |{\bs k}| \leq \Delta E  $ & if $\Delta E/2 \leq |{\bs k}| \leq \Delta E  $\\
& & $ \land \tau \geq \sqrt{\frac{2\Delta E |{\bs k}|-\Delta E^2}{|{\bs k}|^2}} \bigg\} $ & $\land \tau \geq \sqrt{\frac{2\Delta E |{\bs k}|-\Delta E^2}{|{\bs k}|^2}}$\,. \\
& & or $ 0 \leq |{\bs k}| \leq \Delta E/2 \,. $ &  \\ \midrule
$--$ & $\Delta E$ & $|{\bs k}|\tau - \sqrt{{\bs k}^2\tau^2 + \Delta E^2 - 2 \Delta E |{\bs k}|}$ \,, & $|{\bs k}|\tau + \sqrt{{\bs k}^2\tau^2 + \Delta E^2 - 2 \Delta E |{\bs k}|}$ \,, \\  
 &  & if $|{\bs k}|\geq \Delta E \land \tau\geq\sqrt{\frac{2\Delta E |{\bs k}|-\Delta E^2}{|{\bs k}|^2}} $ \,. & if $|{\bs k}|\geq \Delta E \land \tau\geq\sqrt{\frac{2\Delta E |{\bs k}|-\Delta E^2}{|{\bs k}|^2}} $ \,. \\  
 \bottomrule
\end{tabular}
\end{center}
\caption{Summary of the single ($z_{p/m}$) and double ($z_0$) poles, as well as their existence criteria.}
\label{tab:poles}
\end{table*}

For all contributions that can be written in this form, to obtain their contributions to the integrated spectral function $\varrho_E^{++}(p_0=\Delta E)$, we use the three momentum delta function to perform the $\bs q$ integration first. Then, we interchange the order of the loop $\bs k$ integration with the external $\bs p$ integration, and finally we perform the $\bs p$ integration after taking the imaginary part of the retarded correlation function structure shown above, multiplied by $\bs p$-dependent propagators coming from the rest of the diagram. When taking the imaginary part, we apply the residue theorem. These procedures lead to the following schematic arrangement in terms of the contributing poles:
\begin{align}
R^{++}_{0} &\equiv \int_0^{\infty} \text{d} |{\bs k}| \int_{-1}^{1} \text{d} \tau \left[ \text{Res}(G_{0}^{++},z_0) + \text{Res}(G_{0}^{++},z_p) \right], \label{eq:Rpp}\\
R^{+-}_{0} &\equiv  \int_0^{\infty} \text{d} |{\bs k}| \int_{-1}^{1} \text{d} \tau \left[ \text{Res}(G_{0}^{+-},z_0) \right], \label{eq:Rpm} \\
R^{-+}_{0} &\equiv  \int_0^{\Delta E /2} \text{d} |{\bs k}| \int_{-1}^{1} \text{d} \tau \left[ \text{Res}(G_{0}^{-+},z_0) +\text{Res}(G_{0}^{-+},z_p)\right] \label{eq:Rmp}\\
&+ \int_{\Delta E /2}^{\Delta E} \text{d} |{\bs k}|  \bigg\{ \int_{\sqrt{\frac{2\Delta E |{\bs k}|-\Delta E^2}{|{\bs k}|^2}}}^{1} \text{d} \tau  \left[ \text{Res}(G_{0}^{-+},z_0) +\text{Res}(G_{0}^{-+},z_p)-\text{Res}(G_{0}^{-+},z_m)\right]\nonumber  \\
& \phantom{\int_{\Delta E /2}^{\Delta E} \text{d} |{\bs k}| } + \int_{-1}^{\sqrt{\frac{2\Delta E |{\bs k}|-\Delta E^2}{|{\bs k}|^2}}}\text{d} \tau  \left[ \text{Res}(G_{0}^{-+},z_0)\right] \bigg\} \nonumber \\
&+\int_{\Delta E}^{\infty} \text{d} |{\bs k}| \int_{-1}^{1} \text{d} \tau \left[ \text{Res}(G_{0}^{-+},z_0) \right], \nonumber \\
R^{--}_{0}&\equiv  \int_0^{\Delta E} \text{d} |{\bs k}| \int_{-1}^{1} \text{d} \tau \left[ \text{Res}(G_{0}^{--},z_0) \right] \label{eq:Rmm}\\
&+ \int_{\Delta E}^{\infty} \text{d} |{\bs k}|  \bigg\{ \int_{\sqrt{\frac{2\Delta E |{\bs k}|-\Delta E^2}{|{\bs k}|^2}}}^{1} \text{d} \tau  \left[ \text{Res}(G_{0}^{--},z_0) +\text{Res}(G_{0}^{--},z_p)-\text{Res}(G_{0}^{--},z_m)\right]\nonumber\\
& \phantom{\int_{\Delta E /2}^{\Delta E} \text{d} |{\bs k}| } + \int_{-1}^{\sqrt{\frac{2\Delta E |{\bs k}|-\Delta E^2}{|{\bs k}|^2}}}\text{d} \tau  \left[ \text{Res}(G_{0}^{-+},z_0)\right] \bigg\} \,. \nonumber \\
\end{align}
Here we have introduced many new notations that we will explain now. Firstly, we have defined the location of the single ($z_{p/m}$) and double poles ($z_0$) where the integral over ${\bs p}$ gives residue contributions (the double pole comes from having $[D^R(p)]^2$ multiplying the gauge boson self-energy diagrams; in diagrams (5), (5r), (6), (6r) it is only a single pole because there is only a single $D^R(p)$ factor, and there is no corresponding pole in the rest of the diagrams). Their existence criteria, as well as their values as a function of ${\bs k}$ and $\Delta E$ are listed in Table~\ref{tab:poles}. The four $G^{\sigma_1 \sigma_2}_0$ functions originate from the numerator structures that accompany the four single poles in Eq.~(\ref{eq:selfgeneric}), and can be explicitly determined from comparison after performing the integration order as listed above ($\sigma_1$ and $\sigma_2$ denote the different relative sign choices in front of the absolute values $|{\bs q}|$, $|{\bs k}|$ in the four denominators of~\eqref{eq:selfgeneric}). Concretely, for the fermion self-energy these are given by Eq.~(4.15) in Ref.~\cite{Binder:2020efn}. As discussed in the main text, in the case of the gauge boson self-energy the double and single poles are paired, in the sense that they are individually divergent when the momenta become collinear, and only the sum over both residues is collinear-finite. A similar pairing is required for diagrams (5), (5r), (6), (6r), where the sum over poles is also collinear-finite. Because there is no $D^R(p)$ propagator in the rest of the diagrams, their respective spectral functions are safe in the collinear limit (as explained in the main text, the purported collinear divergence appears when two poles become closer, but there is no pair of poles at all in these diagrams, only single poles).

For all terms we then manage to further perform the angular $\tau = \cos\theta$ ($\theta$ denotes the relative angle bewtween ${\bs p}$ and ${\bs k}$) integration analytically. As an illustrative example, for diagram (3) in Figure~\ref{fig:diagrams} we can even perform the last $k$ integral analytically, obtaining a fully analytic expression:
\begin{align}
R_3\left(x=\frac{\Delta E}{T}\right) = - \frac{3 N_c}{2x^2} \big[ & -2 \Re[\text{Li}_2(e^x)]+2 \text{Li}_2[-\cosh(x)+\sinh(x)+1] \\ & + x^2-2 x \ln[e^x-1]+\pi^2 \big] \,.\nonumber
\end{align}
The function is entirely negative, monotonically increasing for increasing $x$, and at large $x$ it satisfies $\lim_{x\rightarrow \infty} R_3(x)=0$.

We note that, for diagrams (5) and (5r) in Figure~\ref{fig:diagrams}, which involve contributions coming from Wilson lines, the remaining $k$ integration needs to be performed with a principal part prescription at the location $k=\Delta E$, in the same way as the result is described in the main text.

\subsection{Integration order with regulator}

In this Appendix, we carry out the calculations of the momentum integrals including an explicit regulator for the collinear limit.

Starting from equation~\eqref{eq:integrated-gluon-self-energy-before-integration}, we can work with the full expression (before taking the real part) for the temperature-dependent pieces. We define
\begin{equation}
\int_{\p} L_T = i\int_{\k,\q} \sum_{\sigma_1,\sigma_2}  \frac{\sigma_2 2 n_B(|\k|) }{2|\k|2|\q|} \frac{(-1) N((p_0,\q - \k),(-\sigma_1 |\k|,-\k)) }{( (p_0 + i\epsilon)^2 - (\q - \k)^2 )^2 (p_0 + \sigma_1 |\k| - \sigma_2 |\q| + i0^+)} \,,
\end{equation}
which satisfies
\be
{\rm Re} \{g^2 N_c (N_c^2-1) \int_{\p} L_T\} = \left. \varrho_E^{++}(p_0) \right|_{\rm NLO}^{\rm gauge\ boson+ghost} - \left. \varrho_E^{++}(p_0) \right|_{{\rm NLO}, \, T=0 }^{\rm gauge\ boson+ghost}  \,.\quad\ 
\ee

First, we do the integral over the angle between $\k $ and $\q $, $u \equiv \frac{\k \cdot \q}{kq}$, which only involves the numerator and the retarded propagator with momentum $p$. Explicitly, writing
\begin{equation}
    N((p_0,\q - \k),(-\sigma_1 |\k|,-\k)) = c(p_0,-\sigma_1 k, q, k) + d(p_0,-\sigma_1 k, q, k) u + e(p_0,-\sigma_1 k, q, k) u^2\,,
\end{equation}
we have
\be
\int_{\p} L_T &=& \frac{(-i) (2\pi) (4\pi) }{ (2\pi)^6 } \nn\\
&\times& \int_{0}^\infty \diff k \diff q \sum_{\sigma_1,\sigma_2}  \frac{  \sigma_2 n_B(k)  }{8 k q (p_0 + \sigma_1 k - \sigma_2 q + i0^+)} \int_{-1}^1 \diff u \frac{ c + du + eu^2 }{  ( u - u_0 + i{\rm sgn}(p_0) 0^+)^2  } \,,\quad\quad\quad
\ee
with $u_0 = u_0(p_0,q,k) = (k^2 + q^2 - p_0^2)/(2kq)$. The integral over $u$ can be done straightforwardly. If we introduce a regulator $\delta$ to control the collinear divergence,
\begin{equation}
\begin{split}
    &\int_{-1+\delta}^{1-\delta} \diff u \frac{ c + du + eu^2 }{  ( u - u_0 + i{\rm sgn}(p_0) 0^+)^2  } \\ &=   \bigg[ 2 e - ( e u_0^2 + d u_0 + c  ) \left( \frac{1}{1 - u_0+ i0^+ {\rm sgn}(p_0) } - \frac{1}{-1 - u_0+ i0^+ {\rm sgn}(p_0) } \right) \\
& \quad  + (2e u_0 + d ) \bigg( \ln \left| \frac{1 - \delta - u_0}{-1 + \delta - u_0} \right| - i \pi \Theta(1-\delta -u_0) \Theta(u_0 - (-1 + \delta))   \bigg)  \bigg] \,,
\end{split}
\end{equation}
where we have let $\delta \to 0$ in the non-problematic terms; namely, in every place except the logarithms and the terms needed to compensate for them. For notational simplicity we define
\begin{align}
    {\rm NC} = 2e/(8kq)\,, & & {\rm NL} = (2 e u_0 + d)/(8 k q)\,, & & {\rm ND} = -(e u_0^2 + du_0 + c)/(8p_0) \,.
\end{align}
Then we have
\be
\int_{\p} L_T^{(1)} &=& \frac{(-i) }{ 8\pi^4 } \int_{0}^\infty \!\! \diff k  \diff q \, n_B(k) \sum_{\sigma_1,\sigma_2} \bigg[  \frac{  \sigma_2  {\rm NC}  }{ p_0 + \sigma_1 k - \sigma_2 q + i0^+} \nn\\
&& \quad \quad +  \frac{  \sigma_2 {\rm NL}   }{ p_0 + \sigma_1 k - \sigma_2 q + i0^+} \bigg( \ln \left| \frac{1 - \delta - u_0}{1 - \delta + u_0} \right| - i \pi \Theta(1-\delta -u_0) \Theta(u_0 +1 - \delta)   \bigg) \nn\\
&& \quad \quad +  \sum_{\sigma_1' \sigma_2'} \frac{  \sigma_2 {\rm ND}   }{ p_0 + \sigma_1 k - \sigma_2 q + i0^+} \frac{\sigma_1' \sigma_2'}{p_0 + \sigma_1' k - \sigma_2' q + i0^+}  \bigg] \,.
\ee
Numerically the strategy is to take the real part of this expression and integrate over $q$. The terms proportional to ${\rm NC}$ and ${\rm ND}$ may be evaluated directly using the residue theorem, and then we take $\delta \to 0$ at the end.

For the terms proportional to ${\rm NL}$, there are two cases:
    \begin{enumerate}
        \item $\sigma_1 \sigma_2 = 1$, in which case the logarithmic term can be written as
        \be
        \label{eqn:log}
            && \ln \left| \frac{ (|\k| - |\q|)^2 - p_0^2 + 2\delta |\k| |\q| }{(|\k| + |\q|)^2 - p_0^2 - 2\delta |\k| |\q| } \right|_{|\q| = |\k| + \sigma_2 p_0} \Theta(|\k| + \sigma_2 p_0) \nn\\
            &=& \ln \left| \frac{ 2\delta |\k| |\q| }{(|\k| + |\q|)^2 - p_0^2 - 2\delta |\k| |\q| } \right|_{|\q| = |\k| + \sigma_2 p_0} \Theta(|\k| + \sigma_2 p_0) \,,
        \ee
        where the theta functions come from the fact that the integration variable $|\q|$ is positive. This diverges as $\delta \to 0$, so it must be compensated by the other term with the $\Theta$ functions. For purposes of evaluating the divergent terms, we can add and subtract a term that treats ${\rm NL}$ as a constant, where all $|\q|$ dependence in NL is set to be the value where the singularity occurs $|\q| = \sigma_1 \sigma_2 |\k| + \sigma_2 p_0$. So, we have to evaluate
        \begin{equation}
        \int_0^\infty \diff |\q| \frac{\Theta\big( (|\k| + |\q|)^2 - p_0^2 - 2\delta |\k| |\q| \big) \Theta\big( p_0^2 - (|\k| - |\q|)^2 - 2\delta |\k| |\q| \big) }{ p_0 + \sigma_1|\k| - \sigma_2|\q| } \,.
        \end{equation}
        The Heaviside step functions contain essential information. They bound either momentum in terms of the other as
        \begin{equation}
            \left| (1-\delta) k - \sqrt{p_0^2 - 2\delta k^2 + \delta^2 k^2} \right| < q < (1-\delta) k  + \sqrt{p_0^2 - 2\delta k^2 + \delta^2 k^2} \,,
        \end{equation}
        which means that our integral is actually
        \begin{equation}
        \begin{split}
        \label{eqn:angle}
            \int_{\big| (1-\delta) |\k| - \sqrt{p_0^2 - 2\delta |\k|^2 + \delta^2 |\k|^2} \big|}^{(1-\delta) |\k| + \sqrt{p_0^2 - 2\delta |\k|^2 + \delta^2 |\k|^2}} \frac{\diff |\q|}{p_0 + \sigma_1 |\k| - \sigma_2 |\q|} \\
            = -\sigma_2 \ln \left|\frac{ -\sigma_2 p_0 - \sigma_1 \sigma_2 |\k| +  (1-\delta) |\k| + \sqrt{p_0^2 - 2\delta |\k|^2 + \delta^2 |\k|^2}}{ -\sigma_2 p_0 - \sigma_1 \sigma_2 |\k| + |  (1-\delta) |\k| - \sqrt{p_0^2 - 2\delta |\k|^2 + \delta^2 |\k|^2}| } \right| \\
            = -\sigma_2 \ln \left|\frac{ -\sigma_2 p_0 - \delta |\k| + \sqrt{p_0^2 - 2\delta |\k|^2 + \delta^2 |\k|^2}}{ -\sigma_2 p_0 - |\k| + |(1-\delta) |\k| - \sqrt{p_0^2 - 2\delta |\k|^2 + \delta^2 |\k|^2}| } \right| \,,
        \end{split}
        \end{equation}
        where we have used $\sigma_1 \sigma_2 = 1$. The sign of the integral~(\ref{eqn:angle}) depends on $\sigma_2$ (actually, on $\sigma_2 \,{\rm sgn}(p_0)$, but we take $p_0>0$ throughout), which is an external overall factor. All that remains now is to take the limit. We first consider the case with $\sigma_2 = -1$. It is now clear that Eq.~(\ref{eqn:angle}) is divergent as $\delta \to 0$ when $|\k| > |p_0|$, and finite if the converse inequality is true (we can always choose $\delta$ small enough so that it doesn't affect the inequalities, because $|\k|$ and $|p_0|$ are fixed at this step). Therefore, we can erase the absolute value in the denominator of Eq.~(\ref{eqn:angle}) and obtain the sum of Eqs.~(\ref{eqn:log}) and~(\ref{eqn:angle}) for $|{\bs k}_1|>|p_0|$
        \begin{equation}
        \begin{split}
            \lim_{\delta \to 0} & \ln \left| \frac{ 2\delta  |\k| (|\k| + \sigma_2 p_0) }{( 2|\k| + \sigma_2 p_0)^2 - p_0^2 - 2\delta |\k|(|\k| + \sigma_2 p_0) } \right|  \\ & -  \sigma_2 \ln \left|\frac{ -\sigma_2 p_0 - \delta |\k| + \sqrt{p_0^2 - 2\delta |\k|^2 + \delta^2 |\k|^2}}{ -\sigma_2 p_0 -\delta |\k| - \sqrt{p_0^2 - 2\delta |\k|^2 + \delta^2 |\k|^2}} \right| \,,
        \end{split}
        \end{equation}
        and we can take $\delta \to 0$ in all terms that are not going to 0, obtaining
        \begin{equation}
        \begin{split}
            \lim_{\delta \to 0} & \ln \left| \frac{ 2\delta  |\k| (|\k| - |p_0|) }{ 4 |\k|^2 - 4  |\k| |p_0| } \right|  +  \ln \left|\frac{ 2|p_0| }{ |p_0| -\delta |\k| - \sqrt{p_0^2 - 2\delta |\k|^2 + \delta^2 |\k|^2}} \right| \\
            =\lim_{\delta \to 0} & \ln \left| \frac{ \delta}2 \right| +  \ln \left|\frac{ 2|p_0| }{ |p_0| -\delta |\k| - |p_0| (1 - \delta |\k|^2/p_0^2 )} \right| \\
            = \lim_{\delta \to 0} & \ln \left| \frac{ \delta}2 \right| +  \ln \left|\frac{ 2 }{  \delta (|\k|^2/p_0^2 - |\k|/|p_0|) } \right| = - \ln \left| \frac{|\k|^2}{|p_0|^2} - \frac{|\k|}{|p_0|} \right| \,,
        \end{split}
        \end{equation}
        which is finite. Then we consider $\sigma_2 = 1$. That means that the divergence comes from the numerator, and the limit of the sum is equal to
        \begin{equation}
        \begin{split}
            \lim_{\delta \to 0} & \ln \left| \frac{ 2\delta  |\k| (|\k| + \sigma_2 p_0) }{( 2|\k| + \sigma_2 p_0)^2 - p_0^2 - 2\delta |\k|(|\k| + \sigma_2 p_0) } \right|  \\ &-  \sigma_2  \ln \left|\frac{ -\sigma_2 p_0 - \delta |\k| + \sqrt{p_0^2 - 2\delta |\k|^2 + \delta^2 |\k|^2}}{ -\sigma_2 p_0 - |\k| + | |\k| - |p_0|| } \right| \\
            = \lim_{\delta \to 0} & \ln \left| \frac{ 2\delta  |\k| (|\k| + |p_0|) }{ 4|\k|^2 + 4 |\k| |p_0|  } \right| -  \ln \left|\frac{ -|p_0| - \delta |\k| + \sqrt{p_0^2 - 2\delta |\k|^2 + \delta^2 |\k|^2}}{ -|p_0| - |\k| + | |\k| - |p_0|| } \right| \\
            =\lim_{\delta \to 0} & \ln \left| \frac{ \delta}2 \right| -  \ln \left|\frac{ - \delta |\k| - \delta |\k|^2/|p_0| }{ -|p_0| - |\k| + | |\k| - |p_0|| } \right| = - \ln \left| \frac{ |\k|^2 + |\k| |p_0| }{|p_0| \min(|\k|,|p_0|) } \right|\,,
        \end{split}
        \end{equation}
        which again, is finite. So we see that this should work for all terms.
        \item $\sigma_1 \sigma_2 = -1$ is the remaining case. Now the first term (the term with the logarithm) after integrating over $|\q|$ goes to
        \begin{equation}
        \label{eqn:case2_log}
        \ln \left| \frac{  (|\k| - |\q|)^2 - p_0^2 + 2\delta |\k| |\q| }{(|\k| + |\q|)^2 - p_0^2 - 2\delta |\k| |\q| } \right| \to \ln \left| \frac{ 4 |\k|^2 - 4\sigma_2 p_0 |\k| }{ 2\delta |\k| (|\k| - \sigma_2 p_0) } \right| \Theta(-|\k| + \sigma_2 p_0)  \,.
        \end{equation}
        The term with the $\Theta$-functions becomes after the integral over $|\q|$
        \begin{equation}
        \label{eqn:case2_theta}
            -\sigma_2  \ln \left|\frac{ -\sigma_2 p_0 + |\k| +  (1-\delta) |\k| + \sqrt{p_0^2 - 2\delta |\k|^2 + \delta^2 |\k|^2}}{ -\sigma_2 p_0 + |\k| + |  (1-\delta) |\k| - \sqrt{p_0^2 - 2\delta |\k|^2 + \delta^2 |\k|^2}| } \right| \,,
        \end{equation}
        which only diverges in the denominator as $\delta \to 0$ if $\sigma_2 = 1$ and $|p_0| > |\q|$. This regime is exactly where the Heaviside step function of the first term with the logarithm~(\ref{eqn:case2_log}) is supported. In this regime, the term~(\ref{eqn:case2_theta}) can be simplified as
        \begin{equation}
            - \ln \left|\frac{ 2|\k| }{ \delta |\k| - \delta |\k|^2/|p_0| } \right|\,,
        \end{equation}
        and so the limit of the sum of these two terms is
        \begin{equation}
            \lim_{\delta \to 0} \left( \ln \frac{2}{\delta} - \ln \left| \frac{ 2|\k| }{ \delta |\k| - \delta |\k|^2/|p_0|} \right| \right) = \ln \left| 1 - \frac{|\k|}{|p_0|}\right| \,.
        \end{equation}
        This shows that the collinear divergence cancels out exactly. Therefore, this guarantees that if we evaluate the integrals (numerically or analytically) with $\delta > 0$ and then take $\delta \to 0$ at the end, we will get a well-defined result. 
    \end{enumerate}

An equivalent treatment can be done for diagrams $(5)$, $(5r)$, $(6)$, $(6r)$, by simply modifying the number of propagators $D^R(p)$ that appear. However, we will not apply the methods shown in this section for these diagrams, since we want to evaluate them in dimensional regularization to extract their UV divergence in vacuum, which will be performed in the next section of this Appendix.

\subsection{Contribution from diagrams \texorpdfstring{$(5)$}{(5)}, \texorpdfstring{$(5r)$}{(5r)}, \texorpdfstring{$(6)$}{(6)}, and \texorpdfstring{$(6r)$}{(6r)}} \label{app:3-prop-detail}

Here we give the details of how we evaluate the contribution of diagrams $(5)$, $(5r)$, $(6)$, $(6r)$, as outlined in Section~\ref{sec:3-propagator-evaluation}, to the integrated spectral function
\begin{equation}
\begin{split}
\left. \varrho_E^{++}(p_0) \right|_{\rm NLO}^{5-6} &= i g^2 N_c (N_c^2-1) \tilde{\mu}^\epsilon \int_{\k,\p} \frac{1+2n_B(k)}{2k} \sum_{\sigma_1}   N_{3p}((p_0,\p),(\sigma_1 k, \k)) \\
  & \quad \quad \quad \quad \quad \quad \quad \times {\rm Re}\left\{ \frac{i}{((p_0+i0^+)^2 -\p^2 )((p_0-k_0 + i0^+)^2 - (\p-\k)^2 )} \right\} \\
  & \quad + g^2 N_c (N_c^2 -1) \pi \int_{\k,\p}   \frac{ \left[ k_0 N^{(5),(6)}(p,k) \right]_{k_0=0} }{\k^2} \ml{P} \left( \frac{1}{p_0^2 - (\p - \k)^2} \right) \delta(p^2)  \,,
\end{split}
\end{equation}
where $N_{3p}(p,k)$ includes the sum of all numerators of diagrams (5), (5r), (6), (6r) and $\tilde{\mu}^2 = \mu^2e^{\gamma_E}/(4\pi)$. Here we have factored out of the numerators the factor of $N_c$. All integrals, $\int_\k$ and $\int_\p$, are in $d = 3-\epsilon$ dimensions.

{Let us first evaluate the last piece. To that end, note that 
\begin{align}
    \left[ k_0 N^{(5),(6)}(p,k) \right]_{k_0=0} &= p_0 \left[ 2\k^2 - 2 \p \cdot \k - 2(d-1) p_0^2 + 2\p^2 \right] \nonumber \\ 
    &=  p_0 \left[ \k^2 +(\p-\k)^2 - p_0^2 - (2d-3) p_0^2 + \p^2 \right] \, ,
\end{align}
and therefore
\begin{align}
    &\pi \int_{\k,\p}   \frac{ \left[ k_0 N^{(5),(6)}(p,k) \right]_{k_0=0} }{\k^2} \ml{P} \left( \frac{1}{p_0^2 - (\p - \k)^2} \right) \delta(p^2) \nonumber \\
    &= \pi p_0 \int_\p \delta(p^2) \int_\k \left[ \ml{P} \frac{1}{p_0^2 - (\p-\k)^2} - \frac{1}{\k^2} + \ml{P} \frac{\p^2 - (2d-3)p_0^2}{\k^2 (p_0^2 - (\p-\k)^2)} \right] \, .
\end{align}
The first two terms in the integrand vanish in dimensional regularization, and the last one decays as $\k^4$, which means it is convergent in $d=3$ dimensions for the $\k$ integral. Consequently,
\begin{align}
    &\pi \int_{\k,\p}   \frac{ \left[ k_0 N^{(5),(6)}(p,k) \right]_{k_0=0} }{\k^2} \ml{P} \left( \frac{1}{p_0^2 - (\p - \k)^2} \right) \delta(p^2) \nonumber \\
    &= \pi p_0 \frac{(4\pi)(2\pi)}{(2\pi)^6} \int_0^\infty \!\! d|\p| \, \p^2 \delta(p_0^2 - \p^2) \int_0^\infty \!\! d|\k| \, \k^2 \int_{-1}^1 \! du \, \ml{P} \frac{-4p_0^2}{ \k^2 (2 |\p| |\k| u - \k^2 ) } \nonumber \\
    &= - \frac{4p_0^3}{(2\pi)^3} \frac{p_0^2}{2p_0} \int_0^\infty \!\! d|\k| \int_{-1}^1 \! du \frac{1}{2 p_0 |\k|} \ml{P} \frac{1}{u - \k^2/(2 p_0 |\k|)} \nonumber \\
    &= - \frac{p_0^3}{(2\pi)^3} \int_0^\infty \frac{dk}{k} \ln \left| \frac{1 - \frac{k}{2p_0}}{1 + \frac{k}{2p_0}} \right| =  \frac{p_0^3}{(2\pi)^3} \int_0^\infty \frac{dk}{k} \ln \left| \frac{1 + k}{1 - k} \right| = \frac{p_0^3}{(2\pi)^3} \frac{\pi^2}{2} \,.
\end{align}
}

Now, {for the remaining piece,} instead of splitting the $\p$ integral into angular and radial components, we proceed using the standard QFT machinery to evaluate the $\p$ integral. We define $\tilde{p}_0 \equiv p_0 + i0^+$, and $\bar{N}(p,k) = {\rm Im}\{N_{3p}(p,k)\}$. Then we have
\begin{equation}
\begin{split}
&\left. \varrho_E^{++}(p_0) \right|_{\rm NLO}^{5-6} {- \frac{g^2 N_c (N_c^2-1)\pi^2}{2(2\pi)^3}} \\
&= (-1) g^2 N_c (N_c^2-1) \tilde{\mu}^\epsilon \int_{\k} \frac{1+2n_B(k)}{2k} \sum_{\sigma_1} {\rm Re}\left\{ \int_0^1 \diff x \int_\p \frac{i \bar{N}((p_0,\p+x\k),k) }{ ( \p^2 -  (\tilde{p}_0 - x k_0 )^2 ) } \right\} \\
 &= - g^2 N_c (N_c^2-1) \tilde{\mu}^\epsilon \frac{\Omega_{3-\epsilon}}{(4\pi)^{(3-\epsilon)/2}} \Gamma \left( \frac{-1+\epsilon}{2} \right) \int_0^\infty \frac{\diff k k^{1 - \epsilon}}{(2\pi)^{d-1}} \frac{1+2n_B(k)}{2} \\
 & \times \sum_{\sigma_1} {\rm Re}\bigg\{i \int_0^1 \diff x \bigg[ \frac{3-\epsilon}{2} A(p_0,k_0,x) D^{(1-\epsilon)/2} + \frac{-1+\epsilon}{2} B(p_0,k_0,x) D^{(-1-\epsilon)/2} \bigg] \bigg\}\,,
\end{split}
\end{equation}
where $D = - (\tilde{p}_0 - x k_0 )^2$ and 
\be
\bar{N}((p_0,\p+x\k),k) = A(p_0,k_0) \p^2 + B(p_0,k_0,x) + ({\rm terms\ linear \ in \ } \p)  \,.
\ee
Explicit inspection of the coefficients reveals that $A$ does not depend on $x$.

Next we do the integrals over the Feynman parameter $x$. We can define
\begin{align}
    I_1(p_0,k_0) &= p_0^{-3} {\rm Re}\bigg\{ i \int_0^1 \diff x \, D^{(1-\epsilon)/2} \bigg\} \\
    I_2(p_0,k_0) &= p_0^{-3} {\rm Re}\bigg\{ i \int_0^1 \diff x \, D^{(-1-\epsilon)/2} \bigg\} \\
    I_3(p_0,k_0) &= p_0^{-3} {\rm Re}\bigg\{ i \int_0^1 \diff x \, x \, D^{(-1-\epsilon)/2} \bigg\} \\
    I_4(p_0,k_0) &= p_0^{-3} {\rm Re}\bigg\{ i \int_0^1 \diff x \, x^2 \, D^{(-1-\epsilon)/2} \bigg\} \,,
\end{align}
all of which can be done analytically. If we further decompose $B(p_0,k_0,x) = B_2(p_0,k_0) + B_3(p_0,k_0) x + B_4(p_0,k_0) x^2 $, we can write the final result as
\begin{equation}
\begin{split}
&\left. \varrho_E^{++}(p_0) \right|_{\rm NLO}^{5-6} { - \frac{g^2 N_c (N_c^2-1)\pi^2}{2(2\pi)^3}} \\
&= - g^2 p_0^3 \tilde{\mu}^\epsilon \frac{N_c (N_c^2-1)}{(2\pi)^{d-1}} \frac{\Omega_{3-\epsilon}}{(4\pi)^{(3-\epsilon)/2}} \Gamma \left( \frac{-1+\epsilon}{2} \right) \int_0^\infty \diff k k^{1 - \epsilon} \frac{1+2n_B(k)}{2} \\
 & \quad \quad \quad \quad \times \sum_{\sigma_1} \bigg[ \frac{3-\epsilon}{2} A I_1 + \frac{-1+\epsilon}{2} \big( B_2 I_2 + B_3 I_3 + B_4 I_4 \big)  \bigg] \,. \label{eq:D-36}
\end{split}
\end{equation}

Now let us set $T = 0$ and study the integrand as $k \to \infty$. For notational simplicity, let
\begin{equation}
    K(k;\epsilon) =   \frac{1}{2} k^{1 - \epsilon} \sum_{\sigma_1} \bigg[ \frac{3-\epsilon}{2} A I_1 + \frac{-1+\epsilon}{2} \big( B_2 I_2 + B_3 I_3 + B_4 I_4 \big) \bigg] \,. \label{eq:K-appendix}
\end{equation}
One can then show that for $0< \epsilon < 1$
\begin{align}
    &\lim_{k \to \infty} K(k;\epsilon) k^{1+\epsilon} = - \cos \left( \frac{\pi \epsilon}{2} \right) p_0^{-\epsilon} \\
    &\lim_{k \to \infty} \left(K(k;\epsilon) + \cos \left( \frac{\pi \epsilon}{2} \right) p_0^{-\epsilon} k^{-1-\epsilon} \right) k^{1+2\epsilon} = (2 - 2\epsilon + \epsilon^2) \cos \left( \frac{\pi \epsilon}{2} \right) \\
    &\lim_{k \to \infty} \left(K(k;\epsilon) + \cos \left( \frac{\pi \epsilon}{2} \right) p_0^{-\epsilon} k^{-1-\epsilon} - (2 - 2\epsilon + \epsilon^2) \cos \left( \frac{\pi \epsilon}{2} \right) k^{-1-2\epsilon} \right) k^{2} = 0 \,,
\end{align}
effectively demonstrating that we need to extract the possible divergences as $k\to \infty$ at two different rates (but only two). In practice, we add and subtract two explicitly calculable integrals with the same degree of divergence as $K(k;\epsilon)$ so that we can take the limit $\epsilon \to 0$ before performing the integral. Concretely, we calculate
\begin{equation}
\begin{split}
    & \lim_{\epsilon \to 0} \int_0^\infty \diff k\, K(k;\epsilon) \\ &= \lim_{\epsilon \to 0} \int_0^\infty \diff k \bigg[ K(k;\epsilon) + \cos \left( \frac{\pi \epsilon}{2} \right) \frac{k^{1-\epsilon} p_0^{-\epsilon}}{k^2 + p_0^2} - (2 - 2\epsilon + \epsilon^2) \cos \left( \frac{\pi \epsilon}{2} \right) \frac{k^{1-2\epsilon}}{k^2 + p_0^2} \bigg] \\
    & \quad + \lim_{\epsilon \to 0} \int_0^\infty \diff k \bigg[ - \cos \left( \frac{\pi \epsilon}{2} \right) \frac{k^{1-\epsilon} p_0^{-\epsilon}}{k^2 + p_0^2} + (2 - 2\epsilon + \epsilon^2) \cos \left( \frac{\pi \epsilon}{2} \right) \frac{k^{1-2\epsilon}}{k^2 + p_0^2} \bigg] \,,
\end{split}
\end{equation}
so that the first term of the right-hand side is now absolutely convergent for any $0 \leq \epsilon < 1$, and we can simply take $\epsilon \to 0$ there. If there is any divergence whatsoever, it must be in the second term. We evaluate
\begin{equation}
\begin{split}
    \int_0^\infty \diff k \frac{k^{1-\epsilon}}{k^2 + p_0^2} &= \frac{(2\pi)^{2-\epsilon}}{\Omega_{2-\epsilon}} \int \frac{\diff^{2-\epsilon} k_E}{(2\pi)^{2-\epsilon}} \frac{1}{k_E^2 + p_0^2} \\
    &= \frac{(2\pi)^{2-\epsilon}}{\Omega_{2-\epsilon}} \frac{\Gamma(1-(2-\epsilon)/2 )  }{(4\pi)^{1-\epsilon/2}} (p_0^2)^{-(1-(2-\epsilon)/2)} \\
    &= \frac{(2\pi)^{2-\epsilon}}{\Omega_{2-\epsilon}} \frac{\Gamma(\epsilon/2 )  }{(4\pi)^{1-\epsilon/2}} p_0^{-\epsilon}\,,
\end{split}
\end{equation}
which shows that
\begin{equation}
    \begin{split}
        &\int_0^\infty \diff k \bigg[ - \cos \left( \frac{\pi \epsilon}{2} \right) \frac{k^{1-\epsilon} p_0^{-\epsilon}}{k^2 + p_0^2} + (2 - 2\epsilon + \epsilon^2) \cos \left( \frac{\pi \epsilon}{2} \right) \frac{k^{1-2\epsilon}}{k^2 + p_0^2} \bigg] \\
        &= \cos \left( \frac{\pi \epsilon}{2} \right) p_0^{-2\epsilon} \left[ - \frac{(2\pi)^{2-\epsilon}}{\Omega_{2-\epsilon}} \frac{\Gamma(\epsilon/2 )  }{(4\pi)^{1-\epsilon/2}} + (2 - 2\epsilon + \epsilon^2) \frac{(2\pi)^{2-2\epsilon}}{\Omega_{2-2\epsilon}} \frac{\Gamma(\epsilon )  }{(4\pi)^{1-\epsilon}} \right] \\
        &= -p_0^{-2\epsilon} + \mathcal{O}(\epsilon).
    \end{split}
\end{equation}
In a nutshell, all we get from the second term as $d \to 4$ is just $-1$. 

For the first term, since the integral is absolutely convergent, we can take the limit $\epsilon\to0$ in the integrand, and obtain
\begin{equation}
\begin{split}
    & \lim_{\epsilon \to 0} \int_0^\infty \diff k \bigg[ K(k;\epsilon) + \cos \left( \frac{\pi \epsilon}{2} \right) \frac{k^{1-\epsilon} p_0^{-\epsilon}}{k^2 + p_0^2} - (2 - 2\epsilon + \epsilon^2) \cos \left( \frac{\pi \epsilon}{2} \right) \frac{k^{1-2\epsilon}}{k^2 + p_0^2} \bigg] \\
    &= \int_0^\infty \diff k \left[ K(k;0) - \frac{k}{k^2 + p_0^2} \right] \\
    &= \frac{1}{12} \left[ -\pi^2 - 6 \left( -4 + (\ln(2))^2 - 2 {\rm Li}_2(-1/2) + {\rm Li}_2(1/4) \right) \right] { = 2 - \frac{\pi^2}{6}} \,.
\end{split}
\end{equation}

Using
\begin{equation}
    -\frac{\Omega_{3-\epsilon}}{(4\pi)^{(3-\epsilon)/2}} \Gamma \left( \frac{-1+\epsilon}{2} \right) = 1 + \mathcal{O}(\epsilon)\,,
\end{equation}
we find the full contribution at $T=0$ from the 3-propagator diagrams is
\begin{equation}
\begin{split}
    &\left. \varrho_E^{++}(p_0) \right|_{{\rm NLO}, \, T=0 }^{5-6} \\
    &= \frac{ g^2 N_c (N_c^2-1) p_0^3}{(2\pi)^3} \left[ -1 + { 2 - \frac{\pi^2}{6} + \frac{\pi^2}{2} } \right] + \mathcal{O}(\epsilon)\, \\
    &= \frac{ g^2 N_c (N_c^2-1) p_0^3}{(2\pi)^3} \left[ {1 + \frac{\pi^2}{3}} \right] + \mathcal{O}(\epsilon)\,
\end{split}
\end{equation}

Since there are no UV divergences in the terms that come from purely temperature-dependent contributions, the finite $T$ contribution is obtained by taking the $\epsilon \to 0$ limit of $K(k;\epsilon)$, defined in~\eqref{eq:K-appendix}, and then plug it in~\eqref{eq:D-36}. The result is given in the main text.

\section{Time-ordered correlator in vacuum} \label{app:T-ordered-vacuum-5-5r}

{
To explore further aspects of similar-looking correlators and compare the finite vacuum constant piece of our result, which is a Wightman correlation function, with the time-ordered vacuum NLO electric field correlator calculated in Ref.~\cite{Eidemuller:1997bb}, we will show here that by taking the time-ordered version of our correlator on the Schwinger-Keldysh contour and integrating over the momentum ${\bs p}$ (which is equivalent to setting ${\bs y}={\bs x}$), we reproduce the results of Ref.~\cite{Eidemuller:1997bb}. To this end, we calculate
\begin{align}
 \big[g_E^{++}\big]_{I=J=1}(y,x) = \Big\langle \ml{T} & \big[{E}_i(y) \ml{W}_{[( y^0, {\bs y}), (+\infty, {\bs y})]} \big]^a
\big[ \ml{W}_{[(+\infty, {\bs x}),(x^0, {\bs x})]} {E}_i(x) \big]^a \Big\rangle_T \,,  
\end{align}
and compare it with
\begin{align}
\big[g_E^{++}\big]^{\ml{T}}(y,x) \equiv \theta(y^0 - x^0) \big[g_E^{++}\big]^>(y,x) + \theta(x^0 - y^0) \big[g_E^{++}\big]^<(y,x) \,,
\end{align}
in order to give an explicit assessment of Eq.~\ref{eq:discrepancy-T-ordered}.}

{
If there were no Wilson lines, then we would simply have an equality between the two, i.e., $ \big[g_E^{++}\big]_{I=J=1}(y,x) = \big[g_E^{++}\big]^{\ml{T}}(y,x)$, because the time-ordering symbol would amount exactly to reordering the electric field insertions. Therefore, perturbatively, any diagram that does not involve Wilson lines will not generate any difference. So we only need to investigate diagrams that involve gauge boson insertions from the Wilson lines. Furthermore, we will restrict ourselves to looking at the real part of the correlation functions in momentum space to see if a difference appears, because this is the contribution that would appear in the spectral function.

\subsection{2-propagator diagrams}

Following our conventions, the time-ordered correlation function as given by $I=J=1$ for the 2-propagator diagrams. After substituting the corresponding expressions for the vertex factors and propagator structures in Tables~\ref{tab:2prop-structure} and~\ref{tab:2prop-vertex}, we find
\begin{align}
\big[g_E^{++}\big]_{I=J=1}^{\rm 2-propagator}(p_0) &= g^2 N_c (N_c^2-1) \int_{\bs p} \int_k \frac{1}{k_0^2} \big[ \left((\p-\k)^2 - (d-1) p_0^2 \right) D^{\ml T}(p-k) D^{\ml T}(k) \nonumber \\ & \quad \quad \quad \quad \quad \quad \quad \quad \quad \quad \quad + \left( (d-1) p_0^2 - \p^2  \right) D^{\ml T}(p) D^{\ml T}(k)  \big] \,,
\end{align}
which by direct calculation in $d=4-\epsilon$ up to $O(\epsilon)$ terms gives (for $p_0>0$)
\begin{align}
& {\rm Re} \left\{ \big[g_E^{++}\big]_{I=J=1}(p_0) \right\} \\ 
&= \frac{\pi N_c (N_c^2 - 1)}{8} \left(\frac{\Omega_{d-1}}{(2\pi)^{d-1}} \right)^2 p_0^{2d-5} \tilde{\mu}^{4-d} \nonumber \\ & \quad \times \int_0^\infty \diff \tilde{k} \, \tilde{k}^{d-5} \left[ |\tilde{k}-1|^{d-3} \left((1-\tilde{k})^2-(d-1)\right) + |\tilde{k}+1|^{d-3} \left((1+\tilde{k})^2-(d-1)\right) + (d-2) \right] \nonumber \\
&= (N_c^2 -1) \frac{(d-2) \pi \Omega_{d-1}}{2 (4\pi^2) (2\pi)^{d-1}} g^2 p_0^{d-1}  N_c \bigg[ \frac{1}{\epsilon} + \frac{7}{12}  + \frac{1}{2}  \ln \left( \frac{\mu^2}{p_0^2} \right) + \frac{\gamma_E + \psi(3/2)}{2} \bigg] \,,
\end{align}
where $\tilde{k}$ is $k/p_0$. This result is
consistent with the finite pieces we obtained in the calculation of $\big[g_E^{++}\big]^>(y,x)$.

\subsection{3-propagator diagrams}

Here the focus will be on diagrams $(5)$ and $(5r)$, since these are the only remaining diagrams that involve Wilson lines. However, and again for convenience, we will include diagrams $(6)$ and $(6r)$ in the relevant numerator. Starting from the propagator structures and vertex factors in Tables~\ref{tab:3prop-structure} and~\ref{tab:3prop-vertex}, we obtain that in vacuum for $p_0>0$, 
\begin{align}
\big[g_E^{++}\big]_{I=J=1}^{\rm 3-propagator}(p_0) &= \frac{g^2 N_c (N_c^2-1)}{2} \nn \\ & \quad \times \int_{\bs p} \int_k \left[ \frac{1}{k_0+i0^+} + \frac{1}{k_0-i0^+} \right] \frac{\bar{N}(p,k)}{(k^2 + i0^+) ((p-k)^2 + i0^+) (p^2 + i0^+) } \,,
\end{align}
where $\bar{N} = 2p_0 \left[ (d-1)k_0 p_0 + 2\k^2 - 2\p \cdot \k + 2\p^2 - 2(d-1)p_0^2 \right]$. It is convenient for our purposes to rearrange $\bar{N}$ as follows:
\begin{align}
    \bar{N}(p,k) &= \tilde{N}(p,k) - 4 p_0 (p_0^2 - \p^2) \\
    \tilde{N}(p,k) &=  2p_0 \left[ (d-1)k_0 p_0 + 2\k^2 - 2\p \cdot \k - 2(d-2)p_0^2 \right] \,.
\end{align}
Then, the contribution from the piece of the numerator that is proportional to $p^2 = p_0^2 - \p^2$ is simply given by 
\begin{align}
    & \frac12 \int_{\bs p} \int_k \left[ \frac{1}{k_0+i0^+} + \frac{1}{k_0-i0^+} \right] \frac{-4 p_0 p^2}{(k^2 + i0^+) ((p-k)^2 + i0^+) (p^2 + i0^+) } \nn \\
    &= \int_{\bs p} \int_k \mathcal{P} \frac{1}{k_0} \frac{-4 p_0}{(k^2 + i0^+) ((p-k)^2 + i0^+)} \nn \\ &= -4p_0 \frac{\Gamma\left(\frac{3-d}{2} \right)^2}{(4\pi)^{d-1}} \int_{-\infty}^{\infty} \frac{\diff k_0}{2\pi} \frac{(-k_0^2-i0^+)^{\frac{d-3}{2}} (-(p_0-k_0)^2 - i0^+)^{\frac{d-3}{2}} }{k_0} \nn \\
    &= -4 p_0 (-i)^{2d-6} \frac{\Gamma\left(\frac{3-d}{2} \right)^2}{2\pi (4\pi)^{d-1}} \int_{0}^{\infty} \diff k_0 \, k_0^{d-4} \left[ |p_0-k_0|^{d-3} - |p_0+k_0|^{d-3} \right] \nn \\
    &{}_{{\rm DR}, \, d \to 4} = \frac{p_0^3}{(2\pi)^3} \,,
\end{align}
where in the last line we have taken the result of the previous line in dimensional regularization, and then set $d=4$.

The other contribution requires a more complicated treatment. Using Georgi and Feynman parameters, we can write
\begin{align}
   &\frac12 \int_{\bs p} \int_k \left[ \frac{1}{k_0+i0^+} + \frac{1}{k_0-i0^+} \right] \frac{\tilde{N}(p,k)}{(k^2 + i0^+) ((p-k)^2 + i0^+) (p^2 + i0^+) } \nn \\
   &= \int_{\p} \frac{1}{p^2 + i0^+}  \int_0^1 \diff x \int_0^\infty \!\! \diff \lambda \, \int_k \Bigg( \frac{\tilde{N}(p,k)}{(\lambda k_0 + (1-x)k^2 + x(p-k)^2 + i0^+)^3} \nn \\ & \quad \quad \quad \quad \quad \quad \quad \quad \quad \quad \quad \quad \quad \quad \quad \quad - \frac{\tilde{N}(p,k)}{(-\lambda k_0 + (1-x)k^2 + x(p-k)^2 + i0^+)^3} \Bigg) \,.
\end{align}
The integral over $k$ is now a textbook loop integral, and the resulting integral over $\lambda$ can be done explicitly in terms of polynomial and Hypergeometric functions.
One then obtains
\begin{align}
   &\frac12 \int_{\bs p} \int_k \left[ \frac{1}{k_0+i0^+} + \frac{1}{k_0-i0^+} \right] \frac{\tilde{N}(p,k)}{(k^2 + i0^+) ((p-k)^2 + i0^+) (p^2 + i0^+) } \nn \\
   &= \int_{\p} \frac{1}{p^2 + i0^+}  \int_0^1 \diff x \Bigg[ \frac{i p_0 (1-d)}{(4\pi)^{d/2}} \Gamma \! \left( \frac{4-d}{2} \right) 4 x p_0 \frac{{}_2F_1 \! \left( \frac12, \frac{4-d}2,\frac32, \frac{-x^2 p_0^2}{x(1-x) \p^2-xp_0^2-i0^+} \right) }{(x(1-x)\p^2 - xp_0^2 - i0^+)^{2-d/2}} \nn \\
   & \quad \quad \quad \quad \quad \quad \quad \quad + \frac{i p_0^2 (d-1) }{2 (4\pi)^{d/2}} \Gamma \! \left( \frac{4-d}{2} \right) \frac{4}{(x(1-x)\p^2 - xp_0^2 - i0^+)^{2-d/2} } \nn \\
   & \quad \quad \quad \quad \quad \quad \quad \quad + \frac{i p_0^2 (d-1) }{2 (4\pi)^{d/2}} \Gamma \! \left( \frac{6-d}{2} \right) 8 x^2 p_0^2 \frac{{}_2F_1 \! \left( \frac12, \frac{6-d}2,\frac32, \frac{-x^2 p_0^2}{x(1-x) \p^2-xp_0^2-i0^+} \right) }{(x(1-x)\p^2 - xp_0^2 - i0^+)^{3-d/2}} \nn \\ 
   & \quad \quad \quad \quad \quad \quad \quad \quad + \frac{(-i)}{2(4\pi)^{d/2}} \Gamma \! \left( \frac{6-d}{2} \right) 4 x p_0 \left[ \left( 2(d-1)(x-2)+4 \right) p_0^3 - 4x(1-x) p_0 \p^2 \right] \nn \\ &  \quad \quad \quad \quad \quad \quad \quad \quad  \quad \quad \quad \quad \quad \quad \quad \quad  \quad \quad \quad \quad \quad \quad \times \frac{{}_2F_1 \! \left( \frac12, \frac{6-d}2,\frac32, \frac{-x^2 p_0^2}{x(1-x) \p^2-xp_0^2-i0^+} \right) }{(x(1-x)\p^2 - xp_0^2 - i0^+)^{3-d/2}} \Bigg] \,.
\end{align}
Using the Pfaff transformations of the Hypergeometric functions, and taking $d \to 4$, we arrive at
\begin{align}
   &\frac12 \int_{\bs p} \int_k \left[ \frac{1}{k_0+i0^+} + \frac{1}{k_0-i0^+} \right] \frac{\tilde{N}(p,k)}{(k^2 + i0^+) ((p-k)^2 + i0^+) (p^2 + i0^+) } \nn \\
   &= \int_{\p} \frac{i}{p^2 + i0^+}  \int_0^1 \diff x \frac{1}{(4\pi)^2} \bigg\{ -12 x p_0^2 \frac{\partial}{\partial a} \left[ {}_2F_1 \! \left( a,1,\frac32,\frac{-x^2 p_0^2}{x(1-x)(p^2+i0^+)} \right)  \right]_{a=0} \nn \\
   & \quad \quad \quad \quad \quad \quad \quad \quad  \quad \quad - \frac{16 x p_0^4 + 8 p_0^2 \p^2 x^2 (1-x)}{x(1-x) (p^2 + i0^+) } {}_2F_1 \! \left( 1,1,\frac32,\frac{-x^2 p_0^2}{x(1-x)(p^2+i0^+)} \right) \bigg\} \nn \\
   &= \frac{1}{16 \pi^4} \int_0^\infty \!\! \diff |\p| \frac{i |\p|^2 p_0^2 }{p_0^2 - |\p|^2 + i0^+} \nn \\
   & \quad \quad \quad \quad \quad \quad   \times \Bigg[ 3 \int_0^1 \!\! \diff x \int_0^1 \!\! \diff y \frac{x}{\sqrt{1-y}} \ln \left( \frac{x(1-x) (p_0^2 - |\p|^2 + i0^+) + y x^2 p_0^2 }{x(1-x)(p_0^2 - |\p|^2 + i0^+)} \right) \nn \\
   & \quad \quad \quad \quad \quad \quad \quad \quad \quad  - 2  \int_0^1 \!\! \diff x \int_0^1 \!\! \diff y \frac{x}{\sqrt{1-y}} \frac{2 p_0^2 + x(1-x) |\p|^2}{x(1-x) (p_0^2 - |\p|^2 + i0^+) + y x^2 p_0^2 } \Bigg] \,,
\end{align}
where in the last line we have used an integral representation of the hypergeometric function:
\begin{align}
    {}_2F_1 \! \left(a,1,\frac32,D \right) = \frac12 \int_0^1 \frac{\diff y}{\sqrt{1-y}} (1 - Dy)^{-a} \,.
\end{align}

Now we take the real part of this expression, as it is all we need to compare with our results from the Wightman correlations. The remaining integrals in $|{\bs p}|$ can then be carried out by complex contour integration, yielding
\begin{align}
    &{\rm Re} \left\{ \int_0^\infty  \!\! \diff |\p| \frac{i |\p|^2}{p_0^2-|\p|^2+i0^+} \frac{2p_0^2 + x(1-x) |\p|^2 }{x(1-x)(p_0^2 - |\p|^2 + i0^+) + y x^2 p_0^2} \right\} \nn \\
    &\quad \quad = \frac{\pi p_0}{2} \left( \frac{2+x(1-x)}{yx^2} - \sqrt{\frac{1-x+yx}{1-x}} \frac{2+x(1-x)+yx^2}{yx^2} \right) \\
    &{\rm Re} \left\{ \int_0^\infty  \!\! \diff |\p| \frac{i |\p|^2}{p_0^2-|\p|^2+i0^+} \ln \left( - \frac{x(1-x) (p_0^2 - |\p|^2 + i0^+) + y x^2 p_0^2 }{p_0^2} \right) \right\} \nn \\
    &\quad \quad= \frac{\pi p_0}{2} \left( \ln \left| y x^2 \frac{\sqrt{\frac{1-x+yx}{1-x}}+1}{\sqrt{\frac{1-x+yx}{1-x}}-1} \right| - 2 \sqrt{\frac{1-x+yx}{1-x}} \right) \\
    &{\rm Re} \left\{ \int_0^\infty  \!\! \diff |\p| \frac{i |\p|^2}{p_0^2-|\p|^2+i0^+} \ln \left( - \frac{x(1-x) (p_0^2 - |\p|^2 + i0^+) }{p_0^2} \right) \right\} \nn \\
    &\quad \quad= \frac{\pi p_0}{2} \left( \ln \left|4 x (1-x) \right| - 2 \right) \,,
\end{align}
where we have split the logarithmic term into two pieces and integrate them separately.
Putting everything together, we find that
\begin{align}
    & {\rm Re} \left\{ \big[g_E^{++}\big]_{I=J=1}^{\rm 3-propagator}(p_0) \right\} \nn \\ &= g^2 N_c (N_c^2 - 1) \frac{p_0^3}{(2\pi)^3} \Bigg[ 1 + \frac{1}{2} \int_0^1 \!\!\! \diff x \int_0^1 \!\!\! \diff y \frac{x}{\sqrt{1-y}} \nn \\
    & \quad \quad \quad \quad \quad \quad \quad \quad   \quad \quad \quad \quad  \times \bigg( 3 - 3 \sqrt{\frac{1-x+yx}{1-x}}  + \frac32 \ln \left| \frac{y x}{1-x} \frac{\sqrt{\frac{1-x+yx}{1-x}}+1}{\sqrt{\frac{1-x+yx}{1-x}}-1} \right| \nn \\ 
    & \quad \quad \quad \quad \quad \quad \quad \quad  \quad \quad \quad \quad \quad   + \sqrt{\frac{1-x+yx}{1-x}} \frac{2+x(1-x)+yx^2}{yx^2} - \frac{2+x(1-x)}{yx^2} \bigg) \Bigg] \nn \\
    &= g^2 N_c (N_c^2 - 1) \frac{p_0^3}{(2\pi)^3} \Bigg[ 1 + \frac{\pi^2}{3} \Bigg] \,.
\end{align}
This reproduces the result of Ref.~\cite{Eidemuller:1997bb}, and that of our main text given in Eq.~\eqref{eq:to-checkApp52}.

\subsection{Conversion of previous results in position space to momentum space}

To reassure the reader that the result of the previous Appendix section matches that of Ref.~\cite{Eidemuller:1997bb}, we convert their results into momentum space. In their work, they calculated
\begin{align}
\mathcal{D}_{\mu\nu\rho\sigma}(z) &= \left\langle 0 \left| \ml{T} \left( F_{\mu \nu}^a(z)  \ml{W}_{[z,0]}^{ab} F_{\rho \sigma}^b(0) \right)  \right| 0 \right\rangle \nn \\
& = \left[ g_{\mu \rho} g_{\nu \sigma} - g_{\mu \sigma} g_{\nu \rho} \right] \left( D(z^2) + D_1(z^2) \right) \nn \\ & \quad + \left[ g_{\mu \rho} z_{\nu} z_{\sigma} - g_{\mu \sigma} z_{\nu} z_{\rho} - g_{\nu \rho} z_{\mu} z_{\sigma} + g_{\nu \sigma} z_{\mu} z_{\rho} \right] \frac{\partial D_1}{\partial z^2} \,,
\end{align}
for an arbitrary spacetime separation $z^{\mu}$. Up to NLO, Ref.~\cite{Eidemuller:1997bb} found
\begin{align}
    D(z^2) &= \frac{N_c^2-1}{\pi^2 z^4} \left[ \frac{\alpha N_c}{\pi} \left(-\frac{1}{4} L + \frac38 \right) \right] \\
    D_1(z^2) &= \frac{N_c^2-1}{\pi^2 z^4} \left[ 1 + \frac{\alpha N_c}{\pi} \left( \left( \frac{\beta_1}{2 N_c} - \frac14 \right) L + \frac{\beta_1}{3N_c} + \frac{29}{24} + \frac{\pi^2}{3} \right) \right] \,,
\end{align}
where $L = \ln (e^{2\gamma_E} \mu^2 z^2/4 )$, $\beta_1 = (11 N_c - 4 N_f)/6$, and $\mu$ is the $\overline{\rm MS}$ renormalization scale.

Our correlator of interest is $\mathcal{D}_{0i0i}(z_0,{\bs z}=0)$. After some algebra, one arrives at
\begin{align}
   \mathcal{D}_{0i0i}(z_0) = 3\frac{N_c^2-1}{\pi^2 z_0^4} \left[ 1 + \frac{g^2}{4\pi^2}\left( \left(  \frac{11}{12} N_c - \frac{1}{3} N_f \right) L + N_c \left( \frac79 + \frac{\pi^2}{3} \right) + \frac19 N_f \right) \right] \,,
\end{align}
where now $L = \ln (e^{2\gamma_E} \mu^2 z_0^2/4 )$. One can then use the Fourier transforms
\begin{align}
    \int_{-\infty}^\infty \diff t\, e^{i p_0 t} \frac{1}{t^4} &= \frac{\pi}{6} p_0^3 \,{\rm sgn}(p_0) \\
    \int_{-\infty}^\infty \diff t\, e^{i p_0 t} \frac{\ln(t^2)}{t^4} &= -\frac{\pi}{18} p_0^3 \,{\rm sgn}(p_0) \left( -11 + 6 \gamma_E + 6 \ln |p_0| \right) \,,
\end{align}
which, upon substitution into our correlator of interest, yield (for $p_0>0$):
\begin{align}
   \mathcal{D}_{0i0i}(p_0) = \frac{ N_c^2-1}{(2\pi)^3} p_0^3 \left[ 4\pi^2 + g^2 \! \left( \! \left( \frac{11}{12} N_c - \frac{1}{3} N_f \right) \! \ln \! \left( \frac{\mu^2}{4 p_0^2}\right) + \left( \frac{149}{36} + \frac{\pi^2}{3} \right) \! N_c  - \frac{10}{9} N_f \right)  \right] \,,
\end{align}   
which exactly matches our vacuum result in~\eqref{eq:spectral-full-result}.
}

\section{Different expressions for the quarkonium correlator present in the literature} \label{app:diff-literature}

In this Appendix we review the different definitions present in the literature~\cite{Brambilla:2016wgg,Brambilla:2017zei,Yao:2020eqy} that appear in the quantum transport equations for quarkonium. The following discussion will further illuminate the nature of the difference between the correlators~\eqref{HQ-corr} and~\eqref{QA-corr}.

We start from the definition of the chromoelectric correlator for the quarkonium transport equation as given in the main text~\eqref{QA-corr}, which appears in the formulation of the open quantum system for quarkonium in the quantum optical limit~\cite{Yao:2020eqy}:
\be
\label{app:QA}
g_E^{\rm Q\bar{Q}}(t) = g^2 T_F \left\langle  {F}_{0i}^a(t) \ml{W}_{[t, 0]}^{ab}
   {F}_{0i}^b(0) \right\rangle \, .
\ee
In the quantum optical limit, it is the correlator~\eqref{app:QA} at finite frequency that contributes to the quarkonium dissociation and recombination rates:
\be
g_E^{Q\bar{Q}}(p_0) = g^2 T_F \int_{-\infty}^{+\infty} \!\! \diff t \, e^{ip_0 t} \left\langle  {F}_{0i}^a(t) \ml{W}_{[t, 0]}^{ab} {F}_{0i}^b(0) \right\rangle \,.
\ee
On the other hand, in the quantum Brownian motion limit studied in Refs.~\cite{Brambilla:2016wgg,Brambilla:2017zei}, it is the zero frequency limit of $g_E^{Q\bar{Q}}$ that matters in the quarkonium transport (see also Ref.~\cite{Yao:2021lus}). However, the expression given in Refs.~\cite{Brambilla:2016wgg,Brambilla:2017zei} (see also Eq.~(2.13) of Ref.~\cite{Brambilla:2020qwo}) is
\begin{align}
\kappa^{Q\bar{Q}} \equiv \frac{g^2T_F}{N_c} {\rm Re} \int_{-\infty}^{+\infty}\!\!\diff t \, \Big\langle \ml{T}\Big( \widetilde{E}_i^a(t) \widetilde{E}_i^a(0) \Big) \Big\rangle\,,
\end{align}
where $\ml{T}$ denotes time-ordering and $\widetilde{E}_i^a(t) = U_{[-\infty,t]} E_i^a(t) U_{[t,-\infty]} = U_{[-\infty,t]} F_{0i}^a(t) U_{[t,-\infty]}$ with $U$ representing a fundamental Wilson line. The expression of $\kappa^{Q\bar{Q}}$ looks different from our expression $g_E^{Q\bar{Q}}(p_0=0)$ here. In the following, we will show they are equivalent for $p_0=0$. First we find
\begin{align}
& \int_{-\infty}^{+\infty} \!\! \diff t \left\langle \ml{T} \left( {F}_{0i}^a(t) \ml{W}_{[t, 0]}^{ab} {F}_{0i}^b(0) \right) \right\rangle \\
& = \int_{0}^{+\infty} \!\! \!\! \diff t \left\langle  {F}_{0i}^a(t) \ml{W}_{[t, 0]}^{ab} {F}_{0i}^b(0) \right\rangle + \int_{-\infty}^{0} \!\! \diff t \left\langle {F}_{0i}^a(0) \ml{W}_{[0, t]}^{ab}  {F}_{0i}^b(t) \right\rangle \nonumber \\
& = \int_{0}^{+\infty} \!\! \!\! \diff t \left\langle  {F}_{0i}^a(t) \ml{W}_{[t, 0]}^{ab} {F}_{0i}^b(0)  \right\rangle + \int_{0}^{+\infty} \!\! \!\! \diff t \left\langle {F}_{0i}^a(0) \ml{W}_{[0, -t]}^{ab}  {F}_{0i}^b(-t)  \right\rangle \nn\\
& = 2 \int_{0}^{+\infty} \!\! \!\! \diff t \left\langle {F}_{0i}^a(t) \ml{W}_{[t, 0]}^{ab} {F}_{0i}^b(0) \right\rangle \,,\nn
\end{align} 
where we have relabeled the color indexes $a$ and $b$ in the second term on the first line, flipped the sign of $t$ in the second term on the second line and used translational invariance in $t$ on the last line. Then we can show
\begin{align}
& {\rm Re} \int_{-\infty}^{+\infty} \!\! \diff t \left\langle \ml{T} \left( {F}_{0i}^a(t) \ml{W}_{[t, 0]}^{ab} {F}_{0i}^b(0) \right) \right\rangle \\ &= 2\,{\rm Re} \int_{0}^{+\infty} \!\! \diff t \left\langle  {F}_{0i}^a(t) \ml{W}_{[t, 0]}^{ab} {F}_{0i}^b(0)  \right\rangle \nonumber \\
& = \int_{0}^{+\infty} \!\! \diff t \left\langle  {F}_{0i}^a(t) \ml{W}_{[t, 0]}^{ab} {F}_{0i}^b(0)  \right\rangle + \int_{0}^{+\infty} \!\! \diff t \left\langle {F}_{0i}^a(0) \ml{W}_{[0, t]}^{ab} {F}_{0i}^b(t) \right\rangle \nn\\
& = \int_{0}^{+\infty} \!\! \diff t \left\langle {F}_{0i}^a(t) \ml{W}_{[t, 0]}^{ab} {F}_{0i}^b(0) \right\rangle + \int_{-\infty}^{0} \!\! \diff t \left\langle {F}_{0i}^a(0) \ml{W}_{[0, -t]}^{ab} {F}_{0i}^b(-t)  \right\rangle \nn\\
& = \int_{0}^{+\infty} \!\! \diff t \left\langle {F}_{0i}^a(t) \ml{W}_{[t, 0]}^{ab} {F}_{0i}^b(0)  \right\rangle + \int_{-\infty}^{0} \!\! \diff t \left\langle {F}_{0i}^a(t) \ml{W}_{[t, 0]}^{ab} {F}_{0i}^b(0)  \right\rangle \nn\\[4pt]
& \propto g_E^{Q\bar{Q}}(p_0=0) \,, \nn
\end{align}
where we have relabeled the color indexes $a$ and $b$ in the second term on the second line, flipped the sign of $t$ in the second term on the third line and used translational invariance in $t$ in the second term on the second-to-last line. Finally we study the relation between $\langle \ml{T} ( {F}_{0i}^a(t) \ml{W}_{[t, 0]}^{ab} {F}_{0i}^b(0) ) \rangle$ and $\langle \ml{T} ( \widetilde{E}_i^a(t) \widetilde{E}_i^a(0) ) \rangle$, both of which can be studied using the closed-time path integral methods~\cite{keldysh1965diagram}. Since both correlators are time-ordered, we can insert all the fields contained in the correlators on the time-ordered branch of the Schwinger-Keldysh contour, i.e., the upper branch with type-$1$ fields. Then we can use the standard SU$(N_c)$ Wilson line algebra to derive
\begin{align}
\label{eq:difference-correlators-explicit}
& g^2 T_F \left\langle \ml{T} \left( {F}_{0i}^a(t) \ml{W}_{[t, 0]}^{ab} {F}_{0i}^b(0) \right) \right\rangle \\
&= g^2 T_F \int \! DA_E DA_1 DA_2 \, e^{iS[A_1]-iS[A_2]-S_E[A_E]} F_{0i}^a[A_1](t) \, \ml{W}_{[t, 0]}^{ab}[A_1] \, F_{0i}^b[A_1](0) \nonumber \\
    &= g^2 \int \! DA_E DA_1 DA_2 \, e^{iS[A_1]-iS[A_2]-S_E[A_E]} \, \nonumber \\ & \quad\quad\quad\quad\quad\quad\quad\quad\quad\quad \times {\rm Tr}_{c} \! \left\{ F_{0i}[A_1](t) \, U_{[t, 0]}[A_1] \, F_{0i}[A_1](0) \,  U_{[0, t]}[A_1] \right\}  \nonumber \\
    &= g^2 \int \! DA_E DA_1 DA_2 \, e^{iS[A_1]-iS[A_2]-S_E[A_E]} \, \nonumber \\ & \quad\quad\quad\quad\quad\quad\quad\quad\quad\quad \times {\rm Tr}_{c} \! \left\{ U_{[-\infty, t]}[A_1] \, F_{0i}[A_1](t) \, U_{[t, 0]}[A_1] \, F_{0i}[A_1](0) \, U_{[0, -\infty]}[A_1] \right\}  \nonumber \\
    &= g^2 \left\langle \mathcal{T} \, {\rm Tr}_c \left( U_{[-\infty,t]} F_{0i}(t) U_{[t,0]} F_{0i}(0) U_{[0,-\infty]} \right) \right\rangle \nonumber \\[4pt]
& = g^2T_F \langle \ml{T} ( \widetilde{E}_i^a(t) \widetilde{E}_i^a(0) ) \rangle \,,
 \nn
\end{align}
where the subscripts $1$, $2$ and $E$ denote the type-1, type-2 and Euclidean fields. 
Putting everything together, we have proved that $\kappa^{Q\bar{Q}}$ defined in Refs.~\cite{Brambilla:2016wgg,Brambilla:2017zei} and our expression $g_E^{Q\bar{Q}}(p_0=0)$ are the same, up to a trivial normalization factor.

We want to emphasize that the second-to-last line of this expression~\eqref{eq:difference-correlators-explicit} does not match the correlator that defines the heavy quark diffusion coefficient $\kappa^Q$~\cite{Casalderrey-Solana:2006fio}, which is given by
\be
\kappa^Q \propto g^2 {\rm Re} \int_{-\infty}^{+\infty} \!\!\diff t  \left\langle {\rm Tr}_c \left( U_{[-\infty,t]} F_{0i}(t) U_{[t,0]} F_{0i}(0) U_{[0,-\infty]} \right) \right\rangle  = 
{\rm Re} \int_{-\infty}^{+\infty} \!\!\diff t\, g_E^Q(t) \, .
\ee
The key difference between the second-to-last line of Eq.~\eqref{eq:difference-correlators-explicit} and $g_E^Q$ is the operator ordering: in the former case the operators are time-ordered while in the latter they are ordered in the sequence as shown. Conceptually they are different in the sense of Figure~\ref{fig:correlator-rep}: the Wilson loop in~\eqref{HQ-corr} is interrupted by the (thermal) trace over states, whereas~\eqref{QA-corr} can be written in a way such that the Wilson lines only appear between the two chromoelectric field operators. In the original formulation of the heavy quark diffusion coefficient~\cite{Casalderrey-Solana:2006fio}, the Wilson line configuration wraps around the closed-time Schwinger-Keldysh contour with a winding number equal to one. The Euclidean calculation of the heavy quark diffusion coefficient~\cite{Burnier:2010rp} also has this feature (see~\cite{Eller:2019spw} for an explicit proof that the Minkowski formulation~\cite{Casalderrey-Solana:2006fio} and the Euclidean formulation~\cite{Burnier:2010rp} give the same result). The Wilson line configuration in the correlator for quarkonium has a winding number equal to zero.
This mathematical difference has physical origin as discussed in the main text. 
Therefore, these two quantities $\kappa^{Q\bar{Q}}$ and $\kappa^Q$ (or more generally $g_E^{\rm Q\bar{Q}}$ and $g_E^Q$) cannot be used interchangeably, which has also been noted in Ref.~\cite{Eller:2019spw}.

The first verification that $g_E^{\rm Q\bar{Q}}$ and $g_E^Q$ are different was achieved in~\cite{Burnier:2010rp}, which can be seen by comparing the results obtained therein for $g_E^Q$ to the results of~\cite{Eidemuller:1997bb}, which first computed the correlator $g_E^{\rm Q\bar{Q}}$ in vacuum. Our result~\eqref{eq:spectral-full-result} further verifies this difference, by determining $g_E^{\rm Q\bar{Q}}$ both in vacuum and at finite temperature. Furthermore, the imaginary part of Eq.~\eqref{eq:difference-correlators-explicit} at zero frequency differs from the imaginary part of $g_E^{Q}$ at zero frequency already at $\ml{O}(g^4)$~\cite{Eller:2019spw}. In the present paper we further study their difference with a more general gauge choice, and show the breakdown of a naive axial gauge calculation.

\section{Chromoelectric correlator in axial gauge} \label{app:axial-NLO}
Here we study the chromoelectric field correlators for heavy quarks and quarkonia in axial gauge, in which the Wilson lines become identities and thus can be neglected. This is one of the necessary steps to arrive at the resolution of the puzzle outlined in Sec.~\ref{sec:axial-resolution}. The time-ordered chromoelectric field correlator is 
\be
g_{E,\,T}^{\rm Axial}(p_0) = g^2 \int_{-\infty}^{+\infty} \! \diff t\, e^{i p_0 t} \langle 0 |  \mathcal{T} ( {E}^a_{i}(t,{\bs x})  {E}^a_{i}(0,{\bs x}) ) | 0 \rangle\,,
\ee
and we want to calculate it in axial gauge at next-to-leading order (NLO). We will focus on the gluon polarization diagram, which contributes to the correlator at NLO. First we work out the gluon propagator in axial gauge. The free part of the gauge boson Lagrangian plus the gauge-fixing term in momentum space can be written as
\begin{align}
\frac{i}{2} \int \diff^4 k \, A^{\mu a}(-k) \Big( - g_{\mu\nu}(k^2 +i\varepsilon) +k_\mu k_\nu - \frac{1}{\xi}n_\mu n_\nu \Big) A^{\nu a}(k) \,,
\end{align}
where $\varepsilon$ comes from the boundary condition of the path integral at $t=\pm\infty$ and $\xi$ is a gauge-fixing parameter to be set later. Inverting $i g_{\mu\nu}(k^2 +i\varepsilon) -i k_\mu k_\nu +i n_\mu n_\nu/\xi$ gives the time-ordered gluon propagator
\begin{align}
[D_T(k)]_{\mu\nu}^{ab} = \frac{i\delta^{ab}}{k^2+i\epsilon} \left[ -g_{\mu\nu} + \frac{ n\cdot k(k_\mu n_\nu + k_\nu n_\mu) - [\xi(k^2+i\varepsilon)+n^2] k_\mu k_\nu +i\varepsilon n_\mu n_\nu }{i\varepsilon[\xi(k^2+i\varepsilon)+n^2] + (n\cdot k)^2} \right] \,.
\end{align}
Setting $\xi=0$ and $n_{\mu}=(1,0,0,0)$ for temporal axial gauge and neglecting terms proportional to $\varepsilon$ in the numerator lead to
\begin{align}
[D_T(k)]_{\mu\nu}^{ab} = \frac{i\delta^{ab} P_{\mu\nu}(k) }{k^2+i\epsilon} \equiv D_T(k) \delta^{ab} P_{\mu\nu}(k) \,,
\end{align}
where
\begin{align}
    P_{\mu \nu}(k) = -g_{\mu \nu} + \frac{k_0(k_\mu n_\nu + n_\mu k_\nu)}{k_0^2 + i\varepsilon} - \frac{k_\mu k_\nu}{k_0^2 + i\varepsilon} \, .
\end{align}
Then, using the Feynman rules of non-Abelian gauge theory we find the contribution of the gluon polarization diagrams (with its two external legs connected with the two chromoelectric fields) to the time-ordered chromoelectric correlator
\begin{align}
\left. g_{E,\,T}^{\rm Axial}(p_0) \right|_{\rm NLO} &= g^2 \int_{\bs p} D_T(p)^2 (i p_0 g_i^{\,\sigma'} - i p_i g_0^{\,\sigma'}) P_{\sigma' \sigma}(p) (- i p_0 g_i^{\,\rho'} + i p_i g_0^{\,\rho'}) P_{\rho' \rho}(p) \nonumber \\
& \quad \frac12 \bigg( \delta^{cd} \int_k D_T(k) \delta^{ab} P_{\mu \nu}(k) (-i g^2) \left[ f^{abe} f^{cde} (g^{\mu \rho} g^{\nu \sigma} - g^{\mu \sigma} g^{\nu \rho}) \right.  \nonumber \\
    & \quad \quad \quad \quad \quad \quad \quad \quad \quad \quad \quad \quad \quad \quad + \left. f^{ace} f^{dbe} (g^{\mu \sigma} g^{\nu \rho} - g^{\mu \nu} g^{\rho \sigma}  ) \right.  \nonumber \\
    & \quad \quad \quad \quad \quad \quad \quad \quad \quad \quad \quad \quad \quad \quad + \left. f^{ade} f^{bce} (g^{\mu \nu} g^{\rho \sigma} - g^{\mu \rho} g^{\nu \sigma}) \right] \nonumber \\
& + \delta^{cc'} \int_k D_T(k) D_T(p-k) \delta^{aa'} \delta^{bb'} P_{\mu\mu'}(k) P_{\nu\nu'}(p-k) \nonumber \\
    & \quad \quad \quad \quad \quad \times g f^{abc} \left[ g^{\mu \nu} (p - 2k)^{\sigma} + g^{\nu \sigma} (k-2p)^{\mu} + g^{\sigma \mu}(p+k)^{\nu} \right] \nonumber \\
    & \quad \quad \quad \quad \quad \times g f^{a'b'c'} \left[ g^{\mu' \nu'} (2k-p)^{\rho} + g^{\nu' \rho} (2p-k)^{\mu'} + g^{\rho \mu'}(-p-k)^{\nu'} \right] \bigg) \,,
\end{align}
where $\int_{\bs p} \equiv \int\frac{\diff^d p}{(2\pi)^d}$ and $\int_k \equiv \int\frac{\diff^D k}{(2\pi)^D}$.
Our strategy to evaluate these integrals is to do the ${\bs p}$ and ${\bs k}$ integrals first, using dimensional regularization in $d = 3 - \tilde{\epsilon}$ spatial dimensions for both of them. $D = 4 - \tilde{\epsilon}$ is the total number of spacetime dimensions. (The calculation is only consistent if we use the same dimensionality for both ${\bs p}$ and ${\bs k}$ integrals.) We leave the integral over $k_0$ to be done at the end of the calculation.
We proceed by reducing the integral into a handful of integral structures $\tilde{I}_i(p,k)$ and their respective numerators $N_i(p_0,k_0)$ that do not depend on spatial momenta, which gives
\be
g_{E,\,T}^{\rm Axial}(p_0) = \frac{N_c (N_c^2-1) g^4 p_0^2 }{2} \int_{-\infty}^{+\infty} \! \frac{\diff k_0} {2\pi} \int \frac{\diff^d {\bs p} \diff^d {\bs k} }{(2\pi)^{2d}} \sum_i N_i(p_0,k_0) \tilde{I}_i(p,k) \, .
\ee
Also, we denote $I_i(p_0,k_0) = \int \frac{\diff^d {\bs p} \diff^d {\bs k} }{(2\pi)^{2d}} \tilde{I}_i(p,k)$.

Below we list the resulting integral structures, accompanied by their respective numerators:
\begin{enumerate}
    \item \begin{align}
    I_1 &= \int_{\bs p} \int_{\bs k} \frac{1}{(p^2+i\varepsilon)^2 (k^2 + i\varepsilon) ((p-k)^2 + i\varepsilon) } \, , \\ 
    N_1 &= 0 \, , \end{align}
    \item \begin{align}
    I_2 &= \int_{\bs p} \int_{\bs k} \frac{1}{(p^2+i\varepsilon) (k^2 + i\varepsilon) ((p-k)^2 + i\varepsilon) } \nonumber \\ &= - \frac{\Gamma(4-D)}{(4\pi)^{D-1} } \int_0^1 \diff x \diff y \frac{ \left[ - y (k_0 - x p_0)^2 - (1 - y + y x(1-x)) p_0^2 - i\varepsilon \right]^{D-4} }{y^{\frac{D-3}{2}} (1-y+yx(1-x))^{\frac{D-1}{2}}} \, , \\ 
    N_2 &=  \frac{4 (D-2) (k_0^2 - k_0 p_0 + p_0^2)^2 }{k_0 (k_0 - p_0) p_0^2 } \, , \end{align}
    \item \begin{align}
    I_3 &= \int_{\bs p} \int_{\bs k} \frac{1}{(p^2+i\varepsilon)^2 (k^2 + i\varepsilon)} = - (-i)^{2D-8} \frac{\Gamma \! \left( \frac{3-D}{2} \right) \Gamma \! \left( \frac{5-D}{2} \right) }{(4\pi)^{D-1}} \frac{|k_0 p_0|^{D-3}}{p_0^2} \, , \\ 
    N_3 &=  \frac{2 (D-2) \left[ 2(D-1) k_0^2 - (D-2) p_0 k_0 + 2(D-1) p_0^2 \right] }{(D-1) k_0 p_0 } \, , \end{align}
    \item \begin{align}
    I_4 &= \int_{\bs p} \int_{\bs k} \frac{1}{(p^2+i\varepsilon) (k^2 + i\varepsilon)}  = (-i)^{2D-6} \frac{\Gamma\!\left(\frac{3-D}{2} \right)^2}{(4\pi)^{D-1}} |k_0 p_0|^{D-3} \, , \\ 
    N_4 &=  \frac{2 \left[ 2(D-2) k_0^4 - 3(D-2) k_0^3 p_0 + D\, k_0^2 p_0^2 - (D-2) k_0 p_0^3 + (D-2) p_0^4 \right]  }{ k_0^2 p_0^3 (k_0 - p_0) } \, , \end{align}
    \item \begin{align}
    I_5 &= \int_{\bs p} \int_{\bs k} \frac{1}{((p-k)^2+i\varepsilon) (k^2 + i\varepsilon)} = (-i)^{2D-6} \frac{\Gamma\!\left(\frac{3-D}{2} \right)^2}{(4\pi)^{D-1}} |k_0 (p_0-k_0) |^{D-3} \, , \\ 
    N_5 &=  \frac{D-2}{k_0^2} + \frac{D-2}{(k_0-p_0)^2} + \frac{2}{p_0^2} \, , \end{align}
\end{enumerate}
All the other integral structures give vanishing contributions in dimensional regularization. Note that the term 1.~vanishes because the numerator happens to be zero, and the term 3.~also vanishes upon integration over $k_0$ because the integrand is just a polynomial in $k_0$. (In dimensional regularization the limit $\tilde{\epsilon} \to 0$ cannot be taken before performing all integrals that involve $d$, which means the limit should be taken after the $k_0$ integral.) Then, one can show that for $p_0>0$
\begin{enumerate}
    \item \begin{align}
    \frac{N_c (N_c^2-1) g^4 p_0^2 }{2}  \int_{-\infty}^{+\infty} \frac{\diff k_0}{2\pi}  N_1(p_0,k_0) I_1(p,k) = 0
     \, , \end{align}
    \item \begin{align}
    & \frac{N_c (N_c^2-1) g^4 p_0^2 }{2}  \int_{-\infty}^{+\infty} \frac{\diff k_0}{2\pi}  N_2(p_0,k_0) I_2(p,k) \nonumber \\ &= \frac{N_c(N_c^2-1)g^4 p_0^3}{(2\pi)^3} \left[ \frac{11}{6 \tilde{\epsilon} } + \frac{11}{6} \ln \! \left( \frac{\mu^2}{4p_0^2} \right) + \frac{167}{36} + \frac{\pi^2}{3} \right]
     \, , \end{align}
    \item \begin{align}
    \frac{N_c (N_c^2-1) g^4 p_0^2 }{2} \int_{-\infty}^{+\infty} \frac{\diff k_0}{2\pi}  N_3(p_0,k_0) I_3(p,k) = 0
     \, , \end{align}
    \item \begin{align}
    & \frac{N_c (N_c^2-1) g^4 p_0^2 }{2} \int_{-\infty}^{+\infty} \frac{\diff k_0}{2\pi}  \left[ N_4(p_0,k_0) I_4(p,k) + N_5(p_0,k_0) I_5(p_0,k_0) \right] \nonumber \\ &= \frac{N_c (N_c^2 - 1) g^4 p_0^3}{ (2\pi)^3} \frac5{12}
     \, . \end{align}
\end{enumerate}

The final contribution to evaluate is from the coupling constant counterterm, since the definition of the chromoelectric correlator contains $g^2$. The contribution for $p_0>0$ reads
\begin{align}
    & (Z_g - 1 ) (N_c^2 - 1) g^2 \int_{\bs p}  D_T(p) (i p_0 g_i^{\,\sigma} - i p_i g_0^{\,\sigma}) P_{\sigma \sigma'}(p) (- i p_0 g_i^{\,\sigma'} + i p_i g_0^{\,\sigma'}) \nonumber \\
    &= \frac{g^4}{8\pi^2 (D-4)} \frac{11}{3} N_c (N_c^2 - 1) p_0^2 \int_{\bs p} \frac{i (\delta_{ii} - {\bs p}^2/p_0^2 )}{p_0^2 - {\bs p}^2 + i\varepsilon} \nonumber \\
    &= \frac{g^4 N_c (N_c^2 - 1)}{(2\pi)^3} p_0^2 \left[ \frac{11}{24 \pi^2 (D-4)} \frac{1}{(2\pi)^{D-4}} \pi \frac{1}{2p_0} \Omega_{D-1} ( D-2 ) p_0^{D-2}  \right] \nonumber \\
    &= \frac{g^4 N_c (N_c^2 - 1)}{(2\pi)^3} p_0^3 \frac{11}{48\pi} \left[ \frac{1}{ (D-4)} \frac{\Omega_{D-1} ( D-2 ) (p_0/\tilde{\mu} )^{D-4}}{(2\pi)^{D-4}}   \right] \nonumber \\
    &= \frac{g^4 N_c (N_c^2 - 1)}{(2\pi)^3} p_0^3 \left[ - \frac{11}{6\tilde{\epsilon}} + \frac{11}{48\pi} \frac{\partial}{\partial D} \left( \frac{\Omega_{D-1} ( D-2 ) (p_0/\tilde{\mu} )^{D-4}}{(2\pi)^{D-4}} \right)_{D=4} \right] \nonumber \\
    &= \frac{g^4 N_c (N_c^2 - 1)}{(2\pi)^3} p_0^3 \left[ - \frac{11}{6\tilde{\epsilon}} + \frac{11}{12} \left( 1 + \ln( \pi (p_0/(2\pi \tilde{\mu}))^2 ) - ( 2 - \gamma_E - 2\ln(2) ) \right) \right] \nonumber \\
    &=  \frac{g^4 N_c (N_c^2 - 1)}{(2\pi)^3} p_0^3 \left[ - \frac{11}{6\tilde{\epsilon}} - \frac{11}{12} - \frac{11}{12} \ln \! \left( \frac{\mu^2}{4p_0^2} \right)  \right] \,,
\end{align}
where $\mu^2 = 4\pi e^{-\gamma_E} \tilde{\mu}^2$.
Adding everything up, one obtains for $p_0>0$
\begin{equation}
    \left. g_{E,\,T}^{\rm Axial}(p_0) \right|_{\rm NLO}  = \frac{g^4 N_c (N_c^2 - 1)}{(2\pi)^3} p_0^3 \left[ \frac{11}{12}\ln \! \left( \frac{\mu^2}{4p_0^2} \right) + \frac{149}{36} + \frac{\pi^2}{3} \right],
\end{equation}
which is what we give in the main text.

The corresponding Euclidean correlator can be evaluated in exactly the same way. In the same notation (but with the understanding that $k_0$ and $p_0$ are now Euclidean quantities), the relevant integral structures are
\begin{enumerate}
    \item \begin{align}
    I_1 &= \int_{\bs p} \int_{\bs k} \frac{1}{(p^2)^2 k^2 (p-k)^2 } \, , \\ 
    N_1 &= 0 \, , \end{align}
    \item \begin{align}
    I_2 &= \int_{\bs p} \int_{\bs k} \frac{1}{p^2 k^2  (p-k)^2 } \nonumber \\ &=  \frac{\Gamma(4-D)}{(4\pi)^{D-1} } \int_0^1 \diff x \diff y \frac{ \left[  y (k_0 - x p_0)^2 + (1 - y + y x(1-x)) p_0^2 \right]^{D-4} }{y^{\frac{D-3}{2}} (1-y+yx(1-x))^{\frac{D-1}{2}}}  \, , \\ 
    N_2 &= - \frac{4 (D-2) (k_0^2 - k_0 p_0 + p_0^2)^2 }{k_0 (k_0 - p_0) p_0^2 } \, , \end{align}
    \item \begin{align}
    I_3 &= \int_{\bs p} \int_{\bs k} \frac{1}{(p^2)^2 k^2} = \frac{\Gamma \! \left( \frac{3-D}{2} \right) \Gamma \! \left( \frac{5-D}{2} \right) }{(4\pi)^{D-1}} \frac{|k_0 p_0|^{D-3}}{p_0^2} \, , \\ 
    N_3 &= - \frac{2 (D-2) \left[ 2(D-1) k_0^2 - (D-2) p_0 k_0 + 2(D-1) p_0^2 \right] }{(D-1) k_0 p_0 } \, , \end{align}
    \item \begin{align}
    I_4 &= \int_{\bs p} \int_{\bs k} \frac{1}{p^2 k^2 }  =  \frac{\Gamma\!\left(\frac{3-D}{2} \right)^2}{(4\pi)^{D-1}} |k_0 p_0|^{D-3} \, , \\ 
    N_4 &=  -\frac{2 \left[ 2(D-2) k_0^4 - 3(D-2) k_0^3 p_0 + D k_0^2 p_0^2 - (D-2) k_0 p_0^3 + (D-2) p_0^4 \right]  }{ k_0^2 p_0^3 (k_0 - p_0) } \, , \end{align}
    \item \begin{align}
    I_5 &= \int_{\bs p} \int_{\bs k} \frac{1}{(p-k)^2 k^2 } = \frac{\Gamma\!\left(\frac{3-D}{2} \right)^2}{(4\pi)^{D-1}} |k_0 (p_0-k_0) |^{D-3} \, , \\ 
    N_5 &= - \frac{D-2}{k_0^2} - \frac{D-2}{(k_0-p_0)^2} - \frac{2}{p_0^2} \, , \end{align}
\end{enumerate}
all of which give the same contributions to the correlator as in the time-ordered case.

\section{Calculation Details of Spectral Function Difference}
\label{app:spectral}

To firmly establish the importance of the nonperturbative property $\rho^{++}_{\rm adj}(\omega) \neq -\rho^{++}_{\rm adj}(-\omega)$ of the spectral function for quarkonium transport, as required to formulate a lattice QCD calculation in Section~\ref{sec:euclidean}, in this Appendix we provide the details of the calculation of the difference between the spectral function for single heavy quark transport and the spectral function for quarkonium transport.
As explained in Section~\ref{sect:non-odd}, the difference between the spectral function for quarkonium transport and that for single heavy quark transport is given by the diagrams (j) in Refs.~\cite{Burnier:2010rp,Eller:2019spw}, or (5), (5r) in Fig.~\ref{fig:diagrams}. The diagrammatic representation of their difference in real time in terms of Wightman functions was given in Section~\ref{sec:axial-gauge}, where gauge invariance was also examined.

Following the calculation details of Section~\ref{sec:3-propagator-evaluation}, we find that the difference between these two spectral functions stemming from these diagrams is given by
\begin{align}
    \rho^{++}_{\rm adj}(\omega) -  \rho_{\rm fund}(\omega) &= \int_{{\bs p}, k}  \frac{T_F}{3N_c} g^4 N_c (N_c^2 - 1)  2\pi \delta(k_0)   \\ 
    &\times  \big[ g_{\mu \nu} (p - 2k)_\delta + g_{\nu \delta} (k - 2p)_{\mu} + g_{\delta \mu} (p+k)_\nu \big] \nonumber \\
    &\times   (p_0 g_{i\delta'} - p_i g_{0\delta'} ) \big((p_0-k_0) g_{i\nu'} - (p_i - k_i) g_{0\nu'} \big) \nonumber \\
    &\times   {\rm Re} \Big\{  [\rho(p)]^{\delta' \delta} [D_T(p-k)]^{\nu \nu'} [D_T(k)]^{\mu 0}  \nonumber \\
    & \quad - [D_{T}(p)]^{\delta' \delta } \big( [D_>(p-k)]^{\nu' \nu } [D_>(k)]^{ 0 \mu}  - [D_<(p-k)]^{\nu' \nu } [D_<(k)]^{\mu 0}  \big) \Big\} \, , \nn
\end{align}
where $p_0 = \omega$. By using the thermal (KMS) relations between the free propagators $D_>, D_<, D_T$ and $\rho$, this can be further simplified to
\begin{align}
\label{eq:general-spectral-diff-NLO}
    \rho^{++}_{\rm adj}(\omega) -  \rho_{\rm fund}(\omega) &= \int_{{\bs p}, k}  \frac{T_F}{3N_c} g^4 N_c (N_c^2 - 1)  2\pi \delta(k_0)    \\ 
    & \quad \times \big[ g_{\mu \nu} (p - 2k)_\delta + g_{\nu \delta} (k - 2p)_{\mu} + g_{\delta \mu} (p+k)_\nu \big] \nonumber \\
    & \quad \times (p_0 g_{i\delta'} - p_i g_{0\delta'} ) \big((p_0-k_0) g_{i\nu'} - (p_i - k_i) g_{0\nu'} \big) \nonumber \\
    & \quad \times (-1) [\rho(p)]^{\delta' \delta} {\rm Im} \{ [D_R]^{\nu \nu'}(p-k) \} {\rm Im} \{ [D_R]^{\mu 0}(k) \} \, .  \nn
\end{align}
In our convention, the free propagators in Feynman gauge are given by
\begin{align}
    [\rho(p)]^{\mu \nu} = (- g^{\mu \nu}) (2\pi) {\rm sgn}(p_0) \delta(p^2) & & [D_R(p)]^{\mu \nu} = \frac{ - i g^{\mu \nu}}{p^2 + i 0^+ {\rm sgn}(p_0) } \, ,
\end{align}
and using them to calculate the difference, one arrives at
\begin{align}
    &\rho^{++}_{\rm adj}(\omega) -  \rho_{\rm fund}(\omega) \\ &= \int_{{\bs p}, k}  \frac{T_F}{3N_c} g^4 N_c (N_c^2 - 1)  (2\pi) \delta(k_0)   (2\pi) {\rm sgn}(\omega) \delta(p^2) \mathcal{P} \left( \frac{2 d \omega^3 - 2 \omega ({\bs p} - {\bs k})^2}{k^2 (p-k)^2 } \right)  \, . \nonumber
\end{align}

In dimensional regularization, $({\bs p} - {\bs k})^2$ may be exchanged by $\omega^2$ because $\int_{\bs k} \frac{1}{{\bs k}^2}$ vanishes. Then, setting $d = 3$, this integral becomes
\begin{align}
    \rho^{++}_{\rm adj}(\omega) -  \rho_{\rm fund}(\omega) =  \frac{T_F}{3N_c} g^4 N_c (N_c^2 - 1) |\omega|^3   \int_{{\bs p}, {\bs k}}   (2\pi) \delta(p^2) \mathcal{P} \left( \frac{(-4) }{{\bs k}^2 [ \omega^2 - ({\bs p} - {\bs k})^2 ] } \right)  \, .
\end{align}
The explicit calculation of this integral is equivalent to the one presented in Eq.~\eqref{eq:result-diff-wightman}. The final result is
\begin{align}
    \rho^{++}_{\rm adj}(\omega) -  \rho_{\rm fund}(\omega) &=  \frac{T_F}{3N_c} g^4 N_c (N_c^2 - 1) |\omega|^3   \frac{\pi^2}{(2\pi)^3}  = \frac{g^4 T_F (N_c^2 - 1) \pi^2}{3 (2\pi)^3} |\omega|^3 \, ,
\end{align}
as claimed in the main text.

It is noteworthy that the difference between the spectral functions, as given in Eq.~\eqref{eq:general-spectral-diff-NLO} may also be used in conjunction with HTL-resummed propagators to explore the value of the difference (a modification to the gluon 3-vertex is also necessary, according to the HTL effective theory Feynman rules. They can be found in Ref.~\cite{Andersen:2002ey}.). However, as discussed in Section~\ref{sec:fitting-ansatz-rho}, a full fixed-order calculation at $\mathcal{O}(g^6)$, which is the leading contribution to the difference in the small frequency domain, also requires considering 2-loop diagrams, which we will not pursue here.

%% file: appendixb-v2.tex
\chapter{Appendix: Calculation of the Chromoelectric Field Correlator at Strong Coupling in $\mathcal{N}=4$ Yang-Mills Theory}

\section{Operator ordering aspects of Wilson loops} \label{sec:App-W-ordering}

In this Appendix we discuss operator ordering aspects that are necessary to rigorously define observables in terms of Wilson loops, which become most apparent in the strongly coupled calculation of Section~\ref{sec:strong-coupling}, and specifically concern the setup described in Section~\ref{sect:setup_begin}.

\subsection{Time-ordered products of Wilson lines}

In this work, we deal with the calculation of the time-ordered correlator
\begin{align}
    [g_E^{{\mathcal{T}}}](t) = \langle \hat{\mathcal{T}} E_i^a(t) \mathcal{W}^{ab}_{[t,0]} E_j^b(0) \rangle_T \, ,
\end{align}
where $\mathcal{W}^{ab}_{[t,0]}$ is an adjoint Wilson line, that is written in terms of fundamental Wilson lines as
\begin{equation} \label{eq:app-adj-as-fund}
    \W^{ab}_{[t_2,t_1]} = \frac{1}{T_F} {\rm Tr}_{\rm color}  \left[ \hat{\T} \, T^a_F U_{[t_2,t_1]} T^b_F U_{[t_2,t_1]}^\dagger \right] \, ,
\end{equation}
where the time-ordering symbol is necessary to preserve the explicit ordering of operators in an adjoint Wilson line.

For concreteness, we write both adjoint and fundamental lines below:
\begin{align}
    \W^{ab}_{[t_2,t_1]} &= \left[ {P} \exp \left( i g \int_{t_1}^{t_2} dt\, A_0^c(t) T_{\rm Adj}^c \right) \right]^{ab} \, , \\
    U_{[t_2,t_1],ij} &= \left[ {P} \exp \left( i g \int_{t_1}^{t_2} dt\, A_0^c(t) T_{\rm Fund}^c \right) \right]_{ij} \, ,
\end{align}
where the difference is the representation of the SU($N_c$) generator matrices. $[T_{\rm Fund}^a]_{ij} \equiv [T_{ F}^a]_{ij}$ are the generators of the fundamental representation, normalized in the conventional way ${\rm Tr}[T_{F}^a T_{F}^b] = T_F \delta^{ab} $ with $T_F = 1/2$, and $[T_{\rm Adj}^a]^{bc} = -i f^{abc}$, where $f^{abc}$ are the structure constants of the group $[T^a , T^b] = i f^{abc} T^c$\@.

Note that the operator ordering in the adjoint Wilson line is not the ``natural'' one in terms of fundamental Wilson lines, because the operator products are not ordered in the same way as the matrix products that contract color indices. In this sense, Eq.~\eqref{eq:app-adj-as-fund} without the time-ordering symbol is only indicative of the color product structure, but is not an explicit expression in terms of how the gauge field operators $A_0^a(t)$ are ordered. This is specified by the symbol $\hat{\T}$, but by itself does not provide an explicit method to calculate it. One explicit way to evaluate it is given by its path integral representation, to which we will return in a moment. Before doing that, however, it is useful to discuss how this time ordering of fundamental Wilson lines appears from the dynamics of two (coincident) point color charges, which we take to be in the fundamental and anti-fundamental representations of SU($N_c$)\@.

To accomplish this, let us collectively denote the ``colors'' of the $Q \bar{Q}$ pair by $\big( Q\bar{Q} \big)_{ij}$, with $i$ being the index of the quark in the fundamental representation, and $j$ the index of the anti-quark in the anti-fundamental representation. The dynamics are given by
\begin{align}
    \frac{d}{dt} \big( Q\bar{Q} \big)_{ij} &= \left[ i g A_0^a(t) \big[ T^a_{\rm Fund} \big]_{ii'} \delta_{jj'} + i g A_0^a(t) \big[ T^a_{\rm Anti-Fund} \big]_{jj'} \delta_{ii'} \right] \big( Q\bar{Q} \big)_{i'j'} \nonumber \\
    &= \left[ i g A_0^a(t) \big[ T^a_{F} \big]_{ii'} \delta_{j'j} - i g A_0^a(t) \big[ T^a_{F} \big]_{j'j} \delta_{ii'} \right] \big( Q\bar{Q} \big)_{i'j'} \, ,
\end{align}
where we have used that $ \big[T^a_{\rm Anti-Fund}\big]_{ij} = - [T^a_{\rm Fund}]_{ji}$\@.

Formally, we can write the solution to this equation as
\begin{equation}
    \big( Q\bar{Q} \big)_{ij} (t) = \mathcal{W}_{i i_0,j_0 j}(t) \big( Q\bar{Q} \big)_{i_0j_0}(t=0) \, ,
\end{equation}
where $\mathcal{W}_{i i_0,j_0 j}(t)$ obeys the same equation as $\big( Q\bar{Q} \big)_{ij}$
\begin{equation} \label{eq:App-def-W-eqn}
    \frac{d}{dt} \mathcal{W}_{i i_0,j_0 j} = \left[ i g A_0^a(t) \big[ T^a_{F} \big]_{ii'} \delta_{j'j} - i g A_0^a(t) \big[ T^a_{F} \big]_{j'j} \delta_{ii'} \right] \mathcal{W}_{i' i_0,j_0 j'} \, ,
\end{equation}
with $\mathcal{W}_{i i_0,j_0 j}(t=0) = \delta_{i i_0} \delta_{j j_0}$ as the initial condition. Note that, by construction, we have
\begin{equation}
    \mathcal{W}_{i i_0,j_0 j} = \hat{\T} \big( \big[U_{[t,0]}\big]_{i i_0} \big[ U_{[t,0]}^\dagger \big]_{j_0 j} \big) \, .
\end{equation}
A quick way to see this is to note that if $A_0^a$ were ordinary numbers, then the time ordering would be irrelevant and the Wilson lines in the last expression would be completely decoupled from each other. However, because $A_0^a(t)$ are in principle non-commuting operators, we have to keep track of the fact that $A_0^a(t)$ is always inserted to the left of operators $A_0^a(t')$ when $t > t'$\@. This is due to $A_0^a(t)$ being to the left of $\W$ in its defining differential equation~\eqref{eq:App-def-W-eqn}\@. Therefore, what we get out of $\mathcal{W}_{i i_0,j_0 j}$ is a fundamental and an anti-fundamental Wilson line put together, with their operators time-ordered.

A consistency check is to verify that we can get the adjoint Wilson line from $\mathcal{W}_{i i_0,j_0 j}$\@. Indeed, we can consider
\begin{align}
    \frac{1}{T_F} [T^a_F]_{j i} \mathcal{W}_{i i_0,j_0 j} [T^b_F]_{i_0 j_0} = \frac{1}{T_F} {\rm Tr}_{\rm color}  \left[ \hat{\T} \, T^a_F U_{[t_2,t_1]} T^b_F U_{[t_2,t_1]}^\dagger \right] \, ,
\end{align}
and get, after contracting the color indices and using the Lie Algebra of the group,
\begin{align}
    \frac{d}{dt} \big( [T^a_F]_{j i} \mathcal{W}_{i i_0,j_0 j} [T^b_F]_{i_0 j_0} \big) = i g A_0^c(t) \big( - i f^{cad} \big) \big( [T^d_F]_{j' i'} \mathcal{W}_{i' i_0,j_0 j'} [T^b_F]_{i_0 j_0} \big) \, ,
\end{align}
which is exactly the defining equation for an adjoint Wilson line:
\begin{equation}
    \frac{d}{dt} \W^{ab} = i g A_0^c(t) \big( - i f^{cad} \big) \W^{db} \, .
\end{equation}

It is also direct to see from here that the Wilson loop we consider in Section~\ref{sect:setup_begin} satisfies
\begin{align}
    \langle \hat{\T} W[\mathcal{C}] \rangle &= \frac{1}{N_c} \langle {\rm Tr}_{\rm color} \left[ \hat{\T} U_{[t,0]} U_{[t,0]}^\dagger \right] \rangle \nonumber \\
    &= \frac{1}{N_c} \langle \hat{\T} \big( \big[U_{[t,0]} \big]_{i i_0}  \big[ U_{[t,0]}^\dagger \big]_{i_0 i} \big) \rangle \nonumber \\
    &= \frac{1}{N_c} \langle \W_{i i_0 , i_0 i} \rangle \, ,
\end{align}
and contracting the indices in the previous equations, we see that
\begin{equation}
    \frac{d}{dt} \W_{i i_0 , i_0 i} = 0 \, ,
\end{equation}
and therefore, given the initial condition $\mathcal{W}_{i i_0,j_0 j}(t=0) = \delta_{i i_0} \delta_{j j_0}$, we conclude that
\begin{align}
    \langle \hat{\T} W[\mathcal{C}] \rangle &= \frac{1}{N_c} \langle \W_{i i_0 , i_0 i}(t=0) \rangle \nonumber \\
    &= \frac{1}{N_c} \delta_{i i_0} \delta_{i_0 i} = 1 \, ,
\end{align}
as claimed in Section~\ref{sec:setup}.

It is worth noting that this is self-evident from the path integral formulation. Indeed, collectively denoting the field content of the theory by $\varphi$, one can calculate time-ordered (vacuum) correlation functions as
\begin{equation}
    \langle \hat{\T} {O}_1 \ldots {O}_n \rangle = \frac{1}{{Z}} \int D\varphi\, e^{i S[\varphi]} {O}_1[\varphi] \ldots {O}_n[\varphi] \, ,
\end{equation}
which means that for our pair of fundamental Wilson lines we have
\begin{align}
    \langle \hat{\T} \big( \big[U_{[t,0]} \big]_{i i_0}  \big[ U_{[t,0]}^\dagger \big]_{i_0 i} \big) \rangle &= \frac{1}{{Z}} \int D\varphi\, e^{i S[\varphi]} \big[U_{[t,0]} \big]_{i i_0}  \big[ U_{[t,0]}^\dagger \big]_{i_0 i} \nonumber \\
    &= \frac{1}{{Z}} \int D\varphi\, e^{i S[\varphi]} = 1 \, .
\end{align}
The step to the last line is achieved because, inside the path integral, the Wilson lines are just SU$(N_c)$ unitary matrices that are inverses of each other. The fact that we get one out of this is evidently consistent with our previous discussion.

\subsection{Standard products of Wilson lines}

In other contexts, it is also possible for the Wilson lines to have different operator orderings. For instance, we could consider a different kind of Wilson loop, without a time-ordering symbol:
\begin{equation}
    \langle W[\mathcal{C}] \rangle = \frac{1}{N_c} \langle U_{[t,0]} U^\dagger_{[t,0]} \rangle = 1\,,
\end{equation}
which also equals unity. The reason for that here, however, is that the operators $U_{[t,0]}$ and $U_{[t,0]}^\dagger$ are inverses of each other, and should be interpreted as written, with the operator products appearing in the same way as the color products.

Interestingly, the path integral description of this object is less simple than for the time-ordered loop. The reason for this is that inserting complete bases of states along the operator products to convert the expectation value into a path integral requires following the time contour defined by the explicit operator ordering in the correlation function. In the time-ordered case, operators are, by definition, arranged further to the left at later times, and hence it is sufficient to insert complete bases of states that span the $[0,t]$ time interval once. However, for the loop considered in this section, one has to insert complete bases of states along the interval $[0,t]$ once in the forward direction (for $U$), and once in the backward direction (for $U^\dagger$)\@.

In the context of the AdS/CFT correspondence, this type of operator ordering is realized by using the gauge/gravity duality for each segment of the path integral contour and imposing appropriate matching conditions~\cite{Skenderis:2008dg,Skenderis:2008dh,vanRees:2009rw}\@. Both the heavy quark diffusion coefficient calculation~\cite{Casalderrey-Solana:2006fio} and the jet quenching parameter claculation~\cite{DEramo:2010wup} implicitly have this feature. This is in contrast to the calculation presented herein, which does not require to match the background solution across manifolds that have different segments of the Schwinger-Keldysh contour as their boundaries.

\subsection{The timelike adjoint Wilson line in AdS/CFT and the role of the $S_5$} \label{app:proof-bound}

In the main text, in Eq.~\eqref{eq:claim-to-show-AppB}, we claim that
\begin{equation}
    \left| \frac{1}{ \mathcal{Z}} {\rm Tr}_{\mathcal{H}} \! \left(  e^{-\beta H} \hat{\mathcal{T}} W_S[\mathcal{C}_0,\hat{n}] \right) \right| \leq 1 \, .
\end{equation}
for a timelike path $\mathcal{C}_0$ that goes over a straight segment of length $\mathcal{T}$ and then backtracks to its starting point (in what follows, it will be clear that it is not essential for the path to be straight, but it does have to be timelike).

Showing the bound is straightforward once the notation is made explicit. The main ingredient that has to be dealt with carefully is time-ordering. The simplest way to proceed is to define the time-ordered version of the Wilson loop by introducing a more general object that contains it through the differential equation satisfied by the color degrees of freedom of the heavy quarks. Let $\mathcal{W}_{i i_0,j_0 j}$ be such that (in this expression we use the convention that repeated indices are summed; the rest of summations in this section will be made explicit)
\begin{align} \label{eq:App-def-W-eqn-2}
    &\frac{\rm d}{{\rm d}t} \mathcal{W}_{i i_0,j_0 j} \\ &= \left[ i g \left( A_0^a(t) + \hat{n}_1(t) \cdot \phi^a(t) \right) \big[ T^a_{F} \big]_{ii'} \delta_{j'j} - i g \left( A_0^a(t) + \hat{n}_2(t) \cdot \phi^a(t) \right) \big[ T^a_{F} \big]_{j'j} \delta_{ii'} \right] \mathcal{W}_{i' i_0,j_0 j'} \, , \nonumber
\end{align}
with $\mathcal{W}_{i i_0,j_0 j}(t = -\mathcal{T}/2) = \delta_{i i_0} \delta_{j j_0}$. The $S_5$ coordinates are given by $\hat{n}_1(t)$ and $-\hat{n}_2(t)$, representing their values on each side of the contour $\mathcal{C}_0$. The minus sign is necessary to be consistent with the definition~\eqref{eq:W-loop-S}, where there is no sign flip in the prefactor of the scalars caused by the sign flipping of $\dot{x}^{\mu}$. Then, one has $\hat{\mathcal{T}} W_S[\mathcal{C}_0,\hat{n}] = \frac{1}{N_c} \sum_{i,i_0}\mathcal{W}_{i i_0,i_0 i}(t = \mathcal{T}/2)$.

More importantly, $\mathcal{W}$ is a unitary operator on the Hilbert space $\mathcal{H}_{\rm ext} = \mathcal{H} \otimes {\rm Fund}_{N_c} \otimes \overline{\rm Fund}_{N_c}$, which describes the Hilbert spaces of the QGP without any external charge, the heavy quark and the heavy antiquark respectively. As such, we can write
\begin{equation} \label{eq:starting-proof}
    \frac{1}{ \mathcal{Z}} {\rm Tr}_{\mathcal{H}} \! \left(  e^{-\beta H} \hat{\mathcal{T}} W_S[\mathcal{C}_0,\hat{n}] \right) = \frac{1}{\mathcal{Z} N_c} \sum_n e^{-\beta E_n} \sum_{i,j = 1}^{N_c} \langle n , i , i | \mathcal{W} | n, j , j \rangle \, ,
\end{equation}
where we have labeled states in $\mathcal{H}_{\rm ext}$ as $|n, i, j \rangle$, in which $n$ labels the energy eigenstates of $\mathcal{H}$, $i$ labels the color index of ${\rm Fund}_{N_c}$, and $j$ labels the color index of $\overline{\rm Fund}_{N_c}$.
Generally, the action of an operator can be written in terms of its matrix elements. Inserting an identity as a complete set of states yields
\begin{equation}
    \mathcal{W} | n, i , j \rangle = \sum_{m} \sum_{k,l = 1}^{N_c} | m, k, l \rangle [\mathcal{W}]_{mkl,nij}   \, ,
\end{equation}
and the fact that $\mathcal{W}$ is a unitary operator means that we can write its matrix elements in terms of its eigenstates' components $v_{nij}^{(L)}$ as
\begin{equation}
    [\mathcal{W}]_{mkl,nij} = \sum_L v_{mkl}^{(L)*} e^{i \phi_L} v_{nij}^{(L)}
\end{equation} 
where the eigenstates are labelled by $L$.
We then have
\begin{align}
    \frac{1}{ \mathcal{Z}} {\rm Tr}_{\mathcal{H}} \! \left(  e^{-\beta H} \hat{\mathcal{T}} W_S[\mathcal{C}_0,\hat{n}] \right) &= \frac{1}{\mathcal{Z} N_c} \sum_n e^{-\beta E_n} \sum_{i,j = 1}^{N_c} [\mathcal{W}]_{nii,njj} \\ &= \frac{1}{\mathcal{Z} N_c} \sum_n e^{-\beta E_n} \sum_{i,j = 1}^{N_c} \sum_L v_{nii}^{(L)*} e^{i \phi_L} v_{njj}^{(L)} \nonumber \\
    &= \frac{1}{\mathcal{Z} N_c} \sum_n \sum_L e^{-\beta E_n} \left| \sum_{i=1}^{N_c} v_{nii}^{(L)} \right|^2 e^{i \phi_L} \, . \nonumber
\end{align}
Whatever the eigenvectors $v_{nij}^{(L)}$ are, this sum is largest in absolute value if all of the phases $e^{i \phi_L}$ are equal. However, if this is the case, then it follows that $\mathcal{W} = e^{i \phi} \mathbbm{1}$. Therefore, from Eq.~\eqref{eq:starting-proof} we have
\begin{align}
    \left| \frac{1}{ \mathcal{Z}} {\rm Tr}_{\mathcal{H}} \! \left(  e^{-\beta H} \hat{\mathcal{T}} W_S[\mathcal{C}_0,\hat{n}] \right) \right| &\leq \frac{1}{\mathcal{Z} N_c} \sum_n e^{-\beta E_n} \sum_{i,j = 1}^{N_c} \langle n , i , i | \mathbbm{1} | n, j , j \rangle \\ &= \frac{1}{\mathcal{Z} N_c} \sum_n e^{-\beta E_n} \sum_{i,j = 1}^{N_c} \delta_{ij} \nonumber \\
    &= \frac{\sum_n e^{-\beta E_n}}{\mathcal{Z}} \frac{\sum_{i= 1}^{N_c} 1}{ N_c} = 1 \, , \nonumber
\end{align}
as initially claimed in Eq.~\eqref{eq:claim-to-show-AppB}.

Furthermore, this bound is saturated by configurations where $\hat{n}$ takes antipodal positions on the $S_5$. This is easy to see from the defining equation~\eqref{eq:App-def-W-eqn-2}, because, noting that $\hat{\mathcal{T}} W_S[\mathcal{C}_0,\hat{n}] = \frac{1}{N_c} \mathcal{W}_{i i_0,i_0 i}(t = \mathcal{T}/2)$, it suffices to inspect this differential equation for $i = j$ and $i_0 = j_0$. Explicitly, we have
\begin{align} \label{eq:App-work-W-eqn}
    &\frac{\rm d}{{\rm d}t} \mathcal{W}_{i i_0,i_0 i} \\ &= \left[ i g \left( A_0^a(t) + \hat{n}_1(t) \cdot \phi^a(t) \right) \big[ T^a_{F} \big]_{ii'} \delta_{j'i} - i g \left( A_0^a(t) + \hat{n}_2(t) \cdot \phi^a(t) \right) \big[ T^a_{F} \big]_{j'i} \delta_{ii'} \right] \mathcal{W}_{i' i_0,i_0 j'} \nonumber \\
    &= \left[ i g \left( A_0^a(t) + \hat{n}_1(t) \cdot \phi^a(t) \right) \big[ T^a_{F} \big]_{j'i'} - i g \left( A_0^a(t) + \hat{n}_2(t) \cdot \phi^a(t) \right) \big[ T^a_{F} \big]_{j'i'} \right] \mathcal{W}_{i' i_0,i_0 j'} \, , \nonumber
\end{align}
which vanishes if $\hat{n}_1 = \hat{n}_2$. As discussed below Eq.~\eqref{eq:App-def-W-eqn-2}, this corresponds to taking antipodal positions on the $S_5$ for the generalized Wilson loop~\eqref{eq:W-loop-S}. The bound is then saturated because 
\begin{align}
    \frac{\rm d}{{\rm d}t} \mathcal{W}_{i i_0,i_0 i} = 0 \implies \hat{\mathcal{T}} W_S[\mathcal{C}_0,\hat{n}] &= \frac{1}{N_c} \mathcal{W}_{i i_0,i_0 i}(t = \mathcal{T}/2) \nonumber \\ &= \frac{1}{N_c} \mathcal{W}_{i i_0,i_0 i}(t = -\mathcal{T}/2) = \frac{1}{N_c} \delta_{i i_0} \delta_{i i_0} = 1 \, . \nonumber
\end{align}
Any other configuration will give a highly oscillatory contribution to the trace over $\mathcal{H}$, and thus its numerical value would be suppressed. Therefore, the dominant contribution indeed comes from the configurations we just described.

One can then also verify on the gravity side of the duality that the extremal worldsheet associated with this configuration is stable and allows for a calculation of the correlator~\eqref{eq:EE-corr-from-variations} by solving a set of linear differential equations for the path variations in the dual gravitational description, as we do in Section~\ref{sec:QQ-setup}.

As a final comment, we note that the above argument relies crucially on $W_S[\mathcal{C}_0,\hat{n}]$ being constructed from unitary operators. This is true for timelike Wilson loops, but if the path $\mathcal{C}$ is spacelike, then the prefactor $\sqrt{\dot{x}^2}$ of the scalars in the exponential of Eq.~\eqref{eq:W-loop-S} becomes imaginary. Consequently, there is no unitarity bound for such Wilson loops, and thus our preceding argument does not follow through.

\section{Analysis of the worldsheet turnaround region and its consequences for the $i\epsilon$ prescription} \label{app:turnaround-worldsheet}

In this Appendix we verify the $i\epsilon$ prescription we arrived at in Section~\ref{sec:ends-matching} by analyzing the behavior of the worldsheet around the turnaround times $t = \pm \T/2$. Specifically, we study whether one can get extra imaginary terms in the equations of motion by having a transition where the induced metric on the worldsheet goes from having Minkowski signature (i.e., timelike) to having Euclidean signature (i.e., spacelike)\@.

To have control over the behavior of the worldsheet at the turnaround times $t = \pm \T/2$, we need to regulate the backtracking of the loop in a way that its tangent vector is continuous throughout, such that if we look closely enough, the extremal surface will still be smooth. Our choice of regulator for the present purpose is to introduce a small spatial separation $L$ between the two lines, which is compatible with the discussion in the previous sections. Conversely, the only way to smoothly turn from a timelike tangent vector $\dot{x}^\mu$ going in the future direction to one going in the past direction is by having a segment where it is spacelike. As such, our choice for a regulator is actually generic.

This motivates studying the behavior of a worldsheet close to a spacelike boundary segment. To gain intuition, let us first discuss a few examples. A family of solutions that is easily obtained at $T=0$ is $z(t,x) = \sqrt{t^2 - x^2 - \rho_0^2}$, either for $t \geq \sqrt{x^2 + \rho_0^2}$ or for $t \leq - \sqrt{x^2 + \rho_0^2}$\@. These solutions are spacelike surfaces that satisfy the Euler-Lagrange equations obtained from the Nambu-Goto action that are bounded by the hyperbola $t^2 - x^2 = \rho_0^2$ at $z = 0$\@. Another family of solutions to the Euler-Lagrange equations is given (implicitly) by the integral $\int_0^{z(t)/z_c} \frac{u^2 \diff u}{\sqrt{1+u^4}} = \frac{t}{z_c}$, where $z_c$ is a parameter defining different solutions, all of which are bounded by the line $t=0$ at $z=0$, valid in a small neighborhood of a spatial Wilson line segment with varying $x$ and all else held constant (to fully determine a unique solution it is necessary to specify how the surface is closed, or equivalently, how the Wilson loop path closes itself, far away from the region we just studied)\@. All of these have the crucial property that they are spacelike surfaces, which motivates investigating whether our previous conclusion is affected when we deform the contour slightly by introducing a spatial separation.

Note that the $i\epsilon$ in Eq.~\eqref{eq:bkg-action-general} also provides a prescription to evaluate the action in the case of a spacelike worldsheet. Specifically, it determines that a spacelike worldsheet has a Nambu-Goto action determined by the substitution
\begin{equation} \label{eq:spacelike-prescription}
    \sqrt{- \det \left( \partial_\alpha X^\mu \partial_\beta X^\nu g_{\mu \nu} \right) } \to -i \sqrt{\left| \det \left( \partial_\alpha X^\mu \partial_\beta X^\nu g_{\mu \nu} \right) \right| } \, ,
\end{equation}
which, satisfactorily, is exactly what we would get by demanding that whenever the worldsheet is spacelike, the Nambu-Goto action should be the same as if we had started in Euclidean signature from the beginning.

Now we may ask what happens if we include perturbations on top of a background worldsheet that features a transition from spacelike to timelike and vice-versa. Given that these perturbations are introduced on top of a solution that extremizes the action, the action for the fluctuations in a spacelike region should be real and positive definite. We will verify this explicitly in what follows, as it will be crucial to our results that the $i\epsilon$ prescription would not be modified by contributions from a spacelike region.

When the background worldsheet is spacelike, the argument of the square root in Eq.~\eqref{eq:bkg-action-general} becomes negative, and we must therefore use Eq.~\eqref{eq:spacelike-prescription} to get
\begin{equation}
    i S_{\rm NG}^{(0)}[\Sigma_0] = \! \int \frac{\diff s \,  \diff z}{z^2} \sqrt{ \frac{\dot{x}^2}{f} + \frac{z^2 \dot{\phi}^2 }{f}  - \dot{t}^2 - f \big( \dot{t} x' - t' \dot{x} \big)^2 - f z^2 \big( \dot{t} \phi' - \dot{\phi} t' \big)^2 + z^2 \big( \dot{x} \phi' - x' \dot{\phi} \big)^2 } \,, \label{eq:bkg-action-imag}
\end{equation}
which for the fluctuations read
\begin{equation}
    i S_{\rm NG}^{(2)}[\Sigma_0;y] \!= \!\! \int \frac{\diff s \,  \diff z}{2 z^2} \frac{- f \big(\dot{t} y' - t' \dot{y} \big)^2 + \big( \dot{x} y' - x' \dot{y} \big)^2 + z^2 \big( \dot{y} \phi' - y' \dot{\phi} \big)^2 + \frac{\dot{y}^2}{f} }{\sqrt{ \frac{\dot{x}^2}{f} + \frac{z^2 \dot{\phi}^2 }{f}  - \dot{t}^2 - f \big( \dot{t} x' - t' \dot{x} \big)^2 - f z^2 \big( \dot{t} \phi' - \dot{\phi} t' \big)^2 + z^2 \big( \dot{x} \phi' - x' \dot{\phi} \big)^2 }} \,, \label{eq:fluct-action-imag}
\end{equation}
As written, this is a general expression. However, the expression is explicit enough for us to make generic statements about how the fluctuations $y(s,z)$ behave on a background specified by $X^\mu = (t(s,z), x(s,z),0,0,z,\hat{n}(s,z))$\@. The key observation is that the quadratic form in the numerator of the integrand in Eq.~\eqref{eq:fluct-action-imag} can be written as
\begin{equation}
    \begin{pmatrix}
        y' & \dot{y} 
    \end{pmatrix} 
    \begin{pmatrix}
        \dot{x}^2 + z^2 \dot{\phi}^2 - f \dot{t}^2 & f \dot{t} t' - \dot{x} x' - z^2 \dot{\phi} \phi' \\
        f \dot{t} t' - \dot{x} x' - z^2 \dot{\phi} \phi' &  \frac{1}{f} + x'{}^2 + z^2 \phi'{}^2 - f t'{}^2
    \end{pmatrix}
    \begin{pmatrix}
        y' \\ \dot{y} 
    \end{pmatrix} \, ,
\end{equation}
and noting that the $2 \times 2$ matrix in the middle of this expression is equal, component by component, to $ \partial_\alpha X^\mu \partial_\beta X^\nu g_{\mu \nu}$, where $X^\mu = (t(s,z), x(s,z),0,0,z,\hat{n}(s,z))$ describes the background solution, with the first component of the matrix (for the indices $\alpha, \beta$) corresponding to a derivative with respect to $s$, and the second with respect to $z$. Therefore, if the background worldsheet is spacelike, it follows that both eigenvalues of the induced metric $ \partial_\alpha X^\mu \partial_\beta X^\nu g_{\mu \nu}$ are positive, and hence, that it is a positive definite matrix. Consequently, the action for the fluctuations~\eqref{eq:fluct-action-imag} is positive definite whenever the background worldsheet is spacelike.

Then, extending the boundary contour as $\T \to \infty$, the contributions of these regions will be of the form (for definiteness, consider $\tau = \T/2 \to +\infty$)
\begin{equation}
    i S_{\rm NG}^{(2)}[y]_{\rm spacelike} = \int_0^{(\pi T)^{-1}} \!\!\!\!\!\! \diff z \, \vec{y}\, {}^T(\tau = +\infty,z)  \cdot \Sigma(z,z') \cdot \vec{y}(\tau = +\infty,z)
\end{equation}
for some positive definite quadratic form $\Sigma$ (note that the minus sign in the definition~\eqref{eq:quadratic-action-general} means that the Gaussian integral over the fluctuations $y$ is convergent)\@. The positive definiteness of $\Sigma$ is guaranteed by the fact that it is determined by the induced metric of a spacelike surface, as discussed below Eq.~\eqref{eq:fluct-action-imag}\@. The same is true for the region at $\tau = - \infty$\@. The net effect of this on the action, after decomposing $y$ in terms of the mode functions at intermediate times, is to add an $i\epsilon$ to the $\omega^2$ coming from the temporal derivatives in the action, with $\epsilon > 0$\@.\footnote{This is similar to the derivation of how the $i\epsilon$ appears in the free Feynman propagator in a quantum field theory (see, e.g.,~\cite{Weinberg:1995mt})\@.} This $i\epsilon$ modifies the mode equations by effectively shifting $\omega^2 \to \omega^2(1 + i\epsilon)$, in the same way as the modification induced by the Schwinger-Keldysh contour tilts on the complex plane. As such, the prescription is unambiguously determined.

As a side note, we remark that the above discussion did not address how the solutions on the different sides of the turnaround region are coupled to each other. Continuity of the fluctuations and of their (appropriately normalized) derivatives is a first requirement, but, as hinted from our previous discussions, closer inspection from the field theory perspective reveals that the solutions must actually be more constrained than that. To see this, if instead of considering perturbations as given by Eq.~\eqref{eq:f-antisymm}, we consider
\begin{align} \label{eq:f-symm}
f^\mu(s) = \begin{cases} g^\mu(s) & -\frac{\T}2 < s < \frac{\T}2 \\  g^\mu (\T - s) &  \frac{\T}2 < s < \frac{3\T}2  \end{cases} \, ,
\end{align}
we then obtain $W[\mathcal{C}_{f}] = 1$ for all $f$, since the loop consists of a single line that was traveled back and forth on top of each other. Hence, taking derivatives with respect to this kind of contour deformations gives zero, and as such, the corresponding response kernel for Wilson loop variations evaluated with AdS/CFT techniques must also be identically zero, whenever the loop satisfies $W[\mathcal{C}_f] = 1$ to begin with. We stress that this is the case for the loop with $\hat{n}$ at antipodal positions on the $S_5$, but it will not necessarily be the case when $\hat{n}$ is constant (even though, as we will see later, deformations as in Eq.~\eqref{eq:f-symm} do not contribute in this case as well)\@. That being said, since $W[\mathcal{C}_f] = 1$ is a property of the Wilson loop~\eqref{eq:W-loop}, the correlator we are after is unequivocally determined by the antisymmetric deformations, as presented in Eq.~\eqref{eq:f-antisymm}\@.


\section{The Wilson loop with constant \texorpdfstring{$S_5$} {} coordinate} \label{sec:HQ-setup}

As we discussed in Section~\ref{sec:nhat}, the standard choice to do calculations of Wilson loops using the AdS/CFT correspondence in strongly coupled $\mathcal{N}=4$ SYM is to set a constant value for $\hat{n}$ throughout the Wilson loop. This is indeed the setup used in the celebrated paper by Maldacena~\cite{Maldacena:1998im} to calculate the heavy quark interaction potential at strong coupling. Since our interest is to describe the dynamics of a heavy quark-antiquark pair close together, we find this is a natural starting point that warrants exploration, regardless of our previous observation that this choice for $\hat{n}$ does not preserve all properties we expect from the Yang-Mills Wilson loop. We will consider the situation where the $\hat{n}$ coordinates are at antipodal points on the $S_5$ for the heavy quark and the heavy antiquark respectively in Section~\ref{sec:QQ-setup}\@.

The calculation consists of three steps. First, in Section~\ref{sec:HQ-bkg} we will discuss the ``background'' worldsheet solution that hangs from the unperturbed Wilson loop (i.e., without the deformations that give rise to the field strength insertions), thus establishing the geometry on which fluctuations can propagate. Secondly, in Section~\ref{sec:HQ-fluct} we will discuss the action for the perturbations on top of this background solution, and how to extract the correlation function of interest for the specific geometry we describe in the first step. Finally, in Section~\ref{sec:EE-calculation-HQ} we will calculate the correlation function as prescribed by the previous steps. We will provide more details for fluctuations that are transverse to the worldsheet, and discuss longitudinal fluctuations (to be defined in what follows) in a more succinct way. We will also check our results by a numerical calculation of the background extremal surface and the correlation function in Euclidean signature.

\subsection{Background} \label{sec:HQ-bkg}

The heavy quark interaction potential can be extracted from a rectangular Wilson loop of temporal extent $\T$ and spatial separation $L$, with $\T \gg L$\@. Its calculation using supersymmetric Wilson loops has been discussed many times in the literature.
The original papers discussed this at zero temperature~\cite{Maldacena:1998im,Rey:1998ik}\@. More general setups were later discussed including finite temperature effects and a relative velocity between the heavy-quark pair and the medium, e.g.~\cite{Liu:2006ug,Peeters:2006iu,Chernicoff:2006hi,Avramis:2006em,Liu:2006nn}\@. The same loop has also been considered in AdS/QCD to calculate the characteristic correlation lengths of field strength correlators~\cite{Andreev:2010zg} in the limit $L \gg \T$. In what follows, we review the main features of the extremal surface that appears in the AdS/CFT calculation of the static heavy-quark potential in $\mathcal{N} = 4$ SYM. Our goal is to study it in the limit $L\to 0$, where the two parallel Wilson lines that construct the timelike segments of the loop get pulled close together.
Because the solution for this Wilson loop has been well-studied in the literature, we will only briefly review the results and highlight their most important features for our purposes.

\begin{figure}
    \centering
    \includegraphics[width=0.7\textwidth]{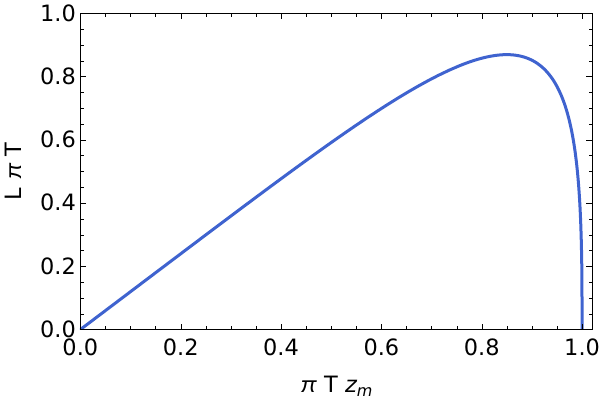}
    \caption{Solutions to Eq.~\eqref{eq:zm-solutions} in the $z_m$-$L$ plane. For each $L$ below a threshold value, there are two solutions $z_m(L)$\@.}
    \label{fig:zm-solutions}
\end{figure}

In the presence of a black hole described through the metric~\eqref{eq:Schwarzschild-AdS}, the minimal area worldsheet configuration that hangs from a rectangular contour of size $\T \times L$ on the boundary, with $\T \gg L$, can be parametrized by
\begin{align}
X^\mu(\tau,\sigma) = (\tau, \sigma, 0 , 0, z(\sigma), \hat{n}_0) \, ,
\end{align}
where $\sigma \in [0,L]$ and $\tau \in [-\T, \T]$\@. For such a parametrization, the Nambu-Goto action reads
\begin{equation}
    \mathcal{S}_{\rm NG} [\Sigma] = - \frac{\sqrt{\lambda}}{2\pi} \int d\tau \, d\sigma \frac{\sqrt{f + z'{}^2}}{z^2} \, .
\end{equation}
Because the action does not depend on $\sigma$ explicitly, there is a conserved quantity, which is the analog of the Hamiltonian $\mathcal{H}$ in standard classical mechanics, with $\mathcal{H} = p \dot{q} - \mathcal{L}$ and $p = \partial \ml{L}/\partial \dot{q}$\@. Using this conserved quantity, one finds that the background worldsheet satisfies
\begin{align} \label{eq:z-eom}
    \sqrt{z'{}^2 + f} = \frac{z_m^2 f }{z^2 \sqrt{f_m}} \iff z' = \pm \sqrt{\frac{f}{f_m}} \sqrt{\frac{z_m^4 - z^4}{z^4}} \, ,
\end{align}
where we have denoted $f_m = f(z_m)$\@. This equation can be integrated to find an implicit solution for $z(\sigma)$, which is given by
\begin{align} \label{eq:implicit-wx}
    1 - \frac{z^3(\sigma)}{z_{m}^3} \frac{\Gamma(5/4) F_1 \! \left( \frac{3}{4}, \frac12, \frac12, \frac74, \frac{z^4(\sigma)}{z_{m}^4}, \pi^4 T^4 z^4(\sigma) \right) }{\sqrt{\pi} \, \Gamma(7/4) {}_2 F_1 \! \left( \frac12, \frac34, \frac54, \pi^4 T^4 z_{m}^4  \right) } = 2\left| \frac{\sigma}{L} - \frac12 \right|,
\end{align}
where $z_m$ is the maximum value of the radial AdS coordinate $z(\sigma)$ and $F_1$ is the Appell hypergeometric function. It is in turn given by
\begin{align} \label{eq:zm-solutions}
    2 \pi T z_{m} \sqrt{1 - \pi^4 T^4 z_{m}^4} \frac{\sqrt{\pi} \, \Gamma(7/4) }{3 \, \Gamma(5/4)} {}_2 F_1 \! \left[ \frac12, \frac34, \frac54, \pi^4 T^4 z_{m}^4 \right] = \pi T L \, .
\end{align}
This equation has two solutions for any given value of $\pi T L < \pi T L_{\rm max} \approx 0.86912$, corresponding to a value of $z_m$ given by $z_m^{\rm crit} \approx 0.84978$\@. This is depicted in Fig.~\ref{fig:zm-solutions}\@. As discussed in Refs.~\cite{Friess:2006rk,Avramis:2006nv}, the solutions with $z_m > z_m^{\rm crit}$ are unstable, and beyond this value the preferred configuration is that of two disconnected, radially infalling surfaces from two parallel Wilson lines. Since we will be interested in the $L \to 0$ limit, namely, $L \pi T \ll 1$, the solution we have to consider is always in the branch with $z_m < z_m^{\rm crit}$\@.

The energy of this configuration\footnote{Because the energy of a configuration is dynamically reflected on the time evolution factor as $e^{-iE (2\T) }$, the energy associated to a Wilson loop configuration in the AdS language after subtracting the mass of the heavy quarks is given by $E = -(\mathcal{S}_{\rm NG}[\Sigma] - \mathcal{S}_0[\mathcal{C},\hat{n}] )/(2\T)$\@.} is given by
\begin{align}
E^c(L) &= - \frac{\sqrt{2\pi \lambda}}{z_m(L) \Gamma(1/4)^2 } (1 - (\pi T z_m(L))^4 ) {}_2 F_1 \! \left[ \frac12, \frac34, \frac14, (\pi T z_{m}(L))^4 \right] + \sqrt{\lambda} T \nonumber \\
&= - \frac{4\pi^2}{\Gamma(1/4)^4} \frac{\sqrt{\lambda}}{L} + \sqrt{\lambda} T + \ml{O}((\pi T L)^3) \, .
\end{align}
This is a Coulomb-like potential, which diverges in the short-distance limit $L \to 0$\@. The constant term proportional to $T$ comes from subtracting the area of the disconnected worldsheet that hangs only down to the horizon $z = (\pi T)^{-1}$, instead of all the way to $z \to \infty$\@. This term $\sqrt{\lambda}T$ corresponds to twice the thermal correction to heavy quark mass. The above equation means that, in the absence of another extra normalization factor in the RHS of Eq.~\eqref{eq:EE-corr-from-variations}, the result of taking the limit $L \to 0$ will be ill-defined. As such, to extract a finite correlation function from here we must also divide by the expectation value of the unperturbed Wilson loop, $\langle W[\mathcal{C}] \rangle_T = \exp ( - i \T E^c(L) )$, as anticipated in Eq.~\eqref{eq:EE-from-Delta}\@.

With the background solution in hand, we can now consider perturbations on top of it. By evaluating the second derivative of the action with respect to the perturbations, we can extract the non-Abelian electric field correlation function we seek.

\subsection{Fluctuations} \label{sec:HQ-fluct}

We now describe the dynamics induced on the worldsheet by small deformations on the contour that  bounds it. Following the discussion we presented in Section~\ref{sec:EE-setup-AdS}, one arrives at the conclusion that this is achieved by introducing small fluctuation fields on the string, which capture how the boundary perturbations propagate into the string. These fields obey second-order partial differential equations that are determined from the Nambu-Goto action. Once one solves the corresponding differential equations, one has to evaluate the Nambu-Goto action ``on-shell,'' i.e., on the solution to the equations of motion, as a function of a general boundary condition $h^i(t)$\@. Then, by taking functional derivatives with respect to $h^i(t)$ of the on-shell action, one can extract the correlation function we are interested in. Because of notational clarity, we will give the results in terms of the kernel $\Delta_{ij}$, which will be different for this configuration than that for the configuration in Section~\ref{sec:QQ-setup}\@. We will denote this section's expression for this kernel by $\Delta_{ij}^c$, where the ``c'' may stand for ``connected'' or ``Coulomb,'' in reference to the shape of the background worldsheet and to the nature of the interaction potential, respectively. We will denote the solution of the next section by $\Delta_{ij}^d$ with ``d'' standing for ``disconnected''\@. Only after we have both of them at hand will we use Eq.~\eqref{eq:EE-Delta} to relate our answer to the non-Abelian electric field correlator of interest.

To evaluate $\Delta_{ij}^c$, the first step to take is introduce perturbations along all of the AdS${}_5$ coordinates, and consider a string parametrized by
\begin{align}
X^\mu(\tau,\sigma) = (\tau + y_0(\tau,\sigma), \sigma + y_1(\tau,\sigma), y_2(\tau,\sigma) , y_3(\tau,\sigma) , z(\sigma) + y_4(\tau,\sigma), \hat{n}) \, .
\end{align}
We do not consider fluctuations on the $S_5$ coordinates because they are decoupled from the rest at the quadratic level in the Nambu-Goto action, which is all we need to evaluate our correlator. However, this parametrization has redundancies in it, because fluctuations that lie on the tangent space to the worldsheet are not physical deformations, but rather a coordinate reparametrization. This means we can choose our worldsheet coordinates such that we can set $y_0(\tau,\sigma) = 0$, as well as set to zero a certain linear combination of $y_1$ and $y_4$\@. To find it, we need to project $y_1$ and $y_4$ along the directions parallel and perpendicular to the worldsheet. Let $\delta(\tau, \sigma)$ parametrize the fluctuations orthogonal to the worldsheet, and $r(\tau,\sigma)$ describe reparametrizations along the worldsheet. Projecting along the parallel and orthogonal directions to the tangent vector of the worldsheet by means of the metric $g_{\mu \nu}$, one finds that the parallel and perpendicular fluctuations are parametrized by
\begin{align}
    \delta_{\parallel} X^\mu(\tau,\sigma) &= (y_0(\tau,\sigma) ,  r(\tau,\sigma), 0, 0, z'(\sigma) r(\tau,\sigma), 0) \\
    \delta_{\perp} X^\mu(\tau,\sigma) &= (0 ,   \frac{z'(\sigma) \delta(\tau, \sigma) }{\sqrt{z'(\sigma)^2 + f(z(\sigma)) } } , y_2(\tau,\sigma) , y_3(\tau,\sigma) , -  \frac{f(z(\sigma)) \delta(\tau, \sigma) }{\sqrt{z'(\sigma)^2 + f(z(\sigma))} }, 0) \, .
\end{align}
As long as we are in the linear response regime, it is a straightforward exercise to show that the Nambu-Goto action only depends on $\delta_\perp X^\mu$, with no dependence on $\delta_\parallel X^\mu$ after using the background equations of motion.\footnote{Of course, the reparametrization invariance of the worldsheet means that there are degrees of freedom absent in the action beyond the linear response regime, but to determine them one would require more information than just the tangent vector to the surface.}
Therefore, we can describe the perturbed worldsheet by
\begin{align} \label{eq:fluct-parametrization}
X^\mu(\tau,\sigma) = (\tau, \sigma + \frac{z'(\sigma) \delta(\tau, \sigma) }{\sqrt{z'(\sigma)^2 + f(z(\sigma)) } } , y_2(\tau,\sigma) , y_3(\tau,\sigma) , z(\sigma) -  \frac{f(z(\sigma)) \delta(\tau, \sigma) }{\sqrt{z'(\sigma)^2 + f(z(\sigma))} }, \hat{n}) \, .
\end{align}
This will be accurate as long as the fluctuations can be treated perturbatively, which is indeed the case of interest because we only need to evaluate the derivative of the action about the background configuration up to quadratic order in the fluctuations, which means that the corresponding equations of motion we will have to solve are linear.

The next step is to write down the action up to quadratic order and derive the equations of motion for the perturbations. After using the background equations of motion and the boundary conditions, one finds that the linear terms in the fluctuations vanish, and it is then straightforward to show Eq.~\eqref{eq:NG-action} becomes
\begin{align}
    \mathcal{S}_{\rm NG}[\Sigma] = - \frac{\sqrt{\lambda}}{2\pi} \left( S_{\rm NG,c}^{(0)}[z] + S_{\rm NG,c}^{(2),\parallel}[\delta] + S_{\rm NG,c}^{(2),\perp}[y_2] + S_{\rm NG,c}^{(2),\perp}[y_3]  \right) \, ,
\end{align}
where each term is given by
\begin{align}
    S^{(0)}_{\rm NG,c}[z] &= \T \int_{0}^L \!\! d\sigma \frac{\sqrt{z'{}^2 + f} }{z^2} \, , \\
    S^{(2),\parallel}_{\rm NG,c}[\delta] &= \int_{-\T/2}^{\T/2} \!\! d\tau \! \int_{0}^L \!\! d\sigma \left[ \frac{f }{2 z^2 \sqrt{z'{}^2 + f} } \delta'{}^2 - \frac{ \sqrt{z'{}^2 + f}}{2 z^2 f} \dot{\delta}{}^2 \right. \nonumber \\ & \quad \quad \quad \quad \quad \quad \quad \left. + \frac{2 z z' f (\pi T z)^4}{z^4 (z'{}^2 + f )^{3/2}} \delta \delta' -  \frac{ f  ( z'{}^2 f + 1 + 5 (\pi T z)^4  )  }{ z^4 (z'{}^2 + f )^{3/2} } \delta^2 \right] \, , \label{eq:action-delta} \\
    S^{(2),\perp}_{\rm NG,c}[y] &= \int_{-\T/2}^{\T/2} \!\! d\tau \! \int_{0}^L \!\! d\sigma \left[ \frac{f}{2 z^2 \sqrt{z'{}^2 + f}} y'{}^2 - \frac{ \sqrt{z'{}^2 + f}}{2 z^2 f} \dot{y}{}^2 \right] \, . \label{eq:action-y}
\end{align}
In these equations, $\parallel$ and $\perp$ should be distinguished from the meanings of being tangent and perpendicular to the background worldsheet. Rather, they indicate whether the perturbations on the boundary ($z=0$) are in the same plane as the Wilson loop or perpendicular to it.

From the action $S^{(2)}_{{\rm NG},c}$, one can derive the equations of motion for the fluctuations. 
We want to emphasize that if we had kept the redundant fluctuations $y_0(\tau,\sigma), r(\tau,\sigma)$, we would have obtained an action containing them up to quadratic order. However, upon calculating their equations of motion, one finds that they are trivial (they vanish identically), and also do not enter the equations of motion for the rest of the fluctuations up to the linear response level.

Before proceeding to the calculation of the kernel $\Delta_{ij}^c$, we note that from here we can already give formal expressions for the on-shell action at quadratic order in the perturbations. As one can always do for a quadratic action of dynamical variables and their first derivatives, we can integrate the action density by parts to obtain the equation of motion plus a total derivative, which reduces to a boundary term. Using Eq.~\eqref{eq:z-eom} and considering nonzero boundary conditions at the timelike segments of the Wilson loop only, we find (in the limit $\T \to \infty$)
\begin{align}
    S^{(2),\parallel}_{\rm NG,c}[\delta]_{\rm on-shell} &= \frac{\sqrt{f_m}}{2 z_m^2} \int_{-\infty}^\infty \!\! d\tau \left[ \delta'(\tau,\sigma = L) \delta(\tau,\sigma = L) - \delta'(\tau,\sigma = 0) \delta(\tau,\sigma = 0) \right] \, , \\
    S^{(2),\perp}_{\rm NG,c}[y]_{\rm on-shell} &= \frac{\sqrt{f_m}}{2 z_m^2} \int_{-\infty}^\infty \!\! d\tau \left[ y'(\tau,\sigma = L) y(\tau,\sigma = L) - y'(\tau,\sigma = 0) y(\tau,\sigma = 0) \right] \, ,
\end{align}
which conveniently are of the same form. The relative simplification of these expressions is due to the fact that all of the coefficients of $\delta$ and $\delta'$ are evaluated at the boundary $z=0$\@. 

The remaining task is to find the derivatives $\delta'$ and $y'$ that solve the equations of motion derived from Eqs.~\eqref{eq:action-delta} and~\eqref{eq:action-y}, in terms of general boundary conditions on the timelike segments of the Wilson loop, of the form implied by Eq.~\eqref{eq:f-antisymm}\@. Concretely, we seek $\delta'$ and $y'$ whose boundary conditions at the timelike segments of the Wilson loop are given by
\begin{align}
    \delta(\tau,\sigma=L) &= \delta(\tau,\sigma=0) = h^\parallel(\tau) \, , \\
    y(\tau,\sigma=L) &= - y(\tau, \sigma = 0) = h^\perp(\tau) \, .
\end{align}
There is no sign flip in the boundary condition for $\delta$ relative to that of $y$ because the parametrization~\eqref{eq:fluct-parametrization} already takes it into account.

Because the corresponding equations of motion are linear, it is possible to write down the derivative of the solutions at the boundaries in terms of linear response kernels $K^c_\parallel(\tau,\tau')$, $K^c_\perp(\tau,\tau')$\@. Because of how we have parametrized the longitudinal fluctuations, $\delta$ will be an even function of $\sigma$ around $\sigma = L/2$, and $y$ will be odd. Then, we can write
\begin{align}
    \delta'(\tau,\sigma=L) = - \delta'(\tau,\sigma=0) &= \int_{-\infty}^\infty \!\! d\tau' K^c_\parallel(\tau,\tau';L) h^\parallel(\tau') \, , \\
    y'(\tau,\sigma=L) = y'(\tau,\sigma=0) &= \int_{-\infty}^\infty \!\! d\tau' K^c_\perp(\tau,\tau';L) h^\perp(\tau') \, ,
\end{align}
with which the on-shell actions can be written as
\begin{align}
    S^{(2),\parallel}_{\rm NG,c}[h]_{\rm on-shell} &= \frac{\sqrt{f_m}}{z_m^2} \int_{-\infty}^\infty \!\! d\tau \,d\tau' h^\parallel(\tau) K^c_\parallel(\tau,\tau';L) h^\parallel(\tau') \, , \\
    S^{(2),\perp}_{\rm NG,c}[h]_{\rm on-shell} &= \frac{\sqrt{f_m}}{ z_m^2} \int_{-\infty}^\infty \!\! d\tau \,d\tau' h^\perp(\tau) K^c_\perp(\tau,\tau';L) h^\perp(\tau') \, .
\end{align}
Because of the time translational symmetry in the limit $\T \to \infty$, we also have $K^c_\parallel(\tau,\tau';L) = K^c_\parallel(\tau-\tau';L)$, $K^c_\perp(\tau,\tau';L) = K^c_\perp(\tau-\tau';L)$\@. From here, it is clear that by evaluating $K^c_\parallel, K^c_\perp$ we will have all the information we need to evaluate the contribution of each type of fluctuations to $\Delta_{ij}^c$:
\begin{align}
 \Delta^c_{ij}(t_2 - t_1;L) =  \frac{\sqrt{\lambda}}{\pi}  \frac{\sqrt{f_m}}{z_m^2} \left[ \delta_{i1} \delta_{1j} K^c_\parallel(t_2-t_1;L) + (\delta_{i2} \delta_{2j} + \delta_{i3} \delta_{3j}) K^c_\perp(t_2 - t_1;L) \right] \, . \label{eq:Delta-in-terms-of-K}
\end{align}

Finally, to evaluate all the different response kernels $K^c(\tau-\tau')$, because of the time translational invariance, it is most helpful to introduce their Fourier transforms, which for the fluctuations are given by
\begin{align}
     \delta_\omega(\sigma) &= \int_{-\infty}^\infty d\tau \,e^{i\omega \tau} \delta(\tau,\sigma)\, , \\
    y_\omega(\sigma) &= \int_{-\infty}^\infty d\tau \,e^{i\omega \tau} y(\tau,\sigma) \, ,
\end{align}
and for the response kernels by
\begin{align}
     K_{\parallel}^c(\omega, \omega') &= \int_{-\infty}^\infty d\tau \, d\tau' \, e^{i \omega \tau - i \omega' \tau'} K_{\parallel}^c(\tau-\tau') \equiv (2\pi) \delta(\omega - \omega') K_{\parallel}^c(\omega)\, , \\
     K_{\perp}^c(\omega, \omega') &= \int_{-\infty}^\infty d\tau \, d\tau' \, e^{i \omega \tau - i \omega' \tau'} K_{\perp}^c(\tau-\tau') \equiv (2\pi) \delta(\omega - \omega') K_{\perp}^c(\omega)\, ,
\end{align}
where $K_{\perp/\parallel}^c(\omega) = \int_{-\infty}^\infty d\tau e^{i\omega \tau} K_{\perp/\parallel}^c(\tau)$ is the Fourier representation of the response kernel for either type of perturbation. With this, we can simply write down 
\begin{align}
      \delta_\omega'(\sigma = L) &= K_\parallel^c(\omega) \delta_\omega(\sigma=L)\, , \\ 
    y_\omega'(\sigma = L) &= K_\perp^c(\omega) y_\omega(\sigma=L)\, ,
\end{align}
and the problem is reduced to finding the respective response function $K_{\perp/\parallel}^c(\omega)$ at each frequency.

\subsection{Calculation of the time-ordered non-Abelian electric field correlator} \label{sec:EE-calculation-HQ}

After setting up all of the machinery, we now describe the calculation of the response kernels for fluctuations in the configuration where the two timelike segments of the SYM Wilson loop have the same $S_5$ coordinates. We proceed with a greater level of detail for transverse fluctuations, which is arguably the simpler case, in the hope that it will make the longitudinal discussion less cumbersome. We also provide an increased level of detail in the hope that future calculations of fluctuations on top of extremal worldsheets to extract correlation functions from holography may benefit from this discussion.

Furthermore, in anticipation of obtaining results that might require careful regularization, we will also carry out the calculation allowing for more flexibility in the fluctuations than using a single perturbation $h^i$\@. Specifically, we will set boundary conditions on the two timelike segments of the contour $\mathcal{C}$ independently. The purpose of this will be to verify that the contributions to the response kernel proportional to a Dirac delta function in time are not due to (omitted) contact terms in the RHS of Eq.~\eqref{eq:E-insertion-2}\@.

\subsubsection{Transverse fluctuations}  \label{sec:transverse-v-fluct}

As we just discussed, our goal now is to solve for $y_\omega(\sigma)$ and extract its derivatives on the boundary. 
Varying $S_{\rm NG,c}^{(2),\perp}$ with respect to $y$ and transforming to the frequency domain, the equation we have to solve is
\begin{equation}
\frac{\partial^2 y_\omega}{\partial \sigma^2}(\sigma) + \frac{z_m^4}{z^4(\sigma)} \frac{\omega^2}{f_m} y_\omega(\sigma) = 0 \, ,
\end{equation}
where $z(\sigma)$ is determined by solving
\begin{equation}
\frac{f}{z^2 \sqrt{f + z'^2}} = \frac{\sqrt{f_{m}}}{z_{m}^2} \iff z' = \pm \sqrt{f} \sqrt{\frac{z_m^4}{z^4} \frac{f}{f_m} -1}
\end{equation}
subject to $z'=0 \iff z = z_m$ and $z(\sigma = L) = z(\sigma = 0)  = 0$\@. In the interval $\sigma \in (0,L/2)$ we take the plus sign (as the worldsheet goes into AdS${}_5$), and the minus sign when $\sigma \in (L/2,L)$\@. The most useful form of the above expression is
\begin{equation} \label{eq:eom-z}
z' = \pm \sqrt{ \frac{f}{f_m} \frac{z_m^4 - z^4}{z^4} } \, .
\end{equation}

To avoid introducing unnecessary numerical uncertainties, the best alternative is to transform the equation for $y_\omega(\sigma)$ into an equation for $y_\omega(z)$, because then we will not need to solve for $z(\sigma)$ explicitly.\footnote{It is actually possible to do this, but because we are able to perform the change of variables, the explicit form of the solution $z(\sigma)$ becomes unimportant.} Performing the transformation, we have
\begin{align}
&\sqrt{ \frac{f}{f_m} \frac{z_m^4 - z^4}{z^4} } \frac{\partial}{\partial z} \left( \sqrt{ \frac{f}{f_m} \frac{z_m^4 - z^4}{z^4} } \frac{\partial y_\omega}{\partial z} \right) + \frac{z_m^4}{z^4} \frac{\omega^2}{f_m} y_\omega = 0 \, , \nonumber \\
\implies & f\frac{z_m^4 - z^4}{z^4} \frac{\partial^2 y_\omega}{\partial z^2} - \frac{2[ (z_m^4-z^4) + z^4 f ] }{z^5}  \frac{\partial y_\omega}{\partial z} + \frac{z_m^4}{z^4} \omega^2 y_\omega = 0 \, , \nonumber \\
\implies & \frac{\partial^2 y_\omega}{\partial z^2} - \frac{2}{z} \left[ \frac{1}{f} + \frac{z^4}{z_m^4 - z^4} \right]  \frac{\partial y_\omega}{\partial z} + \frac{\omega^2 z_m^4}{(z_m^4 - z^4) f } y_\omega = 0 \, .
\end{align}

At this point, it is useful to introduce a rescaling of the radial AdS coordinate: $\xi = z/z_m \in (0,1)$\@. In terms of this variable, we have
\begin{equation} \label{eq:eom-y}
 \frac{\partial^2 y_\omega}{\partial \xi^2} - \frac{2}{\xi} \left[ \frac{1}{1 - (\pi T z_m)^4 \xi^4 } + \frac{\xi^4}{1 - \xi^4} \right]  \frac{\partial y_\omega}{\partial \xi} + \frac{\omega^2 z_m^2}{(1 - \xi^4) (1 - (\pi T z_m)^4 \xi^4) } y_\omega = 0 \, .
\end{equation}

The same equation applies for both copies of the transformed intervals $\sigma \in (0,L/2)$ and $\sigma \in (L/2,L)$\@. Let us denote the corresponding solutions as a function of $z$ by $y_\omega^L(z)$ and $y_\omega^R(z)$, respectively, where $L,R$ stand for ``left'' and ``right'' sides of the worldsheet. All that we need to specify in order to close the system are the boundary conditions. In terms of the original coordinate $\sigma$, we have
\begin{align}
y_\omega(\sigma = [L/2]^-) = y_\omega(\sigma = [L/2]^+) \, , & & \frac{\partial y_\omega}{\partial \sigma}(\sigma = [L/2]^-) = \frac{\partial y_\omega}{\partial \sigma}(\sigma = [L/2]^+)  \, ,
\end{align}
which, in terms of the $z$ coordinate, transform to
\begin{align}
\lim_{z \to z_m} y_\omega^L(z) = \lim_{z \to z_m} y_\omega^R(z) \, , & & \lim_{z \to z_m}\sqrt{\frac{f}{f_m} \frac{z_m^4 -z^4}{z^4} }  \frac{\partial y_\omega^L}{\partial z} = - \lim_{z \to z_m} \sqrt{\frac{f}{f_m} \frac{z_m^4 -z^4}{z^4} } \frac{\partial y_\omega^R}{\partial z} \, ,
\end{align}
or, equivalently,
\begin{align}
\lim_{\xi \to 1} y_\omega^L(\xi) = \lim_{\xi \to 1} y_\omega^R(\xi) \, , & & \lim_{\xi \to 1} \sqrt{1 - \xi^4} \frac{\partial y_\omega^L}{\partial \xi} = - \lim_{\xi \to 1} \sqrt{1 - \xi^4} \frac{\partial y_\omega^R}{\partial \xi} \, .
\end{align}

To implement these matching conditions, the best way is to do a WKB-type analysis to extract the leading/possibly singular behavior of the solution near the horizon. We can do that analytically by writing
\begin{align}
y_\omega^L(\xi) &= A_L \exp \left( i  \int_0^\xi \frac{\omega z_m \xi'^3 d\xi'}{\sqrt{(1-\xi'^4)(1-(\pi T z_m \xi')^4)}} \right) F^{-}_\omega(\xi) \nonumber \\ & \quad + B_L \exp \left( -i  \int_0^\xi \frac{\omega z_m \xi'^3 d\xi'}{\sqrt{(1-\xi'^4)(1-(\pi T z_m \xi')^4)}} \right) F^{+}_\omega(\xi) \nonumber \\
&\equiv A_L \, y^-_\omega(\xi) + B_L \, y^+_\omega(\xi) \, , \label{eq:yL} \\
y_\omega^R(\xi) &= A_R \exp \left( i  \int_0^\xi \frac{\omega z_m \xi'^3 d\xi'}{\sqrt{(1-\xi'^4)(1-(\pi T z_m \xi')^4)}} \right) F^{-}_\omega(\xi) \nonumber \\ & \quad + B_R \exp \left( -i  \int_0^\xi \frac{\omega z_m \xi'^3 d\xi'}{\sqrt{(1-\xi'^4)(1-(\pi T z_m \xi')^4)}} \right) F^{+}_\omega(\xi) \nonumber \\
&\equiv A_R \, y^-_\omega(\xi) + B_R \, y^+_\omega(\xi) \, , \label{eq:yR}
\end{align}
where $\partial F^{\pm}_\omega/\partial \xi$ is finite at the turning point $\xi = 1$\@. With this decomposition, the matching conditions translate into
\begin{align}
A_L \, y_\omega^-(\xi=1) + B_L \, y_\omega^+(\xi=1) &= A_R \, y_\omega^-(\xi=1) + B_R \, y_\omega^+(\xi=1) \, , \\
i A_L \, y_\omega^-(\xi=1) - i B_L \, y_\omega^+(\xi=1) &= -i A_R \, y_\omega^-(\xi=1) + i B_R \, y_\omega^+(\xi=1) \, .
\end{align}

To solve the system in terms of one of the amplitudes, we need to specify the boundary conditions. To extract the correlation function we are interested in, the natural choice is to prescribe $y^L_\omega(\xi = 0) = - y^R_\omega(\xi = 0)$\@. However, in order to illustrate the nature of contact divergences that will appear in this calculation, we will instead consider the boundary condition $y^R_\omega(\xi = 0) = 0$\@. We can obtain the boundary condition that defines our correlation function (i.e., $y^L_\omega(\xi = 0) = - y^R_\omega(\xi = 0)$) by taking linear superpositions of the boundary condition $y^{L/R}_\omega(\xi = 0) = 0$ and using that the equations we are looking at are symmetric under the exchange of the $L,R$ labels.  With this, it is appropriate to define the response functions $K^{AB}(\omega)$ as the derivative responses $y_\omega'(\sigma)$ on side $A$ due to a unit perturbation on side $B$\@.

Then, the null boundary condition $y^R_\omega(\xi = 0) = 0$ at $\sigma = L$ translates into
\begin{equation} \label{eq:just-after-KAB}
A_R \, y_\omega^-(\xi=0) + B_R \, y_{\omega}^+(\xi=0) = 0 \, ,
\end{equation}
which, together with the matching conditions at $\xi=1$, fully determine the solution up to an overall constant:
\begin{align}
A_L &= - \frac{y^+_\omega(\xi=1)}{y_\omega^+(\xi=0)} \frac{y^-_\omega(\xi=0)}{y^-_\omega(\xi=1)} A_R \, , \\
B_R &= - \frac{y^-_\omega(\xi=0)}{y^+_\omega(\xi=0)} A_R \, , \\
B_L &= \frac{y^-_\omega(\xi=1)}{y^+_\omega(\xi=1)} A_R \, .
\end{align}

Formally, all that remains is to evaluate the response kernels $K^{RL,c}_{\perp}$ and $K^{LL,c}_{\perp}$\@. We remind the reader that the superscripts are there to make it explicit that they represent partial contributions to the response kernel we want to evaluate, each coming from specific boundary conditions and response locations. These are determined by
\begin{align}
K^{RL,c}_{\perp}(\omega,L) &= -\frac{1}{y_\omega^L(\xi=0)} \lim_{\xi \to 0} \sqrt{\frac{ 1- (\pi T z_m \xi)^4 }{f_m} \frac{1 - \xi^4}{\xi^4} } \frac{1}{z_m} \frac{\partial y_\omega^R}{\partial \xi} \nonumber \\
&= - \frac{1}{z_m \sqrt{f_m}} \frac{1}{y_\omega^L(\xi=0)} \lim_{\xi \to 0} \frac{1}{\xi^2} \frac{\partial y_\omega^R}{\partial \xi} \, , \label{eq:KRLc-def} \\
K^{LL,c}_{\perp}(\omega,L) &= \frac{1}{y_\omega^L(\xi=0)} \lim_{\xi \to 0} \sqrt{\frac{ 1- (\pi T z_m \xi)^4 }{f_m} \frac{1 - \xi^4}{\xi^4} } \frac{1}{z_m} \frac{\partial y_\omega^L}{\partial \xi} \nonumber \\
&= \frac{1}{z_m \sqrt{f_m}} \frac{1}{y_\omega^L(\xi=0)} \lim_{\xi \to 0} \frac{1}{\xi^2} \frac{\partial y_\omega^L}{\partial \xi} \, . \label{eq:KLLc-def} 
\end{align}
Furthermore, choosing the normalization of the mode functions such that $y^{\pm}_\omega(\xi=0) = 1$, and writing the result in terms of the regular functions $F_\omega^{\pm}(\xi)$ whenever possible, we have the response kernels in each case, before subtractions and regularizations, given by
\begin{align}
K^{RL,c}_{\perp}(\omega,L) &=  \frac{1}{z_m \sqrt{f_m}} \frac{\lim_{\xi\to 0} \frac{1}{\xi^2} \left[ \frac{\partial F_\omega^-}{\partial \xi} - \frac{\partial F_\omega^+}{\partial \xi}  \right] }{- \frac{y^{+}_\omega(\xi=1)}{y^-_\omega(\xi=1)} + \frac{y^{-}_\omega(\xi=1)}{y^+_\omega(\xi=1)}  } \, , \\
K^{LL,c}_{\perp}(\omega,L) &=  \frac{1}{z_m \sqrt{f_m}} \frac{\lim_{\xi\to 0} \frac{1}{\xi^2} \left[ \frac{y^{+}_\omega(\xi=1)}{y^-_\omega(\xi=1)} \frac{\partial F_\omega^-}{\partial \xi} - \frac{y^{-}_\omega(\xi=1)}{y^+_\omega(\xi=1)} \frac{\partial F_\omega^+}{\partial \xi}  \right] }{- \frac{y^{+}_\omega(\xi=1)}{y^-_\omega(\xi=1)} + \frac{y^{-}_\omega(\xi=1)}{y^+_\omega(\xi=1)}  } \, .
\end{align}

Note that all of the above expressions are valid for arbitrary $L >0$, and furthermore, all of the discussion in this section holds for arbitrary, complex $\omega$\@.\footnote{This will be useful to enforce the time-ordering prescription.} No approximations have been made. All that remains is to solve the equation for the modes $y_\omega^\pm(\xi)$, or equivalently $F_\omega^\pm(\xi)$, and with that the above expressions can be calculated explicitly.

It is useful to note that the denominators in the expressions for $K_{\perp}^{AB,c}$ can be written in terms of the Wronskian of the differential equation for $y^\pm_\omega$\@. First we observe that
\begin{equation}
    y^+_\omega \frac{\partial y^-_\omega}{\partial \xi} - y^-_\omega \frac{\partial y^+_\omega}{\partial \xi} = W(\xi) \, , 
\end{equation}
where
\begin{align} \label{eq:Wronsk}
    W(\xi) = C \exp \left( 2 \int^\xi \frac{d\xi}{\xi} \left[ \frac{1}{1 - (\pi T z_m \xi)^4} + \frac{\xi^4}{1-\xi^4}  \right] \right) = \frac{C \xi^2}{\sqrt{(1-\xi^4)(1 - (\pi T z_m \xi)^4)}} \, .
\end{align}
The constant $C$ can be fixed by looking at the behavior of $y^{\pm}_\omega$ when $\xi \to 1$\@. The result is
\begin{equation} \label{eq:Wronsk-C}
    C = 2 i \omega z_m y^+_\omega(\xi=1) y^-_\omega(\xi=1) \, ,
\end{equation}
where we have worked under the normalization $y^\pm_\omega(\xi=0) = 1$.

Finally, one can integrate the equation for the Wronskian to derive that
\begin{align}
    \frac{y^-_\omega(\xi=1)}{y^+_\omega(\xi=1)} &= \exp \left( \int_0^1 d\xi \frac{W(\xi)}{y^+_\omega(\xi) y^-_\omega(\xi)} \right) \nonumber \\
    &= \exp \left( 2 i \omega z_m \int_0^1 d\xi \frac{y^+_\omega(\xi=1) y^-_\omega(\xi=1)}{y^+_\omega(\xi) y^-_\omega(\xi)} \frac{ \xi^2}{\sqrt{(1-\xi^4)(1 - (\pi T z_m \xi)^4)}} \right) \, .
\end{align}
Also, note that by construction we have $y^+_\omega(\xi) y^-_\omega(\xi) = F^+_\omega(\xi) F^-_\omega(\xi)$, and all contributions to the response kernels $K_\perp^{RL/LL,c}$ can be written entirely in terms of the regular functions $F^\pm_\omega$\@.
We then find the denominators in the expressions for $K_\perp$ are given by:
\begin{align}
    &- \frac{y^{+}_\omega(\xi=1)}{y^-_\omega(\xi=1)} + \frac{y^{-}_\omega(\xi=1)}{y^+_\omega(\xi=1)} \nonumber \\ &= 2i \sin \left( 2 \omega z_m \int_0^1 d\xi \frac{F^+_\omega(\xi=1) F^-_\omega(\xi=1)}{F^+_\omega(\xi) F^-_\omega(\xi)} \frac{ \xi^2}{\sqrt{(1-\xi^4)(1 - (\pi T z_m \xi)^4)}} \right) \, .
\end{align}
This is consequential because it provides a clean expression to implement the time-ordering prescription to evaluate the correlator.
Concretely, the time-ordering prescription is implemented by taking $\omega \to \omega (1 + i\epsilon)$\@. It is then convenient to define $\phi_\omega(z_m)$ as
\begin{equation}
    \phi_\omega(z_m) \equiv 2 \omega z_m \int_0^1 d\xi \frac{F^+_\omega(\xi=1) F^-_\omega(\xi=1)}{F^+_\omega(\xi) F^-_\omega(\xi)} \frac{ \xi^2}{\sqrt{(1-\xi^4)(1 - (\pi T z_m \xi)^4)}} \, ,
\end{equation}
with which the $i\epsilon$ prescription implies that we can write\footnote{Strictly speaking, one also has to analyze the mode functions and determine explicitly whether $F_\omega^+ F_\omega^-$ gives another $\ml{O}(\epsilon)$ contribution when introducing the prescription. As it turns out, this contribution adds up with the naive one, giving the same overall effect.}
\begin{equation}
    - \frac{y^{+}_\omega(\xi=1)}{y^-_\omega(\xi=1)} + \frac{y^{-}_\omega(\xi=1)}{y^+_\omega(\xi=1)} = 2i \left[ \sin (\phi_\omega(z_m)) + i\epsilon \phi_\omega(z_m) \cos (\phi_\omega(z_m)) \right] \, .
\end{equation}

Using this, and the fact that the mode functions $F_\omega^{\pm}$ satisfy $\frac{\partial^2 F_\omega^{\pm}}{\partial \xi^2}(\xi = 0) = \omega^2 z_m^2 F^{\pm}_\omega(\xi=0)$, an equality that follows from Eq.~\eqref{eq:eom-y}, we get
\begin{align}
K^{RL,c}_{\perp}(\omega,L) &= -i \frac{1}{2z_m \sqrt{f_m}} \frac{\lim_{\xi\to 0} \frac{1}{\xi^2} \left[ \frac{\partial F_\omega^-}{\partial \xi} - \frac{\partial F_\omega^+}{\partial \xi}  \right] }{\sin (\phi_\omega(z_m)) + i\epsilon \phi_\omega(z_m) \cos (\phi_\omega(z_m))  } \nonumber \\
&=  \frac{1}{z_m \sqrt{f_m} } \frac{ \omega z_m F_\omega^+(\xi=1) F_\omega^-(\xi=1) }{\sin(\phi_\omega(z_m)) + i\epsilon \phi_\omega(z_m) \cos(\phi_\omega(z_m))  } \, , \label{eq:Delta-yy-RL-c} \\
K^{LL,c}_{\perp}(\omega,L) &= -i \frac{1}{2z_m \sqrt{f_m}} \frac{\lim_{\xi\to 0} \frac{1}{\xi^2} \left[ e^{-i\phi_\omega(z_m)} \frac{\partial F_\omega^-}{\partial \xi} - e^{i \phi_\omega(z_m) } \frac{\partial F_\omega^+}{\partial \xi}  \right] }{\sin (\phi_\omega(z_m)) + i\epsilon \phi_\omega(z_m) \cos (\phi_\omega(z_m))  } \nonumber \\ 
&= \frac{1}{z_m \sqrt{f_m}} \frac{ \omega z_m F_\omega^+(\xi=1) F_\omega^-(\xi=1) \cos (\phi_\omega(z_m) ) }{\sin(\phi_\omega(z_m)) + i\epsilon \cos(\phi_\omega(z_m))  } \nonumber \\ & \quad -  \frac{1}{4 z_m \sqrt{f_m}} \left[ \frac{\partial^3 F_\omega^-}{\partial \xi^3} +  \frac{\partial^3 F_\omega^+}{\partial \xi^3} \right]_{\xi=0} -  \frac{\omega^2 z_m^2}{\sqrt{f_m}} \lim_{z \to 0} \frac{1}{z} \, , \label{eq:Delta-yy-LL-c}
\end{align}
where we have used our expression for the Wronskian as given by Eqs.~\eqref{eq:Wronsk} and~\eqref{eq:Wronsk-C}\@. The Wronskian is what allowed us to write the difference of the derivatives of $F^+_\omega$ and $F^-_\omega$ purely in terms of $F^{\pm}_\omega$ with no derivatives.

We clearly see that $K_{\perp}^{RL,c}$, $K_{\perp}^{LL,c}$ are different functions. But this is fine, since we only expect them to be equal in the limit $L \to 0$, and up to contact terms. Indeed, the last term in the expression for $K_{\perp}^{LL,c}$ is a divergent term that is exactly of this form. On the other hand, by construction, $K_{\perp}^{RL,c}$ will feature no such contact term contributions, because the variations of the Wilson loop, in the language of Section~\ref{sec:W-loop-AdS}, are always at different values of the parameter $s$\@. While this means that the $RL$ setup to extract the correlator gives a cleaner signal than the $LL$ kernel, where no subtraction for contact terms is required, we shall still calculate both as a consistency check. In what follows, since we have isolated its origin, we will omit the contact term $\frac{\omega^2 z_m^2}{\sqrt{f_m}} \lim_{z \to 0} \frac{1}{z}$ as it does not enter the definition of the correlator from the variations of the Wilson loop~\eqref{eq:EE-corr-from-variations} at $t_1 \neq t_2$\@.

All that remains now is to evaluate the mode functions and substitute the result into the expressions for the response kernels. The equation for the mode functions $F_\omega^{\pm}$ can be found by explicitly substituting $y^\pm_\omega = \exp(\mp i (\cdots) ) F_\omega^\pm$ into the equation of motion for $y$, given by Eq.~\eqref{eq:eom-y}\@. To optimize the notation, we introduce $h \equiv \pi T z_m$ and $\Omega \equiv \omega/(\pi T)$\@.
The equation for $F^\pm_{\omega}$ then reads
\begin{align} 
    \frac{\partial^2 F^\pm_\omega}{\partial \xi^2} &- 2 \left[ \frac{1 - h^4 \xi^8}{\xi(1-\xi^4)(1-h^4\xi^4)} \pm \frac{i \Omega h \xi^3}{\sqrt{(1-\xi^4)(1-h^4 \xi^4)}} \right] \frac{\partial F^\pm_\omega}{\partial \xi} \nonumber \\ & \quad \quad \quad \quad + \left[ \mp \frac{i \Omega h \xi^2}{\sqrt{(1-\xi^4)(1-h^4 \xi^4)}} + \frac{\Omega^2 h^2 (1 - \xi^6) }{(1-\xi^4)(1-h^4 \xi^4)} \right] F^\pm_\omega = 0 \, . \label{eq:F-transverse-eom}
\end{align}
Now we can proceed to study the behavior of the solutions. The defining condition we have to impose is the regularity of $\partial F_\omega^\pm/\partial \xi$ at the turning point, which, in terms of $\partial^2 F^\pm_\omega / \partial \xi^2$ being finite as $\xi \to 1$ requires $\partial F_\omega^\pm(\xi = 1) / \partial \xi  = 0$, which can be seen by analyzing the most divergent pieces of Eq.~\eqref{eq:F-transverse-eom} when $\xi\to1$\@. The near-boundary behavior of the mode functions also requires
\begin{align}
    \frac{\partial F^\pm_\omega}{\partial \xi}(\xi=0) = 0 \, , & & \frac{\partial^2 F^\pm_\omega}{\partial \xi^2}(\xi=0) = \Omega^2 h^2 F^\pm_\omega(\xi=0) \,,
\end{align}
which can be obtained by expanding Eq.~\eqref{eq:F-transverse-eom} in a power series in $\xi$ near $\xi=0$\@.
While the regularity condition at the midpoint is in principle enough to determine the mode function up to an overall normalization, these near-boundary conditions may also be used in a numerical solution of Eq.~\eqref{eq:F-transverse-eom}\@.

\paragraph{Taking the limit $L \to 0$} \hspace{\fill} \label{sec:App-transverse-connected}

In this section we evaluate the response kernels $K_{\perp}^{AB,c}$ introduced in the paragraph above Eq.~\eqref{eq:just-after-KAB}, in the limit of small $L$\@. The first step is to take the limit $\xi \to 0$, as this is part of the definition of the correlator at any $L$\@. We use the notation $F_\omega^\pm$ with the understanding that its dependence on $\omega$ will often be through $\Omega = \omega/ (\pi T)$\@. We start from the expressions~\eqref{eq:Delta-yy-RL-c} and~\eqref{eq:Delta-yy-LL-c}\@.

The relevant limit corresponds to taking $z_m \to 0$, which is equivalent to taking $L\to 0$ at fixed $T$\@. Furthermore, because we have an explicit factor of $z_m^{-3}$, we have to calculate the series expansion of the rest of the expression up to order $z_m^3$, so that in the end we get a result of the form
\begin{equation}
    \frac{c_3}{L^3} + \frac{c_2}{L^2} + \frac{c_1}{L} + c_0 \, .
\end{equation}

Let us examine this term by term. We first note that we need to evaluate all terms in this expression up to cubic power in $z_m$\@. In particular, we have to evaluate
\begin{equation}
    \phi_\omega(z_m) \equiv 2 \omega z_m \int_0^1 d\xi \frac{F^+_\omega(\xi=1) F^-_\omega(\xi=1)}{F^+_\omega(\xi) F^-_\omega(\xi)} \frac{ \xi^2}{\sqrt{(1-\xi^4)(1 - (\pi T z_m \xi)^4)}}
\end{equation}
at least up to cubic order in $z_m$\@. As with the rest of this expression, this requires to solve for $F^{\pm}_\omega$ up to $\ml{O}(h^3)$, where $h = \pi T z_m$\@. However, by simple inspection, one quickly realizes that the structure of the solution, up to $\mathcal{O}(h^3)$, is of the form
\begin{equation}
    F^-_\omega(\xi) = F^{(0)}_\omega(\xi) + i \Omega h F^{(1)}_\omega(\xi) + (i \Omega h)^2 F^{(2)}_\omega(\xi) + (i \Omega h )^3 F^{(3)}_\omega(\xi) + \mathcal{O}(h^4) \, ,
\end{equation}
where $F^{(i)}_\omega(\xi)$ are real functions of $\xi$\@. This means that
\begin{equation}
    F^-_\omega(\xi) F^+_\omega(\xi) = |F^-_\omega(\xi)|^2 = \left(F_\omega^{(0)}(\xi) \right)^2 + \Omega^2 h^2 \left( \left(F_\omega^{(1)}(\xi) \right)^2 - 2 F_\omega^{(0)}(\xi) F_\omega^{(2)}(\xi) \right) + \mathcal{O}(h^4) \, ,
\end{equation}
which implies that the product $F^-_\omega(\xi) F^+_\omega(\xi)$ is an even function of $h$ (similarly for $F^+_\omega + F^-_\omega$)\@. This in turn implies that the whole object is an odd function of $h$\@. Moreover, since no terms in the power series up to $\mathcal{O}(h^3)$ involve the temperature explicitly, we have that both correlators are of the form
\begin{equation}
    \frac{c_3}{z_m^3} + \frac{c_1 \omega^2}{z_m} \, 
\end{equation}
and because of Eq.~\eqref{eq:zm-solutions}, $z_m = \frac{3 \Gamma(5/4)}{2\sqrt{\pi} \Gamma(7/4) } L + \mathcal{O}(L^5)$, with which
\begin{equation}
    D_{AB}^{yy}(\omega,L\to 0) \approx \frac{\tilde{c}_3}{L^3} + \frac{\tilde{c}_1 \omega^2}{L} \, .
\end{equation}
In terms of the time coordinate, both terms only contribute to the infinitesimal neighborhood of $t-t' = 0$, and therefore we anyways expect a divergence. The leading nontrivial dependence on $\omega/T$ appears at linear order in $L$, which is to say, from the $\mathcal{O}(h^4)$ corrections from each term. 

For future reference, note that $F_{\omega}^{(0)}$, $F_{\omega}^{(1)}$, and $F_{\omega}^{(2)}$ can be determined explicitly:
\begin{align}
    F_{\omega}^{(0)} &= 1 \, ,\\
    F_{\omega}^{(1)} &= \frac{\xi^3}{3} {}_2 F_1 \! \left( \frac12 , \frac34 , \frac74, \xi^4  \right) + \frac{\sqrt{1-\xi^4}-1}2 \, ,\\
    F_{\omega}^{(2)} &= \frac{\xi^7}{21} {}_p F_q \! \left( \left\{ 1, \frac54 \right\}, \left\{ \frac{11}4 \right\} , \xi^4 \right) + \frac{ \Gamma(5/4) \left( -3\xi^4 + 4\xi^3 -12 \xi^2 + 6 - 6\sqrt{1-\xi^4} \right)}{6 \Gamma(1/4)} \nonumber \\ & \quad + \frac{  E({\rm asin}(\xi), -1) - F({\rm asin}(\xi), -1 )  }{2 \Gamma(1/4) \Gamma(5/4) }   \Big[ \sqrt{2 \pi^3} - 4  \Gamma(5/4)^2 \left( 1 +  E({\rm asin}(\xi), -1) \right) \nonumber \\ & \quad \quad\quad \quad\quad \quad\quad \quad\quad \quad\quad \quad \quad \quad\quad \quad\quad \quad\quad \quad\quad  + 4 \Gamma(5/4)^2  F({\rm asin}(\xi), -1 )    \Big] \, ,
\end{align}
where ${\rm asin}(x) = \arcsin(x)$, $E(x,y) = {\rm EllipticE}(x,y)$, and $F(x,y) = {\rm EllipticF}(x,y)$\@.

Let us evaluate these numbers, and the first nontrivial correction from $T$-dependent effects. Let us then organize the calculation in powers of $h$, up to $\mathcal{O}(h^4)$\@. We define
\begin{equation}
    G_\omega(\xi) = |F^-_\omega(\xi)|^2 = G_\omega^{(0)}(\xi) + \Omega^2 h^2 G_\omega^{(2)}(\xi) + \Omega^4 h^4 G_\omega^{(4)}(\xi) + \mathcal{O}(h^6) \, ,
\end{equation}
where
\begin{align}
    G_\omega^{(0)}(\xi) &= \left(F^{(0)}(\xi) \right)^2 = 1 \, , \\
    G_\omega^{(2)}(\xi) &= \left(F_\omega^{(1)}(\xi) \right)^2 - 2 F_\omega^{(0)}(\xi) F_\omega^{(2)}(\xi) \, , \\
    G_\omega^{(4)}(\xi) &= \left(F_\omega^{(2)}(\xi) \right)^2 - 2 F_\omega^{(1)}(\xi) F_\omega^{(3)}(\xi) + 2 F_\omega^{(0)}(\xi) F_\omega^{(4)}(\xi) \, ,
\end{align}
For notational brevity, it is also useful to define
\begin{equation}
    H_\omega(\xi) = \frac{F^+_\omega(\xi=1) F^-_\omega(\xi=1)}{F^+_\omega(\xi) F^-_\omega(\xi)} = H_\omega^{(0)}(\xi) + \Omega^2 h^2 H_\omega^{(2)}(\xi) + \Omega^4 h^4 H_\omega^{(4)}(\xi) + \mathcal{O}(h^6) \, ,
\end{equation}
where, in terms of $G^{(n)}$, we have
\begin{align}
    H_\omega^{(0)}(\xi) &= \frac{G_\omega^{(0)}(\xi=1)}{G_\omega^{(0)}(\xi)} = 1 \, , \\
    H_\omega^{(2)}(\xi) &= G^{(2)}_\omega(\xi=1) - G^{(2)}_\omega(\xi) \, , \\
    H_\omega^{(4)}(\xi) &= \left( G^{(2)}_\omega(\xi) \right)^2 - G^{(2)}_\omega(\xi) G^{(2)}_\omega(\xi=1) - G^{(4)}_\omega(\xi) + G^{(4)}_\omega(\xi=1) \, .
\end{align}
This means that we can expand $\phi_\omega(z_m)$ as
\begin{equation}
    \phi_\omega(z_m) = h \Omega \phi_\omega^{(1)} + h^3 \Omega^3 \phi_\omega^{(3)} + h^5 \Omega^5 \phi_\omega^{(5)} \,,
\end{equation}
where
\begin{align}
    \phi_\omega^{(1)} &= 2 \int_0^1 \frac{\xi^2 d\xi}{\sqrt{1-\xi^4}} =  \frac{2 \sqrt{\pi} \, \Gamma(7/4)}{3\, \Gamma(5/4)} \, , \\
    \phi_\omega^{(3)} &= 2 \int_0^1 \frac{\xi^2 d\xi}{\sqrt{1-\xi^4}} H^{(2)}_\omega(\xi) \, , \\
    \phi_\omega^{(5)} &= 2 \int_0^1 \frac{\xi^2 d\xi}{\sqrt{1-\xi^4}} \left( \frac{\xi^4}{2 \Omega^4} + H_\omega^{(4)}(\xi) \right) \, .
\end{align}
Here $G_\omega^{(4)}(\xi)$, $H_\omega^{(4)}(\xi)$ and $\phi_\omega^{(5)}$ have a nontrivial dependence on $\omega$\@.

We can absorb the $i\epsilon$ into the definition of $\omega$ and rotate $\omega \to \omega(1+i\epsilon)$ at the end. It means we can expand the trigonometric functions in the definitions of the correlators and proceed without obstacle. However, because the structures are slightly different, we proceed separately for each correlator.

\paragraph{Calculating \texorpdfstring{$K^{RL,c}_{\perp}(\omega,L)$} {} } \hspace{\fill}

Recall that
\begin{align}
    K^{RL,c}_{\perp}(\omega,L) = \frac{\omega}{\sqrt{f_m}} \frac{ F_\omega^+(\xi=1) F_\omega^-(\xi=1) }{\sin(\phi_\omega(z_m)) } \, .
\end{align}
This means that, expanding up to $\mathcal{O}(h)$, we have
\begin{align}
    & K^{RL,c}_{\perp}(\omega,L) \nonumber \\ & = \frac{\omega}{\sqrt{f_m} }  \left[ \frac{1}{h \Omega \phi^{(1)}_\omega} + h \Omega \left( \frac{G^{(2)}_\omega(\xi=1)}{\phi^{(1)}_\omega} + \frac{\left(\phi^{(1)}_\omega \right)^3 - 6 \phi^{(3)}_\omega}{6 \left(\phi^{(1)}_\omega \right)^2} \right) \right. \nonumber \\
    & \quad + \left. h^3 \Omega^3 \left( \frac{ 360 G^{(4)}_\omega(\xi=1) \left(\phi^{(1)}_\omega \right)^2 + 60 G^{(2)}_\omega(\xi=1) \left(\phi^{(1)}_\omega \right)^4 + 7 \left(\phi^{(1)}_\omega \right)^6}{360 \left(\phi^{(1)}_\omega \right)^3} \right. \right. \nonumber \\
    &\quad + \left. \left. \frac{- 360 G^{(2)}_{\omega}(\xi=1) \phi^{(1)}_\omega \phi^{(3)}_\omega  + 60 \left(\phi^{(1)}_\omega \right)^3 \phi^{(3)}_\omega + 360 \left(\phi^{(3)}_\omega \right)^2 - 360 \phi^{(1)}_\omega \phi^{(5)}_\omega }{360 \left(\phi^{(1)}_\omega \right)^3 } \right) + \mathcal{O}(h^5) \right] 
\end{align}

As a function of $L$, there will be one further contribution coming from the mapping $z_m(L)$, which receives corrections of $\mathcal{O}(h^4)$ at small $h$. These will only contribute at $\mathcal{O}(L)$ from the $1/h^3$ term.

We can get the terms proportional to $1/h^3$ and $1/h$ explicitly, because we can solve for $F_\omega^-$ up to $\mathcal{O}(h^2)$ explicitly. Then, writing
\begin{equation}
    K^{RL,c}_{\perp}(\omega,L) = \frac{z_m^2}{\sqrt{f_m}} \left( \frac{c_3^{RL}}{L^3} + \frac{c_1^{RL} \omega^2}{L} + L \, \omega^4 f_{RL}(\omega/T) + \mathcal{O}(L^3) \right) \, ,
\end{equation}
we have
\begin{equation}
    c_3^{RL} = \frac{1}{\left(\frac{3\Gamma(5/4)}{2\sqrt{\pi}\Gamma(7/4)} \right)^3 \times 2 \int_0^1 \frac{\xi^2 d\xi}{\sqrt{1-\xi^4}}} = \left( \frac{2 \sqrt{\pi} \, \Gamma(7/4)}{3\, \Gamma(5/4)} \right)^2 \approx 1.43554002209 \, ,
\end{equation}
and
\begin{align}
    c_1^{RL} &= \frac{1}{\frac{3\Gamma(5/4)}{2\sqrt{\pi}\Gamma(7/4)}} \left[ \frac{1 - \frac{4\pi^3}{\left(3 \Gamma(-3/4) \Gamma(5/4)\right)^2 } }{\frac{2 \sqrt{\pi} \Gamma(7/4)}{3 \Gamma(5/4)}} + \frac16 \frac{2 \sqrt{\pi} \, \Gamma(7/4)}{3\, \Gamma(5/4)} - \frac{2 \int_0^1 \frac{\xi^2 d\xi}{\sqrt{1-\xi^4}} H^{(2)}_\omega(\xi)}{\left(  \frac{2 \sqrt{\pi} \Gamma(7/4)}{3 \Gamma(5/4)} \right)^2}  \right] \nonumber \\
    &\approx 0.49022320139 \, .
\end{align}

The last term has two new contributions that would have to be determined numerically: $G_\omega^{(4)}(\xi=1)$ and $\phi_\omega^{(5)}$\@. We could continue this process forever as well.

Now, because all of the $1/L$ divergences come from the vacuum part, we can calculate the vacuum-subtracted contribution to the correlator without expanding in powers of $z_m$ to get the $T$-dependent part. We achieve this by numerically solving for $F_\Omega$ using the methods discussed in Section~\ref{sec:EE-calculation-HQ-numerics}\@. This extraction gives
\begin{equation}
    K^{RL,c}_{\perp}(\omega,L) - K^{RL,c}_{\perp}(\omega,L)_{T=0} = (-0.20898059) \times \frac{z_m^2}{\sqrt{f_m}} \left[ (\pi T)^4 L + \mathcal{O}((\pi T L)^3) \right] \, ,
\end{equation}
with no frequency dependence at this order in $\pi T L$\@.
    
\paragraph{Calculating \texorpdfstring{$K^{LL,c}_{\perp}(\omega,L)$} {} } \hspace{\fill}

Let us introduce the notation $\bar{K}^{LL,c}_{\perp}= K^{LL,c}_{\perp} + \frac{\omega^2 z_m^2}{\sqrt{f_m}} \lim_{z\to 0} \frac{1}{z^2}$, so that the divergent piece from the contact term is automatically subtracted. Then
\begin{align}
    \bar{K}^{LL,c}_{\perp}(\omega,L) = \frac{\omega}{\sqrt{f_m}} \frac{ F_\Omega^+(\xi=1) F_\Omega^-(\xi=1) }{\tan(\phi_\omega(z_m)) } -  \frac{1}{4 z_m \sqrt{f_m}} \left[ \frac{\partial^3 F^-_\omega}{\partial \xi^3} + \frac{\partial^3 F^+_\omega}{\partial \xi^3} \right]_{\xi=0} \, .
\end{align}

We again proceed to expand up to $\mathcal{O}(h)$\@. The result is
\begin{align}
    & \bar{K}^{LL,c}_{\perp}(\omega,L) \nonumber \\ & = \frac{\omega}{\sqrt{f_m}}  \left[ \frac{1}{h \Omega \phi^{(1)}_\omega} + h \Omega \left( \frac{G^{(2)}_\omega(\xi=1)}{\phi^{(1)}_\omega} - \frac{\left(\phi^{(1)}_\omega \right)^3 + 3 \phi^{(3)}_\omega}{3 \left(\phi^{(1)}_\omega \right)^2} \right) \right. \nonumber \\
    & \quad + \left. h^3 \Omega^3 \left( \frac{ 45 G^{(4)}_\omega(\xi=1) \left(\phi^{(1)}_\omega \right)^2 - 15 G^{(2)}_\omega(\xi=1) \left(\phi^{(1)}_\omega \right)^4 - \left(\phi^{(1)}_\omega \right)^6}{45 \left(\phi^{(1)}_\omega \right)^3} \right. \right. \nonumber \\
    &\quad + \left. \left. \frac{- 45 G^{(2)}_{\omega}(\xi=1) \phi^{(1)}_\omega \phi^{(3)}_\omega  - 15 \left(\phi^{(1)}_\omega \right)^3 \phi^{(3)}_\omega + 45 \left(\phi^{(3)}_\omega \right)^2 - 45 \phi^{(1)}_\omega \phi^{(5)}_\omega }{45 \left(\phi^{(1)}_\omega \right)^3 } \right) + \mathcal{O}(h^5) \right] \nonumber \\
    & \quad - \frac{1}{2 z_m \sqrt{f_m}} \left[ - h^2 \Omega^2 \frac{\partial^3 F^{(2)}_\omega}{\partial \xi^3} + h^4 \Omega^4 \frac{\partial^3 F^{(4)}_\omega}{\partial \xi^3} + \mathcal{O}(h^6) \right]_{\xi=0} \, .
\end{align}
We need the same numbers and functions as before to evaluate this, and we can similarly write
\begin{equation}
    \bar{K}^{LL,c}_{\perp}(\omega,L) = \frac{z_m^2}{\sqrt{f_m}} \left( \frac{c_3^{LL}}{L^3} + \frac{c_1^{LL} \omega^2}{L} + L \, \omega^4 f_{LL}(\omega/T) + \mathcal{O}(L^3) \right) \, .
\end{equation}
The resulting numbers are again calculable. We have
\begin{equation}
    c_3^{LL} = \frac{1}{\left(\frac{3\Gamma(5/4)}{2\sqrt{\pi}\Gamma(7/4)} \right)^3 \times 2 \int_0^1 \frac{\xi^2 d\xi}{\sqrt{1-\xi^4}}} = \left( \frac{2 \sqrt{\pi} \, \Gamma(7/4)}{3\, \Gamma(5/4)} \right)^2 \approx 1.43554002209 \, ,
\end{equation}
and
\begin{align}
    c_1^{LL} &= \frac{1}{\frac{3\Gamma(5/4)}{2\sqrt{\pi}\Gamma(7/4)}} \left[ \frac{1 - \frac{4\pi^3}{\left(3 \Gamma(-3/4) \Gamma(5/4)\right)^2 } }{\frac{2 \sqrt{\pi} \Gamma(7/4)}{3 \Gamma(5/4)}} - \frac13 \frac{2 \sqrt{\pi} \, \Gamma(7/4)}{3\, \Gamma(5/4)} - \frac{2 \int_0^1 \frac{\xi^2 d\xi}{\sqrt{1-\xi^4}} H^{(2)}_\omega(\xi)}{\left(  \frac{2 \sqrt{\pi} \Gamma(7/4)}{3 \Gamma(5/4)} \right)^2}  \right] \nonumber \\ & \quad + \frac12 \frac{1}{\frac{3\Gamma(5/4)}{2\sqrt{\pi}\Gamma(7/4)}} \left[ 1 + \frac{ 2\sqrt{\pi} \Gamma(7/4) + 3(-1 + 2 E(-1) - 2K(-1) ) \Gamma(5/4)}{3 \Gamma(5/4) } \right] \nonumber \\
    &\approx 1.20799321244 \, ,
\end{align}
where $E(-1) = {\rm EllipticE}(-1)$ and $K(-1) = {\rm EllipticK}(-1)$ are Elliptic integrals.

We can calculate the vacuum-subtracted contribution to the correlator as before. The result is explicitly the same as for the $RL$ case up to leading order in $\pi T L$:
\begin{equation}
    \bar{K}^{LL,c}_{\perp}(\omega,L) - \bar{K}^{LL,c}_{\perp}(\omega,L)_{T=0} = (-0.20898059) \frac{z_m^2}{\sqrt{f_m}} \left[ (\pi T)^4 L + \mathcal{O}((\pi T L)^3) \right] \, .
\end{equation}
Higher order terms in $L$ may differ because the $RL$ and $LL$ response functions are only guaranteed to agree up to contact terms in the limit $L \to 0$\@.

\paragraph{Result for the transverse fluctuation response kernels} \hspace{\fill}

We now put together the results we obtained in the preceding calculations. As we just showed, taking the $L \to 0$ limit, we obtain that
\begin{align}
    K^{RL,c}_{\perp}(\omega,L) = \frac{z_m^2}{\sqrt{f_m}} \left( \frac{c_3^{RL}}{L^3} + \frac{c_1^{RL} \omega^2}{L} + \ml{O}(L) \right) \, , \nonumber \\
    K^{LL,c}_{\perp}(\omega,L) = \frac{z_m^2}{\sqrt{f_m}} \left( \frac{c_3^{LL}}{L^3} + \frac{c_1^{LL} \omega^2}{L} + \ml{O}(L) \right) \, ,
\end{align}
where the dominant contribution in the limit $L\to 0$ is determined by
\begin{align}
    c_3^{RL} = c_3^{LL} = \left( \frac{2 \sqrt{\pi} \, \Gamma(7/4)}{3\, \Gamma(5/4)} \right)^2 \approx 1.43554 \, .
\end{align}
We have kept outside the definition of $c_3$ an overall factor of $ \frac{z_m^2}{\sqrt{f_m}}$ (which does depend on $L$) for convenience to translate to the result for $\Delta^c_{ij}$, which has the inverse of this factor in the front (see Eq.~\eqref{eq:Delta-in-terms-of-K} for comparison)\@.
The subleading $1/L$ contribution is different for each case (see Section~\ref{sec:App-transverse-connected}), i.e., $c_1^{RL} \neq c_1^{LL}$, but this presents no issue because we anyways do not expect the results to agree beyond the leading term as a function of $L$\@. On the other hand, the leading contribution is the same for both procedures, and it does not receive contributions from contact terms, because the construction of the $RL$ kernel explicitly prevents this.

Having done the above, one finds that when we introduce anti-symmetric perturbations $h_\perp(t)$ as discussed in Section~\ref{sec:W-loop-AdS}, the sum of the response kernels gives a total of
\begin{equation}
    K^{c}_{\perp}(\omega,L) = 2 \frac{z_m^2}{\sqrt{f_m}} \left( \frac{2 \sqrt{\pi} \, \Gamma(7/4)}{3\, \Gamma(5/4)} \right)^2 \left[ \frac{1}{L^3} + \ml{O}(L^{-1}) \right] \, ,
\end{equation}
and therefore the kernel that determines the two-point function for transverse deformations is given by
\begin{equation}
    \Delta^c_{22}(\omega,L) = \Delta^c_{33}(\omega,L) = \frac{2 \sqrt{\lambda}}{\pi} \left( \frac{2 \sqrt{\pi} \, \Gamma(7/4)}{3\, \Gamma(5/4)} \right)^2 \frac{1}{L^3} + \ml{O}(L^{-1}) \, .
\end{equation}
The main feature of this result is that it diverges as $L^{-3}$ when $L \to 0$\@. 

This concludes our calculation of the response functions that determine the linear response of the Nambu-Goto action to transverse deformations on the boundary countour that defines the Wilson loop, for the background configuration that describes a Coulomb-type potential between the heavy quarks. To complete the result, we now move on to calculate the longitudinal one, following the same steps.

\subsubsection{Longitudinal fluctuations}

Having gone through all the machinery in the previous section, we shall give an abbreviated discussion of the calculation for the case of longitudinal fluctuations. First, we note that to obtain the leading behavior that we got in the previous section, it is sufficient to work in the $T=0$ case. This is clear by looking at how $T$ appears in the solution for $z(\sigma)$ and in the action for the fluctuations $\delta(\tau,\sigma)$: After factoring out the overall scale $L$ from $z(\sigma)$, it is manifest that the leading appearance of $T$ is of the order $(\pi T L)^4$\@. It is then clear that, if we find a $1/L^3$ dependence for $\Delta^{c}_{11}(\omega,L)$ in vacuum, this will be the dominant contribution in the limit $L \to 0$ (note that $\Delta$ has mass dimension three)\@.

When $T=0$, the action for the fluctuations reduces to
\begin{equation}
    S^{(2),\parallel}_{\rm NG,c}[\delta] = \int_{-\T}^{\T} \!\! d\tau \! \int_{0}^L \!\! d\sigma \left[ \frac{1 }{2 z^2 \sqrt{z'{}^2 + 1} } \delta'{}^2 - \frac{ \sqrt{z'{}^2 + 1}}{2 z^2} \dot{\delta}{}^2  -  \frac{ 1  }{ z^4 \sqrt{z'{}^2 + f } } \delta^2 \right] \, .
\end{equation}
Furthermore, if we are only after finding the leading behavior of $\Delta^{AB,c}_{11}$, we can even drop the term with time derivatives in this action, because we will be in the regime $\omega^2 L^2 \ll 1 $\@. This will only modify the result by terms that go as $1/L$\@. 

After using the conservation equation for the background worldsheet in vacuum (i.e., the conserved quantity that appears due to there not being any explicit $\sigma$ dependence in the action), which we can write as $z^2 \sqrt{z'{}^2+1} = z_m^2$, one obtains the following equation of motion for the fluctuations:
\begin{equation}
    \frac{\partial^2 \delta}{\partial \sigma^2}(\sigma) + \frac{2}{z(\sigma)^2} \delta(\sigma) = 0 \, .
\end{equation}
As with the transverse fluctuations, we can change variables from $\sigma$ to $\xi = z/z_m$, and rewrite this equation of motion in terms of two domains, one for $ \sigma \in (0,L/2) $, where we will use $\delta^L$, and the other for $ \sigma \in (L/2, L) $, where we will use $\delta^R$\@. The equation of motion for both of them is
\begin{equation} \label{eq:delta-vacuum}
    (1-\xi^4) \frac{\partial^2 \delta}{\partial \xi^2} - \frac{2}{\xi} \frac{\partial \delta}{\partial \xi} + 2 \xi^2 \delta = 0 \, ,
\end{equation}
subject to matching conditions
\begin{align}
\lim_{\xi \to 1} \delta^L(\xi) = \lim_{\xi \to 1} \delta^R(\xi) \, , & & \lim_{\xi \to 1} \sqrt{1 - \xi^4} \frac{\partial \delta^L}{\partial \xi} = - \lim_{\xi \to 1} \sqrt{1 - \xi^4} \frac{\partial \delta^R}{\partial \xi} \, .
\end{align}

Conveniently, the solutions to Eq.~\eqref{eq:delta-vacuum} can be found explicitly:
\begin{equation}
    \delta^{L,R}(\xi) = A_{L,R} \sqrt{1 - \xi^4} + B_{L,R} \frac{\xi^3 \sqrt{1-\xi^4} }{3} {}_2 F_1 \left[ \frac34 , \frac32 , \frac74, \xi^4 \right] \, .
\end{equation}
Then, the matching conditions set
\begin{align}
    B_L = B_R \, , & & 2 ( A_L + A_R ) =  \frac{ \sqrt{\pi} \, \Gamma(7/4) }{3 \, \Gamma(5/4)} ( B_L + B_R ) \, .
\end{align}
We can then define separate response kernels on either side for perturbations on a given side. Following our discussion of transverse fluctuations, we may define, setting $\delta^L(\xi=0) = A_L = 1$ and $\delta^R(\xi=0) = A_R = 0$,
\begin{align}
    K^{RL,c}_\parallel &= \frac{1}{z_m}  \lim_{\xi \to 0} \frac{1}{\xi^2} \frac{\partial \delta^R}{\partial \xi} =  \frac{1}{2z_m}  \frac{\partial^3 \delta^R}{\partial \xi^3} \, , \\
    K^{LL,c}_\parallel &= - \frac{1}{z_m} \lim_{\xi \to 0} \frac{1}{\xi^2} \frac{\partial \delta^L}{\partial \xi} =  \frac{1}{2z_m}  \frac{\partial^3 \delta^L}{\partial \xi^3} \, ,
\end{align}
where in taking the limit we have used that the mode solutions have vanishing first and second derivatives at $\xi=0$ (this can be seen directly from the mode functions, as their dependence on $\xi$ starts at order $\xi^3$)\@. The result is easily found to be
\begin{equation}
    K^{RL,c}_\parallel = - K^{LL,c}_\parallel = \frac{2}{z_m} \frac{3 \, \Gamma(5/4) }{2 \sqrt{\pi} \Gamma(7/4) } + \ml{O}(L) \, .
\end{equation}
As in the case for transverse fluctuations, these two one-sided response kernels have an equal leading order contribution to the $\Delta_{11}^c$ kernel, and no contact term appears at this order.
The symmetrized response kernel $K_\parallel^c$ is then given by
\begin{equation}
    K_\parallel^c(\omega,L) = \frac{4}{z_m} \frac{3 \Gamma(5/4) }{2 \sqrt{\pi} \, \Gamma(7/4) } + \ml{O}(L) \, ,
\end{equation}
which means that the contribution to the two-point function coming from the linearized fluctuations of the Nambu-Goto action is
\begin{equation}
    \Delta_{11}^c(\omega,L) = \frac{4 \sqrt{\lambda}}{\pi} \left( \frac{2 \sqrt{\pi} \, \Gamma(7/4)}{3\, \Gamma(5/4)} \right)^2 \frac{1}{L^3} + \ml{O}(L^{-1}) \, .
\end{equation}
With this, we have calculated the leading contribution as $L \to 0$ of the longitudinal fluctuations.

We can therefore write the complete leading contribution to the two-point function associated to fluctuations on the extremal worldsheet that gives a Coulomb interaction potential between two heavy quarks:
\begin{align} \label{eq:Delta-c-final}
 \Delta^c_{ij}(\omega,L) =  \frac{16 \pi^2 }{\Gamma(1/4)^4}  \frac{\sqrt{\lambda}}{L^3} \left( 2\delta_{i1} \delta_{1j} + \delta_{i2} \delta_{2j} + \delta_{i3} \delta_{3j} \right) + \ml{O}(L^{-1}) \, .
\end{align}

This completes the calculation for the quadratic fluctuations in this background configuration, and it is all we need in order to compare with the result of the next section. However, because this is highly singular as $L \to 0$, we shall perform a numerical check that our result is not exclusive to the large $\T$ limit, and that the same behavior is obtained in the $L\to 0$ limit for a bounded rectangular Wilson loop at fixed $\T$\@.

\subsubsection{Euclidean numerical calculation for variations on a bounded rectangle} \label{sec:EE-calculation-HQ-numerics}

In what follows, we will verify that the above results continue to hold when the (Euclidean) temporal extent of the loop is finite. This is not in vain, as when $T>0$ the Euclidean time direction is finite in extent, and therefore it is relevant to study the expectation value of a Wilson loop with finite temporal extent $\mathcal{T}_E$, to assess definitively whether the temperature can play a role in the expectation value of interest.

To demonstrate this behavior, we calculate the derivative response to transverse perturbations, solving the analogous problem to that in Section~\ref{sec:transverse-v-fluct}, but in Euclidean signature. The background solution on which the perturbations propagate is specified by the action principle
\begin{align}
    S_{{\rm NG}, c, E}^{(0)}[z] &= \int_{0}^{\T_E} \!\! \diff \tau_E \! \int_0^L \!\! \diff \sigma \frac{\sqrt{\left(\frac{\partial z}{\partial \sigma}\right)^2 + f + \frac{1 }{f} \left(\frac{\partial z}{\partial \tau_E}\right)^2 }}{z^2} \nonumber \\
    &= a b  \int_{0}^{1} \diff \bar{\tau}_E \, \int_0^1 \diff \bar{x} \frac{1}{\xi^2} \sqrt{ 1 - \xi^4 + \frac{\xi'{}^2}{b^2} + \frac{\dot{\xi}{}^2}{a^2 (1 - \xi^4)} } \, , \label{eq:NGactionpseudospectral}
\end{align}
where the dot stands for a derivative with respect to the rescaled imaginary time $\bar{\tau}_E$, the prime stands for a derivative with respect to the rescaled spatial coordinate $\bar{x}$, and we have introduced $\bar{\tau}_E \equiv \tau_E/\T_E$, $\bar{x} \equiv \sigma/L$, $\xi \equiv \pi T z$, $a \equiv  \T_E \pi T$, and $b \equiv L \pi T$\@.
We solve for the background worldsheet at four values of $ b \in \{ 1.0 \times 10^{-2}, 5.0 \times 10^{-3}, 2.5 \times 10^{-3}, 1.0 \times 10^{-3} \}$, holding $ a$ fixed at three different values $a \in \{0.1, 0.3, 1.0 \}$\@. That is to say, the aspect ratio of the rectangle $a/b$ ranges from $10$ to $1000$. We note that because $T$ is nonzero, two parameter choices with the same aspect ratio are not equivalent (however, our analytic calculations for an infinite strip suggest that the effect of $T$ should become smaller and smaller as $a,b \to 0$, asymptotically approaching a regime where the result only depends on the aspect ratio $a/b$). On each of the background surfaces we will obtain, we will solve for the linear response on one of the Euclidean time sides of the rectangle to perturbations in the boundary conditions on the other side.

We obtain the numerical solutions to Eq.~\eqref{eq:NGactionpseudospectral} using the pseudospectral method~\cite{boyd01}\@.
The pseudospectral method is an elegant way to solve boundary value problems such as the one in Eq.~\eqref{eq:NGactionpseudospectral}, which approximates the continuous differential equation by a finite set of coupled equations.
Specifically, we introduce an ansatz for $\xi$ in terms of the Chebyshev polynomials:
\[
\xi(\bar x, \bar{\tau}_E ) = \sum_{i,j=0}^{N_\text{coll}} c_{ij}T_i\left(2\bar x - 1\right)T_j\left(2\bar{\tau}_E - 1\right),
\]
with $c_{ij}$ unknown coefficients for which we need to solve and $T_i$ denoting the Chebyshev polynomial or order $i$\@. We then plug the ansatz into the equations of motion obtained from the action shown in Eq.~\eqref{eq:NGactionpseudospectral}\@. By evaluating these equations of motion at a finite number of points called \emph{collocation points}, we obtain a set of coupled equations. The number of collocation points is given by $N_{\rm coll}$\@. Any collocation points that lie on the boundary of the problem are constrained using the boundary conditions instead.
This immediately highlights a significant advantage of this pseudospectral method, as in this way boundary conditions are automatically satisfied, which is otherwise nontrivial for other approaches to boundary value problems.

For a general choice of such collocation points, the solution obtained in this way will not converge to the solution of the differential equation as we take the number of collocation points $N_\text{coll}$ to infinity.
If, however, we choose our collocation points to lie on the simultaneous zeroes of $T_{N_\text{coll} + 1}(2\bar x - 1)$ and $T_{N_\text{coll} + 1}(2{\bar \tau}_E - 1)$, the solution obtained is guaranteed to converge to the solution of Eq.~\eqref{eq:NGactionpseudospectral}, where the error goes like $\exp\left(-cN_\text{coll}\right)$, with $c$ some positive constant~\cite{boyd01}.
In practice, this means that with this choice of collocation points, the convergence as we take $N_\text{coll} \rightarrow \infty$ is very fast, so that we can achieve impressive precision even with a relatively small number of collocation points.
By varying the number of collocation points, we can also get an estimate of our truncation error.

For linear problems, the procedure described above leads to a set of $(N_\text{coll} + 1)^2$ coupled linear equations, which is exactly the number of unknown coefficients $c_{ij}$ we have, so in this case one can find the solution by matrix inversion.
Here we should note that the matrix that needs to be inverted is often close to singular, requiring us to work with more significant figures than machine precision provides.

A second complication is that the problem defined by Eq.~\eqref{eq:NGactionpseudospectral} is not linear.
Because of this, to find a solution we linearize the equations around a trial solution, and then use the Newton-Raphson method to iteratively update the trial solution until our iteration converges.
This introduces the usual difficulties associated with Newton-Raphson, namely that for certain choices of initial trial solution the iteration might not converge, but if for the first couple of steps in the iteration one uses very small step size in the update, it is generally not hard to reach the correct solution from a reasonably chosen initial trial solution.

\begin{figure}
    \centering
    \includegraphics[width=0.44\textwidth]{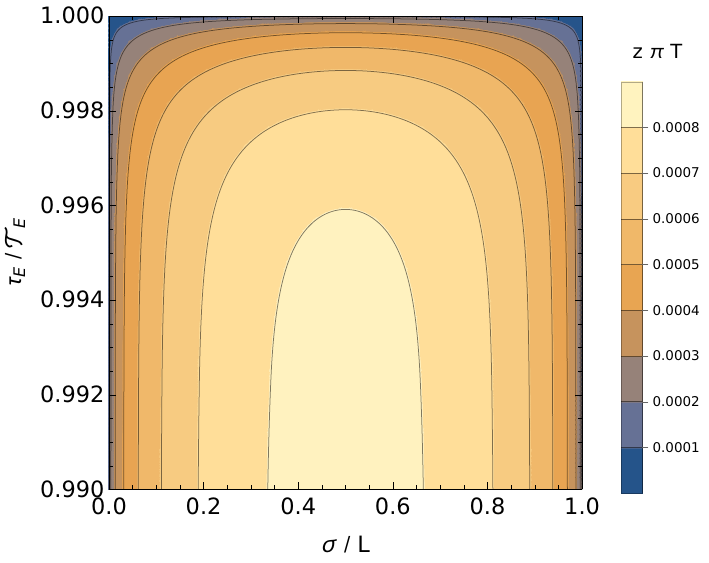}
    \includegraphics[width=0.54\textwidth]{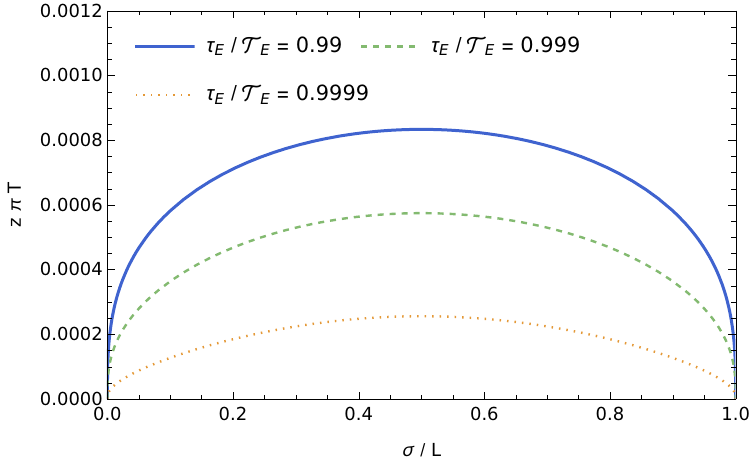}
    \caption{Example of a Euclidean worldsheet configuration hanging from a rectangle of dimensions $\T_E \times L = 0.1 (\pi T)^{-1} \times 10^{-3} (\pi T)^{-1}$ on the boundary of AdS${}_5$ into its radial direction, where a black hole lies at $z = (\pi T)^{-1}$\@. Left panel: contour plot of the solution close to one of the ends along the temporal extent of the Euclidean worldsheet. Right panel: cross sections of the same worldsheet at different values of the Euclidean time coordinate. Away from $\tau_E = 0, \T_E$, the solution approaches the stable solution of the 1D problem given by Eq.~\eqref{eq:implicit-wx} in the small $z_m$ branch. Close to the corners it smoothly interpolates between the solution to the effective 1D problem and the boundary conditions specified by the bounded rectangular Wilson loop.}
    \label{fig:background-sample}
\end{figure}

A sample solution for $a = 0.1$ and $b = 10^{-3}$ can be found in Fig.~\ref{fig:background-sample}\@.
We can see that for $\tau_E$ away from the boundary conditions at $\tau_E = 0$ and $\tau_E = \mathcal{T}_E$, the solution agrees with the 1D problem from Eq.~\eqref{eq:implicit-wx}\@.

Once we obtain the background solution, we can introduce perturbations on the boundary and solve for the response functions.
On each of the background surfaces, we solve for the linear response on one Euclidean time side of the rectangle to perturbations in the boundary conditions on the other side. We use the same decomposition as for the background solution and write the perturbation as 
\begin{equation}
    y(\bar{x},\bar{\tau}_E) = \sum_{i,j=0}^{N_{\rm coll}} d_{ij} T_i(2 \bar{x} - 1) T_j(2 \bar{\tau}_E - 1) \, ,
\end{equation}
where $d_{ij}$ are the unknown coefficients we need to solve for.

Given that the numerical method to solve for the background worldsheet already defines a preferred basis on which to formulate this problem, we will calculate the response functions by mapping a perturbation onto a given basis element $T_n(2\bar{\tau}_E - 1)$, where $T_n$ is a Chebyshev polynomial determining the boundary condition for the transverse fluctuations along the time axis on one side of the rectangular contour, to another basis element $T_m(2\bar{\tau}_E - 1)$ on the other side of the contour. That is to say, given a boundary condition at $\bar{x} = 0$, specified by $y(0,\bar{\tau}_E) = T_n(2\bar{\tau}_E - 1)$, we will want to determine the response at the other side of the contour $y'(\bar{x} = 1, \bar{\tau}_E)$, decomposed in terms of Chebyshev polynomials.

To put this on a concrete mathematical footing, we shall repeat some of the discussions in Section~\ref{sec:HQ-fluct}, keeping in mind that we now work in Euclidean signature with a finite Euclidean time extent. As before, the response kernel of interest is obtained by solving the equations of motion derived from the action for the fluctuations, which can be written as
\begin{align}
    S^{(2),\perp}_{{\rm NG},c,E}[y] &= \int_{0}^{\T_E} \!\! \diff \tau_E \! \int_{0}^L \!\! \diff \sigma  \frac{  \left(\frac{\partial y}{\partial \tau_E}\right)^2  + f \left(\frac{\partial y}{\partial \sigma}\right)^2  + \frac{1}{f} \left(  \frac{\partial z}{\partial \sigma} \frac{\partial y}{\partial \tau_E} - \frac{\partial z}{\partial \tau_E} \frac{\partial y}{\partial \sigma} \right)^2 }{2 z^2 \sqrt{f + \left(\frac{\partial z}{\partial \sigma}\right)^2 + \frac{1 }{f} \left(\frac{\partial z}{\partial \tau_E}\right)^2 }} \,  \nonumber \\
    &= a b \int_{0}^{1} \diff \bar{\tau}_E \int_0^1 \diff \bar{x}  \frac{ \frac{\dot{y}{}^2}{\T_E^2} \left(1 + \frac{\xi'{}^2}{b^2 (1 - \xi^4) } \right) + \frac{y'{}^2}{L^2} \left( 1 - \xi^4 + \frac{\dot{\xi}{}^2}{a^2(1-\xi^4)} \right) - \frac{2 y' \dot{y} \xi' \dot{\xi} }{L \T_E b a (1 - \xi^4) } }{ 2 \xi^2 \sqrt{1 - \xi^4 + \frac{\xi'{}^2}{b^2 } + \frac{\dot{\xi}{}^2}{a^2(1-\xi^4)} } } \, .
\end{align}
Evaluating this action on-shell, with the only non-vanishing boundary conditions being in the temporal segments of the Wilson loop, we get 
\begin{align}
    S^{(2),\perp}_{{\rm NG},c,E}[y]_{\rm on-shell} &= \frac{ab}{2L^2} \int_{0}^{1} \diff \bar{\tau}_E \int_0^1 \diff \bar{x} \frac{\partial}{\partial \bar{x} } \left( \frac{ \left( 1 - \xi^4 + \frac{\dot{\xi}{}^2}{a^2(1-\xi^4)} \right) y y' }{\xi^2 \sqrt{1 - \xi^4 + \frac{\xi'{}^2}{b^2 } + \frac{\dot{\xi}{}^2}{a^2(1-\xi^4)} }} \right) \nonumber \\
    &= \frac{ab}{2L^2} \int_{0}^{1} \diff \bar{\tau}_E \left[ \lim_{\bar{x} \to 1} \frac{y y'}{\xi^2 \sqrt{1 + \frac{\xi'{}^2}{b^2} } } - \lim_{\bar{x} \to 0} \frac{y y'}{\xi^2 \sqrt{1 + \frac{\xi'{}^2}{b^2} } } \right] \, ,
\end{align}
where we have dropped $\xi$ and $\dot{\xi}$ as they vanish at the boundaries defined by the temporal segments of the Wilson loop, where by definition $\xi = 0$ and therefore $\dot{\xi}=0$\@. A posteriori, knowing the solution to the background worldsheet, one can verify that the limits $\ell(\bar{\tau}_E; a,b) \equiv \lim_{{\bar x} \to 0,1} \xi^2 \sqrt{1 + \frac{\xi'{}^2}{b^2} } $ are finite and equal. 
To avoid issues with contact terms, we will extract the quadratic kernel from variations on opposite sides of the contour. That is to say, we will calculate the quadratic kernel $\Delta_{\perp,E}^{c}(\tau_{E1},\tau_{E2};L)$ that appears as a $ \int d\tau_{E1} d\tau_{E2} \, y_L(\tau_{E1}) \Delta(\tau_{E1},\tau_{E2}) y_R(\tau_{E2}) $ contribution to the on-shell action, given by
\begin{equation}
    \Delta_{\perp,E}^{c}(\tau_1,\tau_2;L) = \frac{\sqrt{\lambda}}{\pi} \frac{ab}{ L^2 \T_E^2 } \left( \frac{1}{\ell(\bar{\tau}_{E1})} + \frac{1}{\ell(\bar{\tau}_{E2})} \right) \widetilde{K}^{RL,c}_{\perp, { E}}(\bar{\tau}_{E2},\bar{\tau}_{E1}) \, ,
\end{equation}
where $\widetilde{K}^{RL,c}_{\perp, E}(\tau_{E2},\tau_{E1})$ is defined as the following response kernel:
\begin{equation}
    y'(\bar{x} = 1, \bar{\tau}_E) = - \int_0^1 \diff \bar{\tau}_E' \, \widetilde{K}^{RL,c}_{\perp, { E}}(\bar{\tau}_E,\bar{\tau}_E') y(\bar{x}=0,\bar{\tau}_E') \, .
\end{equation}

\begin{figure}
    \centering
    \includegraphics[width=\textwidth]{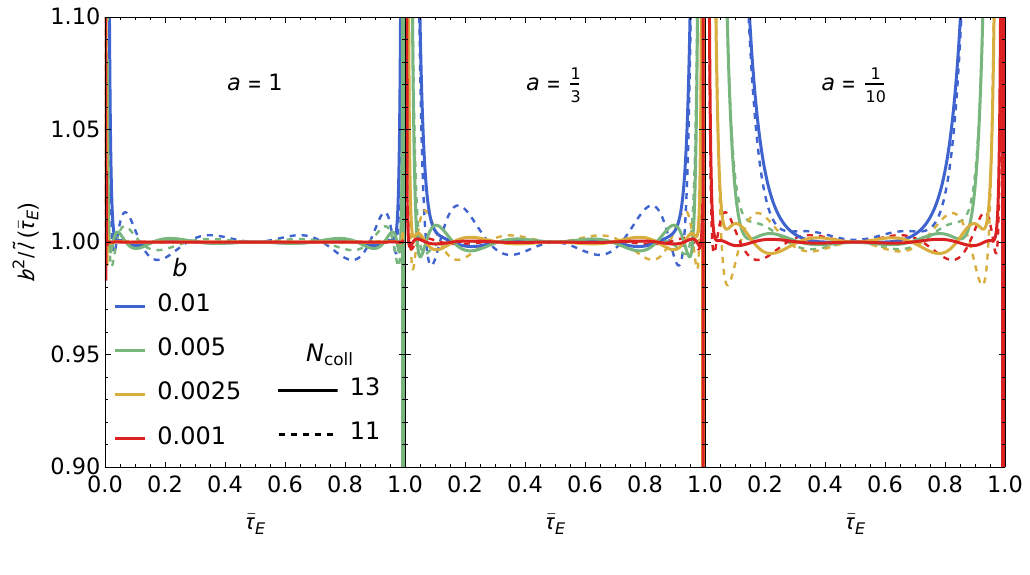}
    \caption{Plots of $b^2/\tilde{\ell}(\bar{\tau}_E)$ for different values of $a, b$ and different number of collocation points $N_{\rm coll}$\@. Here we introduced $\tilde{\ell}(\bar{\tau}_E) = \left( \frac{2 \sqrt{\pi} \, \Gamma(7/4)}{3\, \Gamma(5/4)} \right)^{2} \ell(\bar{\tau}_E)$ so that the limit is rescaled to unity.}
    \label{fig:ell-approach}
\end{figure}

All that remains is to evaluate the response kernel $\widetilde{K}^{RL,c}_{\perp, {E}}$ and the limit $\ell(\bar{\tau}_E)$ in the background solution. Before proceeding, we first discuss what the expected result is. From our analysis of the effective 1D problem (in the limit $\T \to \infty$) in Minkowski signature, we see from Eq.~\eqref{eq:Delta-c-final} that the limit $L \propto b \to 0$ should give us
\begin{equation}
    \Delta_{\perp,E}^{c}(\tau_1,\tau_2;L) = \frac{2\sqrt{\lambda}}{\pi} \left( \frac{2 \sqrt{\pi} \, \Gamma(7/4)}{3\, \Gamma(5/4)} \right)^2 \frac{1}{L^3} \delta(\tau_1 - \tau_2) + \ml{O}(L^{-1}) \, .
\end{equation}
Noting that $\delta(\tau_1 - \tau_2) = \T_E^{-1} \delta(\bar{\tau}_{E1} - \bar{\tau}_{E2})$, and that in the strict limit $\T \to \infty$ we have $\ell = b^2 \left(\frac{2\sqrt{\pi}\, \Gamma(7/4)}{3 \, \Gamma(5/4)} \right)^{-2} $, we expect
\begin{align}
    \lim_{b \to 0} \widetilde{K}^{RL,c}_{\perp, {E}}(\bar{\tau}_E,\bar{\tau}_E') = \delta(\bar{\tau}_{E1} - \bar{\tau}_{E2}) \, \, , & & \lim_{b \to 0} \frac{b^2}{ \ell(\bar{\tau}_E) } = \left( \frac{2 \sqrt{\pi} \, \Gamma(7/4)}{3\, \Gamma(5/4)} \right)^{2} \, \, . \label{eq:numerical-target}
\end{align}
Given these expectations, our numerical evaluation of $\widetilde{K}^{RL,c}_{\perp, { E}}$ and $\ell(\bar{\tau}_E)$ needs only to demonstrate Eq.~\eqref{eq:numerical-target}\@.

\begin{figure}
    \centering
    \includegraphics[width=\textwidth]{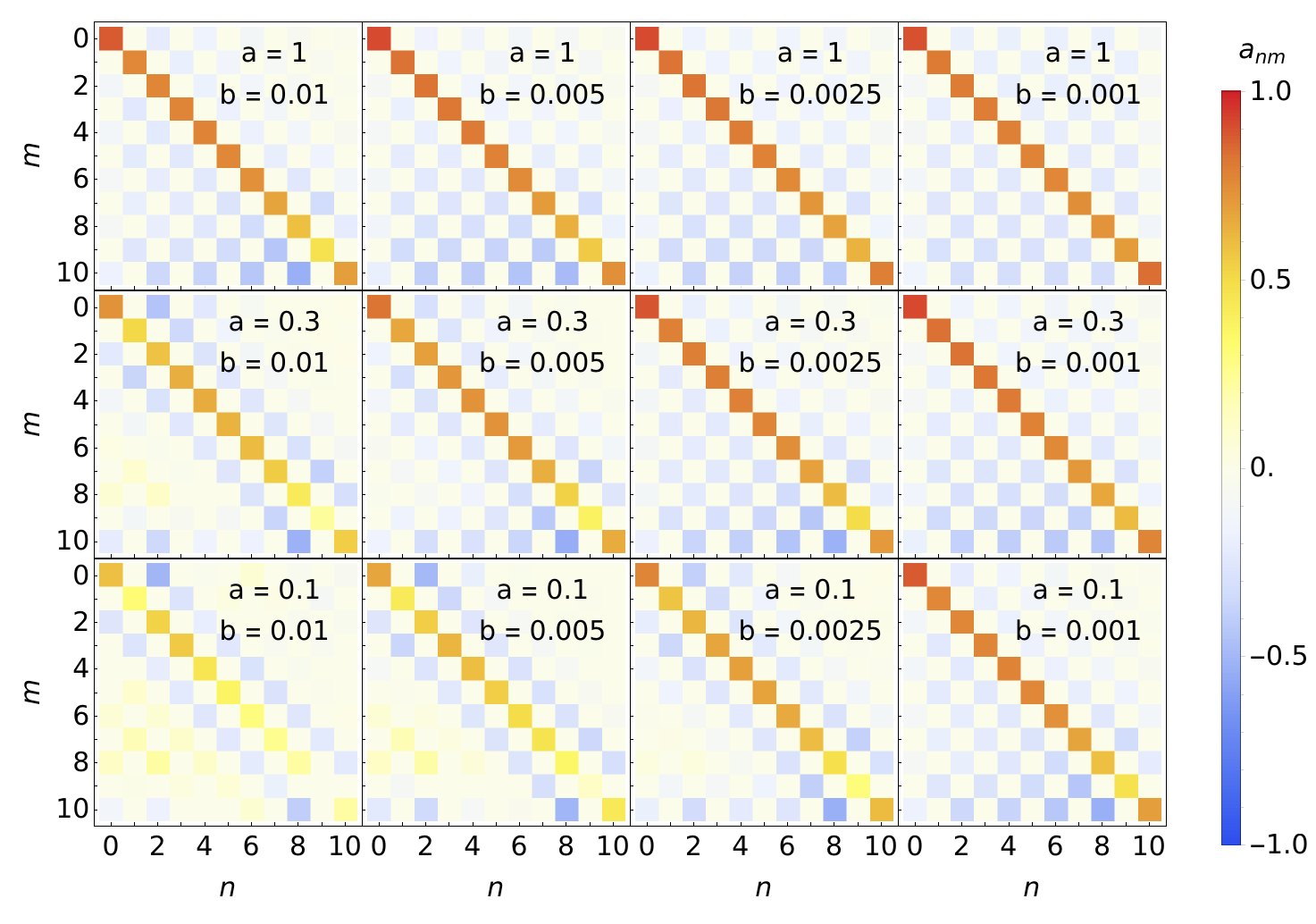}
    \caption{Coefficients $a_m^{(n)}$ as defined in Eq.~\eqref{eq:anm-def}, for $N_{\rm coll} = N_{\rm pols} = 11$\@, displayed as the entries of a matrix with its numerical values described by colors. As discussed in the main text, the expectation is that as $b/a \to 0$, $a_m^{(n)} \to \delta_m^n$, which is to say, the figures at the top right of the above panels should approach the identity matrix, converging to unity on the matrix diagonal and to zero elsewhere.}
    \label{fig:amn-approach-10}
\end{figure}

\begin{figure}
    \centering
    \includegraphics[width=\textwidth]{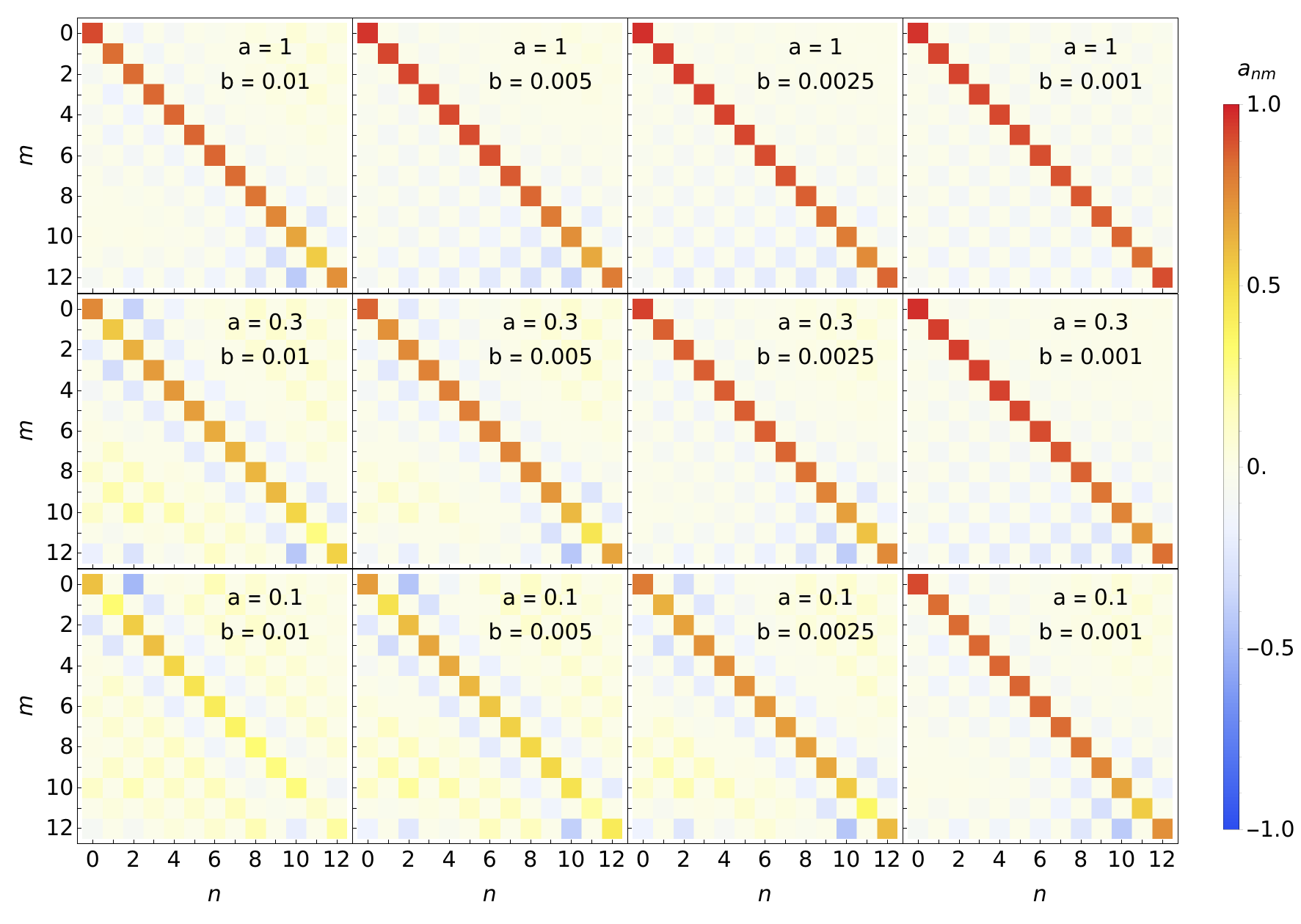}
    \caption{Coefficients $a_m^{(n)}$ as defined in Eq.~\eqref{eq:anm-def}, for $N_{\rm coll} = N_{\rm pols} = 13$\@, displayed as the entries of a matrix with its numerical values described by colors. As for Figure~\ref{fig:amn-approach-10}, the expectation is that as $b/a \to 0$, $a_m^{(n)} \to \delta_m^n$, which is to say, the figures at the top right of the above panels should approach the identity matrix, converging to unity on the matrix diagonal and to zero elsewhere. Note that the convergence here is better due to the larger value of $N_{\rm coll}, N_{\rm pols}$.}
    \label{fig:amn-approach-12}
\end{figure}

We show results for $b^2/\ell(\bar{\tau}_E)$, normalized by its limiting value in Fig.~\ref{fig:ell-approach}\@. We observe two general trends:
\begin{enumerate}
    \item As $b \to 0$, the solution indeed approaches the limiting value, converging first in the middle region $\bar{\tau}_E \sim 1/2$ and later near the corners. For sufficiently small $b$, the convergence is more strongly dependent on the ratio $b/a$ than on $b$ alone. 
    \item The oscillations in the solution, which are artifacts of a truncated basis, get suppressed as we increase the number of collocation points $N_{\rm coll}$, and the convergence of $b^2/\ell(\bar{\tau}_E)$ as $b \to 0$ is observed even more clearly at large $N_{\rm coll}$\@. By increasing the number of collocation points we would reduce the truncation effects, and the solution would approach the exact profile at each value of $b$, with which the limit $b \to 0$ could be examined even more precisely. However, we will refrain to go further, as we deem the plots in Fig.~\ref{fig:ell-approach} as sufficient evidence for the value of the limit we wished to verify.
\end{enumerate}

Next we test the convergence of $\widetilde{K}^{RL,c}_{\perp, {\rm E}}$ by studying the derivative response $y'$ at $\bar{x}=1$ to a boundary condition specified by a Chebyshev polynomial $T_n(2\bar{\tau}_E - 1)$ at $\bar{x}=0$\@. Concretely, we expand the derivative response in terms of Chebyshev polynomials and find the coefficients $a_m^{(n)}$ of the expansion
\begin{equation} \label{eq:anm-def}
    y_{(n)}'(\bar{x}=1, \bar{\tau}_E) = -\sum_{m=0}^{N_{\rm pols}-1} a_m^{(n)} T_m(2\bar{\tau}_E - 1) \, ,
\end{equation}
and we plot these coefficients for different values of $a$ and $b$\@. $N_{\rm pols}$ is the number of Chebyshev polynomials we use in the spectral approach and it is equal to the number of collocation points $N_{\rm coll}$\@.
The expectation is that $a_n^{(n)} \to 1$ and $a_m^{(n)} \to 0$ if $m \neq n$\@. As we can see from Figs.~\ref{fig:amn-approach-10} and~\ref{fig:amn-approach-12}, as the ratio $b/a$ goes to zero, the convergence $a_m^{(n)} \to \delta_{mn}$ is rather good, and qualitatively improves as we refine the set of basis functions in the pseudospectral method.

In a nutshell, we see that everything in the numerical Euclidean approach is consistent with our previous real time analysis in Section~\ref{sec:transverse-v-fluct}, meaning that the quadratic kernel $\Delta^c$ diverges as $L^{-3}$ when we take $L \to 0$\@. As such, we conclude that the limit $L \to 0$ of this SYM Wilson loop does not describe the physics we wish to capture, because it is dominated by UV contributions that are not in the domain of any low-energy effective description. This is also consistent with our previous discussion that we in fact expect $\langle \hat{\T} W[\mathcal{C}] \rangle_T = 1$ for the SU(3) Wilson line configuration that is relevant to quarkonium. As such, we conclude that we must seek other configurations to describe the Wilson loop that is relevant for quarkonium dynamics in a thermal medium.


\section{Analytic aspects of the correlation functions}

In this Appendix we discuss important analytic properties of the chromoelectric correlator we calculated in Section~\ref{sec:QQ-setup}. In Appendix~\ref{sec:App-omega-expansion} we discuss the small frequency expansion that we used in Section~\ref{sec:ads-cft-final-result} to arrive at Eq.~\eqref{eq:small-omega-expansion}. In Appendix~\ref{sec:App-analytic-rot} we prove Eq.~\eqref{eq:analytic-prop-gE}, which is needed as an intermediate step to obtain the spectral function that we calculate in Section~\ref{sec:ads-cft-spectral-eval}.

\subsection{Expansion of \texorpdfstring{$F^{-}_\Omega$} {} in powers of \texorpdfstring{$\Omega$} {}} \label{sec:App-omega-expansion}

Consider the defining equation for $F^-_\omega(\xi)$, given by Eq.~\eqref{eq:F-thermal}:
\begin{align} \label{eq:F-thermal-app}
    \frac{\partial^2 F^-_\omega}{\partial \xi^2} - 2 \left[ \frac{1 + \xi^4}{\xi(1-\xi^4)} - \frac{i \Omega \xi^3}{1-\xi^4} \right] \frac{\partial F^-_\omega}{\partial \xi} + \left[  \frac{i \Omega \xi^2}{1-\xi^4} + \frac{\Omega^2 (1 - \xi^6) }{(1-\xi^4)^2} \right] F^-_\omega = 0 \, ,
\end{align}
and instead of attempting to find a solution for arbitrary $\Omega$, let us expand the solution in powers of $\Omega = \frac{\omega}{\pi T}$\@. To that end, we write
\begin{equation}
    F_\omega^-(\xi) = F^{(0)}(\xi) + i\Omega F^{(1)}(\xi) + (i\Omega)^2 F^{(2)}(\xi) + \mathcal{O}(\Omega^3) \, ,
\end{equation}
and solve Eq.~\eqref{eq:F-thermal-app} order by order in $\Omega$ with boundary conditions determined by
\begin{align}
    F^-_\omega(0) = 1 \, \, , & & \frac{\partial_\xi F_\omega^-(1)}{F_\omega^-(1)} = \frac{i \Omega}{4} \frac{1 - \frac{3i\Omega}{2}}{1 - \frac{i\Omega}{2}} \, .
\end{align}
The solutions can then be found order by order:
\begin{align}
    F^{(0)}(\xi) &= 1 \, , \\
    F^{(1)}(\xi) &= \frac{1}{4} \left[ \ln \left( (1+\xi)^2 (1+\xi^2) \right) - 2 \arctan(\xi) \right] \, , \\
    F^{(2)}(\xi) &= \frac{1}{8} \left[ \left( \arctan(\xi) - \ln(1+\xi) - 4 \right) \left( \arctan(\xi) - \ln(1+\xi) \right) \right. \nonumber \\ & \quad \quad \left. - \left( \arctan(\xi) - \ln(1+\xi) + 2 \right) \ln(1+\xi^2) + \frac{1}{4} \left(\ln(1+\xi^2) \right)^2 \right] \, ,
\end{align}
and with them, one can easily evaluate the input needed for the correlation function:
\begin{equation}
    \frac{-i}{F^-_{|\omega|}(0)} \frac{\partial^3 F^-_{|\omega|}}{\partial \xi^3}(0) = 2 |\Omega| + 2 i \Omega^2 + \mathcal{O}(\Omega^3) \, .
\end{equation}
This verifies Eq.~\eqref{eq:small-omega-expansion}. Higher order terms may be obtained by solving the differential equation~\eqref{eq:F-thermal-app} up to higher powers of $\Omega$\@.

\subsection{Consequences of the pole positions of the time-ordered correlator} \label{sec:App-analytic-rot}

In this section we shall prove that $[g^\T_E]$ satisfies Eq.~\eqref{eq:analytic-prop-gE}\@.

Up to overall factors, and setting the normalization $F^\pm_\omega(\xi=0) = 1$, the time-ordered correlator we obtained is given by
\begin{equation}
    G(\omega) = -i \frac{\partial^3 F^-_{|\omega|} }{\partial \xi^3}\bigg|_{\xi=0} \, ,
\end{equation}
which we obtained by shifting $\omega \to \omega(1 + i\epsilon)$, which is essentially a prescription to avoid potential poles along the real $\omega$ axis.

In order to prove that $[g^\T_E]$ satisfies Eq.~\eqref{eq:analytic-prop-gE}, we will take a seemingly disconnected starting point, which nonetheless will allow us to prove our claim. To begin, consider the integral
\begin{equation}
    I(\omega) = \int_0^\infty d p_0 \frac{2 \omega G(p_0) }{p_0^2 - \omega^2 + i\epsilon} \, .
\end{equation}
Specifically, we will prove that ${\rm Im} \left\{ I(\omega) \right\} = 0 $\@.

Note that this integral only involves $\omega > 0$\@. Then, because we constructed $G(\omega)$ by shifting potential poles on the positive real axis towards the lower half of the complex plane, there is no obstruction to Wick-rotate the integral onto the positive imaginary axis. This is possible because $F_\omega^-$ itself is an analytic function, provided we handle the potential UV divergences properly. We will deal with the potential large $\omega$ divergences in the next subsection.

After doing the Wick rotation, we get
\begin{equation}
    I(\omega) = - i \int_0^\infty dp_E \frac{2 \omega G(i p_E) }{p_E^2 + \omega^2 } \, .
\end{equation}
Then, by observing that $G(i p_E) = - i \frac{\partial^3 F^-_{i p_E} }{\partial \xi^3}(0) $, and inspecting Eq.~\eqref{eq:F-thermal}, we see that $F^-_{i p_E}$ is a real function (it solves a differential equation with real coefficients and real boundary conditions)\@. Therefore, $I(\omega)$ is a real function, and hence
\begin{equation}
    {\rm Im} \left\{ I(\omega) \right\} = 0 \, .
\end{equation}
It is now straightforward to manipulate this expression into what we want to prove:
\begin{align}
    {\rm Im} \left\{ I(\omega) \right\} &= {\rm Im} \left\{ \int_{-\infty}^\infty \frac{G(p_0)  \, \diff p_0}{p_0 - \omega (1 - i\epsilon) } \right\} \nonumber \\
    &=   \int_{-\infty}^\infty \diff p_0 \left[ \frac{{\rm Im} \left\{ G(\omega) \right\}}{p_0 - \omega} - \pi {\rm sgn}(\omega) \delta(\omega - p_0)  {\rm Re} \left\{ G(p_0) \right\}  \right] \, ,
\end{align}
with which
\begin{align}
    {\rm Im} \left\{ I(\omega) \right\} = 0 \quad \implies \quad {\rm sgn}(\omega) {\rm Re} \left\{ G(\omega) \right\} = \int_{-\infty}^\infty \frac{\diff p_0}{\pi p_0}  {\rm Im} \left\{ G(\omega + p_0) \right\} \, . \label{eq:condition-C6}
\end{align}
    
We have also verified this relation numerically for the vacuum-subtracted correlation functions (i.e., for $\Delta G(\omega) = G(\omega) - G(\omega)_{T = 0}$), where all integrals are convergent. For the vacuum part, where the integrals are UV-divergent because of the $\omega^3$ power-law behavior, we present a proof in the next subsection.

\subsection{UV divergent pieces in the Wick rotation} \label{sec:App-UV-div-Wick}

To be sure that we have the correct expression for all contributions in Eq.~\eqref{eq:gE++-from-T-ordered-sec}, we may work out the contributions proportional to $\omega^3$ from the time-ordered correlator $[g_E^{\T}]$ independently. Because the integrals are divergent, and we do not have a natural regulator that respects all of the AdS symmetries, we will use Lorentz covariance of the boundary theory and the fact that $\mathcal{N}=4$ SYM is a conformal field theory (CFT)\@. Once firmly on the side of the boundary theory, we may use all of the standard dimensional regularization machinery to calculate the integrals.

At $T=0$ we have restored Lorentz covariance of the boundary theory, and therefore we can obtain the same correlation function but with the Wilson lines at an angle with the time axis by applying boosts. This, plus the fact that the theory is a CFT means that $G(\omega) \propto |\omega|^3 $ may be derived by integrating a momentum-space two-point function of a massless particle:
\begin{equation}
    G(\omega) = \# \int \frac{d^3 {k}}{(2\pi)^3} \frac{i \omega^2 }{\omega^2 - {\bf k}^2 + i\epsilon} \, .
\end{equation}
Now, we simply verify the right hand side of Eq.~\eqref{eq:condition-C6} mode by mode, i.e.,
\begin{equation}
    {\rm sgn}(\omega) {\rm Re} \left\{ \frac{i \omega^2 }{\omega^2 - {\bf k}^2 + i\epsilon} \right\} = \int_{-\infty}^\infty \frac{\diff p_0}{\pi p_0} {\rm Im} \left\{ \frac{i (\omega + p_0)^2 }{(\omega + p_0)^2 - {\bf k}^2 + i\epsilon} \right\} \, .
\end{equation}
This identity is indeed satisfied, because we may write the numerator of the integrand on the right hand side as $(\omega + p_0)^2 = (\omega + p_0)^2 - {\bf k}^2 + {\bf k}^2$: the first two terms then cancel the denominator, leaving their contribution as the Cauchy principal value integral of $1/p_0$, which vanishes. The last term gives a contribution that can be cast in the form of a Dirac delta function by means of
\begin{equation}
    \int_{-\infty}^\infty dx\,\mathcal{P} \left( \frac{1}{(x-1) (x^2 - a^2)} \right) = \frac{\pi^2}{2} \delta(|a| - 1) \, ,
\end{equation}
and the left hand side may be immediately seen to be proportional to a Dirac delta function $\delta(\omega^2 - {\bf k}^2)$\@. Verifying that the coefficients match is straightforward.

Therefore, the zero-temperature piece of $[g^\T_E]$ satisfies Eq.~\eqref{eq:analytic-prop-gE}, as does the thermal contribution. Thus, we have verified the claim presented in Section~\ref{sec:ads-cft-spectral-eval}.

%% file: appendixc.tex
\chapter{Appendix: Adiabatic Hydrodynamization and Kinetic Theory}

\section {Numerical implementation of kinetic theory solution in the momentum basis}
\label{app:numerics}

Here we discuss the numerical procedure to solve the FP equation in Section~\ref{sec:scaling-num}. 
We follow Ref.~\cite{Tanji:2017suk} and write  the FP equation \eqref{kin} in terms of variables $p = \sqrt{p_T^2 + p_z^2}$ and $\kappa = p_z/p$. 
We shall use
\begin{align}
	\nabla_{\bf p}^2 f &= \frac{\partial^2 f}{\partial p^2} + \frac{1-\kappa^2}{p^2} \frac{\partial^2 f}{\partial \kappa^2} + \frac{2}{p} \frac{\partial f}{\partial p}  - \frac{2 \kappa}{p^2} \frac{\partial f}{\partial \kappa} \, ,
\\
\nabla_{\bf p} \cdot \left( \frac{{\bf p}}{p} (1+f) f \right) &= \frac{1}{p^2} \frac{\partial}{\partial p} p^2 f (1+f)\, ,
\\
\frac{\partial f}{\partial p_z} &= \kappa \frac{\partial f}{\partial p} + \frac{1-\kappa^2}{p} \frac{\partial f}{\partial \kappa}\, .
\end{align}
To ease the numerical implementation, we additionally consider $l_p \equiv \log p$ and evolve the quantity $\log f$. 
The FP equation then becomes
\begin{align*}
	& \frac{\partial \log f}{\partial \tau} + \frac{\kappa}{\tau} \left[(\kappa^2-1) \frac{\partial \log f}{\partial \kappa} - \kappa \frac{\partial \log f}{\partial l_p} \right]\\
	&= \lambda_0 \lcb[f] \left( e^{-l_p} I_b[f] \left[ \frac{\partial \log f}{\partial l_p} + 2 + 2 e^{\log f} \left(1+ \frac{\partial \log f}{\partial l_p} \right) \right] \right.\\
	&\left. \quad\quad\quad\quad + e^{-2 l_p} I_a[f] \left[ -2\kappa \frac{\partial \log f}{\partial \kappa} - (\kappa^2-1)\left( \frac{\partial^2 \log f}{\partial \kappa^2} + \left( \frac{\partial \log f}{\partial \kappa}\right)^2\right) \right. \right. \\ 
    & \quad\quad\quad\quad\quad\quad \left. \left. + \frac{\partial \log f}{\partial l_p} + \frac{\partial^2 \log f}{\partial l_p^2} + \left( \frac{\partial \log f}{\partial l_p}\right)^2 \right] \right)
\end{align*}
$I_a$, $I_b$, and $\lcb$ are integrals that depend on $f$, with $I_a$ and $I_b$ defined through Eq.~\eqref{eq:IaIb} and $\lcb$ through Eq.~\eqref{lcb-0}.
In these coordinates, we note that $\kappa = \cos \theta = p_z/p$ and $\sin \theta = p_T/p$, which give $p_z = e^{l_p} \kappa$ and $p_T = e^{l_p} \sqrt{1-\kappa^2}$. The spherical volume element is $d^3 p = 2 \pi p^2 dp d\kappa = 2 \pi e^{3 l_p} d l_p d\kappa $. The moments \eqref{moment-def} can therefore be written
\begin{equation}
    n_{m,n}(\tau) = \frac{1}{(2\pi)^2} \int d l_p \, d \kappa \, e^{(3+m+n) l_p} (1-\kappa^2)^{m/2} |\kappa|^n f(p_\perp,p_z,\tau)\, .
\end{equation}
The initial condition for the distribution function \eqref{fI} in these coordinates is
\begin{equation}
	\log f(\pT,\pz;\tau=\tau_I) = \log \frac{\sigma_0}{g_s^2} - \frac{e^{2 l_p}(1-(1-\xi_0^2)\kappa^2)}{Q_s^2}\, .
\end{equation}
We use the finite element method in Mathematica's NDSolve to solve the resulting equations in the range $p \in [5 \cdot 10^{-3},4]$, $\kappa \in [0,1]$ (assuming inversion symmetry in $p_z$) and a maximum cell size of $10^{-3}$.

The solutions displayed in Figures~\ref{fig:FP-exp},~\ref{fig:FP_case1}, and~\ref{fig:FP_case2} were calculated using this method.

\section{The simplification of collision integral
\label{app:Ib}
}

In this Appendix, we discuss the simplification of the collision integral~\eqref{eq:small-angle-kernel} in the situation that the ratio of the typical longitudinal momentum to that of transverse momentum, $r=C/B$, is small. 
In addition, we shall justify dropping $\I_{b}$ term for hard gluons under the condition that typical occupancy is large, i.e. $A\gg 1$. These approximations justify our scaling analysis in Sections~\ref{sec:analytic},~\ref{sec:Adiabatic}, and~\ref{sec:evo}.

We begin by substituting Eq.~\eqref{f-w} into the FP equation~\eqref{kin} with Eq.~\eqref{eq:small-angle-kernel} , with which we obtain the equation for the rescaled distribution function $w$ explicitly as
\begin{align}
\label{w-eq-full-1}
    \pd_{y}w=&-\a w + (1-\g)\xi\,w_{\xi}-\beta \zeta w_{\zeta}
    \no 
    \\
    &\quad + \frac{q}{C^{2}}\, 
    \le[
    \le(
    w_{\xi\xi}+r^{2}(\frac{1}{\zeta}w_{\zeta}+w_{\zeta\zeta})
    \ri) \ri. \nonumber  \\ & \quad\quad\quad\quad \le. + 
    \frac{2 c_{b} r^{2}}{(c_{a}\, A+d_{a})}\,\frac{1}{\sqrt{\zeta^{2}+r^{2}\xi^{2}}}\le(2+ \xi\pd_{\xi}+\zeta\pd_{\zeta}\ri)\,\le( w+A\, w^{2}\ri)
    \ri]\, ,
\end{align}
where we have used the relation
\begin{align}
    \frac{I_{b}}{I_{a}}=\frac{2 ABC c_{b}}{\le( c_{a}\, A+d_{a}\ri)\, A B^{2} C}=\frac{2 c_{b}}{B (c_{a}\, A+d_{a})}\, .
\end{align}
Here, we have defined
\begin{align}
  c_{a}&\equiv \,\int_{\xi,\zeta}\, w^{2}\, ,
  \qquad
  d_{a}\equiv\int_{\xi,\zeta}\, w\, ,
  \qquad
  c_{b}\equiv 
   \int_{\xi,\zeta} \frac{w}{\sqrt{\zeta^{2}+r^{2}\xi^{2}}} \,, 
\end{align}
and have introduced short-hand notation for the integration over scaling variables
\begin{align}
    \int_{\xi,\zeta}\equiv \int^{\infty}_{-\infty}\, \frac{d\xi}{2\pi}\int^{\infty}_{0}\, \frac{d\zeta}{2\pi}\, \zeta . 
\end{align}
We shall assume $c_{a},d_{a},c_{b}$ to be order one.

Now, we consider Eq.~\eqref{w-eq-full-1} in the small $r$ limit.
By looking at Eq.~\eqref{q-general}, we can count $\beta$ to be of the order $r^{2}$. 
Therefore at leading order in the small $r$ expansion, we obtain Eq.~\eqref{H}, which is equivalent to using the collision integral~\eqref{CIa-0}. 
The correction due to finite $r$ corresponds to terms proportional to $r^{2}$ in Eq.~\eqref{w-eq-full-1}.

Next, we consider the limit $A\gg 1$. 
In the regime where $A w \gg 1$ is satisfied, 
Eq.~\eqref{w-eq-full-1} reduces to 
\begin{align}
\label{w-eq-2}
\pd_{y}w&=-\a w + (1-\g)w_{\xi}+\frac{q}{C^{2}}w_{\xi\xi}
    \no 
    \\
    & \quad -\beta \zeta w_{\zeta}+\frac{q}{B^2}
    \le[
    (\frac{1}{\zeta}w_{\zeta}+w_{\zeta\zeta})
    + 
    \,\frac{2c_{b}}{c_{a}}\,\frac{1}{\zeta}\le(2+ \xi\pd_{\xi}+\zeta\pd_{\zeta}\ri)\,w^{2}
    \ri]\, . 
\end{align}
For the tail of the distribution, $\zeta,\xi \gg 1$, we have $w\ll 1$, and then
the last term in the bracket of \eqref{w-eq-2} is small compared with the first term in the bracket and can be dropped. 
This corresponds to setting $\I_{b}=0$, i.e., to using the collision kernel given by Eq.~\eqref{CIa-1}.

\section{Numerical implementation of the solution to the kinetic equation in an adiabatic frame} \label{app:numerics-2}

In this Appendix, we discuss the concrete numerical setup we used to obtain the results described in Section~\ref{sec:adiab-beyond-scaling}. Section~\ref{sec:earlynumerical} describes the general setup we used to solve for the evolution of the distribution function when the time-dependent parameters we introduced to find an adiabatic frame corresponded to coordinate rescalings only, i.e., where we take the basis to be time-independent when written as a function of the rescaled coordinates, as in Sections \ref{sec:isotropic} and \ref{sec:expanding}. Section~\ref{sec:r-numerical} describes the additions needed for the case when the basis is not time-independent, as required in the description of Section~\ref{sec:connectstages}.

\subsection{Hamiltonian evolution for scaling regimes} \label{sec:earlynumerical}

Here we will detail the numerical implementation of the scaling solutions described in 
Sects.~\ref{sec:isotropic} and \ref{sec:pre-hydro}, starting with an explanation of the details of the implementation for Sect.~\ref{sec:isotropic}, then the details for the similar case of \ref{sec:pre-hydro}. As discussed in the main text, we choose a basis \eqref{eq:boltzmannbasis}, onto which we project a distribution function and effective Hamiltonian operator \eqref{eq:isotropich}. Matrix elements of the effective Hamiltonian can be expressed in terms of time-independent integrals over the basis functions; using a projection as in Eq.~\eqref{eq:hprojection} the matrix elements can be written in the form
\begin{align}
    H_{kk'} = \alpha+ \delta h^{(1)}_{kk'} - \lambda_0 \ell_{\rm Cb} \frac{I_a}{ D^2} h^{(2)}_{kk'}- \lambda_0 \ell_{\rm Cb} \frac{I_b}{D} h^{(3)}_{kk'}\,,
\end{align}
where
\begin{align}
    h^{(1)}_{kk'} &= \int d^3\chi \psi_k^{(L)} \chi \partial_\chi \psi_{k'}^{(R)}\\
    h^{(2)}_{kk'} &= \int d^3\chi \psi_k^{(L)} \left(\frac{2}{\chi} \partial_\chi + \partial_\chi^2\right) \psi_{k'}^{(R)}\\
    h^{(3)}_{kk'} &= \int d^3\chi \psi_k^{(L)} \left( \frac{2}{\chi} + \partial_\chi \right) \psi_{k'}^{(R)}\,.
\end{align}

The Hamiltonian also in general depends on the non-linear functionals $I_a,I_b,$ and $\ell_{\rm Cb}$. As noted in the text, we write the Coulomb logarithm $\ell_{\rm Cb}$ as
\begin{align}
     \ell_{\rm Cb}[f] =\ln \left( \frac{\sqrt{\langle p^2 \rangle}}{m_D} \right)\, ,
\end{align}
where $m_D$ is the Debye mass $m_D=2N_cg_s^2I_b$. Rather than calculating $\sqrt{\langle p^2 \rangle}$ at each time step, we replace this with an appropriate scaling parameter: $D$ for 
Sects. \ref{sec:isotropic} and \ref{sec:hydro}, and $B$ for 
Sect. \ref{sec:pre-hydro}, which we expect to reflect the characteristic total momentum of the system in each case. While in principle $\ell_{\rm Cb}$ could become negative if $g_s$ is sufficiently large, we don't encounter this issue in any of our solutions. Since $f_I \propto 1/g_s^2$, such problem cannot possibly appear at the initial time for our choice of initial conditions, and we have not encountered a situation where the dynamics modifies this later on.

The functionals $I_a,I_b$ we will treat as independent variables, which we will evolve by writing their time-derivatives in terms of our scaling parameters and time-independent integrals, as we do for the Hamiltonian. For 
Sect.~\ref{sec:isotropic}, this is
\begin{align}
    \partial_\tau I_b &= I_b(\alpha - 2 \delta) + AD^2 \lambda_{k} H_{kk'} w_{k'} \\
    \partial_\tau I_a &= I_a(\alpha-3\delta) +AD^3 q_k H_{kk'} w_{k'}
\end{align}
where
\begin{align}
    \lambda_k &= \int \frac{d^3\chi}{(2\pi)^3}\, \psi_k^{(L)} \frac{2}{\chi} \psi_{k'}^{(R)} \\
    q_k &= \int \frac{d^3\chi}{(2\pi)^3} \psi_k^{(L)}\, \psi_{k'}^{(R)}\,,
\end{align}
although in the specific case of Sect.~\ref{sec:isotropic}, $\partial_\tau I_a=0$ because we have taken the dilute limit in that Section and have no expansion, so $I_a=K_N$ is conserved. 

In Sect.~\ref{sec:pre-hydro}, we express the Hamiltonian \eqref{eq:expandingh} as
\begin{equation}
\begin{split}
H_{ijkl} = \alpha \delta_{ijkl} + \beta h^{(1)}_{ijkl} + (\gamma-1) h^{(2)}_{ijkl} - \tau \lambda_0 \ell_{\rm Cb}  \frac{I_a}{B^2} h^{(3)}_{ijkl} \\- \tau \lambda_0 \ell_{\rm Cb} \frac{I_a}{C^2} h^{(4)}_{ijkl} - \tau \lambda_0 \ell_{\rm Cb}\frac{I_b}{B} h^{(5)}_{ijkl}
\end{split}
\end{equation}
where 
\begin{align}
h^{(1)} &\equiv \zeta \partial_\zeta, \\
h^{(2)} &\equiv \xi \partial_\xi, \\
h^{(3)} &\equiv \frac{1}{\zeta} \partial_\zeta + \partial_\zeta^2, \\
h^{(4)} &\equiv \partial_\xi^2, \;\;\text{and}\\
h^{(5)} &\equiv \frac{2}{\zeta} + \partial_\zeta + \frac{\xi}{\zeta} \partial_\xi
\,.
\end{align}
The functionals $I_a,I_b$ for Sect.~\ref{sec:pre-hydro} are
\begin{align}
    \partial_y I_b &= (\alpha - \beta - \gamma) I_b - ABC \lambda_{ij} H_{ijkl} w_{kl} \\
    \partial_y (\tau I_a) &= (1+2\alpha - 2\beta - \gamma) (\tau I_a-K_{\tilde{N}}) - 2\tau A^2B^2C w_{ij} q_{ijkl} H_{klmn} w_{mn}\,,
\end{align}
where
\begin{align}
    \lambda_{ij} &= \frac{1}{(2\pi)^2} \int_{-\infty}^\infty d\xi \int_0^\infty d\zeta \; \zeta \; \frac{2 \psi_R^{(ij)}}{\zeta} \\
    q_{ijkl} &= \frac{1}{(2\pi)^2} \int_{-\infty}^\infty d\xi \int_0^\infty d\zeta \; \zeta \; \psi_R^{(ij)} \psi_R^{(kl)}
\end{align}
and we have used the approximation $p\approx p_\perp=B\zeta$ in the expression for $\partial_y I_b$.

For convenience, we use the setup described in Appendix~\ref{sec:r-numerical} to calculate the solutions to the kinetic theory in Section~\ref{sec:hydro}. We describe this at the end of the end of that Appendix.

\subsection{Hamiltonian evolution beyond the scaling regime} \label{sec:r-numerical}

Here we describe the setup needed to evolve the system in the time-dependent basis of Section~\ref{sec:connectstages}.
As discussed in the main text, we choose to expand the rescaled distribution function on a basis defined by
\begin{align}
    \psi_{nl}^{(R)} = N_{nl} e^{-\chi} e^{-u^2 r^2/2} L_{n-1}^{(2)}(\chi) Q_l^{(R)}(u;r) \, , & & \psi_{nl}^{(L)} = N_{nl} L_{n-1}^{(2)}(\chi) Q_l^{(L)}(u;r) \, ,
\end{align}
where the polynomials $Q_l^{(R)}(u;r)$, $Q_l^{(L)}(u;r)$ are polynomials on $u$ of degree $l$, constructed such that
\begin{equation}
    \int_{-1}^1 du \, e^{- u^2 r^2/2} Q_l^{(L)}(u;r) Q_k^{(R)}(u;r) = 2 \delta_{lk} \, ,
\end{equation}
which are normalized by setting $Q_0^{(L)} = 1$, $Q_0^{(R)} = 2/J_0(r) $, and $Q_k^{(L)} = J_0(r) Q_k^{(R)}/2$ for $k \geq 1$, and the normalization coefficients $N_{nl}$ are chosen such that
\begin{align}
    \frac{1}{4\pi^2} \int_{-1}^1 du \int_0^\infty d\chi \, \chi^2 \psi_{mk}^{(L)} \psi_{nl}^{(R)} = \delta_{kl} \, .
\end{align}

The precise form of the distribution function we take is then
\begin{equation} \label{eq:f-app-r-explicit}
    f({\bf p},y) = \frac{e^{-y} D_0^3}{D^3(y)} \sum_{k=1}^{N_{\rm states}} c_k(y) \psi_{n(k) \, l(k)}^{(R)}(p/D, u; r) \, ,
\end{equation}
where we have chosen the time-dependent prefactor such that $\partial_y c_1 = 0$ due to the number-conserving property of the collision kernel. Letting $s(k) \equiv \sqrt{1/4 + 2k - 1} - 1/2$, we use
\begin{align}
    n(k) &= k - \frac{\lfloor s(k)  \rfloor \lfloor s(k) + 1  \rfloor }{2} \left\lfloor \frac{k}{\lfloor s(k) \rfloor \lfloor s(k) + 1  \rfloor / 2 + 1} \right\rfloor \\
    l(k) &= \lfloor s(k) + 1  \rfloor - n(k)
\end{align}
as indexing functions for the radial and angular basis functions. What these functions do is to enumerate the $(n,l)$ states as $(1,0), (1,1), (2,0), (1,2), (2,1), (3,0), \ldots$, where the enumeration is ordered by the value of $n+l$, and within each ``layer'' of fixed $n + l$, is goes from the smallest to the largest value of $n$. We do this so that a truncation done at a maximal value of $n+l$ is essentially a truncation on the number of ``nodes'' that the basis functions can accommodate. 

We define the matrix elements
\begin{align}
    a_{kk'} &= \frac{1}{4\pi^2} \int_{-1}^1 \! du \! \int_0^\infty \! d\chi \, \chi^2 \psi_{n(k')\, l(k') }^{(L)} \chi \partial_\chi \psi_{n(k)\, l(k) }^{(R)} \\
    b_{kk'} &= \frac{1}{4\pi^2} \int_{-1}^1 \! du \! \int_0^\infty \! d\chi \left[ \chi^2 e^{-\chi} \partial_\chi \psi_{n(k)\, l(k) }^{(L)}  \partial_\chi \! \left( e^\chi \psi_{n(k')\, l(k') }^{(R)} \right) \right. \nonumber \\ & \quad\quad\quad\quad\quad\quad\quad\quad\quad\quad \left. + (1 - u^2) \partial_u \psi_{n(k)\, l(k) }^{(L)} \partial_u \psi_{n(k')\, l(k') }^{(R)} \right] \\
    d_{kk'} &= \frac{1}{4\pi^2} \int_{-1}^1 \! du \! \int_0^\infty \! d\chi \, \chi^2 \partial_\chi \psi_{n(k')\, l(k') }^{(L)}  \psi_{n(k)\, l(k) }^{(R)} \\
    e_{kk'} &= \frac{1}{4\pi^2} \int_{-1}^1 \! du \! \int_0^\infty \! d\chi \, \chi^2 \psi_{n(k')\, l(k') }^{(L)} \left[ u (1-u^2) \partial_u + u^2 \chi \partial_\chi \right] \psi_{n(k)\, l(k) }^{(R)} \\
    f_{kk'} &= \frac{1}{4\pi^2} \int_{-1}^1 \! du \! \int_0^\infty \! d\chi \, \chi^2 \psi_{n(k')\, l(k') }^{(L)} \partial_r \psi_{n(k)\, l(k) }^{(R)}
\end{align}
and with them the evolution equation for the basis state coefficients is:
\begin{equation}
    \partial_y c_{k} = - [H_{\rm eff}]_{kk'} c_k' \, , \label{eq:app-c-evol}
\end{equation}
where
\begin{equation}
    [H_{\rm eff}]_{kk'} = -(1 - 3 \delta(y) ) \delta_{kk'} - e_{kk'} + \partial_y r(y) f_{kk'} + \delta(y) a_{kk'} + \frac{\tau \lambda_0 \ell_{\rm Cb} I_a}{D^2} \left[ b_{kk'} +  \left( \frac{I_b D}{I_a} - 1 \right) d_{kk'} \right] \, ,
\end{equation}
which we supplement with evolution equations for $D, r$ and $\tilde{\lambda} = I_b / \int_{\bf p} f$:
\begin{align}
    \partial_y D (y) &= - 10 \big( D - D^2 \left\langle \frac{2}{p} \right\rangle  \big) \, , \label{eq:app-d-evol} \\
    \partial_y r &= - \frac{1}{r} \frac{J_0}{J_4 J_0 - J_2^2} \left[ -2 ( J_2 - J_4) + \frac{\tau \lambda_0 \ell_{\rm Cb} I_a }{D^2} (J_0 - 3 J_2) \right] \, , \label{eq:app-r-evol} \\
    \partial_y \tilde{\lambda} &= \tilde{\lambda} + \frac{N_{10} }{c_1} \sum_{k, k'=1}^{N_{\rm states}} \frac{\delta_{0 \, l(k)}}{ N_{n(k)\,0}} \left( e_{k k'} - \partial_y r(y) f_{kk'} - \frac{\tau \lambda_0 \ell_{\rm Cb} I_a}{D^2} \left[ b_{kk'} + \left( \frac{I_b D}{I_a} - 1 \right) d_{kk'} \right] \right) c_{k'}  \, . \label{eq:app-lamb-evol}
\end{align}

The evolution equation for $\tilde{\lambda}$ is derived from the expression of $\tilde{\lambda}$ in terms of $D$ and the basis state coefficients:
\begin{equation}
    \tilde{\lambda} = \frac{  N_{10} }{D c_1} \sum_{k=1}^{N_{\rm states}} c_k \frac{\delta_{0 \, l(k)}}{ N_{n(k)\,0}} \, .
\end{equation}
We use this equation to eliminate $c_{N_{\rm states}}$ from the evolution equations, as this one is the most sensitive to truncation effects. Instead, whenever $c_{N_{\rm states}}$ appears we replace it by its expression in terms of the other basis state coefficients, $\tilde{\lambda}$ and $D$.

When solving the equations, whenever $I_a$ and $\ell_{\rm Cb}$ appear, they are evaluated in terms of the basis state coefficients and $r, D, \tilde{\lambda}$ through the expressions that define them~\eqref{eq:IaIb} and~\eqref{eq:lcb-def} by substituting~\eqref{eq:f-app-r-explicit} into these expressions.


Furthermore, in practice the matrix elements of $H_{\rm eff}$ require evaluating the integral moments $J_n$ introduced in Eq.~\eqref{eq:Jn-moments}. Especially when $r$ grows large (as in the weakly coupled cases we described in Figs.~\ref{fig:scalings-Dr-weak} and~\ref{fig:occupancies-anisotropies-weak}), it is more convenient to numerically solve for
\begin{equation}
    K_n = r^{n+1} \int_{-1}^1 \! du \, u^n e^{-u^2 r^2 /2 } \, .
\end{equation}
If $r$ does not grow large, as in the more strongly coupled systems we considered in Figs.~\ref{fig:scalings-Dr-strong} and~\ref{fig:occupancies-anisotropies-strong}, then it is sufficient to just evaluate these integrals numerically and have them stored for whenever they need to be evaluated. If $r$ grows arbitrarily large, then it is best to evolve $K_n$ as another variable to solve for in the system of ordinary differential equations comprised by Eqs.~\eqref{eq:app-c-evol},~\eqref{eq:app-d-evol},~\eqref{eq:app-r-evol} and~\eqref{eq:app-lamb-evol}. In particular, one can show that
\begin{equation}
    \partial_y K_n = 2 \partial_y r \, r^n \exp(-r^2/2) \, ,
\end{equation}
which we use to evaluate $K_n$ in the more weakly coupled examples of Section~\ref{sec:connectstages}.

For the solutions presented in Section~\ref{sec:hydro}, we used Eqs.~\eqref{eq:app-c-evol},~\eqref{eq:app-d-evol}, and~\eqref{eq:app-lamb-evol} with $r$ set to zero throughout.

\section{Remarks on the omitted $I_b f^2$ term} \label{app:f2}

In this Appendix, we return to the $I_b f^2$ term in the small-angle scattering collision kernel~\eqref{eq:small-angle-kernel} that, as we noted first in Sect.~\ref{sec:AH}, we have omitted throughout this paper. Here we provide a further discussion of the limitations of this approximation, as well as what challenges would need to be faced in order to reinstate the $I_b f^2$ term and treat it within the framework discussed herein.

Quantitatively, we showed in Section~\ref{sec:intro-AH-BSY} that there is no large discrepancy in the dynamics of the hard sector at early times when this term is included/omitted, which we have further verified in Section~\ref{sec:pre-hydro} by comparing to the numerical results that they obtained using the complete small-angle scattering collision kernel, including the $I_b f^2$ term. As such, this approximation does not seem to alter the dynamics significantly at early times. Furthermore, as time goes by, the distribution function becomes more and more dilute, and by the time that the system begins to approach hydrodynamics the condition $f \ll 1$ is certainly satisfied, and therefore dropping the $f^2$ term seems to be a robust approximation in solving the kinetic equation that the small-angle scattering approximation defines.


However, upon closer inspection, one finds that this can only be strictly true for the hard sector, because the equilibrium distribution of the collision kernel is actually the Bose-Einstein distribution if the $I_b f^2$ term is included. 
At small $p$, the Bose-Einstein distribution behaves as $T_{\rm eff}/p$. (Recall that $T_{\rm eff} = I_a/I_b$.) Furthermore, if the condition
\begin{equation} \label{eq:f2-condition}
    I_a \lim_{p \to 0} \int_{-1}^1 du \, p^2 \frac{\partial f}{\partial p} + I_b \lim_{p \to 0} \int_{-1}^1 du \, p^2 f^2 = 0 \, 
\end{equation}
is not satisfied, then the RHS of the kinetic equation specified by the collision kernel~\eqref{eq:small-angle-kernel} generates a term proportional to the delta function $\delta^{(3)}({\bf p})$ due to the action of $\nabla^2_{\bf p}$ on $1/p$ (or equivalently $\nabla_{\bf p} \cdot (\hat{p}/p^2)$), which in turn sources divergent contributions to $f$ and, see \eqref{eq:IaIb}, hence to $I_a$ and thus makes the kinetic equation itself ill-defined. This condition is the same as the one given in 
Ref.~\cite{Blaizot:2014jna} for this collision kernel in a non-expanding geometry. One way to avoid this problem altogether, as discussed in 
Refs.~\cite{Blaizot:2016iir,BarreraCabodevila:2022jhi} for the non-expanding case, is to include $1 \leftrightarrow 2$ number-changing processes to the collision kernel from the full QCD EKT description, which then ensures that the above condition is satisfied. However, this alternative is beyond the scope of the present work. As we note in Section~\ref{sec:outlook}, introducing $1 \leftrightarrow 2$ processes is an important goal for future work.

The preceding argument would seem to indicate that restoring the $I_b f^2$ term in the small-angle scattering collision kernel would introduce pathologies for some initial conditions, but not all of them: note that a Bose-Einstein distribution with temperature given by $T_{\rm eff}$ satisfies~\eqref{eq:f2-condition}.
On the other hand, though, the dynamical evolution that starts from a distribution function that satisfies $\lim_{p \to 0} p f = 0$ (which implies both $\lim_{p \to 0} p^2 f^2 = 0$ and $\lim_{p \to 0} p^2 \partial f/ \partial p = 0$) will not encounter this issue at early times\footnote{We note, however, that in the case of a non-expanding plasma it has been shown that this issue will appear after some finite time if the initial condition is overpopulated~\cite{Blaizot:2011xf,Blaizot:2014jna}, and has been interpreted as the formation of a Bose-Einstein condensate. The formation of such a condensate is forced as the system thermalizes by the simultaneous conservation of number density and energy density. However, it is unclear whether one can engineer initial conditions that would do the same for a longitudinally expanding plasma, as the energy density in this case is not conserved and thus a Bose-Einstein condensate is not guaranteed to form.} because no term proportional to a delta function will be generated on the RHS of the kinetic equation as~\eqref{eq:f2-condition} trivially vanishes. Then, at later times it will approach the shape of a Bose-Einstein distribution and must ultimately become arbitrarily close to it. The evolution toward a Bose-Einstein distribution has to satisfy the condition~\eqref{eq:f2-condition} at all times in order to avoid the emergence of (unphysical) pathologies, and as such, the numerical method one chooses to solve this equation must be equipped to handle this. In particular, if a term proportional to $1/p$ at small $p$ is generated, its coefficient must be such that~\eqref{eq:f2-condition} is satisfied. If the distribution is isotropic, then the integrals over $u$ in~\eqref{eq:f2-condition} are trivial, and
the term that arises is 
$T_{\rm eff}/p$. If, on the other hand, the asymptotic behavior of $f$ as $p \to 0$ is not isotropic, but instead $f \approx g(u)/p$, then Eq.~\eqref{eq:f2-condition} implies that $g(u)$ must satisfy $I_a \int_{-1}^1 g(u) du = I_b \int_{-1}^1 g^2(u) du $.

As a consequence of what we have just described, 
an additional restriction appears when we choose a basis to solve for the dynamics using the AH framework: if one wishes to have a basis state that permits the system to achieve full thermalization, then there must be at least one basis state that accommodates $T_{\rm eff}/p$ as the small $p$ asymptotic behavior. Furthermore, the coefficients in front of such basis states must be such that the physical distribution function satisfies the condition in Eq.~\eqref{eq:f2-condition} at all times, so as to not generate artificial singularities in the evolution of the system. This introduces an explicit additional scale into the basis, which need not be close to the typical hard scale of the distribution function (which one may infer, e.g., from $\langle 2/p \rangle^{-1}$ or $\langle p \rangle$) that we have used to find an approximation to the adiabatic frame in Sections~\ref{sec:hydro} and~\ref{sec:connectstages}. As such, the basis along the radial $p$ component of the distribution function needs to encode two scales, which means that they cannot be accommodated simply by a rescaling, and the situation becomes exactly analogous to the one we successfully dealt with in the angular $u$ direction in Sections~\ref{sec:non-scaling-H} and~\ref{sec:connectstages}. The problem is thus solvable, and whether this is done efficiently or not will crucially depend on whether the choice of basis permits an efficient calculation of the matrix elements of $H_{\rm eff}$. An alternative approach would be to not include any basis function that blows up as $p \to 0$, and simply approximate the Bose-Einstein distribution as best as possible with regular functions, in which case both terms in~\eqref{eq:f2-condition} vanish trivially, but at the cost of never describing fully accurately the IR behavior of the distribution function.

Last, we wish to point out an additional drawback of dropping the $I_b f^2$ term in the kinetic equation, which is that as a consequence the dynamics of $f$ does not satisfy strict energy conservation. This has the practical consequence that the late-time scaling regime with $\alpha = 0$ and $\gamma = \beta = 1/3$ is not guaranteed to last until arbitrarily long times: number conservation and isotropy only fix $\alpha = \gamma + 2\beta - 1$ and $\gamma = \beta$, whereas an energy-conserving collision kernel plus a late-time isotropic scaling regime guarantee that $\delta = \gamma = \beta = 1/3$ as well, because then the conservation properties imply
\begin{align}
    0 &= - \alpha \int_0^\infty d\chi \, \chi^2 w + (1/3 - \delta) \int_0^\infty  d\chi \, \chi^3 \partial_\chi w  \, , \label{eq:AppB-e-cons-1} \\
    0 &= - \alpha \int_0^\infty d\chi \, \chi^3 w + (1/3 - \delta) \int_0^\infty  d\chi \, \chi^4 \partial_\chi w  \, , \label{eq:AppB-e-cons-2}
\end{align}
which fix $\alpha = 0$, $\delta = 1/3$. Without the condition of energy conservation, the late-time scaling regime may take different values for the scaling exponents, and indeed, we observe that if we let our simulations run for arbitrarily long time, eventually a new scaling regime is reached, with $\delta = 1/5$ and $\alpha = -2/5$. This regime can be straightforwardly derived from a scaling analysis of the kinetic equation, by balancing the rate of change of $1/3 - \delta$ (coming from the expansion terms) with $A/D^2 \propto e^{-y}/D^5$ (the scaling behavior of the term that violates energy conservation in $\mathcal{C}[f]$). However, since this regime is clearly unphysical as it would not be present if the collision kernel conserved energy, we do not extend our simulations to such times. Restoring the $I_b f^2$ term would remove this last regime altogether, as energy conservation would then dictate $\alpha = 0$ and $\delta = \gamma = \beta = 1/3$, as prescribed by Eqs.~\eqref{eq:AppB-e-cons-1} and~\eqref{eq:AppB-e-cons-2}. 